\documentclass[a4paper,11pt,titlepage]{article}
\pdfpagewidth
\paperwidth
\pdfpageheight
\paperheight
\usepackage[english]{babel} 
\usepackage{epsfig}
\usepackage{fancyhdr} 
\usepackage{amsmath,amssymb}
\usepackage{amscd} 
\usepackage[T1]{fontenc} 
\usepackage[utf8]{inputenc} 
\usepackage{graphicx,color,listings}
\usepackage{hologo}
\usepackage{geometry}
\usepackage{rotating}
\usepackage{caption}
\usepackage{chemmacros}
\usepackage{upgreek}
 \usepackage{multirow}
\usepackage{picture}
\usepackage{booktabs}
\usepackage{siunitx}
\usepackage{adjustbox}
\usepackage{setspace}

\usepackage{titlesec}
\usepackage{subcaption}

\usepackage[rightcaption]{sidecap}
\graphicspath{ {images/} }
\addto\captionsenglish{}

\title{\textbf{\Huge ASIACO:} \\ \textbf{\LARGE
Asiago Spectroscopy and Imaging Atlas of COmets}}

\author{P. Ochner, F. Manzini, V. Oldani, A. Farina, \\ A. Reguitti, V. Andreoli,  A.C. Mura, I. Albanese, \\ L. Fiaccadori, G. Mocellin, C. Sigismondi}

\vspace{\fill}
\date{\today}

\usepackage{hyperref}
\usepackage[perpage]{footmisc}
\begin{document}

\maketitle
\newpage

\section*{ASIACO}
In the authors intentions, this volume aims to be an atlas of the spectra of all the comets studied in recent years with the spectrograph of the Galileo telescope of the Padua University at the Asiago Astrophysical Observatory. We inaugurated this catalog with the comet C/2012 S1 (ISON) with a spectrum taken on November 7, 2013. \\
The strange title, ASIACO, derives from `Asiago Spectroscopy and Imaging Atlas of COmets'. \\
The creation of this database was possible by the direct contribution of several astronomy students from the
University of Padua.

\section*{The Authors}
\textbf{Paolo Ochner}, Università di Padova, Osservatorio Astronomico di Padova – INAF, Italy (IAU-MPC code 098) - ORCID: 0000-0001-5578-8614
\\
\textbf{Federico Manzini}, Stazione Astronomica di Sozzago, Italy (IAU-MPC code A12) - ORCID: 0000-0002-2393-3618
\\
\textbf{Virginio Oldani}, Stazione Astronomica di Sozzago, Italy (IAU-MPC code A12) - ORCID:\\
\textbf{Andrea Farina}, Università di Padova, Italy - ORCID: 0009-0005-5257-8319
\\
\textbf{Andrea Reguitti}, INAF, Osservatorio Astronomico di Brera, Italy - INAF, Osservatorio Astronomico di Padova,
Italy - ORCID: 0000-0003-4254-2724
\\
\textbf{Vittorio Andreoli}, Università di Trento, Italy \\
\textbf{Alessandra Mura}, Università di Padova, Italy - ORCID: 0009-0009-5174-7765\\
\textbf{Irene Albanese}, Università di Padova, Italy - ORCID: 0009-0006-6223-5018\\
\textbf{Lorenzo Fiaccadori}, Università di Padova, Italy - ORCID: 0009-0000-6092-4887\\
\textbf{Giovanni Mocellin}, Università di Trento, Italy, - ORCID: 0009-0003-7651-5596 \\
\textbf{Costantino Sigismondi}, ICRA/Sapienza, UPRA, ITIS G. Ferraris, Roma, Italy \\ \\

\begin{figure}[h!]
    \centering
    \includegraphics[scale=0.32]{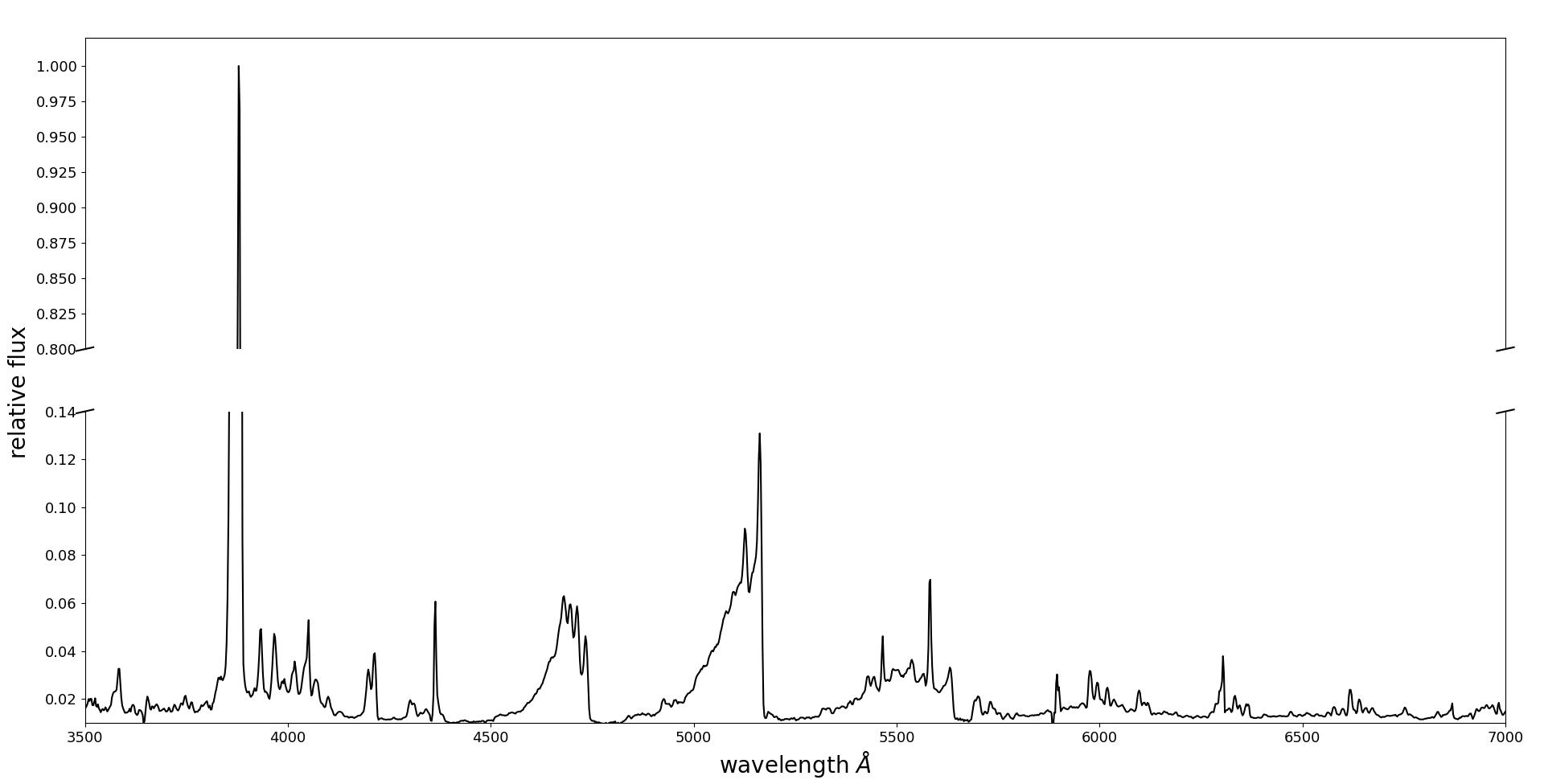}
    \caption{2013-11-07. First spectrum collected for the Atlas project, comet C/2012 S1 (ISON).}
    \label{fig:iimm}
\end{figure}

Corresponding Author: Paolo Ochner (atlantecomete@gmail.com)

\newpage

\section{INTRODUCTION}
Comets have been studied with the Galileo telescope since 1942, in the wake of a long tradition begun in the 17$^{th}$ century by Geminiano Montanari at the Specula of Padua and continued by Antonio Maria Antoniazzi in the same city at the beginning of the 20$^{th}$ century, and by Antonino Gennaro, Salvatore Taffara and Giuseppe Mannino at the Asiago Astrophysical Observatory in the 50s and 60s of the last century.\\ \\
The Galileo telescope is located at the Astrophysical Observatory on the Asiago plateau; it has a diameter of 1.22m ($f/$10) and has given the best contributions to the study of comets in the spectroscopic field (Tab. 1 and 2).\\ \\
Among the many publications on this topic, it is worth mentioning the work of Piero Benvenuti and Karl Wurm (1974) which describes the discovery of a pair of "doublets" in the spectra of comet Kohoutek (C/1973E), later identified by G. Herzberg and H. Lew (1974) as originating from the H$_2$O$^+$ ion. Also worth mentioning, a few years earlier, is a publication by G. Mannino on the spectroscopic analysis of comet Mrkos (1957D), where he announced the rare detection of an intense emission ``doublet'' due to sodium (Na).\\ \\
It is with the desire to celebrate the 80$^{th}$ anniversary of service of this large telescope (inaugurated in 1942) that we present a report of the spectroscopic observations of comets over the last dozen years, accompanied by images obtained almost simultaneously with the other INAF-OAPd telescopes of the Asiago Astronomical Pole (the 1.82m Copernico telescope and the 0.67/0.92m Schmidt telescope) and of the Astronomical Station of Sozzago (0.4m Savonarola telescope).\\ \\
The idea for compiling this volume was born from the authors' desire to make available to the entire scientific community the numerous comet spectra collected with the quality of the Galileo telescope. The pages of this volume are enriched by images obtained in the Johnson-Cousins $UBVRI$ and Sloan $ugriz$ photometric filters (Tab. 4). Everything is in paper format, but also accessible to interested parties in digital form on the dedicated website specifically designed and created to support this volume.\\ \\
The work for the creation of this database, made possible thanks to the direct contribution of several students and PhD students of the astronomy course of the Universities of Padua and Trento, brings with it vast research possibilities, but also an enormous educational value for these collaborators.\\ \\
This publication is related to three goals:
\begin{itemize}
    \item Relaunch the spectroscopic study of comets with the Galileo telescope.
    \item Allow the use of the extensive science data archive created over time.
    \item Continuously expand the database available to the scientific community.
\end{itemize}
The latter goal is strictly connected with the first and requires the constant use of the Galileo telescope. A direct consequence should be the creation of a research group that follows and addresses the needs of the database.
\noindent Finally, it is essential to point out that the study of comets has over time also become a fundamental tool for teaching and for scientific dissemination aimed at the public that increasingly attend the Asiago Astronomical Pole.
\newpage
\section{THE GALILEO TELESCOPE AND ITS INSTRUMENTATION}

The 1.22m telescope of the Asiago Astrophysical Observatory was built by the Officine Galileo in Florence between 1940 and 1942 and dedicated to Galileo Galilei on the occasion of the third centenary of his death. At that time it was the largest telescope in Europe. It was originally designed in two optical configurations, Newton and Cassegrain. However, the telescope is now used only in the Cassegrain configuration with a hyperbolic secondary mirror with a diameter of 0.52m. \\ \\
Since 1998, a Boller \& Chivens (B\&C) spectrograph, manufactured by the Perkin Elmer company (model 58770), has been permanently placed at the Cassegrain focus. The B\&C spectrograph features a slit at the Cassegrain focal plane with a variable aperture up to 1 mm and a length of 28 mm. \\ \\
The collimator for directing the optical beam onto the grating is an off-axis parabolic mirror with a diameter of 0.090 m and a focal length of 0.810 m. A set of four gratings with dispersions ranging from 42 Å/mm to 339 Å/mm completes the spectrograph equipment. The dispersed light beam is directed towards the Dioptric Blue Galileo Camera, with a focal length of 0.188 m, and works in combination with an Andor iDus DU440 CCD camera with a 2098×512-pixel sensor. The spatial resolution on the CCD is 1"/pixel. Several comparison lamps are permanently installed to enable wavelength calibration of the spectra. \\ \\
The side of the slit that faces the incoming light beam has a reflective surface. The image produced by the telescope is reflected on this surface to be captured by the guiding camera (Andor iXon DV885 with an EMCCD sensor); the field of view is 8.5'$\times$6.4' with a resolution of 0.68"/pixel. \\ \\
The movement of the Galileo telescope is implemented with differential tracking to set the tracking of celestial bodies moving at non-sidereal speed without the use of autoguiding, so that the astronomer only needs to make small manual corrections.
\newpage
\section{TECHNICAL DATA OF THE GALILEO TELESCOPE}
Values referred to the intersection between the telescope's polar and declination axis (Tab. 1):
\begin{table}[h!]
\Large
    \centering
    \begin{tabular}{|l c|}
        \hline
        Longitude: &  E 11$^{\circ}$31$^{'}$35.138$^{"}$ \\
        Latitude: & N 45$^{\circ}$51$^{'}$59.340$^{"}$ \\
        Altitude: & 1044.2 m above sea level \\
        IAU-MPC international code: & 043 \\ 
        Diameter of primary mirror (outer edge): & 1.237 m \\
        Effective diameter of primary mirror:    &    1.200 m \\
        Thickness of primary mirror (at the edge): &  0.208 m \\ 
        Weight of primary mirror:              &      575 kg \\
        Focal length of primary mirror:       &       6.000 m \\ 
        Focal ratio of primary mirror:        &       $f/$5.0 \\
        Diameter of secondary mirror:         &       0.520 m \\
        Equivalent focal length:          &        12.100 m \\
        Focal Ratio:                     &         $f/$10.1 \\
        Scale:                         &           17.05 arcsec/mm \\
       \hline   
    \end{tabular}
    \caption{Tab.1}
    \label{tab:telescope info}
\end{table}

\newpage

\section{OUR TECHNIQUES}
\subsection{Spectra}

In the recent years, equipped with a Boller \& Chivens spectrograph, the Galileo telescope has been effectively used for monitoring dozens of comets (with magnitude $\leq$ 16), taking advantage of all the gratings of the available setup (Tab. 2):\\

\begin{table}[h]
\centering
\begin{tabular}{|>{\centering\arraybackslash}m{2.5cm}|>{\centering\arraybackslash}m{2.5cm}|>{\centering\arraybackslash}m{2.5cm}|>{\centering\arraybackslash}m{2.5cm}|>{\centering\arraybackslash}m{4.5cm}|}
\hline
\textbf{Lines/mm} & \textbf{Blaze wavelength (Å)} & \textbf{Dispersion (Å/pix)} & \textbf{Spectral range (Å)} & \textbf{Observable emissions} \\ \hline
300 & 5000 & 2.25 & 4600 & CN, C$_2$ (Swan bands), NH$_2$ bands \\ \hline
600 & 4500 & 1.15 & 2300 & C$_2$ (Swan bands), NH$_2$ bands, [O I] \\ \hline
1200B & 4000 & 0.6 & 1200 & CN, Swan bands \\ \hline
1200R & 6825 & 0.6 & 1200 & Na, [O I], NH$_2$, H$_2$O$^+$, H$\alpha$ \\ \hline
\end{tabular}
\caption{Tab.2}
\end{table}

\noindent
The back-illuminated ANDOR iDus DU440A E2V 42-10 sensor and the altitude of the Asiago Astrophysical Observatory allow for spectral readings from almost 3300 Å to beyond 7500 Å, where fringing, a typical problem of a thin sensor, comes into play. \\
The observational nights have always followed the same schedule (Fig. 1):
\begin{itemize}
    \item A series of bias frames is captured to correct the zero-level counts of the CCD. Additionally, flat-field frames are acquired using a halogen lamp to eliminate any non-uniform pixel response.
    \item Additional spectra from an emission arc lamp (He-Fe-Ar) are taken to determine the dispersion function of the CCD.
    \item At least one ESO catalog spectrophotometric standard star is observed to perform flux calibration;
    \item A star of GV2 spectral type, a "solar analog" is observed to remove the dust reflectance spectrum;
    \item Finally, one or more spectra of the comet are collected.
    
\end{itemize}
\begin{figure}[h!]
    \centering
    \includegraphics[scale=0.25]{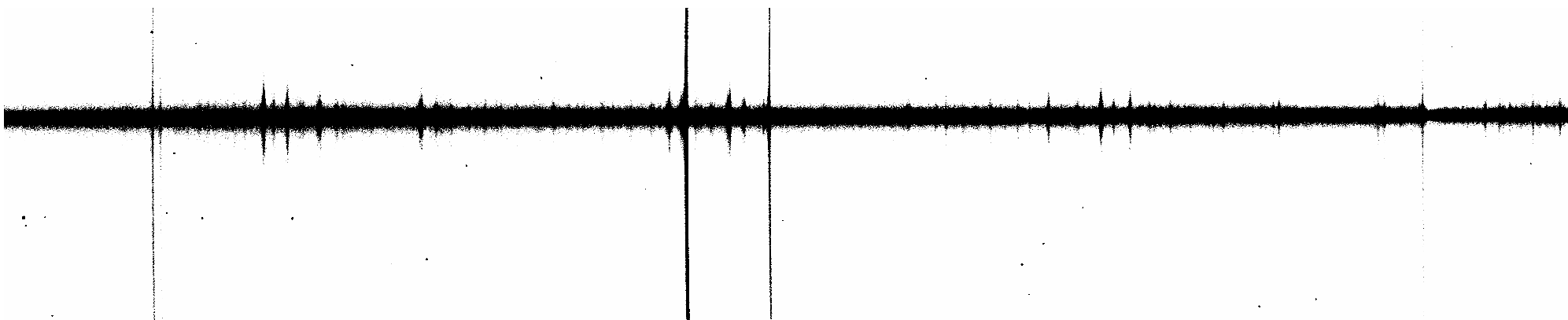}
    \caption{Fig.1 Typical spectrum of a comet (C/2022 E3 - monochromatic image from the Galileo telescope). The radiation from the night sky enters across the entire width of the slit and forms vertical lines throughout its height. The wide field framed by the same slit allows for the extraction of the night sky background far away from the emissions of the nuclear region without being contaminated by the brightness of the outer coma.}
    \label{spec_ex}
\end{figure}
\noindent The length of the slit (8.5' on the celestial plane) has always allowed us to extract the lines of the night sky without "contaminating" the spectrum with the brightness of the comet’s coma.\\
The back-illuminated ANDOR sensor provides excellent performance in the ultraviolet, while
suffering from fringing (photons interference) in the near-infrared due to sensor thinning;
whenever possible, efforts were made to mitigate this effect by taking and averaging different
spectra at different points along the slit. The IRAF v2.6 software was used for the reduction of all spectra, with the standard calibration for dark, bias signal, and flat field. Wavelength calibration was performed using He-Fe-Ar lamps. For proper flux calibration, Bolin
and Lindler’s (1992) ESO Standard Stars were used.  To extract a cometary spectrum, it is necessary to subtract the contribution of the night sky from the raw spectrum (Fig. 2).
\begin{figure}[h!]
    \centering
    \includegraphics[scale=0.6]{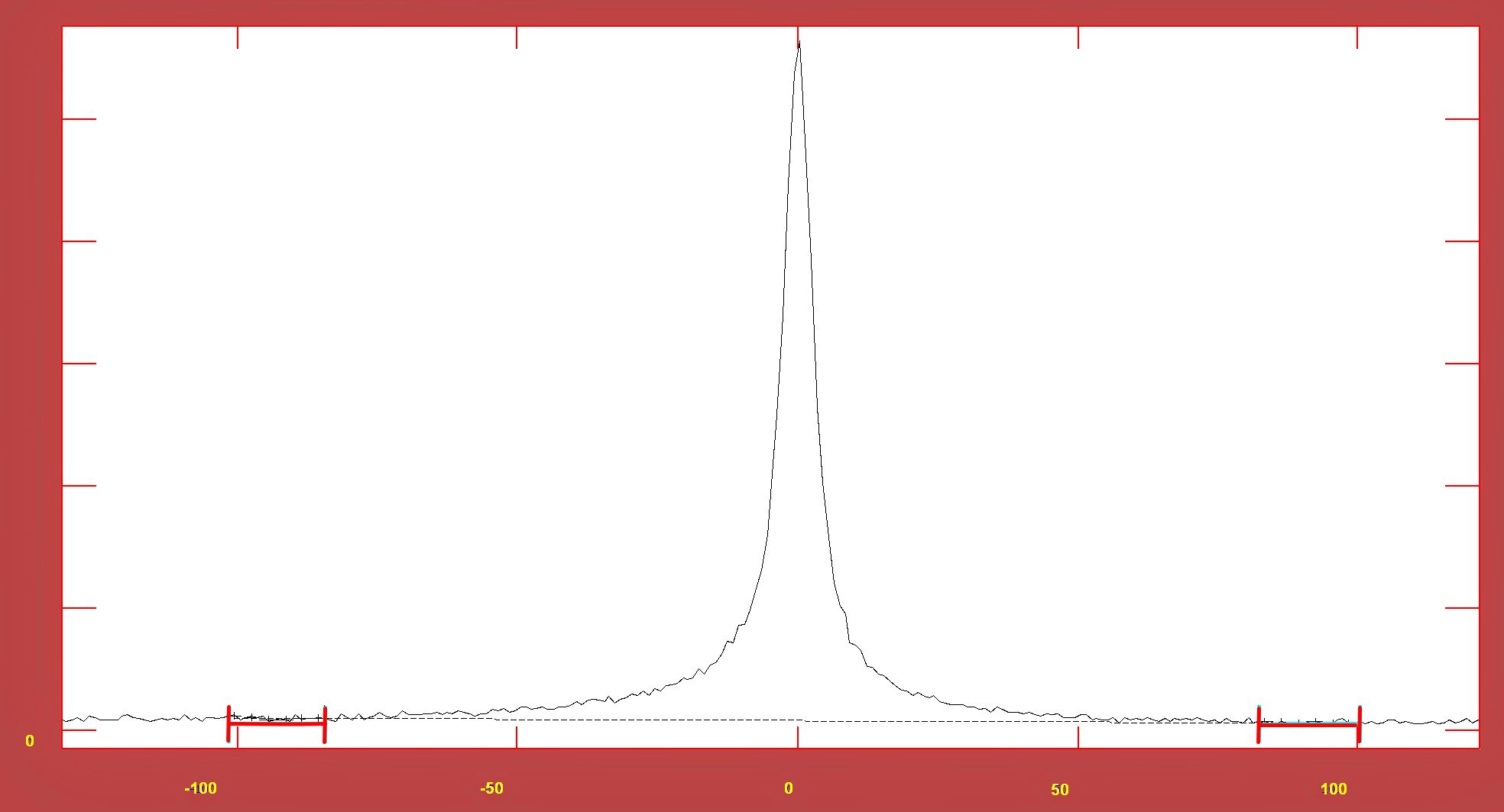}
    \caption{Fig.2 To extract a cometary spectrum, it is necessary to subtract the contribution of the night sky from the raw spectrum. The two side segments adjacent to the peak, indicated in red, represent the areas from which the sky background, far from the contributions of the comet’s nucleus and coma, has been "extracted".}
    \label{fig:apall1}
\end{figure}
\noindent Only later, the calibrated spectra are divided by a spectrum of a star known as a "solar
analog" (spectral classification G2V), also calibrated in the same manner as the comet spectrum (Tab. 3). \\
The results published in this volume, therefore, correspond to a relative flux emitted by each comet. This representation method was chosen because the "analysis" allows for a direct and immediate comparison between the spectra of all comets studied in Asiago: briefly, one can find salient and/or repetitive features in the spectra of the same comet or how these features appear in different comets. Fig. 3 clearly shows the field of view of the Galileo guiding camera with comet C/2020 V2 outside the 200-micron slit.
\begin{figure}[h!]
    \centering
    \includegraphics[scale=0.4]{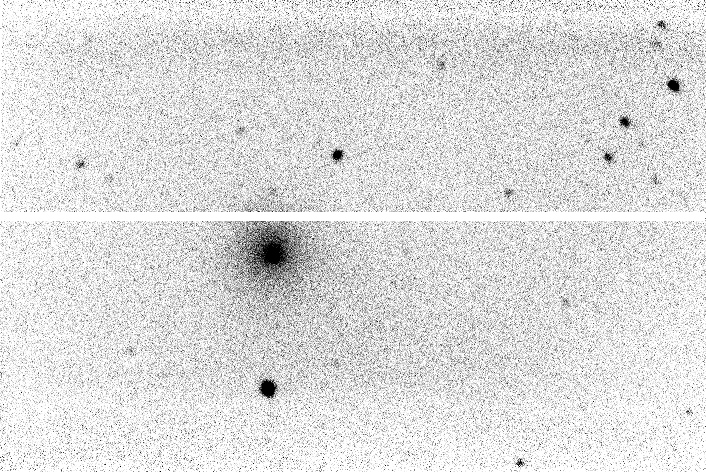}
    \caption{Fig.3 Negative image of the field of view of the guiding camera of the Galileo telescope during a working session on comet C/2020 V2. The white trace corresponds to the position and direction of the spectrograph slit, here set with a width of 200 microns.}
    \label{fig:slit_img}
\end{figure}
\begin{table}[h!]
    \centering
    \begin{tabular}{|c c c c|}
        \hline
         \textbf{name} & \textbf{RA} & \textbf{Dec} & \textbf{V mag} \\ 
          & [hh mm ss] & [$^{\circ}$ ' "] & \\
         \hline 
         \hline
         Land(SA) 93-101 & 01 53 18.0 & +00 22 25 & 9.7 \\ 
         Hyades 64 & 04 26 40.1 & +16 44 49 & 8.1\\
         Land (SA) 98-978 & 06 51 34.0 & -00 11 33 & 10.5 \\
         Land (SA) 102-1081 & 10 57 04.4 & -00 13 12 & 9.9 \\
         Landolt (SA) 107-684 & 15 37 18.1 & -00 09 50 & 8.4 \\
         Land (SA) 107-998 & 15 38 16.4 & +00 15 23 & 10.4 \\
         16 Cyg B & 19 41 52.0 & +50 31 03 & 6.2 \\
         Land (SA) 112-1333 & 20 43 11.8 & +00 26 15 & 10.0 \\
         Land (SA) 115-271 & 23 42 41.8 & +00 45 14 & 9.7 \\
         \hline
         
    \end{tabular}
    \caption{Tab. 3. Stars defined as solar analogs were used as the divisor in the reduction to "relative flux" of all the spectra of this atlas.}
    \label{tab:SolAn}
\end{table}
\newpage
\subsection{THE IMAGING}

This volume is accompanied by numerous images, with at least a couple provided for each comet, along with their respective explanatory captions. \\
The sensors used on the telescopes for imaging were all monochrome, meaning they produced grayscale images. The color figures presented in the volume are the result of the creation of tricolor images composed of individual shots taken through photometric bandpass filters following the Johnson-Cousins (B, V, R, I) or Sloan (u, g, r, i, z) standards (Tab. 4). Other images included in the volume were also captured using standard photometric filters (Fig. 4, 5, 6, 7). \\
The imaging sessions took place during observational nights using the instruments at Cima Ekar, situated a few kilometers southeast of Asiago, at an altitude of 1366 meters, with the IAU-MPC code 098:
\begin{itemize}
    \item the 1.82m Copernico telescope with its Asiago Faint Objects Spectrograph and Camera (AFOSC); it features an Andor iKon-L936 BEX2-DD-9HF camera equipped with a back-illuminated E2V CCD42-40 sensor, providing 2048×2048 pixels with a 13.5 $\mu$m side and a resolution on the sky plane of 0.25"/pixel.
    \item the 0.67/0.92m Schmidt telescope, equipped with a Moravian CCD camera featuring a KAF-16803 sensor. This sensor offers 4096×4096 pixels with a 9x9 $\mu$m size, resulting in a resolution of 0.87"/pixel on the sky plane.
\end{itemize}

\begin{table}[h!]
\centering
\begin{tabular}{|>{\centering\arraybackslash}m{2cm}|>{\centering\arraybackslash}m{2cm}|>{\centering\arraybackslash}m{2cm}|>{\centering\arraybackslash}m{4cm}|}
\hline
\textbf{Filter} & \textbf{CWL [nm]} & \textbf{FWHM [nm]} & \textbf{Remarks} \\ \hline
U & 365 & 60 & Johnson-Cousins \\ \hline
B & 436 & 89 & Johnson-Cousins \\ \hline
V & 544 & 84 & Johnson-Cousins \\ \hline
R & 630 & 120 & Johnson-Cousins \\ \hline
I & 900 & 300 & Johnson-Cousins \\ \hline
u & 354 & 65 & Sloan \\ \hline
g & 477 & 150 & Sloan \\ \hline
r & 623 & 140 & Sloan \\ \hline
i & 762 & 130 & Sloan \\ \hline
z & 913 & 100 & Sloan \\ \hline
\end{tabular}
\caption{Tab. 4. Filters used on both the Copernico and Schmidt telescopes at Cima Ekar (Asiago). The CWL (Central Wavelength) column specifies the central wavelength of the filter's passband.}
\end{table}
\noindent The procedure is slightly different for some images taken with other telescopes in the 0.50 m class; the location is shown in the caption of the relevant images. \\ \\
All original images underwent standard calibration using master bias and dark frames, and flat field corrections. For all imaging sessions, differential tracking was employed, set to the comet’s speed relative to the stars. When multiple images were taken with the same filter in the same observation session, during post-processing we proceeded to co-add them after having realigned them on the optocenter of the cometary nucleus, to ensure a result with sub-pixel precision and raise the signal to background ratio. The processed images show stars as streaks, making it easy to discern the direction of the comet’s motion. \\ \\
Several images accompanying the volume result from the application of a simple photometrically calibrated trichrome in the R channel (Sloan r filter), G channel (Johnson V), and B channel (Johnson B). These images provide a sufficiently realistic representation of how the comet would appear in the sky if one had a calibrated eye sensitive to low light conditions. \\ \\
Other figures result from processing with the Larson-Sekanina mathematical algorithm, which is one of the most used digital filters for the morphological study of comets’ coma (Larson S. \& Sekanina Z., AJ, 1984). The application of this filter highlights micro variations in the brightness within a comet’s inner coma, enabling a detailed study of morphological structures, near the nucleus. In the images processed with this algorithm, star trails (which already appear as bright/white in the original shots) are accompanied by two symmetrical dark/black trails positioned on either side, which are due to the equation applied by the filter (Fig. 6b). \\ \\
These image processing techniques reveal in almost every comet a marked morphology of the inner coma, which can be traced back to its source on the cometary nucleus. They are a useful method for studies aimed at identifying axes of symmetry correlated with the rotation of the nucleus and the orientation of its axis. \\ \\
This volume also contains some figures resulting from isophote visualization, which highlights the extent of the nebulosity (coma+tail) around the comet nucleus along with asymmetries resulting from dust and gas emissions from active areas on the nucleus. The isophote traces have a discrete and precise step in ADU (Analog-to-Digital Units), repeated throughout the image. The visualization of the first isophote corresponds to a brightness just a few ADU above the sky background (Fig. 7b).
\newpage
\section{REPRESENTATION OF IMAGES: THE TRICHROMES}

\begin{figure}[h!]
    \centering
    \includegraphics[width=.5\linewidth]{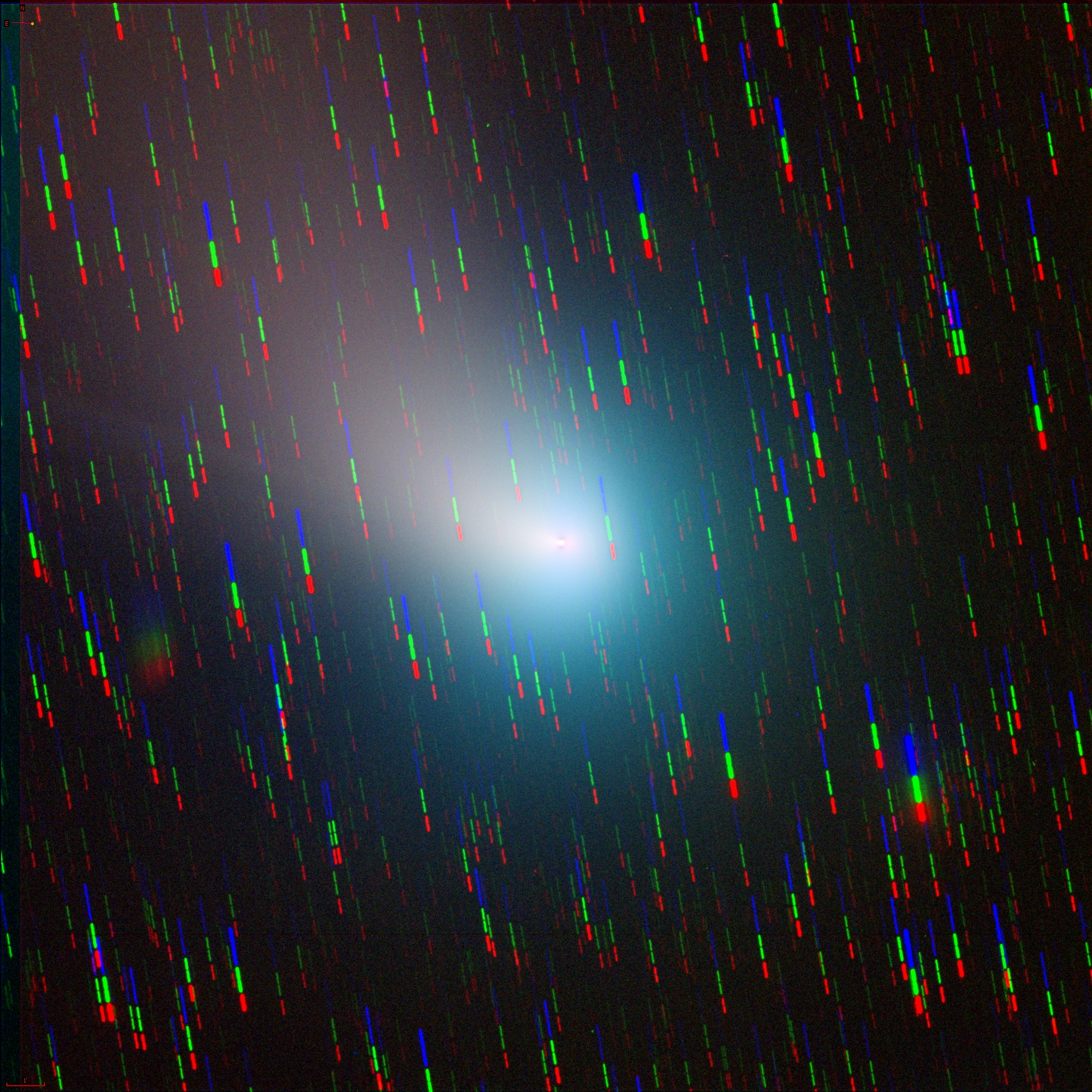}
    \caption{Fig.4 Tri-color image of comet C/2022 E3 (ZTF). The three monochromatic images composing it are derived from shots taken on February 11, 2023, at the 0.67/0.92m Schmidt telescope at the Cima Ekar Observatory (Asiago) using B, V and r photometric filters. The individual frames, with exposure times of 60 and 120 seconds were all aligned using as a reference the optocenter of the comet, which is the brightest photometric point presumed to correspond to the nucleus. A trichrome image immediately highlights the different typical areas of a comet: nucleus, coma and tail. The latter appears reddened due to the presence of micro dust that disperses and reflects sunlight. The blue green of the coma instead derives from the fluorescence of the diatomic carbon atoms ($C_{2}$).}
\end{figure}
\begin{figure}[h!]
\begin{subfigure}{.3\textwidth}
  \centering
  \includegraphics[width=.9\linewidth]{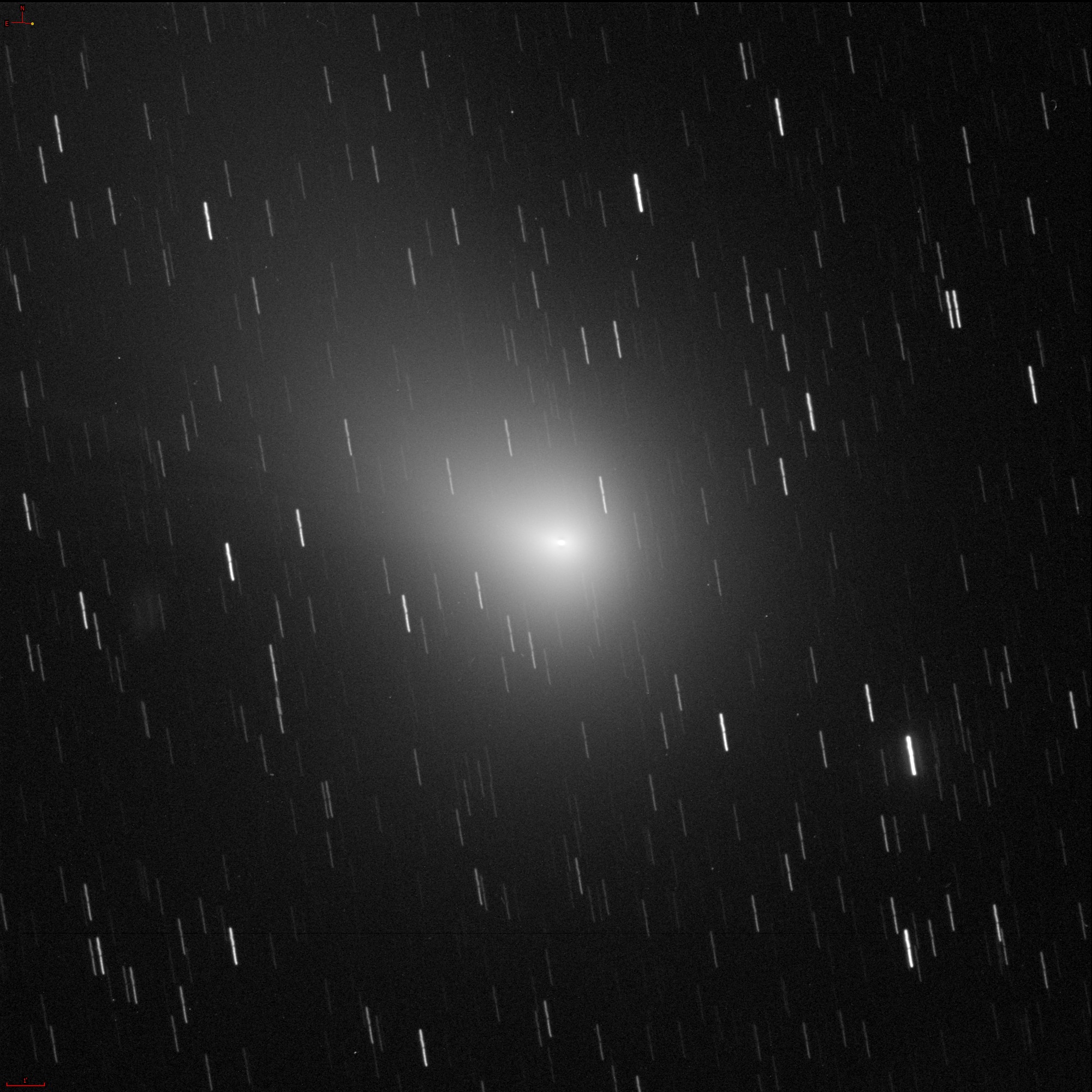}
  \caption{~}

\end{subfigure}%
\begin{subfigure}{.3\textwidth}
  \centering
  \includegraphics[width=.9\linewidth]{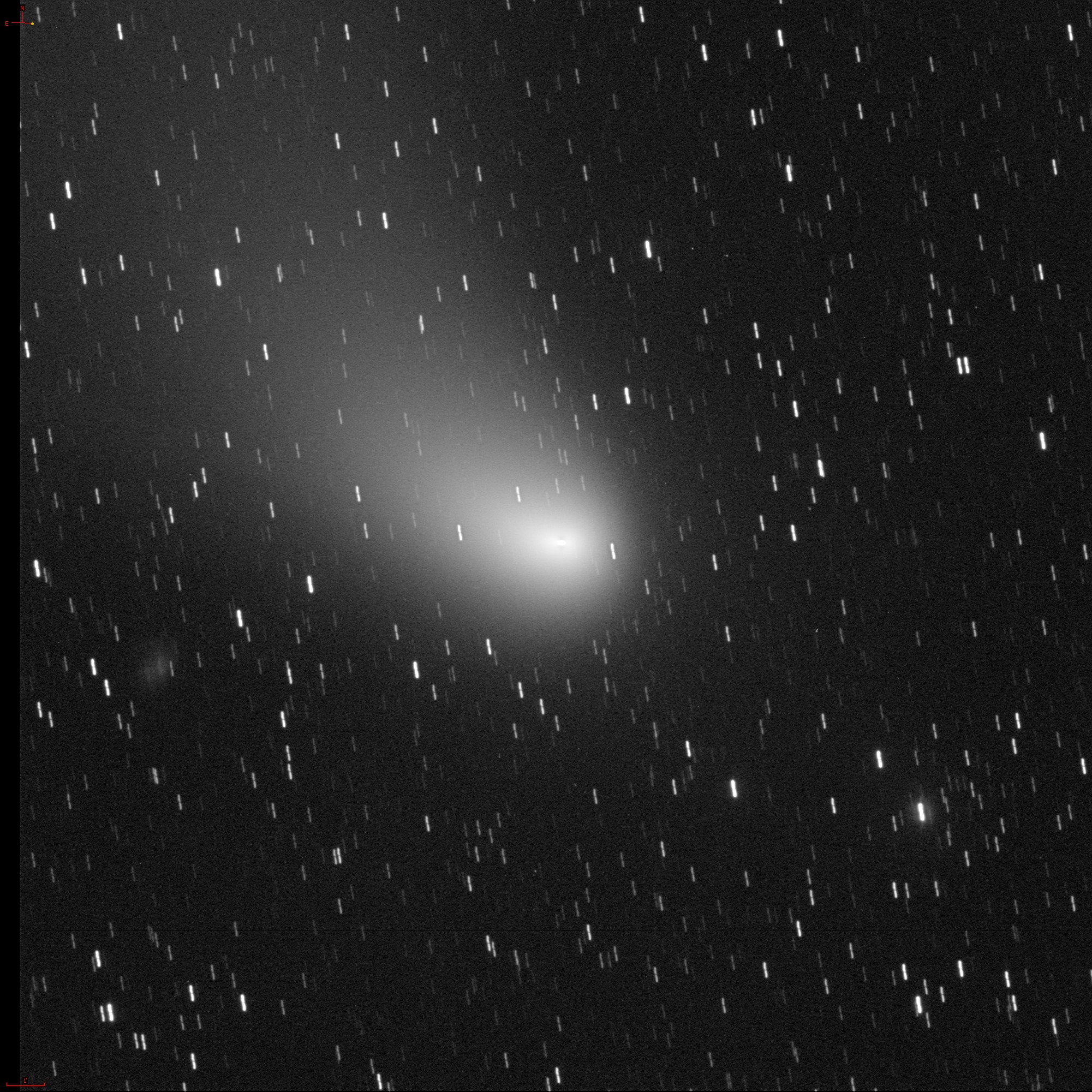}
  \caption{~}

\end{subfigure}
\begin{subfigure}{.3\textwidth}
  \centering
  \includegraphics[width=.9\linewidth]{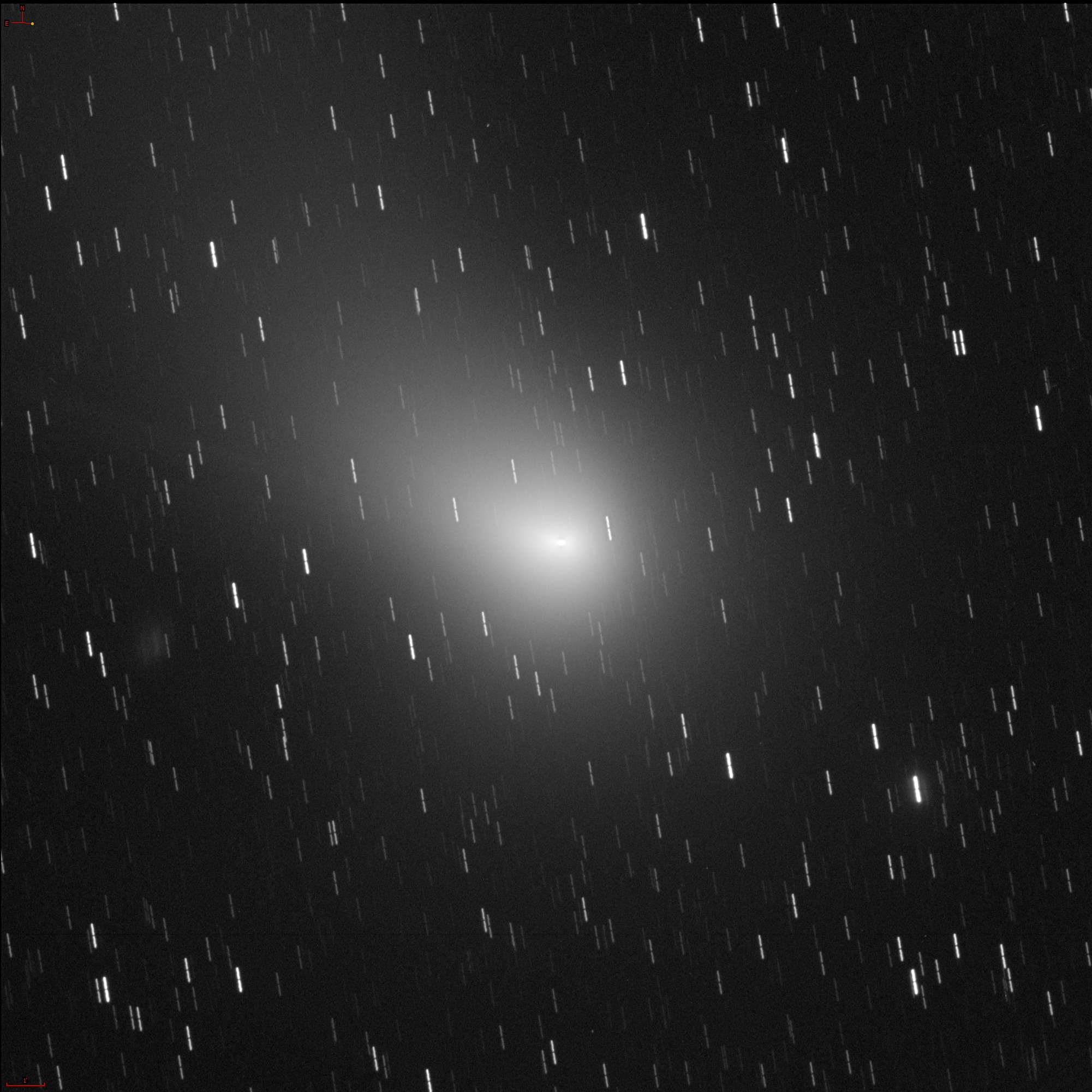}
  \caption{~}

\end{subfigure}
\caption{Fig.5 The trichrome image in Fig. 4 is an RGB composite of the monochromatic shots shown here, respectively taken with B (a), V (b), and r (c) filters. Each of them highlights specific details and features that are only visible in the specific spectral bands used. North is at the top, and east is to the left.}
\end{figure}

\newpage
\section{REPRESENTATION OF IMAGES: THE LARSON-SEKANINA ALGORITHM}

\begin{figure}[h!]
\centering
\begin{subfigure}{\textwidth}
\centering
\includegraphics[width=0.3\textwidth]{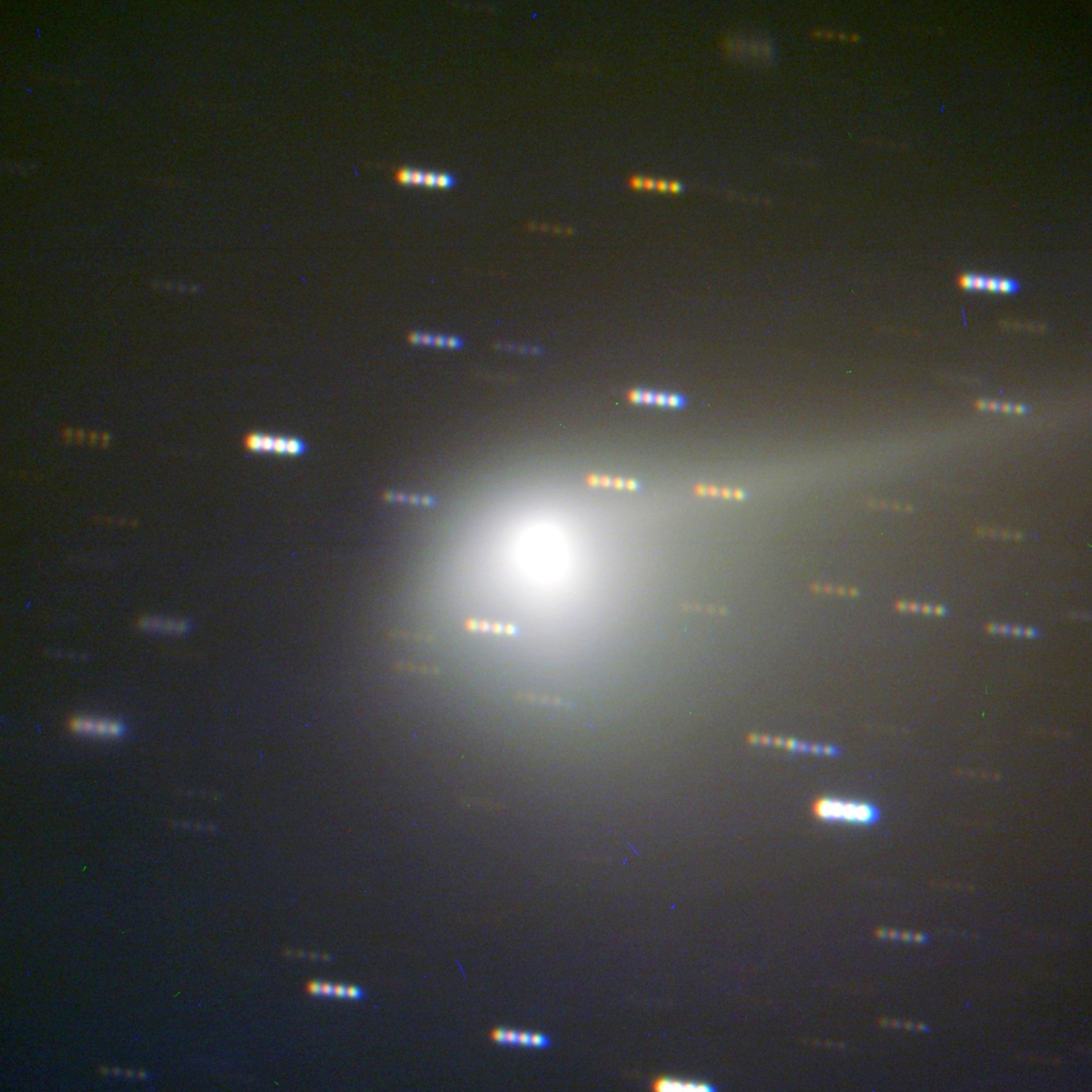}
\caption{~}
\end{subfigure}

\bigskip

\begin{subfigure}{\textwidth}
\centering
\includegraphics[width=0.3\textwidth]{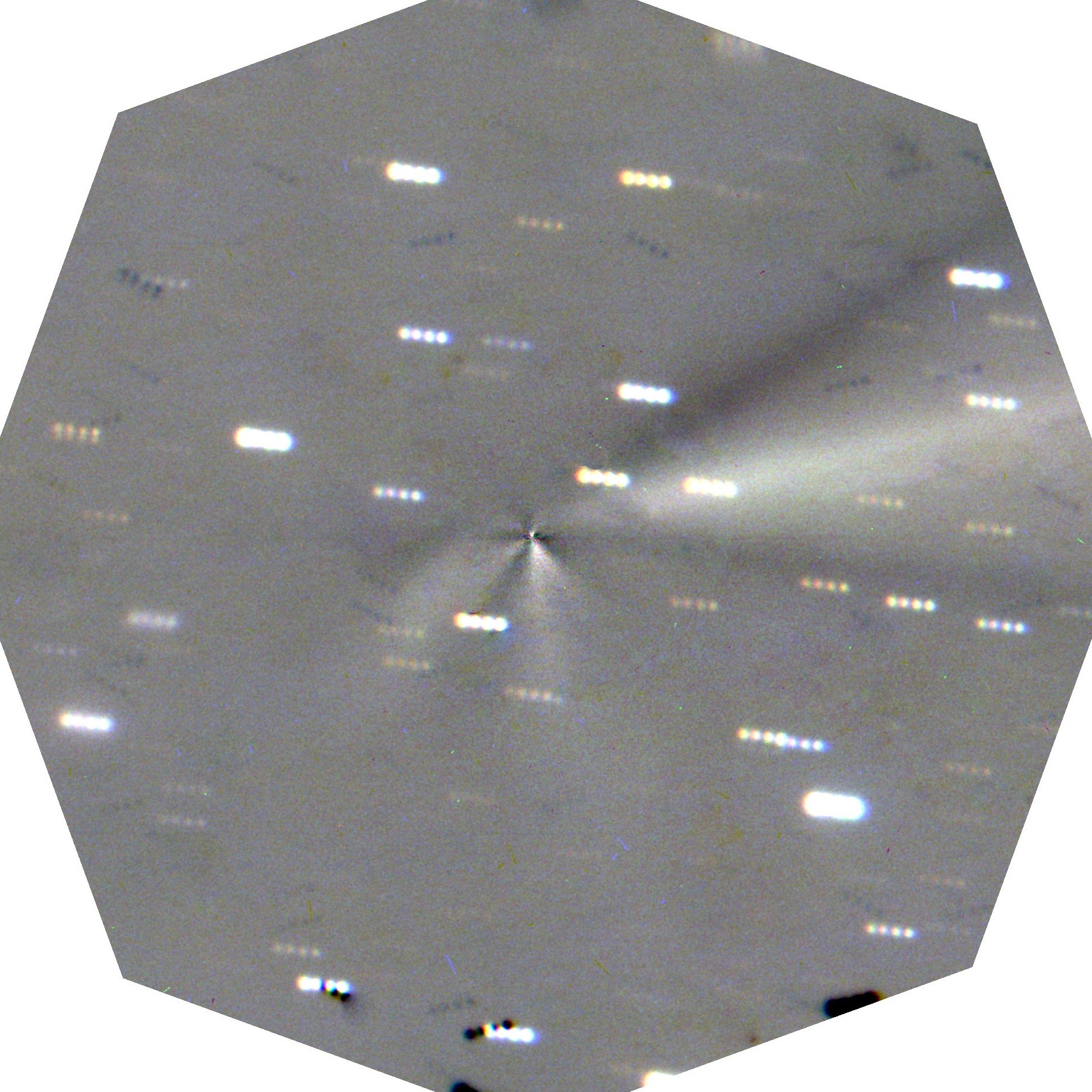}
\caption{~}
\end{subfigure}

\caption{Fig.6 These two images provide a clear illustration of what is "hidden" within a comet’s coma.\\
Panel (a):  a trichrome image of comet 67P (Churyumov-Gerasimenko) taken on January 27, 2022 (Copernico telescope). \\
Panel (b): the application of the Larson-Sekanina algorithm with an angle of $\ 20^\circ$ highlights some morphological structures in the inner coma. These structures directly originate from highly active regions on the nucleus, emitting large quantities of gas and dust. The dust particles maintain their spin and velocity until the solar radiation pressure pushes them back to form the comet’s tail.
In images processed with this algorithm, the star trails, which appear bright in the original shots, are accompanied on both sides by two symmetrical dark trails due to the equation applied by the filter. In this specific case, with the comet in a star-rich field, the dark trails have been partially removed to better reveal the morphological structures. North is at the top, and east is to the left.
}

\end{figure}
\newpage

\section{REPRESENTATION OF IMAGES: THE ISOPHOTE VISUALIZATION}

\begin{figure}[h!]
\centering
\begin{subfigure}{\textwidth}
\centering
\includegraphics[width=0.3\textwidth]{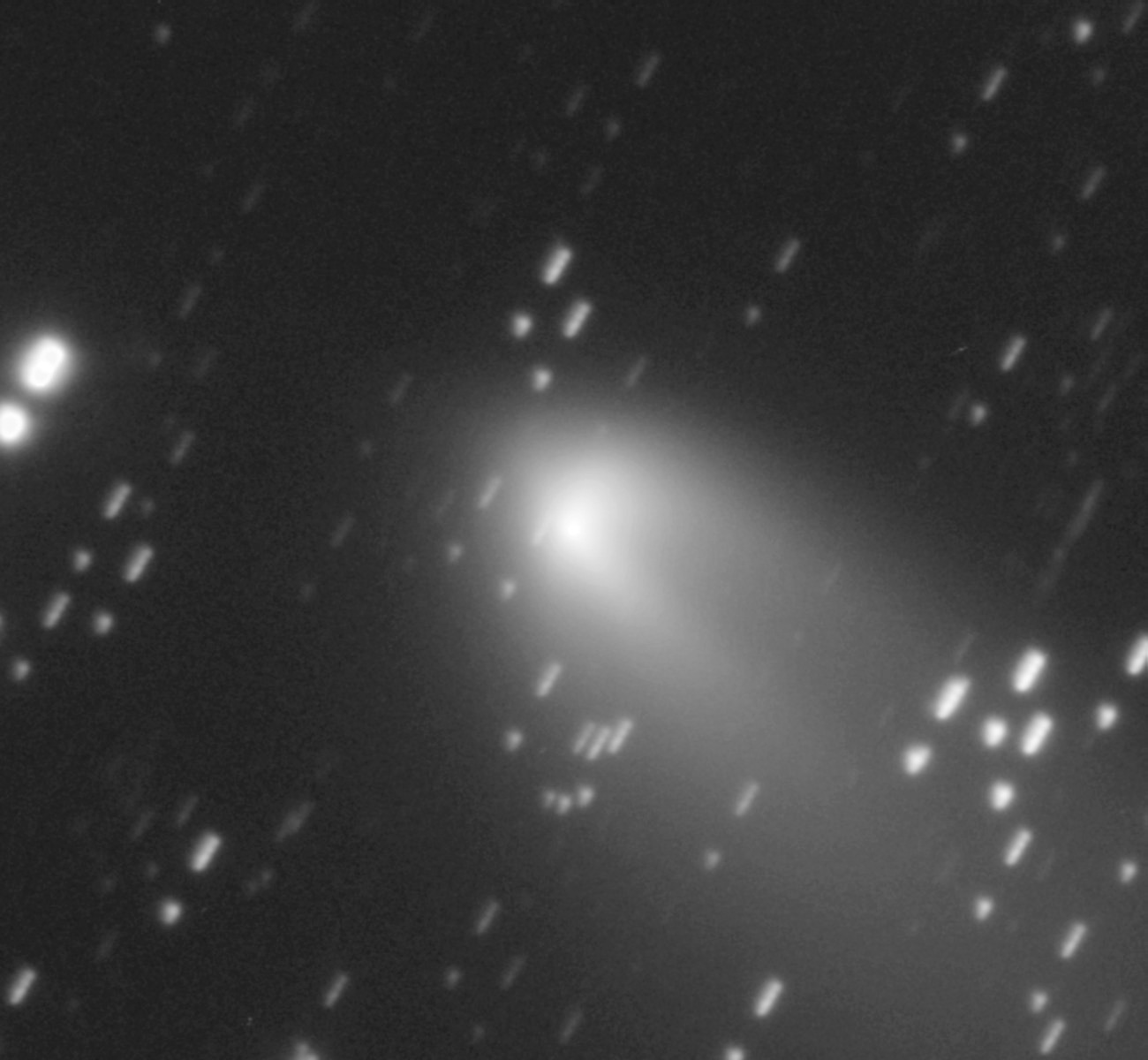}
\caption{~}
\end{subfigure}

\bigskip

\begin{subfigure}{\textwidth}
\centering
\includegraphics[width=0.3\textwidth]{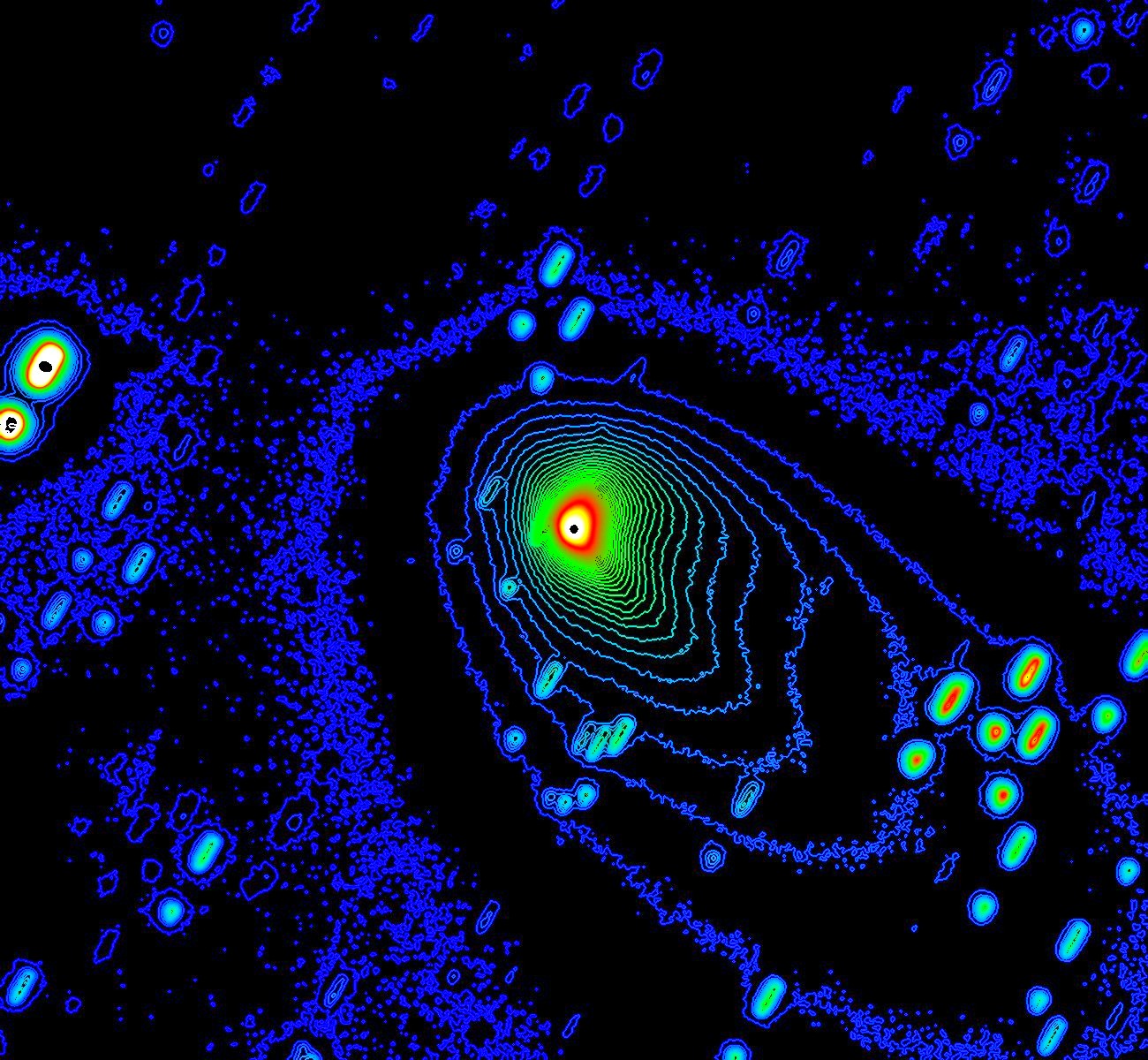}
\caption{~}
\end{subfigure}

\caption{Fig.7
Example of an image of a comet shown with two different graphic modalities.\\
Panel a: a DDP (Digital Development Process) filter has been applied to the original image, sum of single frames of comet 260P. This filter gives the images an appearance similar to those on film. It is useful for identifying low-contrast details within a comet’s coma.\\
Panel b: the isophote visualization involves setting the lowest level isophote to a value a few ADUs higher than the sky background. This highlights the actual extent of the comet’s coma along with asymmetries in the brightness distribution related to emissions of gas and dust from the nucleus. The number of pixels between the minimum and maximum values within which the isophotes are drawn allows for photometry of these diffuse objects. Here, the isophotes are spaced at constant incremental steps of 500 ADUs and are represented in false color, with blue indicating the minimum brightness values in the image.
Image taken with the 1.82m Copernico telescope.
North is at the top, and east is to the left.
}

\end{figure}
\newpage

\section{COMET DATA SHEETS}

In ASIACO, the spectra of 41 comets observed with the Galileo telescope of the Asiago Astrophysical Observatory are presented. For an easier consultation, the sheets dedicated to each comet follow an identical structure.\\ 
The first page of each comet’s sheet features two tables:
the first one (Tab. 4) lists the orbital elements of the comet obtained from the latest data available on the JPL Horizon website (e.g.: \url{https://ssd.jpl.nasa.gov/tools/sbdb_lookup.html#/?sstr=260p}).
\begin{figure}[h!]
    \centering
    \includegraphics[width=0.6\textwidth]{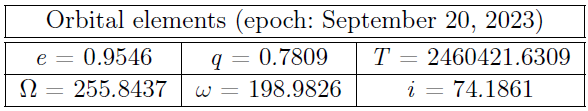}
    \caption{Tab.4 Example of orbital parameter table of comet 12P (Pons-Brooks). Here $e$ is eccentricity, $q$ is the heliocentric distance at perihelion (AU), $T$ is the perihelion date (TDB), $\Omega$ is the argument of pericenter (deg), $\omega$ the longitude of the nodes (deg) and finally $i$ the inclination of the orbit (deg).}
\end{figure}
\noindent The second table provides position and distance ephemerides at 12:00 UT of the two key dates of the transit of the comet: perihelion and nearest approach. It also includes the elongation of the comet from the Sun, the phase angle on the nucleus, and the Earth’s position `above' or `below' the comet's orbital plane. These data allow for the visualization of the comet's position in the sky (Tab. 5).
\begin{figure}[h!]
    \centering
    \includegraphics[width=0.8\textwidth]{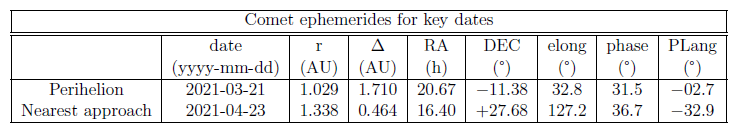}
    \caption{Tab.5 Example of ephemerides of comet 12P (Pons-Brooks) for key dates, perihelion and nearest Earth approach. From left to right: the observation epoch; the heliocentric distance of the target (distance Sun-comet); the distance of the target from the observer (distance Earth-comet); the right ascension of the target; the declination of the target; the solar elongation angle (Sun-Observer-Target); the true phase angle (Sun-Target-Observer); the angle between observer and target orbital plane.}
\end{figure}
\noindent The first page also includes a diagram of the comet's orbit, which size can be directly compared to the orbit of Earth and Jupiter (credits: NASA/JPL-Caltech). The orbit and position of these three celestial bodies are indicated for the date of the comet's perihelion (\url{https://ssd.jpl.nasa.gov/tools/orbit_viewer.html}) (Fig. 8). When necessary, the scale has been enlarged to include the orbits of Saturn and Uranus. Vertical lines are added alongside the comet's orbit to reference its position on the z-axis (the short side of the image) and in relation to the plane of the ecliptic: these lines provide information on when the comet is "above" or "below" the Earth's ecliptic.
\begin{figure}[h!]
    \centering
    \includegraphics[width=0.6\textwidth]{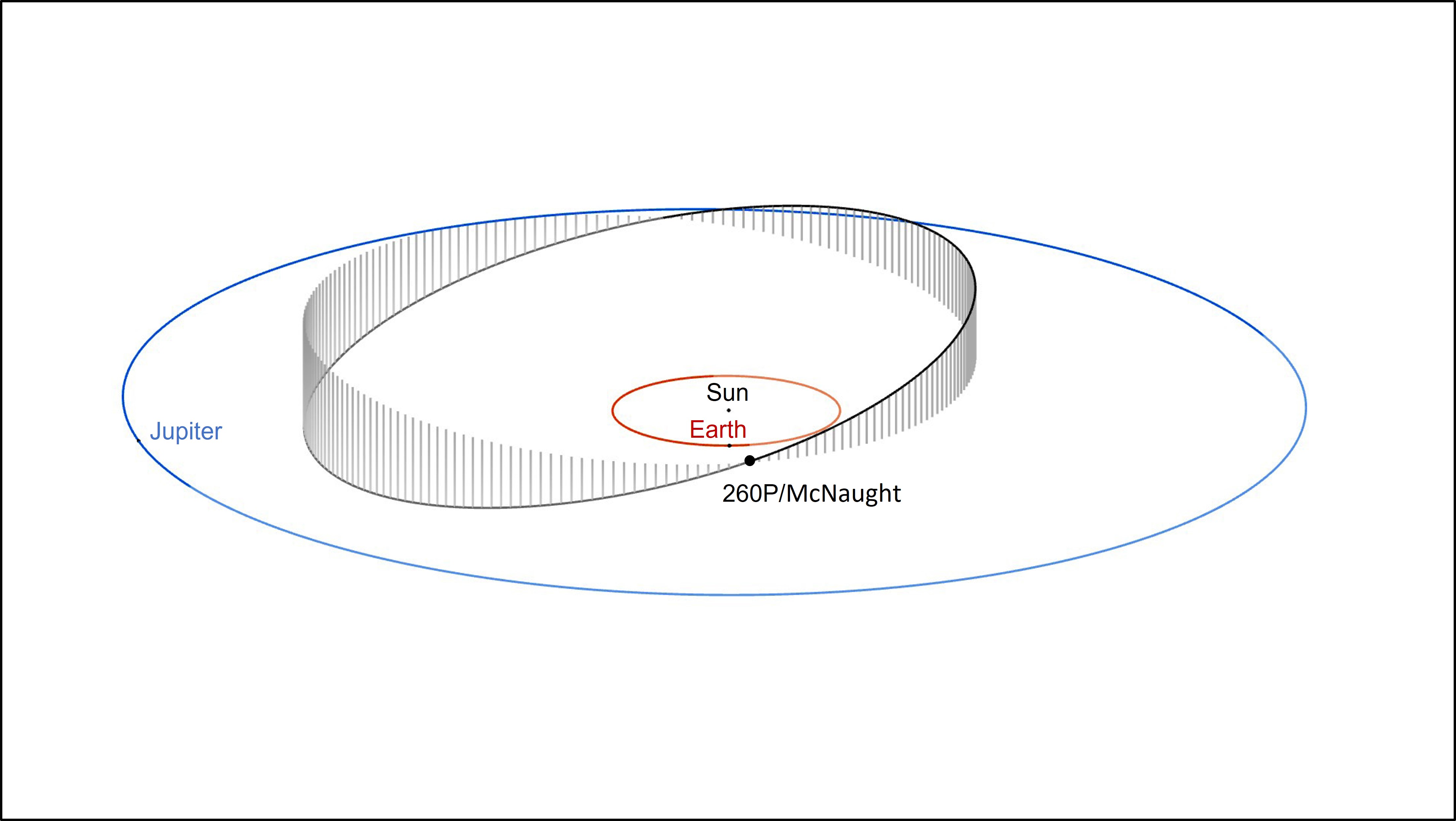}
    \caption{Fig.8 Example of orbit of comet 260P McNaught and position on perihelion date.}
\end{figure}
\noindent Some images of the comet are shown on the second page of each data sheet. Each figure includes the exact orientation and a scale in kilometers measured on the plane of the sky at the comet's distance at the time of imaging. For a more precise description of the visualization methodologies, you can refer to the previous sections.\\
The following pages include a table containing details of all the observations conducted with the spectrograph of the Galileo telescope. Here, you can find the date, the comet's distance from the Sun and the Earth, the equatorial coordinates, and the elongation in degrees from the Sun at the time of each observation. In the following columns, the phase angle on the comet’s nucleus and the angle from which the comet is observed from Earth relative to the comet's orbital plane are reported. Positive values indicate that the observer is located above the object's orbital plane (Tab.6). The last three columns of the table present the spectrograph configuration and the imaging data: respectively the grating used for the observation, the angle of the slit (E-W = 0$^\circ$, N-S = 90$^\circ$) and the exposure time of each spectrum. 
\begin{figure}[h!]
    \centering
    \includegraphics[width=0.8\textwidth]{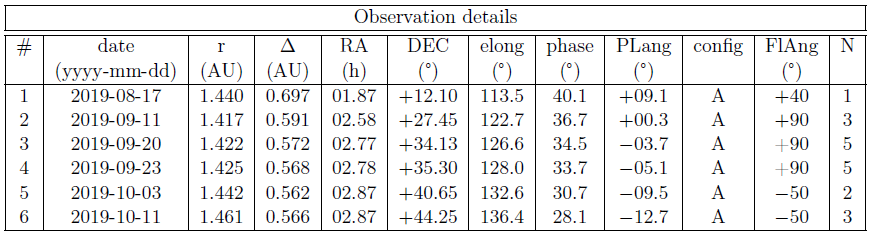}
    \caption{Tab.6 Example of ephemerides table for comet 260P (McNaught). From left to right: The observation run number; the epoch of the observation: the heliocentric distance r of the comet (distance Sun-comet); the distance $\Delta$ of the comet from the observer (distance Earth-comet); the Right Ascension of the target; the Declination of the target; the solar elongation angle (Sun-Observer-Target); the true phase angle (Sun-Target-Observer); the angle between observer and target orbital plane; the kind of grating used; the flange rotation angle of the spectrograph (usually aligned along the target parallactic angle; the number of frames of the target collected during the same run).}
\end{figure}
\noindent The configurations available to the B\&C spectrograph of the Galileo telescope are presented in Tab. 5: the setup may vary for the specific needs of each observation; identification letters (A, B, C, D) have been associated with each configuration and are shown in the table on the third page of the data sheet of each comet (Tab. 7).
\begin{table}[h!]
    \centering
    \begin{tabular}{|c|c|c|c|c|}
        \hline
        \textbf{grating} & \textbf{$\lambda$ of blaze} & \textbf{dispersion (Å/px)} & \textbf{spectral range (Å)} & \textbf{Identify letter} \\ \hline
        300 & 5000 & 2.25 & 3500-7000 & A \\ \hline
        600 & 4500 & 1.15 & 4250-6590 & B \\ \hline
        1200R & 6825 & 0.6 & 5750-7000 & C \\ \hline
        1200B & 4000 & 0.6 & 3820-5020 & D \\ \hline
    \end{tabular}
    \caption{Tab.7 Galileo telescope spectrograph setups. }
    \label{tab:reticoli}
\end{table}
\noindent A representative spectrum of each comet is graphically presented on the last page of the section. The main emission lines due to the molecules and atoms present around the nucleus and in the inner coma of the comet can be identified. The spectra are presented with relative flux as a function of wavelength (\AA). If the flux is too intense, as is typically the case for the CN line, the y-axis has been truncated to better visualize even the weakest emission lines (Fig. 9).
\begin{figure}[h!]
    \centering
    \includegraphics[width=0.7\textwidth]{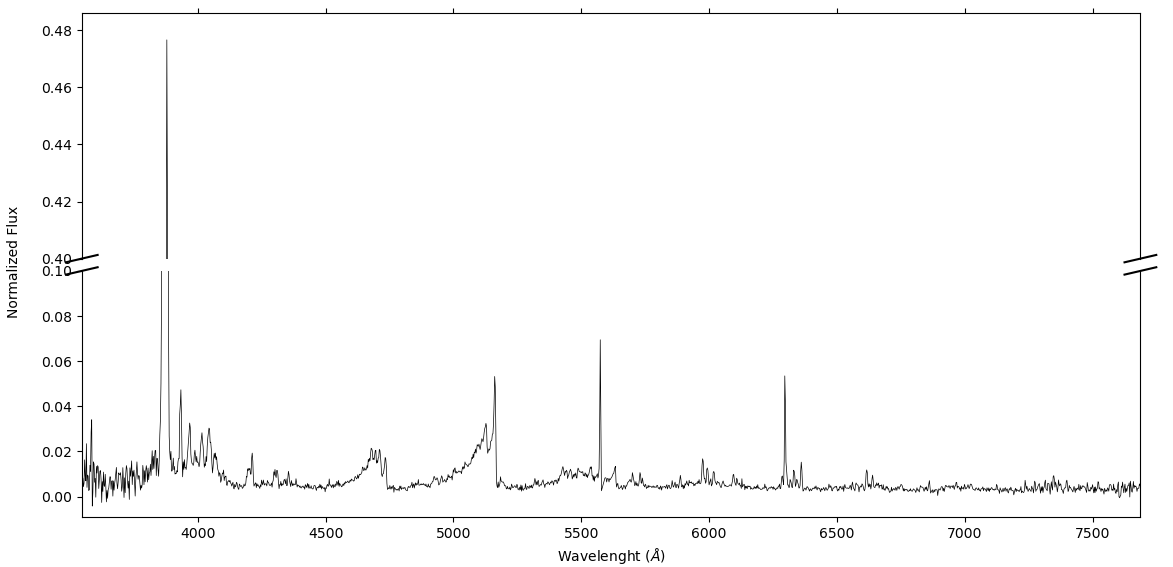}
    \caption{Fig.9 Example of a spectrum of comet C/2023 H2, configuration A}
\end{figure}
\noindent To facilitate a comparison and an initial identification, fig. 10 provides the indication of the relevant regions of the main molecular bands observable on a cometary spectrum. It also shows the three areas of the spectrum where a continuous emission is conventionally identified: the Blue Continuum between 4300 and 4510 Å, the Green Continuum between 5200 and 5320 Å, and the Red Continuum between 7060 and 7180~Å. 
\begin{figure}[h!]
    \centering
    \includegraphics[width=0.75\textwidth]{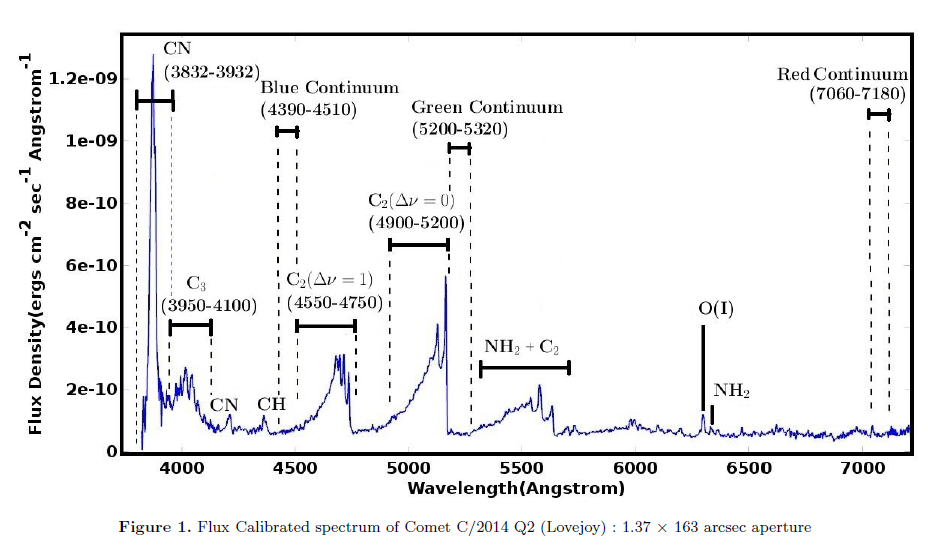}
    \caption{Fig.10 Identification of some chemical species in C/2014 Q2}
\end{figure}
\newpage

\section{THE SPECTRA AT THE GALILEO TELESCOPE}
For a more in-depth study of celestial objects, the setup of the B\&C spectrograph of the Galileo telescope can be substantially varied to suit the observer’s needs. In the specific case of the observations of comets, the possibility of studying them with gratings of different resolutions to investigate the fine structure of the lines in their spectra is of particular interest (Fig. 11, 12).\\
For the sole purpose of producing examples of the possibilities offered by the instrumentation of the Galileo telescope, some spectra of bright comets (therefore with a good S/N) imaged with different gratings and in different areas of the spectrum were chosen for this section. Some graphical sections of spectra, centered at 3880 Å around the main emission line of CN, at 6300 Å for the emission line of oxygen [O I] and at 5890 Å for the presence of the doublet of lines due to sodium (Na), allow to analyze and compare the regions around these lines in detail (Fig. 13, 14).\\
Using the method set by Haser, for each molecule under examination it is possible to calculate a correspondence between the flow observed on the spectrum and the relative quantity of dust and gas emission in kg/sec (Combi, Comets II, 2005). However, this analysis goes beyond the immediate goals of this volume. The authors’ intention is to make dozens of comet spectra available to the scientific community, which can be requested at any time for specific research purposes.\\
It should be kept in mind that the collection and spectral analysis of many comets can favor the observation of important variations in the species and quantities of molecules emitted in possible relation to the internal physical structure of the nucleus and to its distance from the Sun.

\begin{figure}[h!]
\begin{subfigure}{\textwidth}
  \centering
  \includegraphics[width=.7\linewidth]{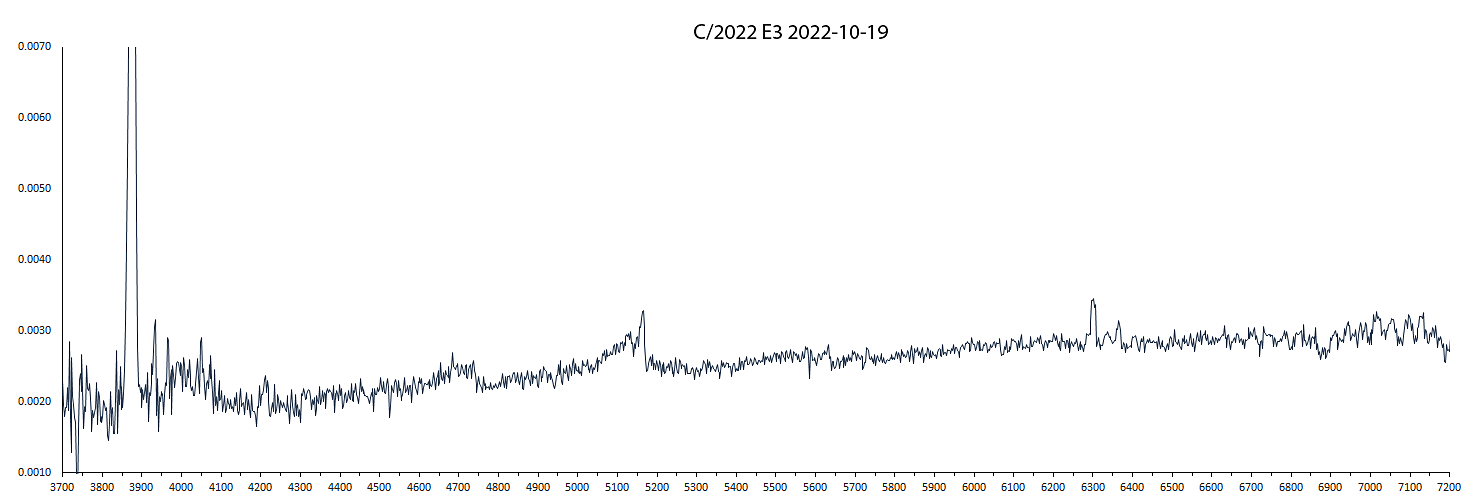}
  \caption{~}

\end{subfigure}
\bigskip
\begin{subfigure}{\textwidth}
  \centering
  \includegraphics[width=.7\linewidth]{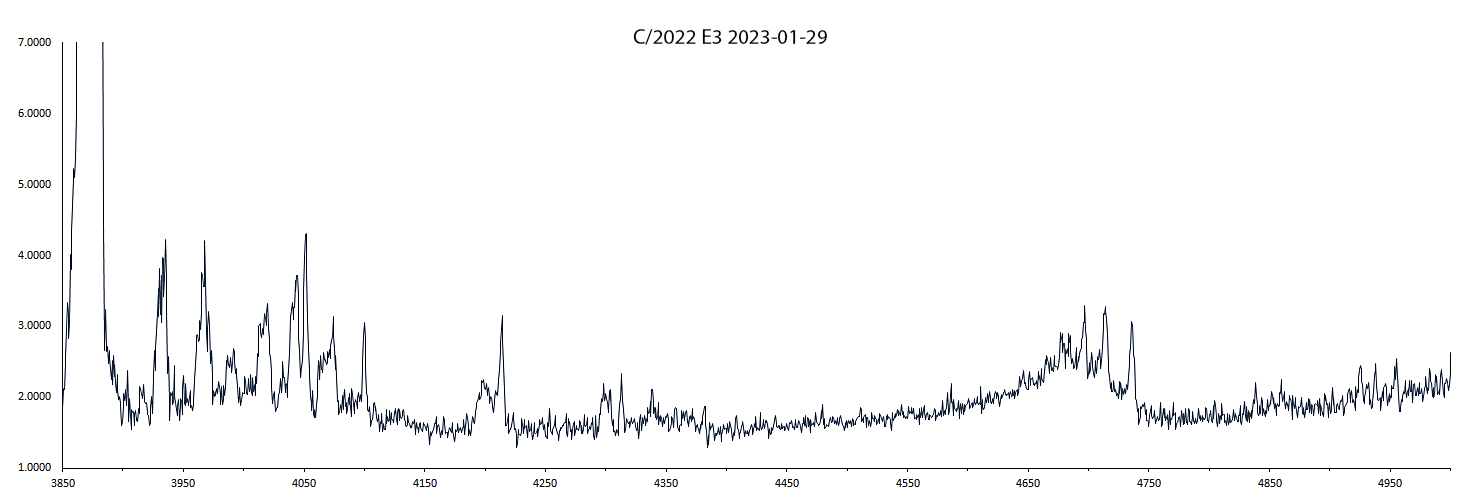}
  \caption{~}

\end{subfigure}
\bigskip
\begin{subfigure}{0.5\textwidth}
  \centering
  \includegraphics[width=.5\linewidth]{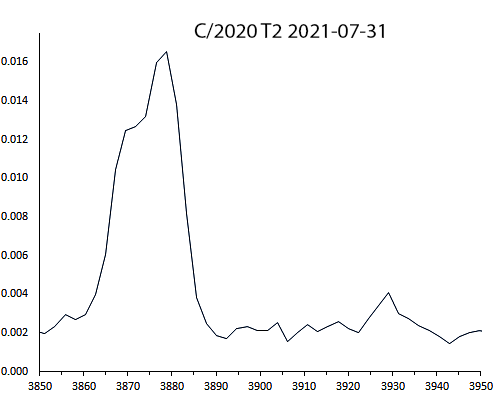}
  \caption{~}

\end{subfigure}
\begin{subfigure}{.5\textwidth}
  \centering
  \includegraphics[width=0.5\linewidth]{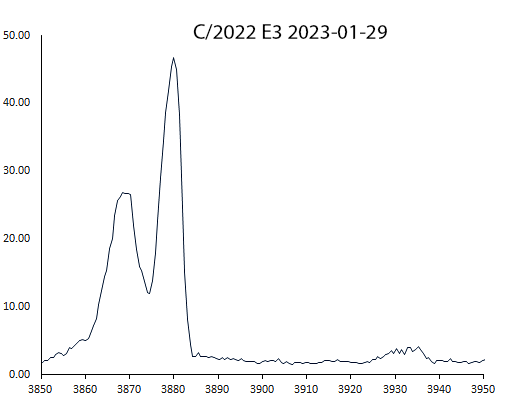}
  \caption{~}

\end{subfigure}
\caption{Fig.11
Comparison of different spectrograph setups applied to the Galileo telescope:\\
Panel (a): comet C/2022 E3, spectrum of October 19, 2022, 300 lines/mm grating, dispersion 2.25 Å/pixel.\\
Panel (b): comet C/2022 E3, spectrum of January 29, 2023, 1200 lines/mm grating, dispersion 0.6 Å/pixel, blue region of the spectrum.\\
Panel (c): comet C/2020 T2, spectrum of July 31, 2021, CN line, 300 lines/mm grating, dispersion 2.25 Å/pixel.\\
Panel (d): comet C/2022 E3, spectrum of January 29, 2023, CN line, 1200 lines/mm grating, dispersion 0.6 Å/pixel.\\
The potential of the higher resolution offered by the grating with 1200 lines is very clear: in c and d 100 angstroms are covered at the same wavelength centered on the CN emissions, in d the line appears split in two.
All spectra result from the division of the recorded comet flux by that of the star defined as Solar Analogue; the scale on the ordinate axis is therefore a relative value.}
\end{figure}
\newpage
\begin{figure}[h!]
\begin{subfigure}{\textwidth}
  \centering
  \includegraphics[width=.8\linewidth]{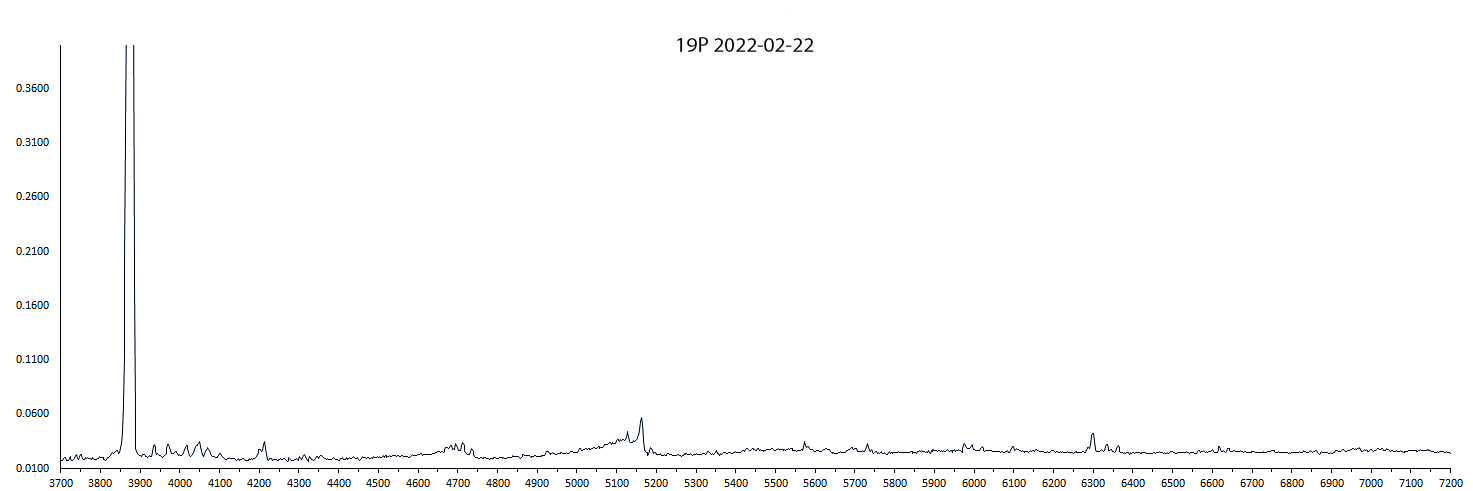}
  \caption{~}

\end{subfigure}
\bigskip
\begin{subfigure}{\textwidth}
  \centering
  \includegraphics[width=.8\linewidth]{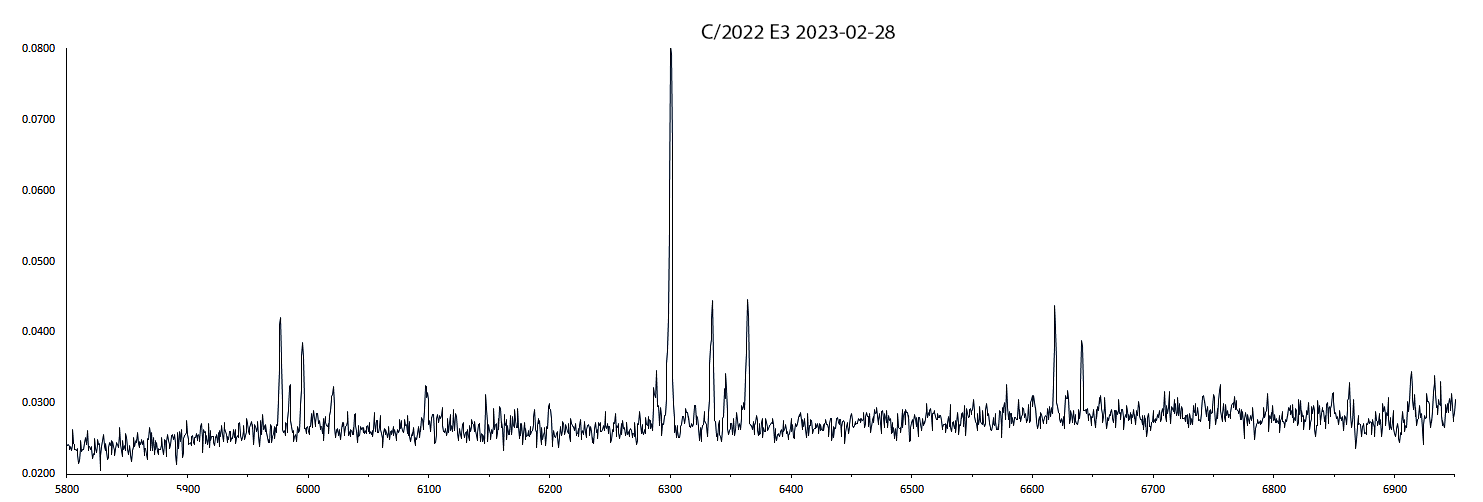}
  \caption{~}

\end{subfigure}
\bigskip
\begin{subfigure}{0.5\textwidth}
  \centering
  \includegraphics[width=.8\linewidth]{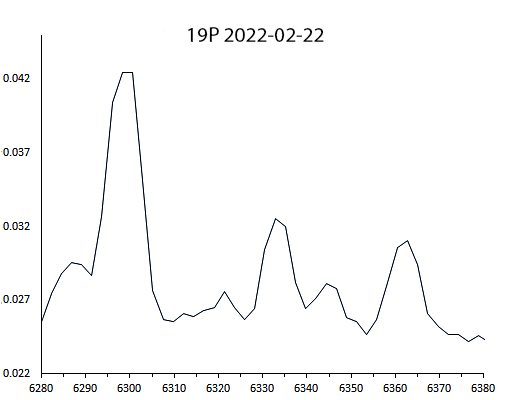}
  \caption{~}

\end{subfigure}
\begin{subfigure}{.5\textwidth}
  \centering
  \includegraphics[width=.8\linewidth]{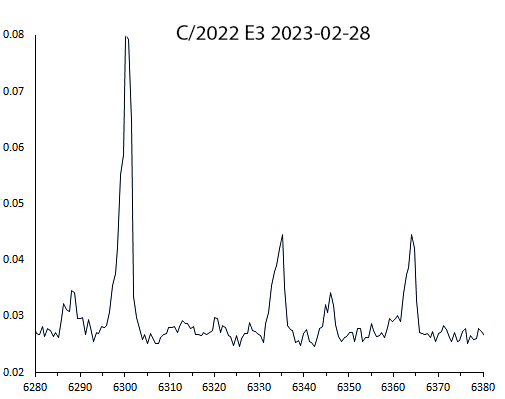}
  \caption{~}

\end{subfigure}
\caption{Fig.12
Comparison of different spectrograph setups applied to the Galileo telescope.\\
Panel (a): comet 19P, spectrum of February 22, 2022, 300 lines/mm grating, dispersion 2.25 Å/pixel.\\
Panel (b): comet C/2022 E3, spectrum of February 28, 2023, 1200 lines/mm grating, dispersion 0.6 Å/pixel, red region of the spectrum.\\
Panel (c): comet 19P, spectrum of February 22, 2022, O I lines, 300 lines/mm grating, dispersion 2.25 Å/pixel; the emission of NH at 6327 Å can be observed in addition to the O I lines (6300 Å+6364 Å).\\
Panel (d): comet C/2020 E3, spectrum of February 28, 2023, O I lines, 1200 lines/mm grating, dispersion 0.6~Å/pixel. The higher resolution also allows the identification of NH$_2$ emissions at 6288 Å and 6345 Å.\\
All spectra result from the division of the recorded comet flux by that of the star defined as Solar Analogue; the scale on the ordinate axis is therefore a relative value.}
\end{figure}
\newpage
\begin{figure}[h!]
\begin{subfigure}{0.45\textwidth}
  \centering
  \includegraphics[width=.8\linewidth]{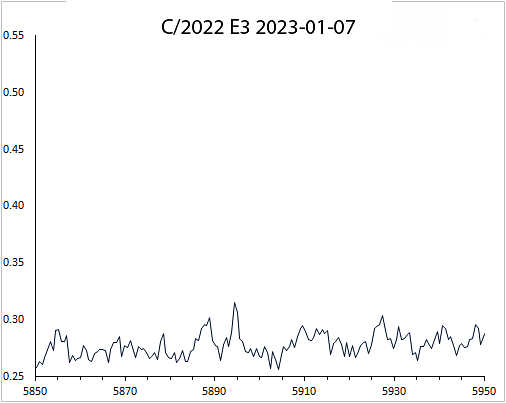}
  \caption{~}

\end{subfigure}
\begin{subfigure}{0.45\textwidth}
  \centering
  \includegraphics[width=.8\linewidth]{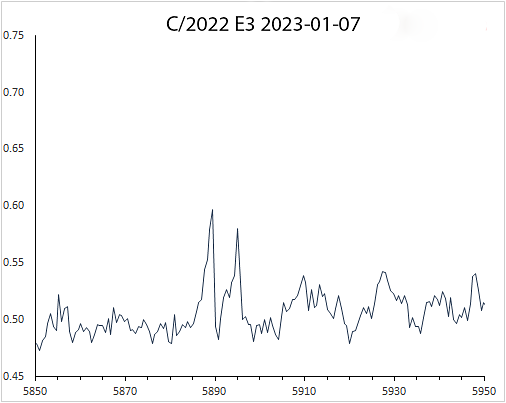}
  \caption{~}

\end{subfigure}
\begin{subfigure}{0.45\textwidth}
  \centering
  \includegraphics[width=.8\linewidth]{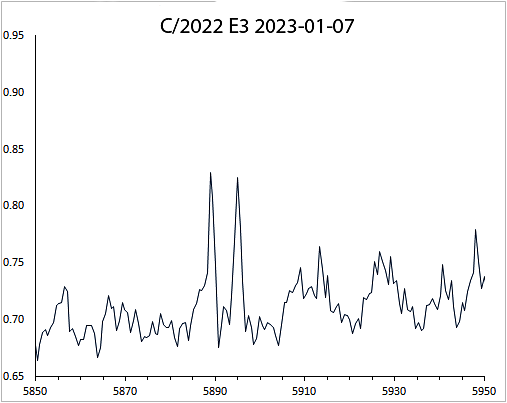}
  \caption{~}

\end{subfigure}
\begin{subfigure}{.45\textwidth}
  \centering
  \includegraphics[width=.8\linewidth]{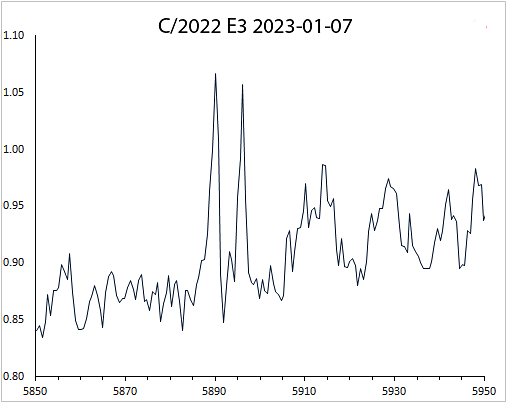}
  \caption{~}

\end{subfigure}
\bigskip
\centering
\begin{subfigure}{.5\textwidth}
  \centering
  \includegraphics[width=.8\linewidth]{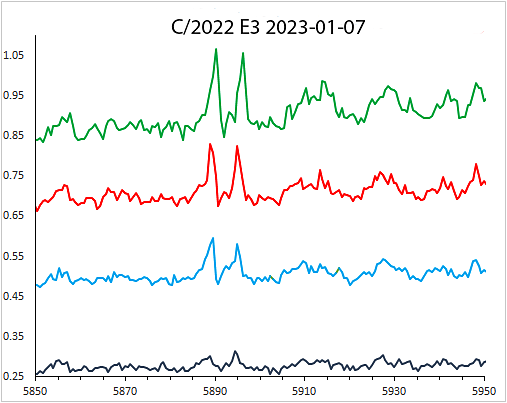}
  \caption{~}

\end{subfigure}
\caption{Fig.13
The emissions due to the Sodium doublet (D1 at 5986 Å and D2 at 5980 Å) are rarely observed and usually in comets approaching the Sun at a distance greater than 1 AU. The presence of sodium emissions on comet C/2022 E3 at 1.115 AU from the Sun was highlighted with the Galileo telescope on January 7, 2023. The spectrograph setup allows to vary the slit width, as shown in the images, at 200$\mu$, 400$\mu$, 600$\mu$, 800$\mu$. Increasing the slit width a larger area of the coma around the nucleus is analyzed. Panel e shows a direct comparison of the results with different slit widths. 
}
\end{figure}
\newpage
\begin{figure}[h!]
\centering
\begin{subfigure}{\textwidth}
\centering
\includegraphics[width=0.4\textwidth]{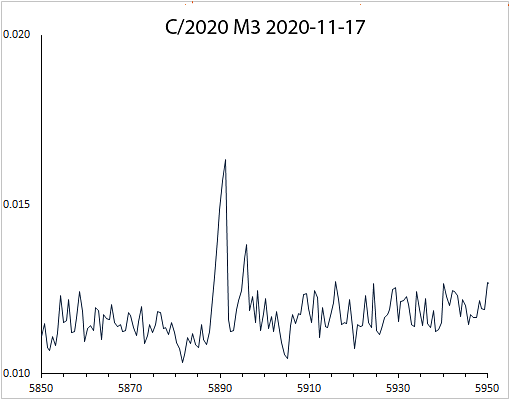}
\caption{~}
\end{subfigure}

\bigskip

\begin{subfigure}{\textwidth}
\centering
\includegraphics[width=0.4\textwidth]{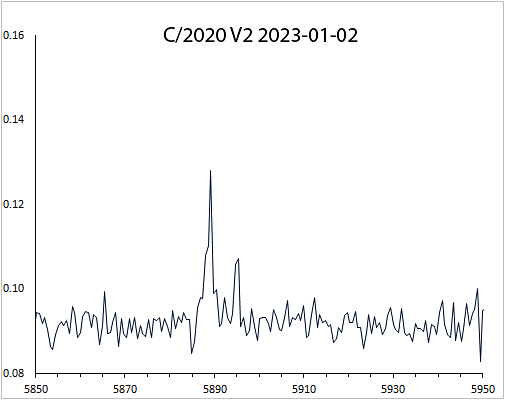}
\caption{~}
\end{subfigure}

\caption{Fig.14
Examples of emission due to Sodium (lines D1 at 5986 Å and D2 at 5980 Å) observed on different comets and at different distances form the Sun with the 1200R grating: \\
Panel (a): comet C/2020 M3: November 17, 2020, at 1.313 AU from the Sun. \\
Panel (b): comet C/2020 V2: January 2, 2023, at 2.646 AU from the Sun. The observation of the sodium emission lines at this distance from the Sun places C/2020 V2 in the small group of comets that show this peculiarity.\\
The scale in ordinate is different in the two panels.
}

\end{figure}
\clearpage
\newpage

\section{REFERENCES}
P. Benvenuti, K. Wurm, A\&A, 1974. Spectroscopic observations of comet Kohoutek.\\
Bolin e Lindler, 1992, Updates to HST Standard-Star Fluxes (\url{https://www.eso.org/sci/observing/tools/standards/spectra/bohlin1992.html}).\\
M. Combi, W. Harris, and W. Smyth, Gas Dynamics and Kinetics in the Cometary Coma: Theory and Observations, in Comets II, 1995.\\
A. Gennaro, S. Taffara, MmSAI, 1945. Posizioni delle comete Oterma … dedotte da fotografie prese con il riflettore di cm. 120 dell’Osservatorio di Asiago.\\
G. Herzberg, H. Lew, A\&A, 1974. Tentative identification of the H2O+ ion in comet Kohoutek.\\
Larson, S. M. \& Sekanina, Z., AJ, 89-1984, p. 571-578.\\
G. Mannino, MmSAI, 1949. Posizioni delle comete Timmers … dedotte da fotografie prese con il riflettore di cm. 120 dell’Osservatorio di Asiago.

\section{WEBSITES}
The Asiago Cima Ekar facility \\
(\url{www.oapd.inaf.it/sede-di-asiago})\\
Telescopes and instrumentation \\
(\url{www.oapd.inaf.it/sede-di-asiago/telescopes-and-instrumentations})\\
The Copernico telescope\\
(\url{www.oapd.inaf.it/sede-di-asiago/telescopes-and-instrumentations/copernico-182cm-telescope})\\
The Schmidt telescope \\
(\url{www.oapd.inaf.it/sede-di-asiago/telescopes-and-instrumentations/schmidt-6792})\\
The Galileo telescope \\
(\url{www.astro.unipd.it/osservatorio/telescopio.html})\\
ESO Standard Stars \\
(\url{https://www.eso.org/sci/observing/tools/standards/spectra/stanlis.html})\\ 
IRAF (Image Reduction and Analysis Facility)\\ 
(\url{https://iraf-community.github.io})\\
JPL Horizon System\\ 
(\url{https://ssd.jpl.nasa.gov/tools/orbit_viewer.html}) 

\clearpage
\newpage
\section*{\Huge GLOSSARY}

\textbf{e:} eccentricity \\
\textbf{q:} perihelion distance [AU] \\
\textbf{T:} time of perihelion passage [TDB]\\
\textbf{$\mathbf{\Omega}$:} longitude of the ascending node [deg]\\
\textbf{$\mathbf{\omega}$:}  argument of perihelion [deg]\\
\textbf{\textit{i}:} inclination, i.e. angle w.r.t. x-y ecliptic plane [deg]\\
\textbf{r:} heliocentric distance [AU]\\
\textbf{$\mathbf{\Delta}$:} distance from target to observer [AU]\\
\textbf{RA:} right ascension [hour]\\
\textbf{DEC:} declination [deg]\\
\textbf{elong:} solar elongation angle [deg]\\
\textbf{phase:} phase angle [deg]\\
\textbf{PLang:} orbit plane angle [deg] \\
\textbf{config:} grating tipe\\
\textbf{FlAng:} spectrograph flange rotation angle [deg]\\
\textbf{N:} number of spectra\\

\newpage

\section*{INDEX}
\begin{itemize}
    \item \hyperref[cometa:12P]{12P (Pons-Brooks)}
    \item \hyperref[cometa:19P]{19P (Borrelly)}
    \item \hyperref[cometa:21P]{21P (Giacobini-Zinner)}
    \item \hyperref[cometa:29P]{29P (Schwassmann-Wachmann 1)}
    \item \hyperref[cometa:38P]{38P (Stephan-Oterma)}
    \item \hyperref[cometa:46P]{46P (Wirtanen)}
    \item \hyperref[cometa:67P]{67P (Churyumov-Gerasimenko)}
    \item \hyperref[cometa:104P]{104P (Koval 2)}
    \item \hyperref[cometa:116P]{116P (Wild 4)}
    \item \hyperref[cometa:123P]{123P (West-Hartley)}
    \item \hyperref[cometa:156P]{156P (Russell-LINEAR)}
    \item \hyperref[cometa:237P]{237P (LINEAR)}
    \item \hyperref[cometa:260P]{260P (McNaught)}
    \item \hyperref[cometa:C2012S1]{C/2012 S1 (ISON)}
    \item \hyperref[cometa:C2013R1]{C/2013 R1 (Lovejoy)}
    \item \hyperref[cometa:C2013US10]{C/2013 US10 (Catalina)}
    \item \hyperref[cometa:C2013X1]{C/2013 X1 (PanSTARRS)}
    \item \hyperref[cometa:C2015O1]{C/2015 O1 (PanSTARRS)}
    \item \hyperref[cometa:C2015V2]{C/2015 V2 (Johnson)}
    \item \hyperref[cometa:C2016N6]{C/2016 N6 (PanSTARRS)}
    \item \hyperref[cometa:C2016R2]{C/2016 R2 (PanSTARRS)}
    \item \hyperref[cometa:C2017K2]{C/2017 K2 (PanSTARRS)}
    \item \hyperref[cometa:C2017T2]{C/2017 T2 (PanSTARRS)}
    \item \hyperref[cometa:C2018N2]{C/2018 N2 (ASASSN)}
    \item \hyperref[cometa:C2018Y1]{C/2018 Y1 (Iwamoto)}
    \item \hyperref[cometa:C2019L3]{C/2019 L3 (ATLAS)}
    \item \hyperref[cometa:C2019T4]{C/2019 T4 (ATLAS)}
    \item \hyperref[cometa:C2019U5]{C/2019 U5 (PanSTARRS)}
    \item \hyperref[cometa:C2019Y4]{C/2019 Y4 (ATLAS)}
    \item \hyperref[cometa:C2020F3]{C/2020 F3 (NEOWISE)}
    \item \hyperref[cometa:C2020M3]{C/2020 M3 (ATLAS)}
    \item \hyperref[cometa:C2020PV6]{C/2020 PV6 (PanSTARRS)}
    \item \hyperref[cometa:C2020R4]{C/2020 R4 (ATLAS)}
    \item \hyperref[cometa:C2020T2]{C/2020 T2 (Palomar)}
    \item \hyperref[cometa:C2020V2]{C/2020 V2 (ZTF)}
    \item \hyperref[cometa:C2021X1]{C/2021 X1 (Maury-Attard)}
    \item \hyperref[cometa:C2022E2]{C/2022 E2 (ATLAS)}
    \item \hyperref[cometa:C2022E3]{C/2022 E3 (ZTF)}
    \item \hyperref[cometa:C2023E1]{C/2023 E1 (ATLAS)}
    \item \hyperref[cometa:C2023H2]{C/2023 H2 (Lemmon)}
    \item \hyperref[cometa:C2023A3]{C/2023 A3 (Tsuchinshan-ATLAS)}
\end{itemize}

\newpage
\clearpage

\titleformat{\section}[display]
  {\normalfont\bfseries}{}{0pt}{\Large}
\titleformat{\subsection}[display]
  {\normalfont\bfseries}{}{0pt}{\large}
\renewcommand{\thesection}{}
\renewcommand{\thesubsection}{}
\section{12P (Pons-Brooks)}
\label{cometa:12P}
\subsection{Description}

12P/Pons-Brooks is a Halley-type comet with a period of 71 years and an absolute magnitude of 5$\pm$0.5.\footnote{\url{https://ssd.jpl.nasa.gov/tools/sbdb_lookup.html\#/?sstr=12P} visited on July 21, 2024} It was first discovered by Jean-Louis Pons, from the Marseille Observatory, on July 12, 1812. Then, it was rediscovered in 1883 by William Robert Brooks from Phelps, NY, and followed until the year after. This object has a history of outbursts, which happened during every perihelion passage. This is also the case for the observations discussed here. A significant outburst event\footnote{\url{http://aerith.net/comet/catalog/0012P/2024.html}, visited on July 20, 2024} \footnote{\url{https://britastro.org/section_news_item/12p-pons-brooks-latest-lightcurve}, visited on July 20, 2024} 
began around 22 UT of July 20, 2023, before the perihelion passage of April 21, 2024 during which the coma of 12P expanded rapidly. 

\noindent
We observed the comet from magnitude 11.5 to 8.\footnote{\url{https://cobs.si/comet/484/ }, visited on July 21, 2024}

\begin{table}[h!]
\centering
\begin{tabular}{|c|c|c|}
\hline
\multicolumn{3}{|c|}{Orbital elements (epoch: September 20, 2023)}                      \\ \hline \hline
\textit{e} = 0.9546 & \textit{q} = 0.7809 & \textit{T} = 2460421.6309 \\ \hline
$\Omega$ = 255.8437 & $\omega$ = 198.9826  & \textit{i} = 74.1861 \\ \hline  
\end{tabular}
\end{table}

\begin{table}[h!]
\centering
\begin{tabular}{|c|c|c|c|c|c|c|c|c|}
\hline
\multicolumn{9}{|c|}{Comet ephemerides for key dates}                      \\ \hline 
\hline
& date         & r    & $\Delta$  & RA      & DEC      & elong  & phase  & PLang  \\
& (yyyy-mm-dd) & (AU) & (AU)      & (h)     & (°)      & (°)    & (°)    & (°) \\ \hline 

Perihelion       & 2024-04-21 & 0.781 & 1.605 & 3.468 & 9.843 & 22.66 & 29.73 & 25.49  \\ 
Nearest approach & 2024-06-02 & 1.101 & 1.546 & 5.883 & 19.653 & 45.30 & 40.88 & 2.73\\ \hline
\end{tabular}

\end{table}

\vspace{0.5 cm}

\begin{figure}[h!]
    \centering
    \includegraphics[scale=0.38]{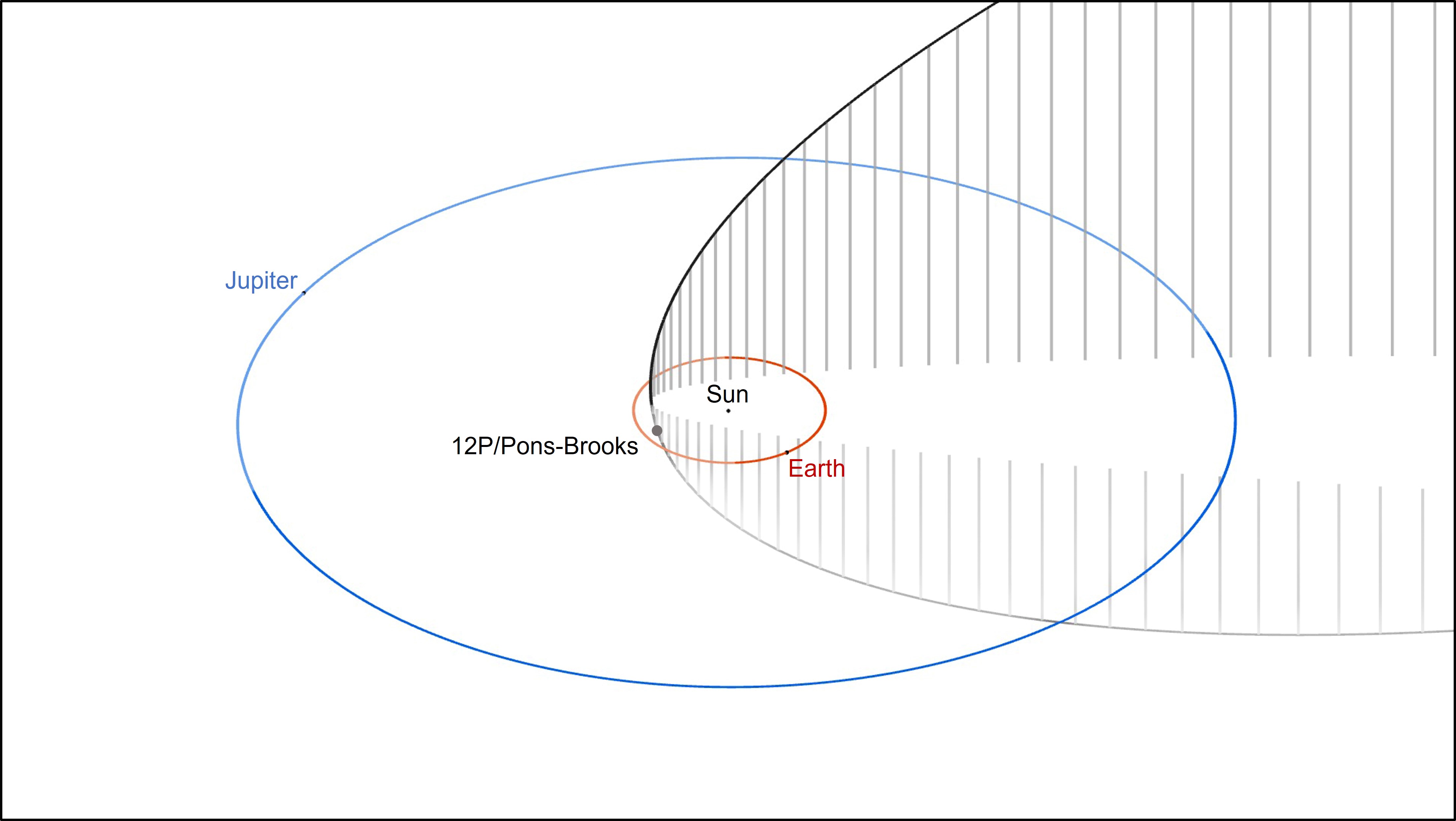}
    \caption{Orbit of comet 12P/Pons-Brooks and position on perihelion date. The field of view is set to the orbit of Jupiter for size comparison. Courtesy of NASA/JPL-Caltech.}
\end{figure}

\newpage

\subsection{Images}

\begin{SCfigure}[0.8][h!]
    \centering
    \includegraphics[scale=0.95]{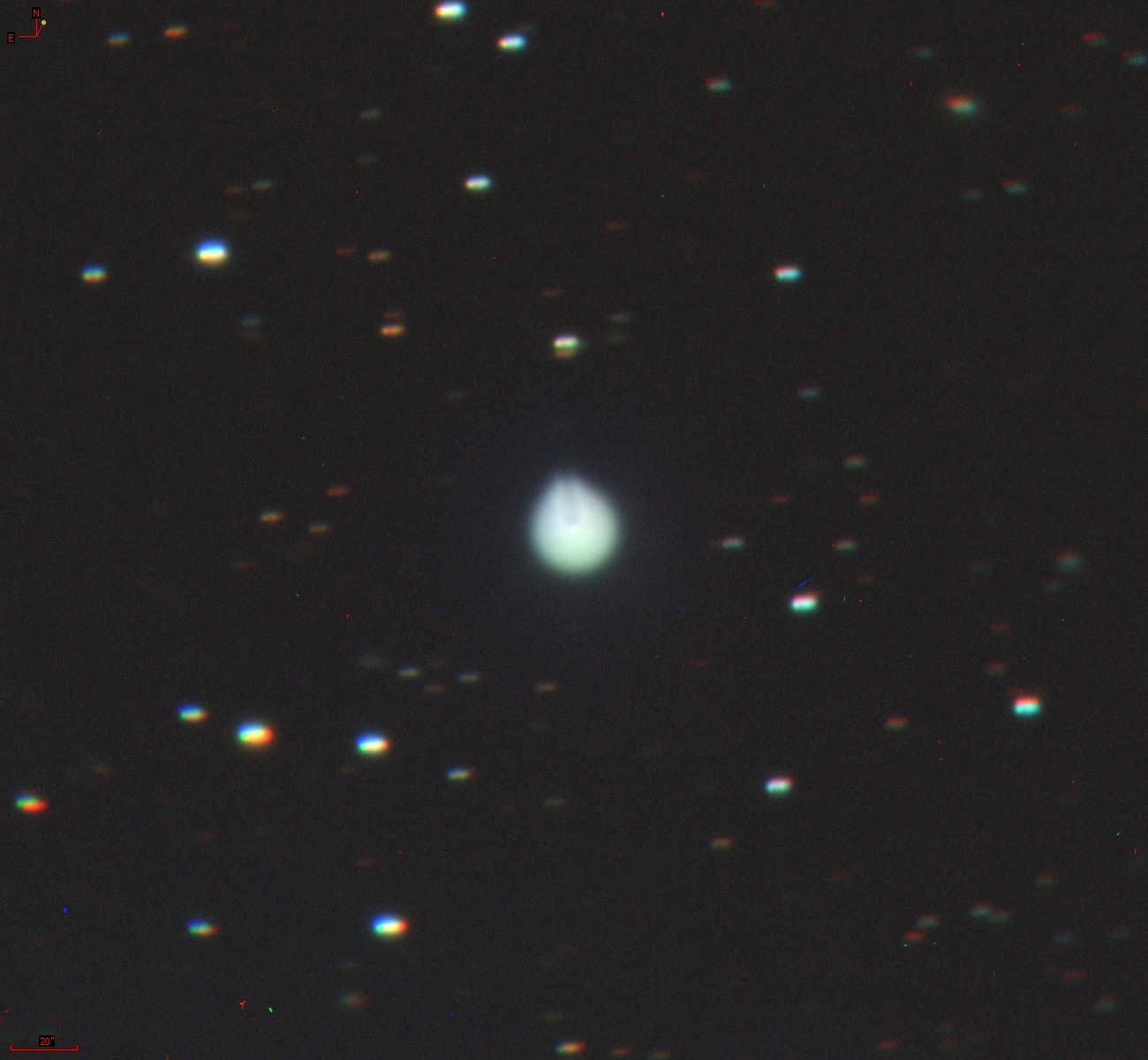}
     \caption{2023-07-22. Composite image in g, r and z filters taken with the 2m Faulkes North Telescope at Las Cumbres Observatory, Hawaii. Initially spherical, the coma assumed a peculiar asymmetric shape caused by radiation pressure from the photons of the Sun acting on the emitted dust.}

\end{SCfigure} 

\begin{SCfigure}[0.8][h!]
    \centering
    \includegraphics[scale=0.228]{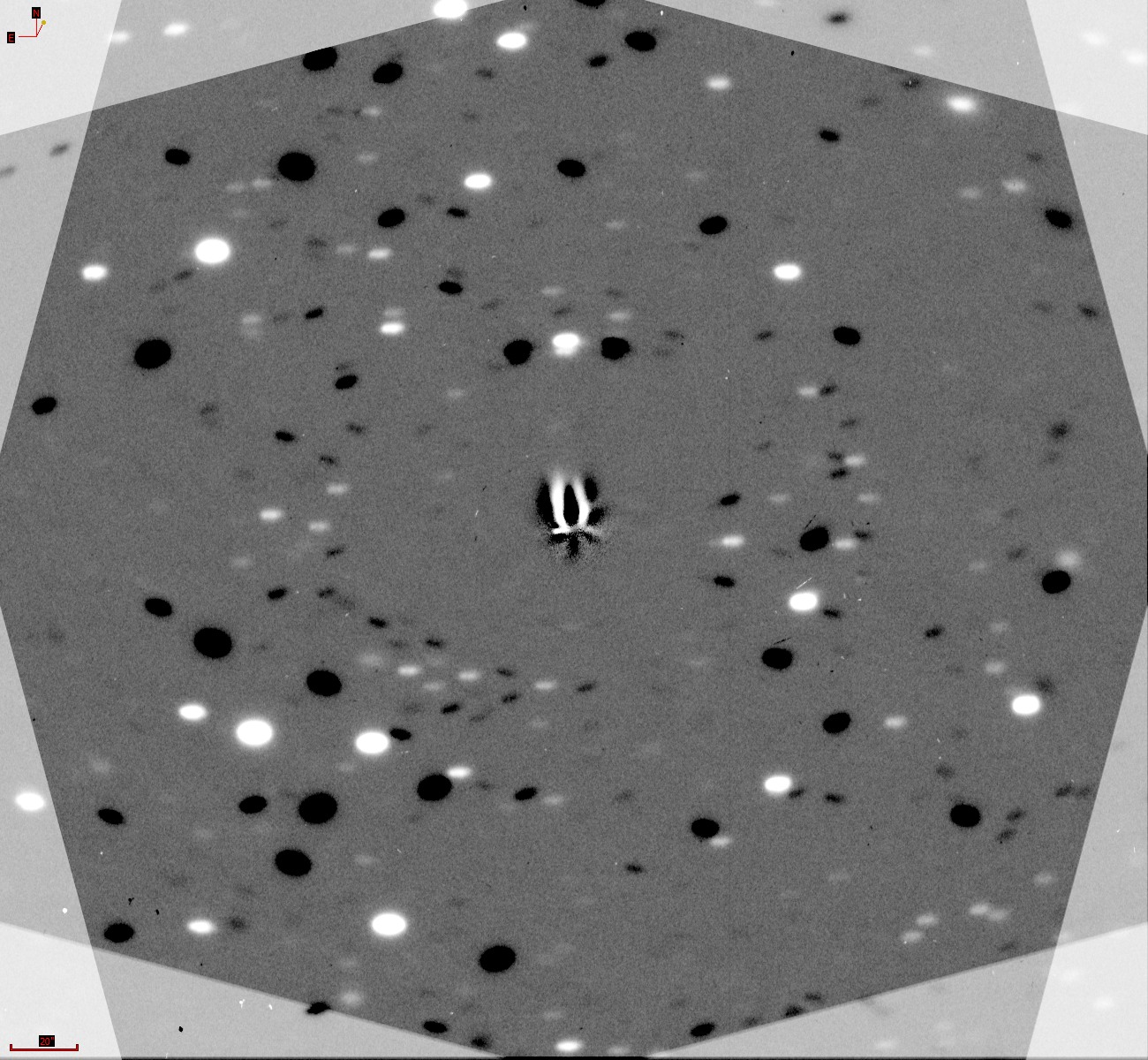}
    \caption{2022-03-09. Larson-Sekanina processing ($\alpha=15$\textdegree) applied to a different image in g, r, i, z filters. There are two clearly noticeable jet-shaped structures where the `horns' in the coma are found. The treatment also reveals two smaller jets in the opposite direction, giving the inner coma a highly symmetric X-shaped structure.}

\end{SCfigure}

\begin{figure}
    \centering
    \includegraphics[scale=0.7]{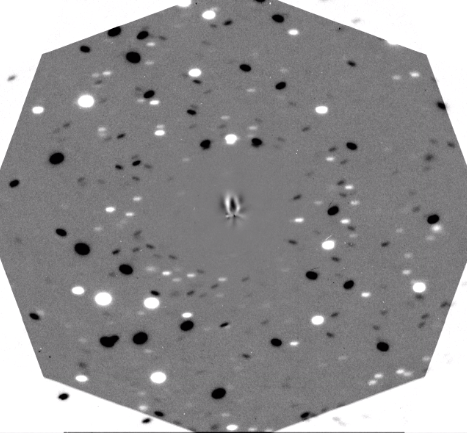}
    \includegraphics[scale=0.7]{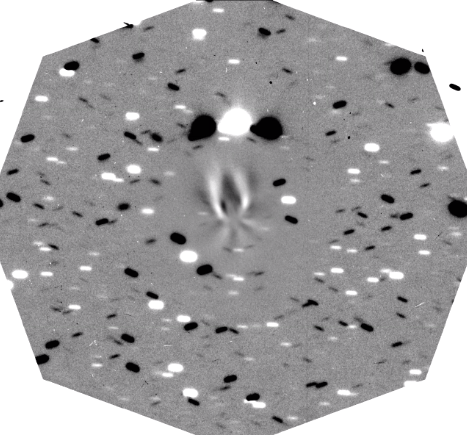} \\
    \includegraphics[scale=0.7]{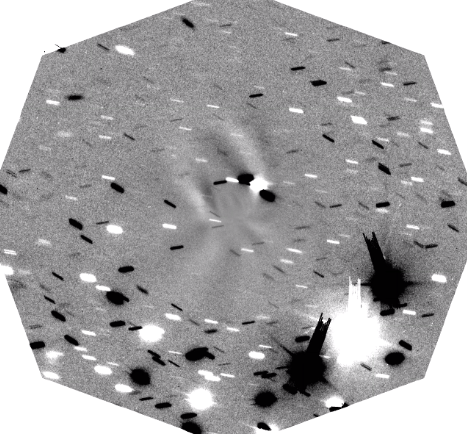}
    \includegraphics[scale=0.7]{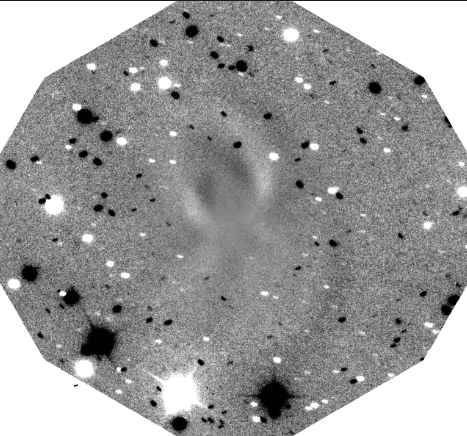}
    \caption{Evolution of the 12P July 2023 outburst over a time-span of 18 days. The Larson-Sekanina filter was applied.}
    \label{fig:enter-label}
\end{figure}
\newpage

\subsection{Spectra}

\begin{table}[h!]
\centering
\begin{tabular}{|c|c|c|c|c|c|c|c|c|c|c|c|}
\hline
\multicolumn{12}{|c|}{Observation details}                      \\ \hline 
\hline
$\#$  & date          & r     & $\Delta$ & RA     & DEC     & elong & phase & PLang & config & FlAng & N \\
      & (yyyy-mm-dd)  &  (AU) & (AU)     & (h)    & (°)     & (°)   & (°)   &  (°)   &       &  (°)  & \\ \hline 

1 & 2023-07-23 & 3.850 & 3.539 & 18.22 & $+$55.79 & 100.14 & 15.06 & $-$11.05 & A & $+$90 & 6 \\
2 & 2023-08-11 & 3.652 & 3.403 & 17.75 & $+$54.77 & 96.02 & 16.02 & $-$14.63 & A & $+$90 & 4 \\
3 & 2023-08-21 & 3.546 & 3.341 & 17.55 & $+$53.60 & 93.28 & 16.55 & $-$15.99 & A & $-$0 & 2 \\
4 & 2023-10-01 & 3.098 & 3.100 & 17.34 & $+$46.71 & 80.63 & 18.59 & $-$16.62 & A & $+$0 & 5 \\
5 & 2023-10-07 & 3.031 & 3.061 & 17.38 & $+$45.65 & 78.85 & 18.88 & $-$15.99 & D & $-$0 & 7 \\
6 & 2023-11-12 & 2.611 & 2.781 & 17.95 & $+$40.34 & 69.86 & 20.85 & $-$08.58 & A & $+$90 & 2 \\
7 & 2023-11-17 & 2.552 & 2.735 & 18.07 & $+$39.79 & 68.89 & 21.19 & $-$07.11 & D & $+$0 & 7 \\
8 & 2023-11-18 & 2.539 & 2.725 & 18.09 & $+$39.69 & 68.70 & 21.27 & $-$06.79 & D & $+$20 & 8 \\
9 & 2023-11-22 & 2.491 & 2.687 & 18.20 & $+$39.31 & 67.98 & 21.57 & $-$05.52 & A & $-$0 & 10 \\
10 & 2023-11-25 & 2.454 & 2.657 & 18.28 & $+$39.04 & 67.46 & 21.81 & $-$04.52 & A & $+$0 & 10 \\ 
11 & 2023-11-26 & 2.442 & 2.647 & 18.31 & $+$38.96 & 67.30 & 21.88 & $-$04.20 & A & $+$20 & 21 \\	
12 & 2023-11-28 & 2.418 & 2.627 & 18.37 & $+$38.82 & 66.96 & 22.06 & $-$03.50 & A & $+$20 & 5 \\
13 & 2023-12-03 & 2.356 & 2.575 & 18.52 & $+$38.48 & 66.18 & 22.50 & $-$01.73 & A & $+$50 & 5 \\ 
14 & 2023-12-06 & 2.319 & 2.543 & 18.62 & $+$38.30 & 65.72 & 22.78 & $-$00.62 & A & $+$59 & 21 \\ 
15 & 2023-12-14 & 2.219 & 2.456 & 18.90 & $+$37.95 & 64.57 & 23.62 & $+$02.51 & A & $-$0 & 10 \\ 
16 & 2023-12-15 & 2.207 & 2.445 & 18.93 & $+$37.93 & 64.44 & 23.73 & $+$02.89 & A & $+$35 & 4 \\ 
17 & 2024-01-04 & 1.951 & 2.217 & 19.79 & $+$37.77 & 61.60 & 26.31 & $+$11.52 & C & $-$0 & 1 \\ 
18 & 2024-01-24 & 1.689 & 1.993 & 20.90 & $+$38.18 & 57.87 & 29.57 & $+$20.82 & B & $+$38 & 3 \\

\hline
\end{tabular}
\end{table}

\begin{figure}[h!]
    \centering
    \includegraphics[scale=0.55]{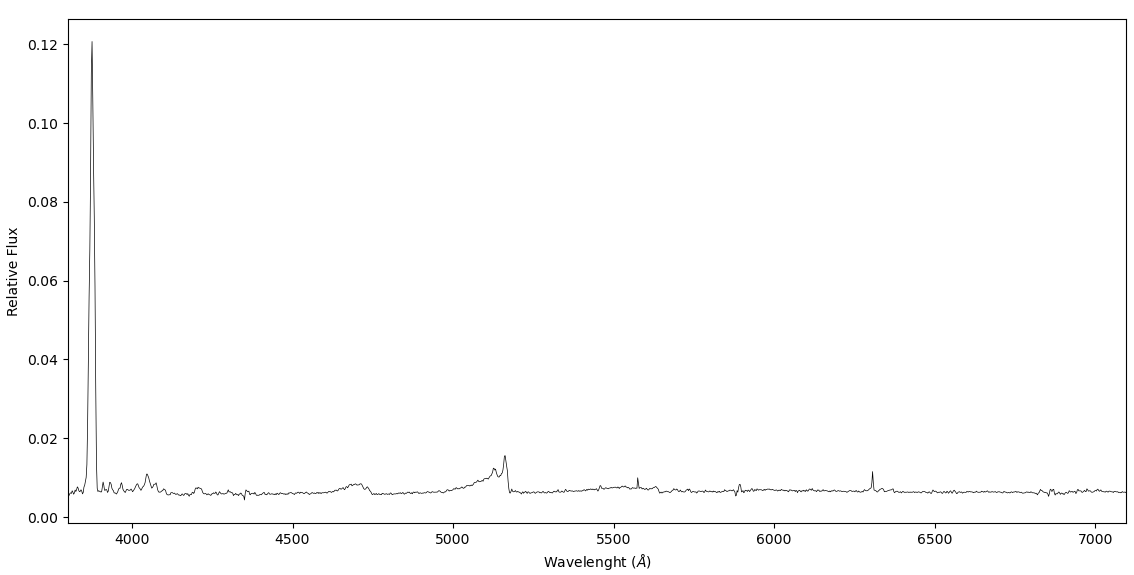}
    \caption{Spectrum of 2023-12-03; configuration A. Flux is normalized to the solar analog Land 107m684.}
\end{figure}

\begin{figure}[h!]
    \centering
    \includegraphics[scale=0.55]{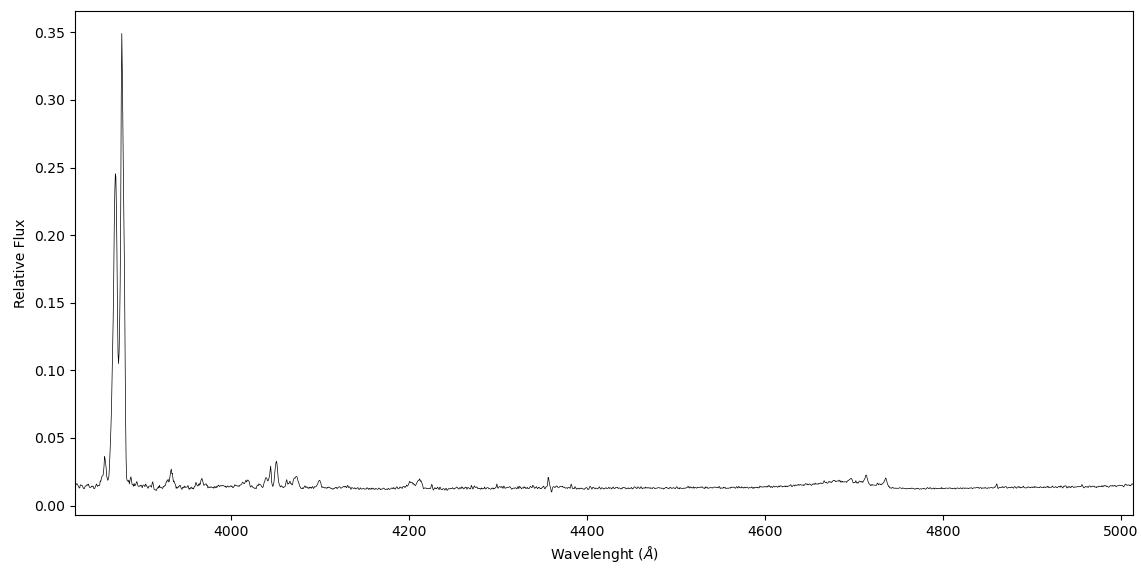}
    \caption{Spectrum of 2024-01-24; configuration B. Flux is normalized to the solar analog Land 107m684.}
\end{figure}

\begin{figure}[h!]
    \centering
    \includegraphics[scale=0.55]{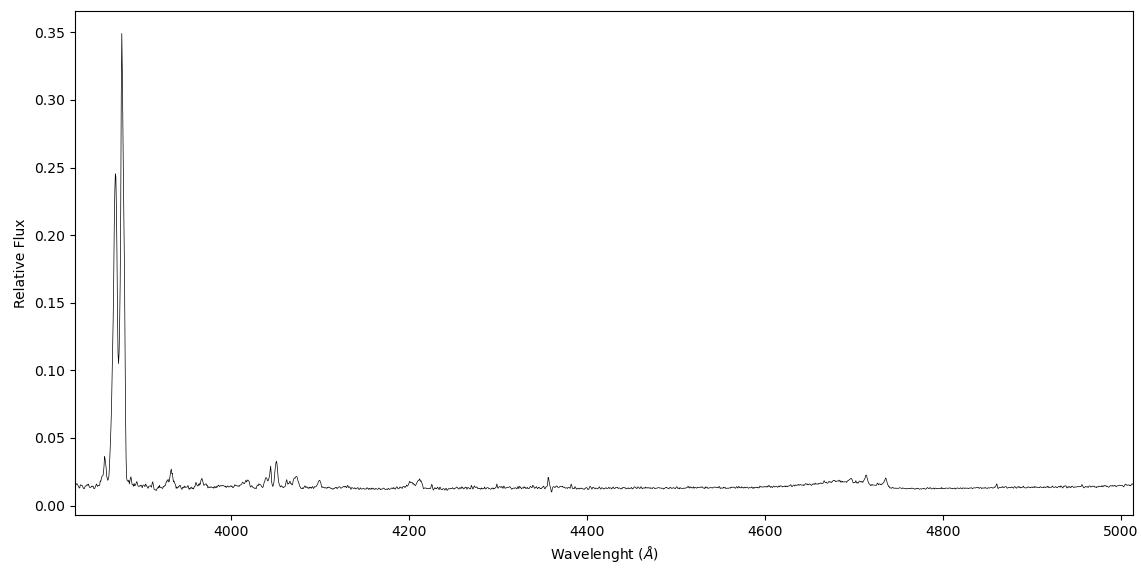}
    \caption{Spectrum of 2023-11-17; configuration D. Flux is normalized to the solar analog Land 107m684.}
\end{figure}

\newpage
\clearpage

\section{19P (Borrelly)}
\label{cometa:19P}
\subsection{Description}

19P/Borrelly is a Jupiter-family comet with a period of 6.85 years and an absolute magnitude of 9.8$\pm$0.8.\footnote{\url{https://ssd.jpl.nasa.gov/tools/sbdb_lookup.html\#/?sstr=19P} visited on July 21, 2024}
It was first discovered by Alphonse Borrelly, from the Marseille Observatory, on December 28, 1904. It was then rediscovered on September 20, 1911 by Harold Knox-Shaw from the Helwan Observatory.
Georges Fayet calculated an elliptical orbit of the new comet taking into account the influences of Jupiter and Saturn.
The Earth crossed the orbital plane of the comet on December 6, 2020, on June 6, 2021, and on December 7, 2021.

\noindent
We observed the comet between magnitude 8 and 10.\footnote{\url{https://cobs.si/comet/335/ }, visited on July 21, 2024} 

\begin{table}[h!]
\centering
\begin{tabular}{|c|c|c|}
\hline
\multicolumn{3}{|c|}{Orbital elements (epoch: March 6, 2021)}                      \\ \hline \hline
\textit{e} = 0.6379 & \textit{q} = 1.3060 & \textit{T} = 2459612.2641 \\ \hline
$\Omega$ = 74.3008 & $\omega$ = 351.8616  & \textit{i} = 29.3187  \\ \hline  
\end{tabular}
\end{table}

\begin{table}[h!]
\centering
\begin{tabular}{|c|c|c|c|c|c|c|c|c|}
\hline
\multicolumn{9}{|c|}{Comet ephemerides for key dates}                      \\ \hline 
\hline
& date         & r    & $\Delta$  & RA      & DEC      & elong  & phase  & PLang  \\
& (yyyy-mm-dd) & (AU) & (AU)      & (h)     & (°)      & (°)    & (°)    & (°) \\ \hline 

Perihelion       & 2022-02-02 & 1.306  & 1.269 & 01.47  & $+$04.95 &  69.5 & 44.9  & $-$19.1  \\ 
Nearest approach & 2021-12-11 & 1.448  & 1.175 & 23.70  & $-$32.03 & 83.8  & 42.5  & $-$02.0 \\ \hline
\end{tabular}

\end{table}

\vspace{0.5 cm}

\begin{figure}[h!]
    \centering
    \includegraphics[scale=0.38]{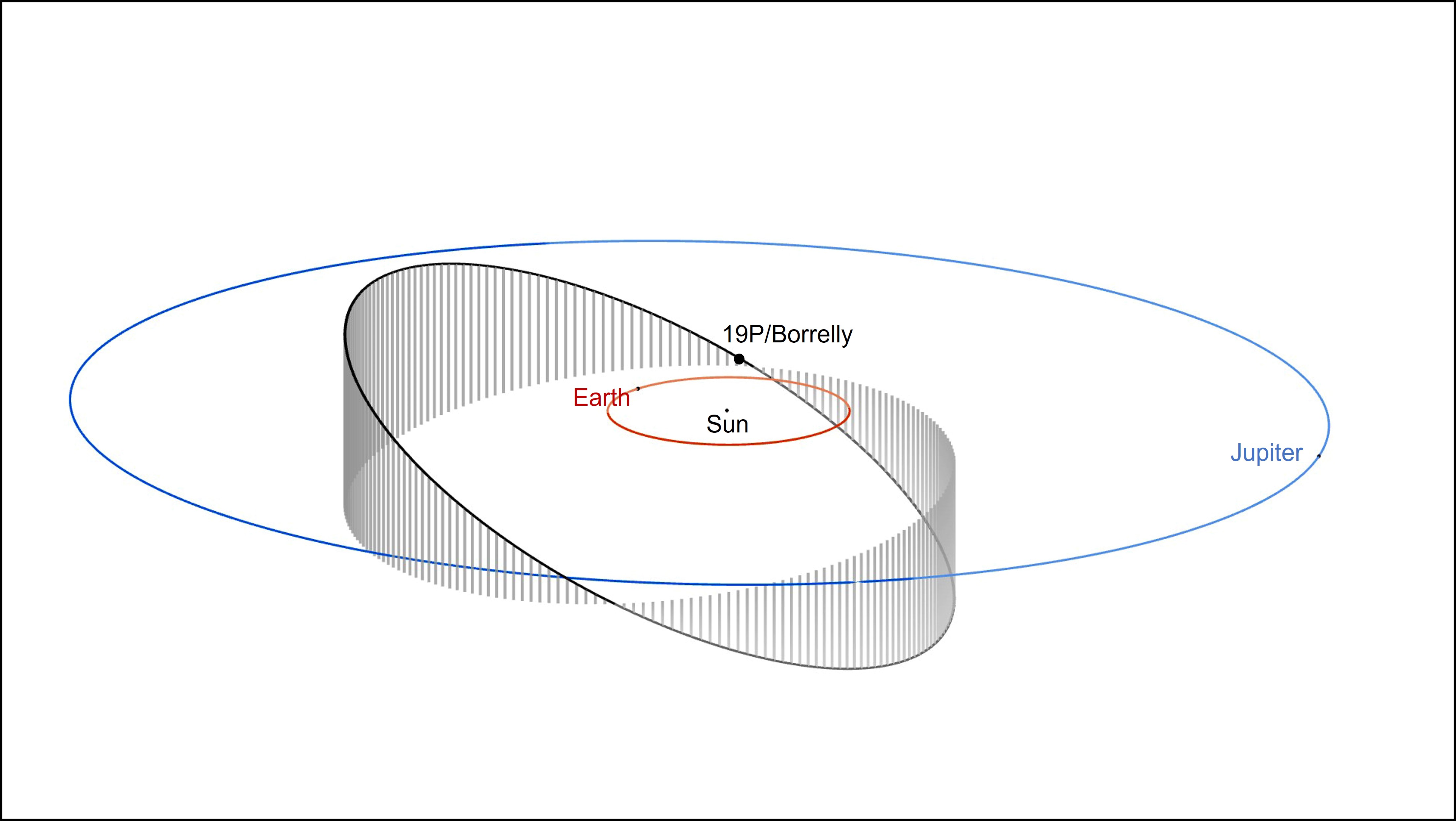}
    \caption{Orbit of comet 19P and position on perihelion date. The field of view is set to the orbit of Jupiter for size comparison. Courtesy of NASA/JPL-Caltech.}
\end{figure} 

\newpage

\subsection{Images}

\begin{SCfigure}[0.8][h!]
    \centering
    \includegraphics[scale=0.4]{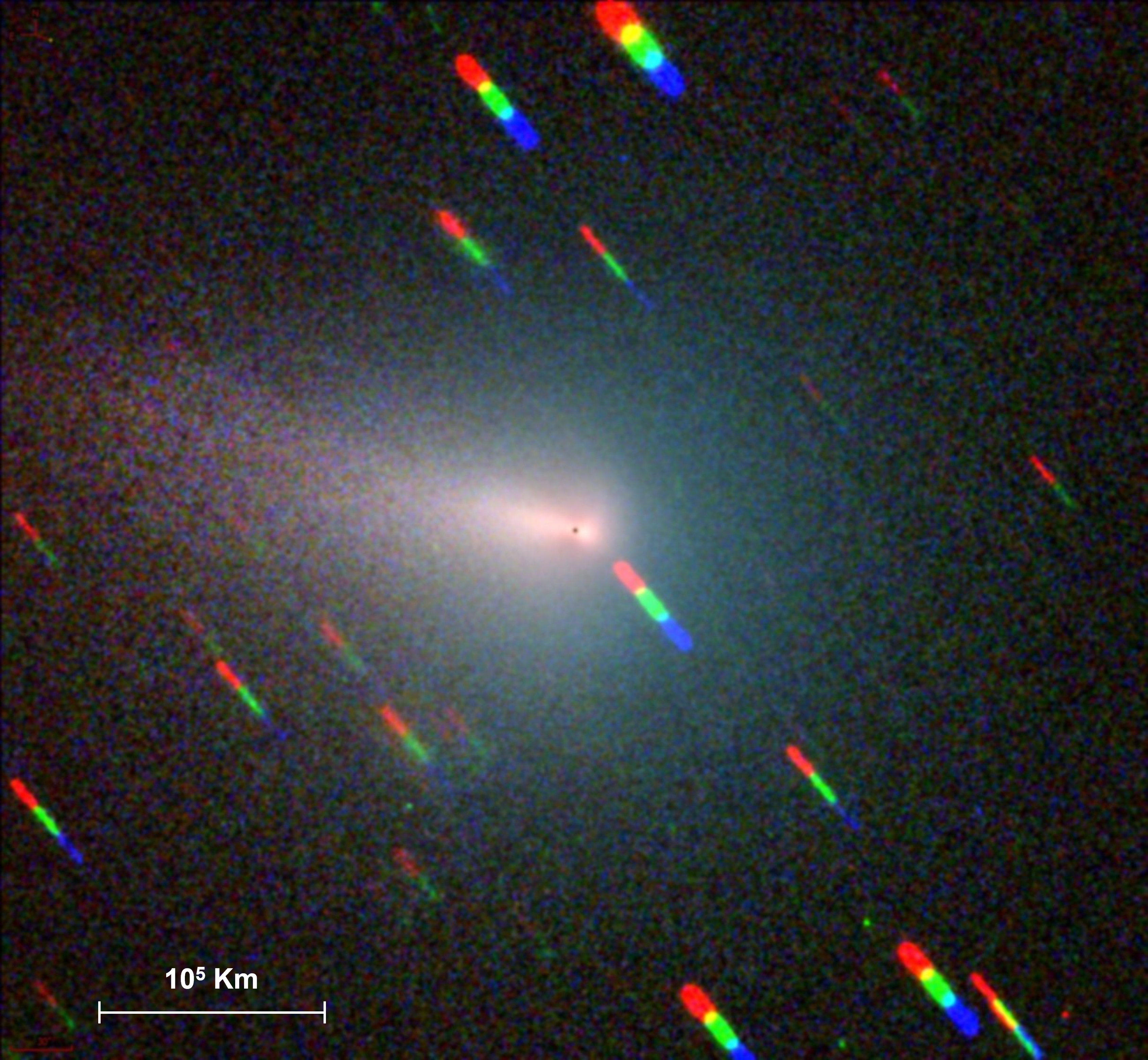}
    \caption{2022-01-30. Three-color BVR composite from images taken with the Asiago Schmidt telescope. The image of the comet was processed with a 1/R attenuator filter to observe the morphological details of the inner coma.}
\end{SCfigure} 

\begin{SCfigure}[0.8][h!]
    \centering
    \includegraphics[scale=0.4]{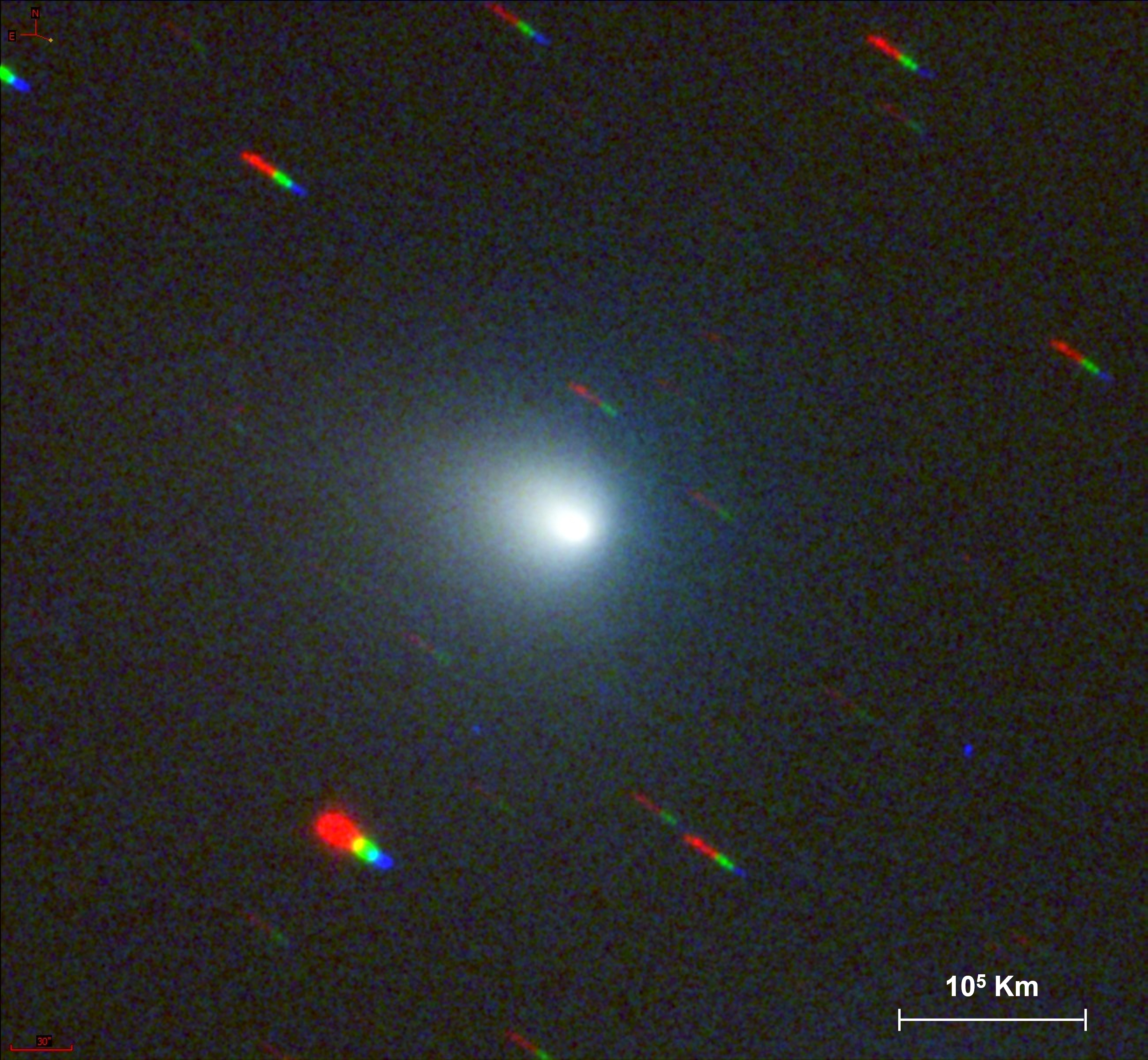}
    \caption{2022-03-12. Images taken with the Asiago Schmidt telescope. Comet Borrelly is moving away from the Sun and the Earth after a very favorable passage. The images, taken with the B, V, R (r+i) filters, have been recalibrated to produce a three-color RGB composite centered on the sensitivity of the human eye. The use of the infrared filter, associated with the red color, accentuates the corresponding component in the trails of the field stars.}
\end{SCfigure}

\newpage

\subsection{Spectra}

\begin{table}[h!]
\centering
\begin{tabular}{|c|c|c|c|c|c|c|c|c|c|c|c|}
\hline
\multicolumn{12}{|c|}{Observation details}                      \\ \hline 
\hline
$\#$  & date          & r     & $\Delta$ & RA     & DEC     & elong & phase & PLang& config  &  FlAng & N \\
      & (yyyy-mm-dd)  &  (AU) & (AU)     & (h)    & (°)     & (°)   & (°)   &  (°)   &       &  (°)  & \\ \hline 

1 & 2022-01-10 & 1.350  & 1.276 & 00.65 & $-$11.02 & 72.1  & 43.8  & $-$12.9 & A & $-$43 & 3 \\
2 & 2022-01-11 & 1.348  & 1.279 & 00.68 & $-$10.30 & 71.8  & 43.9  & $-$13.2 & C & $-$43 & 3 \\
3 & 2022-01-21 & 1.325  & 1.301 & 01.05 & $-$02.98 & 69.3  & 44.0  & $-$15.9 & C & $+40$ & 1 \\
4 & 2022-01-23 & 1.322  & 1.306 & 01.12 & $-$01.53 & 68.9  & 43.4  & $-$16.4 & A & $-$43 & 3\\
5 & 2022-02-22 & 1.325  & 1.437 & 02.40 & $+$18.98 & 63.1  & 41.7  & $-$19.6 & A & $-$45 & 3\\
6 & 2022-03-26 & 1.428  & 1.673 & 04.10 & $+$35.00 & 58.3  & 36.5  & $-$16.0 & A & $+$40 & 3\\
7 & 2022-04-10 & 1.505  & 1.813 & 05.02 & $+$39.58 & 56.1 & 33.5  & $-$12.7 & A & $+$0 & 5 \\
8 & 2022-04-18 & 1.551  & 1.893 & 05.52 & $+$41.22 & 54.9  & 32.0  & $-$10.9 & A & $+$0 & 3\\

\hline
\end{tabular}
\end{table}

\begin{figure}[h!]

    \centering
    \includegraphics[scale=0.55]{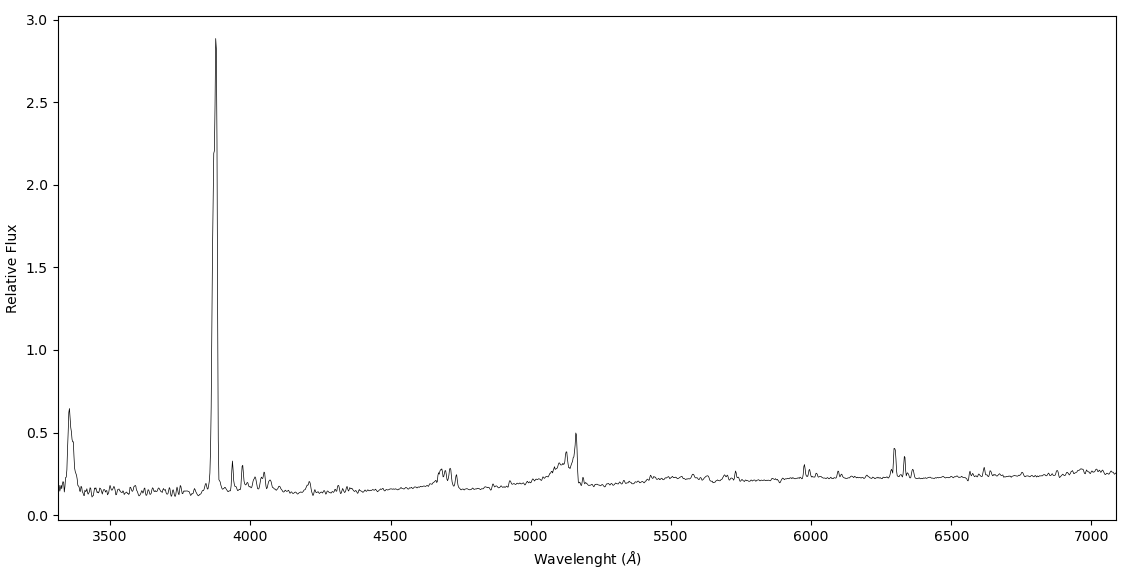}
    \caption{Spectrum of 2022-01-23; configuration A}

\end{figure}

\begin{figure}[h!]

    \centering
    \includegraphics[scale=0.55]{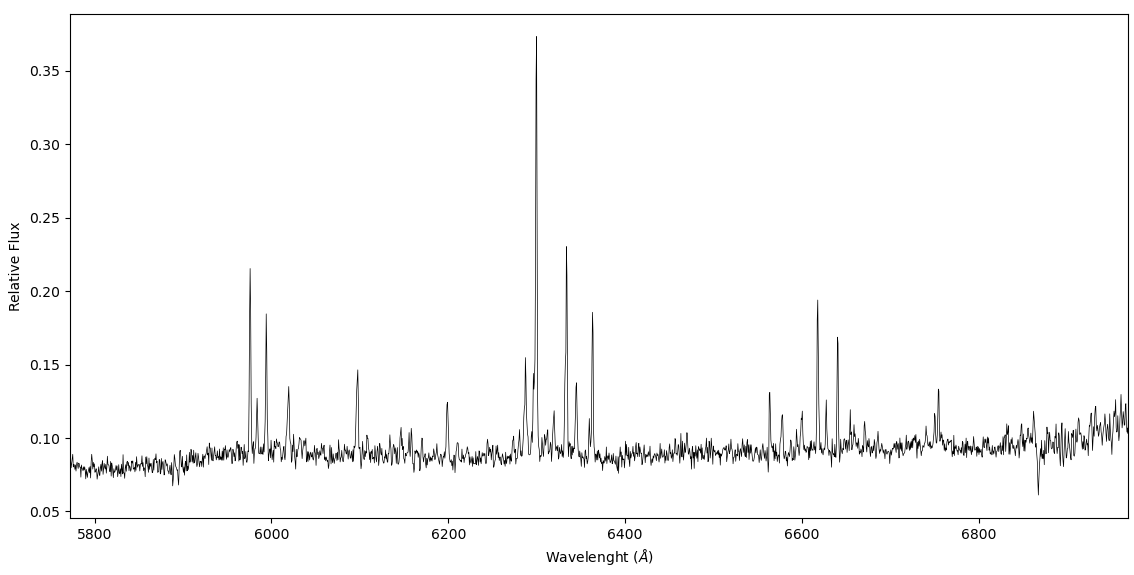}
    \caption{Spectrum of 2022-01-21; configuration C}

\end{figure}

\newpage
\clearpage

\section{21P (Giacobini-Zinner)}
\label{cometa:21P}
\subsection{Description}

21P/Giacobini-Zinner is a Jupiter comet with a period of 6.55 years and an absolute magnitude of 13.3$\pm$0.8.\footnote{\url{https://ssd.jpl.nasa.gov/tools/sbdb_lookup.html\#/?sstr=21P} visited on July 21, 2024} 
It was first discovered by Michael Giacobini from the Nice Observatory on December 20, 1900. It was then rediscovered on October 23, 1913 by Ernst Zinner from the Bamberg Observatory.
The observation was confirmed and the comet was subsequently identified on a photograph taken on November 5, 1942. 
During its apparitions, comet Giacobini–Zinner can reach magnitude 7-8, as during our 2018 observations. In 1946, it underwent a series of outbursts that made it as bright as magnitude 5.
It is the parent body of the Giacobinids meteor shower (also known as the Draconids). 
The comet nucleus is estimated to be 2.0 km in diameter. The comet currently has a minimum orbit intersection distance to Earth of 0.018 AU.

\noindent
We observed the comet around magnitude 7.\footnote{\url{https://cobs.si/comet/403/ }, visited on July 21, 2024}

\begin{table}[h!]
\centering
\begin{tabular}{|c|c|c|}
\hline
\multicolumn{3}{|c|}{Orbital elements (epoch: November 8, 2017)}                      \\ \hline \hline
\textit{e} = 0.7105   &   \textit{q} = 1.0135  &   \textit{T} = 2458371.7657 \\ \hline
$\Omega$ = 195.4050 &   $\omega$ = 172.8102 &    \textit{i} = 32.0030 \\ \hline  
\end{tabular}
\end{table}

\begin{table}[h!]
\centering
\begin{tabular}{|c|c|c|c|c|c|c|c|c|}
\hline
\multicolumn{9}{|c|}{Comet ephemerides for key dates}                      \\ \hline 
\hline
& date & r & $\Delta$ & RA & DEC & elong & phase & PLang \\
& (yyyy-mm-dd) & (AU) & (AU) & (h) & (°) & (°) & (°) & (°) \\ \hline
Perihelion & 2018-09-10 & 1.013 & 0.392 & 05.81 & $+$33.70 & 79.6 & 78.0 & $-$40.3 \\
Nearest approach & 2018-09-11 & 1.013 & 0.392 & 05.88 & $+$31.97 & 79.7 & 78.0 & $-$38.8 \\
\hline
\end{tabular}
\end{table}

\vspace{0.5 cm}

\begin{figure}[h!]
    \centering
    \includegraphics[scale=0.38]{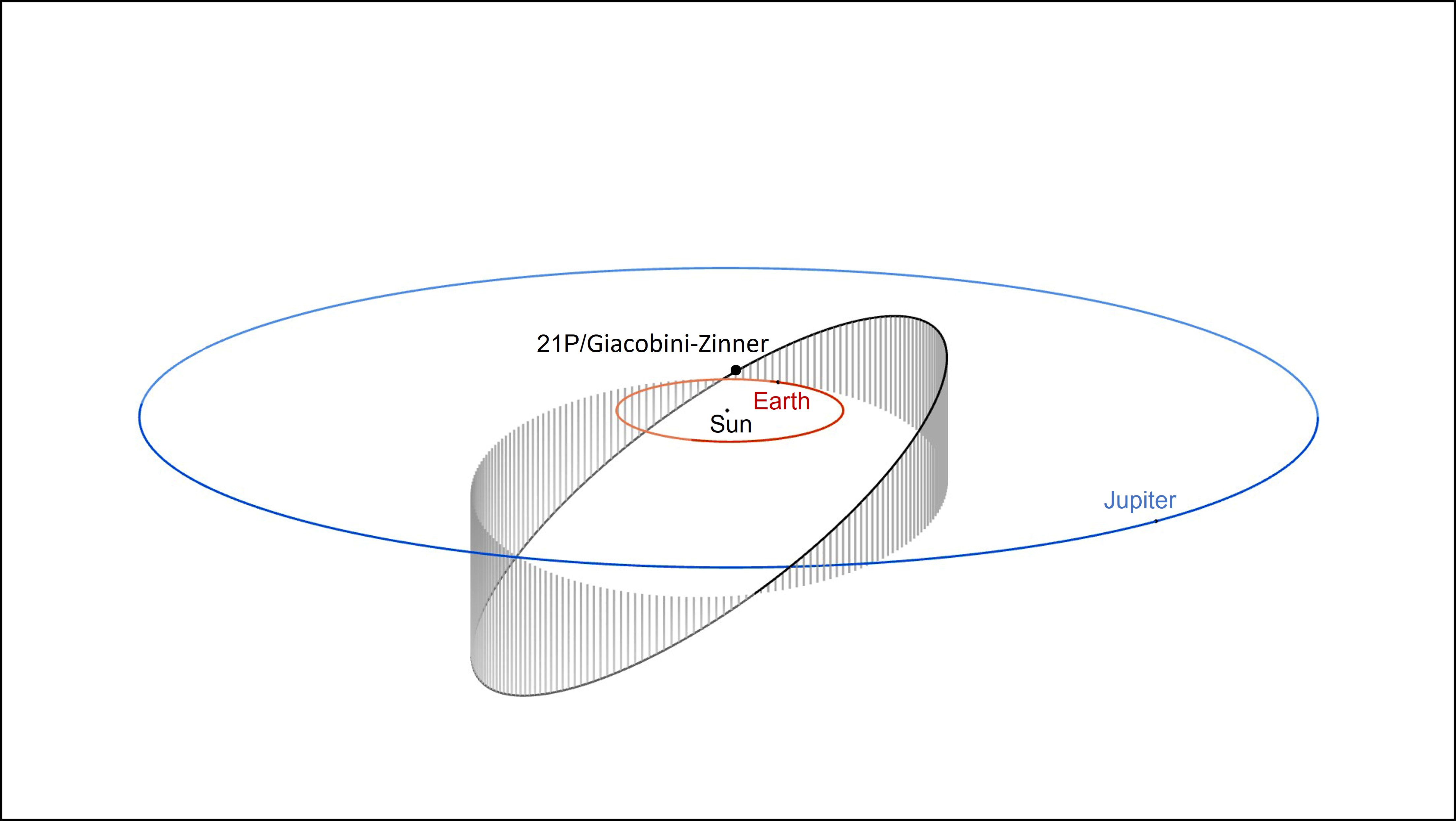}
    \caption{Orbit of comet 21P and position on perihelion date. The field of view is set to the orbit of Jupiter for size comparison. Courtesy of NASA/JPL-Caltech.}
\end{figure} 

\newpage

\subsection{Images}

\begin{SCfigure}[0.8][h!]
    \centering
    \includegraphics[scale=0.4]{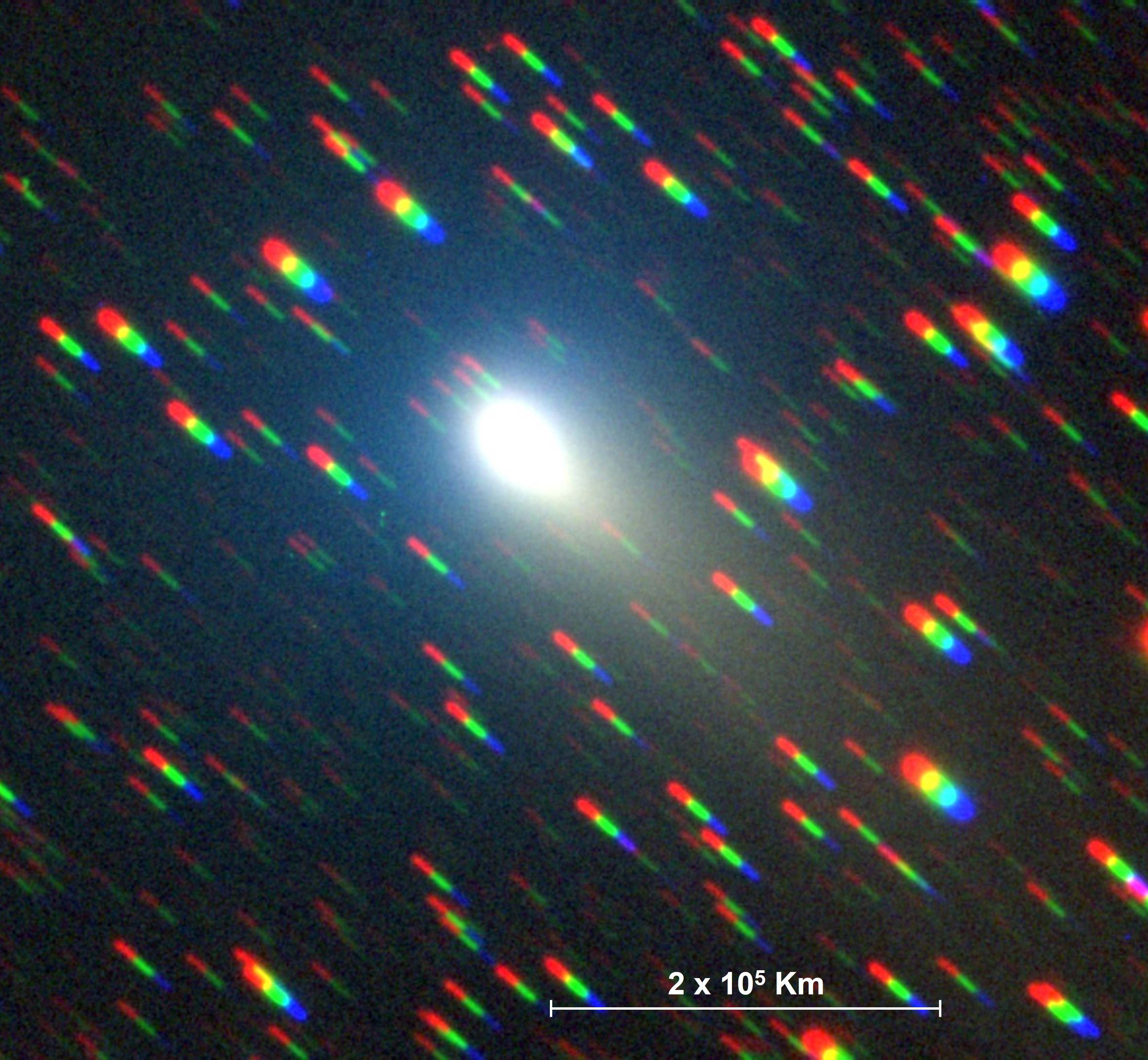}
    \caption{2018-07-17. Image taken with the Asiago Schmidt telescope with B, g, r filters.
Comet Giacobini-Zinner moves fast in a very star-rich field in the constellation Cygnus. The dust tail, which is reddened by the reflection of sunlight, is directed South-West, whereas the blue-green coma develops in a sphere around the nucleus.}
\end{SCfigure} 

\begin{SCfigure}[0.8][h!]
    \centering
    \includegraphics[scale=0.4]{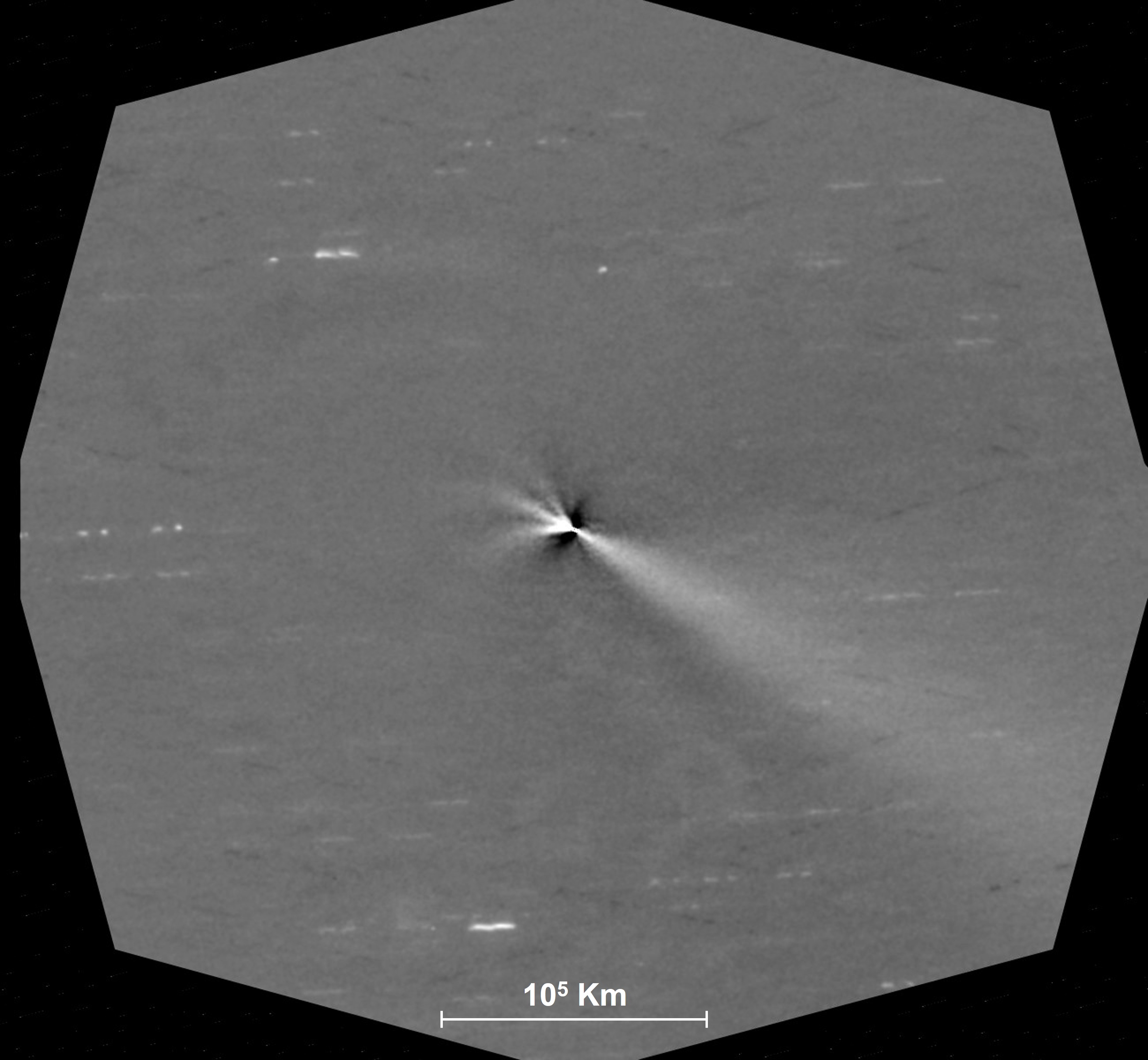}
    \caption{2018-08-08. A mathematical filter was applied to the original image, to highlight the morphology of the inner coma dominated by jet structures directed towards the Sun. These derive from small active areas placed on the nucleus. The tail develops in an anti-solar direction. Image taken with the Asiago Schmidt telescope.}
\end{SCfigure}

\newpage

\subsection{Spectra}

\begin{table}[h!]
\centering
\begin{tabular}{|c|c|c|c|c|c|c|c|c|c|c|c|}
\hline
\multicolumn{12}{|c|}{Observation details}                      \\ \hline 
\hline
$\#$  & date          & r     & $\Delta$ & RA     & DEC     & elong & phase & PLang& config  &  FlAng & N \\
      & (yyyy-mm-dd)  &  (AU) & (AU)     & (h)    & (°)     & (°)   & (°)   &  (°)   &       &  (°)  &  \\ \hline 

1* & 2018-08-28 & 1.028 & 0.417 & 04.70 & $+$52.50 & 80.6 & 75.8 & $-$55.4 & A & $+$0 & 5 \\
2* & 2018-08-29 & 1.025 & 0.414 & 04.82 & $+$51.15 & 80.4 & 76.1 & $-$54.5 & A & $+$0 & 1 \\

\hline
\end{tabular}
\end{table}

\begin{figure}[h!]

    \centering
    \includegraphics[scale=0.368]{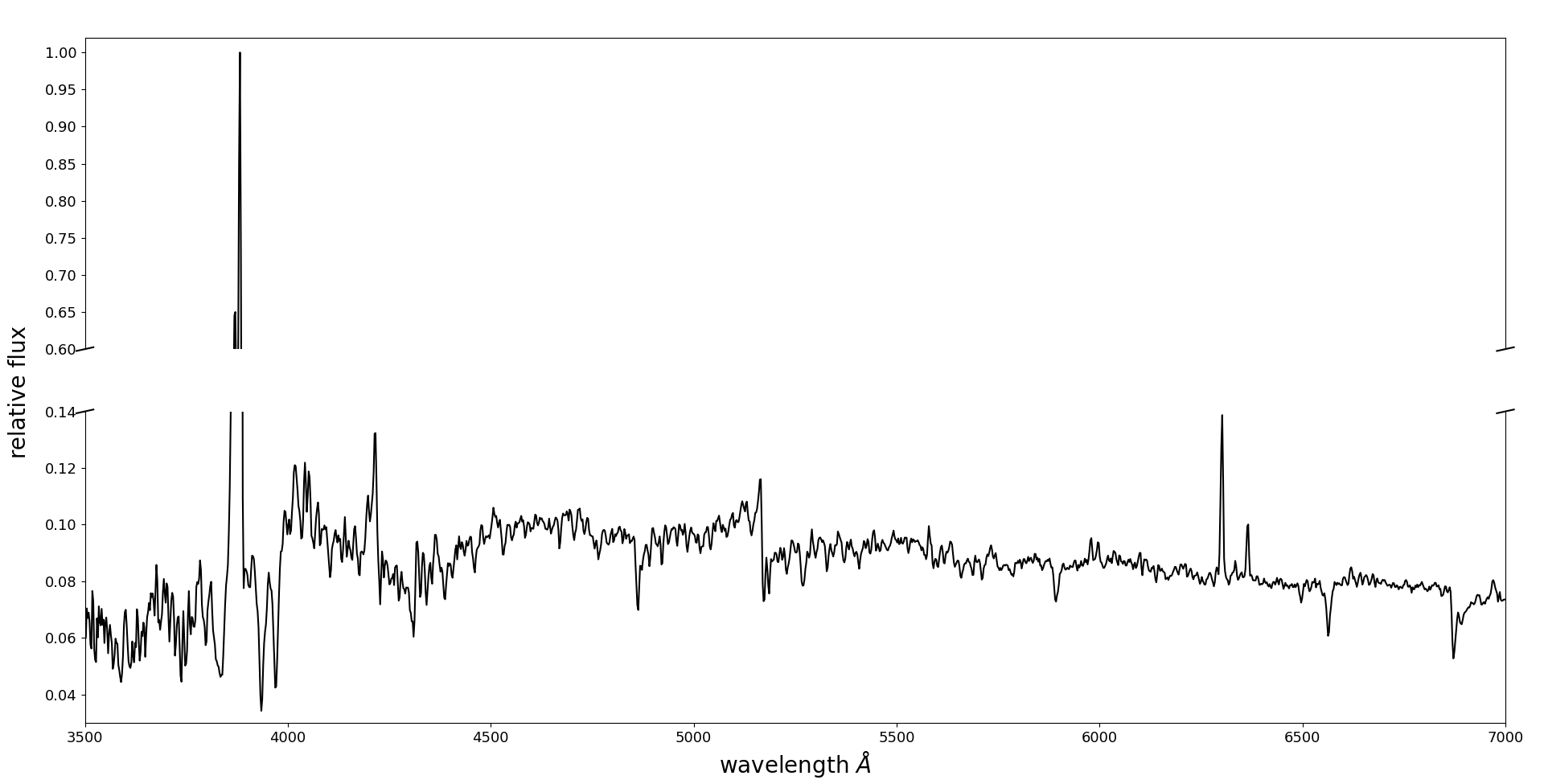}
    \caption{Spectrum of 2018-08-28; configuration A}

\end{figure}

\footnotetext{* No Solar Analog}

\newpage
\clearpage

\section{29P (Schwassmann-Wachmann 1)}
\label{cometa:29P}
\subsection{Description}

29P/Schwassmamm-Wachmann is a Jupiter-family comet with a period of 14.87 years and an absolute magnitude of 10.1$\pm$0.3.\footnote{\url{https://ssd.jpl.nasa.gov/tools/sbdb_lookup.html\#/?sstr=29P}, visited on July 21, 2024} It was first discovered by Friedrich Karl Arnold Schwassmann and Arno Arthur Wachmann from the Hamburg Observatory on November 15, 1927. 
It was actually in 1902 that Karl Wilhelm Reinmuth photographed the comet on March 4, 1902. 
It was Leland Ernest Cunningham who only realized in 1931 that it was the Jupiter-family comet 29P, officially discovered only much later than the first date. 
The comet can, in principle, always be observed because of the low eccentricity of its orbit. 
Comet 29P is known for its numerous bursts of brightness, so it is always worth a look.

\noindent
We observed the comet around magnitude 10.\footnote{\url{https://cobs.si/comet/57/ }, visited on July 21, 2024}

\begin{table}[h!]
\centering
\begin{tabular}{|c|c|c|}
\hline
\multicolumn{3}{|c|}{Orbital elements (epoch: January 1, 2023)}                      \\ \hline \hline
\textit{e} = 0.0448 & \textit{q} = 5.7765 & \textit{T} = 2458593.2922 \\ \hline
$\Omega$ = 312.3941 & $\omega$ =50.9133  & \textit{i} = 9.3641 \\ \hline  
\end{tabular}
\end{table}

\begin{table}[h!]
\centering
\begin{tabular}{|c|c|c|c|c|c|c|c|c|}
\hline
\multicolumn{9}{|c|}{Comet ephemerides for key dates}                      \\ \hline 
\hline
& date         & r    & $\Delta$  & RA      & DEC      & elong  & phase  & PLang  \\
& (yyyy-mm-dd) & (AU) & (AU)      & (h)     & (°)      & (°)    & (°)    & (°) \\ \hline 

Perihelion       & 2019-04-14 & 5.767  & 6.706 & 00.17  & $+$07.91  &  18.9 & 03.2  & $+$01.3  \\ 
Nearest approach & 2019-10-09 & 5.776  & 4.789 & 00.71 & $+$15.58 & 169.8  & 01.8  & $-$01.7 \\ \hline
\end{tabular}

\end{table}

 \vspace{0.5 cm}

\begin{figure}[h!]
    \centering
    \includegraphics[scale=0.38]{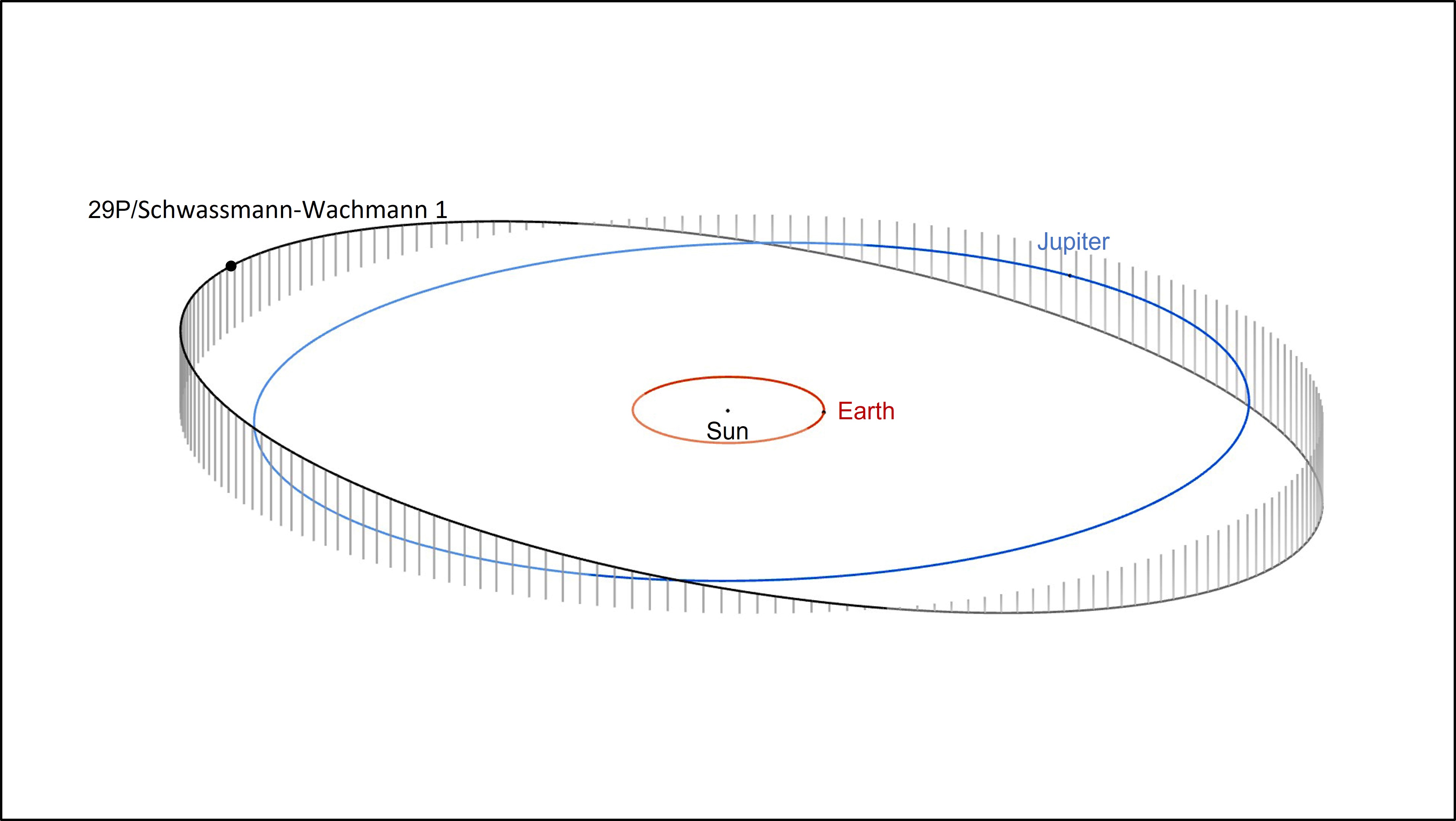}
    \caption{Orbit of comet 29P and position on perihelion date. The field of view is set to the orbit of Jupiter for size comparison. Courtesy of NASA/JPL-Caltech.}
\end{figure}

\newpage

\subsection{Images}

\begin{SCfigure}[0.8][h!]
    \centering
    \includegraphics[scale=0.4]{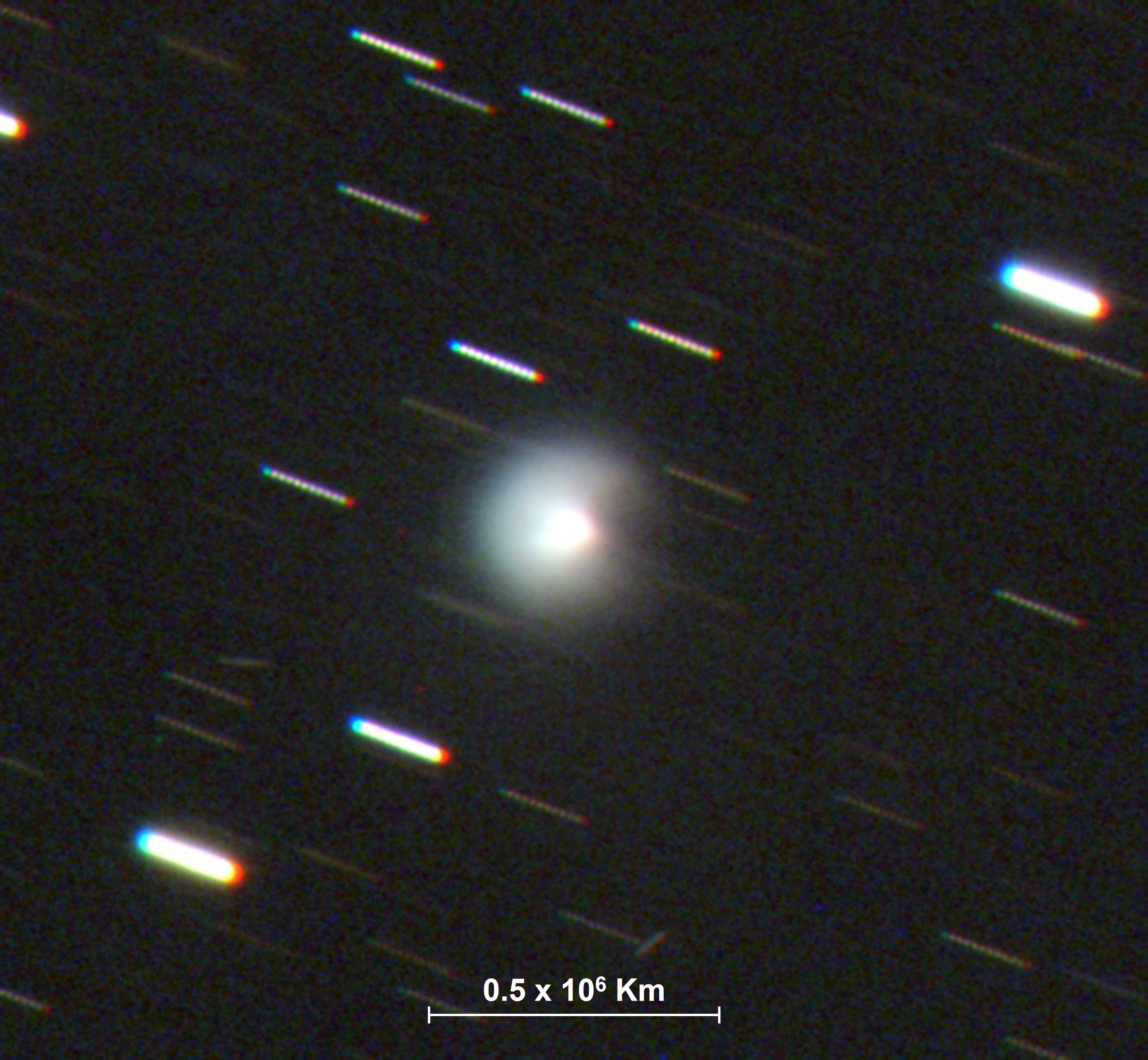}
    \caption{2018-10-03. A three-color Bgr composite was made with images taken with the Asiago Schmidt telescope. The faint traces of two asteroids are also visible in the image: (182342) 2001 QA7 to the left of the comet, (354505) 2004 PX37 to the south.}
\end{SCfigure} 

\begin{SCfigure}[0.8][h!]
    \centering
    \includegraphics[scale=0.4]{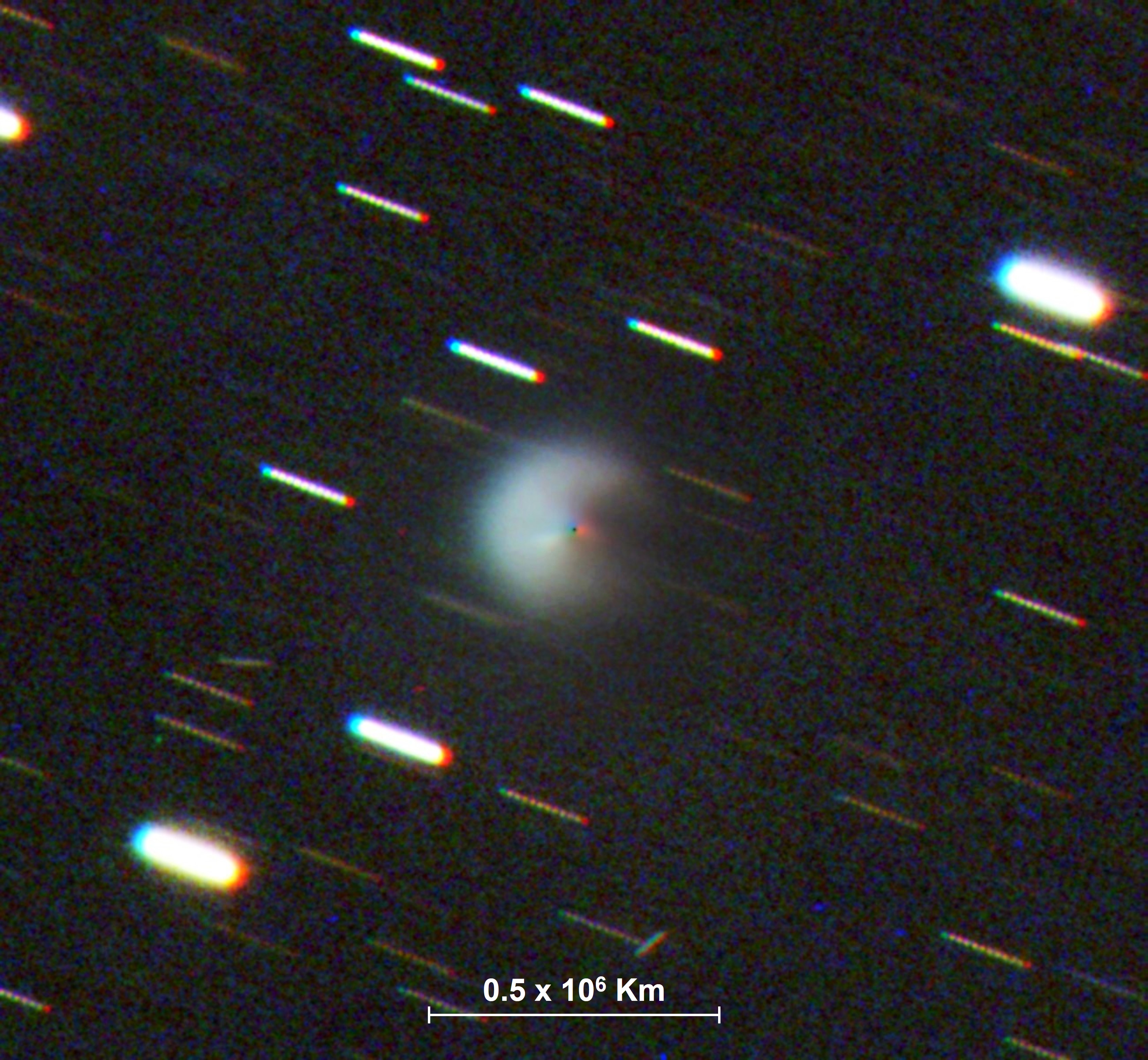}
    \caption{2018-10-03. An algorithm was applied to the original image to subtract the high brightness of the areas close to the nucleus and to show the morphology of the inner coma.
    Several jet structures become visible when the comet encounters an outburst.}
\end{SCfigure}

\newpage

\subsection{Spectra}

\begin{table}[h!]
\centering
\begin{tabular}{|c|c|c|c|c|c|c|c|c|c|c|c|}
\hline
\multicolumn{12}{|c|}{Observation details}                      \\ \hline 
\hline
$\#$  & date          & r     & $\Delta$ & RA     & DEC     & elong & phase & PLang& config  &  FlAng & N \\
      & (yyyy-mm-dd)  &  (AU) & (AU)     & (h)    & (°)     & (°)   & (°)   &  (°)   &       &  (°)  & \\ \hline 

1 & 2018-09-30 & 5.772 & 4.836  & 22.88 & $+$00.10 & 157.1 & 03.15  & $-$1.6   & A & $+$23 & 3 \\
2 & 2022-11-26 & 6.047 & 5.237  & 06.92 & $+$29.30 & 142.1 & 05.80  & $-$1.6   & A & $+$10 & 7 \\

\hline
\end{tabular}
\end{table}

\begin{figure}[h!]

    \centering
    \includegraphics[scale=0.368]{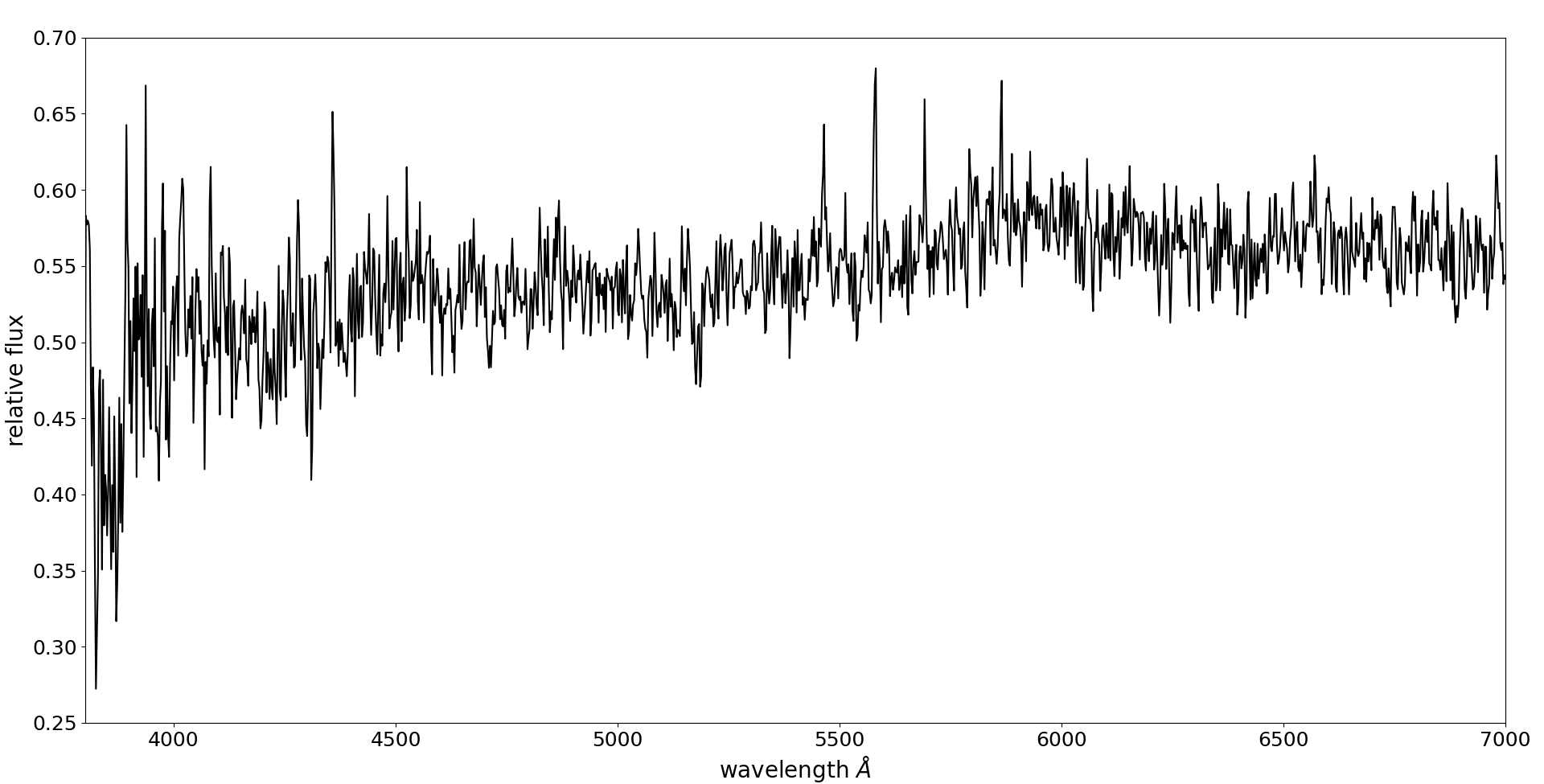}
    \caption{Spectrum of 2018-09-30; configuration A}

\end{figure}

\newpage
\clearpage

\section{38P (Stephan-Oterma)}
\label{cometa:38P}
\subsection{Description}

38P/Stephan-Oterma is a Halley-type comet with a period of 37.93 years and an absolute magnitude of 11.5$\pm$1.0.\footnote{\url{https://ssd.jpl.nasa.gov/tools/sbdb_lookup.html\#/?sstr=38P} visited on July 21, 2024}
It was first discovered by Édouard Stephan from the Marseille Observatory on January 24, 1867. It was then rediscovered on November 6, 1942 by Liisi Oterma from the Iso-Heikkilä Observatory.
\noindent
This comet was actually first observed on January 22, 1867 by Jérôme Eugène Coggia, who initially thought it was a nebula that had not yet been recorded.
Two days later Édouard Stephan checked the data and realized that the nebula had moved and announced the discovery of the comet. 
The first orbital elements were calculated by L. Oterma.
Fred Whipple stated that the orbital elements corresponded to those of Comet 1867 I (Stephan).
So the comet was named Stephan-Oterma.
Earth crossed the comet orbital plane on December 10, 2018.

\noindent
We observed the comet between magnitude 10 and 11.\footnote{\url{https://cobs.si/comet/1024/ }, visited on July 21, 2024}

\begin{table}[h!]
\centering
\begin{tabular}{|c|c|c|}
\hline
\multicolumn{3}{|c|}{Orbital elements (epoch: May 24, 2018)}                      \\ \hline \hline
\textit{e} = 0.8593 & \textit{q} = 1.5885 & \textit{T} = 2458433.4873 \\ \hline
$\Omega$ = 78.0004 & $\omega$ = 359.5887  & \textit{i} = 18.3529  \\ \hline  
\end{tabular}
\end{table}

\begin{table}[h!]
\centering
\begin{tabular}{|c|c|c|c|c|c|c|c|c|}
\hline
\multicolumn{9}{|c|}{Comet ephemerides for key dates}                      \\ \hline 
\hline
& date        & r & $\Delta$ & RA     & DEC    & elong & phase & PLang \\
& (yyyy-mm-dd) & (AU) & (AU)      & (h)     & (°)      & (°)    & (°)    & (°) \\ \hline 
Perihelion       & 2018-11-11 & 1.588 & 0.873 & 07.53 &	$+$21.52 & 116.8 & 33.8 & $+$10.2 \\ 
Nearest approach & 2018-12-17 & 1.652 & 0.766 & 08.47 &	$+$35.05 & 141.2 & 21.9 & $-$2.7 \\ \hline
\end{tabular}

\end{table}

\vspace{0.5 cm}

\begin{figure}[h!]
    \centering
    \includegraphics[scale=0.38]{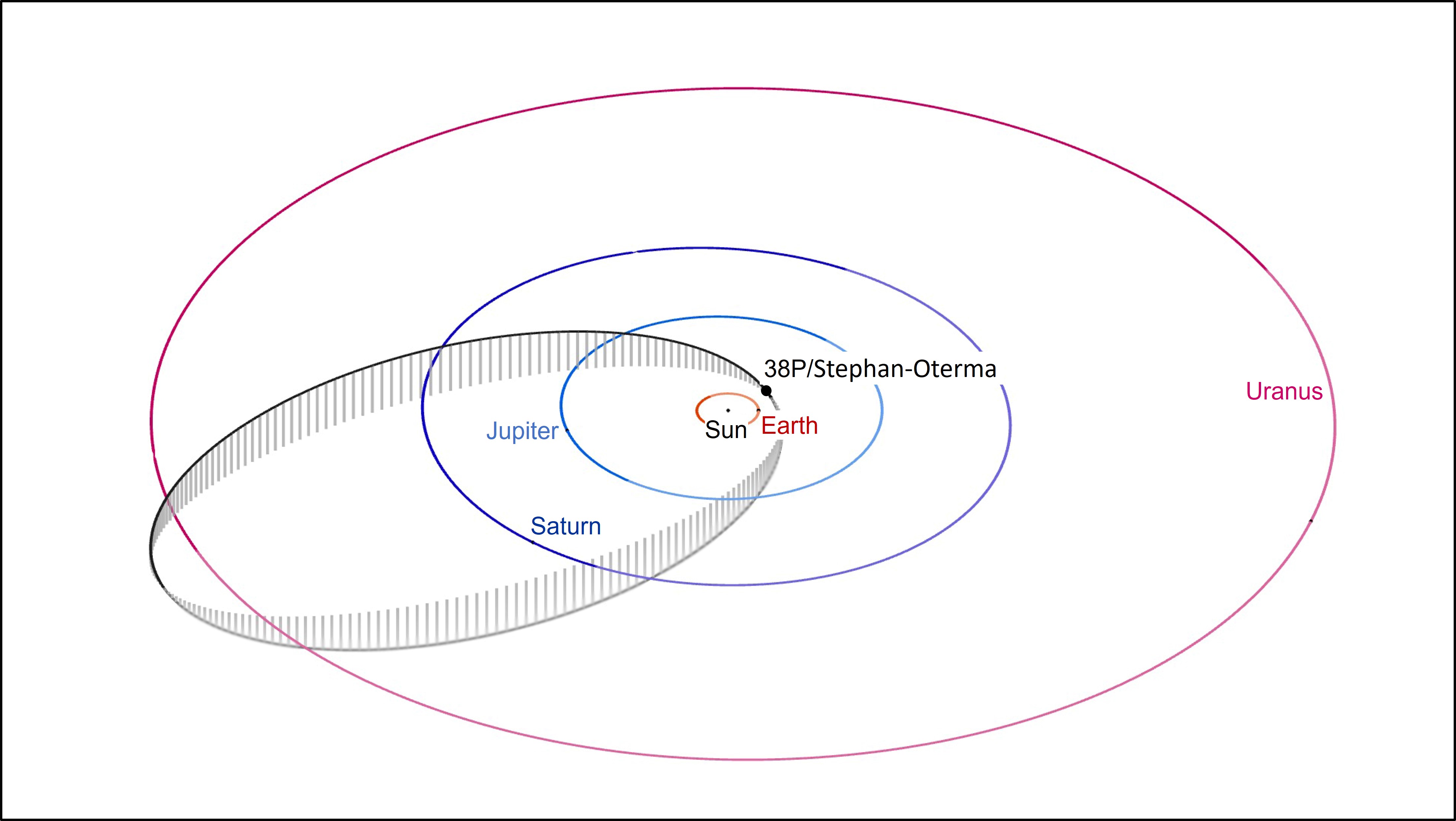}
    \caption{Orbit of comet 38P/Stephan-Oterma and position on perihelion date. The field of view is set to the orbit of Uranus for size comparison. Courtesy of NASA/JPL-Caltech.}
\end{figure} 

\newpage

\subsection{Images}

\begin{SCfigure}[0.8][h!]
    \centering
    \includegraphics[scale=0.4]{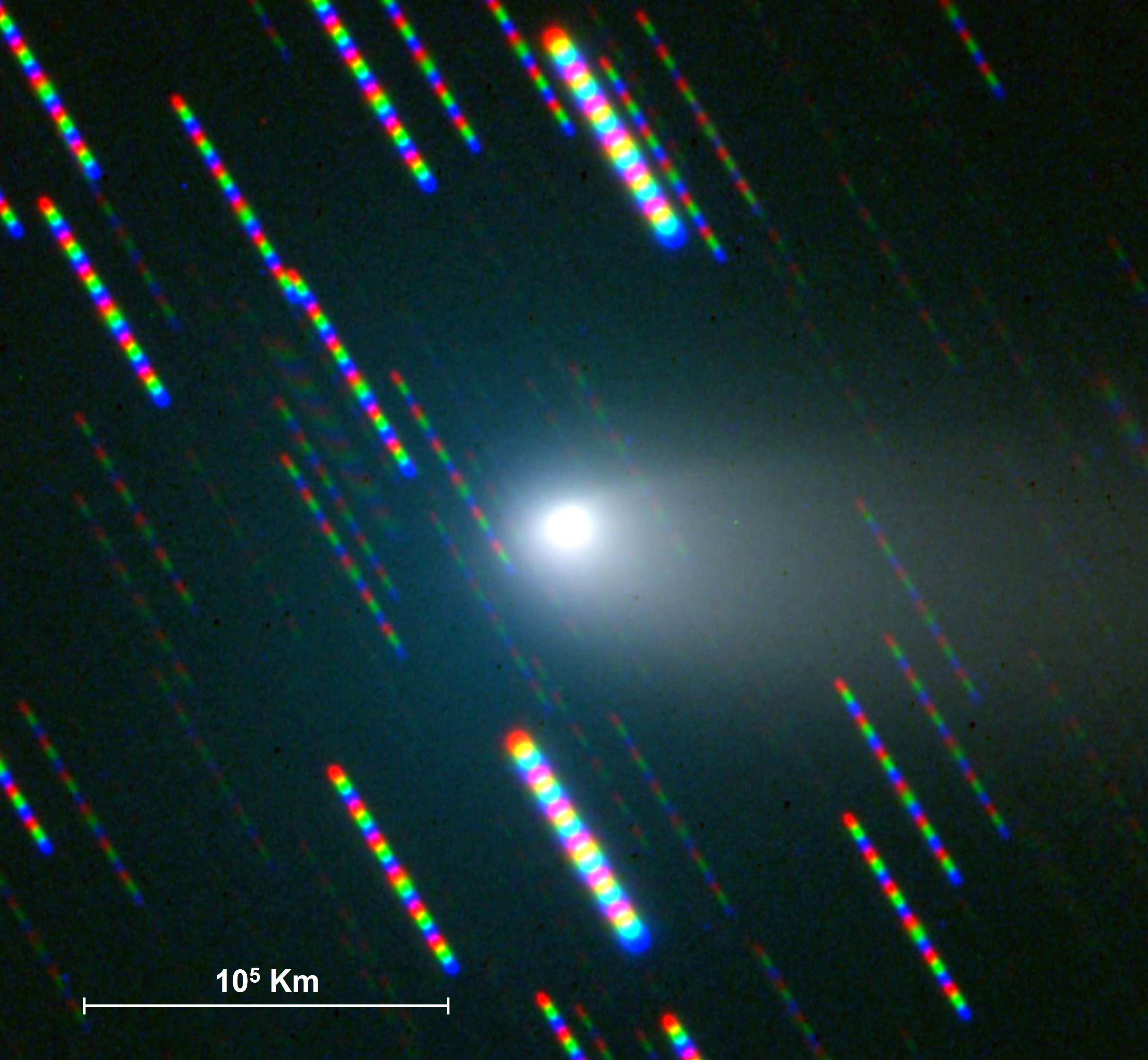}
     \caption{2018-12-08. Three-color (Bgr) composite image taken with the Asiago Schmidt telescope.}
\end{SCfigure} 

\begin{SCfigure}[0.8][h!]
    \centering

    \includegraphics[scale=0.4]{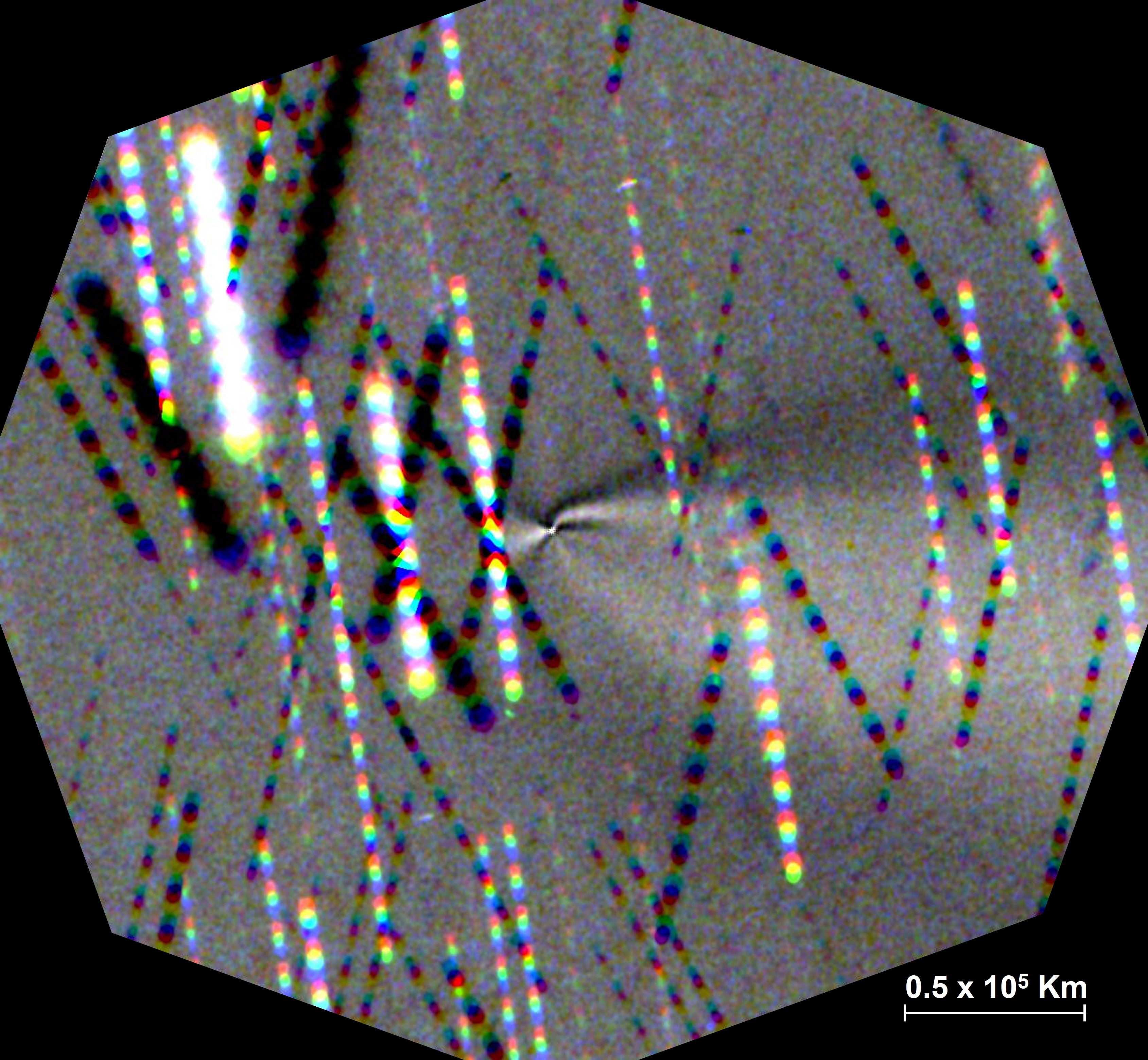}

    \caption{2018-12-27. Image taken with the Asiago Schmidt telescope, processed with a Larson-Sekanina filter ($\alpha=20$\textdegree) to show the morphology of the inner coma, dominated by jet-shaped structures.}
\end{SCfigure}

\newpage

\subsection{Spectra}

\begin{table}[h!]
\centering
\begin{tabular}{|c|c|c|c|c|c|c|c|c|c|c|c|}
\hline
\multicolumn{12}{|c|}{Observation details}                      \\ \hline 
\hline
$\#$  & date          & r     & $\Delta$ & RA     & DEC     & elong & phase & PLang& config  &  FlAng & N \\
      & (yyyy-mm-dd)  &  (AU) & (AU)     & (h)    & (°)     & (°)   & (°)   &  (°)   &       &  (°)  & \\ \hline 

1 & 2018-12-14 & 1.645 & 0.766 & 08.43 & $+$34.25 & 139.8 & 22.7 & $-$01.9  & A & $+$0 & 3 \\
2 & 2018-12-15 & 1.648 & 0.766 & 08.45 & $+$34.65 & 140.5	& 22.3 & $-$02.3  & A & $+$0 & 1 \\
3 & 2018-12-17 & 1.656 & 0.766 & 08.48 & $+$35.45 & 141.9	& 21.5 & $-$03.1  & A & $+$0 & 3 \\
4 & 2018-12-26 & 1.690 & 0.775 & 08.57 & $+$38.98 & 147.8	& 18.0 & $-$06.7  & A & $+$0 & 1 \\
5 & 2019-01-03 & 1.727 & 0.796 & 08.58 & $+$41.47 & 152.0	& 15.5 & $-$09.5  & A & $+$0 & 1 \\
6 & 2019-01-05 & 1.737 & 0.803 & 08.60 & $+$42.38 & 152.7	& 15.0 & $-$10.1 & A & $+$0 & 2 \\
7 & 2019-01-29 & 1.873 & 0.951 & 08.48 & $+$47.02 & 150.6	& 14.9 & $-$14.8 & A & $+$0 & 3 \\
\hline
\end{tabular}
\end{table}

\begin{figure}[h!]

    \centering
    \includegraphics[scale=0.368]{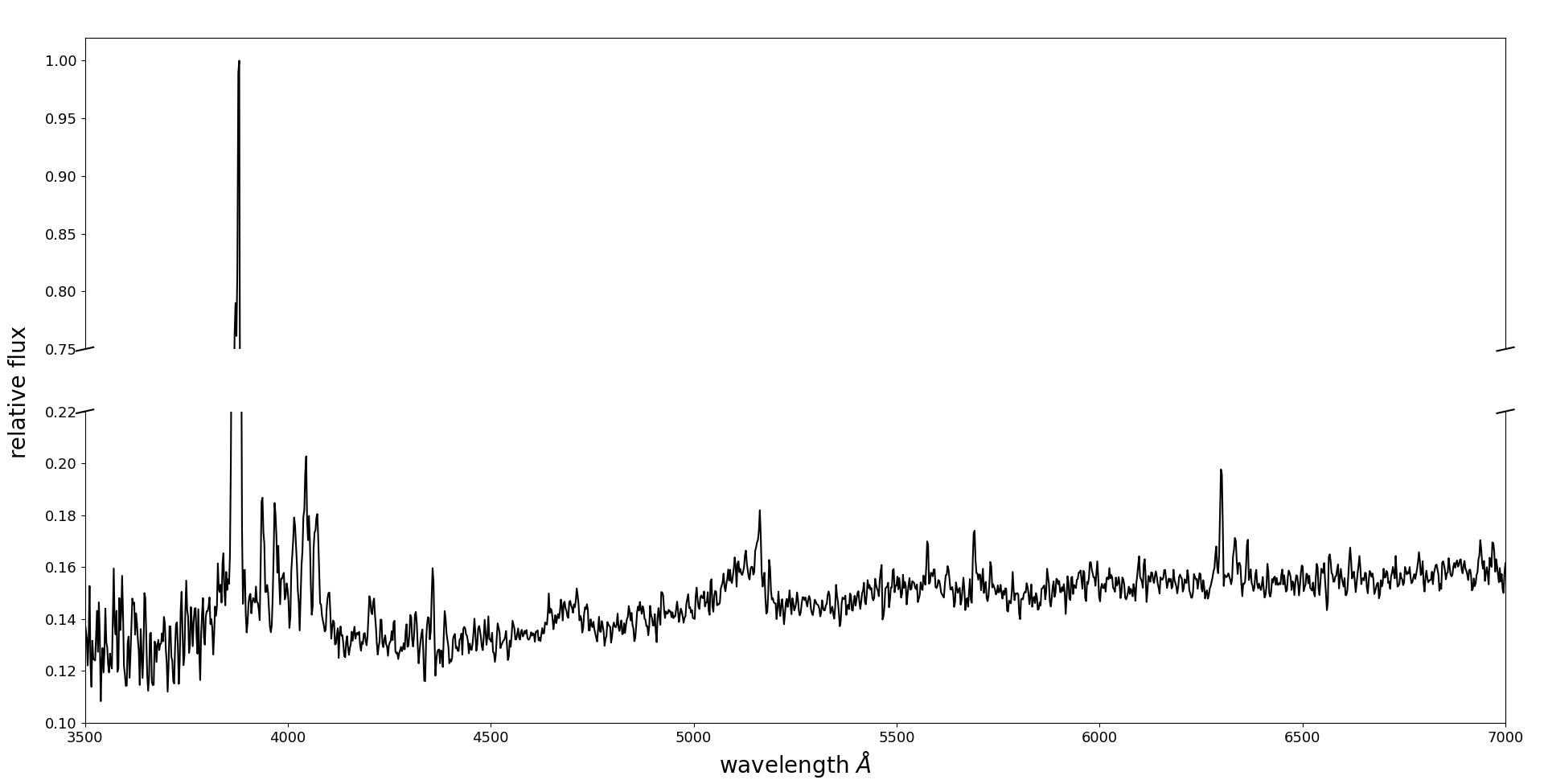}
    \caption{Spectrum of 2018-12-17; configuration A}

\end{figure}

\newpage
\clearpage

\section{46P (Wirtanen)}
\label{cometa:46P}
\subsection{Description}

46P/Wirtanen is a Jupiter-family comet with a period of 5.44 years and an absolute magnitude of 16.6$\pm$0.8.\footnote{\url{https://ssd.jpl.nasa.gov/tools/sbdb_lookup.html\#/?sstr=46P} visited on July 21, 2024}
It was discovered by Carl Alvar Wirtanen on January 17, 1948 from the Lick Observatory. It was then rediscovered on September 8, 1954 by C. A. Wirtanen himself again from the Lick Observatory.
It was initially named 1948b and was finally confirmed on September 26, 1954 by Edgar Roemer and Harold Jeffers, when it was given the provisional designation 1954j. Earth crossed the comet orbital plane on December 14, 2018.

\noindent
We observed the comet between magnitude 7 and 5.\footnote{\url{https://cobs.si/comet/332/ }, visited on July 21, 2024}

\begin{table}[h!]
\centering
\begin{tabular}{|c|c|c|}
\hline
\multicolumn{3}{|c|}{Orbital elements (epoch: December 13, 2018)}                      \\ \hline \hline
\textit{e} = 0.6588 & \textit{q} = 1.0554 & \textit{T} = 2458465.4312 \\ \hline
$\Omega$ = 82.1576 & $\omega$ = 356.3411  & \textit{i} = 11.7475  \\ \hline  
\end{tabular}
\end{table}

\begin{table}[h!]
\centering
\begin{tabular}{|c|c|c|c|c|c|c|c|c|}
\hline
\multicolumn{9}{|c|}{Comet ephemerides for key dates}                      \\ \hline 

& date & r & $\Delta$ & RA & DEC & elong & phase & PLang \\
& (yyyy-mm-dd) & (AU) & (AU) & (h) & (°) & (°) & (°) & (°) \\ \hline

Perihelion       & 2018-12-15 & 1.056  & 0.077 & 03.79 & $+$16.48 & 155.4 & 22.8  & $-$1.4 \\
Nearest approach & 2018-12-16 & 1.057  & 0.075 & 04.20 & $+$27.42 & 159.6  & 19.9  & $-$5.4 \\ \hline
\end{tabular}

\end{table}

\vspace{0.5 cm}

\begin{figure}[h!]
    \centering
    \includegraphics[scale=0.38]{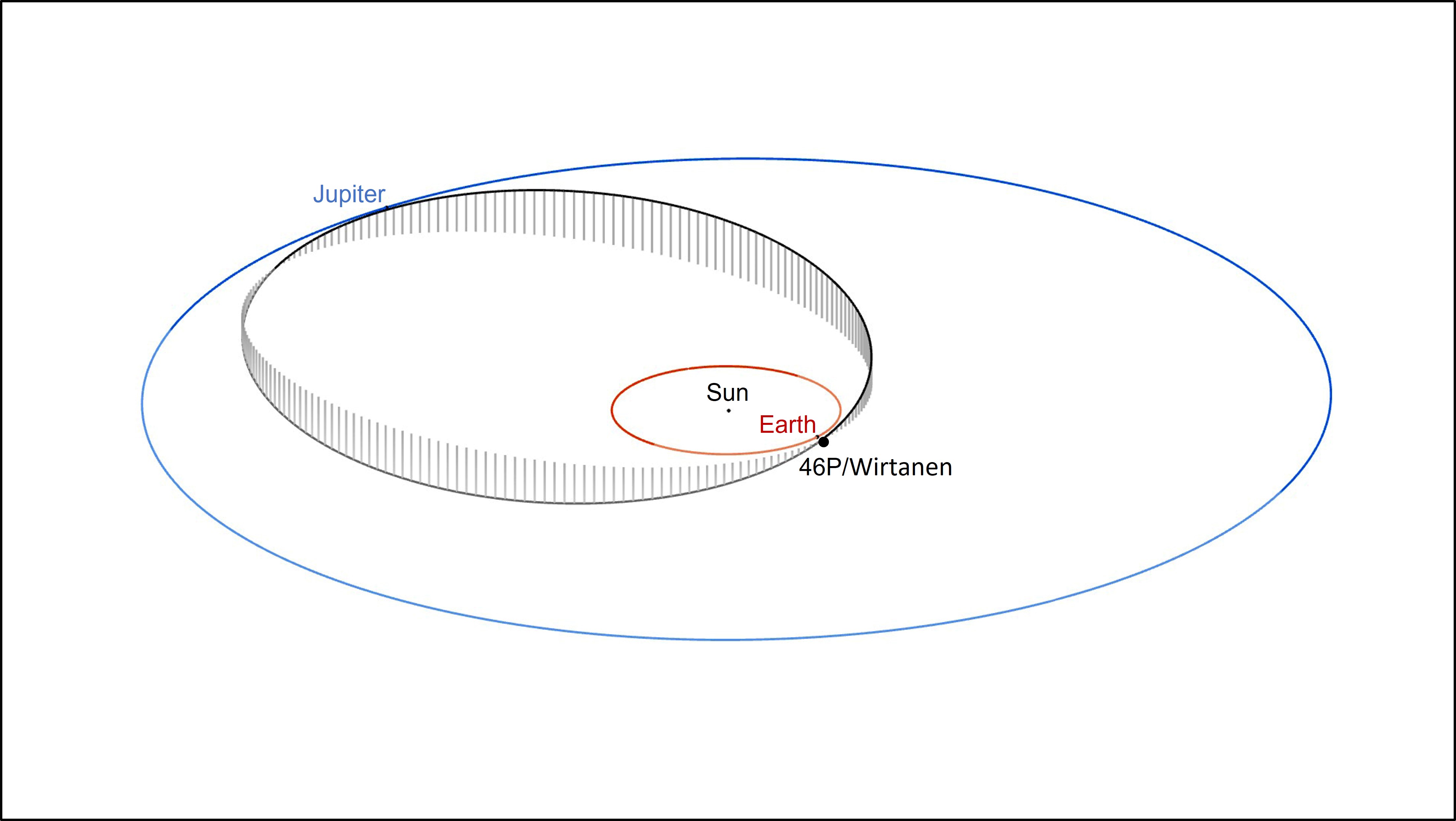}
    \caption{Orbit of comet 46P and position on perihelion date. The field of view is set to the orbit of Jupiter for size comparison. Courtesy of NASA/JPL-Caltech.}
\end{figure} 

\newpage

\subsection{Images}

\begin{SCfigure}[0.8][h!]

 \centering

 \includegraphics[scale=0.4]{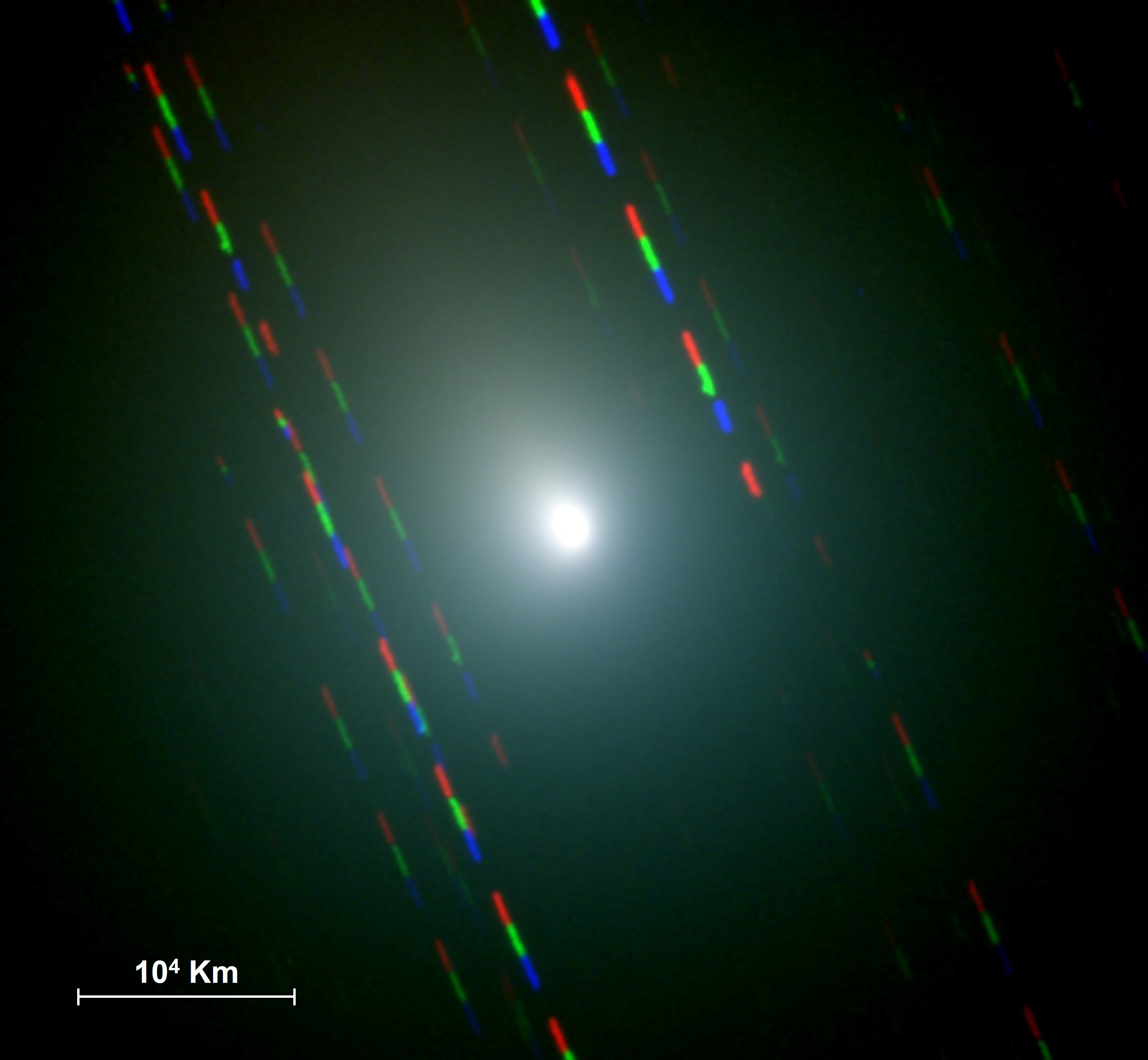}

 \caption{2018-11-28. Three-color (Bgr) composite image taken with the Asiago Copernico telescope. The blue-green coma is mainly due to the fluorescence of the C$_2$ and C$_3$ molecules.}

\end{SCfigure} 

\begin{SCfigure}[0.8][h!]

 \centering

 \includegraphics[scale=0.4]{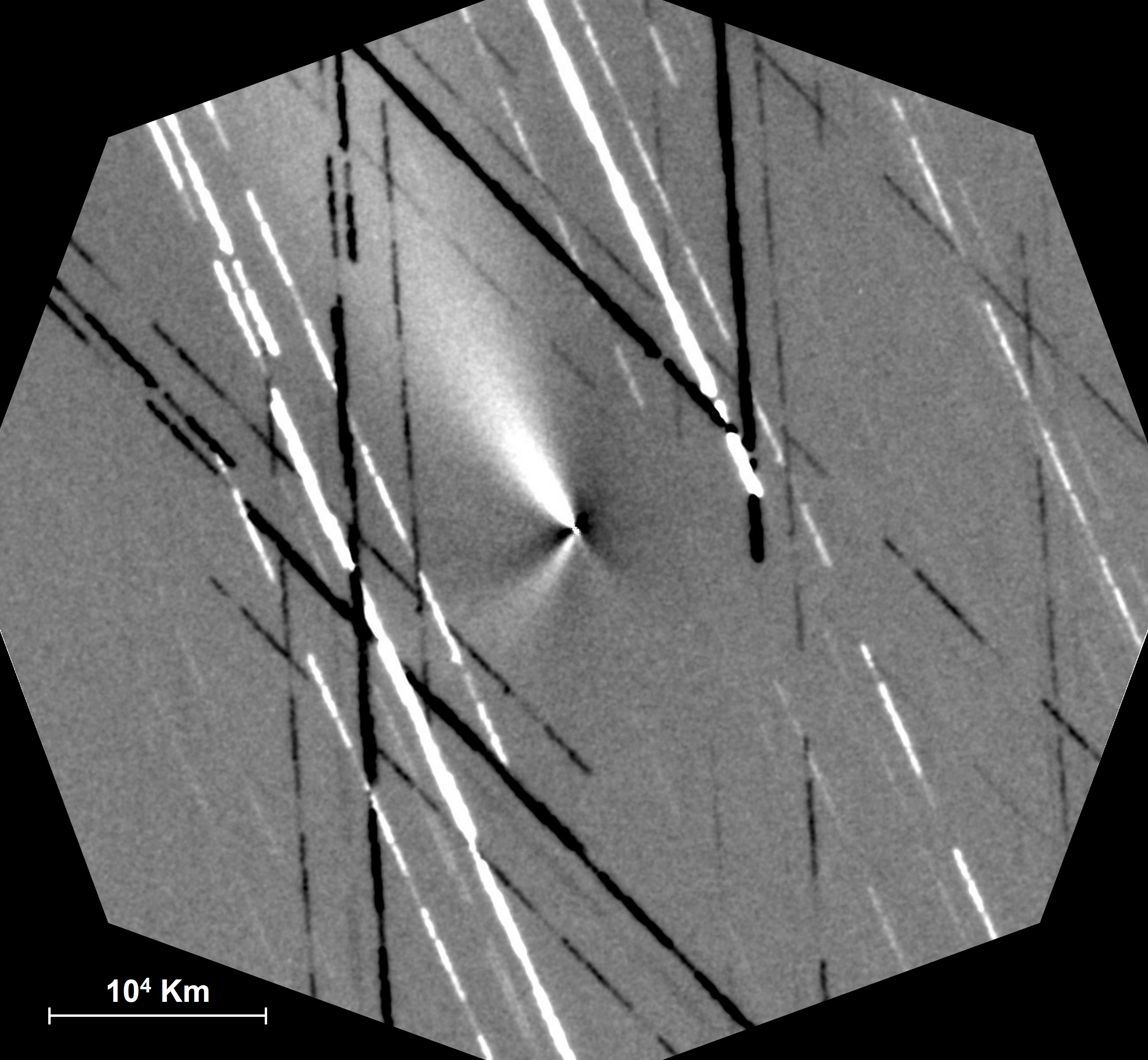}
 
 \caption{2018-11-28. Image taken with the Asiago Schmidt telescope, processed with a Larson-Sekanina filter ($\alpha=20$\textdegree) to show the morphology of the inner coma, dominated by jet-shaped structures.}

\end{SCfigure}

\newpage

\subsection{Spectra}

\begin{table}[h!]
\centering
\begin{tabular}{|c|c|c|c|c|c|c|c|c|c|c|c|}
\hline
\multicolumn{12}{|c|}{Observation details}                      \\ \hline 
\hline
$\#$ & date & r & $\Delta$ & RA & DEC & elong & phase & PLang& config & FlAng & N \\
 & (yyyy-mm-dd) & (AU) & (AU) & (h) & (°) & (°) & (°) & (°) & & (°) & \\ \hline 
 
1 & 2018-11-28 &	1.072 &	0.129 &	02.47  &	$-$22.35  &	  128.7 & 45.9 &	$+$24.8 & A & $+$90 & 6 \\ 
2 & 2018-12-01 &    1.066 & 0.116 & 02.62  &    $-$18.33  &   131.0 & 44.2 &    $+$22.3 & A & +77 & 3 \\ 
3* & 2018-12-08 &    1.057 & 0.089 & 03.01  &    $-$03.45  &  141.6 & 35.4 &    $+$12.5 & A & +75 & 3 \\ 
4 & 2018-12-15 &    1.056 & 0.078 & 03.93  &    $+$20.23  &   157.1 & 21.3 &    $-$03.6 & A & $-$35 & 2 \\ 
5 & 2018-12-17 &	1.058 &	0.078 & 04.02  &    $+$27.41  &   160.0 & 18.6 &	$-$09.1 & A & $+$32 & 1 \\ 
6 & 2018-12-26 &	1.072 & 0.100 & 06.03  &    $+$51.88  &   151.5 & 26.0 &	$-$26.0 & A & $-$80 & 1 \\ 
7 & 2019-01-03 &	1.097 & 0.137 & 07.58  &    $+$58.77  &   143.4 & 32.3 &	$-$31.3 & A & $+$0 & 1 \\ 
8 & 2019-01-05 &	1.104 & 0.147 & 07.88  &    $+$59.27  &   142.4 & 32.9 &	$-$31.8 & A & $-$71 & 1 \\ 
9 & 2019-01-06 &	1.107 & 0.152 & 08.02  &    $+$59.04  &   142.0 & 33.1 &	$-$32.0 & A & $-$71 & 2 \\ \hline
\end{tabular}
\end{table}

\begin{figure}[h!]

    \centering
    \includegraphics[scale=0.368]{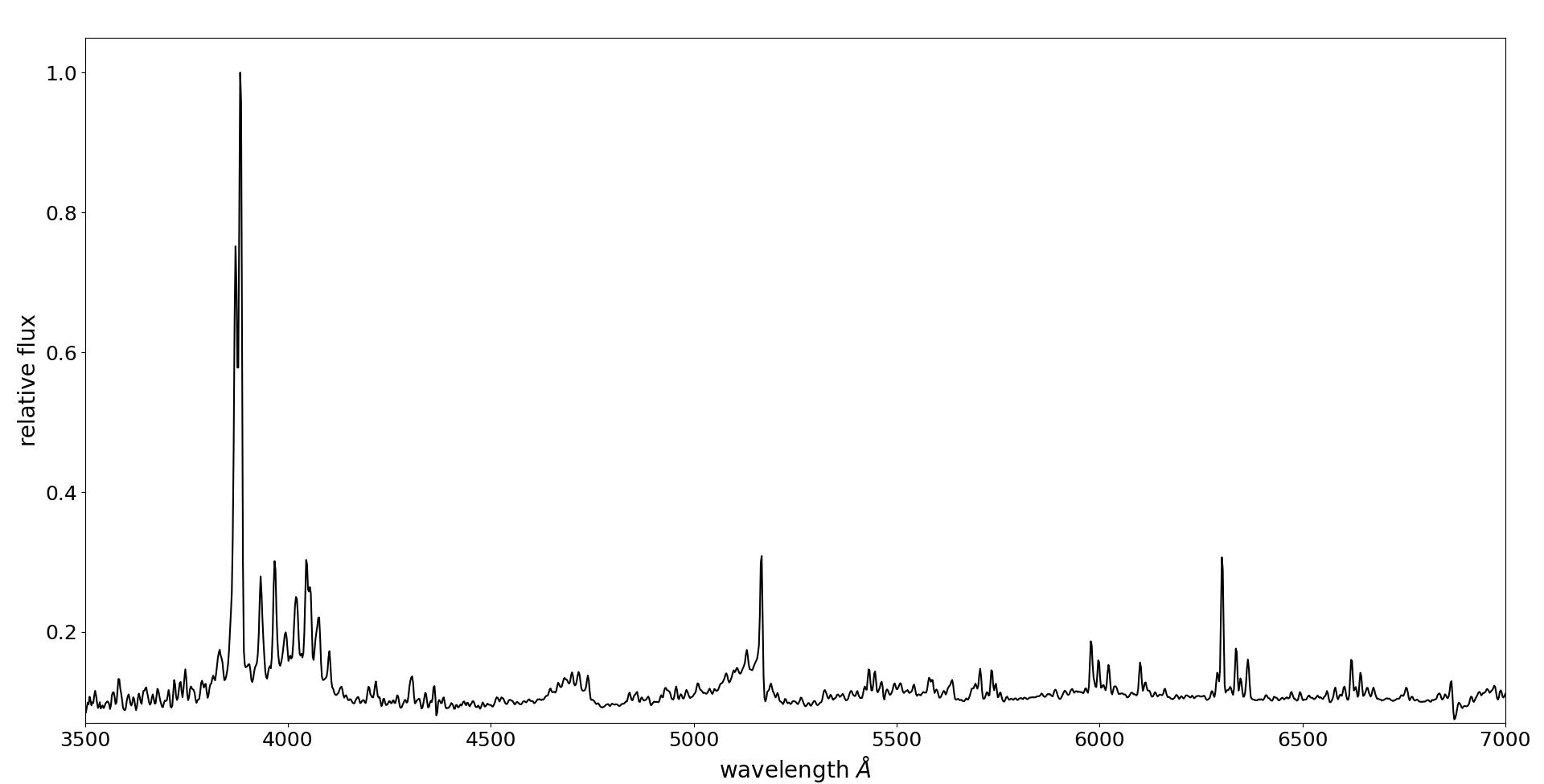}
    \caption{Spectrum of 2018-11-28; configuration A}

\end{figure}

\footnotetext{* No Solar Analog}

\newpage
\clearpage

\section{67P (Churyumov-Gerasimenko)}
\label{cometa:67P}
\subsection{Description}

67P/Churyumov-Gerasimenko is a Jupiter-family comet with a period of 6.44 years and an absolute magnitude of 12.9$\pm$0.8.\footnote{\url{https://ssd.jpl.nasa.gov/tools/sbdb_lookup.html\#/?sstr=67P} visited on July 21, 2024} 
It was first discovered by Klim Ivanovich Churyumov from Kyiv University's Astronomical Observatory on September 20, 1969. It was then rediscovered on August 8, 1975 by Edgar Roamer and Robert McCallister from the Kitt Peak National Observatory. K. Churyumov discovered the comet in a photo taken by Svetlana Ivanovna Gerasimenko on September 11, 1969, which he initially thought to be 32P Comas Sola. In October, it turned out that it was actually a new comet. Further observations by Geoffrey Marsden led to the calculation of an elliptical orbit. The comet is now famous for the visit of the Rosetta mission (ESA).\footnote{\url{https://www.esa.int/Science_Exploration/Space_Science/Rosetta }, visited on July 21, 2024} 

\noindent
We observed the comet between magnitude 9 and 8.\footnote{\url{https://cobs.si/comet/68/ }, visited on July 21, 2024}

\begin{table}[h!]
\centering
\begin{tabular}{|c|c|c|}
\hline
\multicolumn{3}{|c|}{Orbital elements (epoch: October 10, 2015)}                      \\ \hline \hline
\textit{e} = 0.6409 &  \textit{q} = 1.2433 &   \textit{T} = 457247.5887 \\ \hline
$\Omega$ = 50.1356 &   $\omega$ = 12.7982 &    \textit{i} = 7.0403 \\ \hline  
\end{tabular}
\end{table}

\begin{table}[h!]
\centering
\begin{tabular}{|c|c|c|c|c|c|c|c|c|}
\hline
\multicolumn{9}{|c|}{Comet ephemerides for key dates}                      \\ \hline 
\hline
& date         & r    & $\Delta$  & RA      & DEC      & elong  & phase  & PLang  \\
& (yyyy-mm-dd) & (AU) & (AU)      & (h)     & (°)      & (°)    & (°)    & (°) \\ \hline 

Perihelion       & 2021-11-02 & 1.211 & 0.421 & 07.35 & $+$26.35 & 111.2 & 49.8 & $-$0.5 \\ 
Nearest approach & 2021-11-12 & 1.217 & 0.418 & 07.96 & $+$26.71 & 113.2 & 48.4 & $-$2.1 \\ \hline
\end{tabular}

\end{table}
\vspace{0.5 cm}

\begin{figure}[h!]
    \centering
    \includegraphics[scale=0.38]{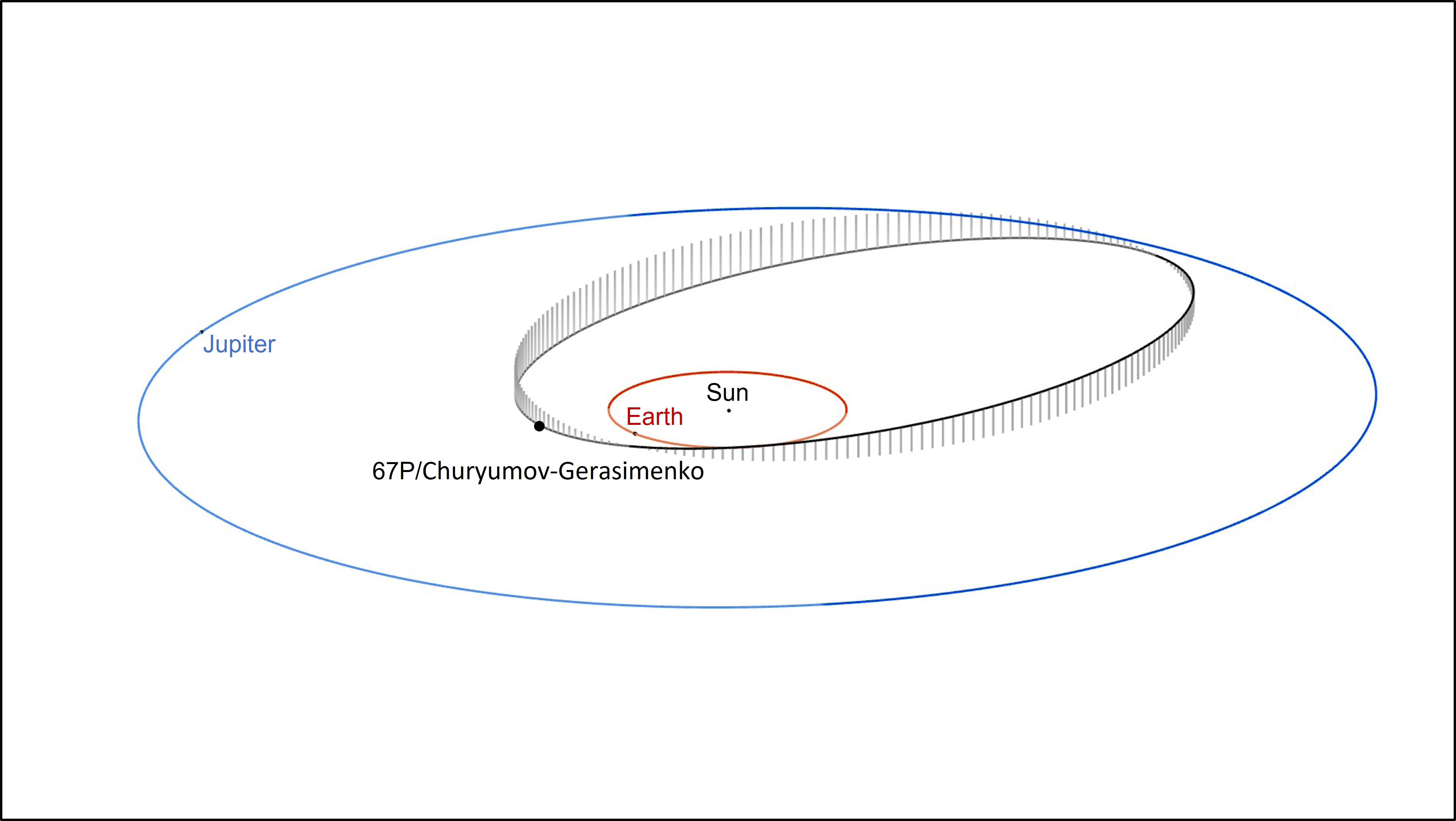}
    \caption{Orbit of comet 67P and position on perihelion date. The field of view is set to the orbit of Jupiter for size comparison. Courtesy of NASA/JPL-Caltech.}
\end{figure} 

\newpage

\subsection{Images}

\begin{SCfigure}[0.8][h!]
    \centering
    \includegraphics[scale=0.4]{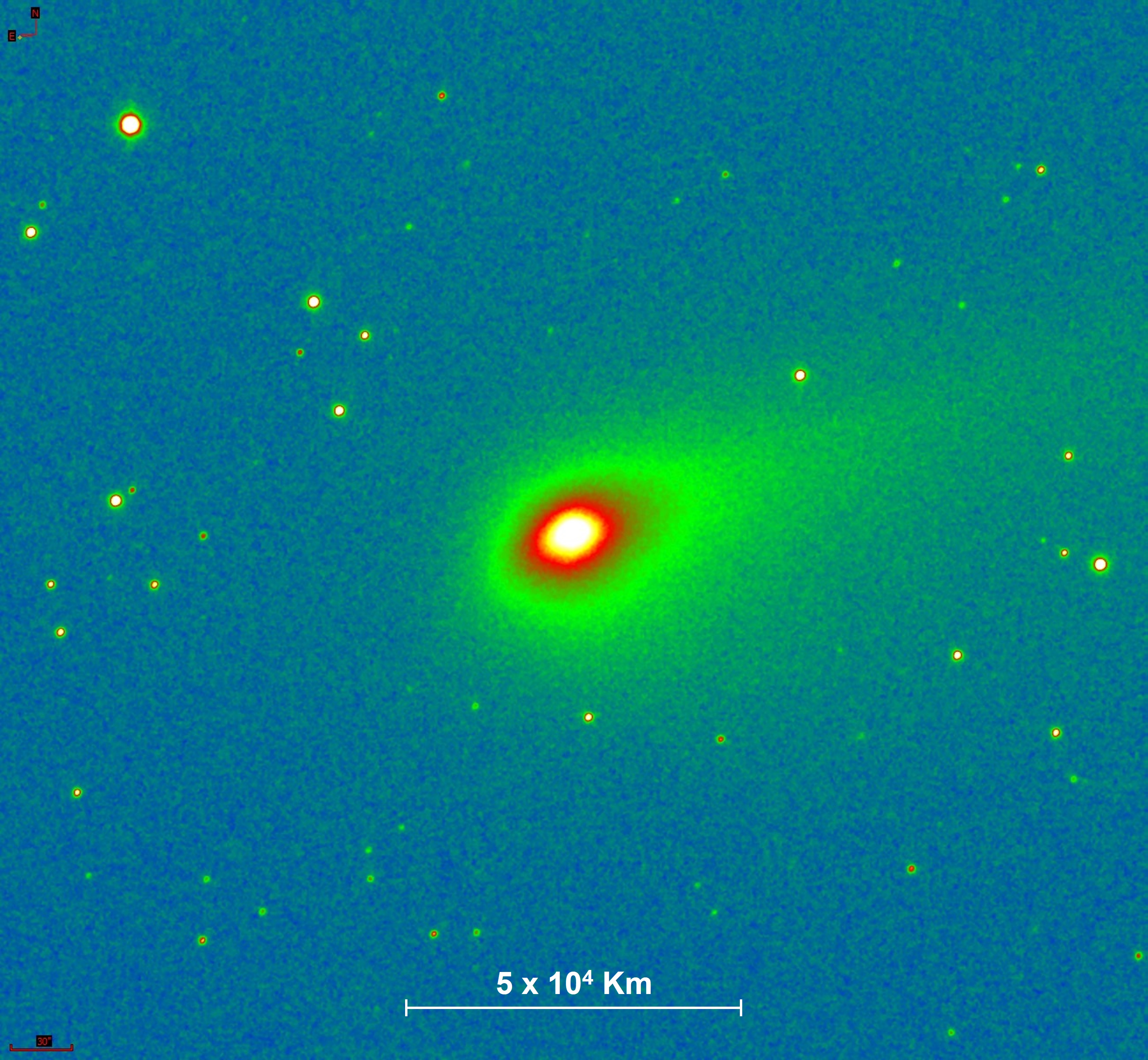}
     \caption{2021-11-19. Image taken with the Asiago Schmidt telescope through a r filter and shown in false colors. The coma appears clearly asymmetrical due to the presence of emissions from discrete active areas on the nucleus.}
\end{SCfigure} 

\begin{SCfigure}[0.8][h!]
    \centering
    
    \includegraphics[scale=0.4]{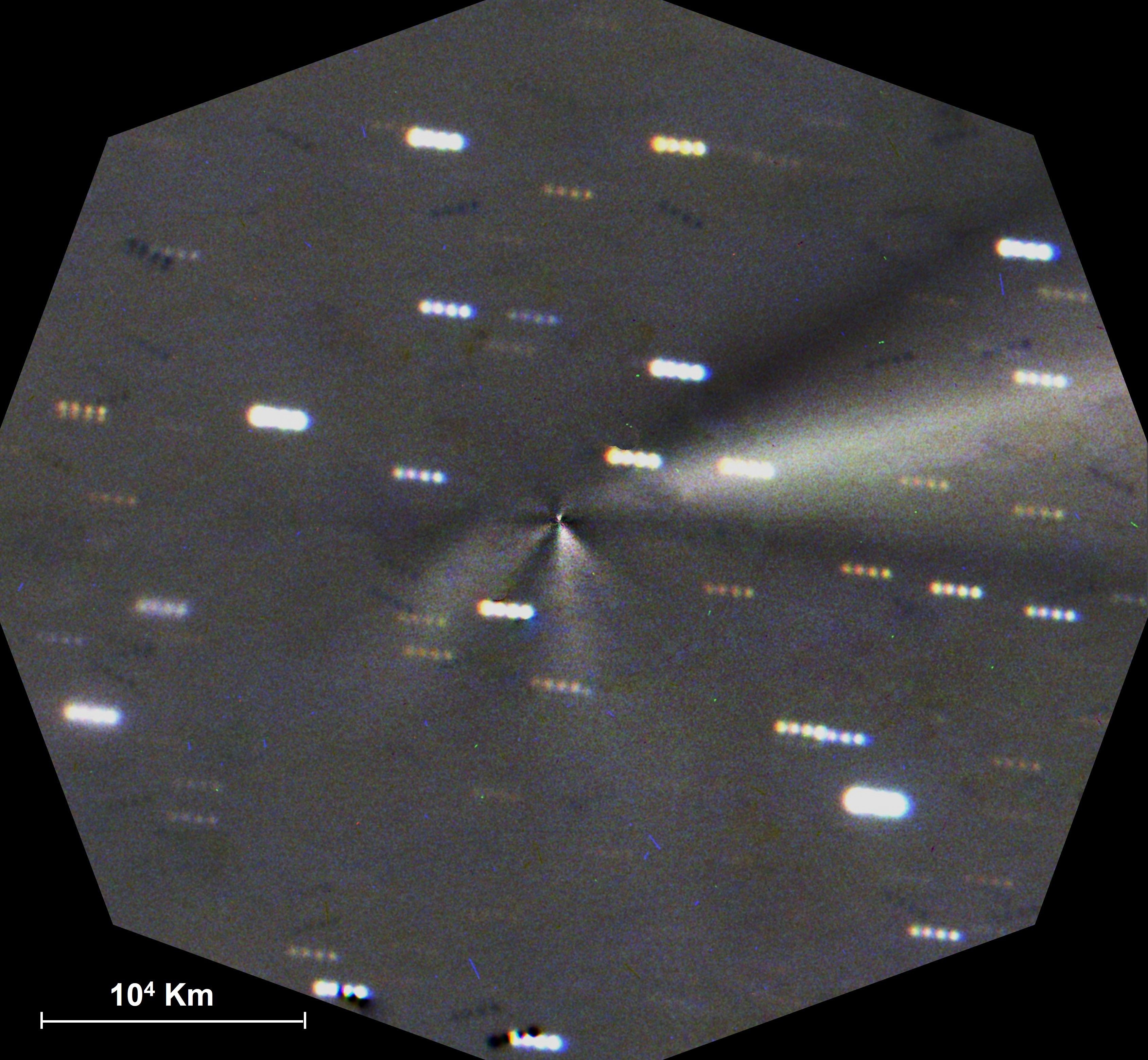}
    
    \caption{2022-01-27. Image taken with the Asiago Copernico Telescope with u, B, V, r filters, and then processed with a Larson-Sekanina algorithm ($\alpha=20$\textdegree) to show the morphology of the inner coma, dominated by jet-shaped structures.}
\end{SCfigure}

\newpage

\subsection{Spectra}

\begin{table}[h!]
\centering
\begin{tabular}{|c|c|c|c|c|c|c|c|c|c|c|c|}
\hline
\multicolumn{12}{|c|}{Observation details}                      \\ \hline 
\hline
$\#$  & date          & r     & $\Delta$ & RA     & DEC     & elong & phase & PLang& config  & FlAng & N \\
      & (yyyy-mm-dd)  &  (AU) & (AU)     & (h)    & (°)     & (°)   & (°)   &  (°)   &       &  (°)  &  \\ \hline 

1 & 2021-11-12 & 1.218  & 0.418 & 08.00 & $+$26.72 & 113.5  & 48.2  & $-$2.2 & A & $-$51 & 2 \\
2 & 2021-12-29 & 1.406  & 0.475 & 09.00 & $+$28.45 & 147.2  & 22.2  & $-$7.1 & A & $+$0 & 3 \\
3 & 2022-01-10 & 1.483  & 0.520 & 08.85 & $+$29.03 & 160.2  & 13.0  & $-$7.1 & C & $+$90  & 3 \\
4 & 2022-01-10 & 1.483  & 0.520 & 08.85 & $+$29.02 & 160.2  & 13.0  & $-$7.1 & D & $+$90 & 3 \\		
\hline
\end{tabular}
\end{table}

\begin{figure}[h!]

    \centering
    \includegraphics[scale=0.368]{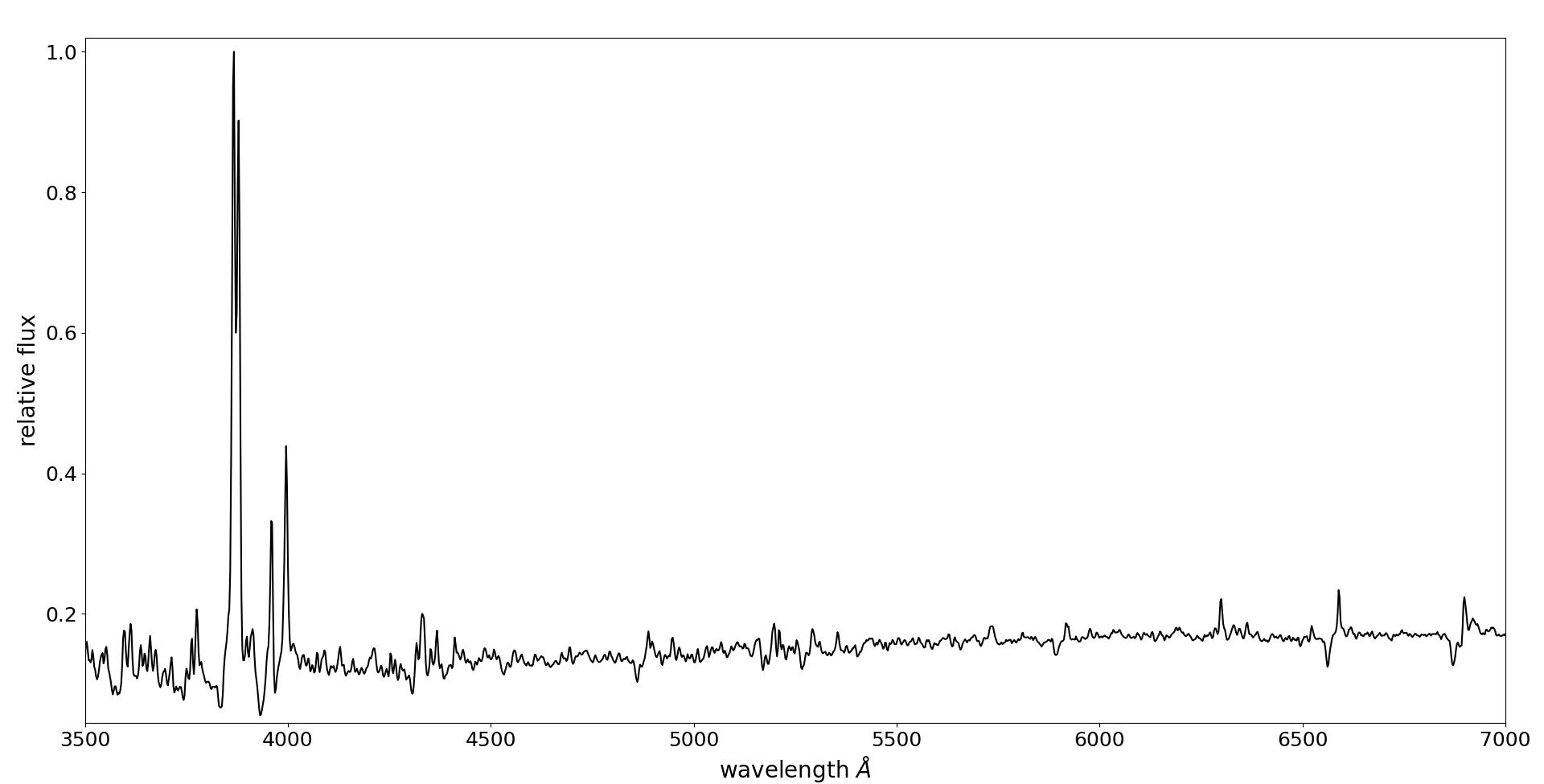}
    \caption{Spectrum of 2021-12-29; configuration A}

\end{figure}

\begin{figure}[h!]

    \centering
    \includegraphics[scale=0.368]{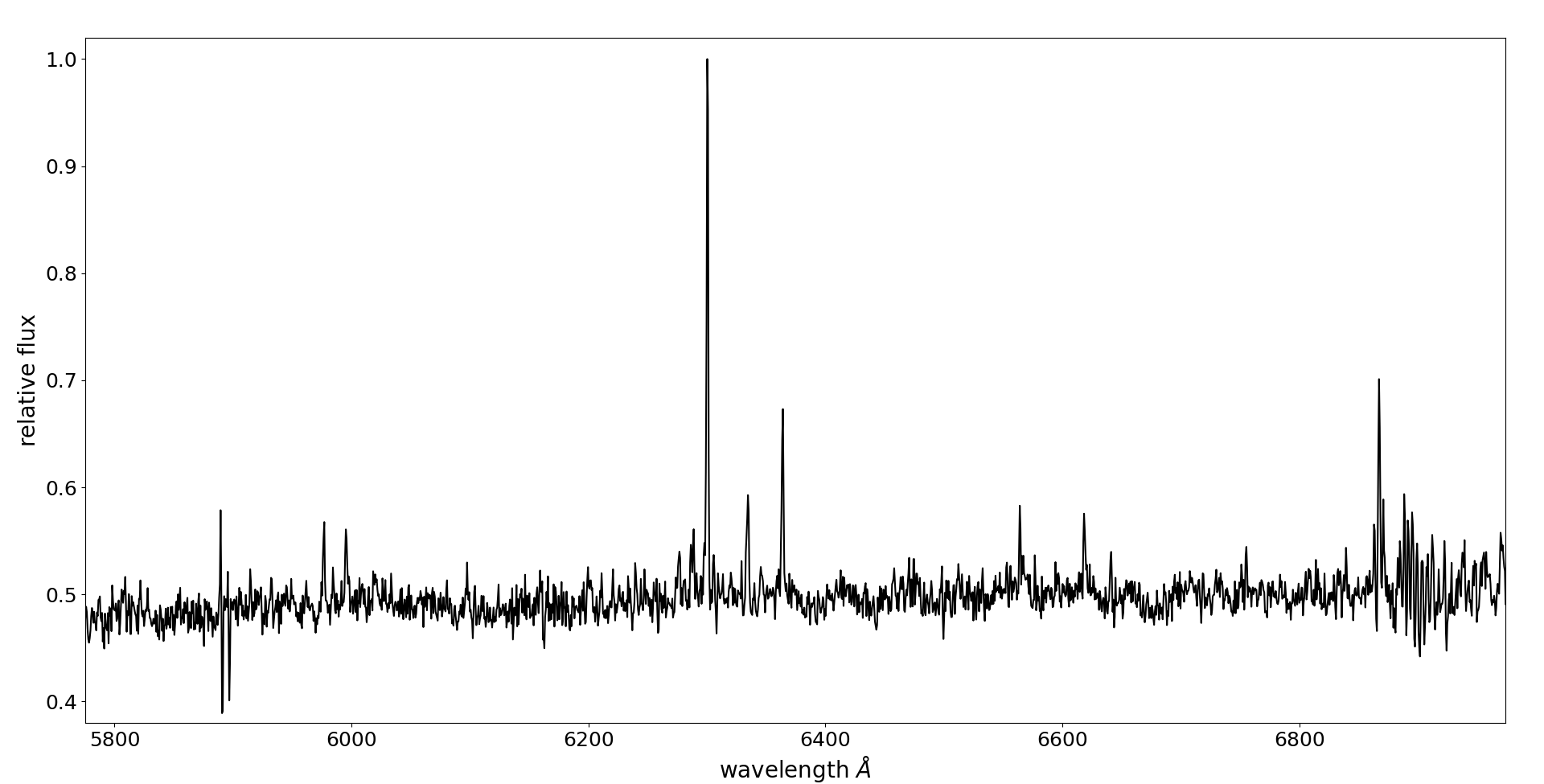}
    \caption{Spectrum of 2022-01-10; configuration C}

\end{figure}
							
\begin{figure}[h!]

    \centering
    \includegraphics[scale=0.368]{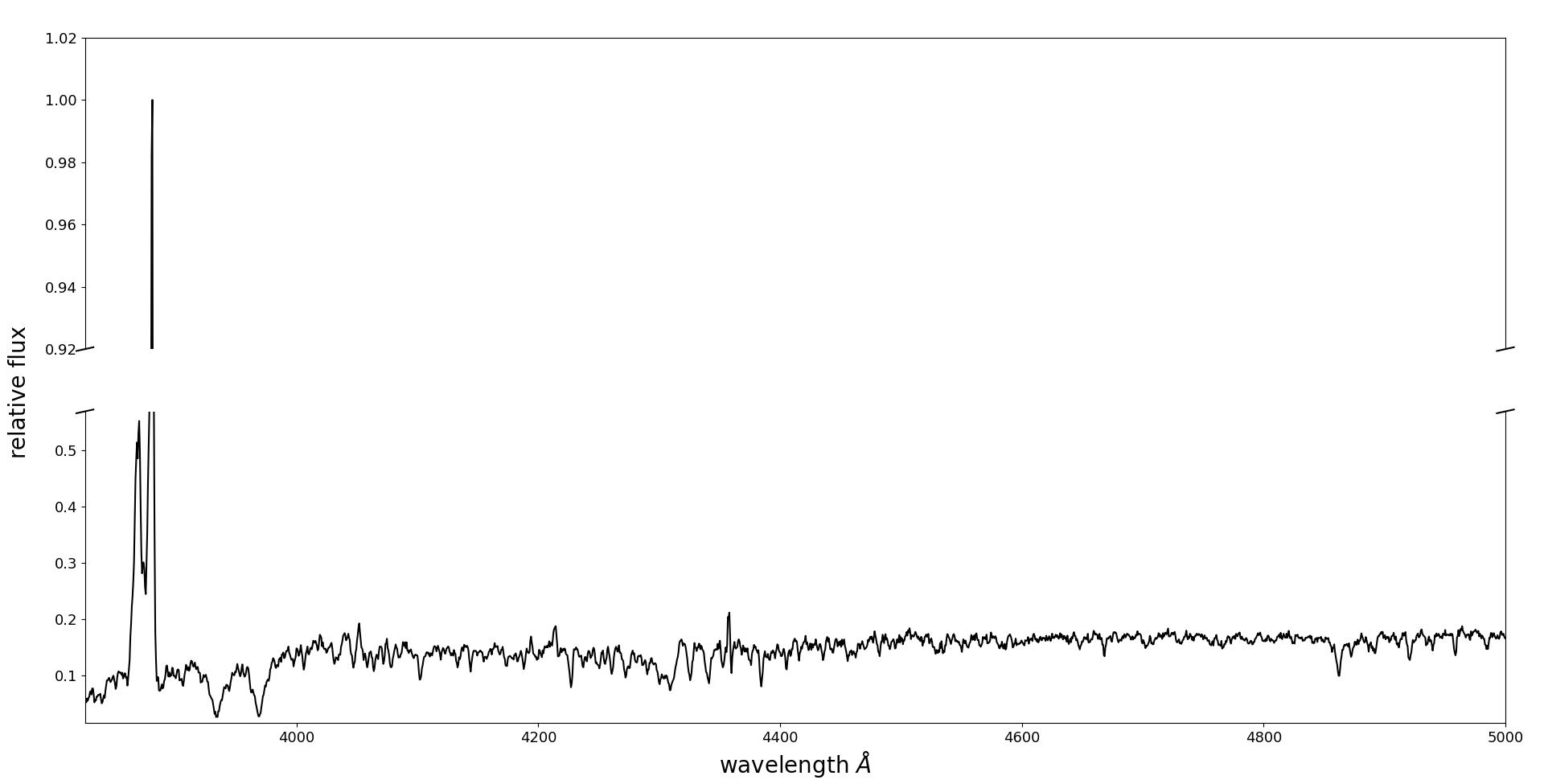}
    \caption{Spectrum of 2022-01-10; configuration D}

\end{figure}

\newpage
\clearpage

\section{104P (Koval 2)}
\label{cometa:104P}
\subsection{Description}

104P/Koval 2 is a Jupiter-family comet with a period of 5.74 years and an absolute magnitude of 14.6$\pm$1.0.\footnote{\url{https://ssd.jpl.nasa.gov/tools/sbdb_lookup.html\#/?sstr=104P} visited on July 21, 2024} 
It was first discovered by Charles Thomas Kowal from the Palomar Observatory on January 27, 1979. It was then rediscovered on December 12, 1991 by Masao Ishikawa from the Kiso Observatory. 

\noindent Actually, Leo Boethin discovered the comet on January 11, 1973 and was able to confirm it on the following two days. Brian Geoffrey Marsden and Syuichi Nakano calculated an elliptical orbit. 
In 2003, B. G. Marsden and Takao Kobayashi independently recognized the identity of the comet with that discovered by Kowal; Gary Kronk finally managed to prove the identity with the object discovered by L. Boethin. However, the comet kept the name Kowal. Earth crossed the comet orbital plane on April 18, 2022 and October 21, 2022. 

\noindent
We observed the comet around magnitude 8.\footnote{\url{https://cobs.si/comet/81/ }, visited on July 21, 2024}

\begin{table}[h!]
\centering
\begin{tabular}{|c|c|c|}
\hline
\multicolumn{3}{|c|}{Orbital elements (epoch: September 29, 2021)}                      \\ \hline \hline
\textit{e} = 0.6655 &   \textit{q} = 1.0731 &   \textit{T} = 2459591.1175 \\ \hline
$\Omega$ = 207.2135  & $\omega$ = 227.2459 &   \textit{i} = 5.7011 \\ \hline
\end{tabular}
\end{table}

\begin{table}[h!]
\centering
\begin{tabular}{|c|c|c|c|c|c|c|c|c|}
\hline
\multicolumn{9}{|c|}{Comet ephemerides for key dates}                      \\ \hline 
\hline
& date        & r & $\Delta$ & RA     & DEC    & elong & phase & PLang \\ 
& (yyyy-mm-dd) & (AU) & (AU)      & (h)     & (°)      & (°)    & (°)    & (°) \\ \hline

Perihelion	& 2022-01-12 & 1.073 & 0.656 & 00.81 &	$-$02.25 &		79.0 &	64.1 &	$+$08.5 \\ 
Nearest approach &	2022-01-28 & 1.096 &	0.638 &	02.04 &	$+$02.93 &		81.9 &	62.9 &	$+$08.7 \\ \hline
\end{tabular}

\end{table}

\vspace{0.5 cm}

\begin{figure}[h!]
    \centering
    \includegraphics[scale=0.38]{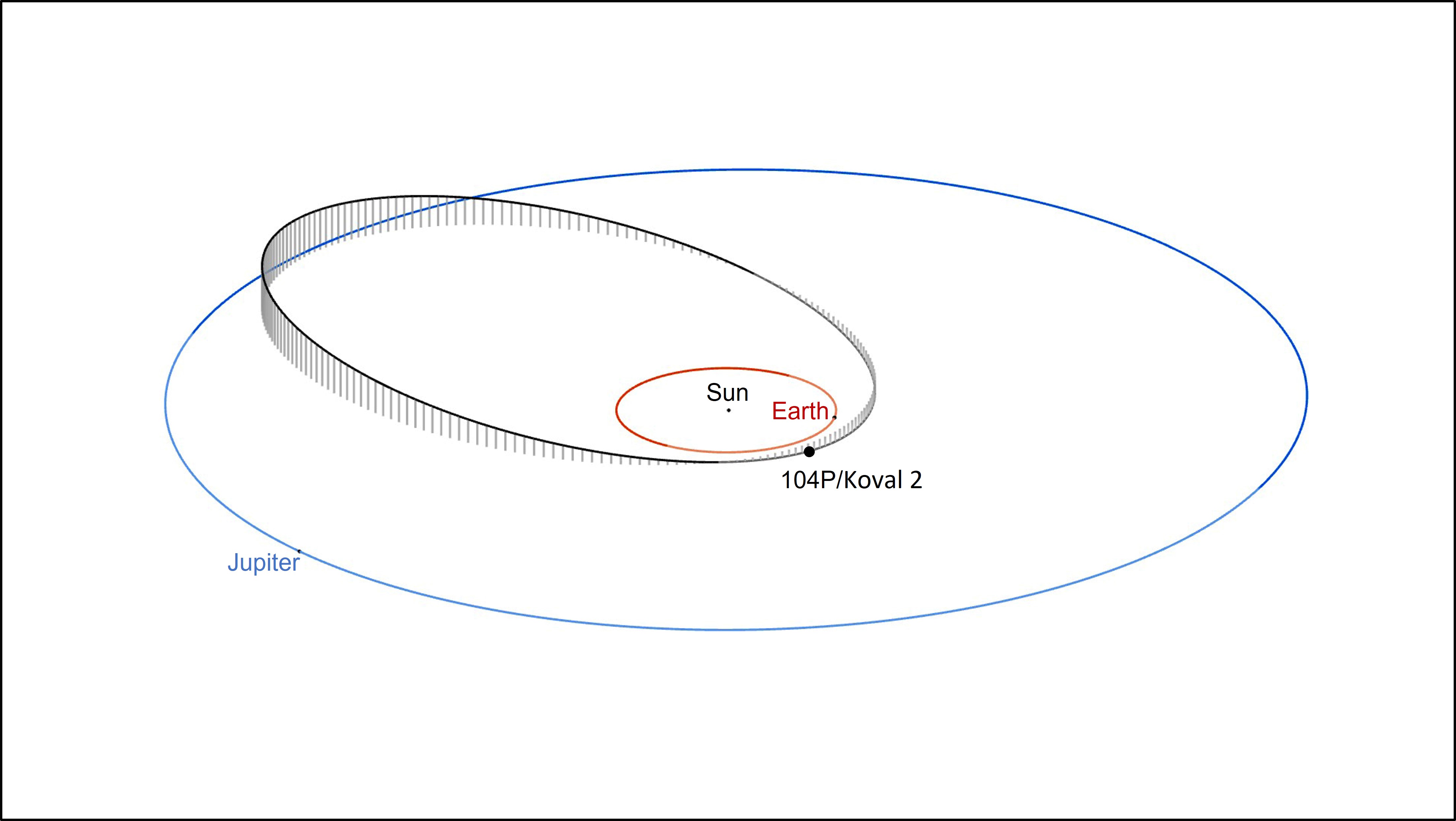}
    \caption{Orbit of comet 104P and position on perihelion date. The field of view is set to the orbit of Jupiter for size comparison. Courtesy of NASA/JPL-Caltech.}
\end{figure} 

\newpage

\subsection{Images}

\begin{SCfigure}[0.8][h!]
    \centering
    \includegraphics[scale=0.4]{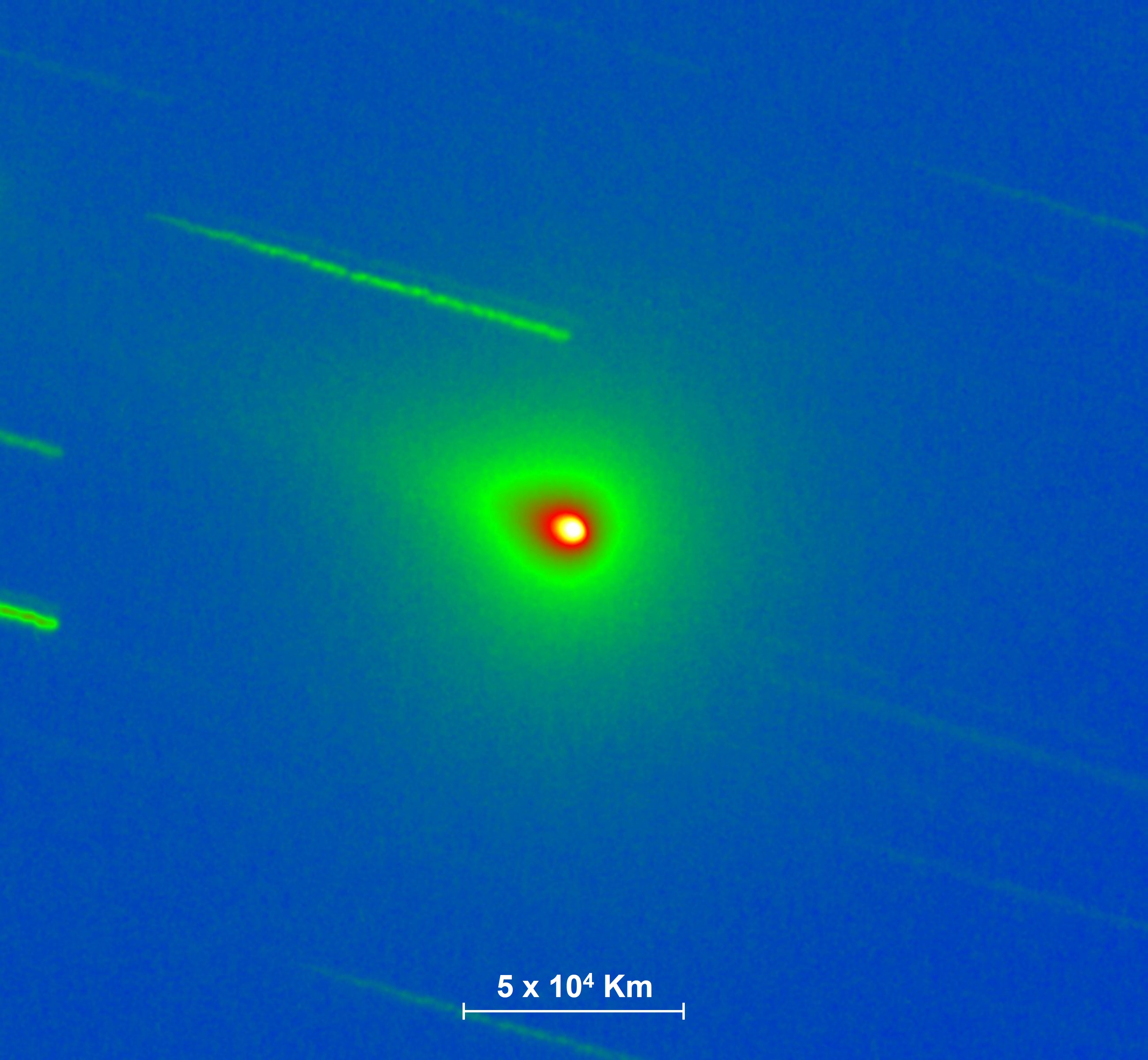}
     \caption{2022-01-30. Stack of unfiltered images taken with the 0.4m Savonarola telescope at the Stazione Astronomica di Sozzago (MPC-A12, Italy). Visualization in false colors. North is up, East is left.}
\end{SCfigure} 

\begin{SCfigure}[0.8][h!]
    \centering
    
    \includegraphics[scale=0.4]{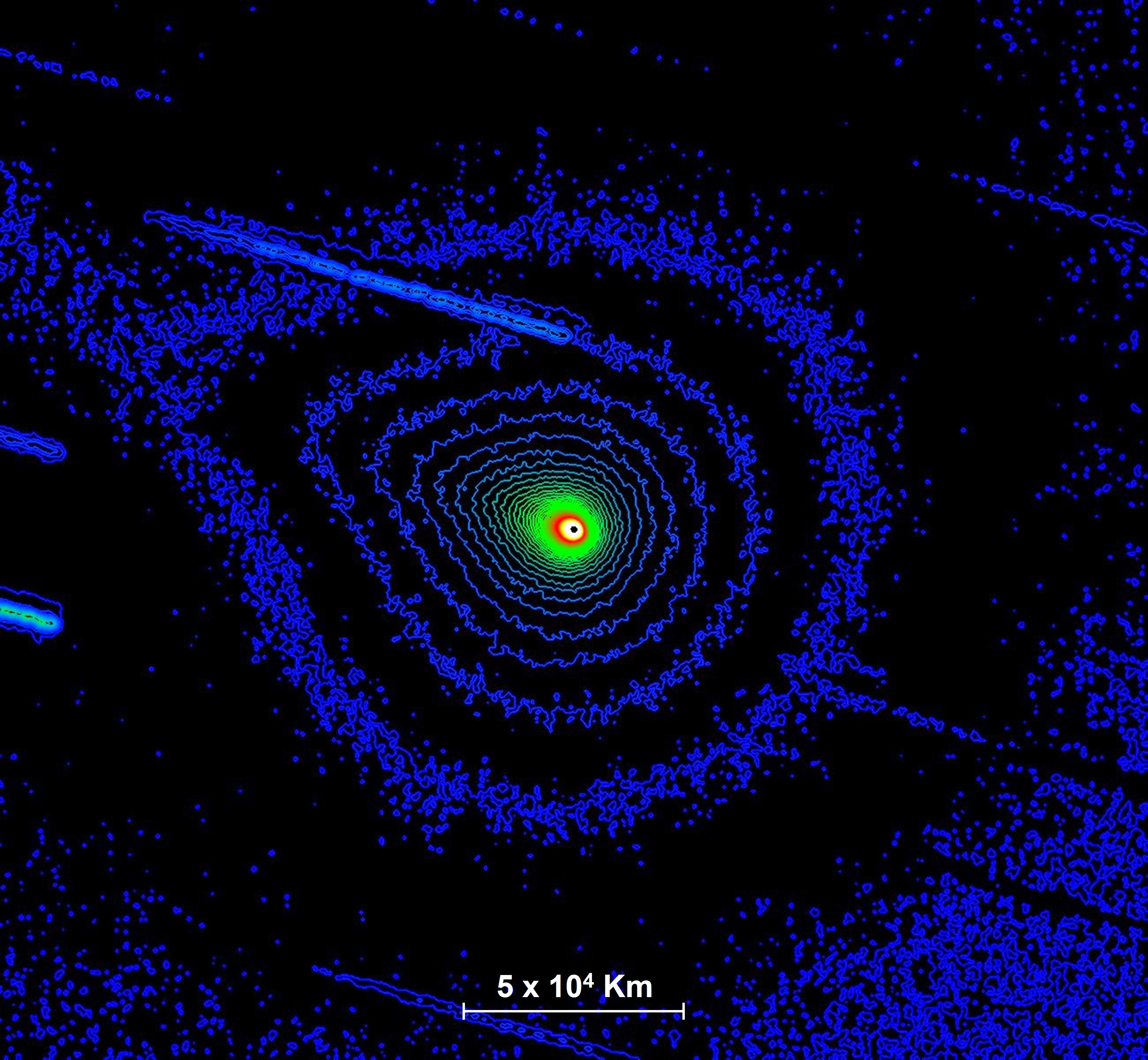}
    
    \caption{2022-01-30. The previous image is displayed in false colors with superimposed isophotes spaced every 3000 ADU, to highlight the general structure of the coma. The nucleus is identified by a black dot.}
\end{SCfigure}

\newpage

\subsection{Spectra}

\begin{table}[h!]
\centering
\begin{tabular}{|c|c|c|c|c|c|c|c|c|c|c|c|}
\hline
\multicolumn{12}{|c|}{Observation details}                      \\ \hline 
\hline
$\#$  & date          & r     & $\Delta$ & RA     & DEC     & elong & phase & PLang& config  &  FlAng & N \\
      & (yyyy-mm-dd)  &  (AU) & (AU)     & (h)    & (°)     & (°)   & (°)   &  (°)   &       &  (°)  & \\ \hline 

1 & 2022-01-30 & 1.104  & 0.639 & 02.28 & $+$03.97 & 82.7  & 62.2  & $+$08.6 & A & $+$0 & 3 \\
\hline
\end{tabular}
\end{table}

\begin{figure}[h!]

    \centering
    \includegraphics[scale=0.368]{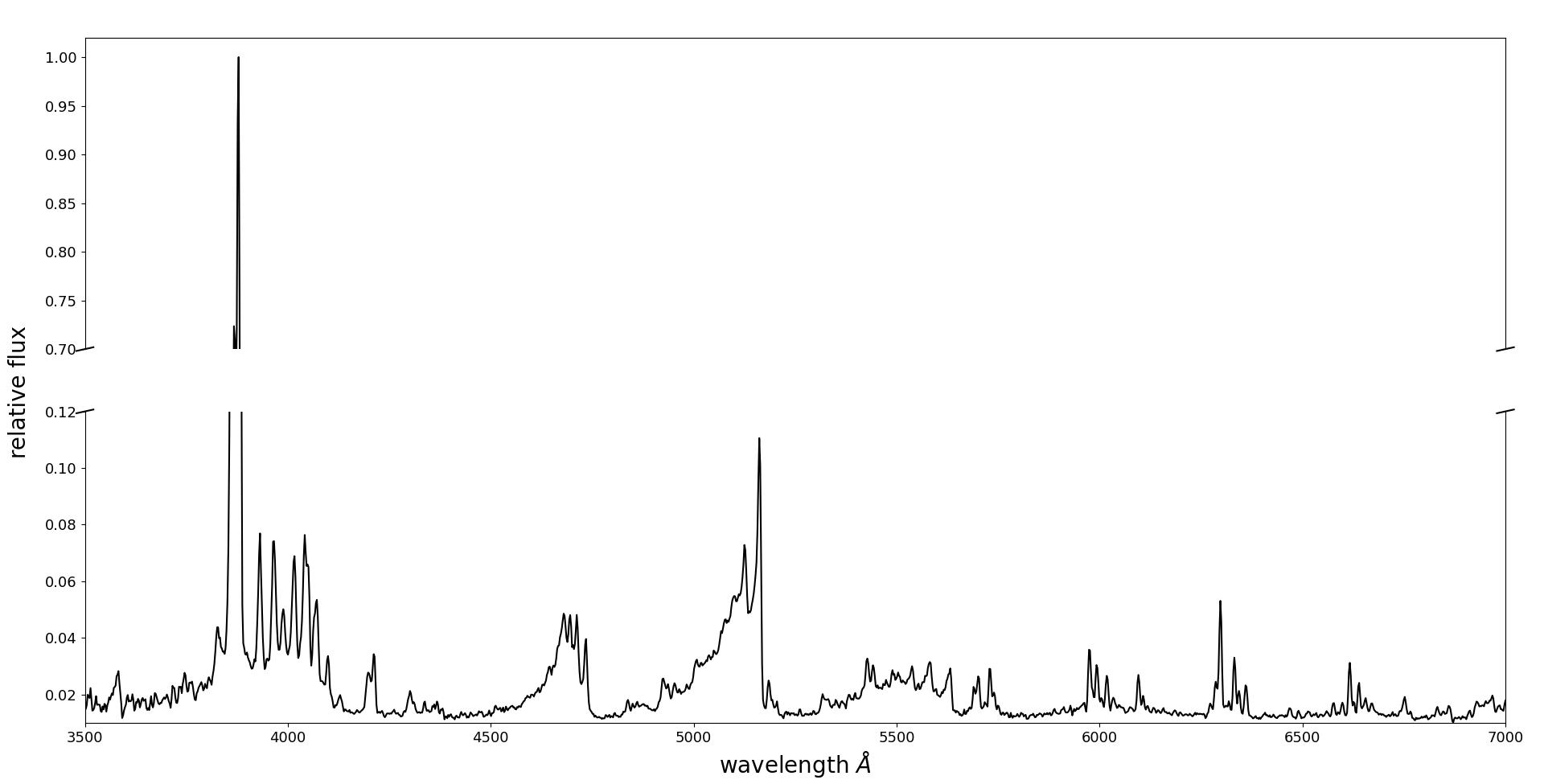}
    \caption{Spectrum of 2022-01-30; configuration A}

\end{figure}

\newpage
\clearpage

\section{116P (Wild 4)}
\label{cometa:116P}
\subsection{Description}

116P/Wild 4 is an Encke-type comet with a period of 6.50 years and an absolute magnitude of  7.6$\pm$0.7.\footnote{\url{https://ssd.jpl.nasa.gov/tools/sbdb_lookup.html\#/?sstr=116P} visited on July 21, 2024} 
It was first discovered by Paul Wild from the Zimmerwald Observatory on January 21, 1990. It was then rediscovered on September 12, 1994 by James Vernon Scotti from the Kitt Peak Observatory.
Earth crossed the orbital plane of the comet on October 14, 2022.

\noindent
We observed the comet around magnitude 13.\footnote{\url{https://cobs.si/comet/84/ }, visited on July 21, 2024}

\begin{table}[h!]
\centering
\begin{tabular}{|c|c|c|}
\hline
\multicolumn{3}{|c|}{Orbital elements (epoch: April 12 2016)}                      \\ \hline \hline
\textit{e} = 0.3723 & \textit{q} = 2.1871 & \textit{T} = 2457399.0268 \\ \hline
$\Omega$ = 20.9919 & $\omega$ = 173.3109  & \textit{i} = 3.6078  \\ \hline  
\end{tabular}
\end{table}

\begin{table}[h!]
\centering
\begin{tabular}{|c|c|c|c|c|c|c|c|c|}
\hline
\multicolumn{9}{|c|}{Comet ephemerides for key dates}                      \\ \hline 
\hline
 & date        & r(AU) & $\Delta$(AU) & RA     & DEC     & elong(°) & phase(°) & PLang(°) \\ \hline 
Perihelion       & 2022-07-17 &	2.197  & 2.573 & 11.48 & $+$03.74 &   57.3 &22.9  & $+$01.4  \\ 
Nearest approach & 2022-02-24 & 2.407  & 1.439 & 09.67 & $+$19.12 & 164.7 & 06.2  & $-$01.8 \\ \hline
\end{tabular}

\end{table}

\vspace{0.5 cm}

\begin{figure}[h!]
    \centering
    \includegraphics[scale=0.38]{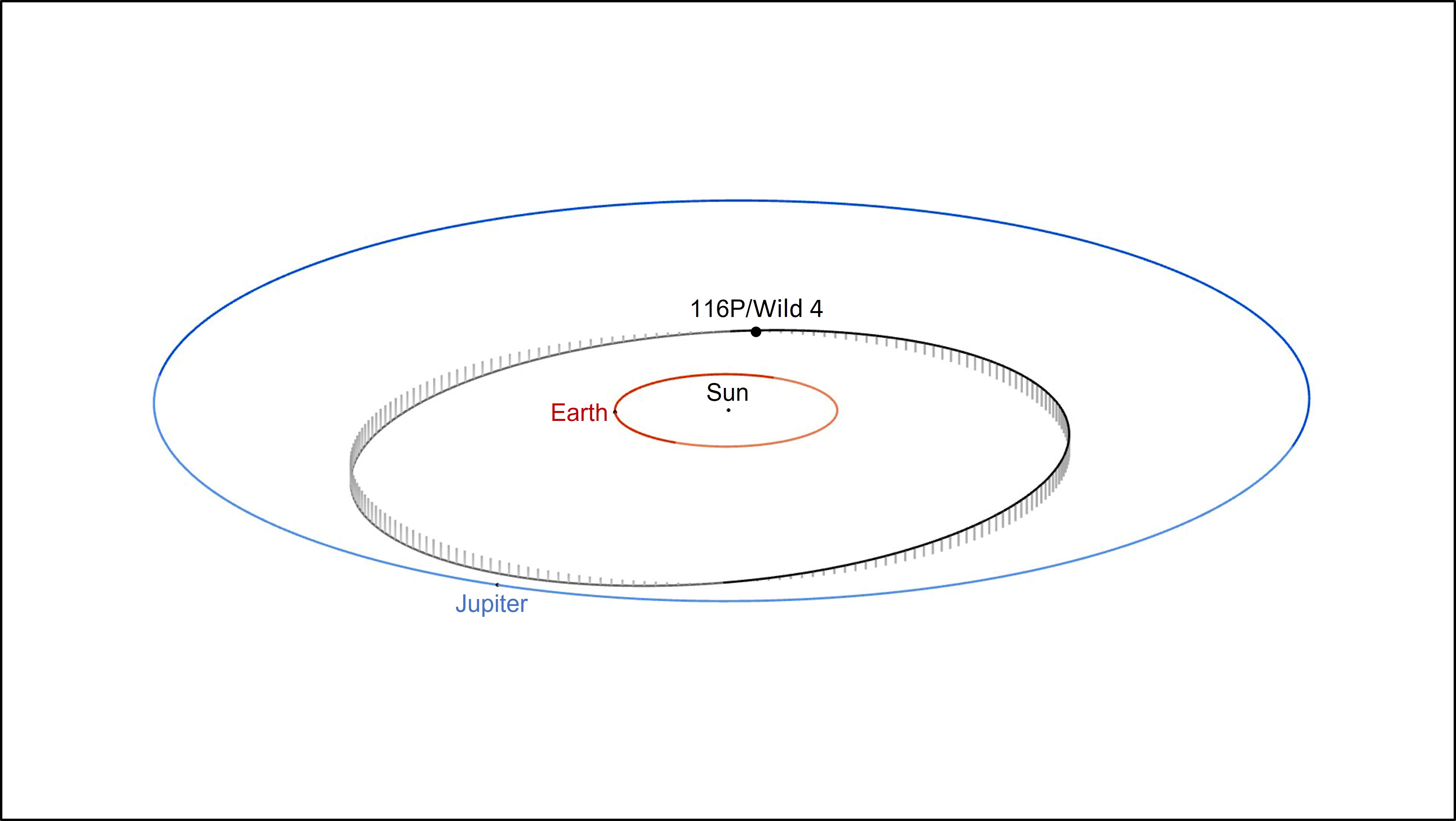}
    \caption{Orbit of comet 116P and position on perihelion date. The field of view is set to the orbit of Jupiter for size comparison. Courtesy of NASA/JPL-Caltech.}
\end{figure} 

\newpage

\subsection{Images}

\begin{SCfigure}[0.8][h!]
    \centering
    \includegraphics[scale=0.4]{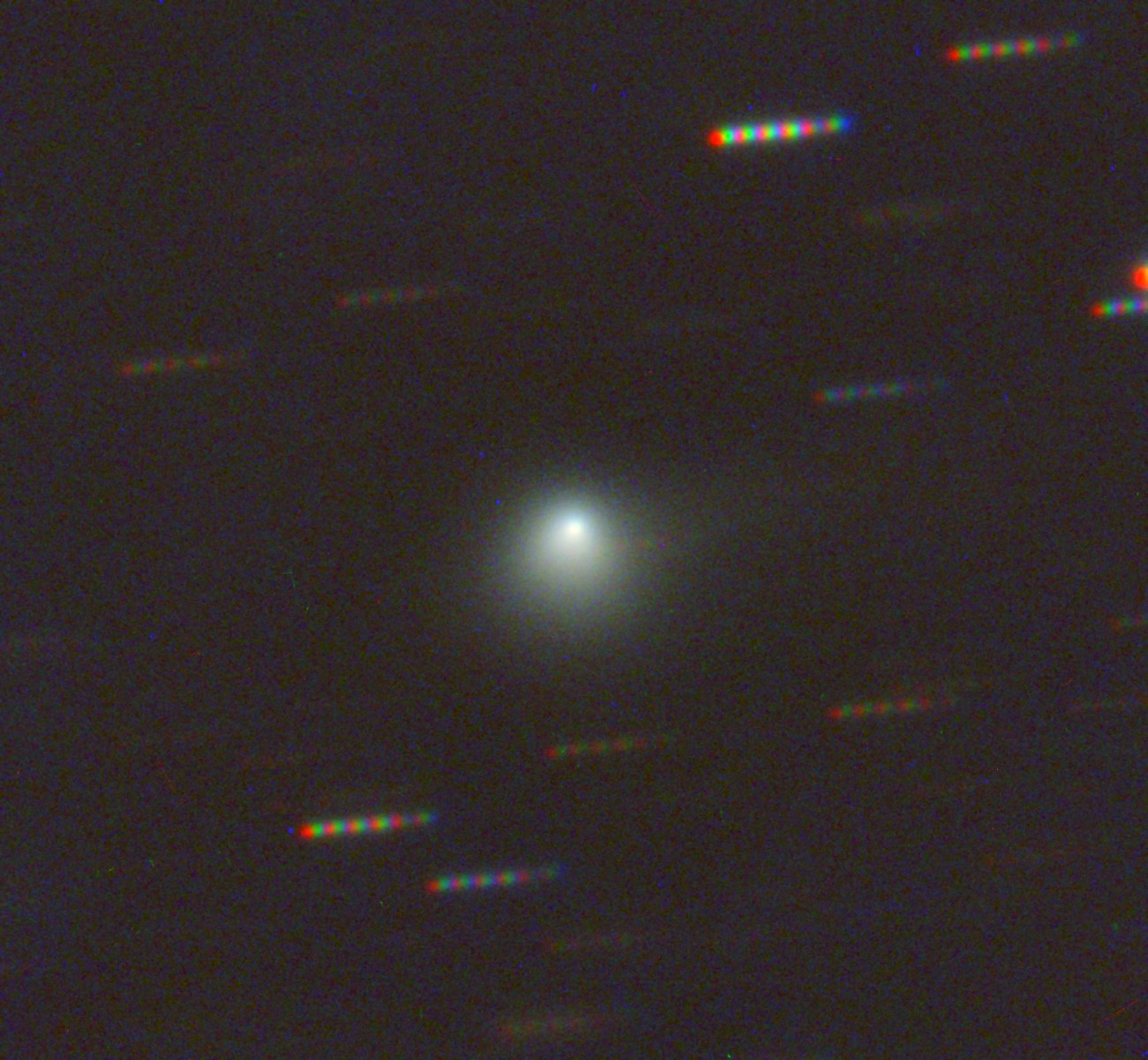}
     \caption{2022-03-06. Three-color BVR composite from images taken with the Asiago Copernico telescope. The comet appears surrounded by a typically green-colored coma, due to the emission of photons from the molecules of diatomic carbon (C$_2$). The tail develops westwards (right in the image).}
\end{SCfigure} 

\begin{SCfigure}[0.8][h!]
    \centering
    \includegraphics[scale=0.4]{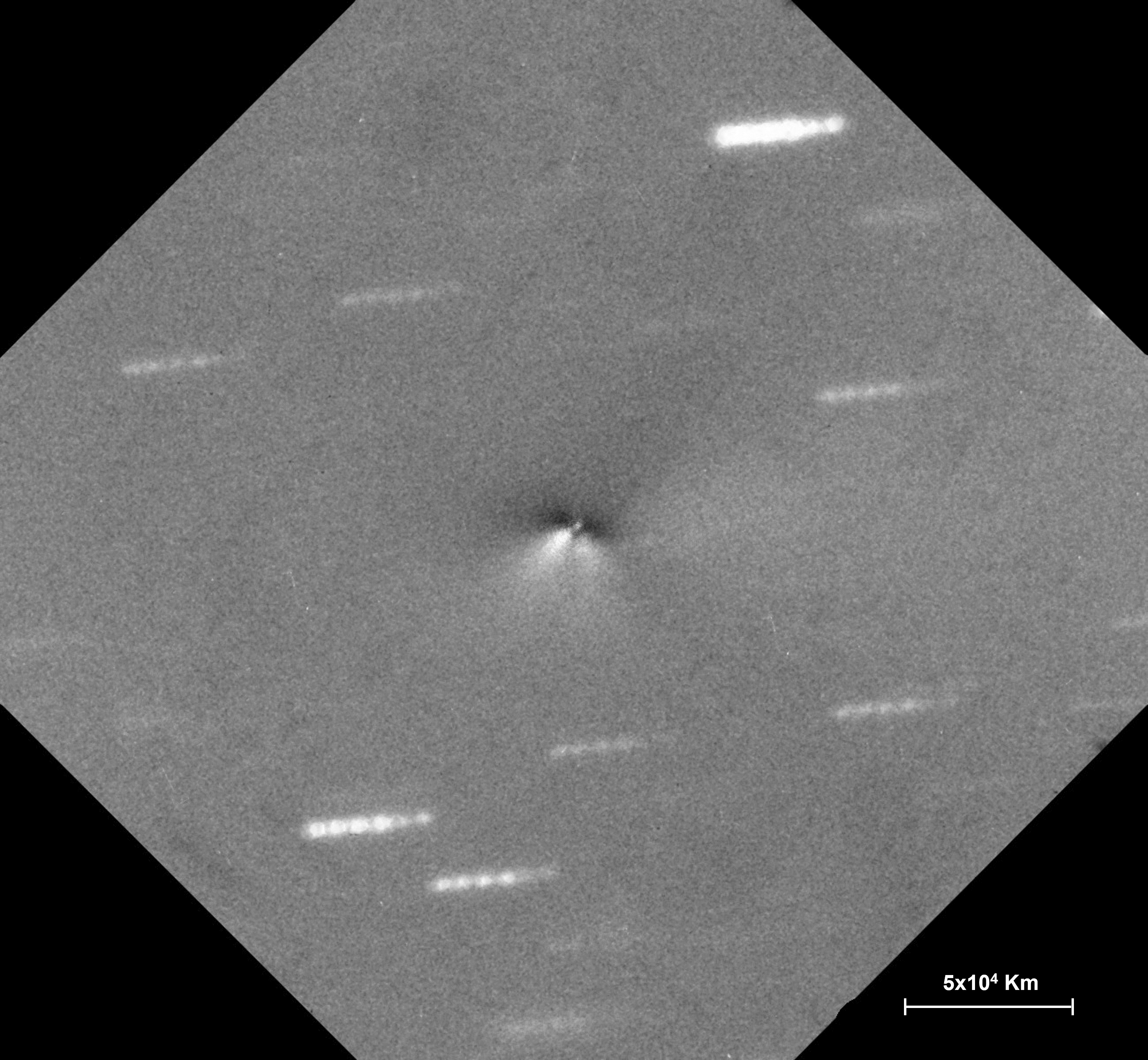}
    \caption{2022-03-06. The previous image was processed with a filter to highlight the details in the inner coma that are closest to the cometary nucleus.
    The observed morphology, with two jets moving away from the nucleus, suggests that there is at least one active area emitting gas and dust.
    For geometric reasons, the tail appears rather short because it develops in the opposite direction to the nucleus with respect to the Earth.}
\end{SCfigure}

\newpage

\subsection{Spectra}

\begin{table}[h!]
\centering
\begin{tabular}{|c|c|c|c|c|c|c|c|c|c|c|c|}
\hline
\multicolumn{12}{|c|}{Observation details}  \\ \hline 
\hline
$\#$ & date        & r & $\Delta$ & RA     & DEC     & elong & phase & PLang & config & FlAng & N \\
  & (yyyy-mm-dd) &  (AU) & (AU)  & (h) &(°)  &  (°) & (°)  &  (°)  & & (°) & \\ \hline 

1 & 2022-03-28 & 2.327 & 1.560 & 09.43 & $+$19.05 & 129.6 & 19.3 & $-$00.5& A & $+$0 & 2 \\

 \hline
\end{tabular}
\end{table}

\begin{figure}[h!]

    \centering
    \includegraphics[scale=0.368]{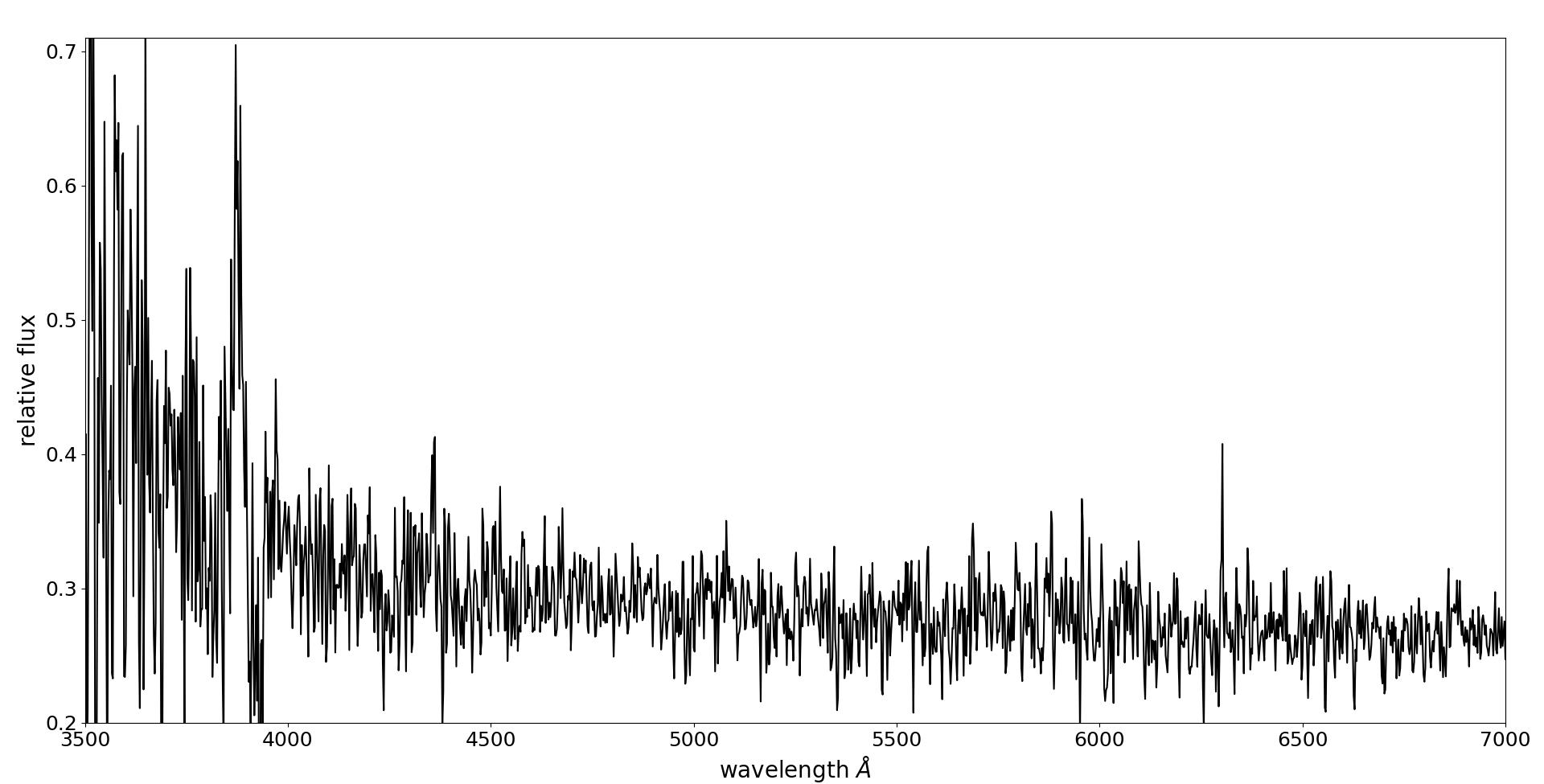}
    \caption{Spectrum of 2022-03-28; configuration A}

\end{figure}

\newpage
\clearpage

\section{123P (West-Hartley)}
\label{cometa:123P}
\subsection{Description}

123P/West-Hartley is a Jupiter-family comet with a period of 7.59 years and an absolute magnitude of 8.2$\pm$0.9.\footnote{\url{https://ssd.jpl.nasa.gov/tools/sbdb_lookup.html\#/?sstr=123P} visited on July 21, 2024} 
It was first discovered by Malcolm Hartley from the Siding Spring Observatory on May 28, 1989. It was then rediscovered on September 21, 1995 by James Vernon Scotti from the Kitt Peak Observatory.
Actually, Richard Martin West discovered the comet from images taken on March 14, 1989 by Guido Pizarro with the 1m Schmidt telescope at the La Silla Observatory.
Confirmation of the observation was initially unsuccessful until M. Hartley discovered the comet on a picture taken on May 28, 1989 with the 1.22m Schmidt telescope at the Siding Spring Observatory, 
which quickly was associated with the discovery of R. West. 
The first elliptical orbit was calculated by Brian Marsden.
The Earth crossed the orbital plane of the comet on May 7, 2019. \\
We observed the comet around magnitude 13.\footnote{\url{https://cobs.si/comet/88/ }, visited on July 21, 2024} 

\begin{table}[h!]
\centering
\begin{tabular}{|c|c|c|}
\hline
\multicolumn{3}{|c|}{Orbital elements (epoch: September 19, 2015)}                      \\ \hline \hline
\textit{e} = 0.4493 & \textit{q} = 2.1261 & \textit{T} = 2458518.8399 \\ \hline
$\Omega$ = 46.5274 & $\omega$ = 102.9184  & \textit{i} = 15.3545  \\ \hline  
\end{tabular}
\end{table}

\begin{table}[h!]
\centering
\begin{tabular}{|c|c|c|c|c|c|c|c|c|}
\hline
\multicolumn{9}{|c|}{Comet ephemerides for key dates}                      \\ \hline 
\hline
& date & r & $\Delta$ & RA & DEC & elong & phase & PLang \\
& (yyyy-mm-dd) & (AU) & (AU) & (h) & (°) & (°) & (°) & (°) \\ \hline 

Perihelion       & 2019-02-05 & 2.127  & 1.253 & 11.67 & $+$30.67  & 143.3 & 16.1  & $-$12.1  \\ 
Nearest approach & 2019-02-28 & 2.135  & 1.200 & 11.46  & $+$31.96  & 154.0 & 11.7  & $-$11.7 \\ \hline
\end{tabular}

\end{table}

\vspace{0.5 cm}

\begin{figure}[h!]
    \centering
    \includegraphics[scale=0.38]{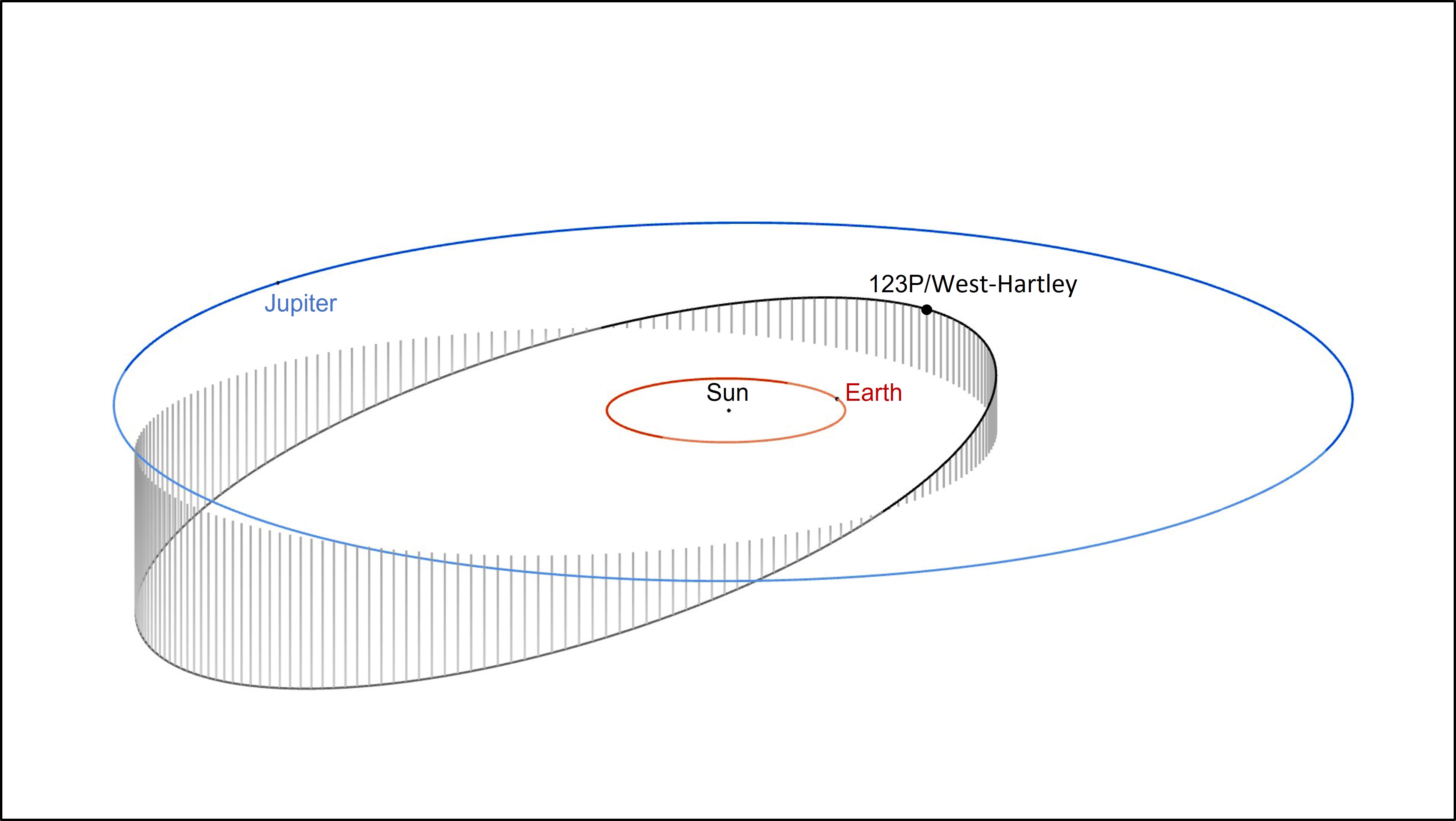}
    \caption{Orbit of comet 123P and position on perihelion date. The field of view is set to the orbit of Jupiter for size comparison. Courtesy of NASA/JPL-Caltech.}
\end{figure} 

\newpage

\subsection{Images}
\begin{SCfigure}[0.8][h!]
 \centering

 \includegraphics[scale=0.4]{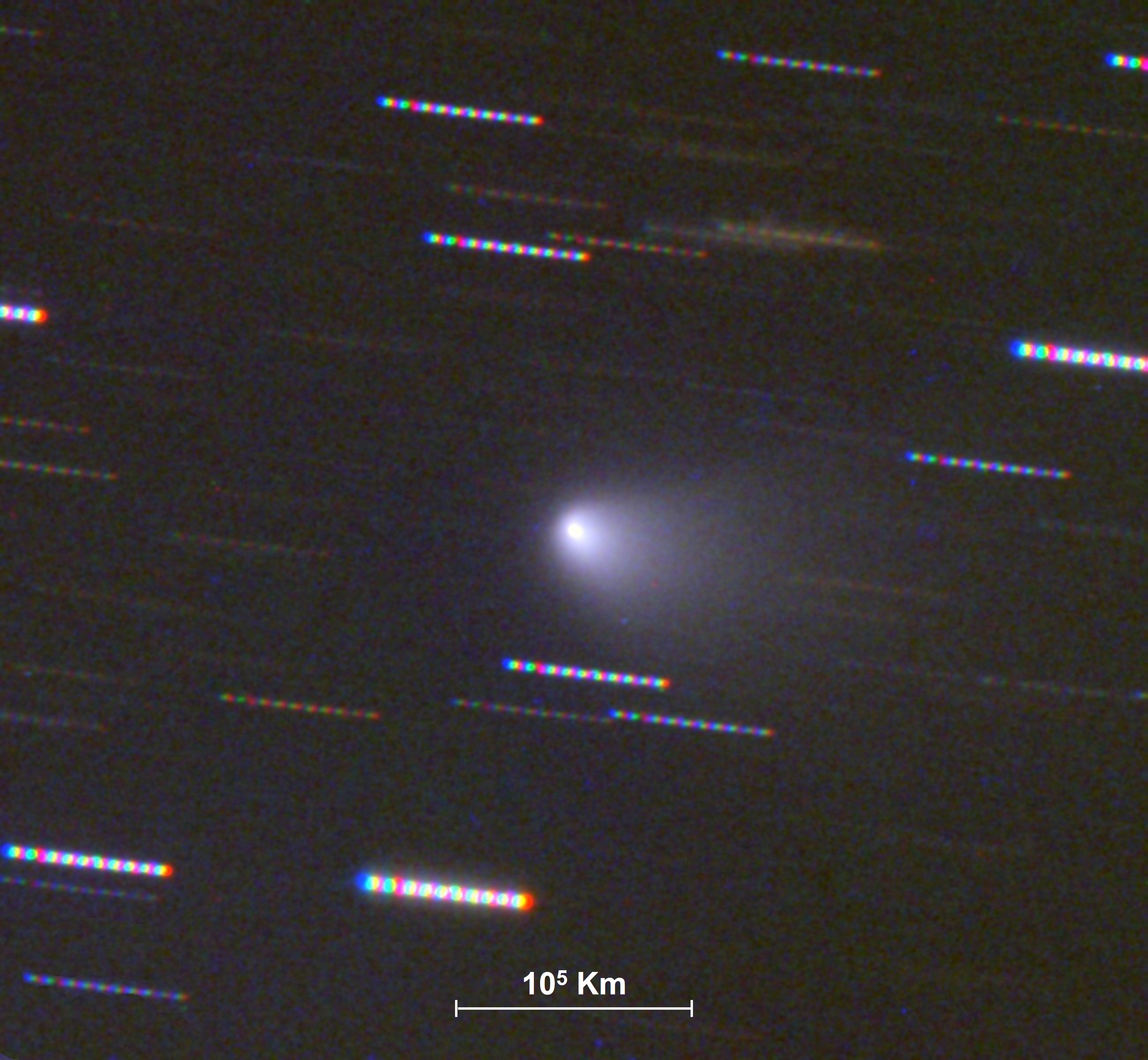}

 \caption{2019-03-05. Three-color (BVR) composite image of comet 123P taken with the Asiago Schmidt telescope.}

\end{SCfigure}

\begin{SCfigure}[0.8][h!]
 \centering
 \includegraphics[scale=0.4]{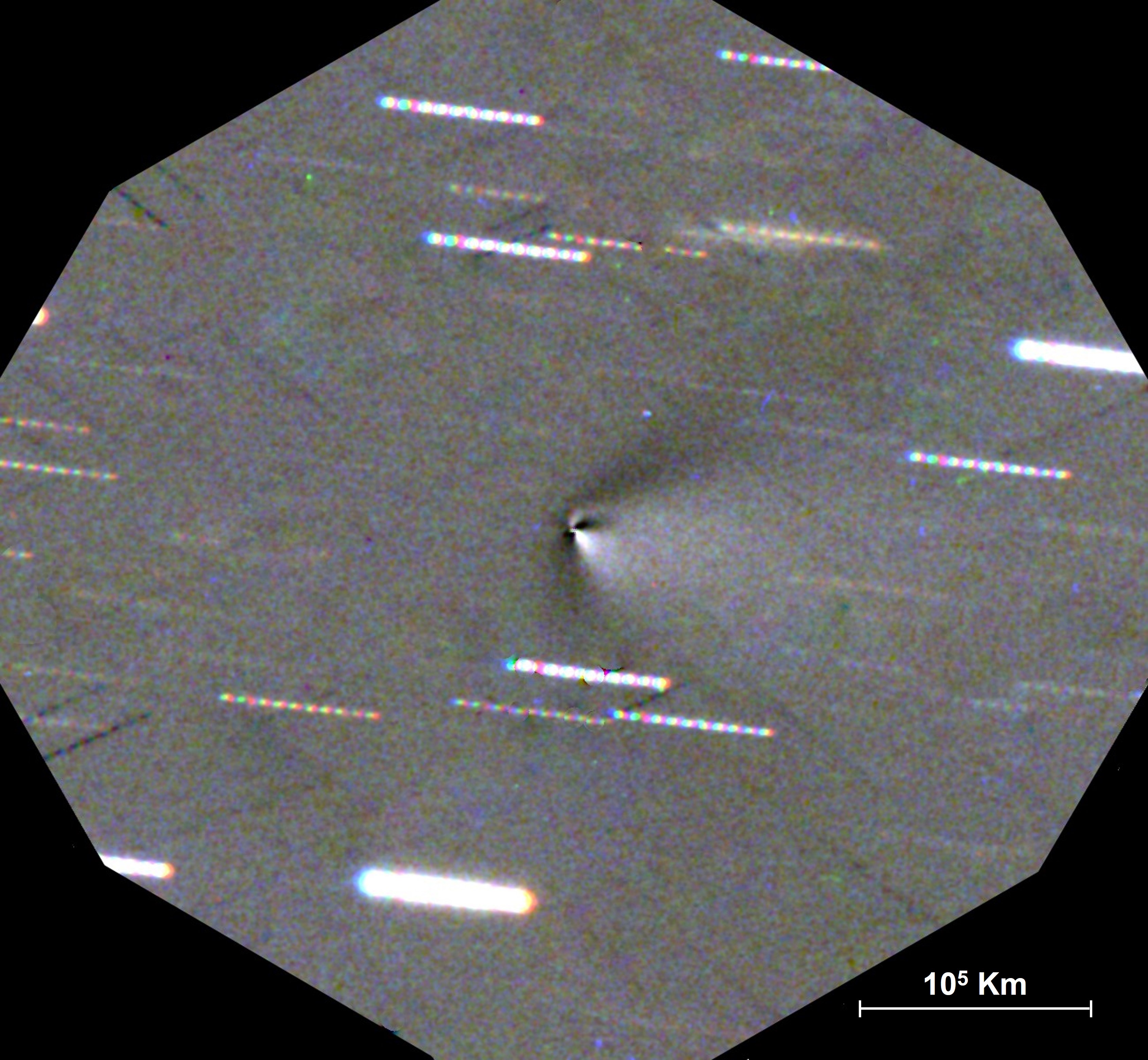}
 \caption{2019-03-05. A mathematical filter was applied to the stacked image to observe the morphology of the inner coma, dominated by a jet structure. The tail develops towards West-South-West.}

\end{SCfigure}

\newpage

\subsection{Spectra}

\begin{table}[h!]
\centering
\begin{tabular}{|c|c|c|c|c|c|c|c|c|c|c|c|}
\hline
\multicolumn{12}{|c|}{Observation details}                      \\ \hline 
\hline
$\#$ & date & r & $\Delta$ & RA & DEC & elong & phase & PLang& config & FlAng & N \\
 & (yyyy-mm-dd) & (AU) & (AU) & (h) & (°) & (°) & (°) & (°) & & (°) & \\ \hline 

1 & 2019-02-16 & 2.129  & 1.211 & 11.57 & $+$31.63 & 150.8 & 13.1 & $-$12.3 & A & $+$0 & 3 \\
2 & 2019-03-05 & 2.139  & 1.204 & 11.37 & $+$31.90 & 153.7 & 11.8 & $-$11.1 & A & $+$0 & 3 \\
 \hline
\end{tabular}
\end{table}

\begin{figure}[h!]

    \centering
    \includegraphics[scale=0.368]{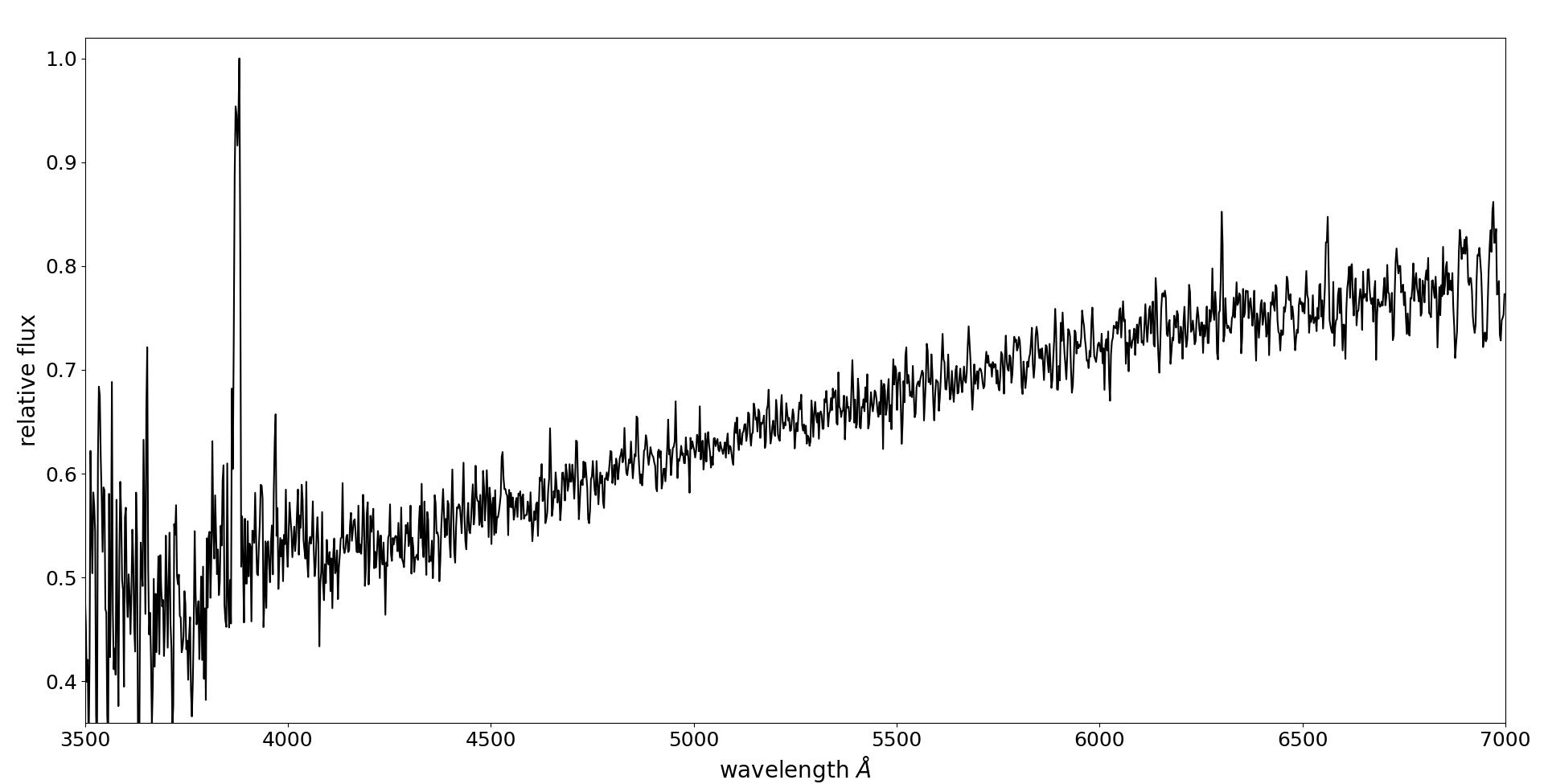}
    \caption{Spectrum of 2019-02-16; configuration A}

\end{figure}

\newpage
\clearpage

\section{156P (Russell-LINEAR)}
\label{cometa:156P}
\subsection{Description}

156P/Russell-LINEAR is a Jupiter-family comet with a period of 6.44 years and an absolute magnitude of 12.7$\pm$0.9.\footnote{\url{https://ssd.jpl.nasa.gov/tools/sbdb_lookup.html\#/?sstr=156P} visited on July 21, 2024} 
It was first discovered by Kenneth Russell from the Siding Spring Observatory in September 1986.

\noindent
Although K. Russell discovered the comet from an image taken by Fred Watson on September 3, 1986, it was no longer observed and was considered lost. 
In 2000, an asteroid object was discovered as part of the LINEAR project and was associated with the comet.
Earth crossed the orbital plane of the comet on October 29, 2020.

\noindent
We observed the comet between magnitude 9 and 10.\footnote{\url{https://cobs.si/comet/537/ }, visited on July 21, 2024}

\begin{table}[h!]
\centering
\begin{tabular}{|c|c|c|}
\hline
\multicolumn{3}{|c|}{Orbital elements (epoch: March 20, 2020)}                      \\ \hline \hline
\textit{e} = 0.6149 & \textit{q} = 1.3331 & \textit{T} = 2459171.3263 \\ \hline
$\Omega$ = 35.3972 & $\omega$ = 0.3781  & \textit{i} = 17.2636  \\ \hline  
\end{tabular}
\end{table}

\begin{table}[h!]
\centering
\begin{tabular}{|c|c|c|c|c|c|c|c|c|}
\hline
\multicolumn{9}{|c|}{Comet ephemerides for key dates}                      \\ \hline 

& date & r & $\Delta$ & RA & DEC & elong & phase & PLang \\
& (yyyy-mm-dd) & (AU) & (AU) & (h) & (°) & (°) & (°) & (°) \\ \hline 

Perihelion       & 2020-11-18 & 1.333  & 0.526 & 23.74  & $-$01.22  & 120.2 & 39.9  & $-$11.3  \\ 
Nearest approach & 2020-10-24 & 1.364  & 0.481 & 23.53  & $-$20.89  & 132.1 & 32.8  & $+$2.7 \\ \hline
\end{tabular}

\end{table}

\vspace{0.5 cm}

\begin{figure}[h!]
    \centering
    \includegraphics[scale=0.38]{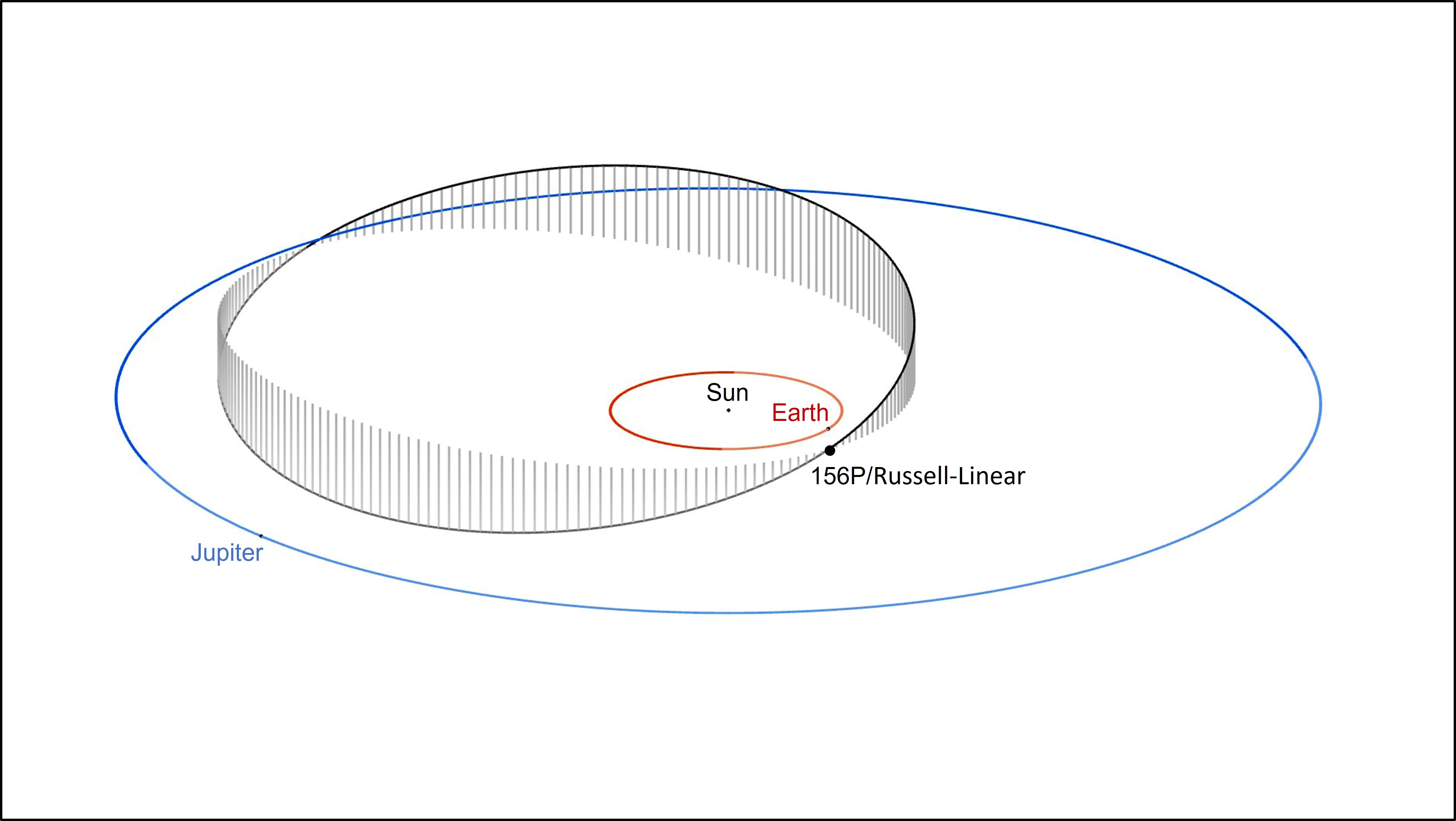}
    \caption{Orbit of comet 156P and position on perihelion date. The field of view is set to the orbit of Jupiter for size comparison. Courtesy of NASA/JPL-Caltech.}
\end{figure}

\newpage

\subsection{Images}

\begin{SCfigure}[0.8][h!]
 \centering
 \includegraphics[scale=0.4]{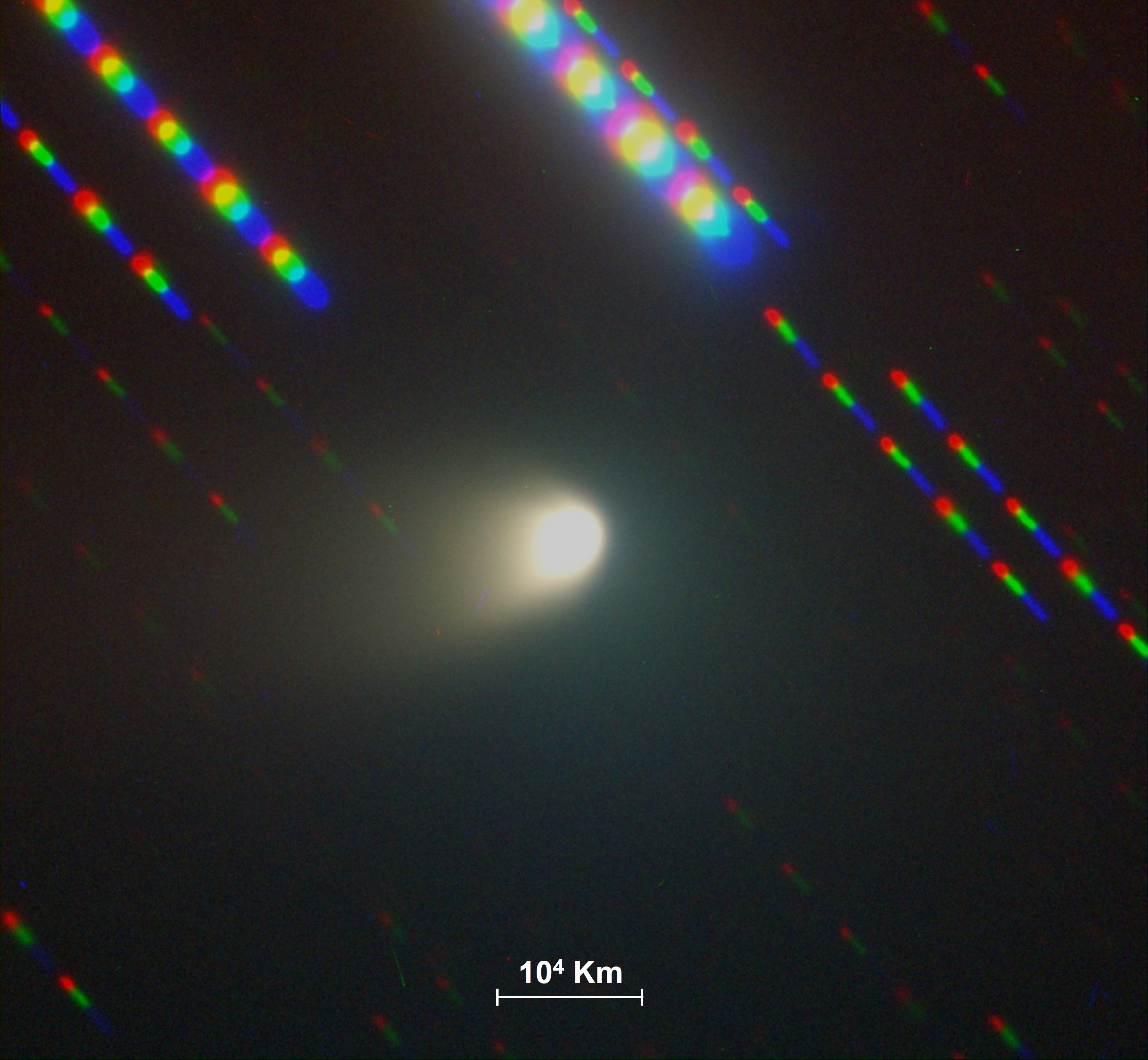}
 \caption{2020-12-16. Three-color (BVr) composite image taken with the Asiago Copernico telescope. The comet is surrounded by a green outer coma due to the fluorescence of C$_2$ and C$_3$ molecules. The dusty tail reflects sunlight. }
\end{SCfigure} 
\begin{SCfigure}[0.8][h!]
 \centering
 \includegraphics[scale=0.4]{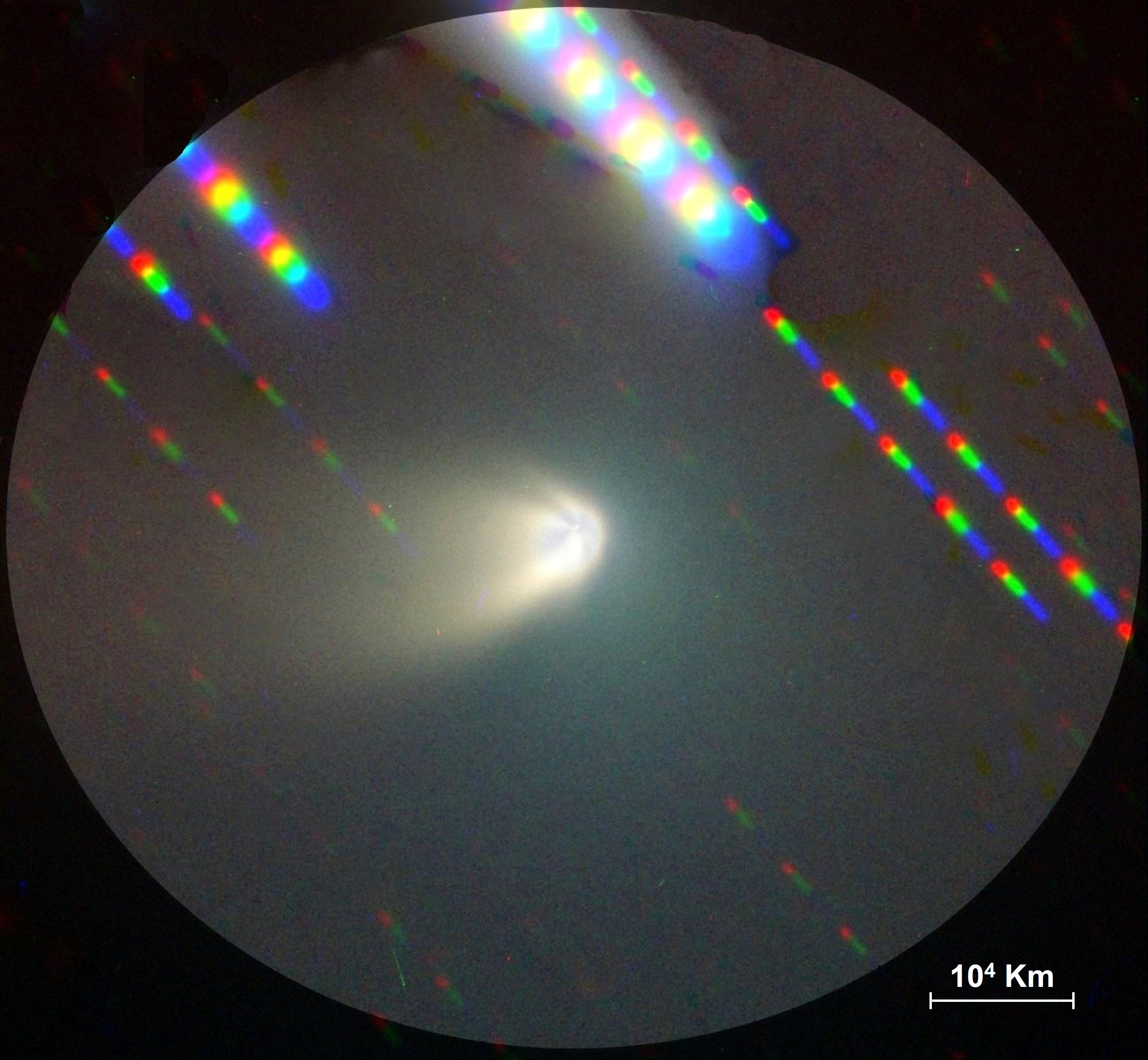}
 \caption{2020-12-16. Three-color (BVr) composite image taken with the Asiago Copernico telescope, processed to highlight the structures in the comet's inner coma.}

\end{SCfigure}

\newpage

\subsection{Spectra}

\begin{table}[h!]
\centering
\begin{tabular}{|c|c|c|c|c|c|c|c|c|c|c|c|}
\hline
\multicolumn{12}{|c|}{Observation details}                      \\ \hline 
\hline

$\#$ & date & r & $\Delta$ & RA & DEC & elong & phase & PLang& config & FlAng & N \\
 & (yyyy-mm-dd) & (AU) & (AU) & (h) & (°) & (°) & (°) & (°) & & (°) &  \\ \hline 
 
1 & 2020-11-20 & 1.333 & 0.537 & 23.82 & $+$01.23  &   118.9 & 40.4 &	$-$12.6 & A & $+$80 & 3 \\ 
2 & 2020-11-24 & 1.335 & 0.553 & 23.92 &    $+$04.28  &   117.4 & 41.0 & $-$14.1 & A & $+$90 & 5 \\ 
3 & 2020-11-27 & 1.338 & 0.568 & 23.98 &    $+$06.52  &   116.2 & 41.4 & $-$15.2 & A & $+$0 & 5 \\ 
4 & 2020-12-13 & 1.367 & 0.662 & 00.48 &    $+$17.18  &   110.7 & 42.4 & $-$18.8 & A & $+$0 & 3 \\ 
5 & 2020-12-26 & 1.407 & 0.760 & 00.98 &    $+$24.22  &   106.9 & 42.0 &	$-$19.5 & A & $+$60 & 5 \\ 
6 & 2020-01-19 & 1.516 & 0.992 & 02.08 &    $+$33.53  &   100.3 & 39.7 &	$-$17.1 & A & $-$0 & 1 \\ 
7 & 2021-01-24 & 1.542 & 1.045 & 02.33 &    $+$34.93  &   98.9  & 39.1 &	$-$16.3 & A & $+$0 & 8 \\ \hline
\end{tabular}
\end{table}

\begin{figure}[h!]

    \centering
    \includegraphics[scale=0.368]{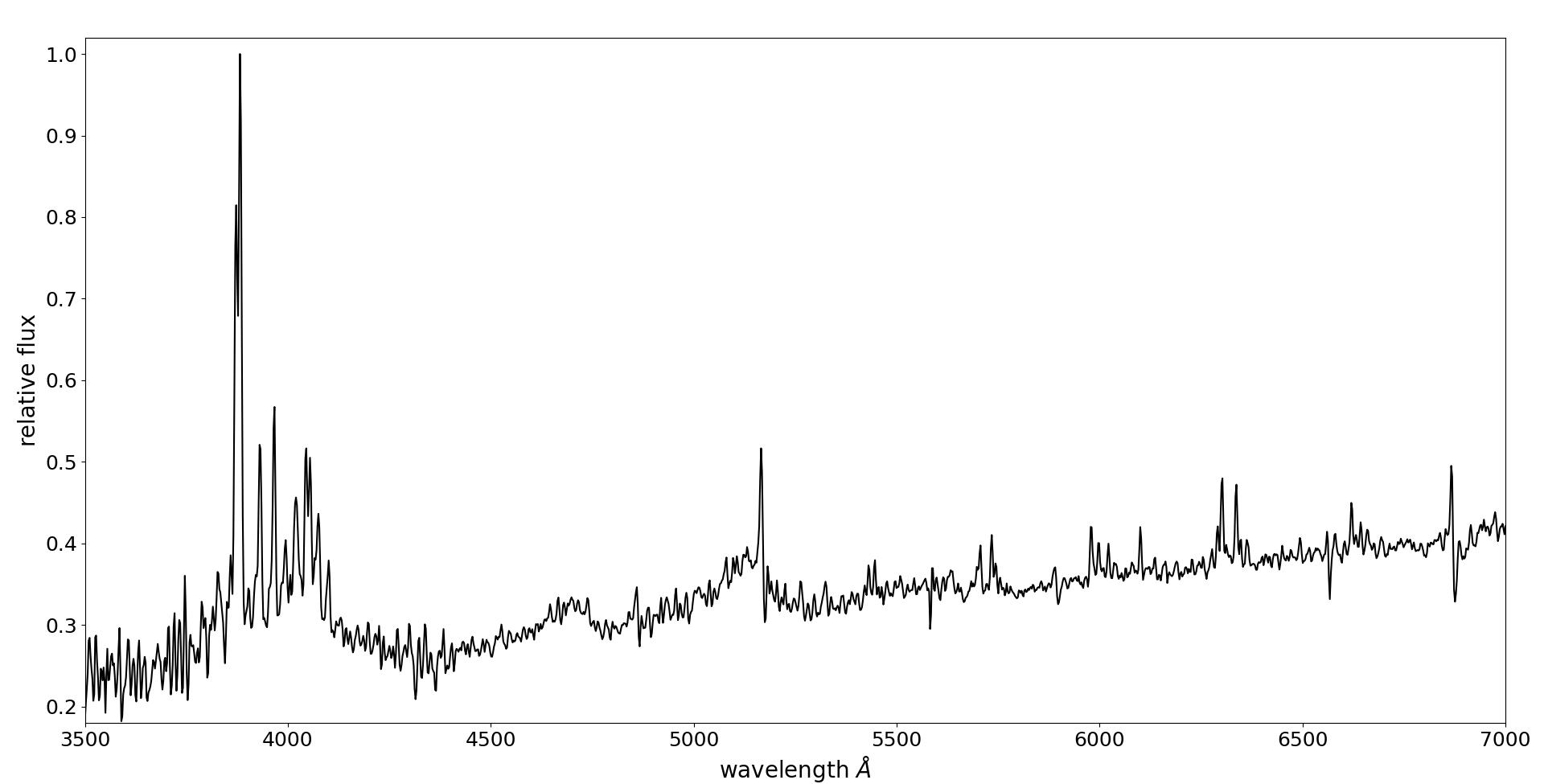}
    \caption{Spectrum of 2020-11-20; configuration A}

\end{figure}




\newpage
\clearpage

\section{237P (LINEAR)}
\label{cometa:237P}
\subsection{Description}

237P/LINEAR is a Jupiter-family comet with a period of 6.58 years and an absolute magnitude of 6.7$\pm$0.8.\footnote{\url{https://ssd.jpl.nasa.gov/tools/sbdb_lookup.html\#/?sstr=237P} visited on July 21, 2024} 
It was first discovered by the automatic LINEAR project on May 17, 2002. In the beginning, it was incorrectly identified as an asteroid by LINEAR and then correctly identified as a comet by the WISE satellite on June 10, 2010. Here are reported the observations of its close approach of 2023, during which the comet reached a distance of 1 AU from Earth and had a magnitude of 12. The spectra show a reddened continuum with no significant emissions, while images highlight an asymmetric shape in the inner coma.

\noindent
We observed the comet around magnitude 12.\footnote{\url{https://cobs.si/comet/302/ }, visited on July 21, 2024} 

\begin{table}[h!]
\centering
\begin{tabular}{|c|c|c|}
\hline
\multicolumn{3}{|c|}{Orbital elements (epoch: December 1, 2021}                      \\ \hline \hline
\textit{e} = 0.4344 & \textit{q} = 1.9866 & \textit{T} = 2460079.0762 \\ \hline
$\Omega$ = 245.3794 & $\omega$ = 25.2355  & \textit{i} = 14.0147 \\ \hline  
\end{tabular}
\end{table}

\begin{table}[h!]
\centering
\begin{tabular}{|c|c|c|c|c|c|c|c|c|}
\hline
\multicolumn{9}{|c|}{Comet ephemerides for key dates}                      \\ \hline 
\hline
& date         & r    & $\Delta$  & RA      & DEC      & elong  & phase  & PLang  \\
& (yyyy-mm-dd) & (AU) & (AU)      & (h)     & (°)      & (°)    & (°)    & (°) \\ \hline 

Perihelion       & 2023-05-15 & 1.987 & 1.326 & 19.85 & $-$11.89 & 115.86 & 27.24 & 2.21  \\ 
Nearest approach & 2023-07-09 & 2.036 & 1.064 & 19.82 & $-$0.51 & 155.12 & 11.67 & $-$8.66\\ \hline
\end{tabular}

\end{table}

\vspace{0.5 cm}

\begin{figure}[h!]
    \centering
    \includegraphics[scale=0.38]{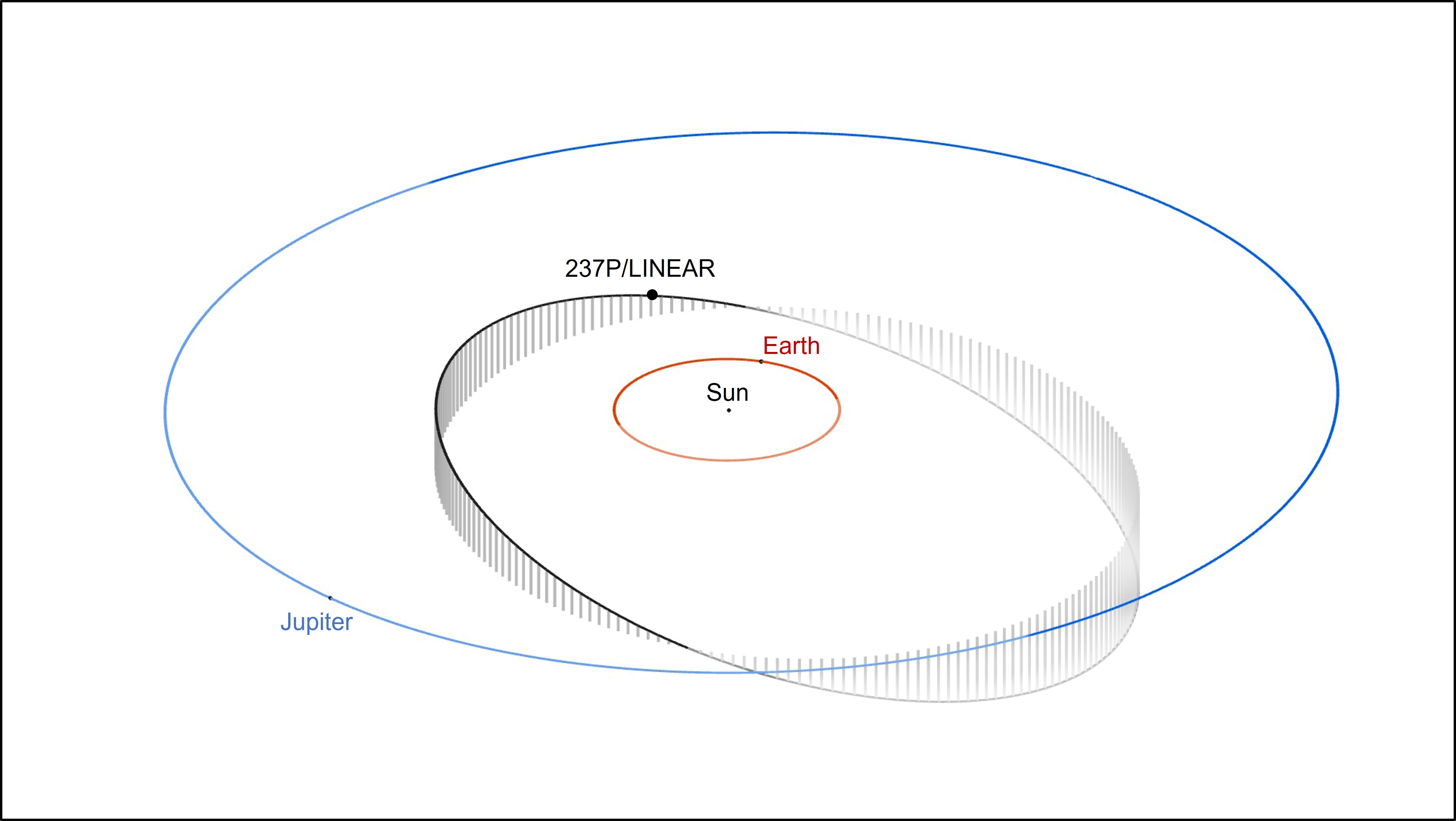}
    \caption{Orbit of comet 237P/Linear and position on perihelion date. The field of view is set to the orbit of Jupiter for size comparison. Courtesy of NASA/JPL-Caltech.}
\end{figure}

\newpage

\subsection{Images}

\begin{SCfigure}[0.8][h!]
    \centering
    \includegraphics[scale=0.23]{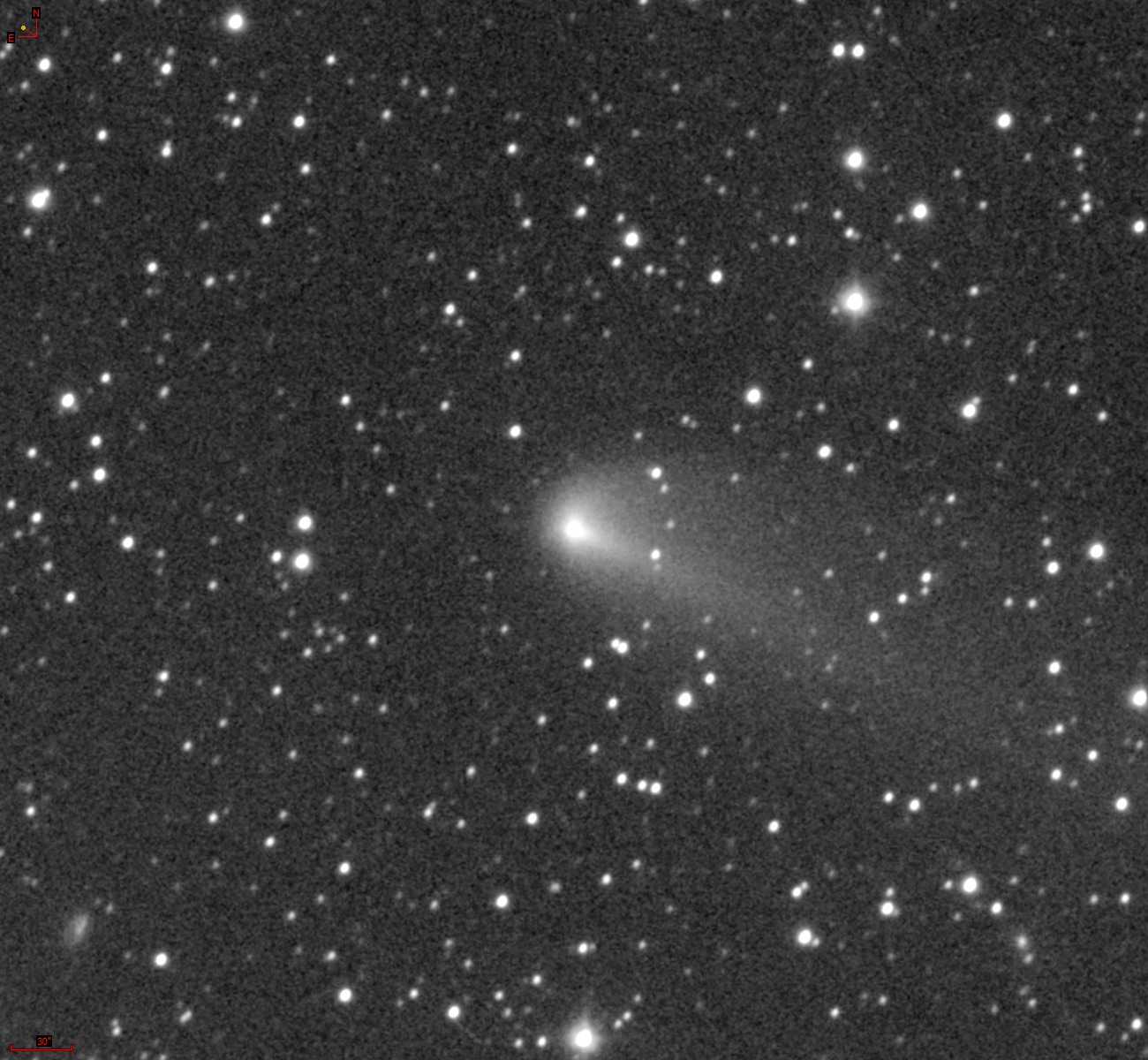}
     \caption{2023-06-18. Unfiltered images of 237P taken at the Asiago Schmidt telescope. An asymmetric morphology of the inner coma is evident.}

\end{SCfigure} 

\begin{SCfigure}[0.8][h!]
    \centering
    \includegraphics[scale=0.23]{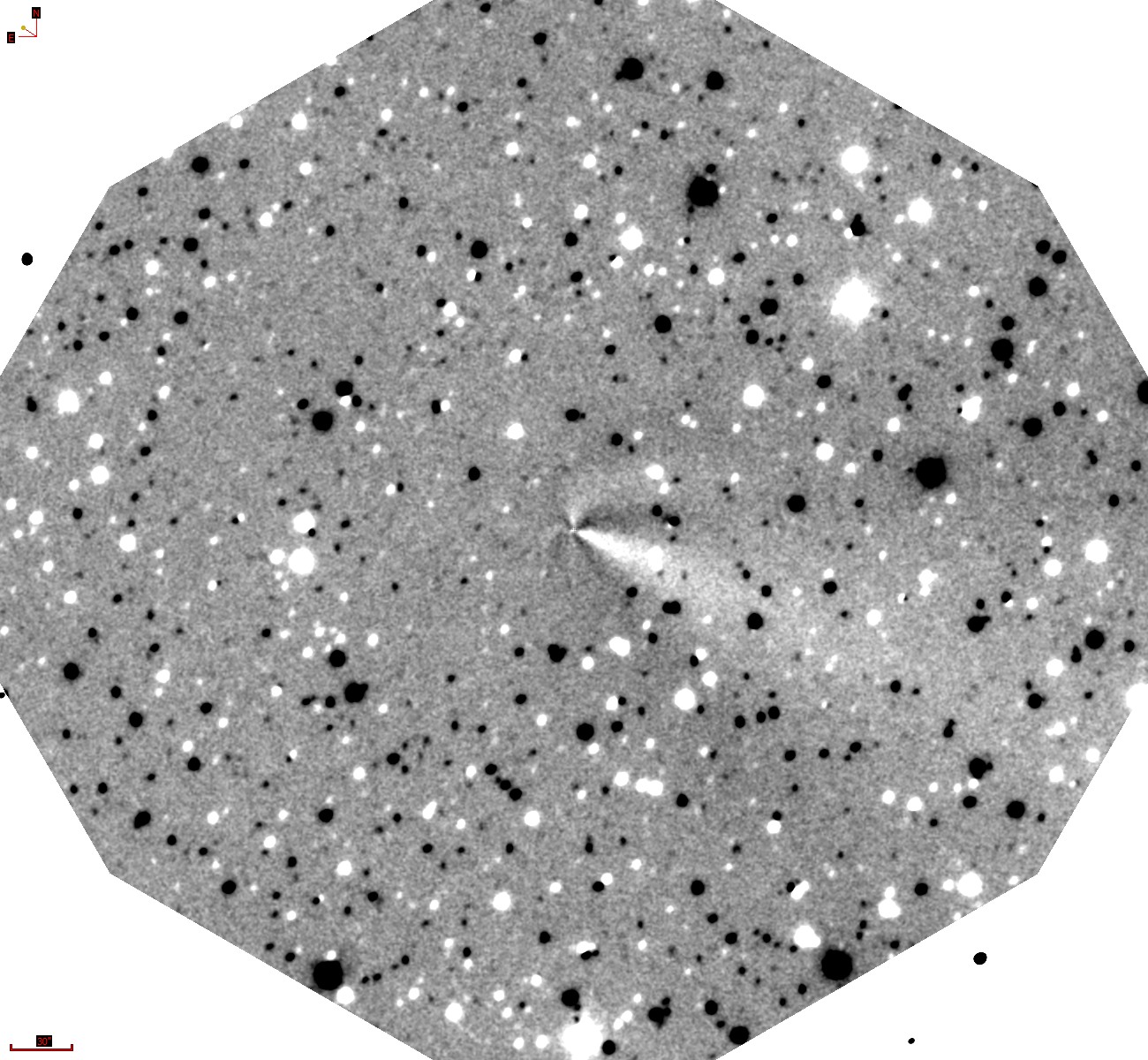}
    \caption{2023-06-18. The previous image was processed with the Larson-Sekanina filter ($\alpha=30$\textdegree). The inner coma is dominated by jet-shaped structures.}

\end{SCfigure}

\newpage

\subsection{Spectra}

\begin{table}[h!]
\centering
\begin{tabular}{|c|c|c|c|c|c|c|c|c|c|c|c|}
\hline
\multicolumn{12}{|c|}{Observation details}                      \\ \hline 
\hline
$\#$  & date          & r     & $\Delta$ & RA     & DEC     & elong & phase & PLang & config  & FlAng & N \\
      & (yyyy-mm-dd)  &  (AU) & (AU)     & (h)    & (°)     & (°)   & (°)   &  (°)   &       &  (°)  &  \\ \hline 

1 & 2023-06-18 & 2.006 & 1.106 & 20.00 & $-$4.00 & 141.9 & 18.2 & $-$4.5 & A & $-$80 & 1 \\

\hline
\end{tabular}
\end{table}

\begin{figure}[h!]
    \centering
    \includegraphics[scale=0.58]{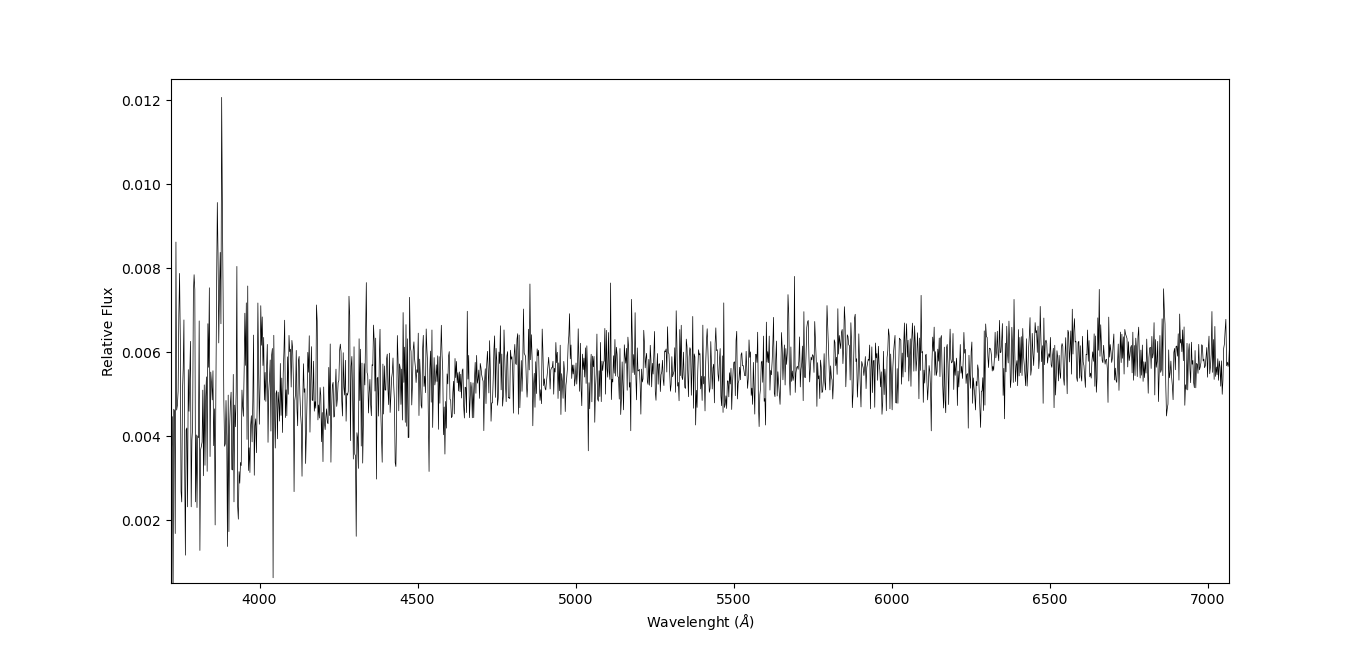}
    \caption{Spectrum of 2023-06-18; configuration A. The flux is normalised to the solar analog Land 107m684.}
\end{figure}

\newpage
\clearpage

\section{260P (McNaught)}
\label{cometa:260P}
\subsection{Description}

260P/McNaught is a Jupiter-family comet with a period of 7.05 years and an absolute magnitude of 12.7$\pm$0.7.\footnote{\url{https://ssd.jpl.nasa.gov/tools/sbdb_lookup.html\#/?sstr=260P} visited on July 21, 2024} 
It was first discovered by Robert McNaught from Siding Spring Observatory on May 20, 2005. It was then rediscovered on May 8, 2012 by Martin Mašek.
The comet was provisionally named P/2005 K3. Earth crossed the orbital plane of the comet on September 12, 2019.

\noindent
We observed the comet around magnitude 11.\footnote{\url{https://cobs.si/comet/754/ }, visited on July 21, 2024}

\begin{table}[h!]
\centering
\begin{tabular}{|c|c|c|}
\hline
\multicolumn{3}{|c|}{Orbital elements (epoch: October 13, 2014)}                      \\ \hline \hline
\textit{e} = 0.5943 &   \textit{q} = 1.4915 &   \textit{T} = 2456183.0058 \\ \hline
$\Omega$ = 351.8632 & $\omega$ = 15.5782 &   \textit{i} = 15.7686 \\ \hline  
\end{tabular}
\end{table}

\begin{table}[h!]
\centering
\begin{tabular}{|c|c|c|c|c|c|c|c|c|}
\hline
\multicolumn{9}{|c|}{Comet ephemerides for key dates}                      \\ \hline 
\hline
& date        & r & $\Delta$ & RA     & DEC    & elong & phase & PLang \\
& (yyyy-mm-dd) & (AU) & (AU)      & (h)     & (°)      & (°)    & (°)    & (°) \\ \hline 
Perihelion       & 2019-09-10 & 1.417  & 0.597 & 02.53 & $+$26.75 & 121.9 & 37.1  & $+$1.1  \\ 
Nearest approach & 2019-10-04 & 1.442 &	0.562 &	02.85 & $+$40.58 &	132.7 &	30.7 &	$-$9.5 \\ \hline
\end{tabular}
\end{table}

\vspace{0.5 cm}

\begin{figure}[h!]
    \centering
    \includegraphics[scale=0.38]{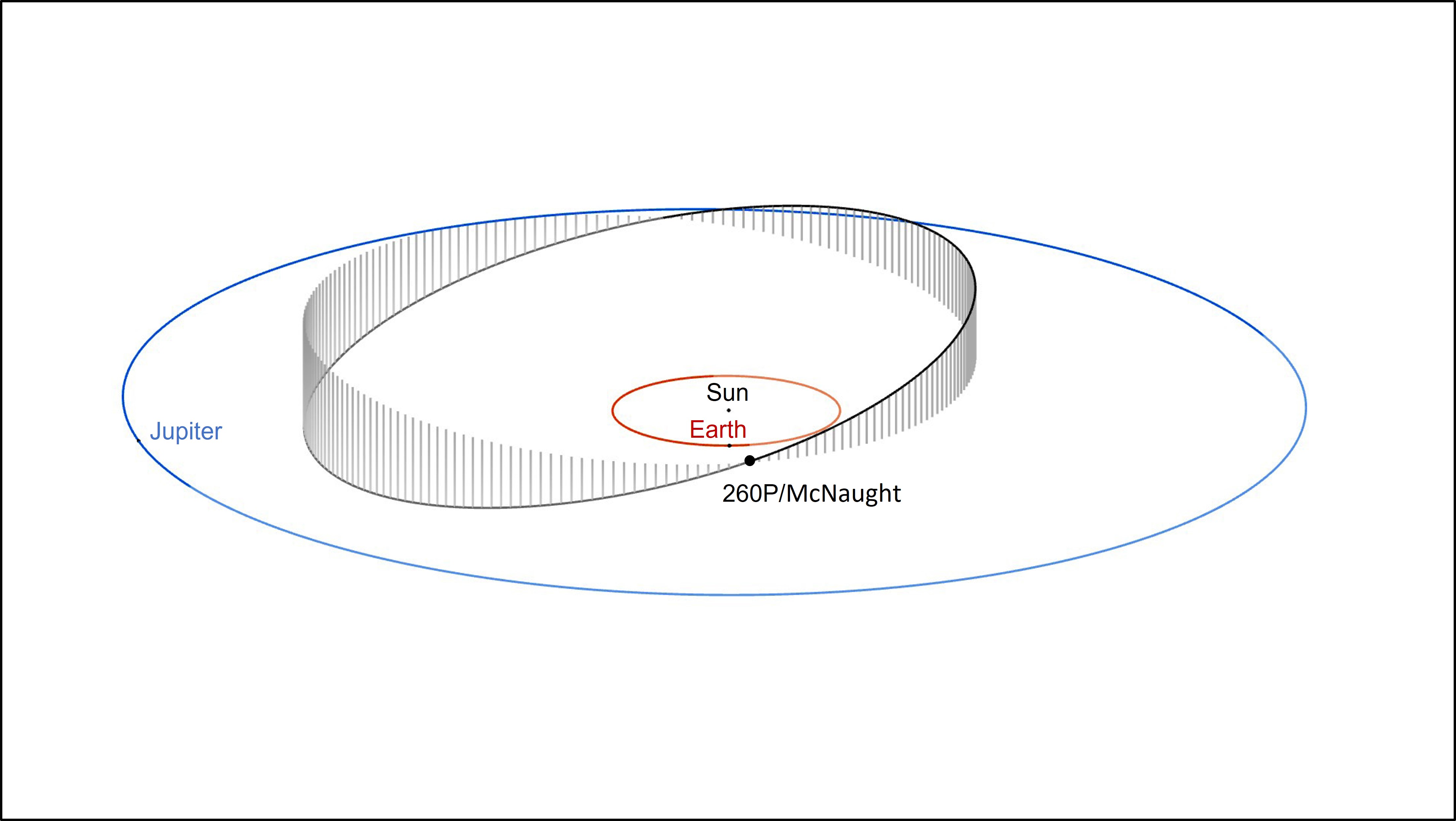}
    \caption{Orbit of comet 260P and position on perihelion date. The field of view is set to the orbit of Jupiter for size comparison. Courtesy of NASA/JPL-Caltech.}
\end{figure} 

\newpage

\subsection{Images}

\begin{SCfigure}[0.8][h!]
    \centering
    \includegraphics[scale=0.4]{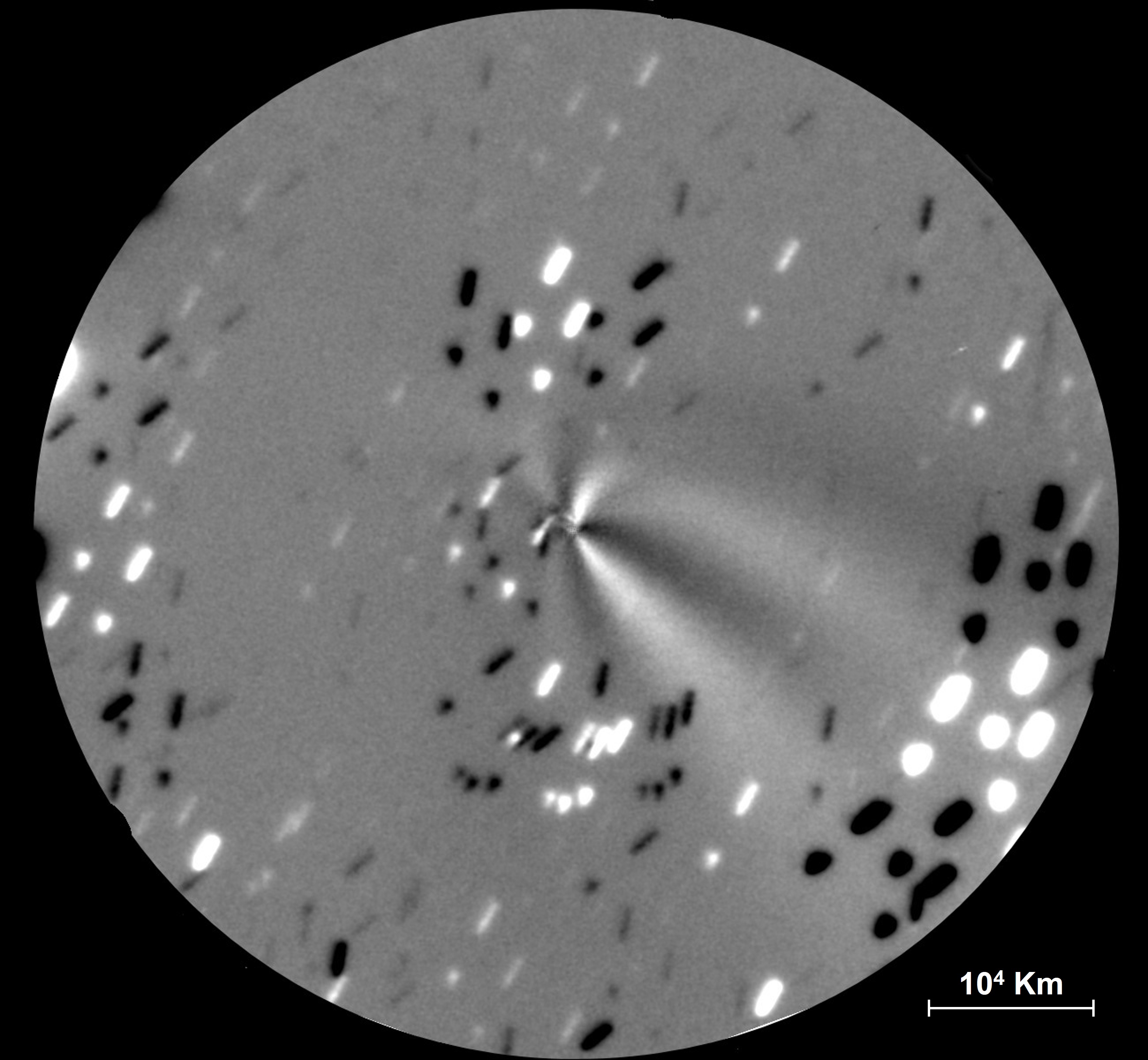}
     \caption{2019-10-23. Image taken with the Asiago Copernico telescope. The processing (Larson-Sekanina algorithm, $\alpha=20$\textdegree) shows a non-isotropic morphology of the comet's inner coma due to the presence of emissions from discrete active areas on the nucleus.}
\end{SCfigure}

\begin{SCfigure}[0.8][h!]
    \centering

    \includegraphics[scale=0.4]{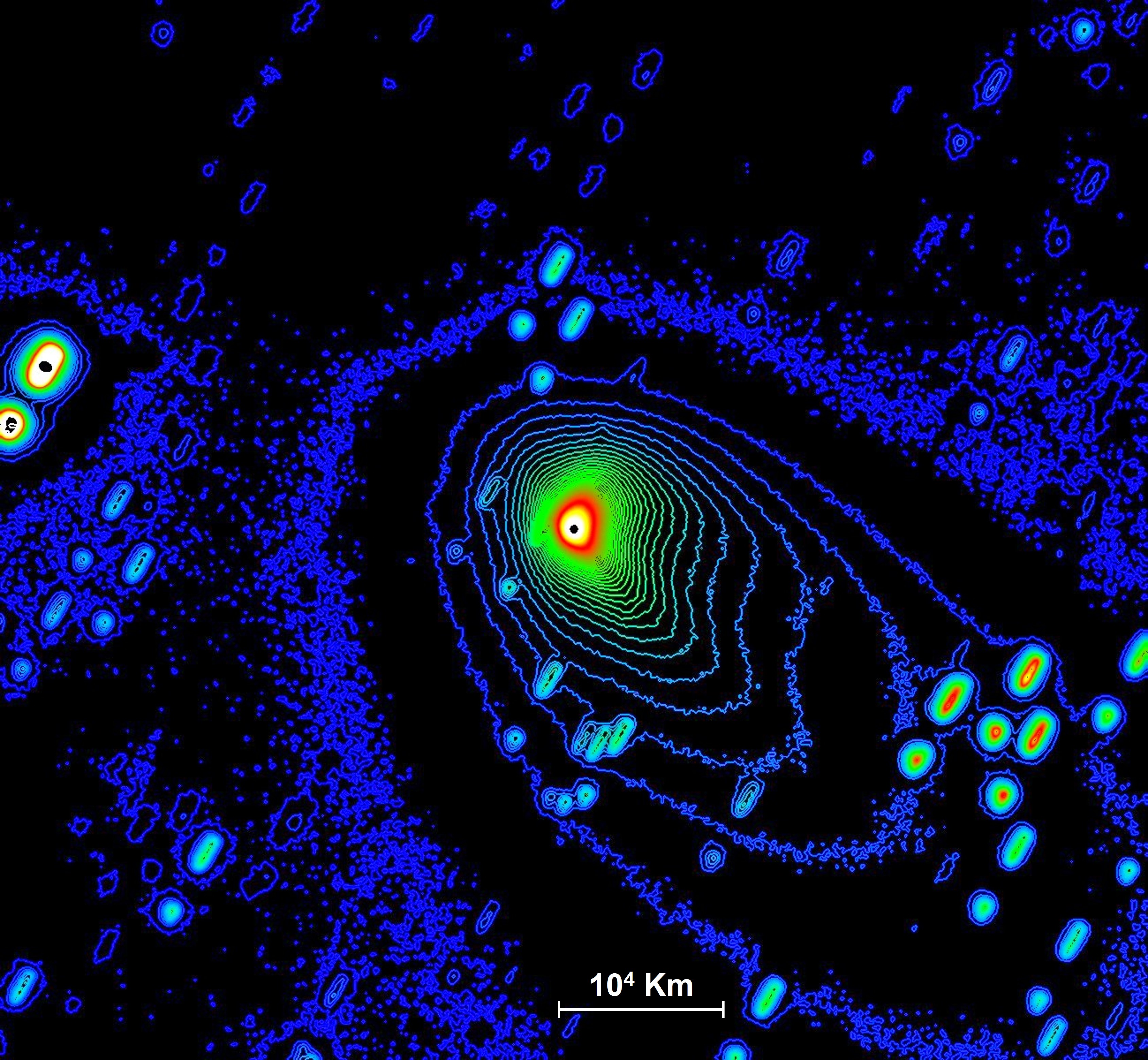}

    \caption{2020-12-16. Image taken with the Asiago Copernico telescope, shown in isophotes in steps of 500 ADU, starting just above the sky background level. An asymmetry of the brightness of the coma is clearly visible.}
\end{SCfigure}

\newpage

\subsection{Spectra}

\begin{table}[h!]
\centering
\begin{tabular}{|c|c|c|c|c|c|c|c|c|c|c|c|}
\hline
\multicolumn{12}{|c|}{Observation details}                      \\ \hline 
\hline
$\#$  & date          & r     & $\Delta$ & RA     & DEC     & elong & phase & PLang& config  & FlAng & N \\
      & (yyyy-mm-dd)  &  (AU) & (AU)     & (h)    & (°)     & (°)   & (°)   &  (°)   &       &  (°)  & \\ \hline 

1 & 2019-08-17 & 1.440 & 0.697 & 01.87 & $+$12.10 & 113.5  & 40.1  & $+$09.1 & A & $+$40 & 1 \\
2 & 2019-09-11 & 1.417  & 0.591 & 02.58 & $+$27.45 & 122.7  & 36.7  & $+$00.3 & A & $+$90 & 3 \\
3 & 2019-09-20 & 1.422  & 0.572 & 02.77 & $+$34.13 & 126.6  & 34.5  & $-$03.7 & A & +90 & 5 \\
4 & 2019-09-23 & 1.425  & 0.568 & 02.78 & $+$35.30 & 128.0  & 33.7  & $-$05.1 & A & +90 & 5 \\
5 & 2019-10-03 & 1.442  & 0.562 & 02.87 & $+$40.65 & 132.6  & 30.7  & $-$09.5 & A & $-$50 & 2 \\
6 & 2019-10-11 & 1.461  & 0.566 & 02.87 & $+$44.25 & 136.4  & 28.1  & $-$12.7 & A & $-$50 & 3 \\
\hline
\end{tabular}
\end{table}

\begin{figure}[h!]

    \centering
    \includegraphics[scale=0.368]{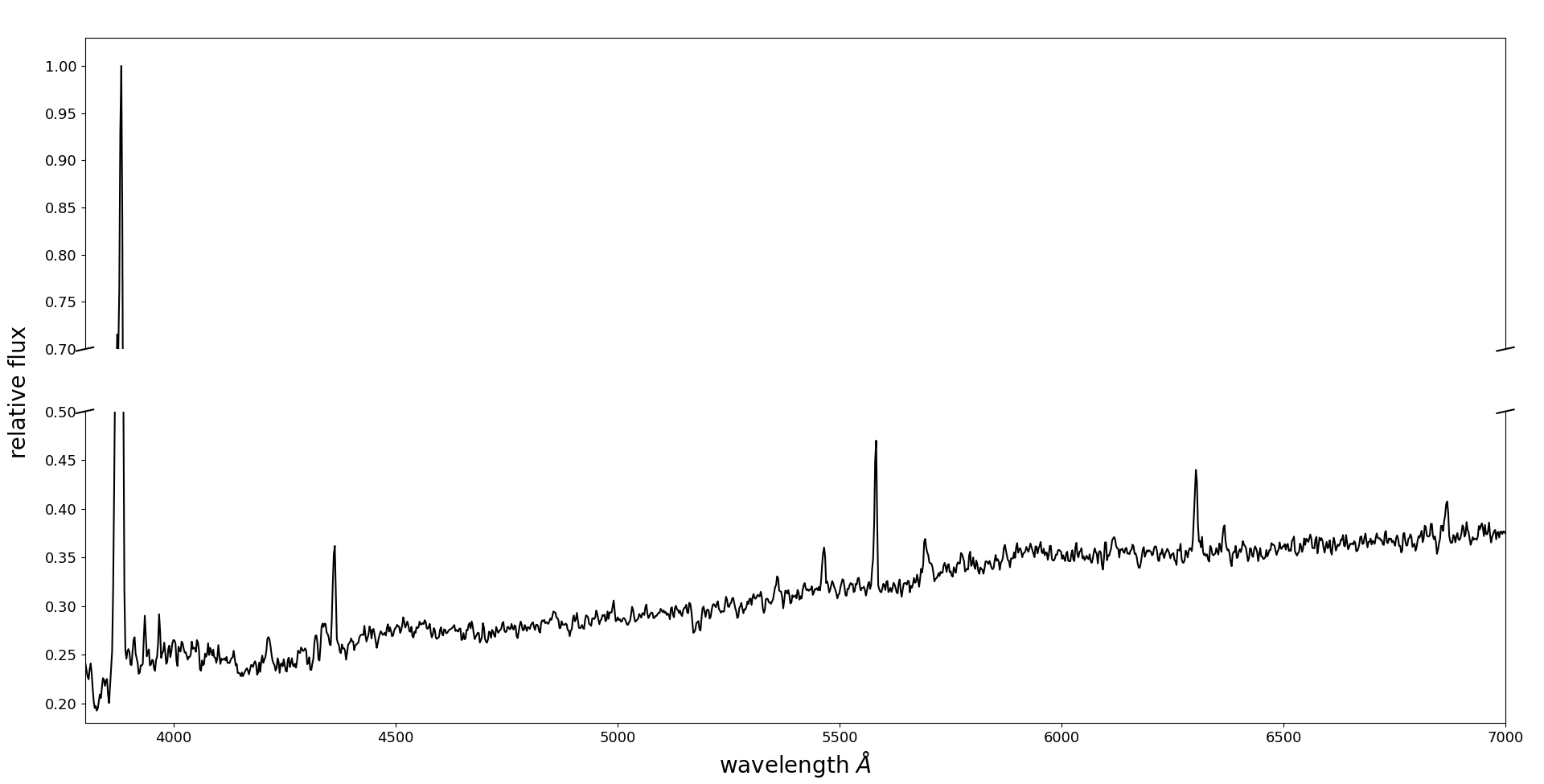}
    \caption{Spectrum of 2019-10-03; configuration A}

\end{figure}

\newpage
\clearpage

\section{C/2012 S1 (ISON)}
\label{cometa:C2012S1}
\subsection{Description}


C/2012 S1 (ISON) is a hyperbolic comet having an absolute magnitude of 10.2$\pm$1.0.\footnote{\url{https://ssd.jpl.nasa.gov/tools/sbdb_lookup.html\#/?sstr=2012\%20S1} visited on July 20, 2024} It was first spotted by Vital Mikalaevič Newski and Artëm Olegovič Novičonok at the International Scientific Optical Network (ISON) in Kislovodsk on September 21, 2012.
\noindent
The great hope of 2013, comet C/2012 S1 (ISON) did not survive its extreme proximity to the Sun, on November 28, 2013. 
In the weeks before it was noticeably weakening and broke a few hours before perihelion. So, nothing came out of the hoped-for spectacular appearance. 
Still, the comet showed many interesting features during its development in the days of the closest proximity to the Sun.
This comet is an example of how the behavior of "new" comets that come directly from the Oort cloud is only partially predictable, as they have not yet been able to probe themselves in the heat of the Sun.
We observed the comet around magnitude 8.\footnote{\url{https://cobs.si/comet/905/ }, visited on July 20, 2024}

\begin{table}[h!]
\centering
\begin{tabular}{|c|c|c|}
\hline
\multicolumn{3}{|c|}{Orbital elements (epoch: May 1, 2013)}                      \\ \hline \hline
\textit{e} = 1.0000051 & \textit{q} = 0.0125 & \textit{T} = 2456625.2645 \\ \hline
$\Omega$ = 295.6865 & $\omega$ =345.5412  & \textit{i} = 62.1629 \\ \hline  
\end{tabular}
\end{table}

\begin{table}[h!]
\centering
\begin{tabular}{|c|c|c|c|c|c|c|c|c|}
\hline
\multicolumn{9}{|c|}{Comet ephemerides for key dates}                      \\ \hline 
\hline
& date         & r    & $\Delta$  & RA      & DEC      & elong  & phase  & PLang  \\
& (yyyy-mm-dd) & (AU) & (AU)      & (h)     & (°)      & (°)    & (°)    & (°) \\ \hline 

Perihelion       & 2013-11-28 & 0.083 & 0.959 & 15.94  & $-$22.73 &  04.6 & 106.9  & $-$44.0  \\ 
\hline
Nearest approach & 2013-11-21 & 0.420 & 0.856 & 14.04  & $-$14.95 &  25.0 & 95.3  & $-$58.9 \\ 
\hline
\end{tabular}

\end{table}

\vspace{0.5 cm}

\begin{figure}[h!]
    \centering
    \includegraphics[scale=0.38]{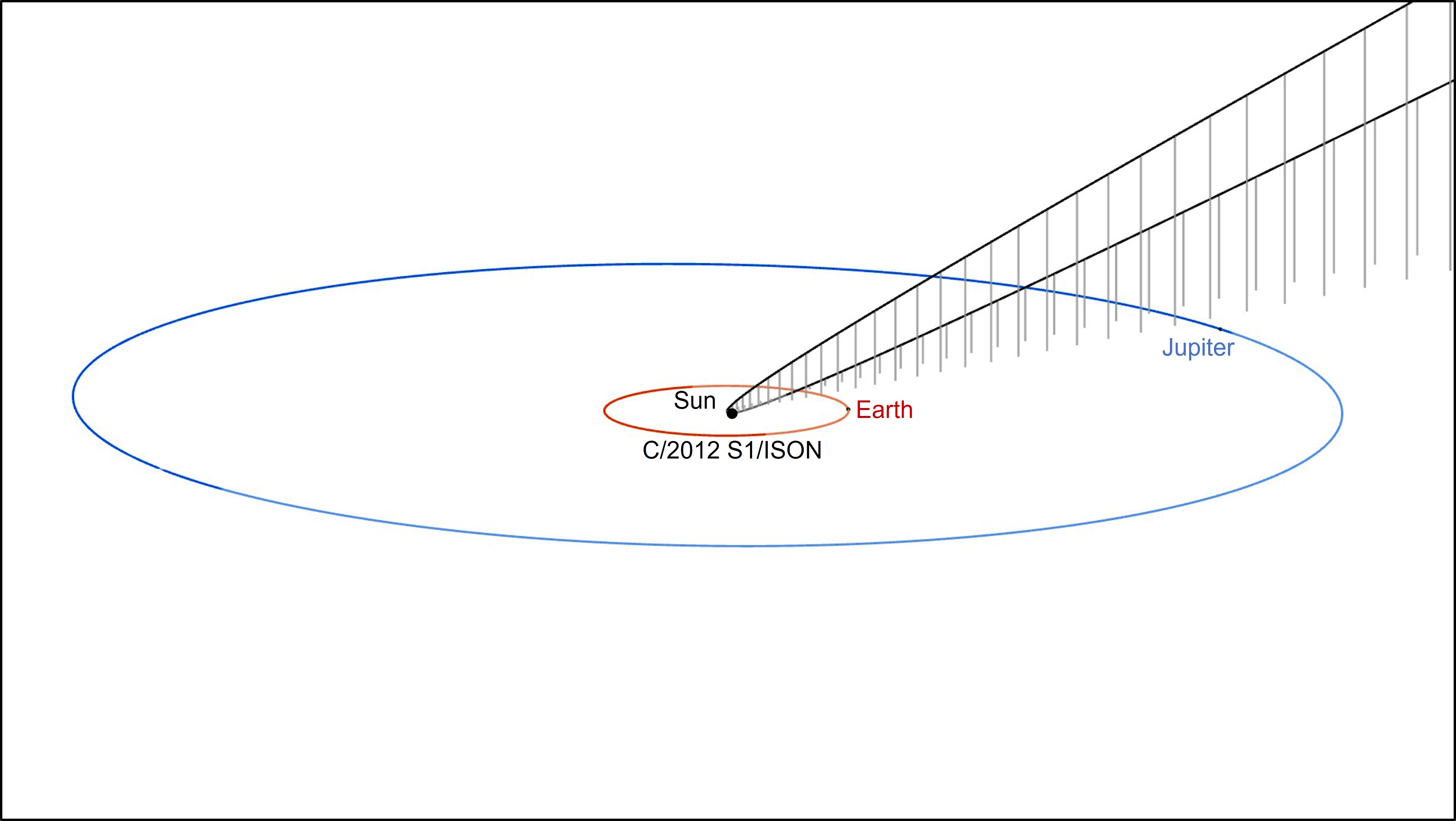}
    \caption{Orbit of comet C/2012 S1 and position on perihelion date. The field of view is set to the orbit of Jupiter for size comparison. Courtesy of NASA/JPL-Caltech.}
\end{figure}

\newpage

\subsection{Images}

\begin{SCfigure}[0.8][h!]
    \centering
    \includegraphics[scale=0.4]{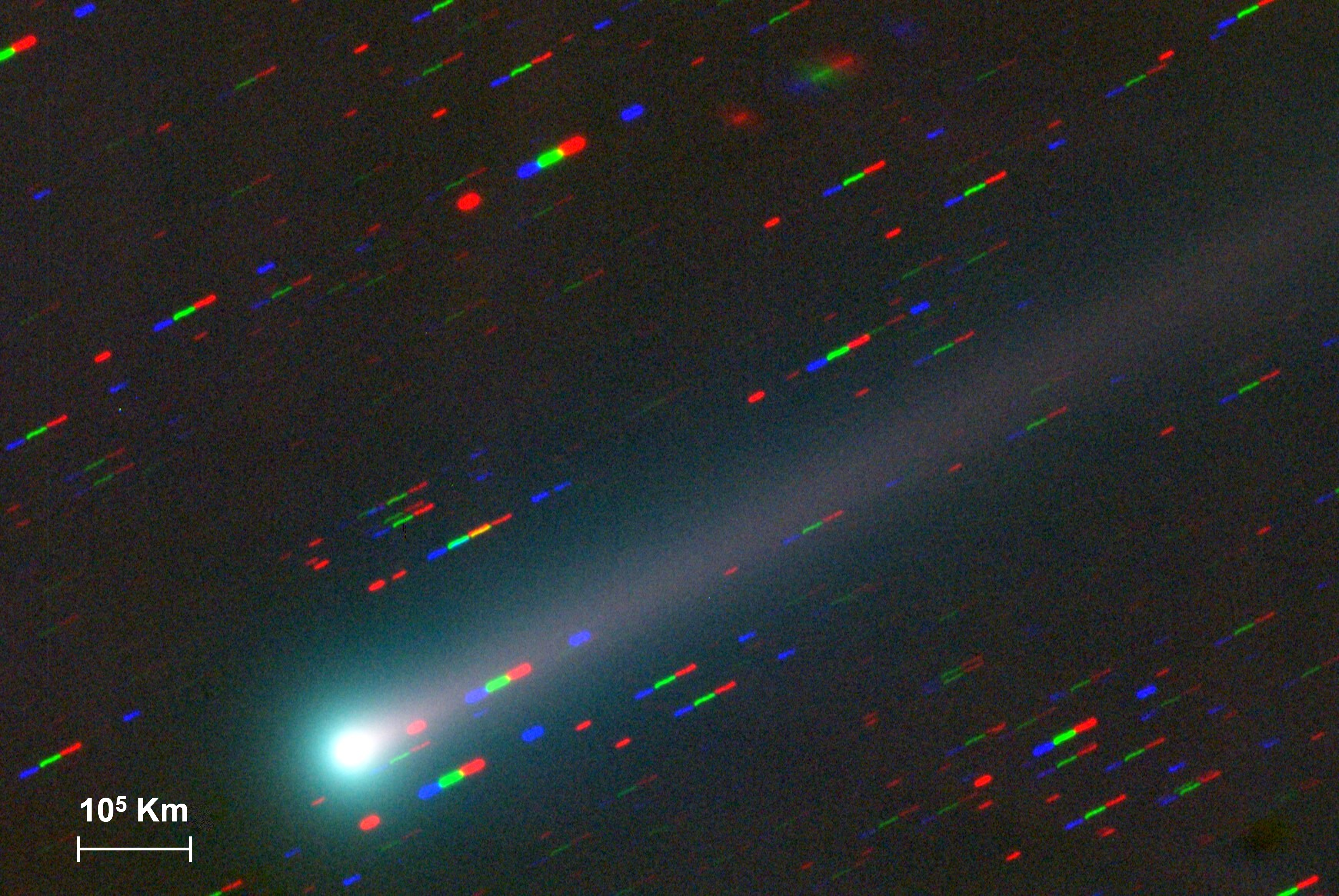}
     \caption{ 2013-11-06. Three-color BVR composite from images taken with the Asiago Schmidt telescope. Comet ISON is projected on the sky with a very long tail whose color is in contrast with that of the coma.
    The tail, which develops towards North-West, is essentially composed of dust that moves away from the nucleus pushed by solar radiation.}
\end{SCfigure} 

\begin{SCfigure}[0.8][h!]
    \centering
    \includegraphics[scale=0.4]{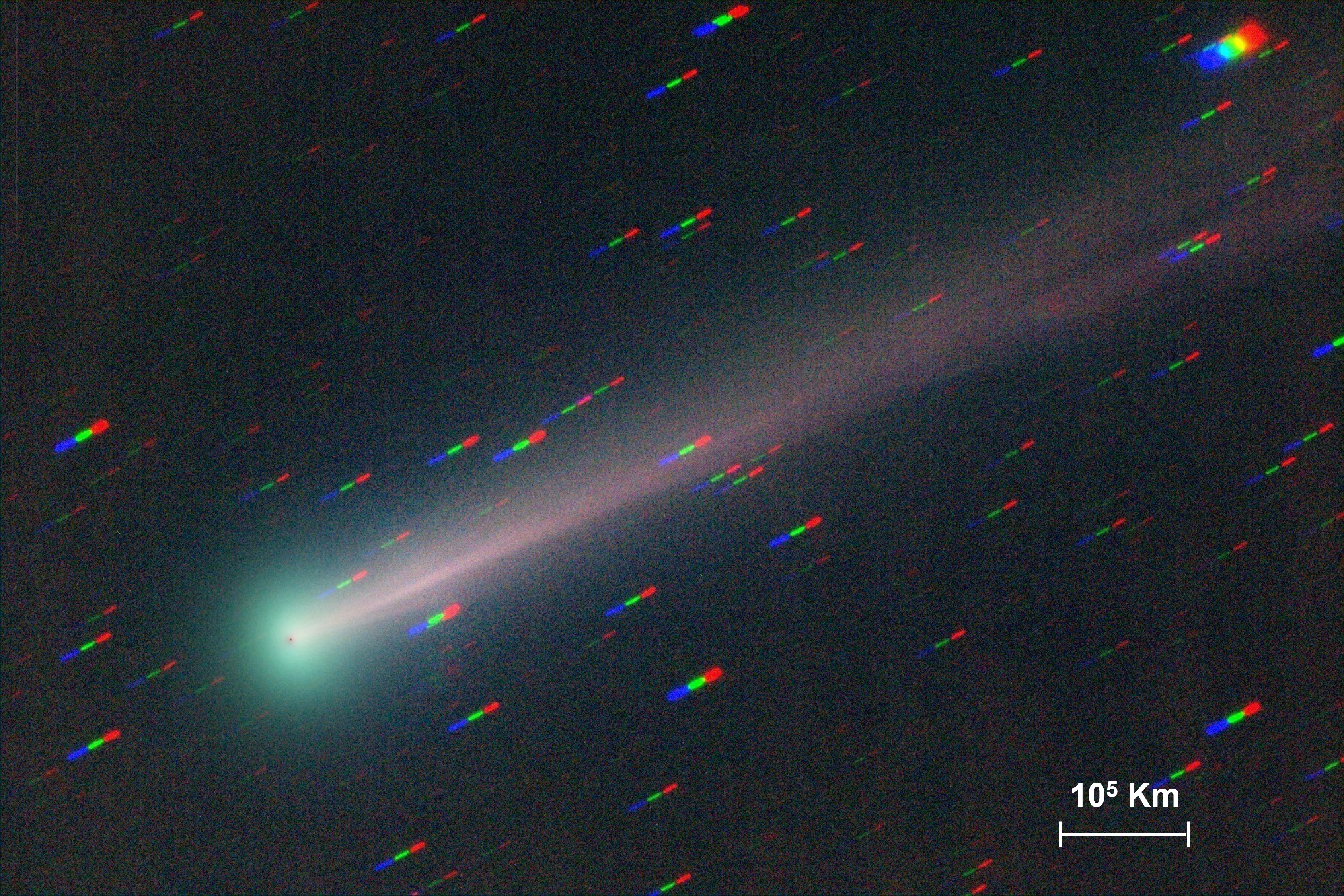}
    \caption{ 2013-11-13. Three-color BVR composite from images taken with the Asiago Schmidt telescope. In less than a week from the image above, comet ISON has approached the Sun by more than 25 million km, thus the nucleus has been further heated.
    The tail appears very long, some tens of millions kilometers. The reddened color is due to the presence of a large amount of micrometer-sized dust particles.
    The coma is green due to the presence of C$_2$ (diatomic carbon) and appears uniform, without a particular morphology.}
\end{SCfigure}

\newpage

\subsection{Spectra}

\begin{table}[h!]
\centering
\begin{tabular}{|c|c|c|c|c|c|c|c|c|c|c|c|}
\hline
\multicolumn{12}{|c|}{Observation details}                      \\ \hline 
\hline
$\#$  & date          & r     & $\Delta$ & RA     & DEC     & elong & phase & PLang& config  & FlAng & N \\
      & (yyyy-mm-dd)  &  (AU) & (AU)     & (h)    & (°)     & (°)   & (°)   &  (°)   &       &  (°)  &  \\ \hline 

1 & 2013-11-07 & 0.841  & 1.067	 & 11.82 & $+$01.55 & 48.1	 & 61.2  & $-$51.0 & A & $+$0 & 4 \\

\hline
\end{tabular}
\end{table}

\begin{figure}[h!]

    \centering
    \includegraphics[scale=0.368]{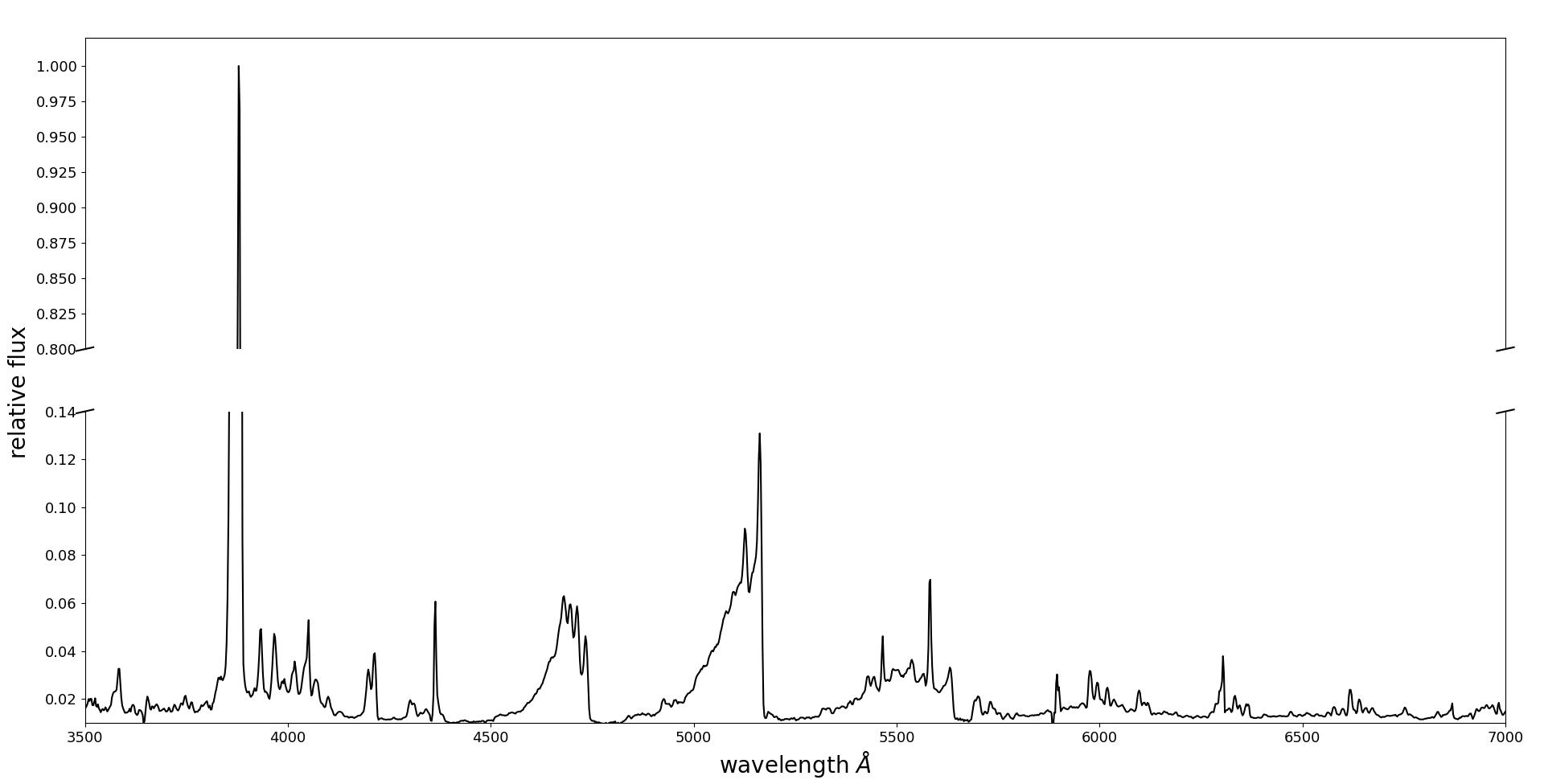}
    \caption{Spectrum of 2013-11-07; configuration A}

\end{figure}

\newpage
\clearpage

\section{C/2013 R1 (Lovejoy)}
\label{cometa:C2013R1}
\subsection{Description}

C/2013 R1 (Lovejoy) is a long period comet with a period of 11705 years and having an absolute magnitude of 11.5$\pm$1.0 \footnote{\url{https://ssd.jpl.nasa.gov/tools/sbdb_lookup.html\#/?sstr=2013\%20R1} visited on July 20, 2024}. It was first spotted by Terry Lovejoy on September 7, 2013 as a magnitude 14 object. He gave the diameter of the condensed coma as 0.5'. 
More observations of Long Period comet C/2013 R1 (Lovejoy) resulted in an overall brightness of magnitude 13.0, which showed a clear central compression and a 1' long, fan-shaped tail at PA=250°. The Earth crossed the comet orbital plane on December 3, 2013.
The comet passed its perihelion, quite close to the Sun, in the Christmas period and approached Earth on November 20, 2013 at 0.40 AU. 
At this point, the comet reached magnitude 7. 

\noindent
We observed the comet around magnitude 4.5.\footnote{\url{https://cobs.si/comet/888/ }, visited on July 20, 2024}

\begin{table}[h!]
\centering
\begin{tabular}{|c|c|c|}
\hline
\multicolumn{3}{|c|}{Orbital elements (epoch: December 29, 2013)} \\ \hline \hline
\textit{e} = 0.9984 & \textit{q} = 0.8118 & \textit{T} = 2456649.2331 \\ \hline
$\Omega$ = 70.7112 & $\omega$ = 67.1664 & \textit{i} = 64.0410 \\ \hline 
\end{tabular}
\end{table}

\begin{table}[h!]
\centering
\begin{tabular}{|c|c|c|c|c|c|c|c|c|}
\hline
\multicolumn{9}{|c|}{Comet ephemerides for key dates} \\ \hline 
\hline
& date & r & $\Delta$ & RA & DEC & elong & phase & PLang \\
& (yyyy-mm-dd) & (AU) & (AU) & (h) & (°) & (°) & (°) & (°) \\ \hline 

Perihelion       & 2013-12-22 & 0.812 & 0.893 & 17.07 & $+$25.71  & 51.0 &	70.3 & $-$19.2  \\
Nearest approach & 2013-11-19 & 1.027 & 0.397 & 11.26 & $+$38.35  & 84.2 & 73.2 & $+$33.1  \\ \hline
\end{tabular}

\end{table}

\vspace{0.5 cm}

\begin{figure}[h!]
    \centering
    \includegraphics[scale=0.38]{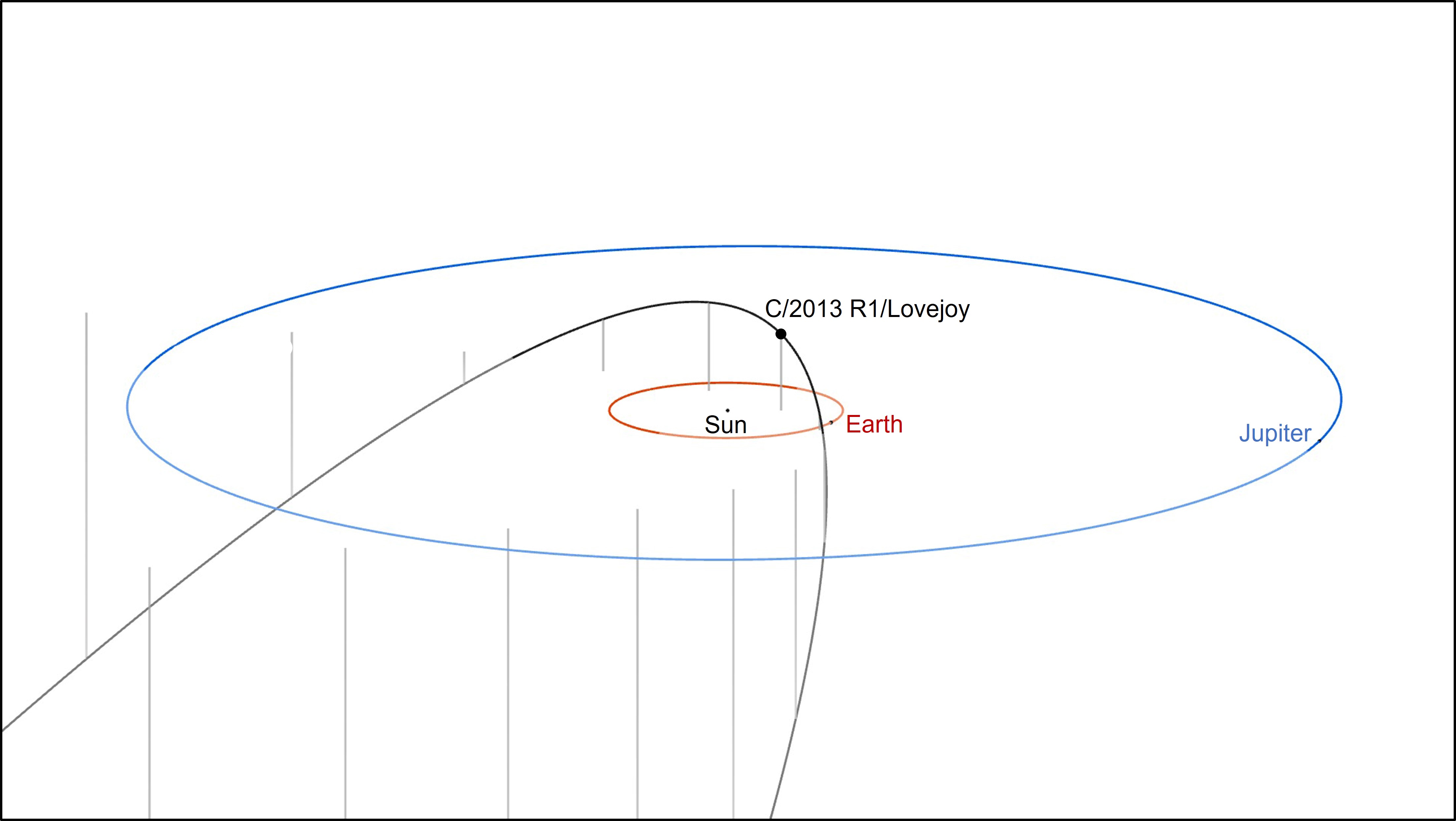}
    \caption{Orbit of comet C/2013 R1 and position on perihelion date. The field of view is set to the orbit of Jupiter for size comparison. Courtesy of NASA/JPL-Caltech.}
\end{figure}

\newpage
\subsection{Images}
\begin{SCfigure}[0.8][h!]
 \centering
 \includegraphics[scale=0.4]{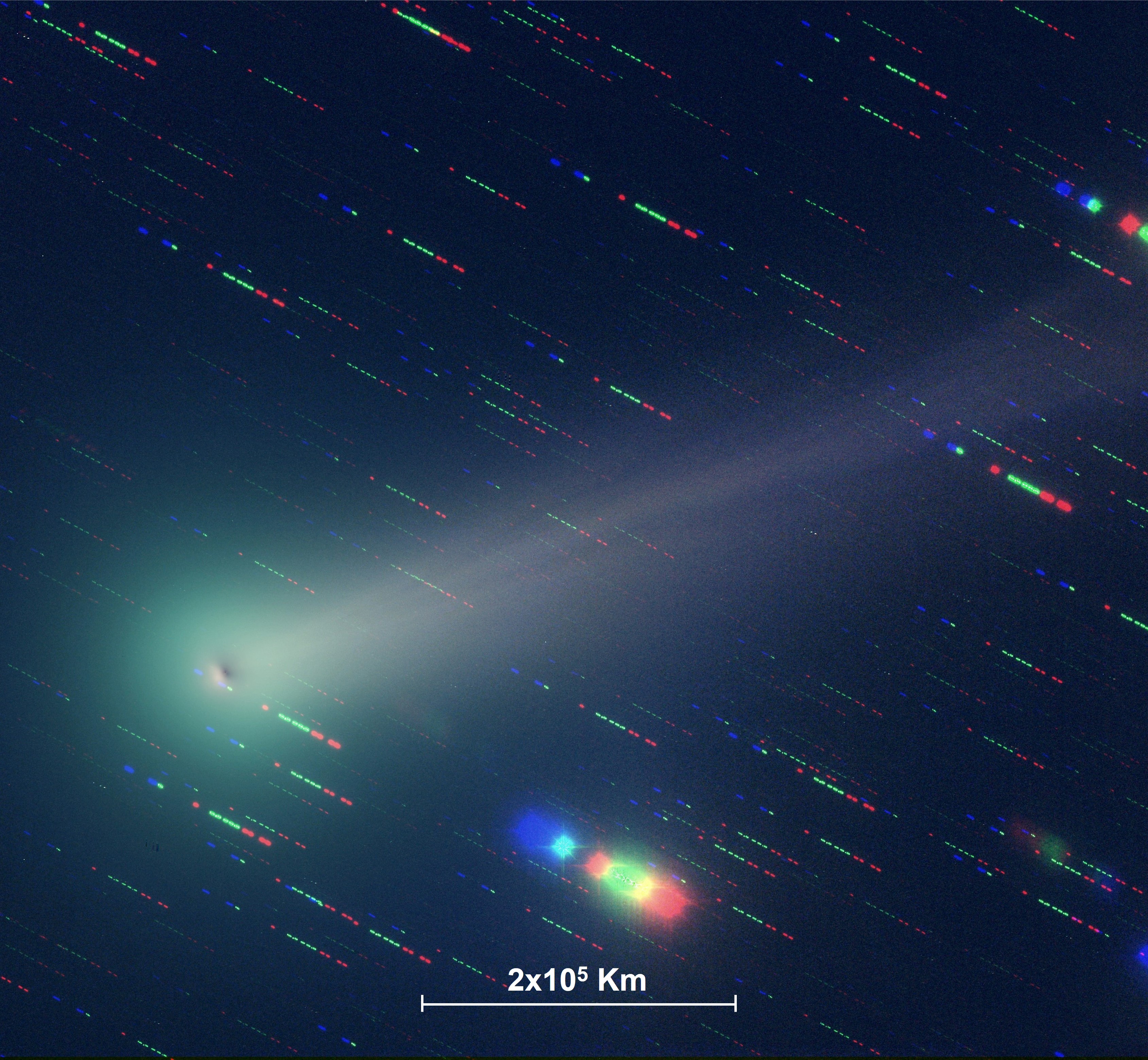}
 \caption{2013.11.16. BVR trichrome from images taken with the Schmidt telescope. Comet Lovejoy was a celestial wonder, with a tail as long as some tens of millions of km, of a contrasting color with that of the coma. The tail is reddened because it is essentially composed of micrometer-sized dust particles pushed away from the nucleus by solar radiation.}
\end{SCfigure} 
\begin{SCfigure}[0.8][h!]
 \centering

 \includegraphics[scale=0.4]{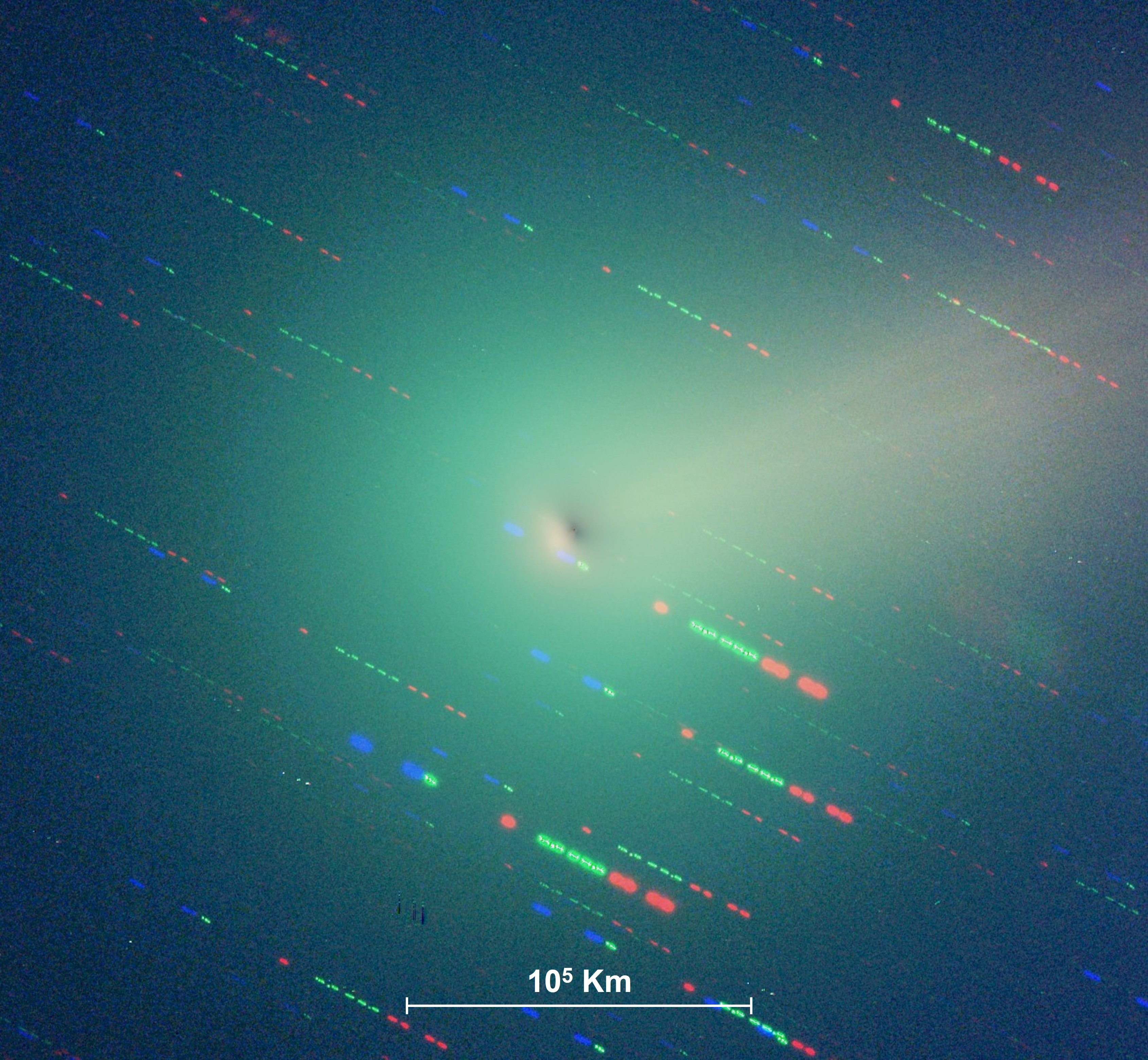}

 \caption{2013.11.16. The coma is green due to the presence of diatomic carbon (C2). An asymmetrical structure can be observed in the sub-solar direction, to the left of the nucleus, which is due to the emissions from active areas placed on the nucleus.}
\end{SCfigure}

\newpage

\subsection{Spectra}
\begin{table}[h!]
\centering
\begin{tabular}{|c|c|c|c|c|c|c|c|c|c|c|c|}
\hline
\multicolumn{12}{|c|}{Observation details} \\ \hline 
\hline
$\#$ & date & r & $\Delta$ & RA & DEC & elong & phase & PLang& config & FlAng & N \\
 & (yyyy-mm-dd) & (AU) & (AU) & (h) & (°) & (°) & (°) & (°) & & (°) &  \\ \hline 
 
1 & 2013-11-29 & 0.925 & 0.466 & 14.40 & $+$41.72 & 68.7  & 83.3 & $+$07.29 & B & $+$0 & 3 \\

\hline
\end{tabular}
\end{table}

\begin{figure}[h!]
 \centering
 \includegraphics[scale=0.368]{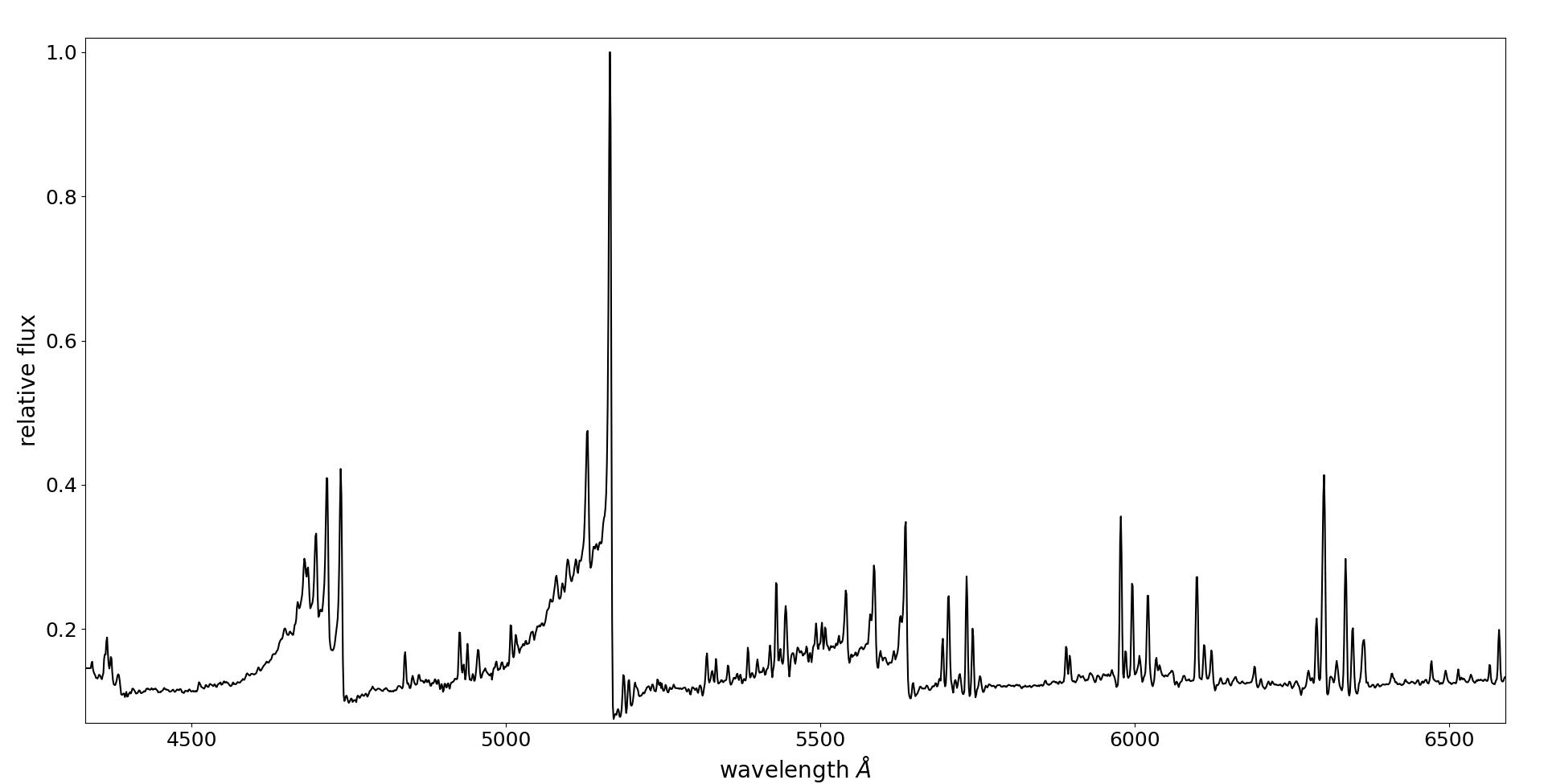}
 \caption{Spectrum of 2013-11-29; configuration B}
\end{figure}

\newpage
\clearpage

\section{C/2013 US10 (Catalina)}
\label{cometa:C2013US10}
\subsection{Description}

C/2013 US10 (Catalina) is an Oort cloud hyperbolic comet having an absolute magnitude of 8.4$\pm$1.0.\footnote{\url{https://ssd.jpl.nasa.gov/tools/sbdb_lookup.html\#/?sstr=2013\%20US10} visited on July 20, 2024} It was first spotted by the Catalina Sky Survey, which uses a 0.68m telescope, on October 31, 2013.

\noindent 
C/2013 US10 is dynamically new. It came from the Oort cloud with a loosely bound chaotic orbit that was easily perturbed by galactic tides and passing stars. 
Before entering the planetary region (around the 1950s), C/2013 US10 had an orbital period of several million years. 
After leaving the planetary region (epoch 2050), it will be on an ejection trajectory. The Earth crossed the comet orbital plane on March 27, 2016.

\noindent
We observed the comet around magnitude 6.\footnote{\url{https://cobs.si/comet/905/ }, visited on July 20, 2024}

\begin{table}[h!]
\centering
\begin{tabular}{|c|c|c|}
\hline
\multicolumn{3}{|c|}{Orbital elements (epoch: November 23, 2015)}                      \\ \hline \hline
\textit{e} = 1.0003 & \textit{q} = 0.8230 & \textit{T} = 2457342.2216 \\ \hline
$\Omega$ = 186.1446 & $\omega$ = 340.3588  & \textit{i} = 148.8783  \\ \hline  
\end{tabular}
\end{table}

\begin{table}[h!]
\centering
\begin{tabular}{|c|c|c|c|c|c|c|c|c|}
\hline
\multicolumn{9}{|c|}{Comet ephemerides for key dates}                      \\ \hline 
\hline
& date & r & $\Delta$ & RA & DEC & elong & phase & PLang \\
& (yyyy-mm-dd) & (AU) & (AU) & (h) & (°) & (°) & (°) & (°) \\ \hline

Perihelion       & 2015-11-16 & 0.823  & 1.743 & 14.37 & $-$18.98  & 14.4  & 17.4  & $-$12.4 \\
Nearest approach & 2016-01-17 & 1.385  & 0.725 & 13.81 & $+$54.32  & 107.4 & 42.7  & $-$41.3 \\ \hline
\end{tabular}

\end{table}

\vspace{0.5 cm}

\begin{figure}[h!]
    \centering
    \includegraphics[scale=0.38]{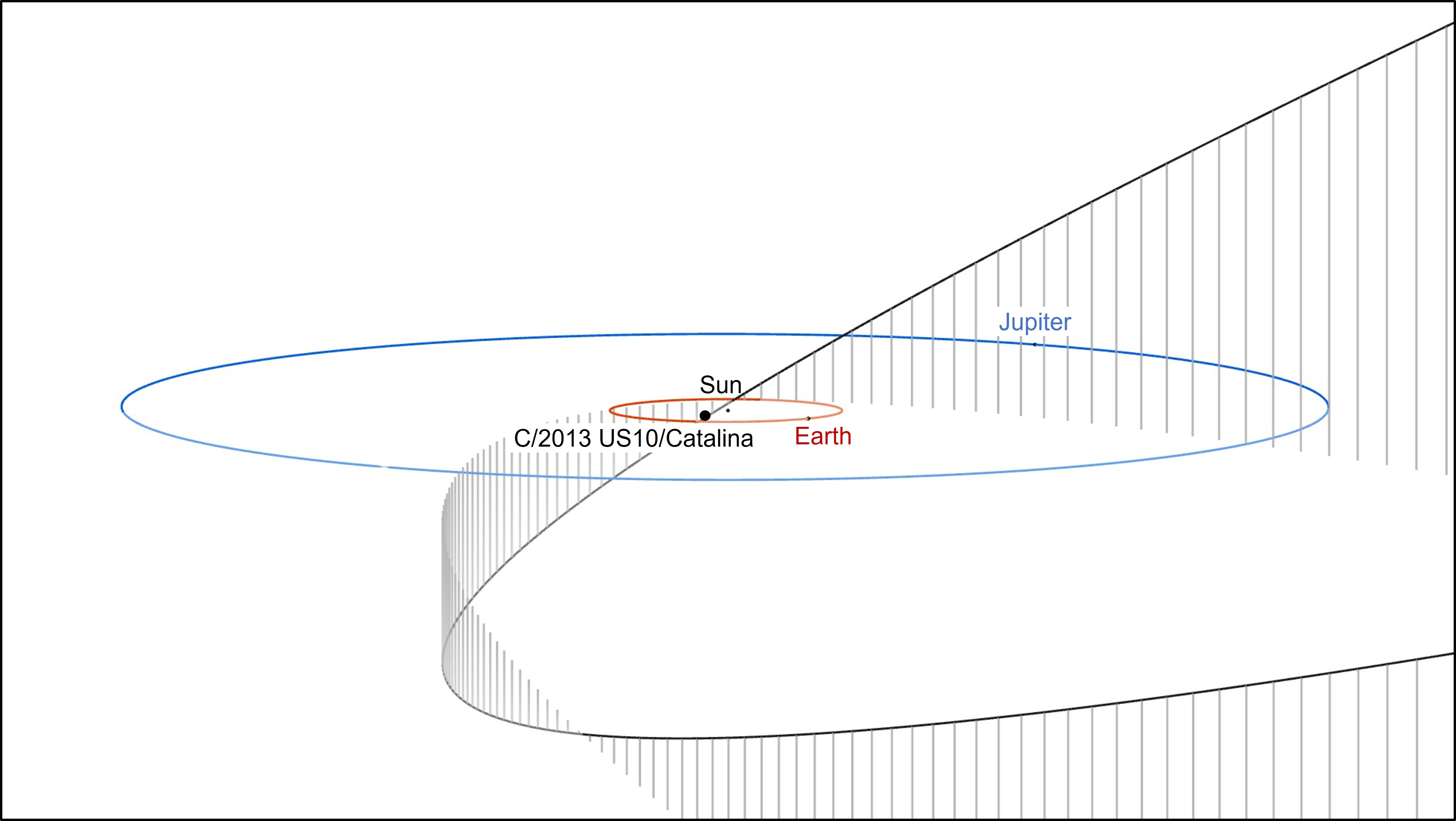}
    \caption{Orbit of comet C/2013 US10 and position on perihelion date. The field of view is set to the orbit of Jupiter for size comparison. Courtesy of NASA/JPL-Caltech.}
\end{figure}

\newpage

\subsection{Images}
\begin{SCfigure}[0.8][h!]
 \centering
 \includegraphics[scale=0.4]{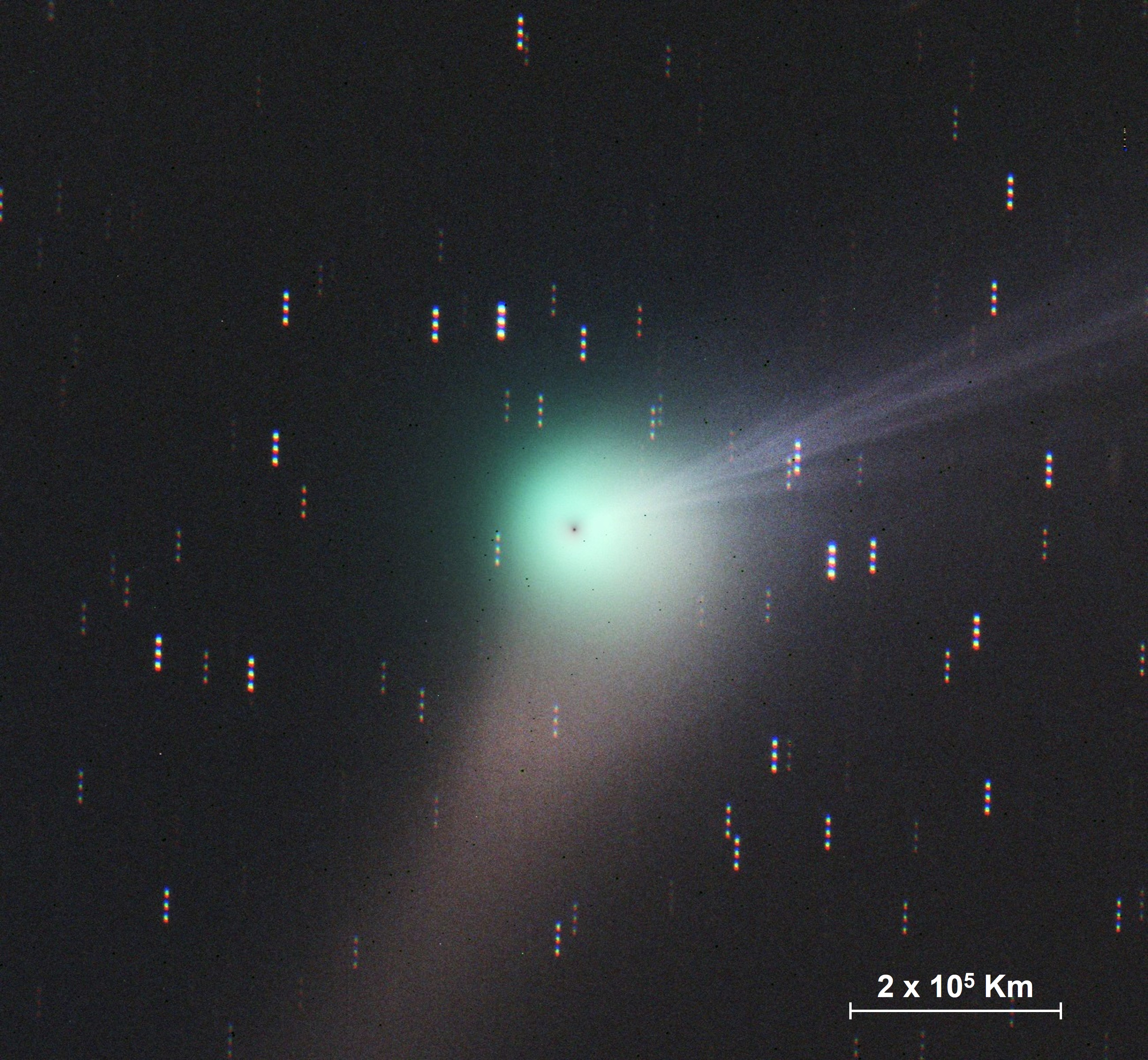}
 \caption{2015-12-10. Three-color BVR composite from images taken with the Asiago Schmidt telescope. Comet Catalina was a real celestial spectacle, because the combination of the geometric conditions of the Earth's orbit and that of the comet made it possible to observe the development of its two tails. Of the two tails, the blue one is composed of molecules ionized by the UV radiation from the Sun, instead the other is reddened because is composed of micrometer-sized dust particles that move away from the nucleus in an anti-solar direction. }
\end{SCfigure} 
\begin{SCfigure}[0.8][h!]
 \centering
 \includegraphics[scale=0.4]{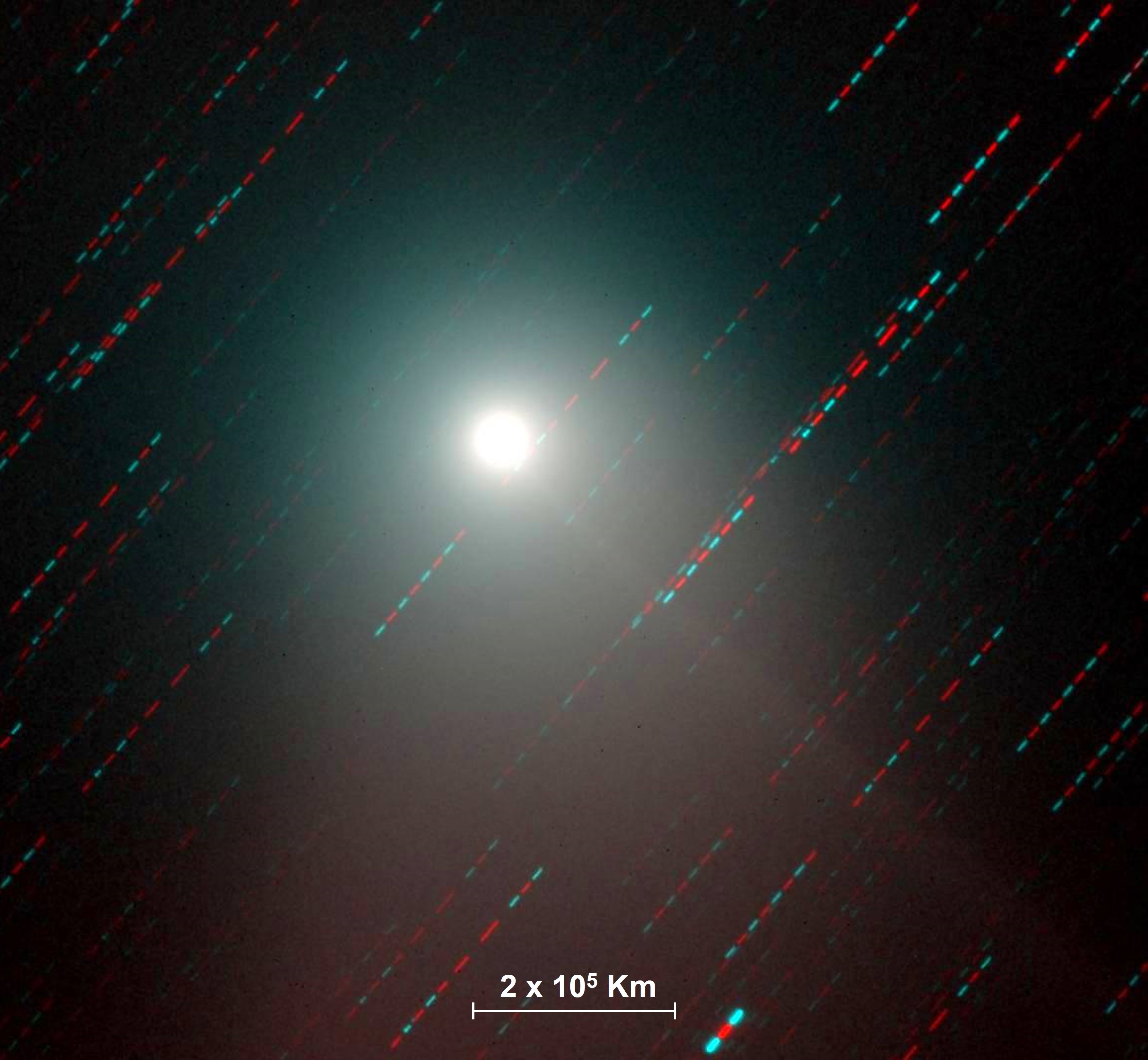}
 \caption{2016-01-26. The coma tend to have a green color due to the presence of emissions by fluorescence of diatomic carbon (C$_2$).}
\end{SCfigure}

\newpage

\subsection{Spectra}

\begin{table}[h!]
\centering
\begin{tabular}{|c|c|c|c|c|c|c|c|c|c|c|c|}
\hline
\multicolumn{12}{|c|}{Observation details}                      \\ \hline 
\hline
$\#$ & date & r & $\Delta$ & RA & DEC & elong & phase & PLang& config & FlAng & N \\
 & (yyyy-mm-dd) & (AU) & (AU) & (h) & (°) & (°) & (°) & (°) & & (°) &  \\ \hline

1*              & 2016-01-20 &	1.441 &	0.737 &	13.45  & $+$64.42  &	112.9 &	39.0 &	$-$39.0 & A & $+$0 & 3\\ \hline 
2*              & 2016-01-25 &	1.493 &	0.775 &	12.67  & $+$73.85  &	115.8 &	36.4 &	$-$35.3 & A & $+$90 & 1\\ \hline

\end{tabular}
\end{table}

\begin{figure}[h!]

    \centering
    \includegraphics[scale=0.368]{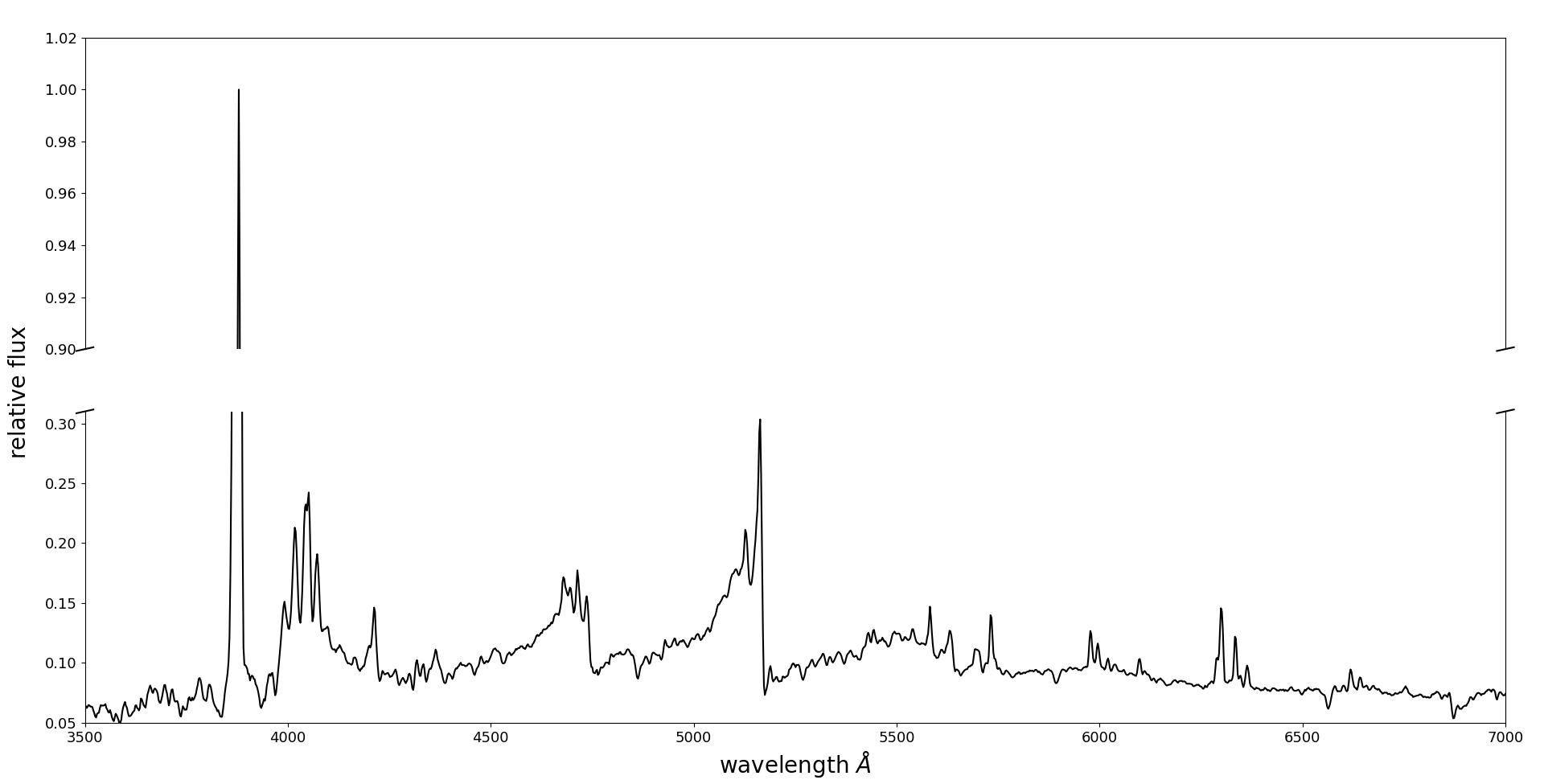}
    \caption{Spectrum of 2016-01-20; configuration A}

\end{figure}

\footnotetext{* No Solar Analog}

\newpage
\clearpage

\section{C/2013 X1 (PanSTARRS)}
\label{cometa:C2013X1}
\subsection{Description}

C/2013 X1 (PanSTARRS) is an hyperbolic Oort Cloud comet with an absolute magnitude of 10.7 $\pm$ 2.0.\footnote{\url{https://ssd.jpl.nasa.gov/tools/sbdb_lookup.html\#/?sstr=2013\%20X1} visited on July 20, 2024} It was first spotted by the 1.8m Panoramic Survey Telescope \& Rapid Response System (Pan-STARRS 1) on December 4, 2013.
It reached the perihelion on April 20, 2016 at a distance of 1.314 AU, and over a period of months many comet chasers and astrophotographers observed and imaged the celestial body. Comet PanSTARRS passed around 0.6 AU from Earth in mid-June 2016. We fully characterized the morphology of the inner coma of the comet with the aid of the Asiago Astrophysical Observatory data. 
The Earth crossed the comet orbital plane on February 2, 2016 and on August 3, 2016.

\noindent
We observed the comet around magnitude 8.\footnote{\url{https://cobs.si/comet/915/}, visited on July 20, 2024}

\begin{table}[h!]
\centering
\begin{tabular}{|c|c|c|}
\hline
\multicolumn{3}{|c|}{Orbital elements (epoch: December 20, 2015)}                      \\ \hline \hline
\textit{e} = 1.0010 &   \textit{q} = 1.3142 &    \textit{T} = 2457499.2287  \\ \hline
$\Omega$ = 130.9549  & $\omega$ = 164.4607 &    \textit{i} = 163.2319  \\ \hline 
\end{tabular}
\end{table}

\begin{table}[h!]
\centering
\begin{tabular}{|c|c|c|c|c|c|c|c|c|}
\hline
\multicolumn{9}{|c|}{Comet ephemerides for key dates}                      \\ \hline 
\hline
& date        & r & $\Delta$ & RA     & DEC    & elong & phase & PLang \\ 
& (yyyy-mm-dd) & (AU) & (AU)      & (h)     & (°)      & (°)    & (°)    & (°) \\ \hline 
Perihelion	 & 2016-04-20 & 1.314 &	1.977 &	23.53 & $+$00.30	& 36.6 & 27.1 &	$+$08.2 \\
Nearest approach & 2016-06-22 & 1.603 &	0.640 & 19.70 & 	$-$46.03 &	149.9 & 18.5 &	$+$17.1 \\ \hline
\end{tabular}

\end{table}

\vspace{0.5 cm}

\begin{figure}[h!]
    \centering
    \includegraphics[scale=0.38]{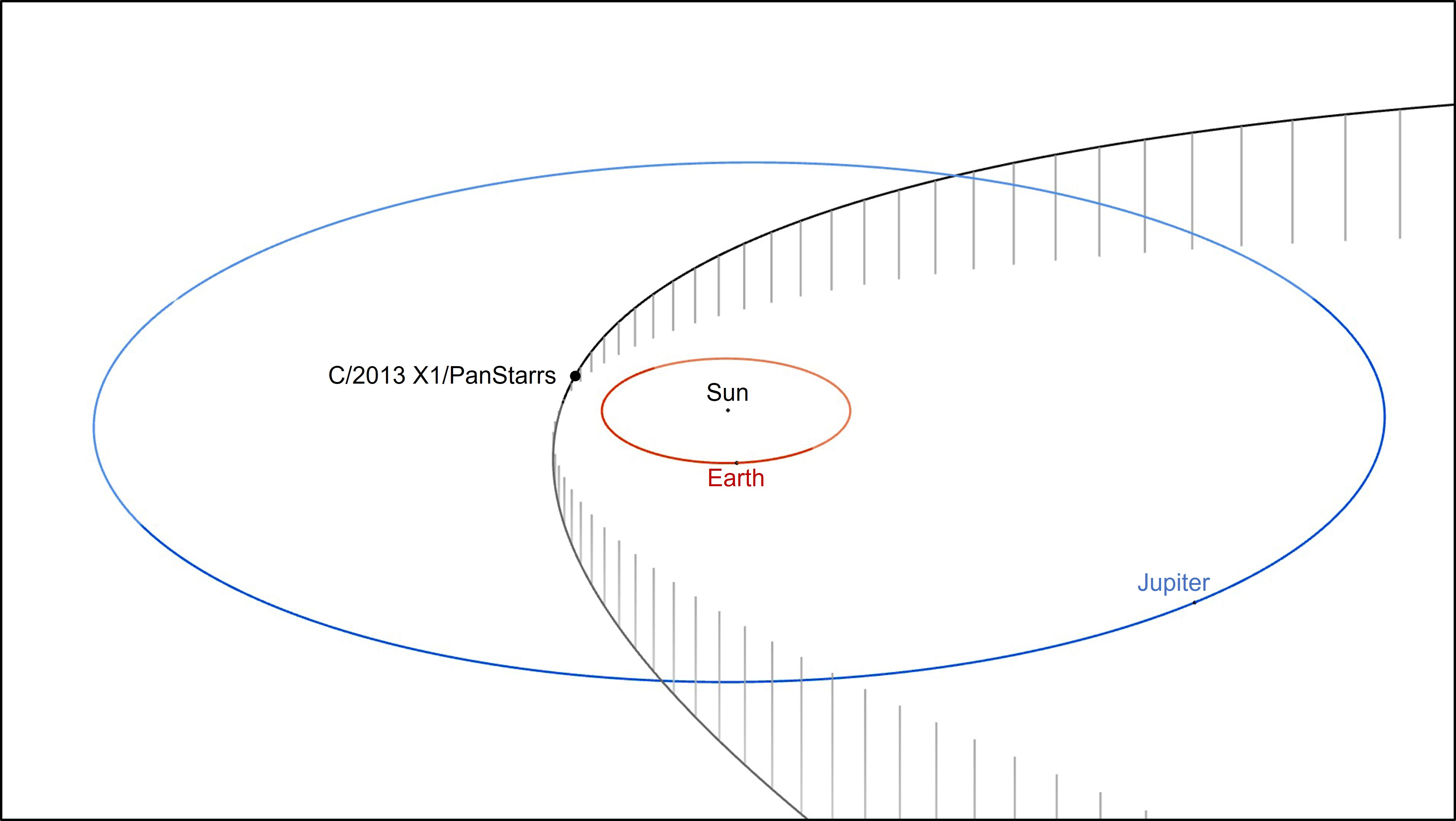}
    \caption{Orbit of comet C/2013 X1 and position on perihelion date. The field of view is set to the orbit of Jupiter for size comparison. Courtesy of NASA/JPL-Caltech.}
\end{figure}

\newpage

\subsection{Images}

\begin{SCfigure}[0.8][h!]
    \centering
    \includegraphics[scale=0.4]{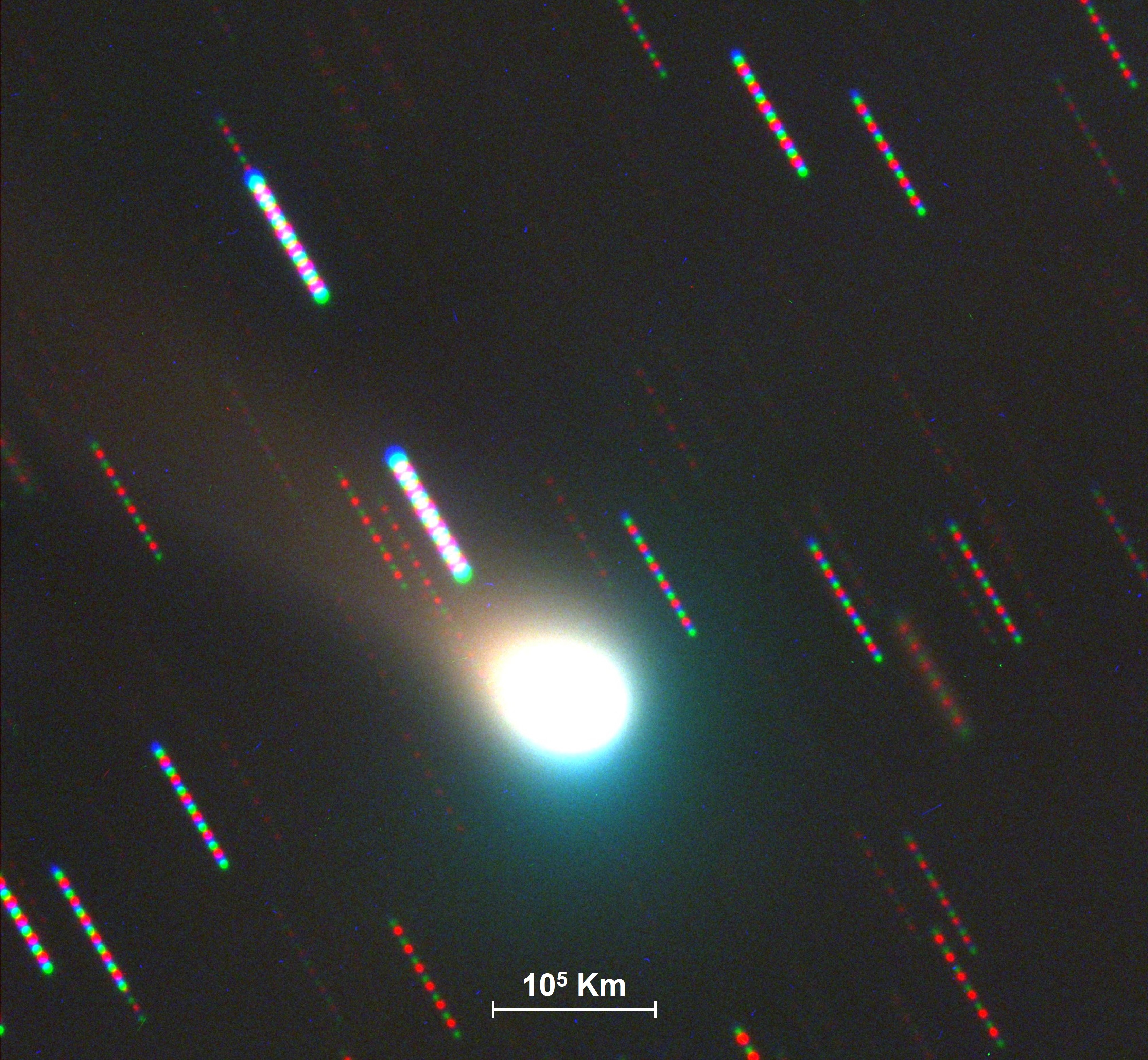}
     \caption{2016-01-20. Three-color BVR composite from images taken with the Asiago Copernico telescope. The tail develops in an antisolar direction, towards North-East; in the treated images, a strong jet can be observed in a subsolar position, which is curved by the radiation pressure.}
\end{SCfigure} 

\begin{SCfigure}[0.8][h!]
    \centering
  
    \includegraphics[scale=0.4]{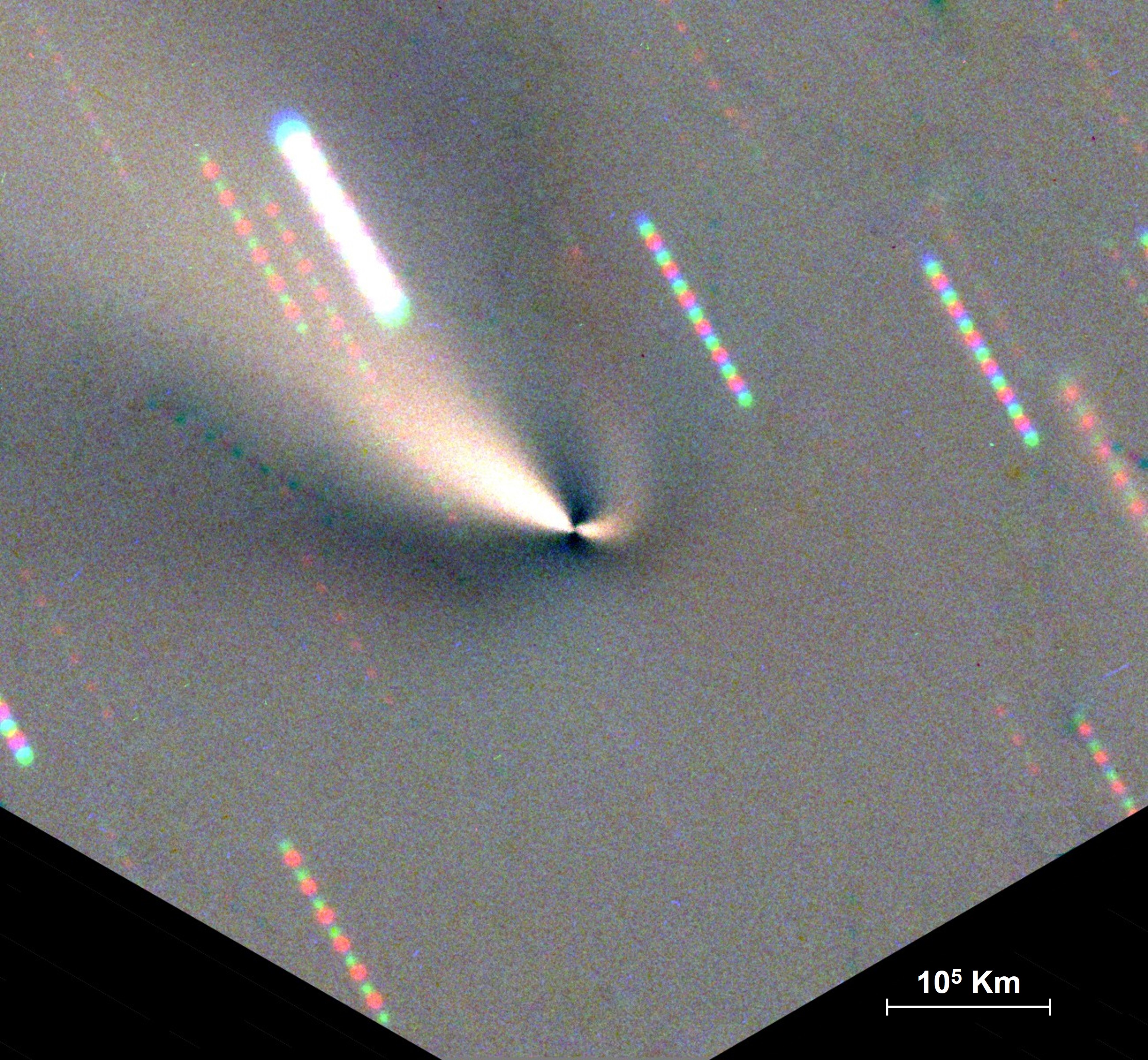}

    \caption{2016-01-20. In the same image, the regions of the inner coma are shown in greater detail, whose structure derives from the active areas present on the nucleus. These active sources emit large quantities of gases, carrying the dust which forms the tail of this comet.}
\end{SCfigure}

\newpage

\subsection{Spectra}

\begin{table}[h!]
\centering
\begin{tabular}{|c|c|c|c|c|c|c|c|c|c|c|c|}
\hline
\multicolumn{12}{|c|}{Observation details}                      \\ \hline 
\hline
$\#$  & date          & r     & $\Delta$ & RA     & DEC     & elong & phase & PLang& config  & FlAng & N \\
      & (yyyy-mm-dd)  &  (AU) & (AU)     & (h)    & (°)     & (°)   & (°)   &  (°)   &       &  (°)  & \\ \hline 

1* & 2016-01-06 & 2.000  & 1.879 & 23.98 & $+$19.70 & 82.3  & 29.2  & $-$3.8 & A & $+$0 & 12 \\
2* & 2016-01-20 & 1.862  & 2.078 & 23.77 & $+$14.83 & 63.6  & 28.3  & $-$1.6 & A & $+$0 & 5 \\
\hline
\end{tabular}
\end{table}

\begin{figure}[h!]

    \centering
    \includegraphics[scale=0.368]{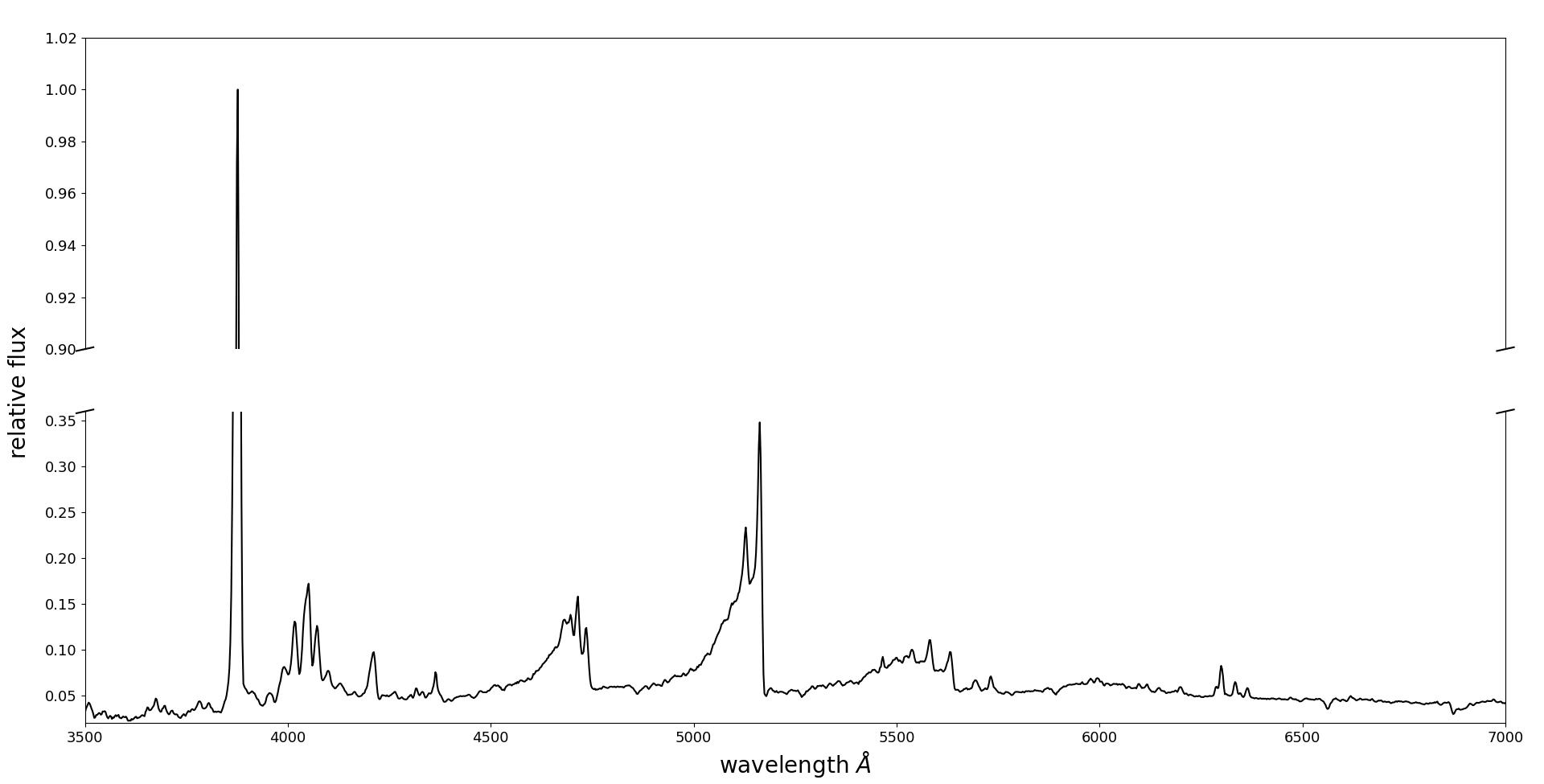}
    \caption{Spectrum of 2016-01-06; configuration A}

\end{figure}

\footnotetext{* No Solar Analog}

\newpage
\clearpage

\section{C/2015 O1 (PanSTARRS)}
\label{cometa:C2015O1}
\subsection{Description}

C/2015 O1 (PanSTARRS) is a Long Period type with a period of 525 million years and an absolute magnitude of 7.2$\pm$1.0.\footnote{\url{https://ssd.jpl.nasa.gov/tools/sbdb_lookup.html\#/?sstr=2015\%20O1} visited on July 20, 2024} 
It was first spotted by Eva Lilly and Richard Wainscoat using the 1.8m Panoramic Survey Telescope \& Rapid Response System (Pan-STARRS 1) on July 19, 2015.
The Earth crossed the comet orbital plane on January 20 and on July 23, 2018.

\noindent
We observed the comet around magnitude 13.\footnote{\url{https://cobs.si/comet/1191/ }, visited on July 20, 2024}

\begin{table}[h!]
\centering
\begin{tabular}{|c|c|c|}
\hline
\multicolumn{3}{|c|}{Orbital elements (epoch: February 12, 2018)}                      \\ \hline \hline
\textit{e} = 0.999994 & \textit{q} = 3.7230 & \textit{T} = 2458168.5294 \\ \hline
$\Omega$ = 299.8558 & $\omega$ = 89.5939  & \textit{i} = 127.2111  \\ \hline  
\end{tabular}
\end{table}

\begin{table}[h!]
\centering
\begin{tabular}{|c|c|c|c|c|c|c|c|c|}
\hline
\multicolumn{9}{|c|}{Comet ephemerides for key dates}                      \\ \hline 
\hline

& date & r & $\Delta$ & RA & DEC & elong & phase & PLang \\
& (yyyy-mm-dd) & (AU) & (AU) & (h) & (°) & (°) & (°) & (°) \\ \hline 

Perihelion       & 2018-02-19 & 3.730  & 3.556 & 16.53  & $+$35.75  & 92.3  & 15.4  & $-$06.4  \\
Nearest approach & 2018-04-05 & 3.752  & 3.233 & 15.37 & $+$52.39 & 113.9 & 14.1  & $-$13.8 \\ \hline
\end{tabular}

\end{table}

\vspace{0.5 cm}

\begin{figure}[h!]
    \centering
    \includegraphics[scale=0.38]{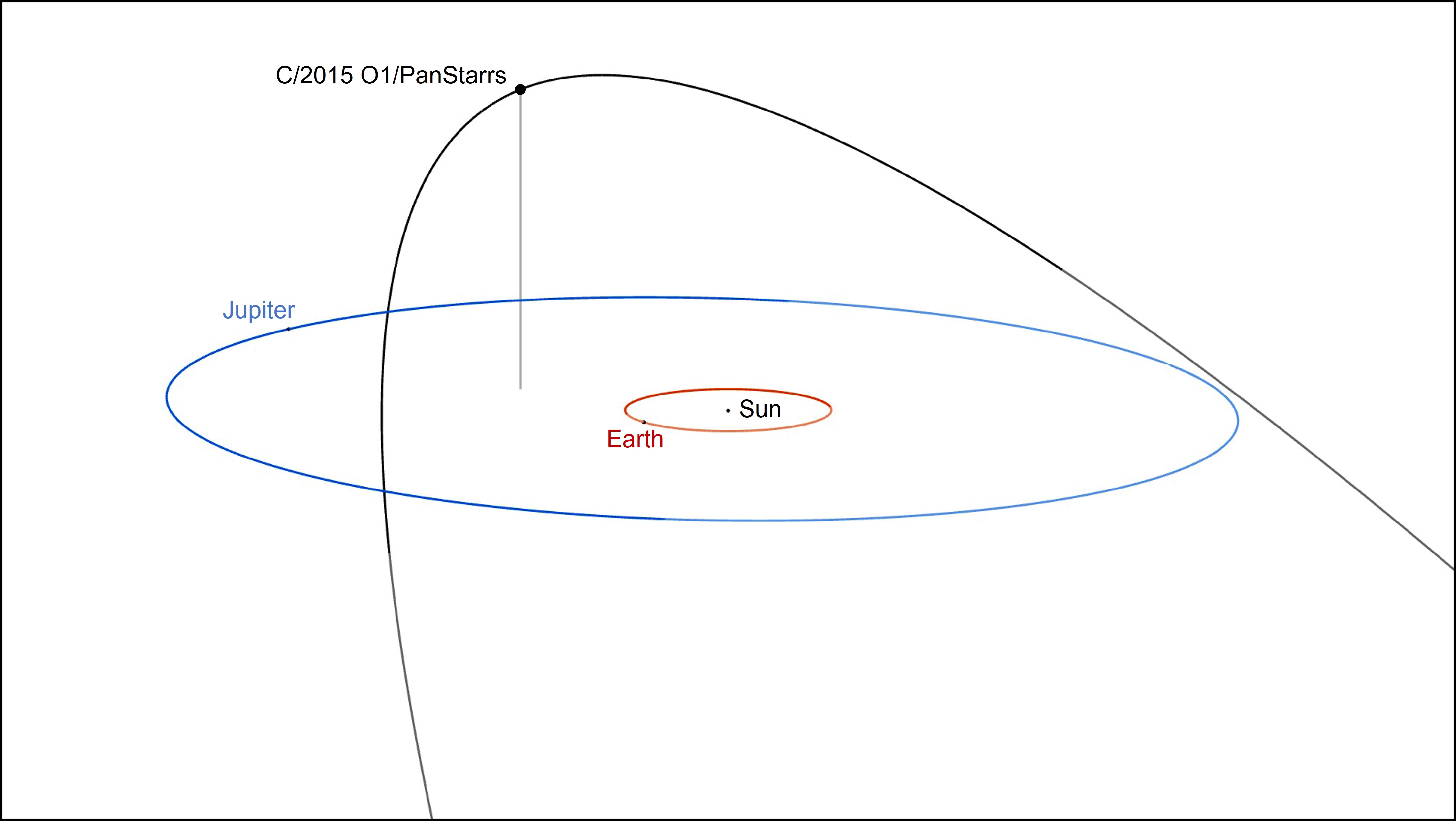}
    \caption{Orbit of comet C/2015 O1 and position on perihelion date. The field of view is set to the orbit of Jupiter for size comparison. Courtesy of NASA/JPL-Caltech.}
\end{figure}

\newpage

\subsection{Images}
\begin{SCfigure}[0.8][h!]
 \centering
 \includegraphics[scale=0.4]{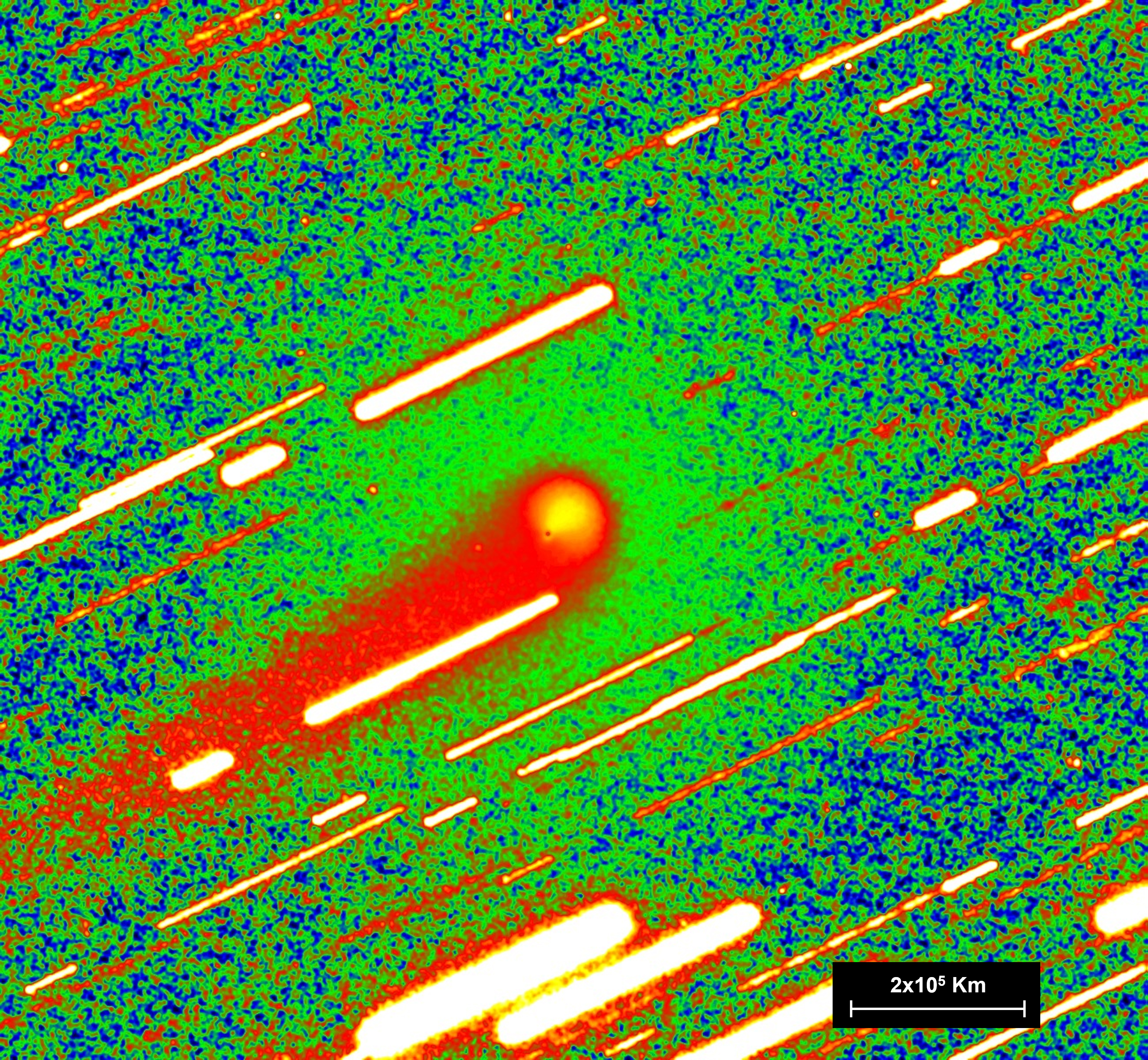}
 \caption{2018-04-13. Comet C/2015 O1 PanSTARRS imaged with the Schmidt telescope. The great distance from the Sun (3.8 AU) did not allow the comet to develop a large tail even shortly after the perihelion transit. To highlight it, special processing is required. In the image, in false colors, a very weak spherical coma around the nucleus can also be observed.}
\end{SCfigure} 
\begin{SCfigure}[0.8][h!]
 \centering
 \includegraphics[scale=0.4]{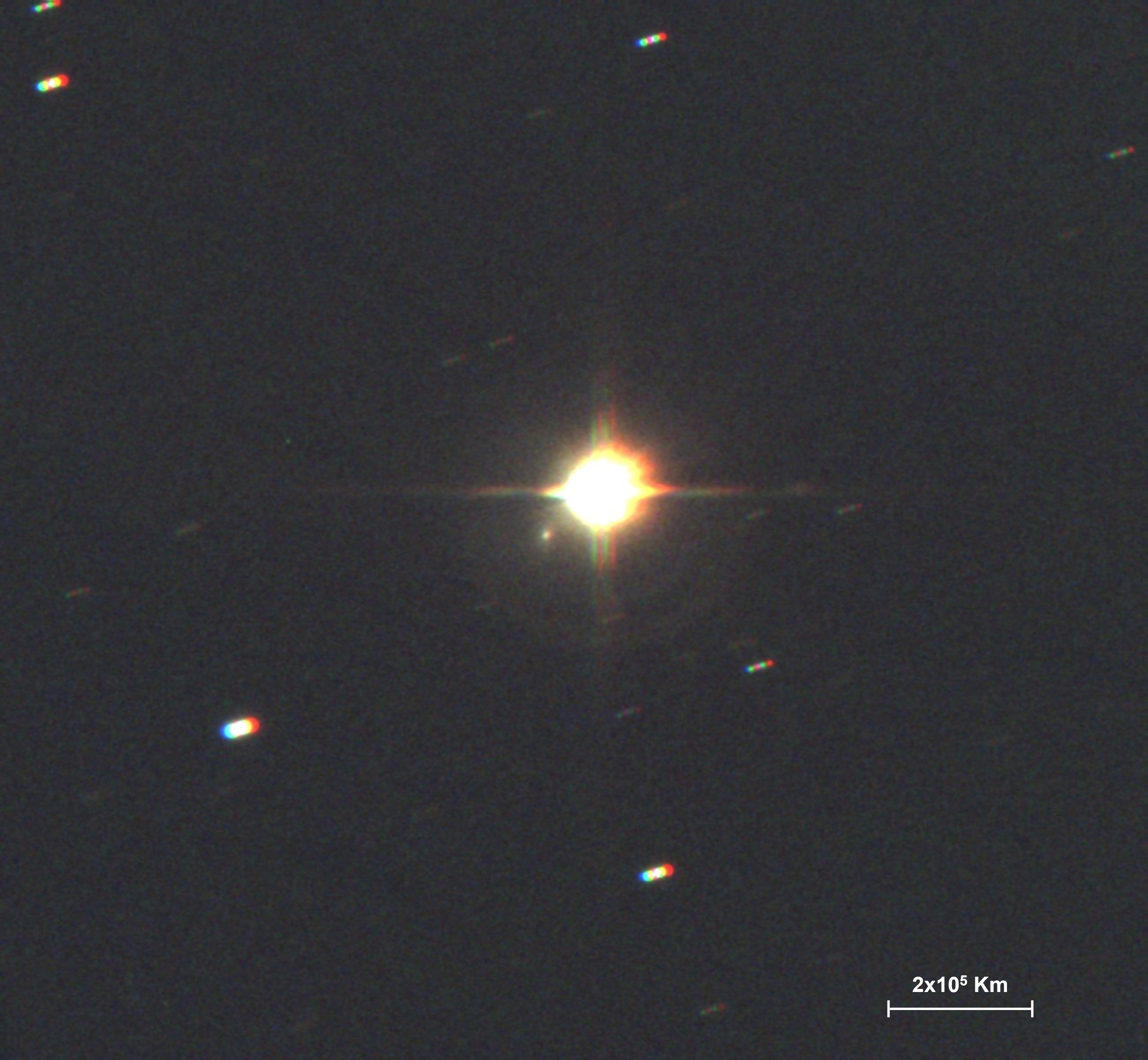}
 \caption{2019-01-24. Ten months after the previous observation, comet C/2015 O1 had moved away at 4.75 AU from the Sun and about 4.0 AU from Earth.
 It was now on a path towards outside the Solar System and its emission was greatly reduced. Here it is observed as a dot with a barely hinted tail, lost in the glow of a red star.
 Image taken with the Asiago Schmidt telescope. }
\end{SCfigure}

\newpage

\subsection{Spectra}

\begin{table}[h!]
\centering
\begin{tabular}{|c|c|c|c|c|c|c|c|c|c|c|c|}
\hline
\multicolumn{12}{|c|}{Observation details}                      \\ \hline 
\hline
$\#$ & date & r & $\Delta$ & RA & DEC & elong & phase & PLang& config & FlAng & N \\
 & (yyyy-mm-dd) & (AU) & (AU) & (h) & (°) & (°) & (°) & (°) & & (°) &  \\ \hline

1               & 2018-04-18 & 3.766  & 3.266 & 14.73  & $+$55.38  & 112.5 & 14.3  & $-$14.1 & A & $-$0 & 2 \\
2               & 2018-04-19 & 3.766  & 3.267 & 14.67  & $+$55.57  & 112.4 & 14.3  & $-$14.1 & A & $-$0 & 4 \\
3               & 2018-04-20 & 3.769  & 3.277 & 14.62  & $+$55.73  & 111.9 & 14.3  & $-$14.1 & A & $+$10 & 4 \\
4               & 2018-04-21 & 3.770  & 3.283 & 14.55  & $+$55.88  & 111.7 & 14.4  & $-$14.1 & A & $+$0 & 4\\
5               & 2018-04-24 & 3.774  & 3.302 & 14.38  & $+$56.28  & 110.5 & 14.5  & $-$13.9 & A & $+$30 & 3 \\
 \hline
\end{tabular}
\end{table}

\begin{figure}[h!]

    \centering
    \includegraphics[scale=0.368]{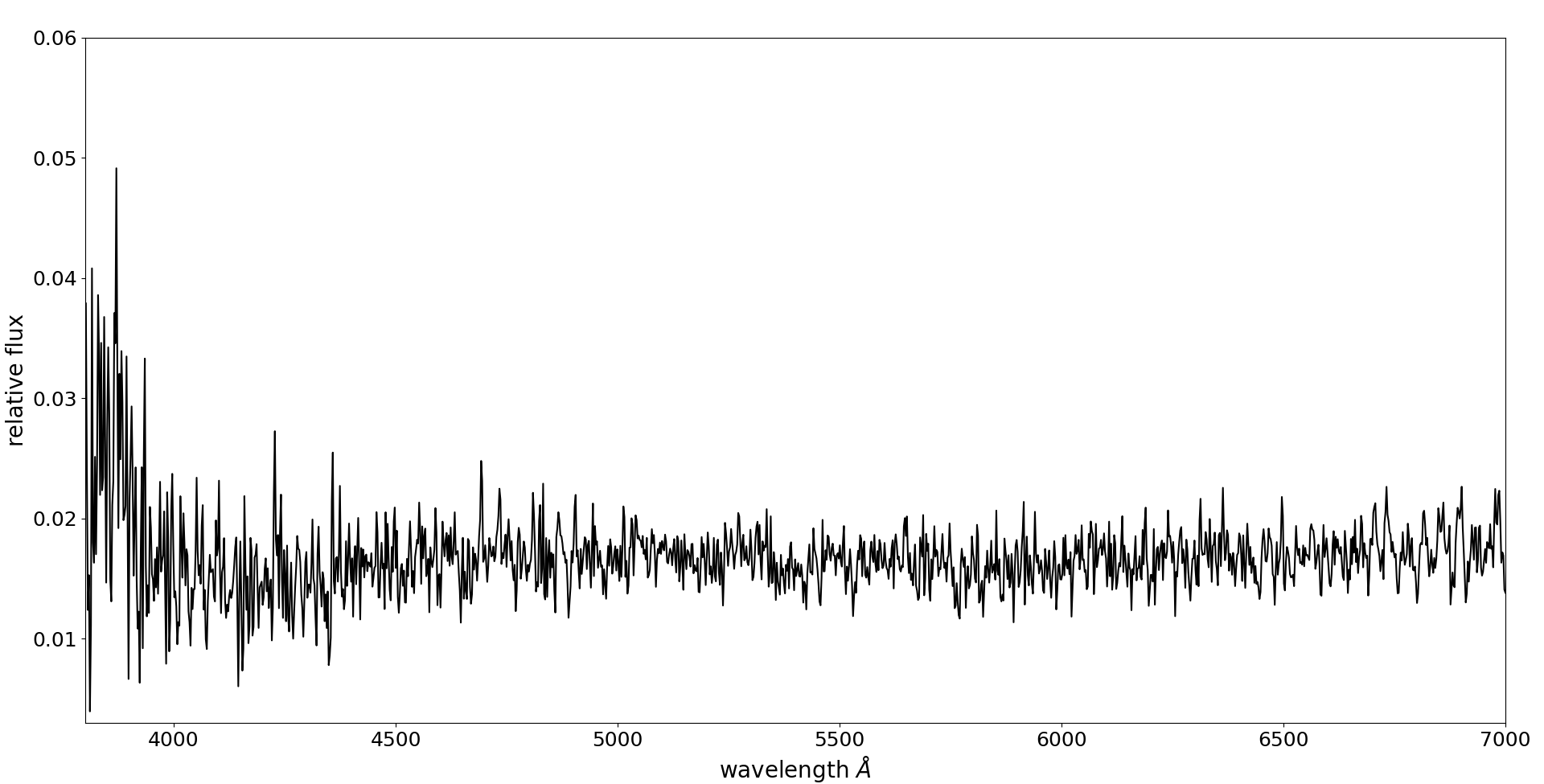}
    \caption{Spectrum of 2018-04-24; configuration A}

\end{figure}

\newpage
\clearpage

\section{C/2015 V2 (Johnson)}
\label{cometa:C2015V2}
\subsection{Description}

C/2015 V2 (Johnson) is a hyperbolic Oort Cloud comet with an absolute magnitude of 10.0$\pm$1.0.\footnote{\url{https://ssd.jpl.nasa.gov/tools/sbdb_lookup.html\#/?sstr=2015\%20V2} visited on July 20, 2024} 
It was first spotted by Jess Johnson with the 0.68m telescope of the Catalina Sky Survey on November 3, 2015.
The comet reached the maximum brightness (magnitude 7.8) in the first week of June 2017. Earth crossed the comet orbital plane on December 16, 2016 and on May 30, 2017.
\noindent
We observed this comet between magnitude 13 and 8.\footnote{\url{https://cobs.si/comet/1347/ }, visited on July 20, 2024}

\begin{table}[h!]
\centering
\begin{tabular}{|c|c|c|}
\hline
\multicolumn{3}{|c|}{Orbital elements (epoch: January 18, 2017)}                      \\ \hline \hline
\textit{e} = 1.0017 & \textit{q} = 1.6370 & \textit{T} = 2457916.8475 \\ \hline
$\Omega$ = 69.8521 & $\omega$ = 164.8981  & \textit{i} = 49.8748  \\ \hline  
\end{tabular}
\end{table}

\begin{table}[h!]
\centering
\begin{tabular}{|c|c|c|c|c|c|c|c|c|}
\hline
\multicolumn{9}{|c|}{Comet ephemerides for key dates}                      \\ \hline 
\hline
& date         & r    & $\Delta$  & RA      & DEC      & elong  & phase  & PLang  \\
& (yyyy-mm-dd) & (AU) & (AU)      & (h)     & (°)      & (°)    & (°)    & (°) \\ \hline 

Perihelion       & 2017-06-12 & 1.637  & 0.823 & 14.44  & $+$10.05 &  125.6 & 30.3  & $+$11.0  \\ 
Nearest approach & 2017-06-04 & 1.640  & 0.811 & 14.63  & $+$18.92 & 127.6  & 29.3  & $+$03.8 \\ \hline
\end{tabular}

\end{table}

\vspace{0.5 cm}

\begin{figure}[h!]
    \centering
    \includegraphics[scale=0.38]{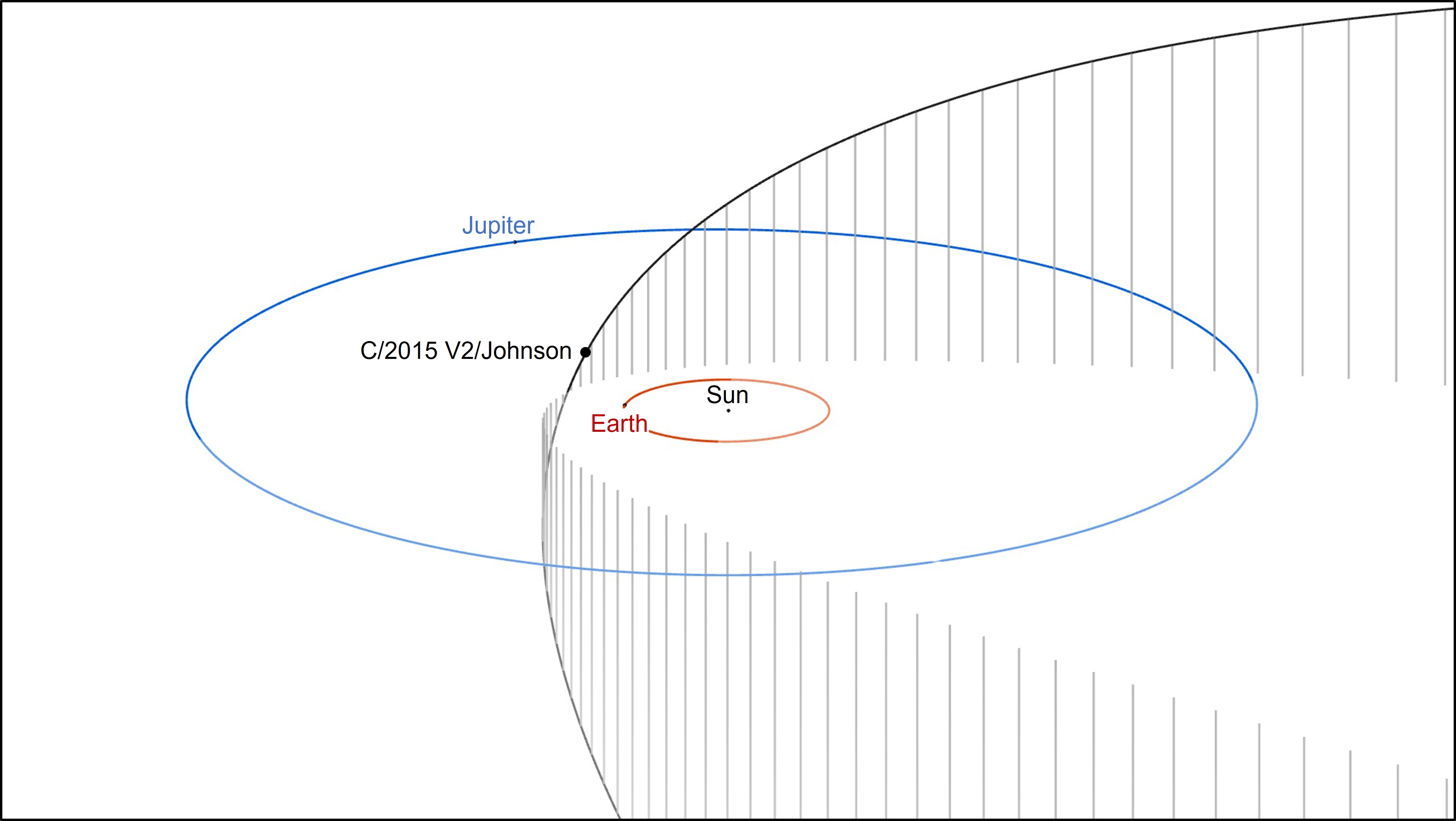}
    \caption{Orbit of comet C/2015 V2 and position on perihelion date. The field of view is set to the orbit of Jupiter for size comparison. Courtesy of NASA/JPL-Caltech.}
\end{figure}


\newpage

\subsection{Images}

\begin{SCfigure}[0.8][h!]
    \centering
    \includegraphics[scale=0.4]{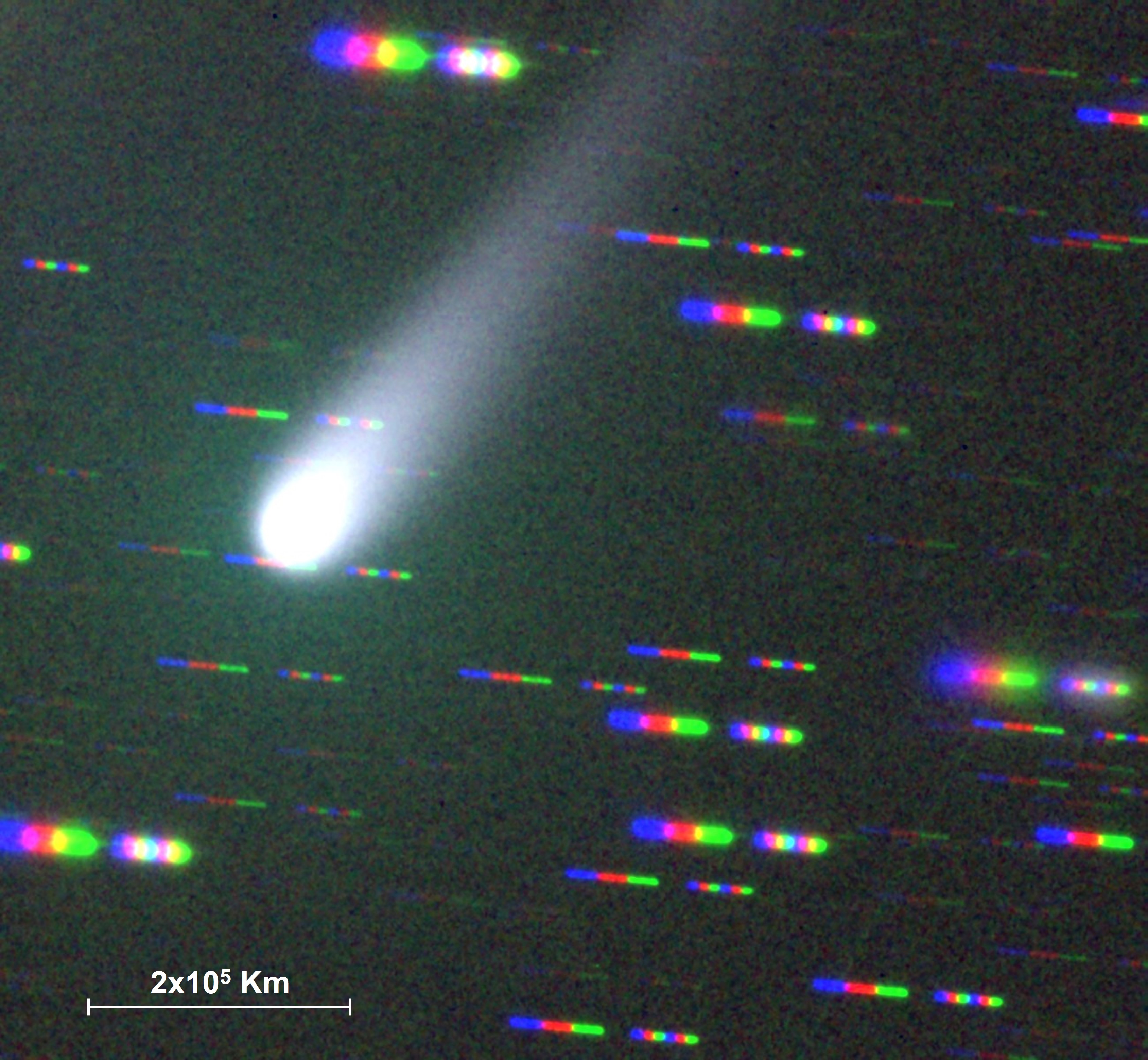}
     \caption{2017-01-24. Three-color BVR composite from images taken with the Asiago Schmidt telescope. The comet shows a very bright nucleus and a very intense inner coma, while the outer coma, spherical, appears greenish due to fluorescence phenomena on the diatomic carbon molecules (C$_2$).}
\end{SCfigure} 

\begin{SCfigure}[0.8][h!]
    \centering

    \includegraphics[scale=0.4]{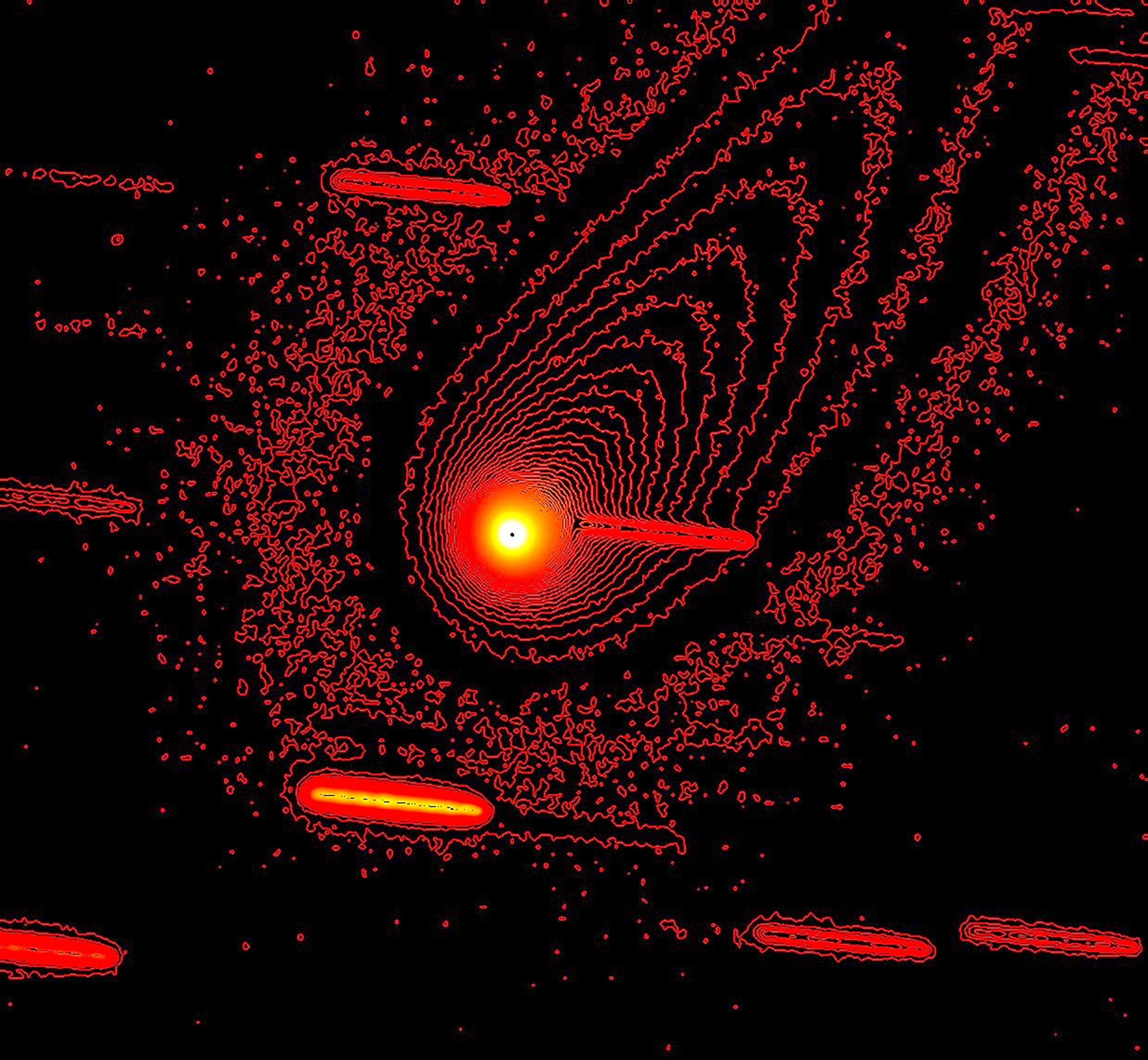}

    \caption{2017-01-27. Image taken with the Schmidt telescope. The false-color isophote visualization shows the greater density of the coma close to the nucleus (indicated by a black dot).}
\end{SCfigure}

\newpage

\subsection{Spectra}

\begin{table}[h!]
\centering
\begin{tabular}{|c|c|c|c|c|c|c|c|c|c|c|c|}
\hline
\multicolumn{12}{|c|}{Observation details}                      \\ \hline 
\hline
$\#$  & date          & r     & $\Delta$ & RA     & DEC     & elong & phase & PLang& config  & FlAng & N \\
      & (yyyy-mm-dd)  &  (AU) & (AU)     & (h)    & (°)     & (°)   & (°)   &  (°)   &       &  (°)  &  \\ \hline 

1 & 2016-11-29 & 2.940 & 2.981 & 13.40 & $+$44.65 & 78.2  & 19.2 & $+$00.5 & A & $+$1 & 5 \\
2 & 2017-06-10 & 1.637  & 0.819 & 14.05 & $+$12.78 & 126.1  & 30.1  & $+$09.7 & A & $-$0 & 1 \\
\hline
\end{tabular}
\end{table}

\begin{figure}[h!]

    \centering
    \includegraphics[scale=0.368]{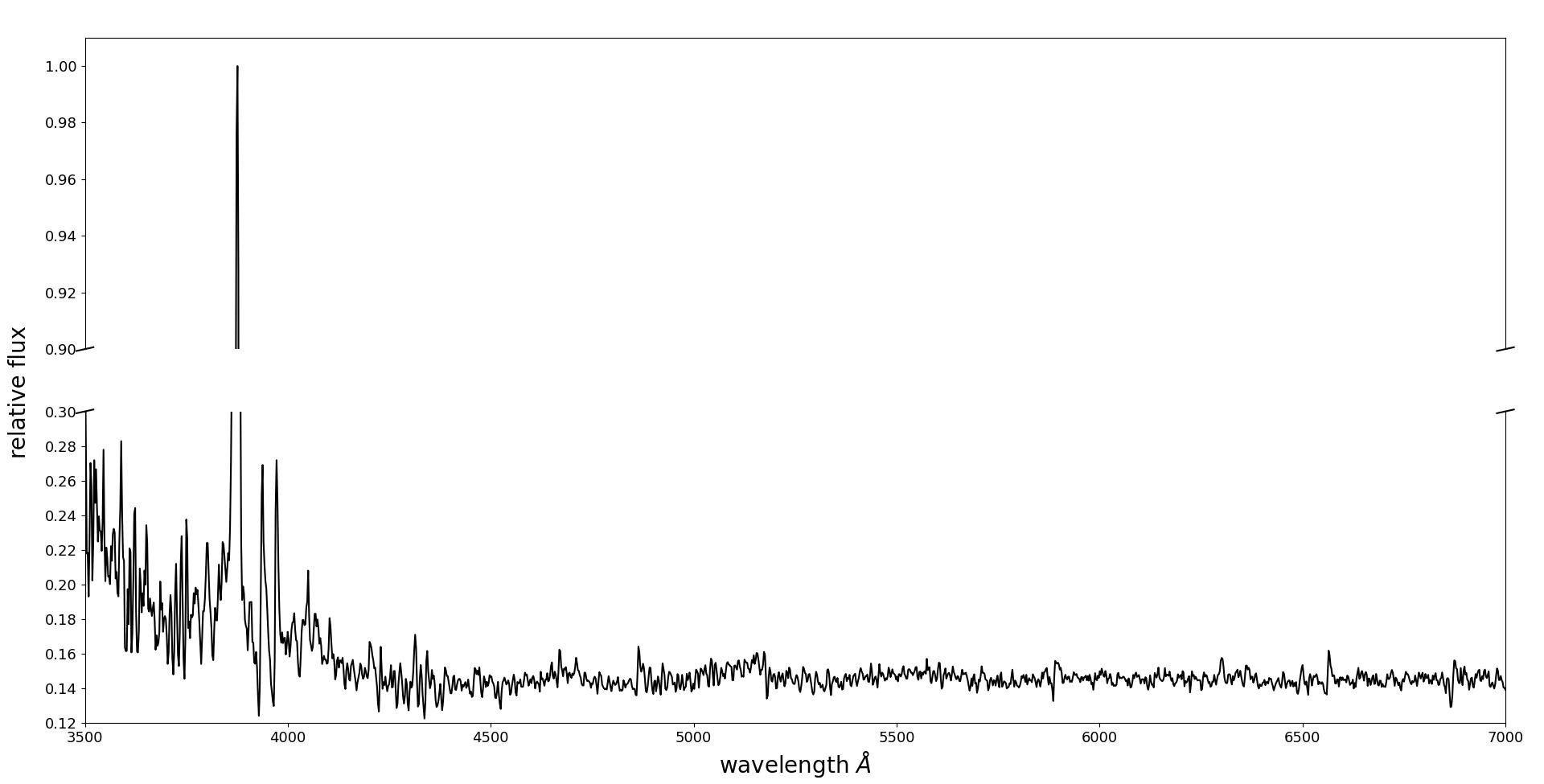}
    \caption{Spectrum of 2016-11-29; configuration A}

\end{figure}

\newpage
\clearpage

\section{C/2016 N6 (PanSTARRS)}
\label{cometa:C2016N6}
\subsection{Description}

C/2016 N6 (PanSTARRS) is a Long Period comet with a period of 88096 years and an absolute magnitude of 6.8$\pm$0.8.\footnote{\url{https://ssd.jpl.nasa.gov/tools/sbdb_lookup.html\#/?sstr=2016\%20N6} visited on July 20, 2024} 
It was first spotted by Richard Wainscoat, Robert Weryk and Eva Lilly using the 1.8m Panoramic Survey Telescope \& Rapid Response System (Pan-STARRS 1) on July 14, 2016.
Its aphelion is at a distance of 3950 AU from the Sun.
The Earth crossed the comet orbital plane on January 19, 2018.

\noindent
We observed the comet around magnitude 12.5.\footnote{\url{https://cobs.si/comet/1619/ }, visited on July 20, 2024}

\begin{table}[h!]
\centering
\begin{tabular}{|c|c|c|}
\hline
\multicolumn{3}{|c|}{Orbital elements (epoch: February 1, 2018)}                      \\ \hline \hline
\textit{e} = 0.9987 & \textit{q} = 2.6691 & \textit{T} = 2458317.7041 \\ \hline
$\Omega$ = 298.9777 & $\omega$ = 162.8032  & \textit{i} = 105.8304  \\ \hline  
\end{tabular}
\end{table}

\begin{table}[h!]
\centering
\begin{tabular}{|c|c|c|c|c|c|c|c|c|}
\hline
\multicolumn{9}{|c|}{Comet ephemerides for key dates}                      \\ \hline 
\hline

& date & r & $\Delta$ & RA & DEC & elong & phase & PLang \\
& (yyyy-mm-dd) & (AU) & (AU) & (h) & (°) & (°) & (°) & (°) \\ \hline 

Perihelion       & 2018-07-18  & 2.669  & 3.648 & 08.45  & $+$31.46  & 13.3  & 05.0  & $-$00.8  \\
Nearest approach & 2018-12-23  & 3.133  & 2.367 & 07.67  & $-$16.05  & 134.3 & 13.0  & $+$10.7 \\ \hline
\end{tabular}

\end{table}

\vspace{0.5 cm}

\begin{figure}[h!]
    \centering
    \includegraphics[scale=0.38]{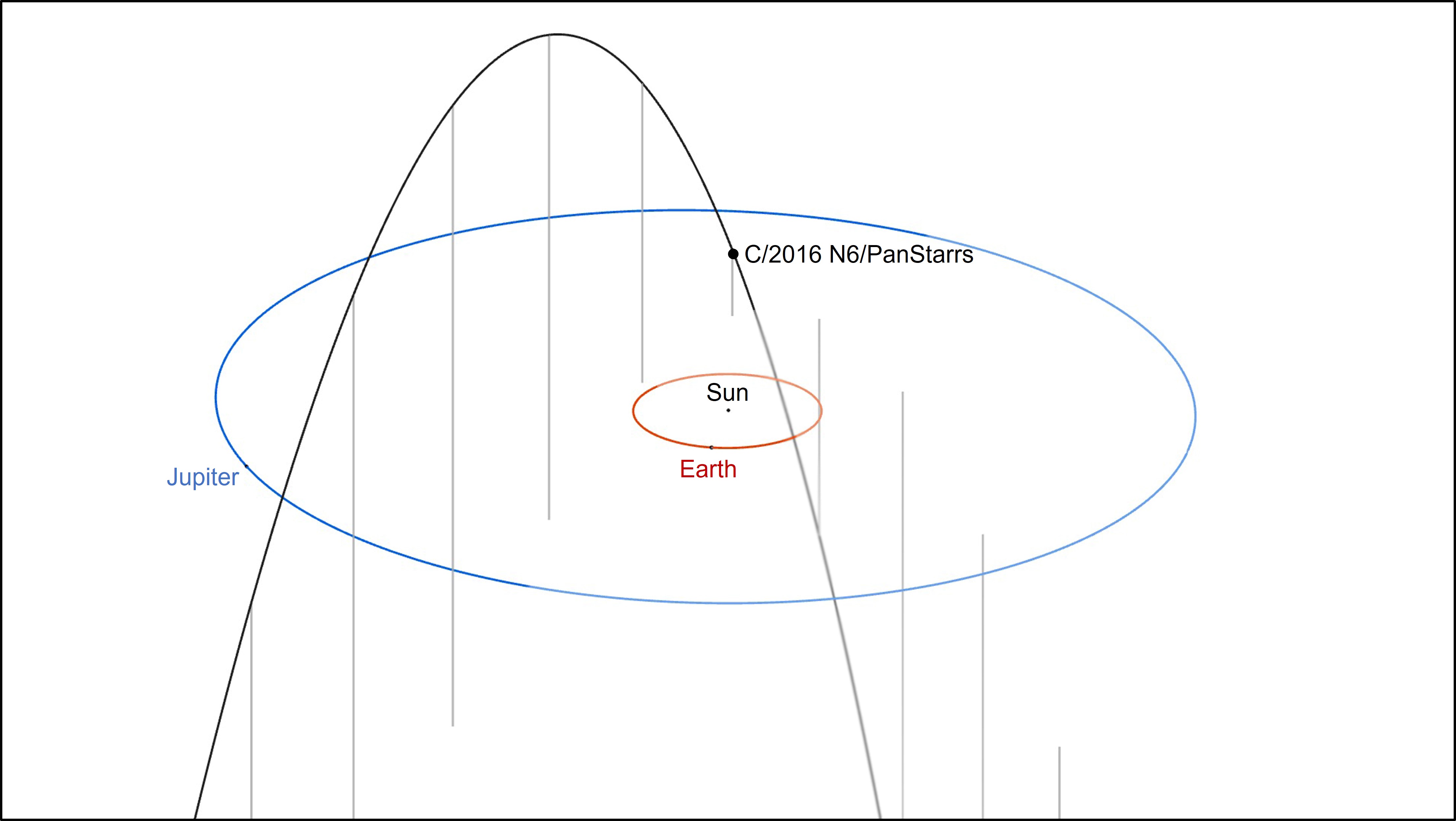}
    \caption{Orbit of comet C/2016 N6 and position on perihelion date. The field of view is set to the orbit of Jupiter for size comparison. Courtesy of NASA/JPL-Caltech.}
\end{figure} 

\newpage


\subsection{Images}
\begin{SCfigure}[0.8][h!]
 \centering
 \includegraphics[scale=0.4]{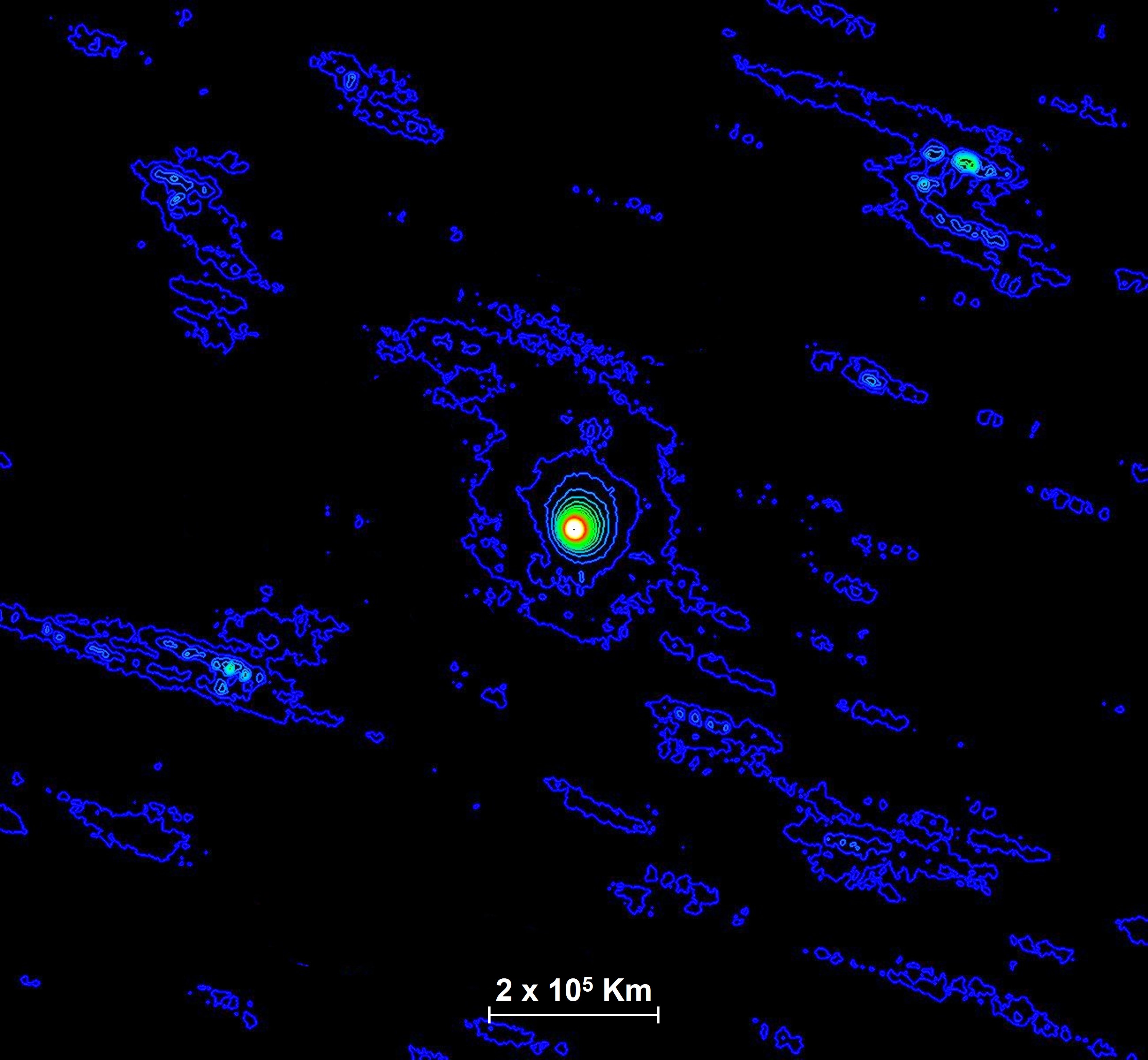}
 \caption{2019-01-06. Image taken with the Asiago Schmidt telescope without filters to acquire the maximum signal from C/2016 N6, which was far away from the Sun and Earth, as an object of magnitude 15. The figure shows a photometric visualization of the image in isophotes. The nucleus is clearly identifiable, as well as a slight asymmetry of the coma.
}
\end{SCfigure} 
\begin{SCfigure}[0.8][h!]
 \centering
 \includegraphics[scale=0.4]{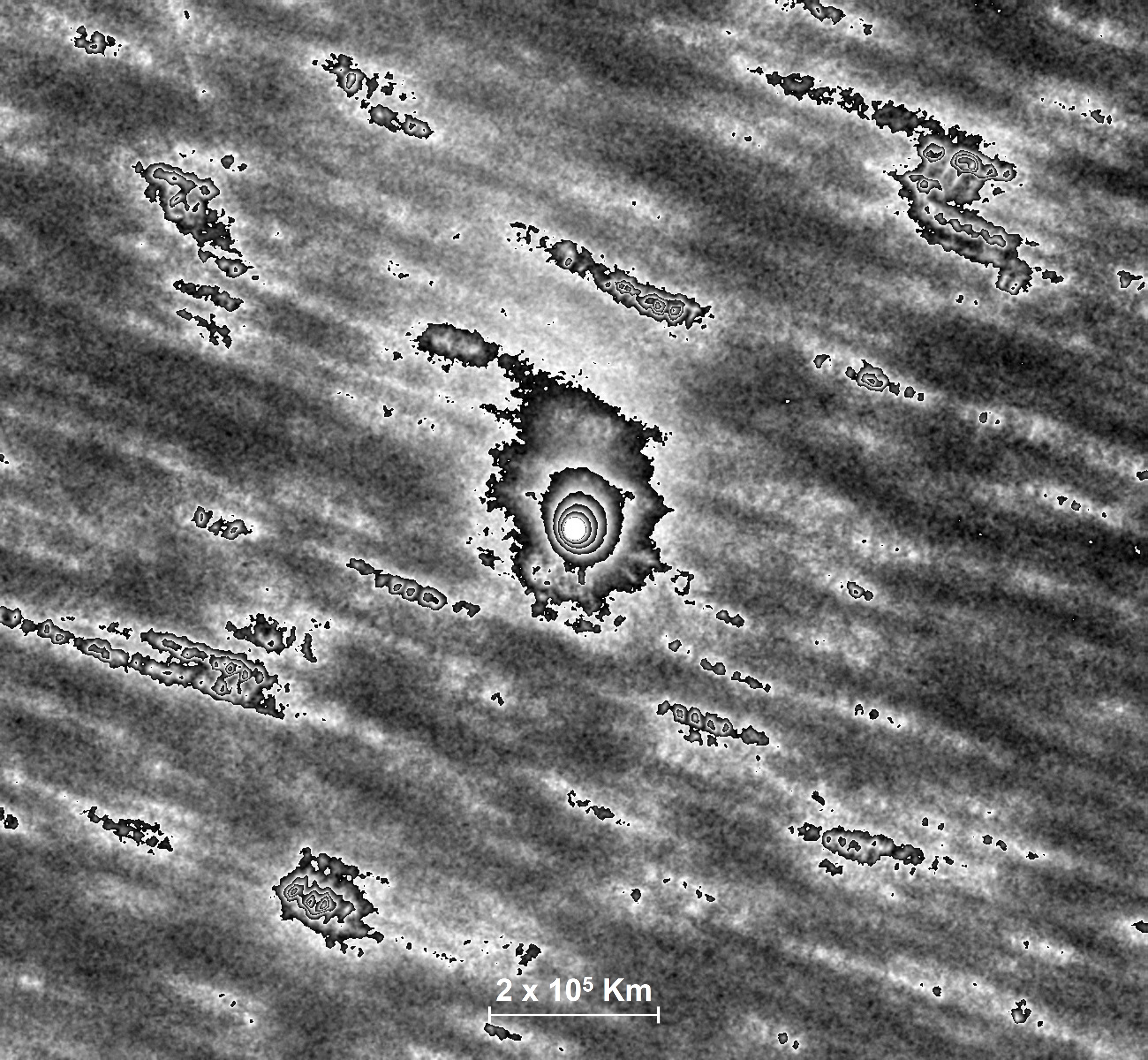}
 \caption{Same image as above shown in isodense}
\end{SCfigure}

\newpage

\subsection{Spectra}

\begin{table}[h!]
\centering
\begin{tabular}{|c|c|c|c|c|c|c|c|c|c|c|c|}
\hline
\multicolumn{12}{|c|}{Observation details}                      \\ \hline 
\hline
$\#$ & date & r & $\Delta$ & RA & DEC & elong & phase & PLang& config & FlAng & N \\
 & (yyyy-mm-dd) & (AU) & (AU) & (h) & (°) & (°) & (°) & (°) & & (°) &  \\ \hline

1               & 2019-01-06 & 3.211 & 2.414 & 07.13  & $-$19.63 & 137.9  & 11.8 & $+$05.1  & A & $+$0 & 2 \\

 \hline
\end{tabular}
\end{table}

\begin{figure}[h!]

    \centering
    \includegraphics[scale=0.368]{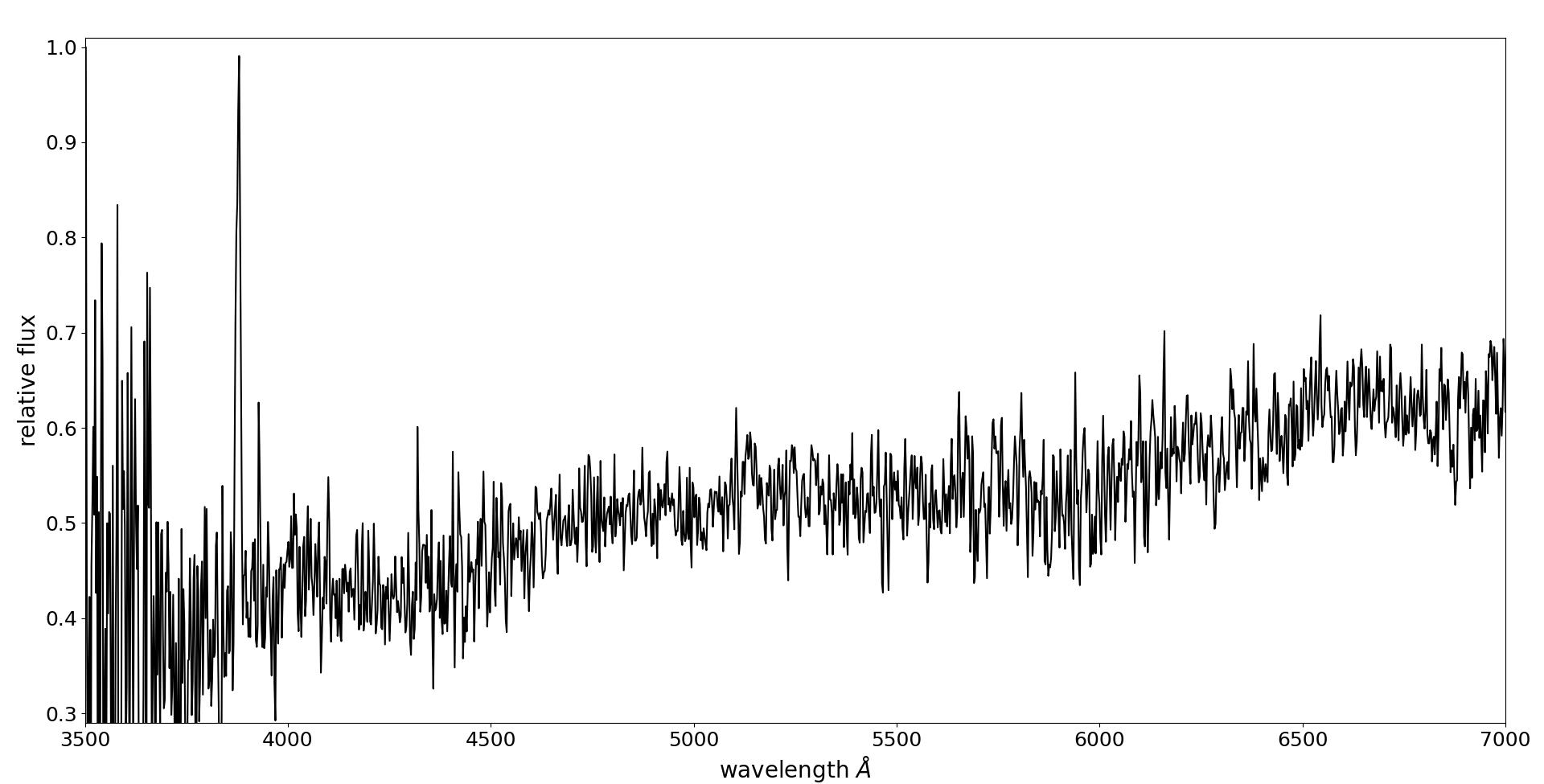}
    \caption{Spectrum of 2019-01-06; configuration A}

\end{figure}

\newpage
\clearpage

\section{C/2016 R2 (PanSTARRS)}
\label{cometa:C2016R2}
\subsection{Description}

C/2016 R2 (PanSTARRS) is a Long Period comet with a period of 18709 years and an absolute magnitude of 7.3$\pm$1.0.\footnote{\url{https://ssd.jpl.nasa.gov/tools/sbdb_lookup.html\#/?sstr=2016\%20R2} visited on July 20, 2024} 
It was first spotted by Robert Weryk and Richard Wainscoat using the 1.8m Panoramic Survey Telescope \& Rapid Response System (Pan-STARRS 1) on July 9, 2016.
The tail of the comet is blue due to the abundance of \ch{N2} molecules in the spectra.
Its aphelion is at a distance of 1407 AU from the Sun.
The Earth crossed the comet orbital plane on June 11 and December 12, 2018.

\noindent
We observed the comet around magnitude 10.5.\footnote{\url{https://cobs.si/comet/1624/ }, visited on July 20, 2024}

\begin{table}[h!]
\centering
\begin{tabular}{|c|c|c|}
\hline
\multicolumn{3}{|c|}{Orbital elements (epoch: May 29, 2018)}                      \\ \hline \hline
\textit{e} = 0.9963 & \textit{q} = 2.6024 & \textit{T} = 2458248.0831 \\ \hline
$\Omega$ = 80.5690 & $\omega$ = 33.1919  & \textit{i} = 58.2241  \\ \hline  
\end{tabular}
\end{table}

\begin{table}[h!]
\centering
\begin{tabular}{|c|c|c|c|c|c|c|c|c|}
\hline
\multicolumn{9}{|c|}{Comet ephemerides for key dates}                      \\ \hline 
\hline

& date & r & $\Delta$ & RA & DEC & elong & phase & PLang \\
& (yyyy-mm-dd) & (AU) & (AU) & (h) & (°) & (°) & (°) & (°) \\ \hline 

Perihelion       & 2018-05-09 & 2.602  & 3.274 & 5.53  & $+$45.00  & 41.3  & 14.8  & $-$08.0  \\
Nearest approach & 2017-12-22 & 2.980  & 2.054 & 4.60 & $+$11.35 & 156.1 & 07.7  & $-$04.0 \\ \hline
\end{tabular}

\end{table}

\vspace{0.5 cm}

\begin{figure}[h!]
    \centering
    \includegraphics[scale=0.38]{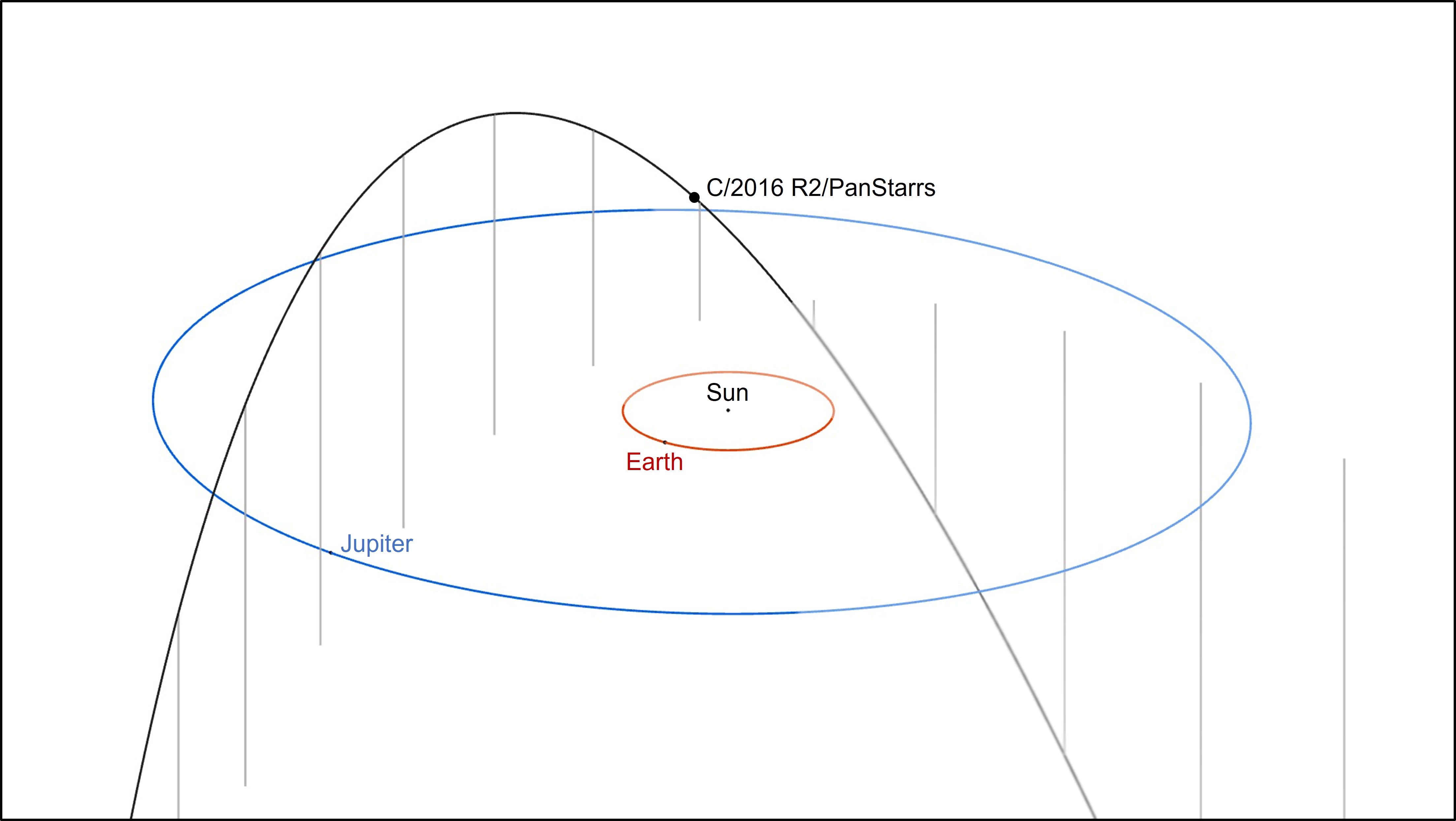}
    \caption{Orbit of comet C/2016 R2 and position on perihelion date. The field of view is set to the orbit of Jupiter for size comparison. Courtesy of NASA/JPL-Caltech.}
\end{figure}

\newpage


\subsection{Images}
\begin{SCfigure}[0.8][h!]
 \centering
 \includegraphics[scale=0.4]{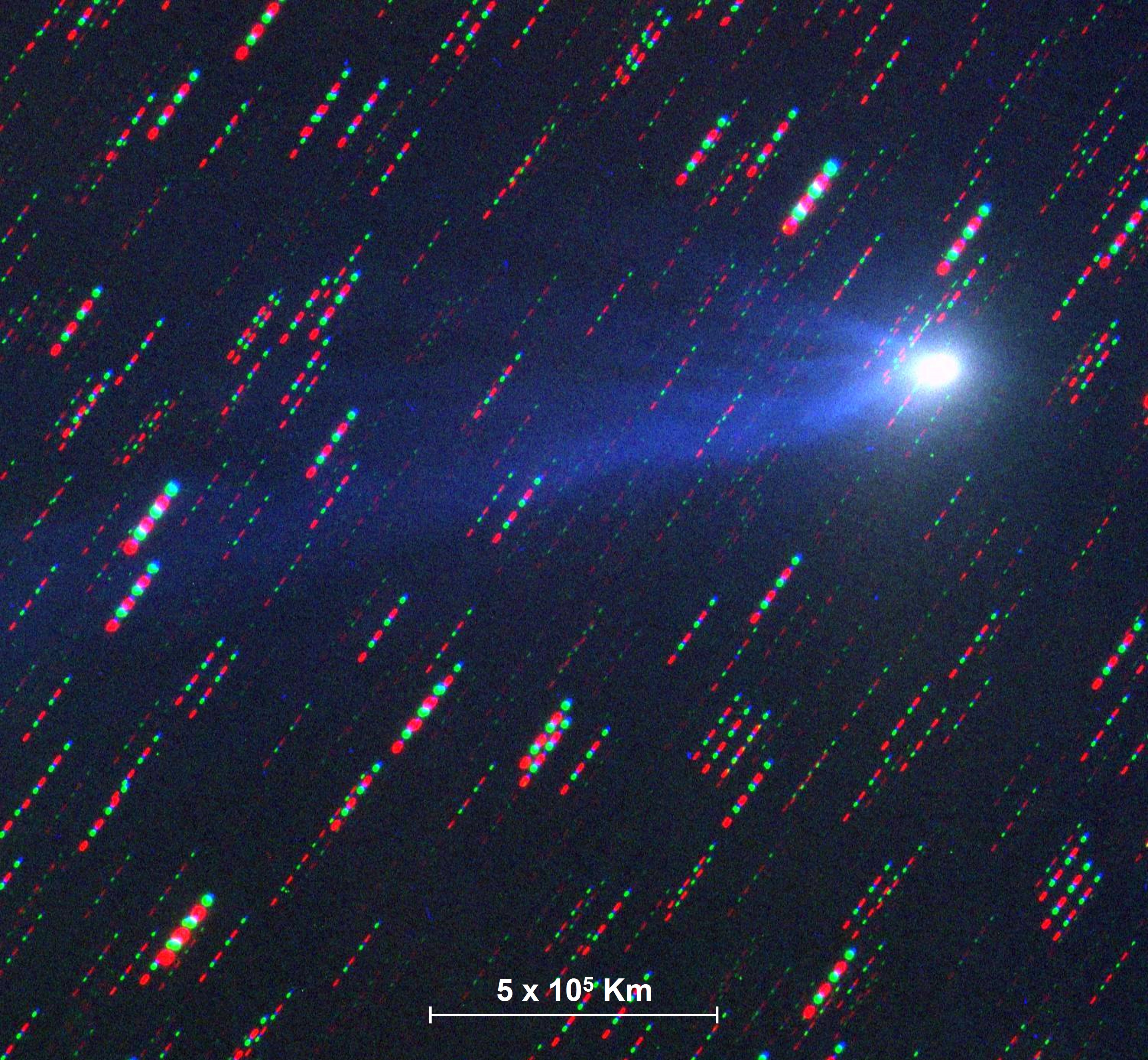}
 \caption{2018-01-13.  Three-color BVR composite from images taken with the Asiago Schmidt telescope. The comet was more than 420 million km from the Sun and 320 million km from Earth and had developed a very long tail that showed structures moving in antisolar direction. The intense blue color of the tail derives from the fluorescence of N2 molecules, of which the comet was very rich. }
\end{SCfigure} 
\begin{SCfigure}[0.8][h!]
 \centering
 \includegraphics[scale=0.4]{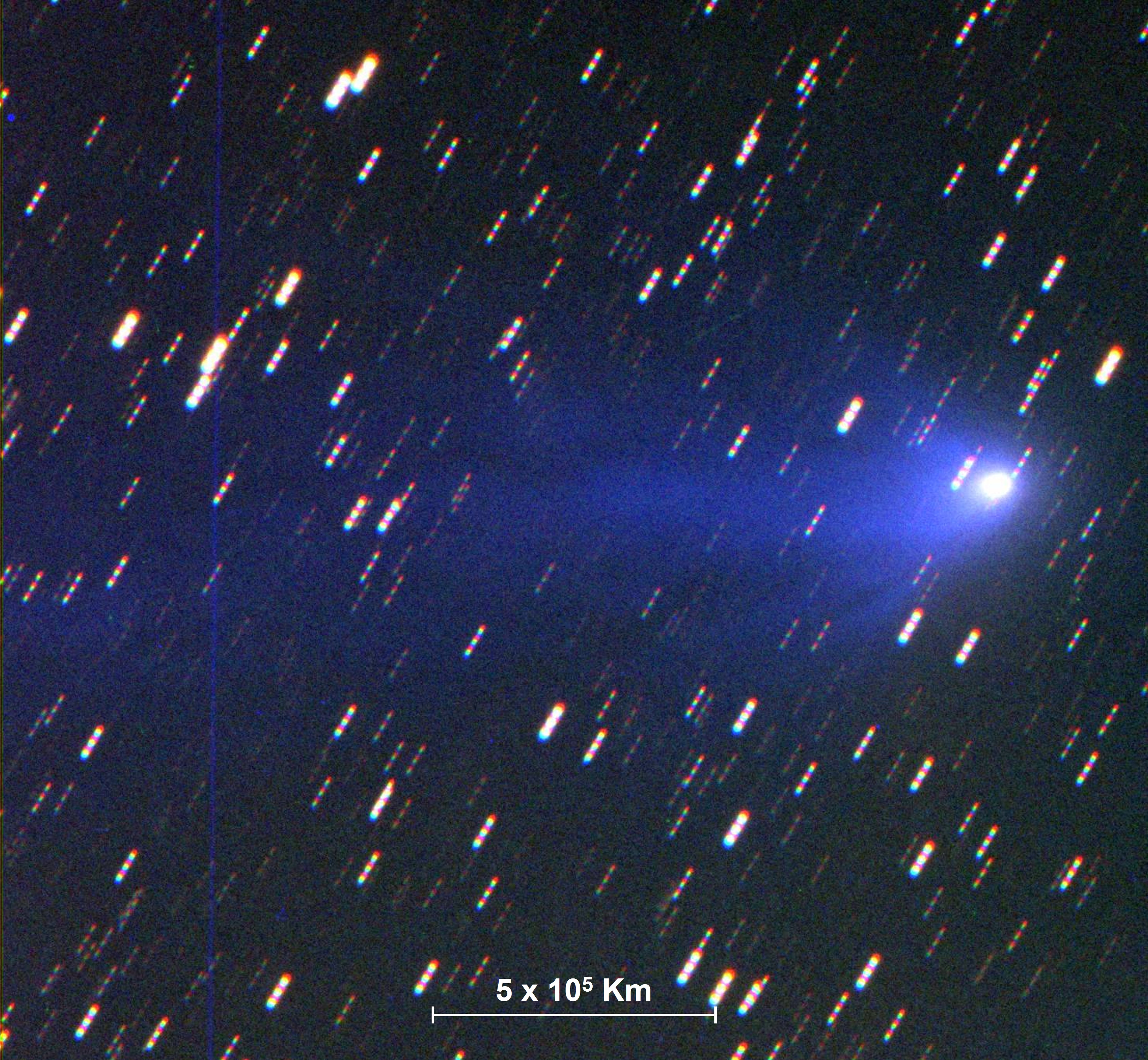}
 \caption{2018-01-18. Three-color BVR composite of the comet  taken a few days later with the Asiago Schmidt telescope. }
\end{SCfigure}

\newpage

\subsection{Spectra}

\begin{table}[h!]
\centering
\begin{tabular}{|c|c|c|c|c|c|c|c|c|c|c|c|}
\hline
\multicolumn{12}{|c|}{Observation details}                      \\ \hline 
\hline
$\#$ & date & r & $\Delta$ & RA & DEC & elong & phase & PLang& config & FlAng & N \\
 & (yyyy-mm-dd) & (AU) & (AU) & (h) & (°) & (°) & (°) & (°) & & (°) & \\ \hline

1               & 2018-01-12 & 2.880 & 2.128 & 04.22  & $+$17.43  & 131.9 & 14.7  & $-$11.9 & A & $+$35 & 3 \\
2               & 2018-01-13 & 2.875  & 2.135 & 04.20  & $+$17.73  & 130.7 & 15.0  & $-$12.2 & A & $-$0 & 1 \\

 \hline
\end{tabular}
\end{table}

\begin{figure}[h!]

    \centering
    \includegraphics[scale=0.368]{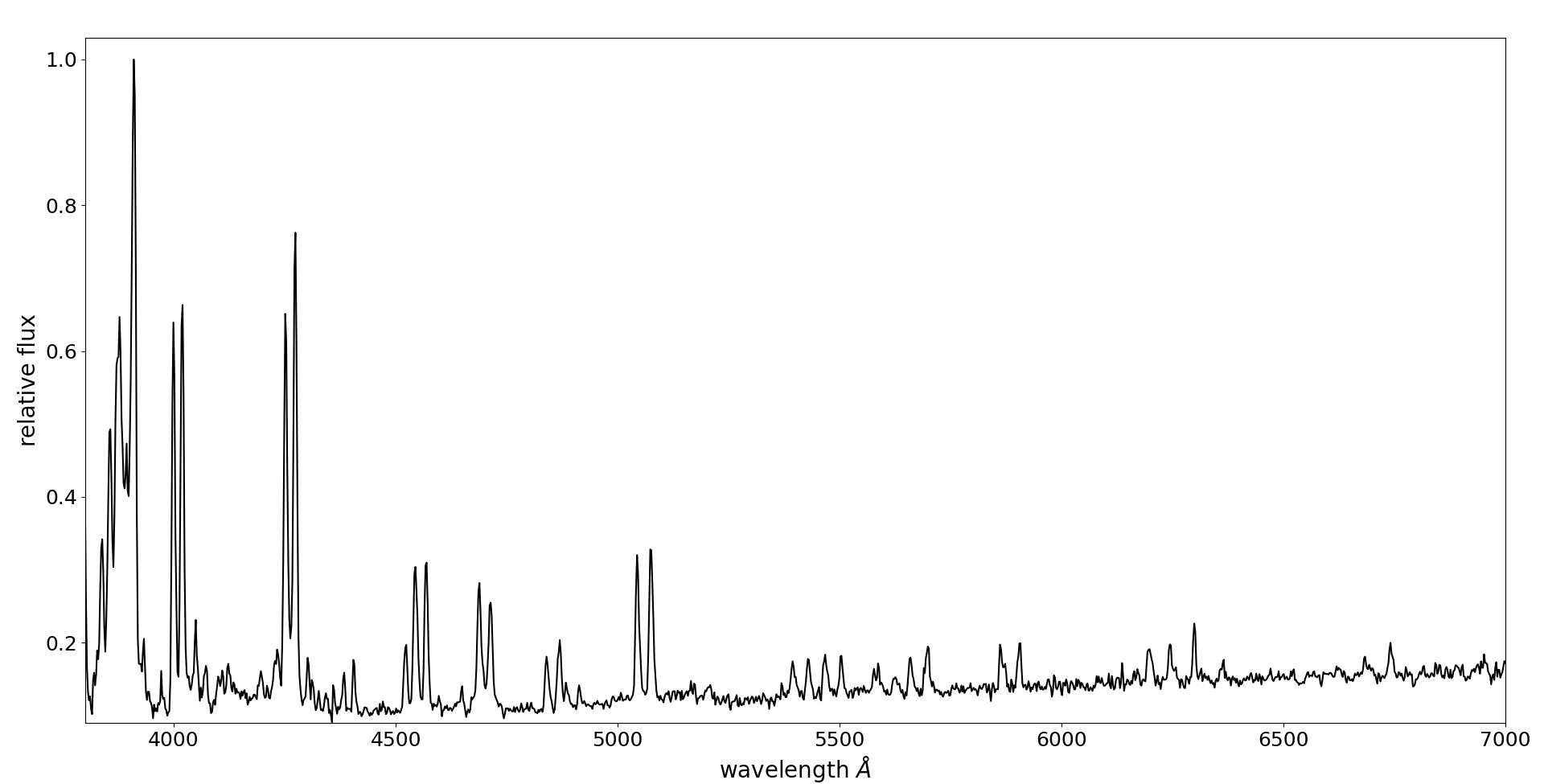}
    \caption{Spectrum of 2018-01-13; configuration A}

\end{figure}

\newpage
\clearpage

\section{C/2017 K2 (PanSTARRS)}
\label{cometa:C2017K2}
\subsection{Description}

C/2017 K2 (PanSTARRS) is a hyperbolic comet with an absolute magnitude of 8.7$\pm$0.8.\footnote{\url{https://ssd.jpl.nasa.gov/tools/sbdb_lookup.html\#/?sstr=2017\%20K2} visited on July 20, 2024}
It was first spotted by the 1.8m Panoramic Survey Telescope \& Rapid Response System (Pan-STARRS 1) on May 21, 2017.
The comet was 16 AU from the Sun when it was discovered. The Earth crossed the comet orbital plane on June 20 and December 20, 2022.

\noindent
We observed the comet from magnitude 14 to 10.\footnote{\url{https://cobs.si/comet/1670/ }, visited on July 20, 2024}

\begin{table}[h!]
\centering
\begin{tabular}{|c|c|c|}
\hline
\multicolumn{3}{|c|}{Orbital elements (epoch: April 26, 2021)}                      \\ \hline \hline
\textit{e} = 1.0005 & \textit{q} = 1.7977 & \textit{T} = 2459933.2794 \\ \hline
$\Omega$ = 88.2464 & $\omega$ = 236.1872  & \textit{i} = 87.5514  \\ \hline  
\end{tabular}
\end{table}

\begin{table}[h!]
\centering
\begin{tabular}{|c|c|c|c|c|c|c|c|c|}
\hline
\multicolumn{9}{|c|}{Comet ephemerides for key dates}                      \\ \hline 
\hline
& date         & r    & $\Delta$  & RA      & DEC      & elong  & phase  & PLang  \\
& (yyyy-mm-dd) & (AU) & (AU)      & (h)     & (°)      & (°)    & (°)    & (°) \\ \hline 

Perihelion       & 2022-12-19 & 1.797  & 2.484 & 17.95  & $-$60.30 &  36.9 & 19.2  & $-$00.1  \\ 
Nearest approach & 2022-07-14 & 2.652  & 1.808 & 16.97  & $-$03.35 & 138.0  & 14.9  & $+$12.8 \\ \hline
\end{tabular}

\end{table}

\vspace{0.5 cm}

\begin{figure}[h!]
    \centering
    \includegraphics[scale=0.38]{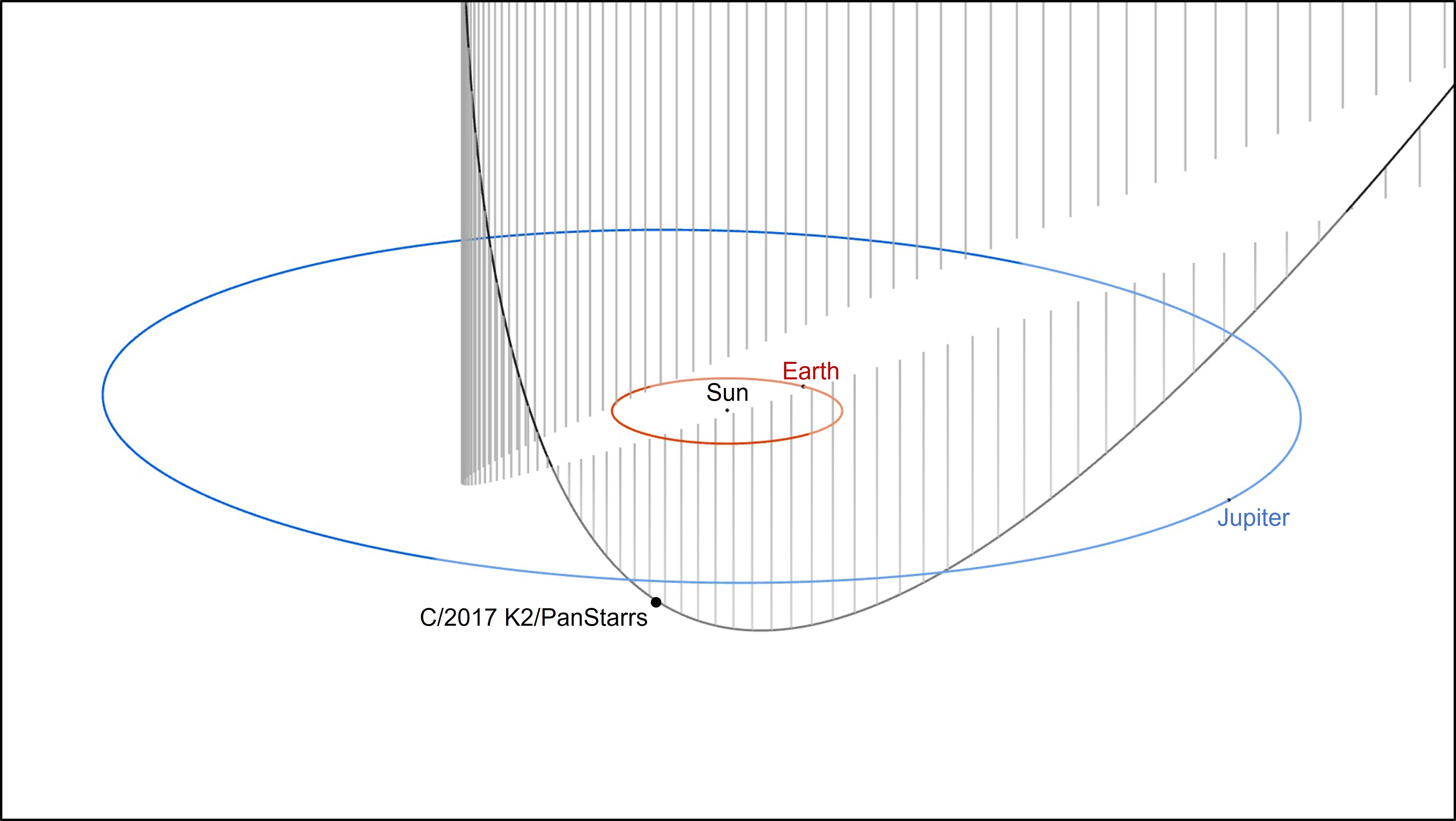}
    \caption{Orbit of comet C/2017 K2 and position on perihelion date. The field of view is set to the orbit of Jupiter for size comparison. Courtesy of NASA/JPL-Caltech.}
\end{figure} 


\newpage

\subsection{Images}

\begin{SCfigure}[0.8][h!]
    \centering
    \includegraphics[scale=0.4]{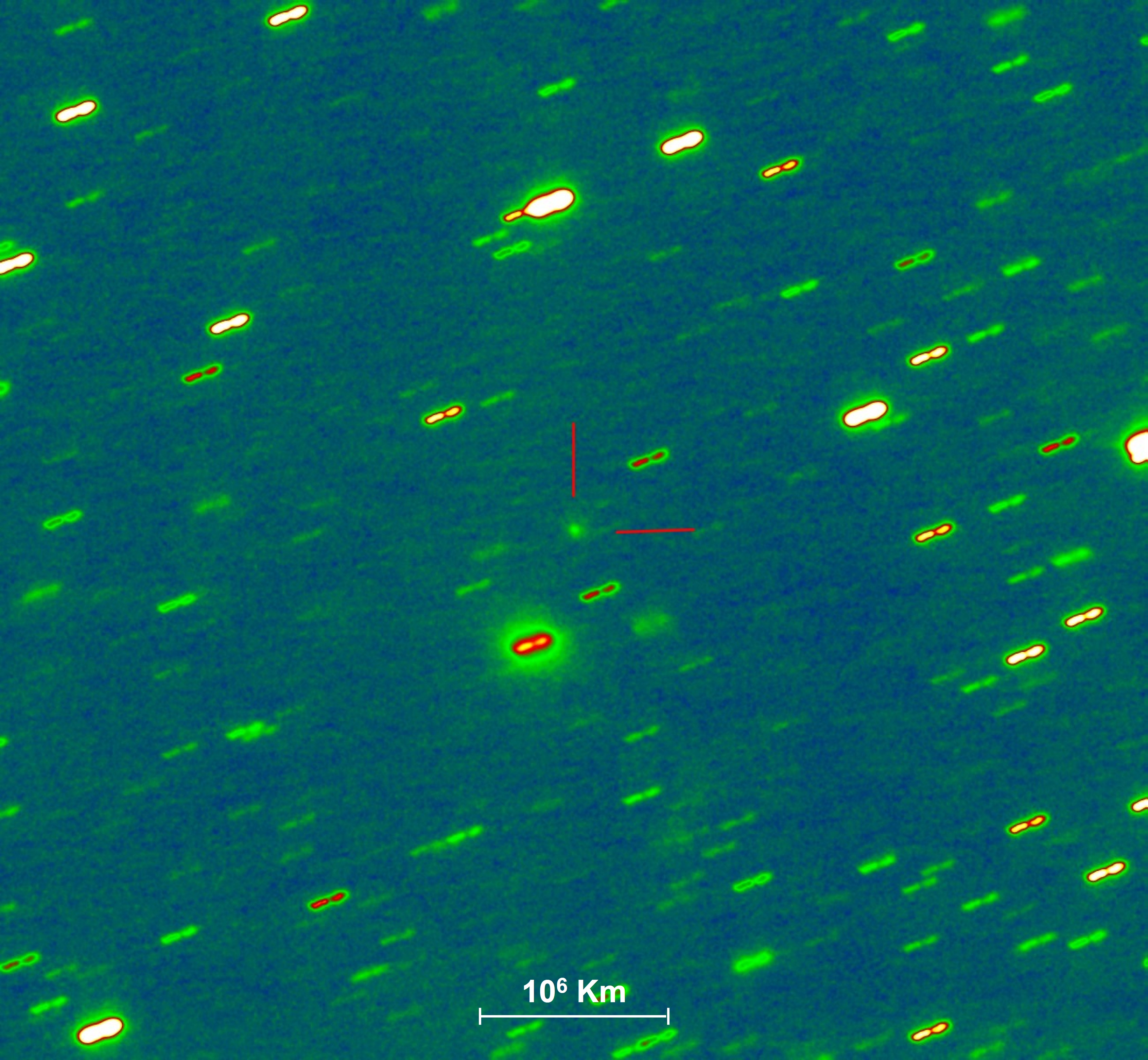}
     \caption{2017-12-03. Comet C/2017 K2 was still very faint at 15 AU from the Sun, where the temperature reaches $-$240 °C, yet there was already a coma composed of super-volatiles. Image taken with the Asiago Schmidt telescope.}
\end{SCfigure}

\begin{SCfigure}[0.8][h!]
    \centering

    \includegraphics[scale=0.4]{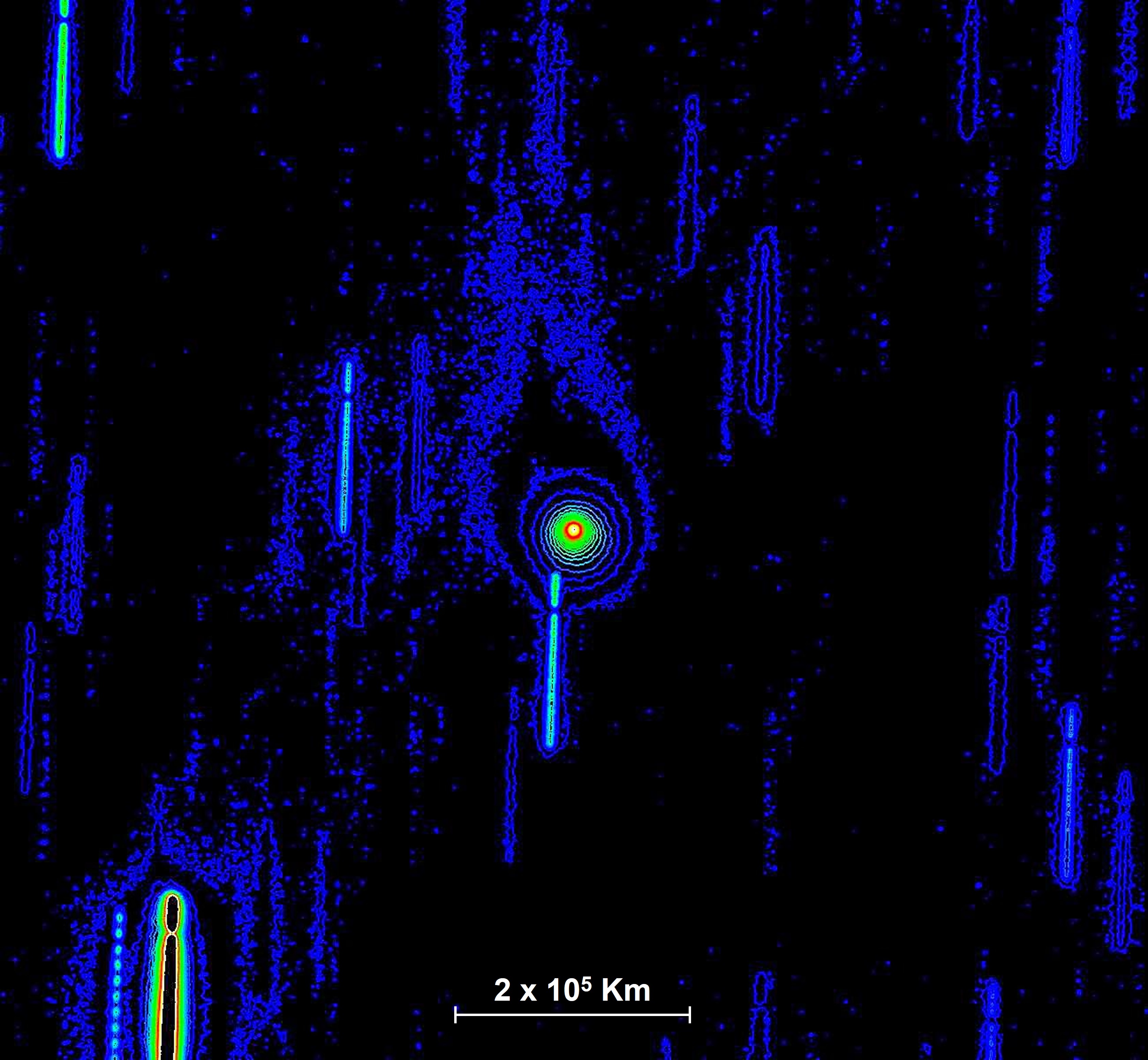}

    \caption{2020-09-12. Approaching the Sun (comet at 8.3 AU), a massive coma is formed, with a hint of a tail pointed in the opposite direction to the Sun and, by geometric effect, also to the Earth. The isophote visualization shows an almost symmetrical coma, with some morphological asymmetry near the nucleus. Image taken with the Asiago Copernico telescope with r and i filters.}
\end{SCfigure}

\newpage

\begin{SCfigure}[0.8][h!]
    \centering
    \includegraphics[scale=0.4]{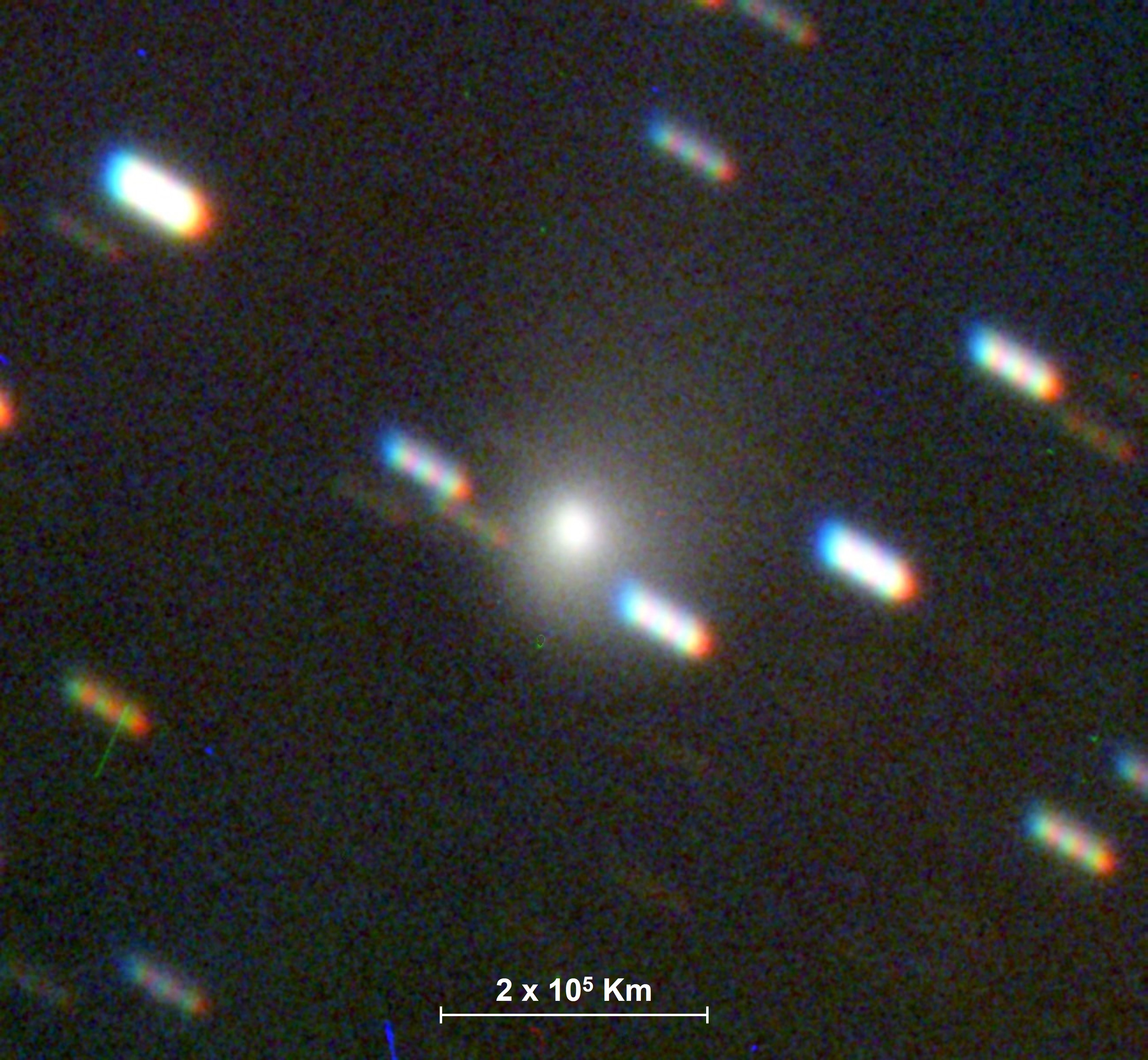}
     \caption{2021-02-14. The comet is increasingly affected by the Sun, to which it is rapidly approaching (comet at 7.1 AU). The coma and the tail are well defined, but a marked asymmetry in the South-West direction with respect to the nucleus is observed. Three-color BVR composite from images taken with the Asiago Copernico telescope.}
\end{SCfigure}

\begin{SCfigure}[0.8][h!]
    \centering
    \includegraphics[scale=0.4]{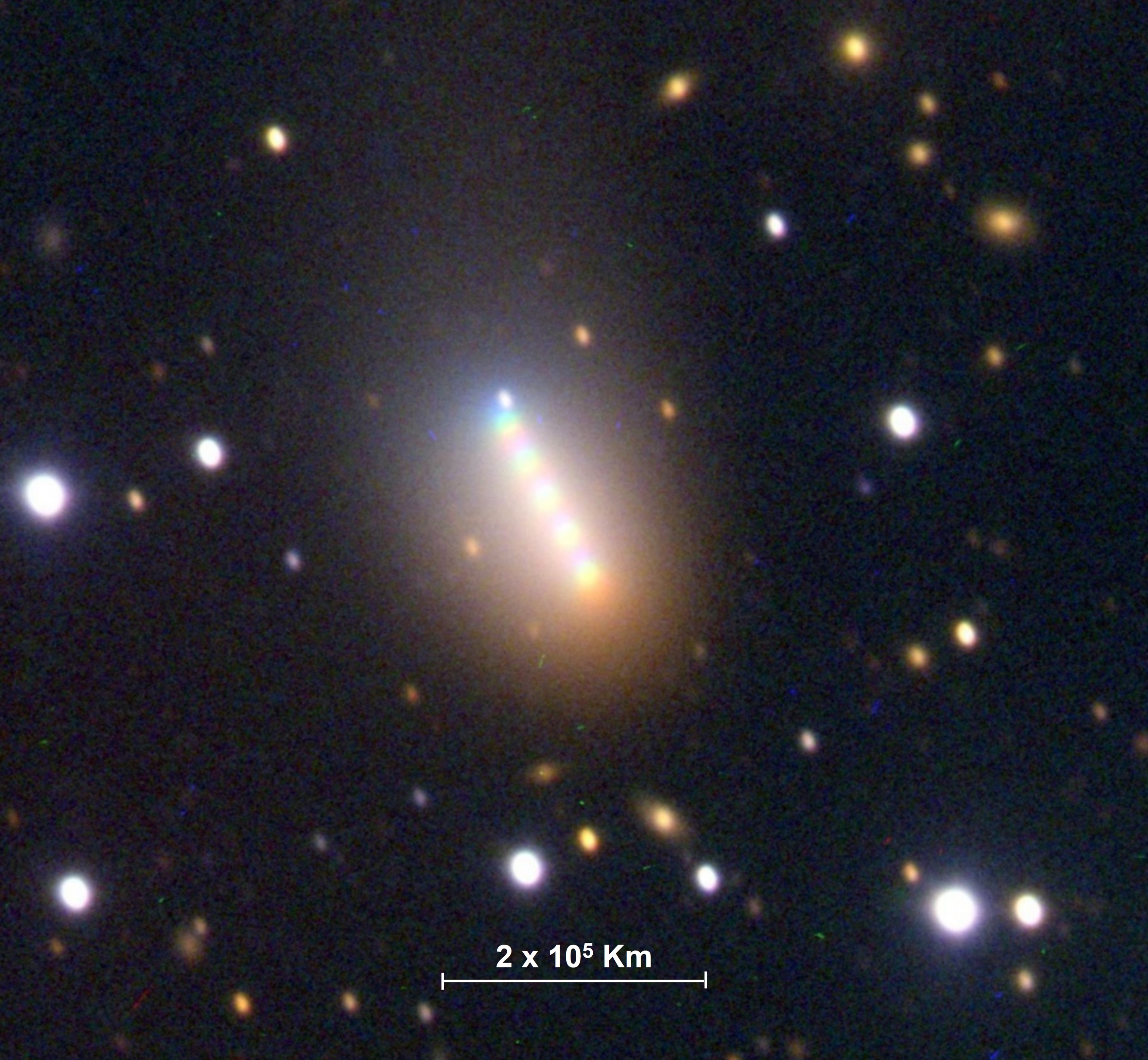} 
    \caption{2021-08-08. The closer they get to the Sun, the faster comets move along their orbit. On August 8, 2021, C/2017 K2 was in the constellation of Hercules surrounded by very distant stars and galaxies, reddened by intergalactic dust. Three-color BVR composite from images taken with the Asiago Copernico telescope.}
\end{SCfigure}

\newpage

\subsection{Spectra}

\begin{table}[h!]
\centering
\begin{tabular}{|c|c|c|c|c|c|c|c|c|c|c|c|}
\hline
\multicolumn{12}{|c|}{Observation details}                      \\ \hline 
\hline
$\#$  & date          & r     & $\Delta$ & RA     & DEC     & elong & phase & PLang& config  & FlAng & N \\
      & (yyyy-mm-dd)  &  (AU) & (AU)     & (h)    & (°)     & (°)   & (°)   &  (°)   &       &  (°)  &  \\ \hline 

1  & 2021-03-16 & 6.855  & 6.912 & 18.45 & $+$38.00 & 82.6  & 08.3  & $-$08.3 & A & $+$0 & 3 \\
2  & 2021-03-21 & 6.814  & 6.832 & 18.47 & $+$38.52 & 84.8  & 08.4  & $-$08.4 & A & $-$0 & 2 \\
3  & 2021-04-08 & 6.662  & 6.528 & 18.47 & $+$40.05 & 93.3  & 08.6  & $-$08.3 & A & $+$90 & 1\\
4  & 2022-01-10 & 4.303	 & 5.009 & 18.08 & $+$12.68 & 40.1  & 08.5 & $-$04.4 & A & $-$70 & 1 \\
5  & 2022-01-22 & 4.205	 & 4.861 & 18.02 & $+$12.07 & 43.7  & 09.3 & $-$06.6 & A & $-$40 & 4  \\ 
6  & 2022-03-26 & 3.627	 & 3.628 & 18.88 & $+$11.55 & 82.1  & 15.8 & $-$15.9 & A & $+$90 & 3 \\
7  & 2022-05-14 & 3.185	 & 2.517 & 18.72 & $+$11.22 & 123.2 & 15.4 & $-$13.2 & A & $+$0 & 6 \\
8 & 2022-05-20 & 3.139	 & 2.415 & 18.63 & $+$10.87 & 127.8 & 14.7 & $-$12.1 & D & $+$0 & 1 \\ 
9 & 2022-05-21 & 3.130	 & 2.395 & 08.62 & $+$10.78 & 128.8 & 14.6 & $-$11.8 & D & $+$79 & 5 \\
10 & 2022-06-16 & 2.890	 & 1.965 & 16.90 & $-$04.42 & 150.2 & 10.1 & $-$01.6 & A & $+$13 & 3 \\
11 & 2022-06-17 & 2.882	 & 1.954 & 16.87 & $-$04.85 & 150.5 & 10.0 & $+$01.1 & A & $+$13 & 6 \\

\hline
\end{tabular}
\end{table}

\begin{figure}[h!]

    \centering
    \includegraphics[scale=0.368]{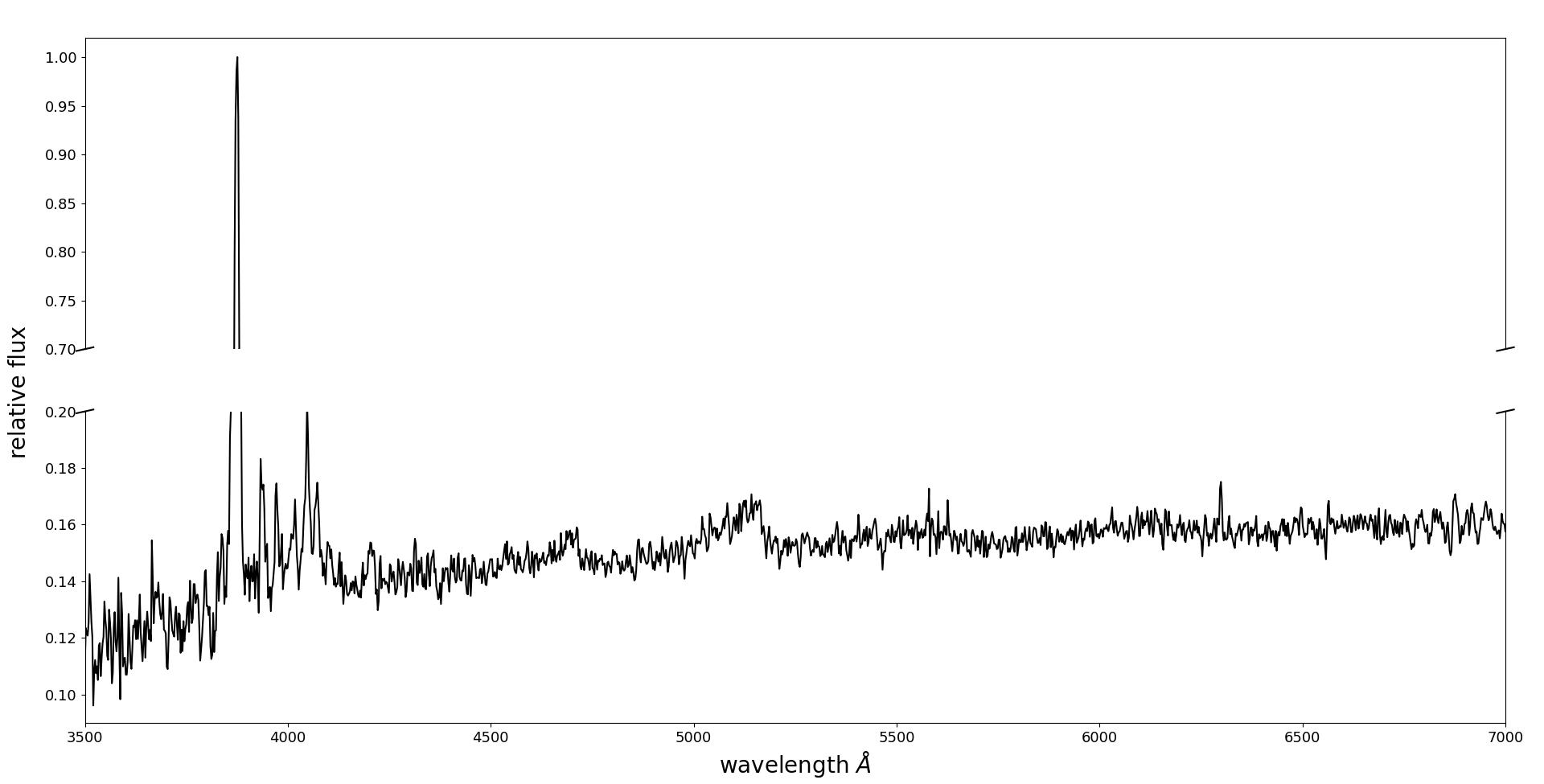}
    \caption{Spectrum of 2022-06-17; configuration A}

\end{figure}

\begin{figure}[h!]

    \centering
    \includegraphics[scale=0.368]{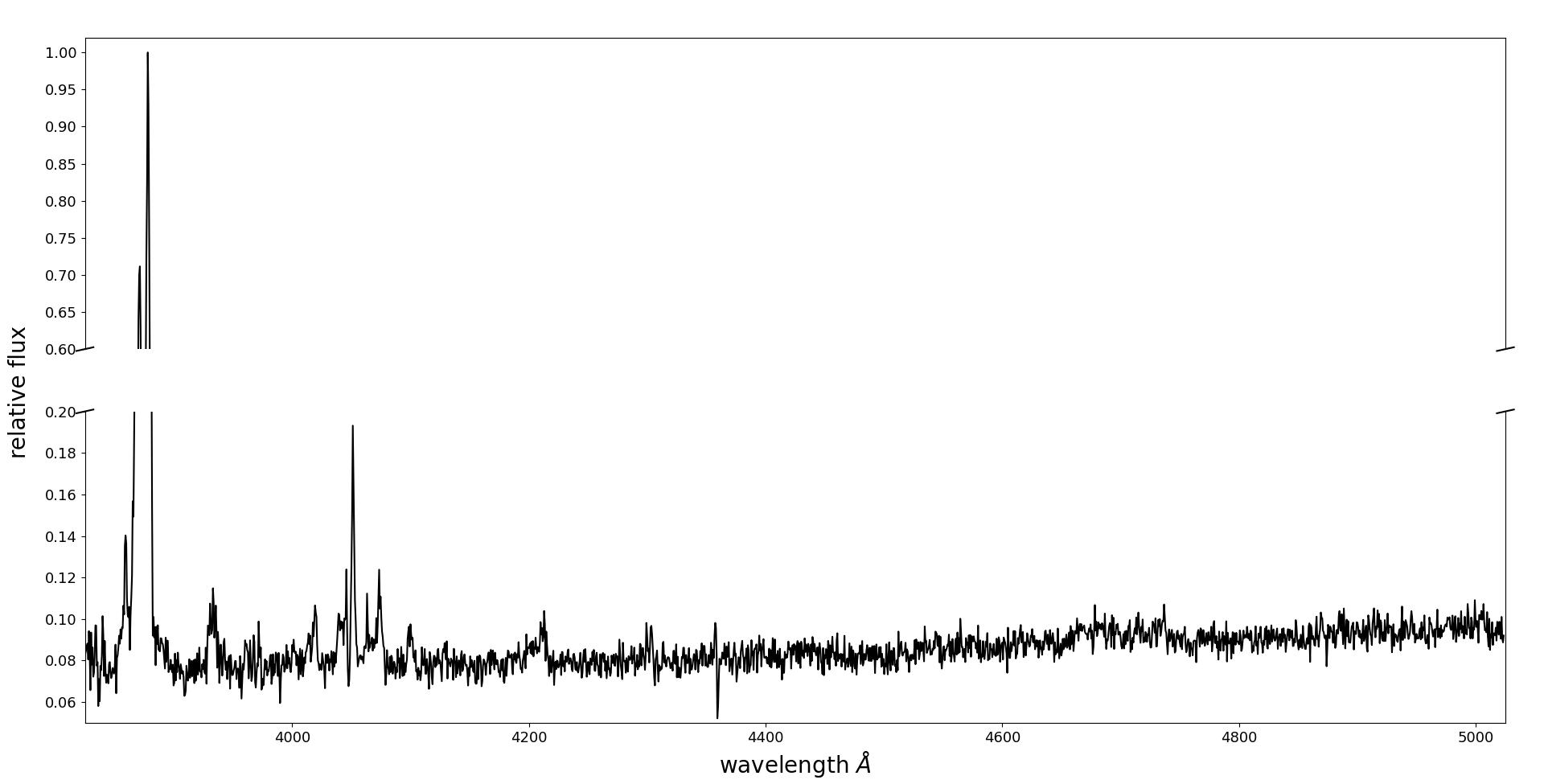}
    \caption{Spectrum of 2022-05-21; configuration D}

\end{figure}

\newpage
\clearpage

\section{C/2017 T2 (PanSTARRS)}
\label{cometa:C2017T2}
\subsection{Description}

C/2017 T2 (PanSTARRS) is a Long Period comet with a period of 354 thousand years and an absolute magnitude of 10.2$\pm$1.0.\footnote{\url{https://ssd.jpl.nasa.gov/tools/sbdb_lookup.html\#/?sstr=2017\%20T2} visited on July 20, 2024}
It was first spotted by Robert Weryk using the 1.8m Panoramic Survey Telescope \& Rapid Response System (Pan-STARRS 1) on October 2, 2017.
The Earth crossed the comet orbital plane on June 11 and December 12, 2018.

\noindent
We observed the comet between magnitudes 10 and 8.\footnote{\url{https://cobs.si/comet/1703/ }, visited on July 20, 2024}

\begin{table}[h!]
\centering
\begin{tabular}{|c|c|c|}
\hline
\multicolumn{3}{|c|}{Orbital elements (epoch: December 13, 2019)}                      \\ \hline \hline
\textit{e} = 0.9997 & \textit{q} = 1.6151 & \textit{T} = 2458974.4499 \\ \hline
$\Omega$ = 64.3788 & $\omega$ = 92.9933  & \textit{i} = 57.2314  \\ \hline  
\end{tabular}
\end{table}

\begin{table}[h!]
\centering
\begin{tabular}{|c|c|c|c|c|c|c|c|c|}
\hline
\multicolumn{9}{|c|}{Comet ephemerides for key dates}                      \\ \hline 
\hline

& date & r & $\Delta$ & RA & DEC & elong & phase & PLang \\
& (yyyy-mm-dd) & (AU) & (AU) & (h) & (°) & (°) & (°) & (°) \\ \hline 

Perihelion       & 2020-05-04 & 1.615  & 1.696 & 6.93  & $+$76.31  & 68.0  & 35.4  & $-$10.1  \\
Nearest approach & 2020-05-27 & 1.644  & 1.659 & 10.48 & $+$67.20 & 71.4 & 35.7  & $+$01.3 \\ \hline
\end{tabular}

\end{table}

\vspace{0.5 cm}

\begin{figure}[h!]
    \centering
    \includegraphics[scale=0.38]{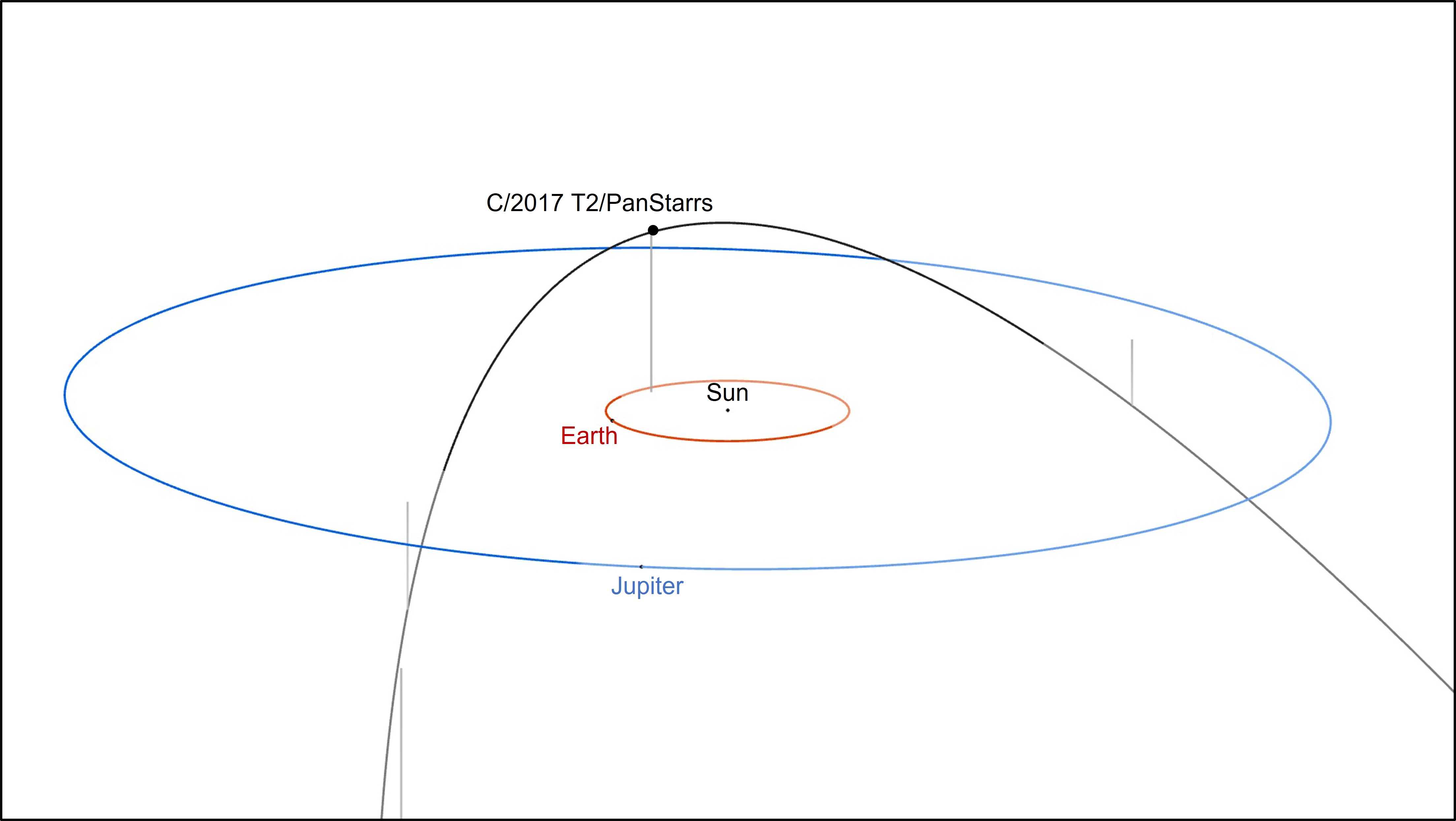}
    \caption{Orbit of comet C/2017 T2 and position on perihelion date. The field of view is set to the orbit of Jupiter for size comparison. Courtesy of NASA/JPL-Caltech.}
\end{figure}

\newpage


\subsection{Images}
\begin{SCfigure}[0.8][h!]
 \centering
 \includegraphics[scale=0.4]{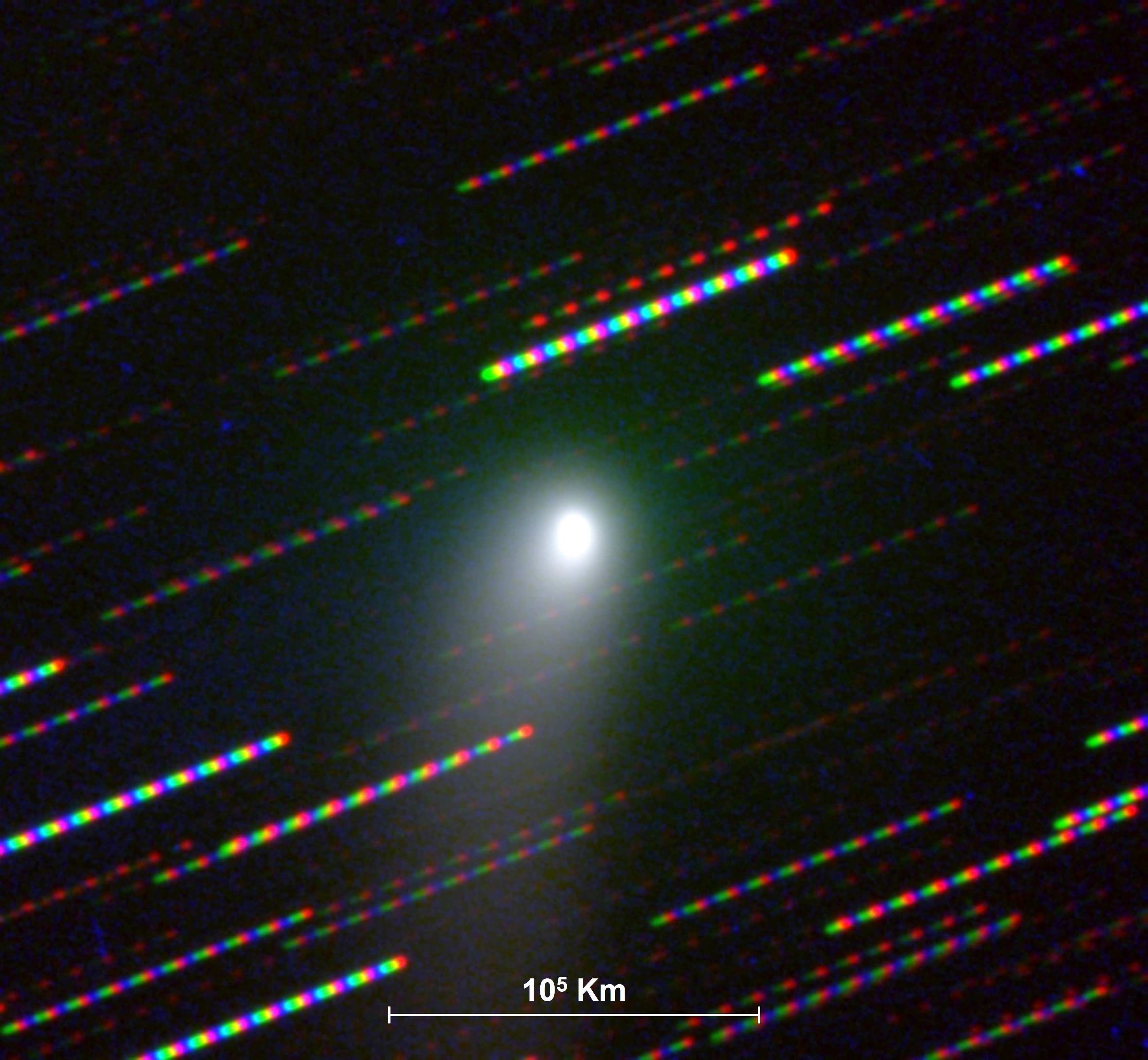}
 \caption{2019.12.29. Three-color uBR composite from images taken with the Asiago Schmidt telescope.
The comet was very bright, hence allowing excellent images in the ultraviolet (u) band and in the blue, where the emissions of CN (cyanogen) and C$_3$ (triatomic carbon) molecules dominate.}
\end{SCfigure}

\begin{SCfigure}[0.8][h!]
 \centering

 \includegraphics[scale=0.4]{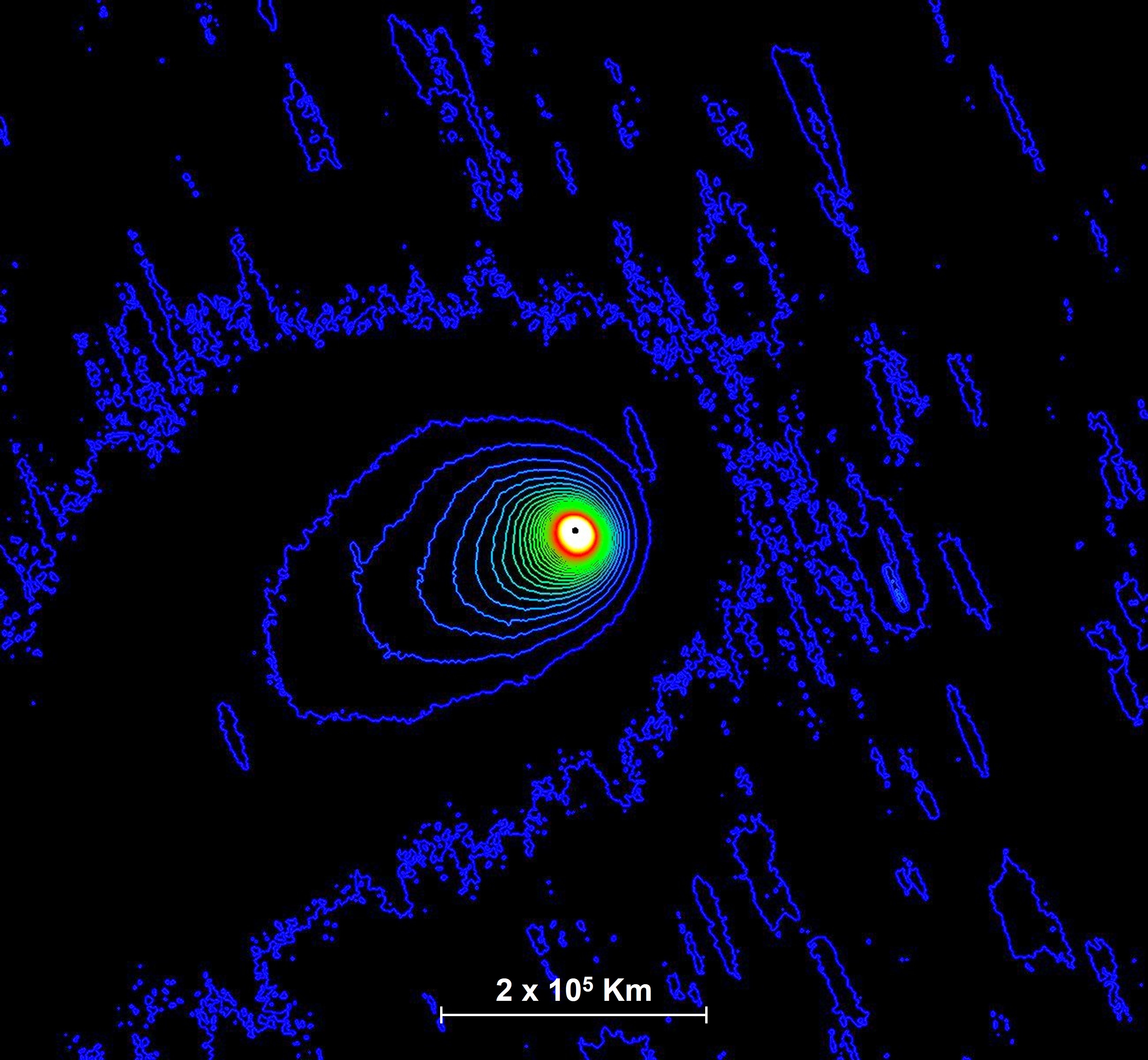}
 \caption{2020-02-28. Composite image taken with r and i filters, visualized in isophotes oriented with a $90^{\circ}$ rotation from the outside towards the nucleus, to show the complexity of the inner coma.}
\end{SCfigure}
\begin{SCfigure}[0.8][h!]
 \centering

 \includegraphics[scale=0.4]{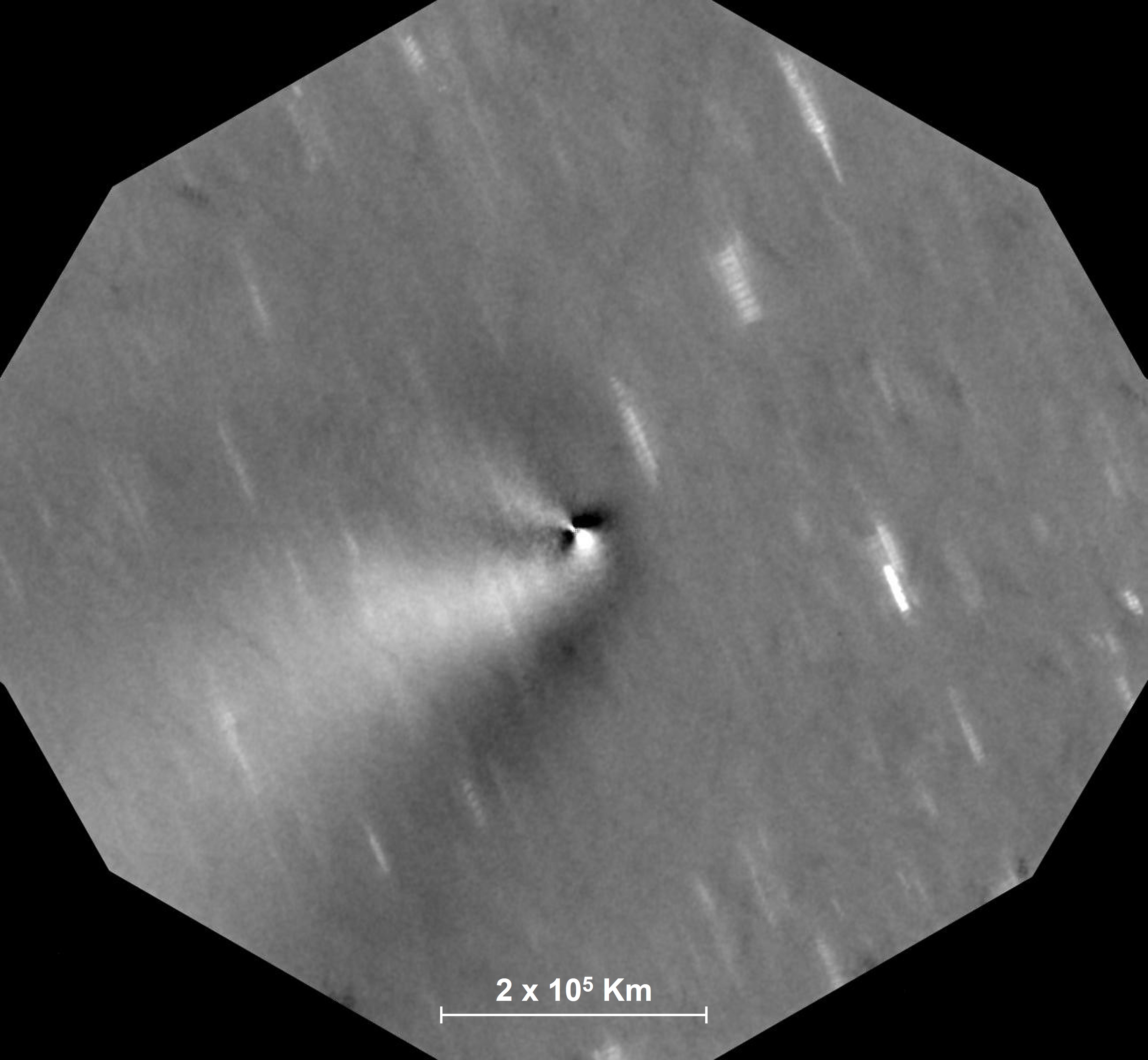}

 \caption{2020-02-28. A more specific mathematical treatment on the same image allows to observe a powerful jet departing from the nucleus in the NE direction. Jets like this have been observed in many comets, and result from the presence of discrete active areas on the surface of the nucleus.}
\end{SCfigure}

\newpage

\clearpage

\subsection{Spectra}

\begin{table}[h!]
\centering
\begin{tabular}{|c|c|c|c|c|c|c|c|c|c|c|c|}
\hline
\multicolumn{12}{|c|}{Observation details}                      \\ \hline 
\hline
$\#$ & date & r & $\Delta$ & RA & DEC & elong & phase & PLang & config & FlAng & N \\
 & (yyyy-mm-dd) & (AU) & (AU) & (h) & (°) & (°) & (°) & (°) & & (°) &  \\ \hline

1               & 2019-12-27 & 2.331  & 1.520 & 03.70 & $+$54.35  & 136.2 & 17.0  & $-$16.5 & A & $+$90 & 2 \\
2               & 2020-01-04 & 2.261  & 1.527 & 03.28 & $+$55.72  & 127.2 & 20.3  & $-$20.3 & A & $+$90 & 3 \\
3               & 2020-01-05 & 2.254  & 1.528 & 03.23 & $+$55.85  & 126.3 & 20.6  & $-$20.6 & A & $+$90 & 7 \\
4               & 2020-01-14 & 2.178  & 1.553 & 02.83 & $+$56.82  & 116.5 & 23.8  & $-$24.0 & A & $+$90 & 2 \\
5               & 2020-01-19 & 2.137  & 1.572 & 02.63 & $+$57.20  & 111.3 & 25.4  & $-$25.5 & A & $+$90 & 1 \\
6               & 2020-01-20 & 2.128  & 1.577 & 02.60 & $+$57.27  & 110.3 & 25.7  & $-$25.8 & A & $+$90 & 3 \\
7               & 2020-01-21 & 2.119  & 1.581 & 02.56 & $+$57.33  & 109.2 & 26.0  & $-$26.1 & A & $+$90 & 1 \\
8               & 2020-02-15 & 1.928  & 1.693 & 02.12 & $+$59.30  & 87.9 & 30.8   & $-$29.2 & A & $+$90 & 1 \\
9               & 2020-02-28 & 1.842  & 1.738 & 02.15 & $+$61.08  & 79.8 & 32.0   & $-$28.7 & A & $+$90 & 1 \\
10              & 2020-05-01 & 1.616  & 1.702 & 06.31 & $+$76.25  & 67.7 & 35.2   & $-$11.3 & A & $-$0 & 2 \\
11              & 2020-05-02 & 1.615  & 1.700 & 06.50 & $+$76.32  & 67.8 & 35.3   & $-$10.9 & A & $+$76 & 4 \\
12              & 2020-05-04 & 1.615  & 1.795 & 06.85 & $+$76.30  & 68.8 & 35.4   & $-$09.9 & A & $+$65 & 3 \\
13              & 2020-05-07 & 1.616  & 1.687 & 07.45 & $+$76.07  & 68.4 & 35.5   & $-$08.6 & A & $+$60 & 3 \\
14              & 2020-05-08 & 1.616  & 1.685 & 07.63 & $+$75.93  & 68.5 & 35.5   & $-$08.1 & A & $+$60 & 1 \\
15              & 2020-05-24 & 1.636  & 1.660 & 10.13 & $+$69.83  & 70.8 & 35.8   & $-$00.9 & A & $+$25 & 3 \\
16              & 2020-05-26 & 1.640  & 1.659 & 10.35 & $+$68.65  & 71.2 & 35.7   & $+$00.5 & A & $+$25 & 5 \\
17              & 2020-06-01 & 1.659  & 1.662 & 10.92 & $+$64.75  & 72.1 & 35.6   & $+$03.5 & A & $+$15 & 3 \\
18              & 2020-06-12 & 1.698  & 1.686 & 11.67 & $+$56.63  & 73.2 & 34.9   & $+$08.6 & A & $-$0 & 5 \\
19              & 2020-06-28 & 1.775  & 1.778 & 12.40 & $+$43.90  & 73.3 & 33.3   & $+$15.0 & A & $-$0 & 1 \\
 \hline
\end{tabular}
\end{table}

\begin{figure}[h!]

    \centering
    \includegraphics[scale=0.368]{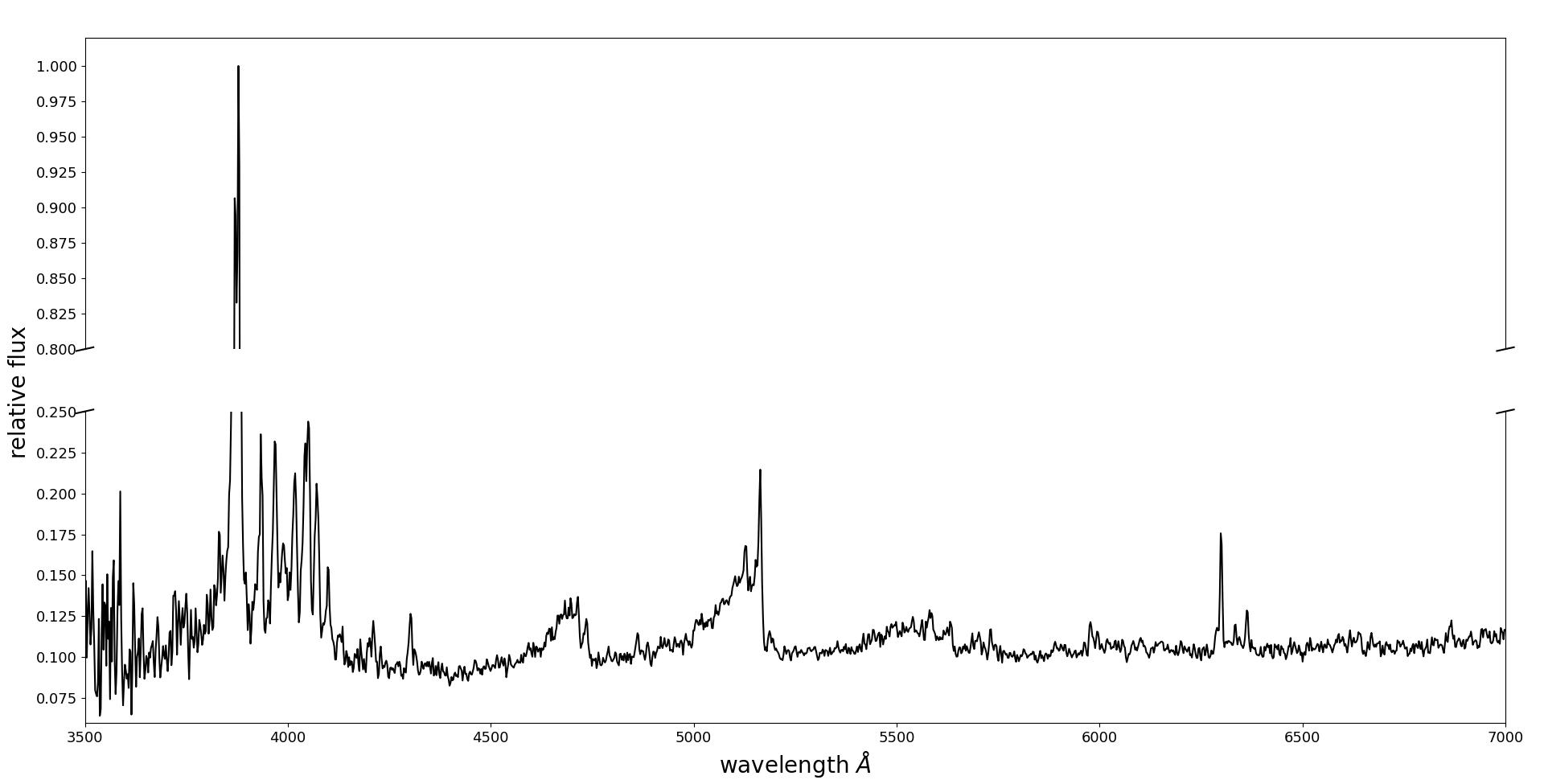}
    \caption{Spectrum of 2020-05-02; configuration A}

\end{figure}

\newpage

\newpage
\clearpage

\section{C/2018 N2 (ASASSN)}
\label{cometa:C2018N2}
\subsection{Description}

C/2018 N2 (ASASSN) is a hyperbolic comet with an absolute magnitude of 9.6$\pm$0.7.\footnote{\url{https://ssd.jpl.nasa.gov/tools/sbdb_lookup.html\#/?sstr=2018\%20N2} visited on July 20, 2024}
It was first spotted by Benjamin Shappee using the 0.14 m telescope as part of the "All-Sky Automated Survey for Supernovae" (ASASSN) on July 7, 2018.
Comet C/2018 N2 has a perihelion at a distance of 3.124 AU from the Sun, but developed a long tail and some morphological structures in the inner coma.
The Earth crossed the comet orbital plane on October 18, 2019.

\noindent
We observed the comet at magnitude 12.\footnote{\url{https://cobs.si/comet/1750/ }, visited on July 20, 2024}


\begin{table}[h!]
\centering
\begin{tabular}{|c|c|c|}
\hline
\multicolumn{3}{|c|}{Orbital elements (epoch: February 26, 2020)}                      \\ \hline \hline
\textit{e} = 1.0002 & \textit{q} = 3.1247 & \textit{T} = 2458798.4642 \\ \hline
$\Omega$ = 25.2578 & $\omega$ = 24.3970  & \textit{i} = 77.5300  \\ \hline  
\end{tabular}
\end{table}

\begin{table}[h!]
\centering
\begin{tabular}{|c|c|c|c|c|c|c|c|c|}
\hline
\multicolumn{9}{|c|}{Comet ephemerides for key dates}                      \\ \hline 
\hline
& date         & r    & $\Delta$  & RA      & DEC      & elong  & phase  & PLang  \\
& (yyyy-mm-dd) & (AU) & (AU)      & (h)     & (°)      & (°)    & (°)    & (°) \\ \hline 

Perihelion       & 2019-11-10 & 3.125 & 2.305 & 00.38 & $+$38.77 & 139.4	& 11.9 & $+$09.3 \\ 
Nearest approach & 2019-10-19 & 3.132 & 2.212 & 01.21 &	$+$36.38 & 153.0	& 08.3	& $+$00.3 \\ \hline
\end{tabular}

\end{table}

\vspace{0.5 cm}

\begin{figure}[h!]
    \centering
    \includegraphics[scale=0.38]{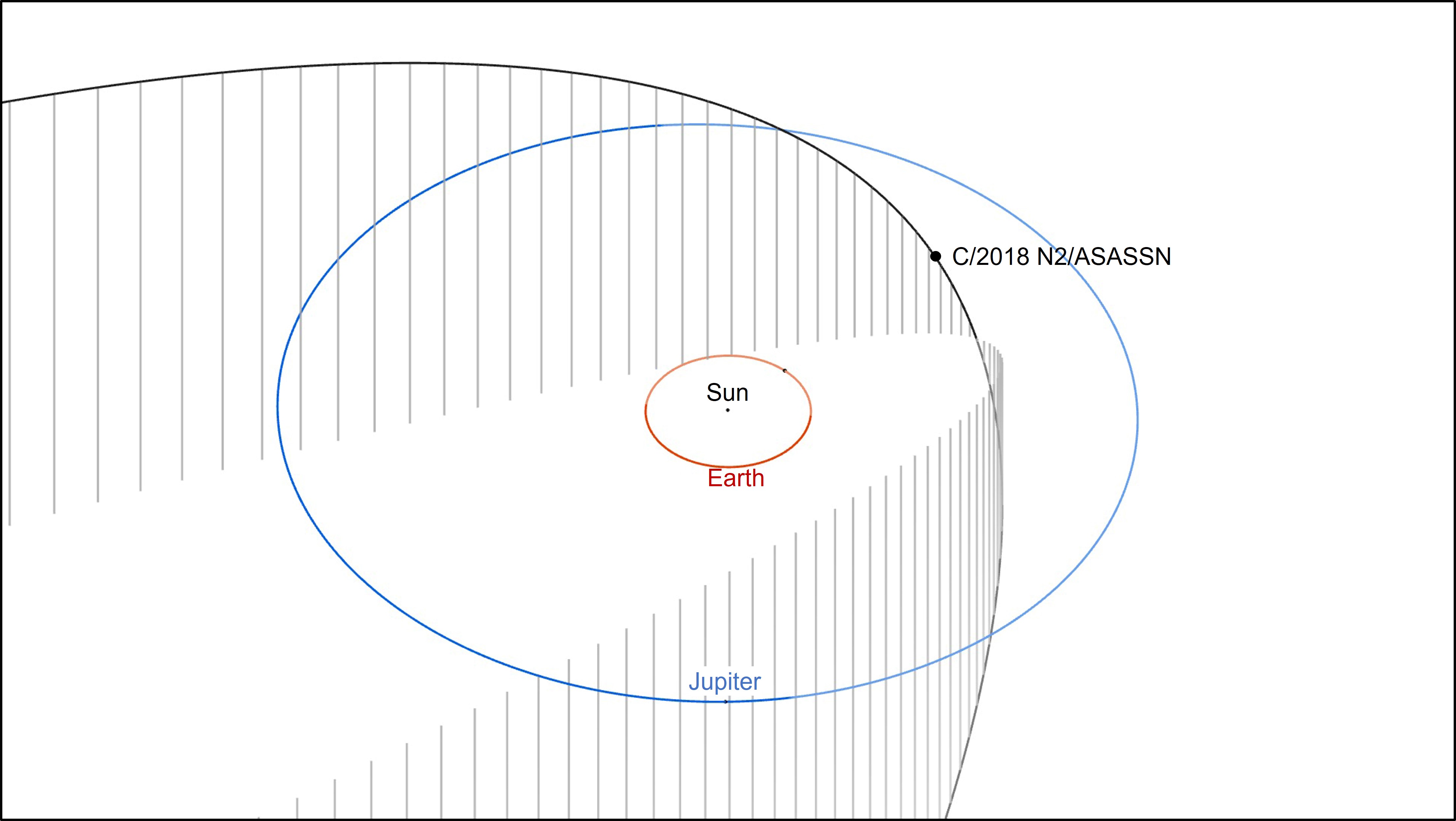}
    \caption{Orbit of comet C/2018 N2 and position on perihelion date. The field of view is set to the orbit of Jupiter for size comparison. Courtesy of NASA/JPL-Caltech.}
\end{figure} 


\newpage

\subsection{Images}

\begin{SCfigure}[0.8][h!]
    \centering
    \includegraphics[scale=0.4]{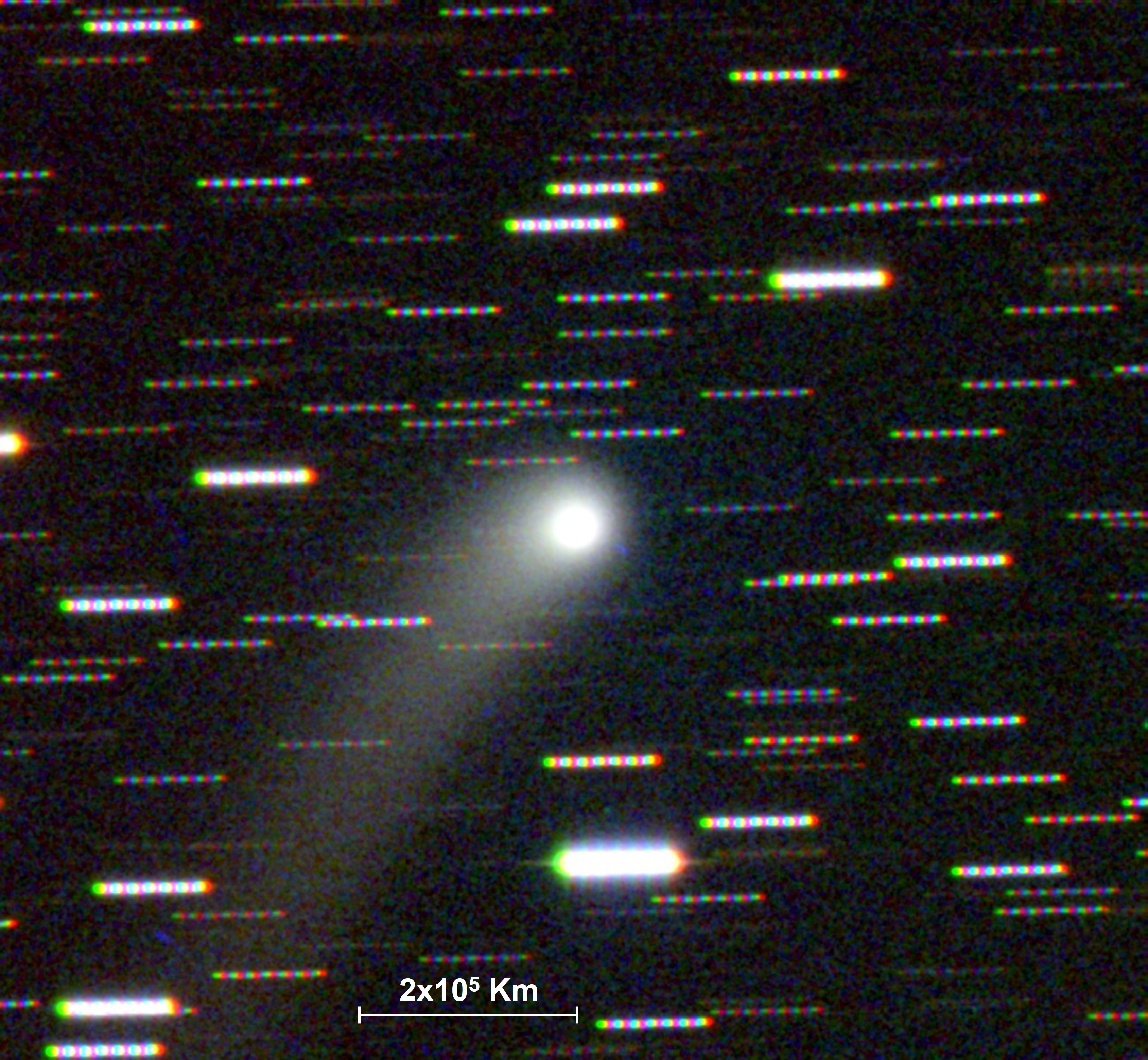}
     \caption{2019-12-05. Three-color BVR composite from images taken with the Asiago Schmidt telescope. The comet was located almost 500 million km from the Sun, and yet it developed a very long tail and a coma in which morphological structures were observed.}
\end{SCfigure} 

\begin{SCfigure}[0.8][h!]
    \centering

    \includegraphics[scale=0.4]{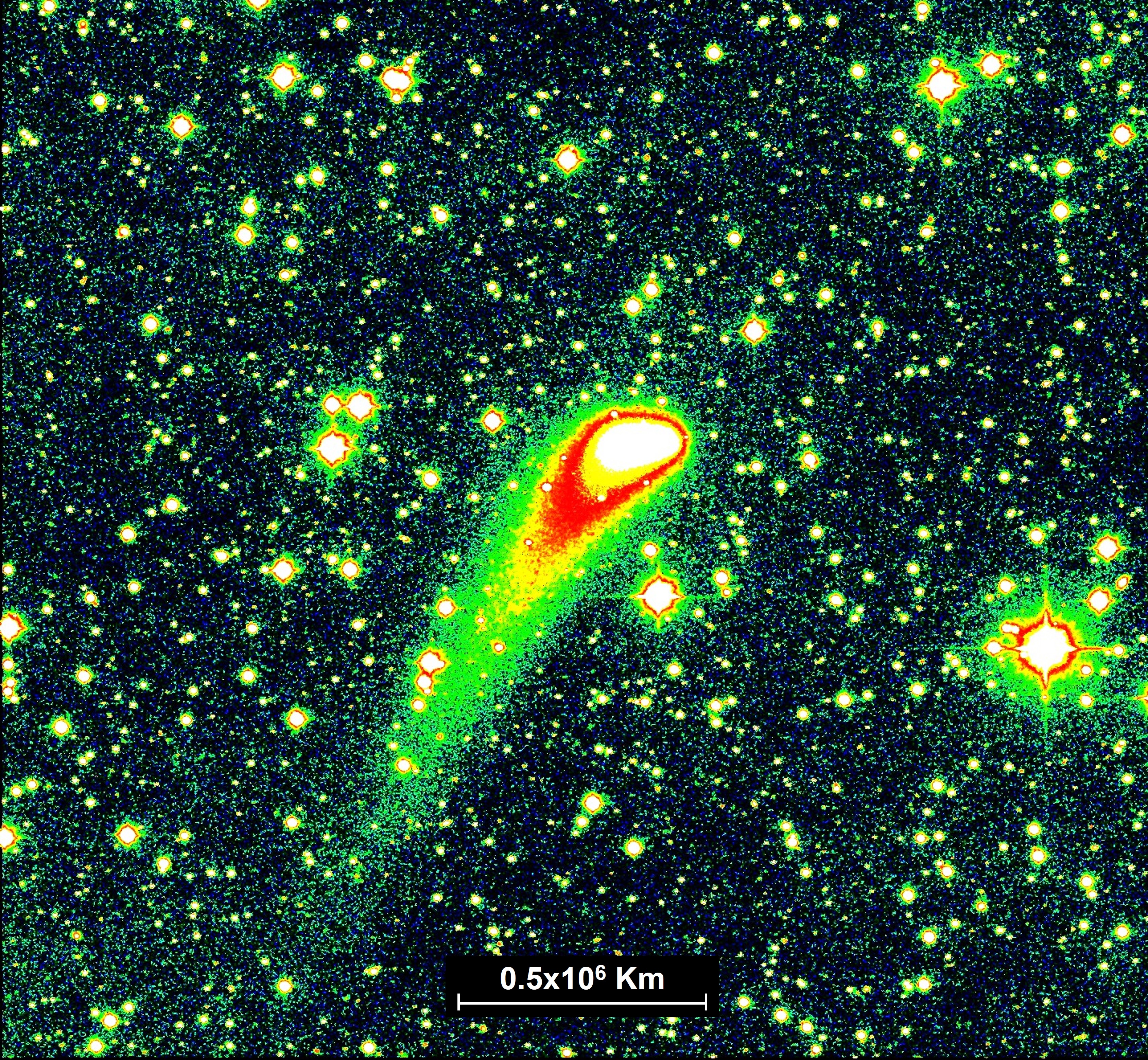}

    \caption{The wide-field false-color view shows almost the entire extension of the tail, which reached several millions km.}
\end{SCfigure}

\newpage

\subsection{Spectra}

\begin{table}[h!]
\centering
\begin{tabular}{|c|c|c|c|c|c|c|c|c|c|c|c|}
\hline
\multicolumn{12}{|c|}{Observation details}                      \\ \hline 
\hline
$\#$  & date          & r     & $\Delta$ & RA     & DEC     & elong & phase & PLang& config  & FlAng & N \\
      & (yyyy-mm-dd)  &  (AU) & (AU)     & (h)    & (°)     & (°)   & (°)   &  (°)   &       &  (°)  & \\ \hline 

1 & 2019-09-23 & 3.159 & 2.343 & 02.12 & $+$30.03 & 137.8	& 12.3 & $-$10.0 & A & $-$49 & 1 \\
\hline
\end{tabular}
\end{table}

\begin{figure}[h!]

    \centering
    \includegraphics[scale=0.368]{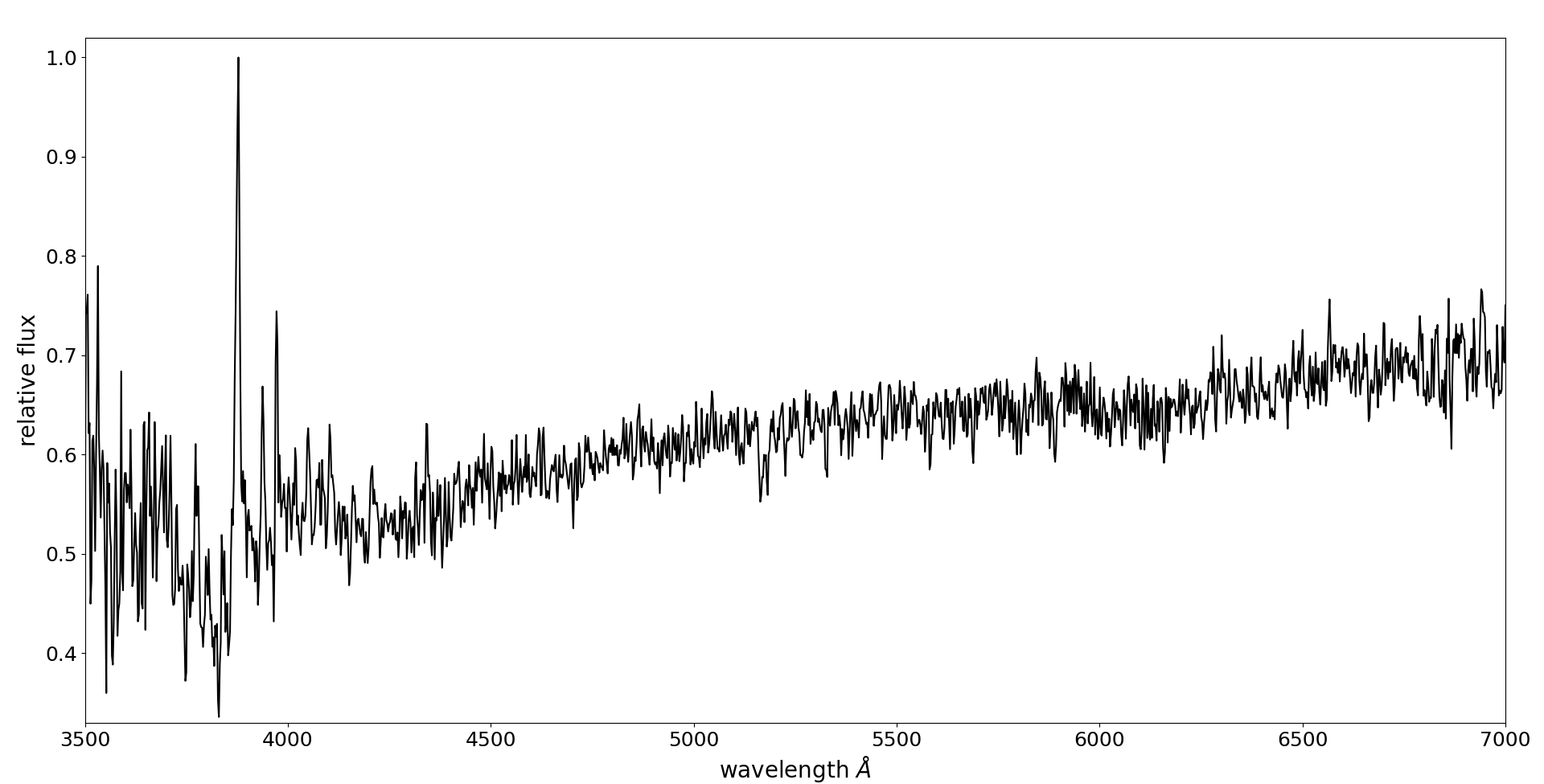}
    \caption{Spectrum of 2019-09-23; configuration A}

\end{figure}

\newpage
\clearpage

\section{C/2018 Y1 (Iwamoto)}
\label{cometa:C2018Y1}
\subsection{Description}

C/2018 Y1 (Iwamoto) is a Long Period comet with a period of 1733 years and an absolute magnitude of 13.6$\pm$1.0.\footnote{\url{https://ssd.jpl.nasa.gov/tools/sbdb_lookup.html\#/?sstr=2018\%20Y1} visited on July 20, 2024}
It was first spotted by the Japanese amateur astronomer Masayuki Iwamoto using its own telescope on December 18, 2018.
Follow-up observations of comet C/2018 Y1 (Iwamoto) revealed a coma as large as 2.9' of magnitude 12.5, with a prominent central condensation but no tail.
Comet Iwamoto has an aphelic distance of more than 287 AU.
The Earth crossed the comet orbital plane on February 16, 2019.

\noindent
We observed the comet at magnitude 7.\footnote{\url{https://cobs.si/comet/1774/ }, visited on July 20, 2024}

\begin{table}[h!]
\centering
\begin{tabular}{|c|c|c|}
\hline
\multicolumn{3}{|c|}{Orbital elements (epoch: March 3, 2019)}                      \\ \hline \hline
\textit{e} = 0.9911 & \textit{q} = 1.2870 & \textit{T} = 2458521.5306 \\ \hline
$\Omega$ = 147.4836 & $\omega$ = 358.0572  & \textit{i} = 160.4035  \\ \hline  
\end{tabular}
\end{table}

\begin{table}[h!]
\centering
\begin{tabular}{|c|c|c|c|c|c|c|c|c|}
\hline
\multicolumn{9}{|c|}{Comet ephemerides for key dates}                      \\ \hline 
\hline
& date         & r    & $\Delta$  & RA      & DEC      & elong  & phase  & PLang  \\
& (yyyy-mm-dd) & (AU) & (AU)      & (h)     & (°)      & (°)    & (°)    & (°) \\ \hline 

Perihelion       & 2019-02-07 & 1.287 & 0.367 & 11.89 & $-$01.18 & 139.2	& 30.0 & $-$08.5 \\ 
Nearest approach & 2019-02-12 & 1.290 & 0.304 & 10.14 &	$+$16.40 & 173.1	& 05.3	& $-$04.7 \\ \hline
\end{tabular}

\end{table}

\vspace{0.5 cm}

\begin{figure}[h!]
    \centering
    \includegraphics[scale=0.38]{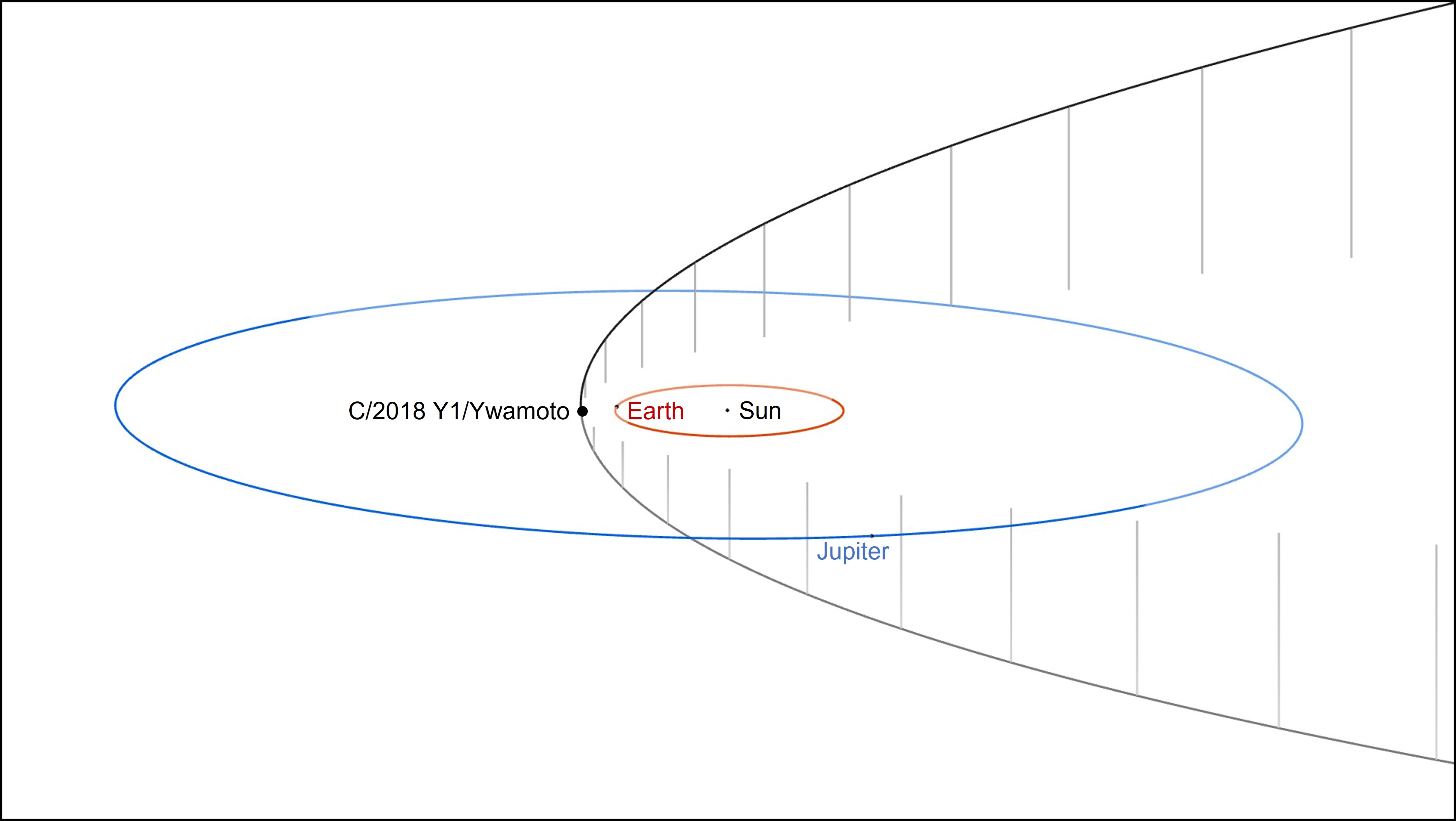}
    \caption{Orbit of comet C2018 Y1 and position on perihelion date. The field of view is set to the orbit of Jupiter for size comparison. Courtesy of NASA/JPL-Caltech.}
\end{figure} 


\newpage

\subsection{Images}

\begin{SCfigure}[0.8][h!]
    \centering
    \includegraphics[scale=0.4]{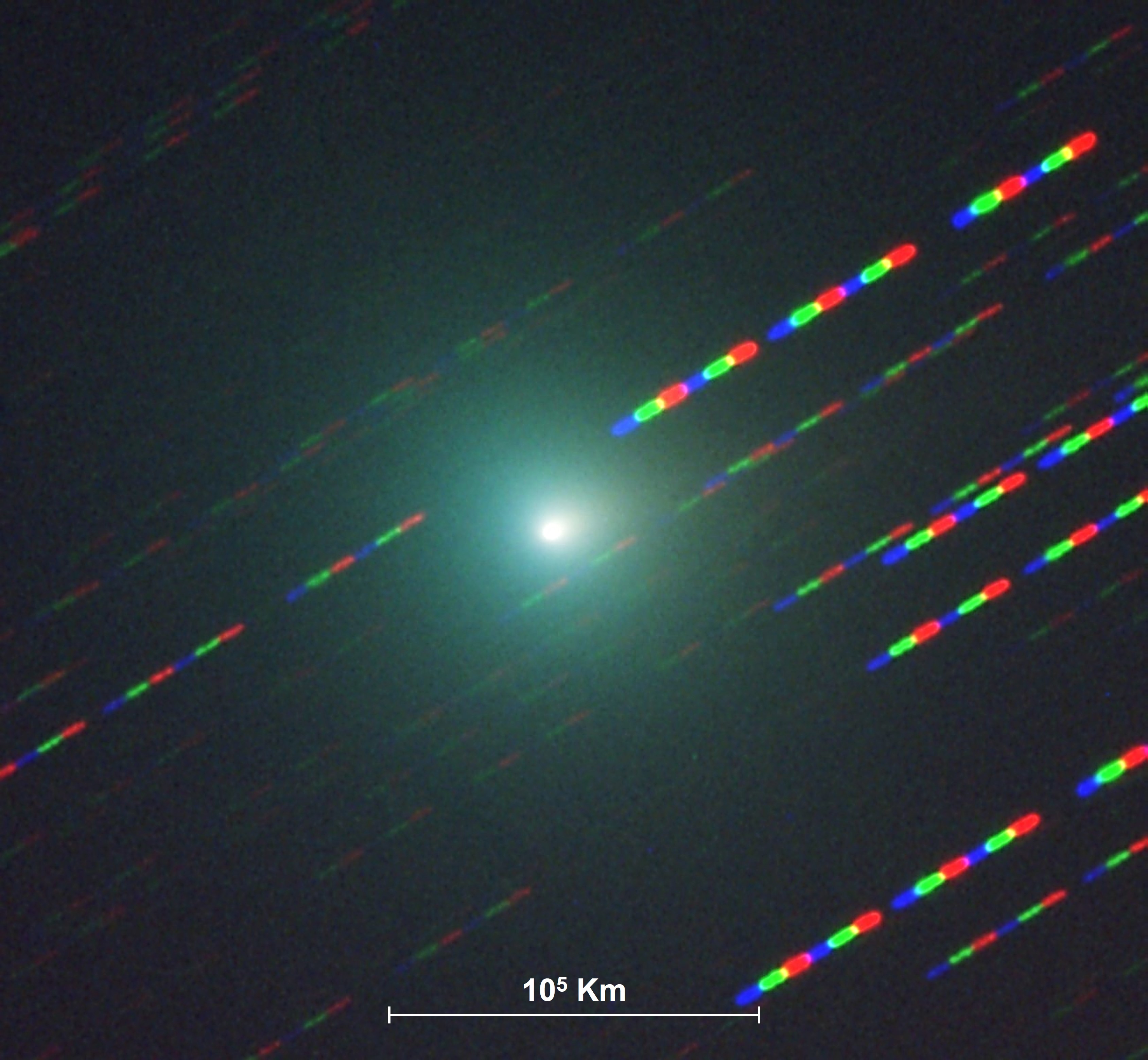}
\caption{2019-01-24. Three-color BVR composite from images taken with the Asiago Schmidt telescope.
At that epoch, the comet was about 80 million km from Earth. Comet Iwamoto has developed a very short tail throughout its appearance, visible as a reddened portion towards West (right). Around the nucleus, a very large gas coma has developed, which appears as a tenuous green haze.}
\end{SCfigure} 

\begin{SCfigure}[0.8][h!]
    \centering
    \includegraphics[scale=0.4]{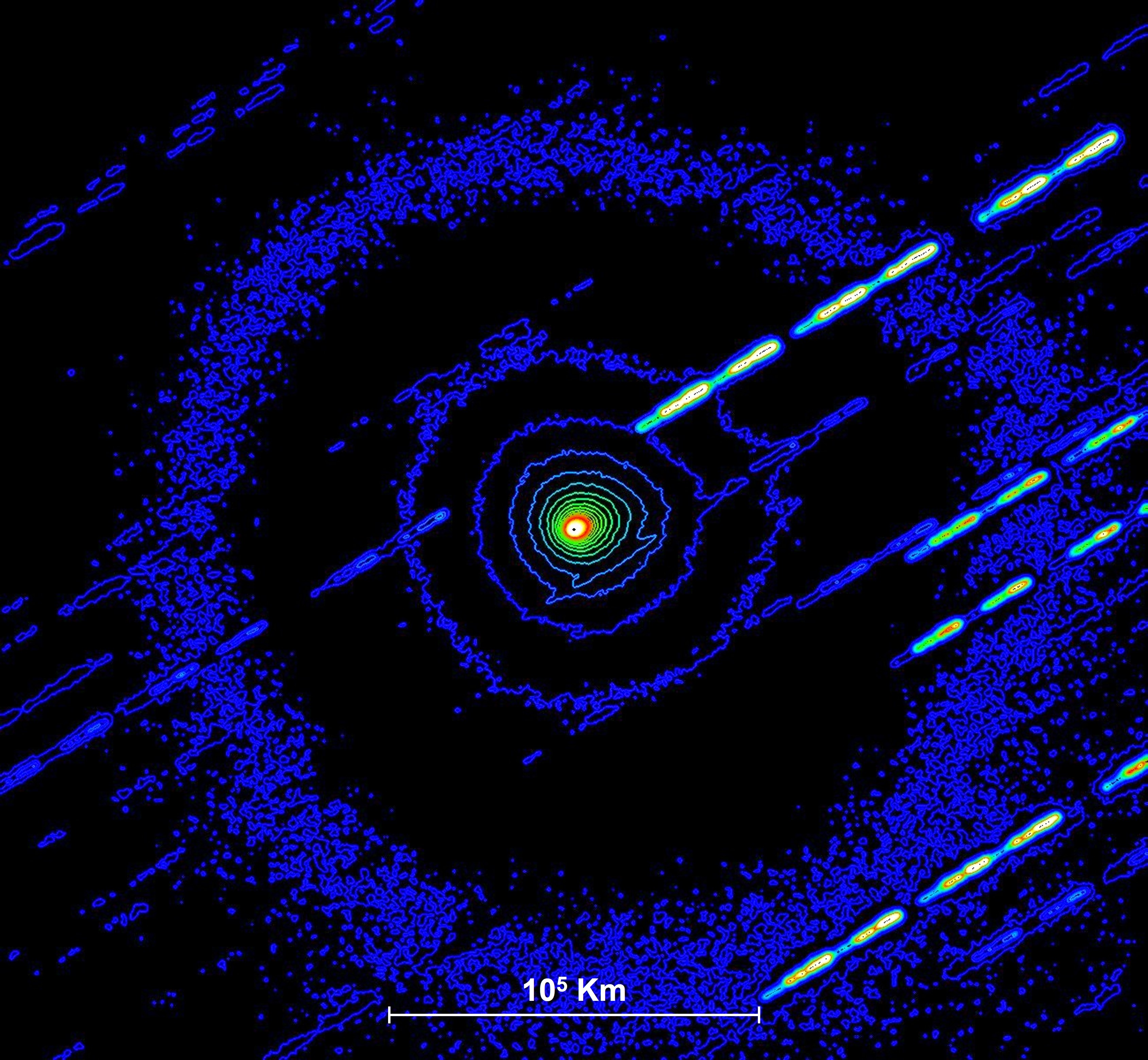}
    \caption{2019-01-24. The isophote visualization shows the cometary nucleus (black dot in the center) surrounded by circles with no apparent morphological structure.}
\end{SCfigure}

\newpage

\subsection{Spectra}

\begin{table}[h!]
\centering
\begin{tabular}{|c|c|c|c|c|c|c|c|c|c|c|c|}
\hline
\multicolumn{12}{|c|}{Observation details}                      \\ \hline 
\hline
$\#$  & date          & r     & $\Delta$ & RA     & DEC     & elong & phase & PLang& config  & FlAng & N \\
      & (yyyy-mm-dd)  &  (AU) & (AU)     & (h)    & (°)     & (°)   & (°)   &  (°)   &       &  (°)  & \\ \hline 

1 & 2019-02-21 & 1.306 & 0.465 & 06.60 & $+$34.63 & 123.9 & 38.9 & $+$03.7 & A & $-$0 & 1 \\
\hline
\end{tabular}
\end{table}

\begin{figure}[h!]

    \centering
    \includegraphics[scale=0.368]{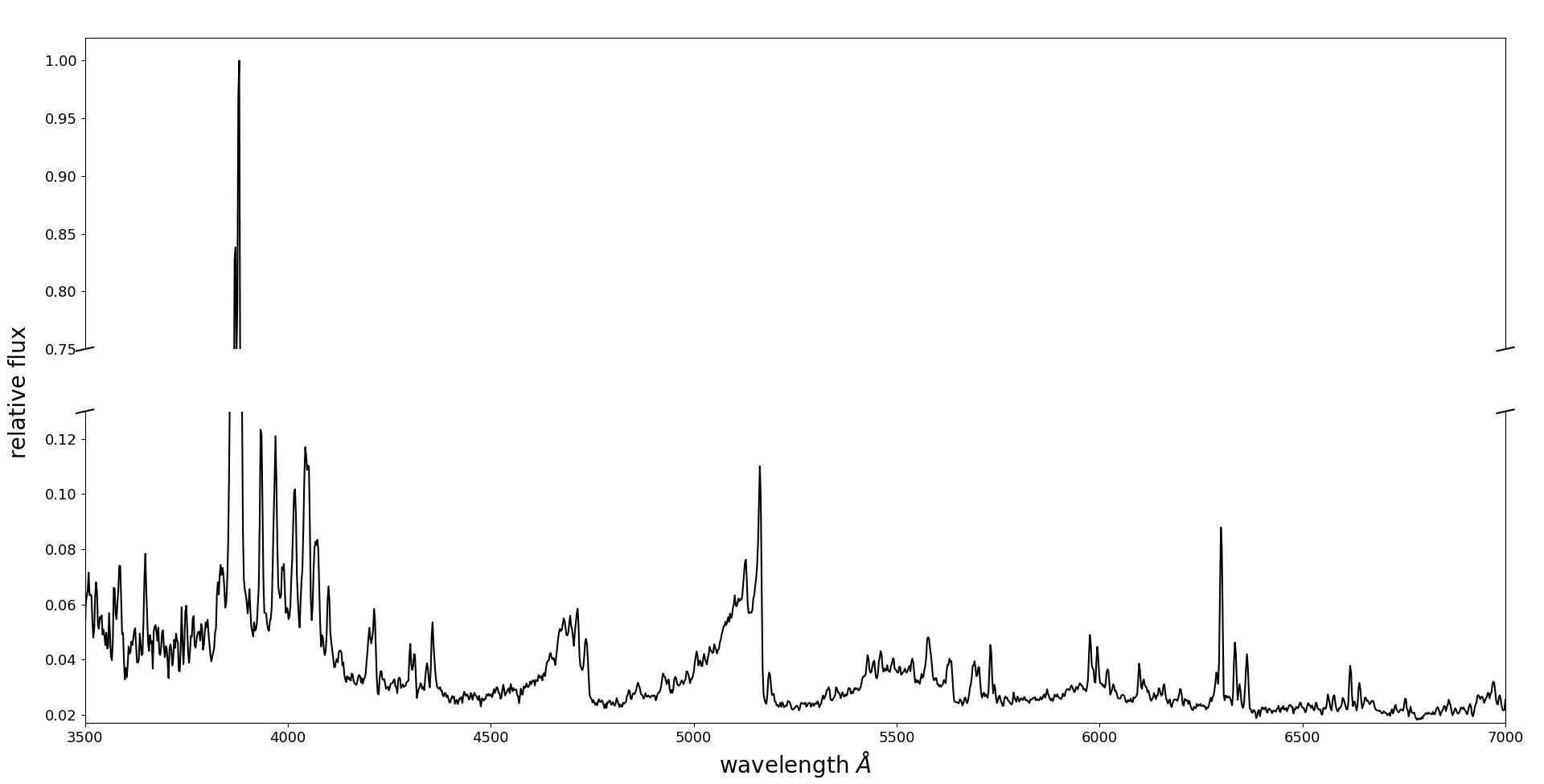}
    \caption{Spectrum of 2019-02-21; configuration A}

\end{figure}

\newpage
\clearpage

\section{C/2019 L3 (ATLAS)}
\label{cometa:C2019L3}
\subsection{Description}

C/2019 L3 (ATLAS) is a hyperbolic comet with an absolute magnitude of 5.5$\pm$0.8.\footnote{\url{https://ssd.jpl.nasa.gov/tools/sbdb_lookup.html\#/?sstr=2019\%20L3} visited on July 20, 2024}
It was first spotted by the 0.5m Asteroid Terrestrial-impact Last Alert System (ATLAS) on June 10, 2019.
The Earth crossed the comet orbital plane on July 13, 2021 and on January 11, 2022.

\noindent
We observed the comet at magnitude 8.6.\footnote{\url{https://ssd.jpl.nasa.gov/tools/sbdb_lookup.html\#/?sstr=2019\%20L3} visited on July 20, 2024}

\begin{table}[h!]
\centering
\begin{tabular}{|c|c|c|}
\hline
\multicolumn{3}{|c|}{Orbital elements (epoch: October 5, 2021)}                      \\ \hline \hline
\textit{e} = 1.0014 &   \textit{q} = 3.5545 &   \textit{T} = 2459589.1341 \\ \hline
$\Omega$ = 290.7888 &    $\omega$ = 171.6140 &    \textit{i} = 48.3639 \\ \hline 
\end{tabular}
\end{table}

\begin{table}[h!]
\centering
\begin{tabular}{|c|c|c|c|c|c|c|c|c|}
\hline
\multicolumn{9}{|c|}{Comet ephemerides for key dates}                      \\ \hline 
\hline
& date        & r & $\Delta$ & RA     & DEC     & elong & phase & PLang \\
& (yyyy-mm-dd) & (AU) & (AU)      & (h)     & (°)      & (°)    & (°)    & (°) \\ \hline
Perihelion	&	2022-01-09 & 3.555 &	2.582 &	07.06 &	$+$31.34 &	169.9 &	02.8 &	$-$00.5  \\
Nearest approach & 2022-01-06 & 3.555 &	2.581 &	07.12 &	$+$31.92 & 170.6 & 02.6 &	$-$01.3 \\ \hline
\end{tabular}

\end{table}

\vspace{0.5 cm}

\begin{figure}[h!]
    \centering
    \includegraphics[scale=0.38]{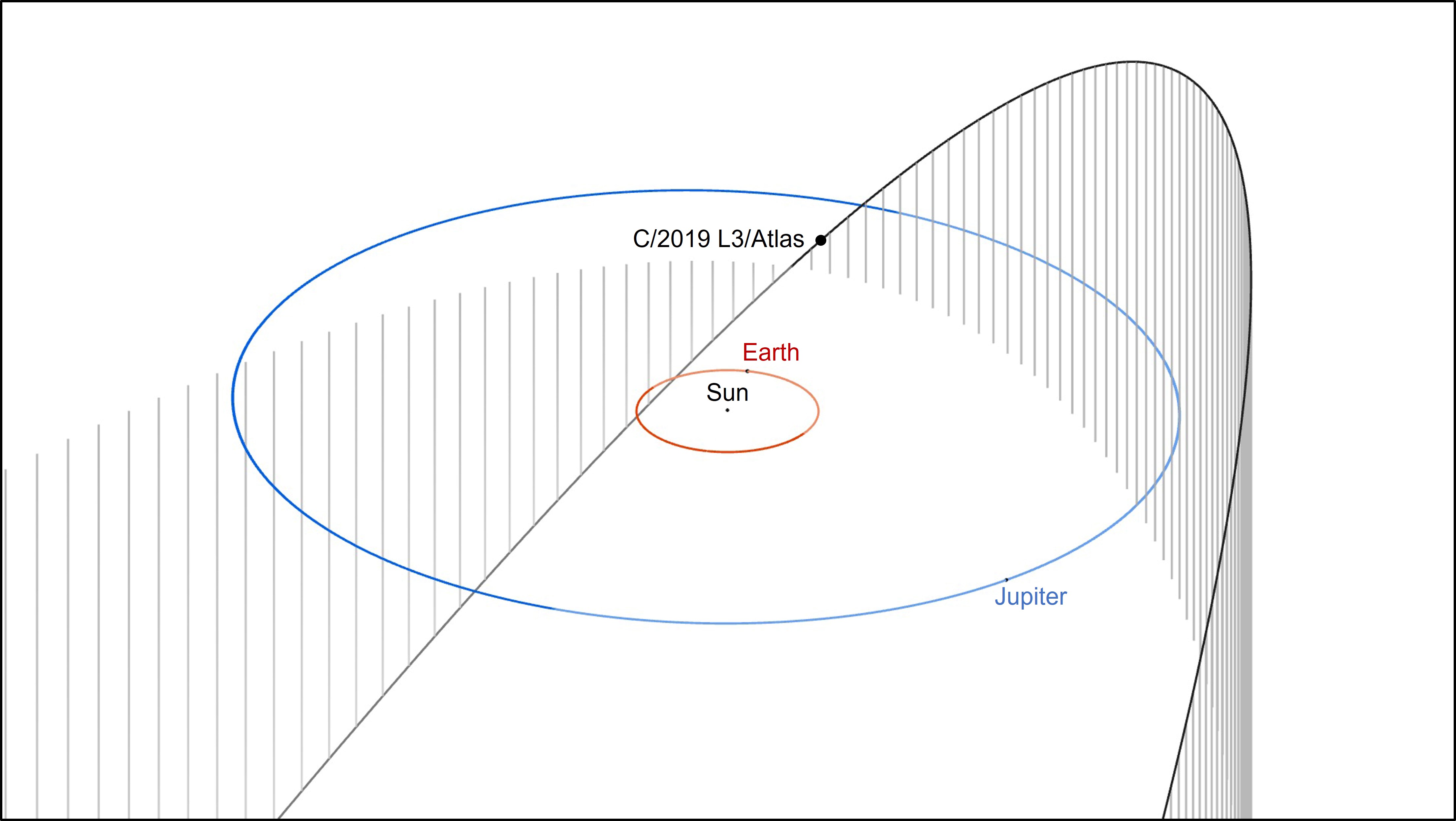}
    \caption{Orbit of comet C/2013 L3 and position on perihelion date. The field of view is set to the orbit of Jupiter for size comparison. Courtesy of NASA/JPL-Caltech.}
\end{figure} 


\newpage

\subsection{Images}

\begin{SCfigure}[0.8][h!]
    \centering
    \includegraphics[scale=0.4]{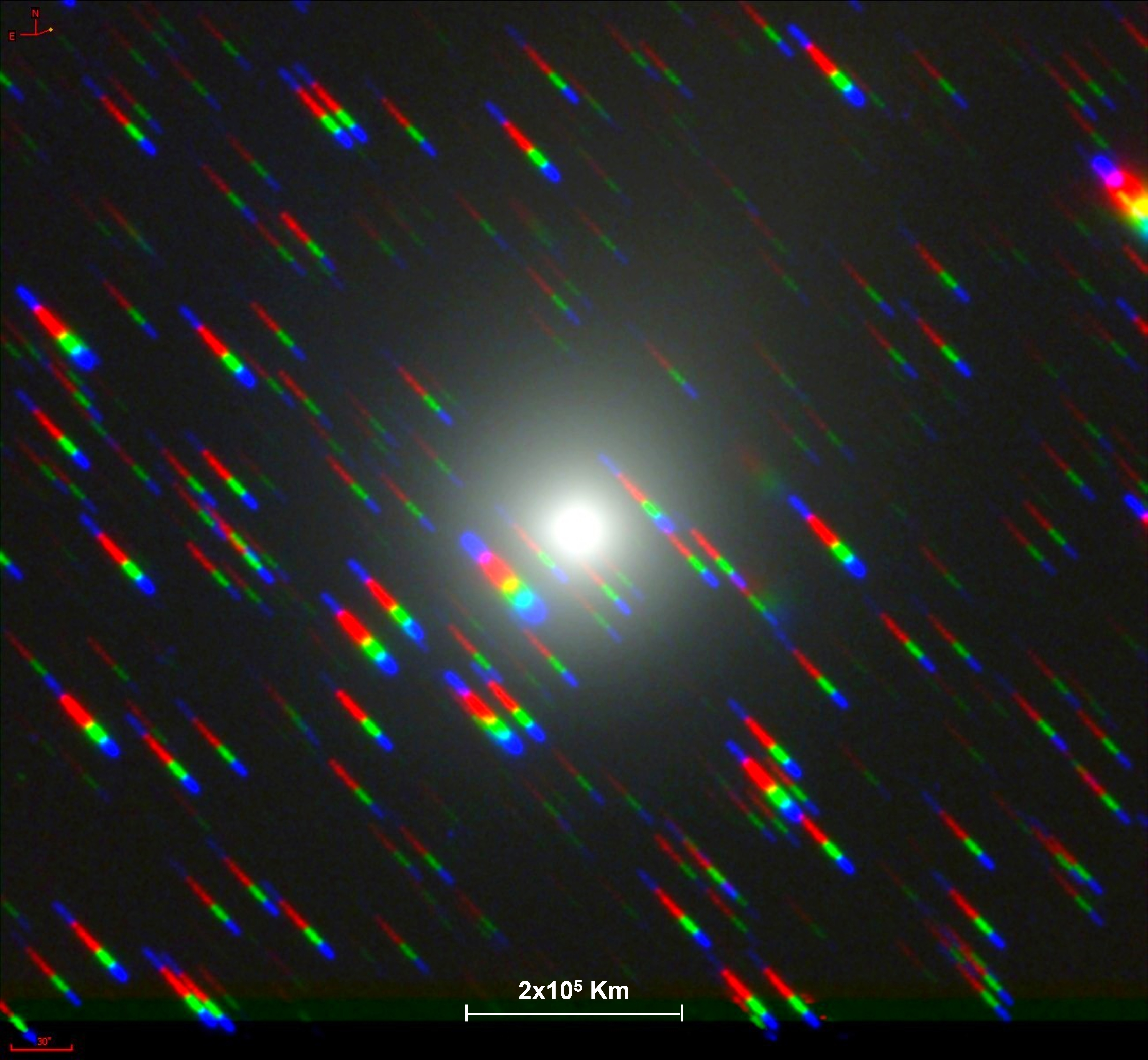}
     \caption{2022-01-23. Three-color BVR composite from images taken with the Asiago Schmidt telescope. UV and IR color channels have also been added to the image. At that epoch, the comet was about 390 million km from Earth.}
\end{SCfigure}

\begin{SCfigure}[0.8][h!]
    \centering
   
    \includegraphics[scale=0.4]{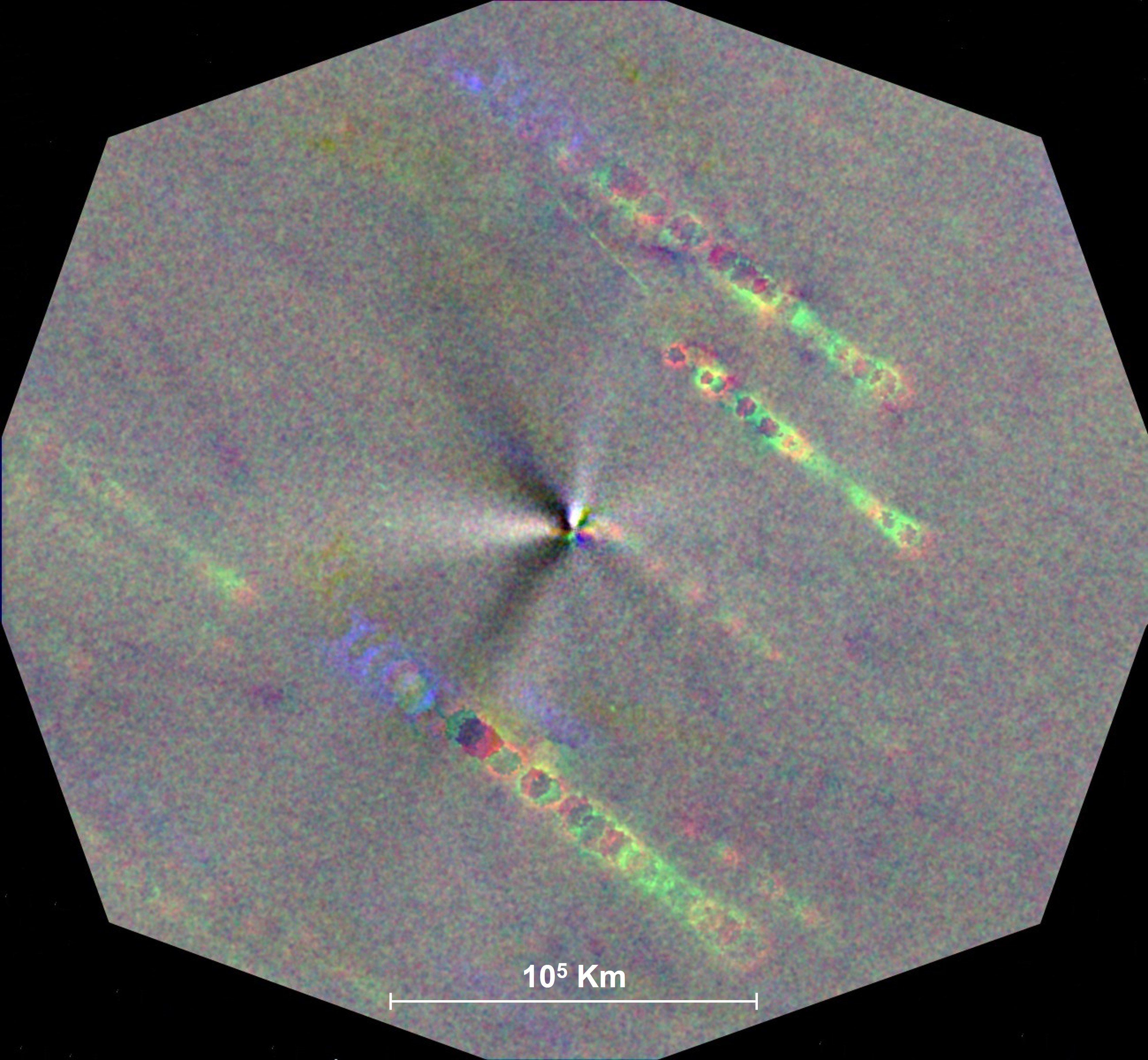}
   
    \caption{2022-12-30. Despite being quite distant from the Sun, comet ATLAS developed an interesting morphology in the inner coma, dominated by the presence of jet-shaped structures, emitted from active areas located on the cometary nucleus. The tail develops towards South.}
\end{SCfigure}

\newpage

\subsection{Spectra}

\begin{table}[h!]
\centering
\begin{tabular}{|c|c|c|c|c|c|c|c|c|c|c|c|}
\hline
\multicolumn{12}{|c|}{Observation details}                      \\ \hline 
\hline
$\#$  & date          & r     & $\Delta$ & RA     & DEC     & elong & phase & PLang& config  & FlAng & N \\
      & (yyyy-mm-dd)  &  (AU) & (AU)     & (h)    & (°)     & (°)   & (°)   &  (°)   &       &  (°)  &  \\ \hline 

1 & 2022-01-06 & 3.555 & 2.581 &	07.12 &	$+$31.78 & 170.6 & 02.6 & $-$01.2 & A & $+$0 & 3 \\

\hline
\end{tabular}
\end{table}

\begin{figure}[h!]

    \centering
    \includegraphics[scale=0.368]{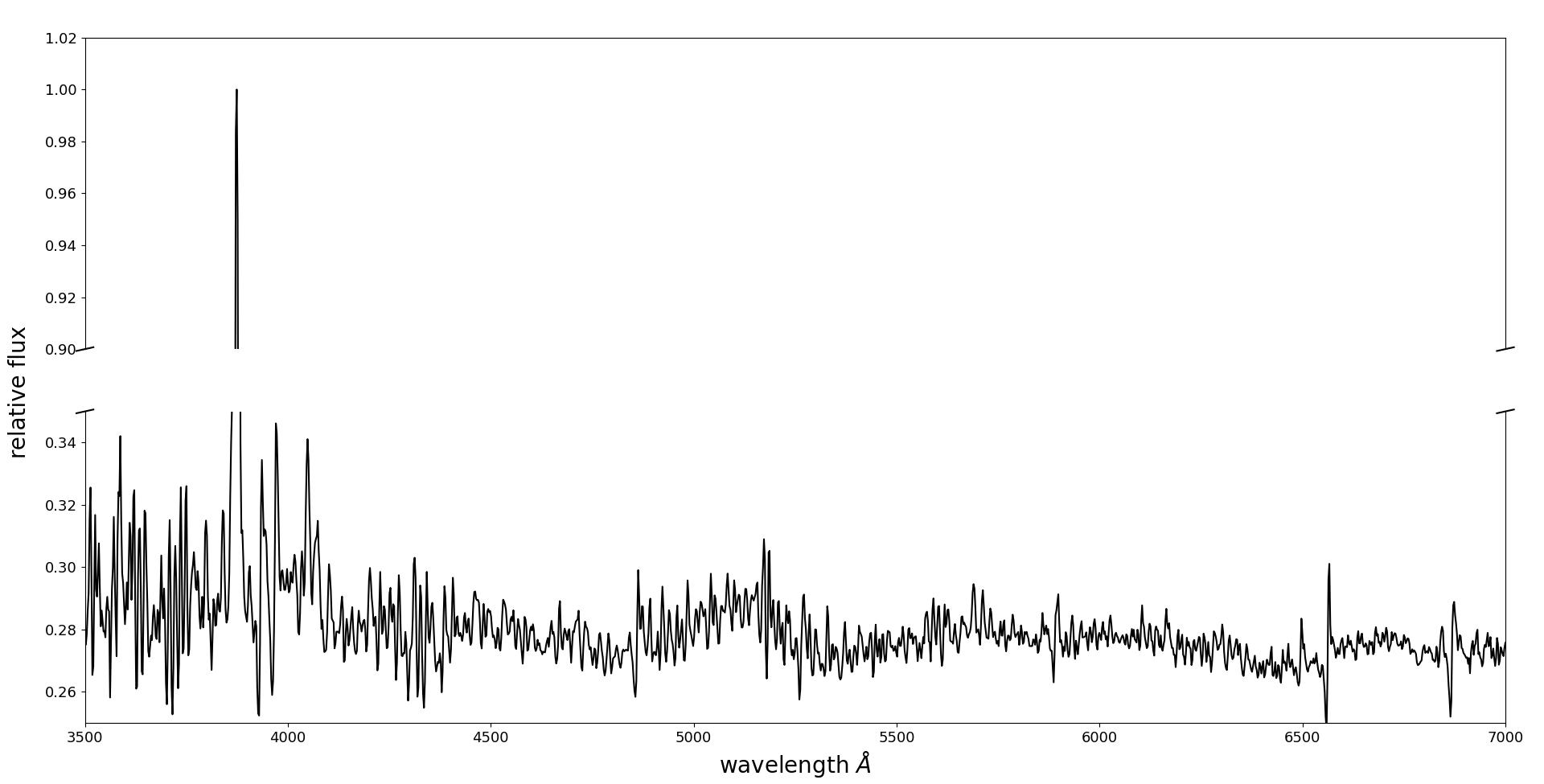}
    \caption{Spectrum of 2022-01-07; configuration A}

\end{figure}

\begin{figure}[h!]

    \centering
    \includegraphics[scale=0.368]{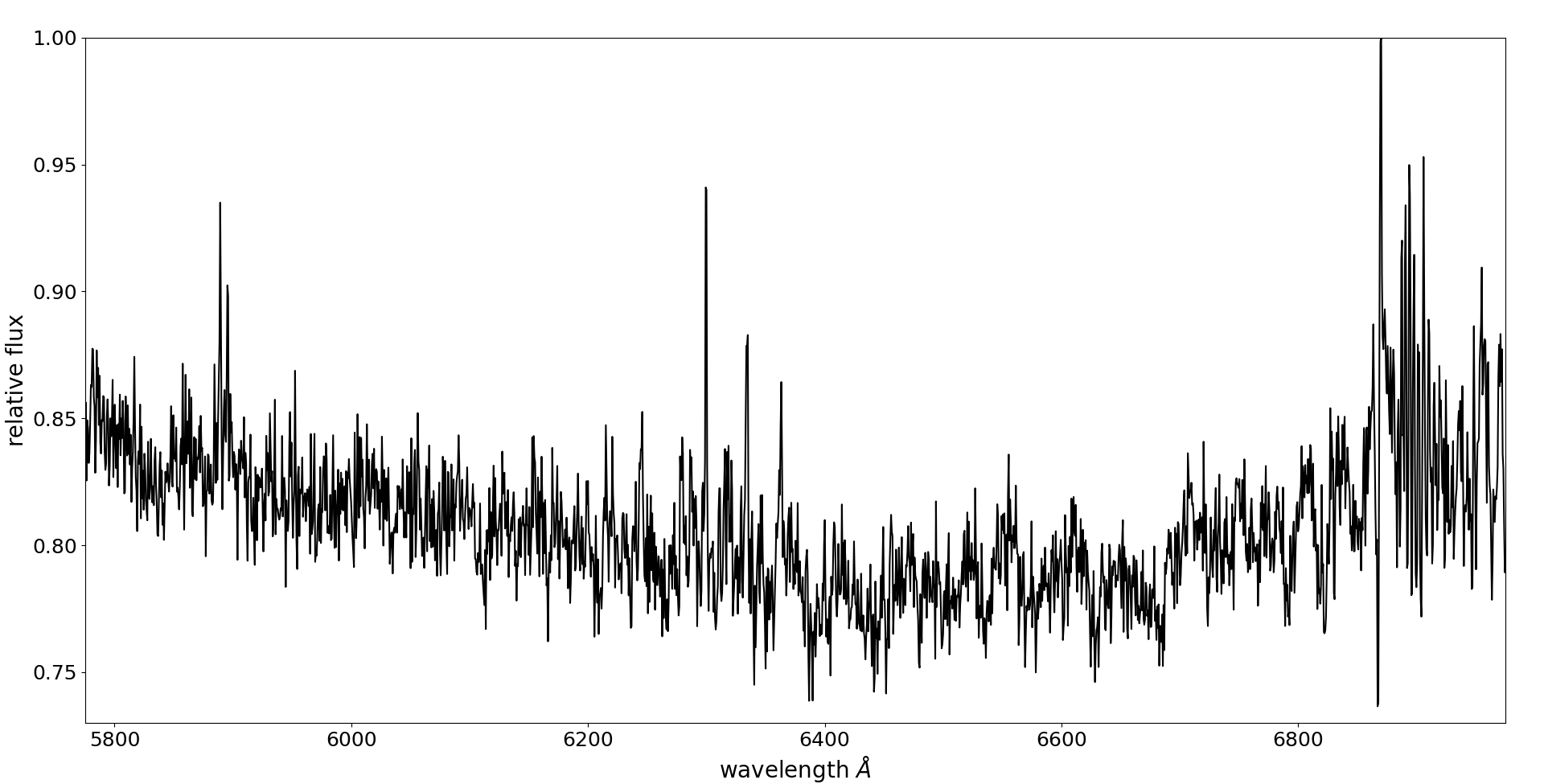}
    \caption{Spectrum of 2022-01-10; configuration C}

\end{figure}

\newpage
\clearpage

\section{C/2019 T4 (ATLAS)}
\label{cometa:C2019T4}
\subsection{Description}

C/2019 T4 (ATLAS) is a Long Period comet with a period of 30637 years and an absolute magnitude of 4.4$\pm$0.7.\footnote{\url{https://ssd.jpl.nasa.gov/tools/sbdb_lookup.html\#/?sstr=2019\%20T4} visited on July 20, 2024}
It was first spotted by the 0.5m Asteroid Terrestrial-impact Last Alert System (ATLAS) on September 10, 2019.
Spectra taken with the Galileo telescope show a strong emission in the \ch{C3} band (triatomic carbon), probably resulting from the dissociation of supervolatiles. The perihelion of C/2019 T4 is beyond 4.2 AU. Since the sublimation of water at this distance is not yet active, the dust from the active areas is conveyed by supervolatiles such as CO, \ch{CO2}, \ch{NH3}.
We observed the comet at magnitude 12.\footnote{\url{https://cobs.si/comet/1822/ }, visited on July 20, 2024}
The Earth crossed the comet orbital plane on February 16, 2019.

\begin{table}[h!]
\centering
\begin{tabular}{|c|c|c|}
\hline
\multicolumn{3}{|c|}{Orbital elements (epoch: December 20, 2022)}                      \\ \hline \hline
\textit{e} = 0.9957 & \textit{q} = 4.2423 & \textit{T} = 2459739.5251 \\ \hline
$\Omega$ = 199.9391 & $\omega$ = 351.1833  & \textit{i} = 53.6319  \\ \hline  
\end{tabular}
\end{table}

\begin{table}[h!]
\centering
\begin{tabular}{|c|c|c|c|c|c|c|c|c|}
\hline
\multicolumn{9}{|c|}{Comet ephemerides for key dates}                      \\ \hline 
\hline
& date         & r    & $\Delta$  & RA      & DEC      & elong  & phase  & PLang  \\
& (yyyy-mm-dd) & (AU) & (AU)      & (h)     & (°)      & (°)    & (°)    & (°) \\ \hline 

Perihelion       & 2022-06-09 & 4.242 & 3.903 & 11.87 & $-$07.46 & 102.7	& 13.5 & $-$10.2 \\ 
Nearest approach & 2022-04-04 & 4.278 & 3.331 & 11.84 &	$-$20.68 & 158.6	& 04.9	& $+$01.4 \\ \hline
\end{tabular}

\end{table}

\vspace{0.5 cm}

\begin{figure}[h!]
    \centering
    \includegraphics[scale=0.38]{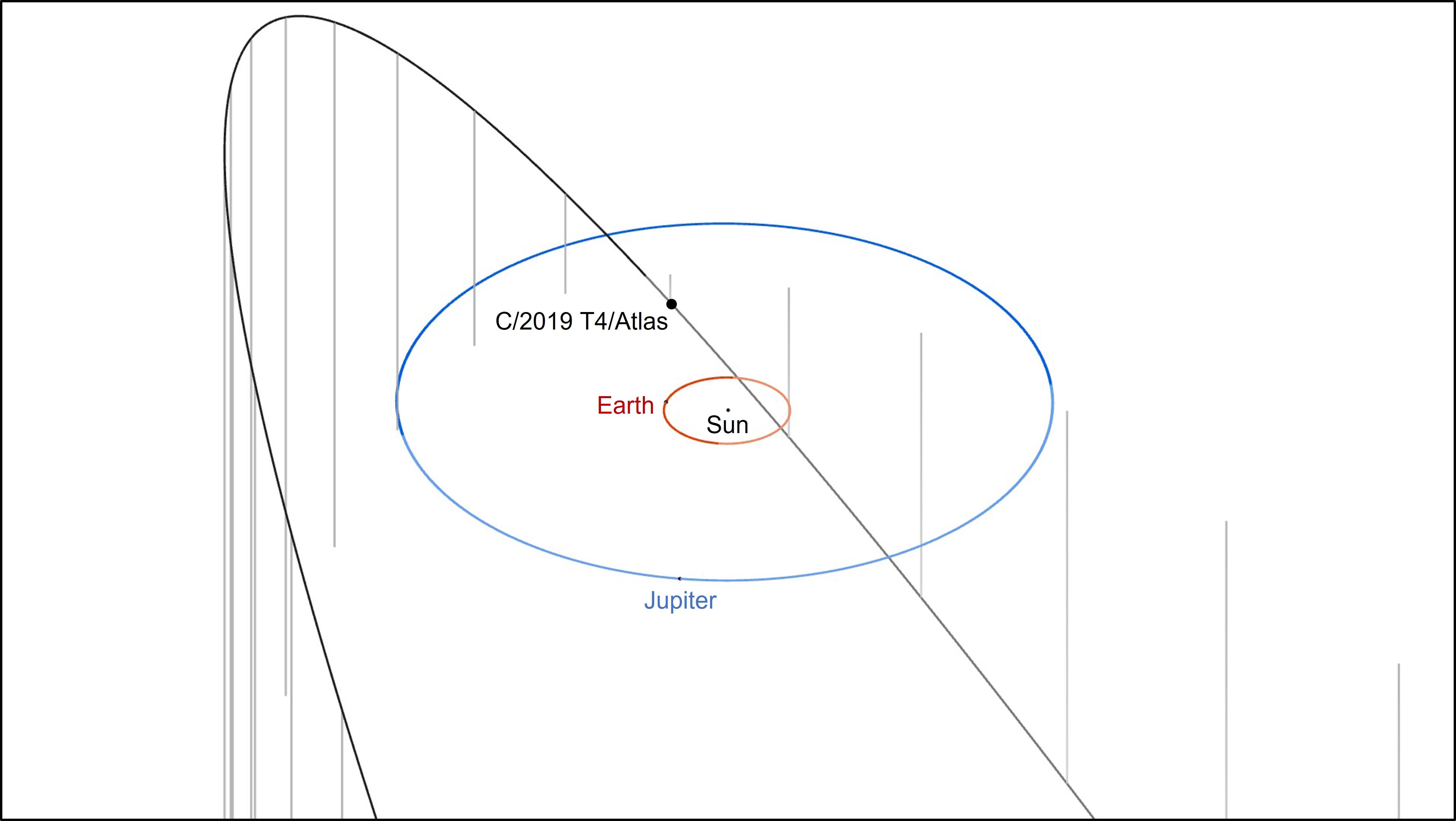}
    \caption{Orbit of comet C/2019 T4 and position on perihelion date. The field of view is set to the orbit of Jupiter for size comparison. Courtesy of NASA/JPL-Caltech.}
\end{figure} 


\newpage

\subsection{Images}

\begin{SCfigure}[0.8][h!]
    \centering
    \includegraphics[scale=0.4]{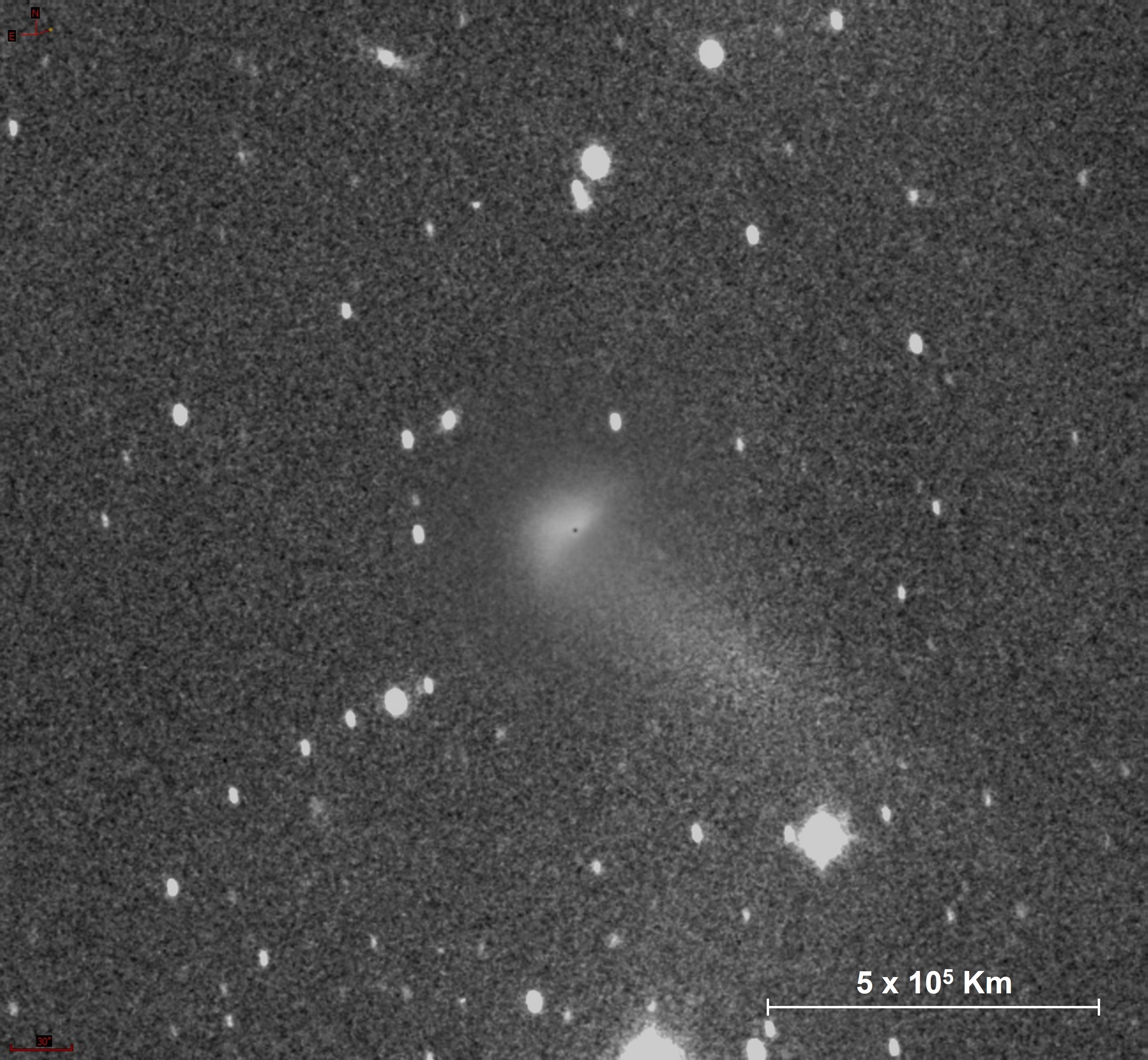}
     \caption{2022-05-21. Image taken with the Schmidt telescope. A simple processing with a 1/R filter shows a clear asymmetry of the inner coma.}
\end{SCfigure} 

\begin{SCfigure}[0.8][h!]
    \centering

    \includegraphics[scale=0.4]{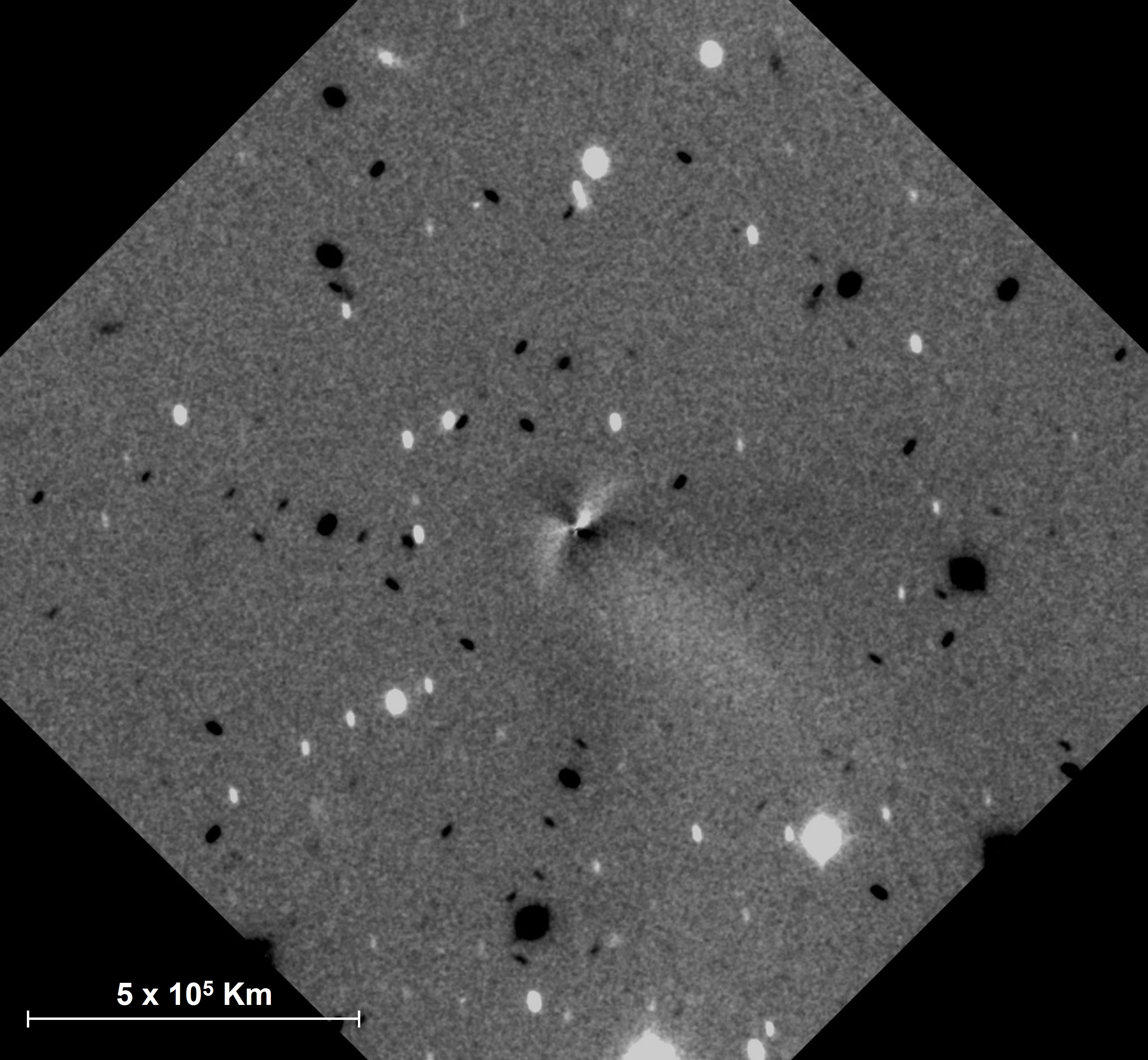}

    \caption{The asymmetry of the inner coma is further evidenced by processing with a Larson-Sekanina filter.}
\end{SCfigure}

\newpage

\subsection{Spectra}

\begin{table}[h!]
\centering
\begin{tabular}{|c|c|c|c|c|c|c|c|c|c|c|c|}
\hline
\multicolumn{12}{|c|}{Observation details}                      \\ \hline 
\hline
$\#$  & date          & r     & $\Delta$ & RA     & DEC     & elong & phase & PLang& config  & FlAng & N \\
      & (yyyy-mm-dd)  &  (AU) & (AU)     & (h)    & (°)     & (°)   & (°)   &  (°)   &       &  (°)  & \\ \hline 

1 & 2022-05-20 & 4.245 & 3.638 & 11.77 & $-$10.68 & 120.7	& 11.8 & $-$08.1 & D & $-$0 & 1 \\
\hline
\end{tabular}
\end{table}

\begin{figure}[h!]

    \centering
    \includegraphics[scale=0.368]{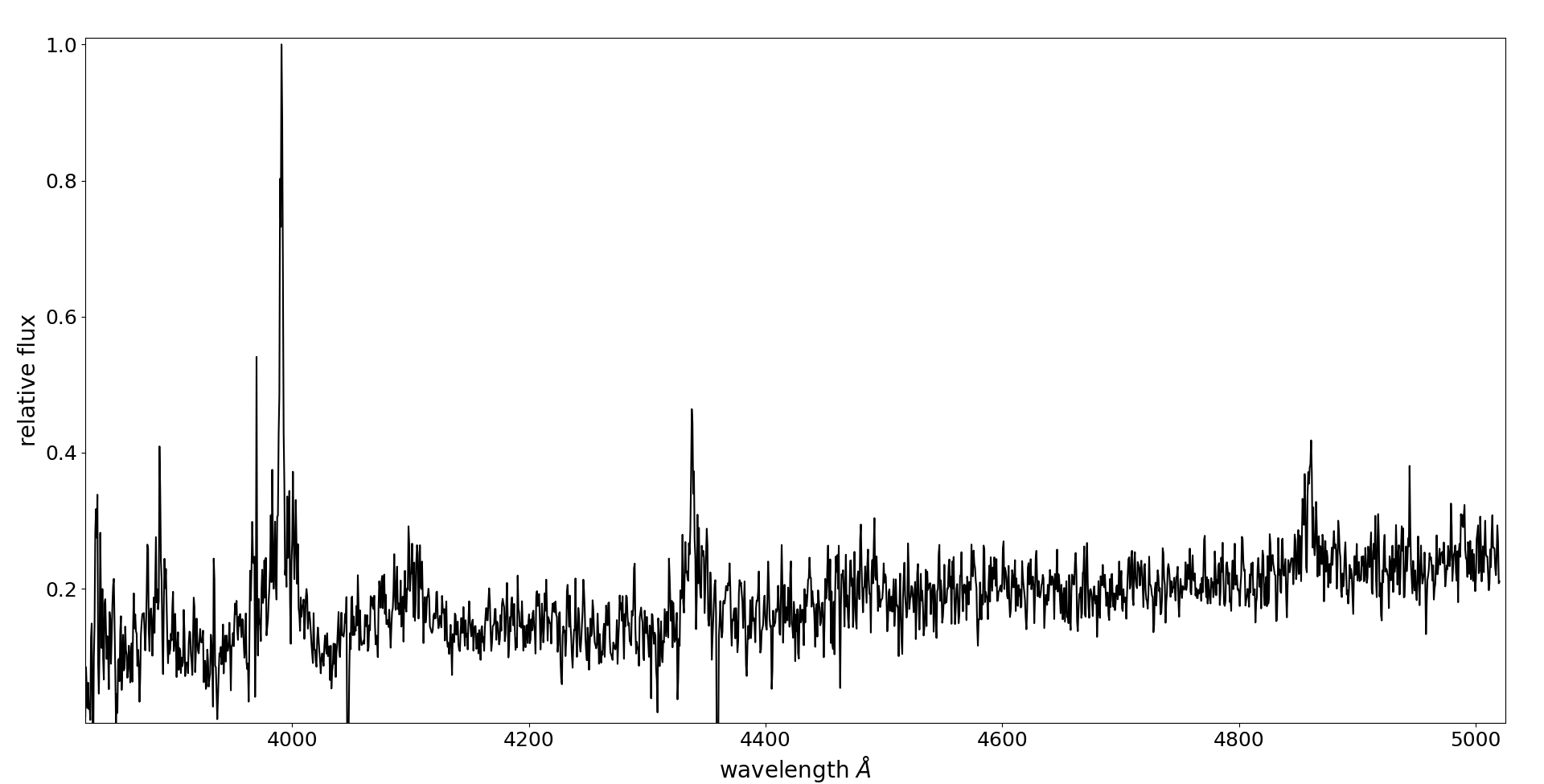}
    \caption{Spectrum of 2022-05-20; configuration D}

\end{figure}

\newpage
\clearpage

\section{C/2019 U5 (PanSTARRS)}
\label{cometa:C2019U5}
\subsection{Description}

C/2019 U5 (PanSTARRS) is a hyperbolic comet with an absolute magnitude of 8.2$\pm$0.7.\footnote{\url{https://ssd.jpl.nasa.gov/tools/sbdb_lookup.html\#/?sstr=2019\%20U5}}
It was discovered by the 1.8m Pan-STARRS 1 telescope located in Haleakala, Hawaii, on October 22, 2019.
C/2019 U5 (PanSTARRS) is a dynamically new comet, which means that this object left the Oort cloud, to reach the inner Solar System for the very first time.
On January 28, 2022, as far as 5.01 AU from the Sun, the comet showed a symmetric coma and a dim tail. Instead, the comet showed evidence of two jets, indicative of active areas on the nucleus.
Looking at the C/2019 U5 spectra, it can be seen that the CN emission line, around $3880$ {\AA} is very faint, probably due to the high perihelion distance, while the \ch{C2} line at $5582$ {\AA} is clearly visible.
We observed the comet at magnitude 12.\footnote{\url{https://cobs.si/comet/2241/ }} 


\begin{table}[h!]
\centering
\begin{tabular}{|c|c|c|}
\hline
\multicolumn{3}{|c|}{Orbital elements (epoch: July 04, 2022)}                      \\ \hline \hline
\textit{e} = 1.0013 & \textit{q} = 3.6239 & \textit{T} = 2460033.4506 \\ \hline
$\Omega$ = 2.6373 & $\omega$ = 181.5172  & \textit{i} = 113.5211 \\ \hline  
\end{tabular}
\end{table}

\begin{table}[h!]
\centering
\begin{tabular}{|c|c|c|c|c|c|c|c|c|}
\hline
\multicolumn{9}{|c|}{Comet ephemerides for key dates}                      \\ \hline 
\hline
& date         & r    & $\Delta$  & RA      & DEC      & elong  & phase  & PLang  \\
& (yyyy-mm-dd) & (AU) & (AU)      & (h)     & (°)      & (°)    & (°)    & (°) \\ \hline 

Perihelion       & 2023-03-29 & 3.624 & 2.634 & 11.94  & $-$1.55 & 171.3 & 2.4 & $-$1.9\\ 
Nearest approach & 2023-03-24 & 3.625 & 2.628 & 12.12 & $-$1.34 & 178.8 & 0.3 & $-$0.2\\ \hline
\end{tabular}

\end{table}

\vspace{0.5 cm}

\begin{figure}[h!]
    \centering
    \includegraphics[scale=0.38]{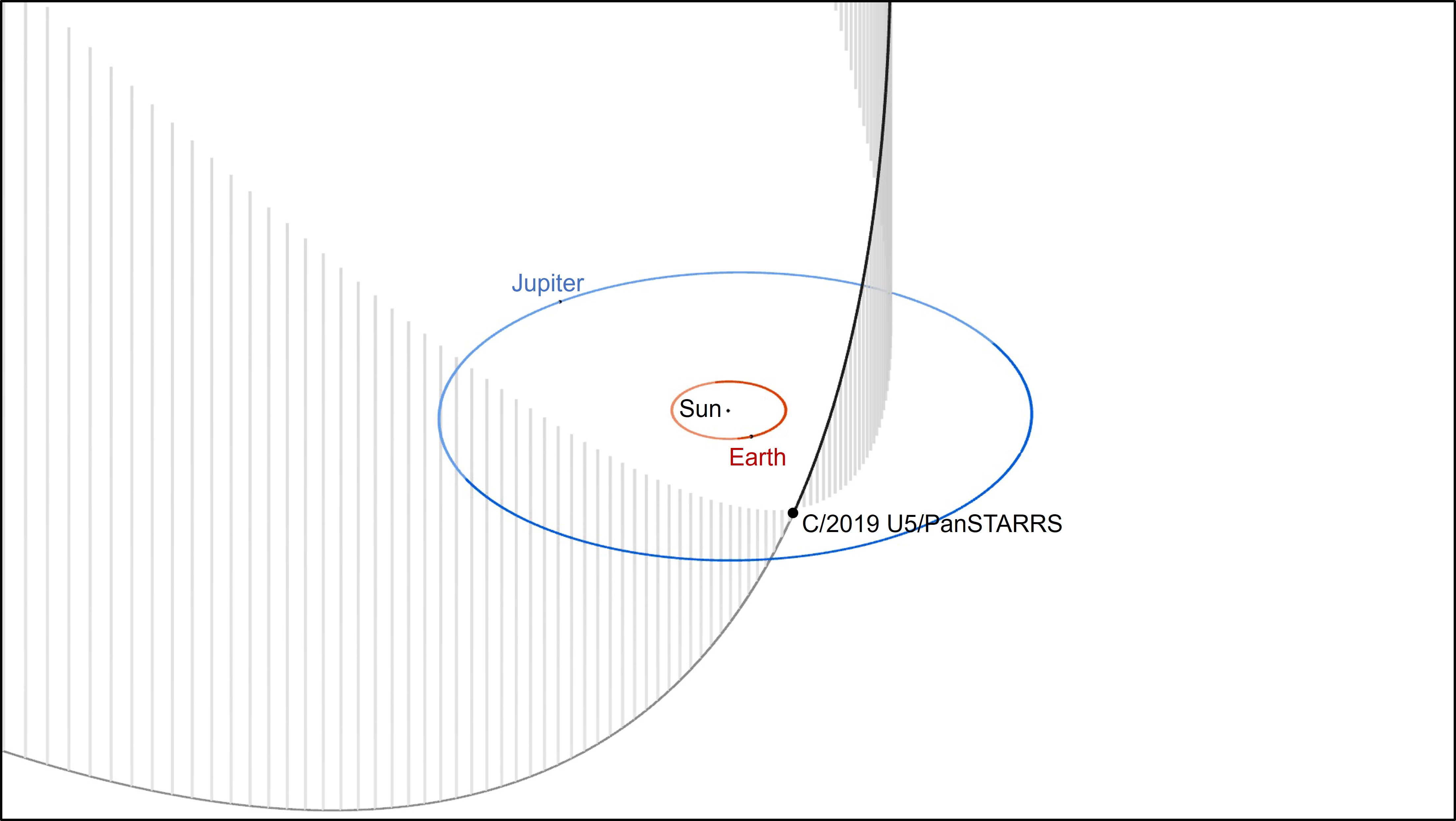}
    \caption{Orbit of comet C/2019 U5 and position on perihelion date. The field of view is set to the orbit of Jupiter for size comparison. Courtesy of NASA/JPL-Caltech.}
\end{figure} 


\newpage

\subsection{Images}

\begin{SCfigure}[0.8][h!]
    \centering
    \includegraphics[scale=0.935]{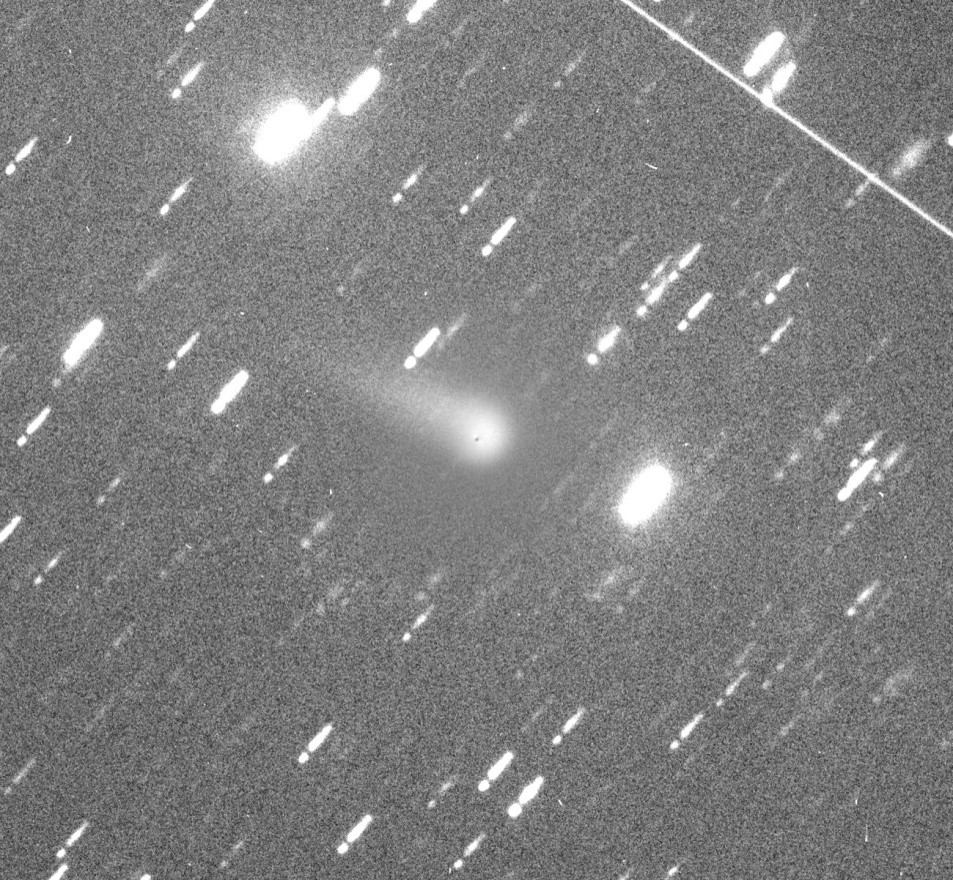}
     \caption{2022-03-09. Image taken with the Asiago Copernico Telescope with the V, r, i filters. A small tail is visible. Its limited extension is due to the long distance of the comet from the Sun ($r=4.96$ AU).}

\end{SCfigure} 

\begin{SCfigure}[0.8][h!]
    \centering
    \includegraphics[scale=0.935]{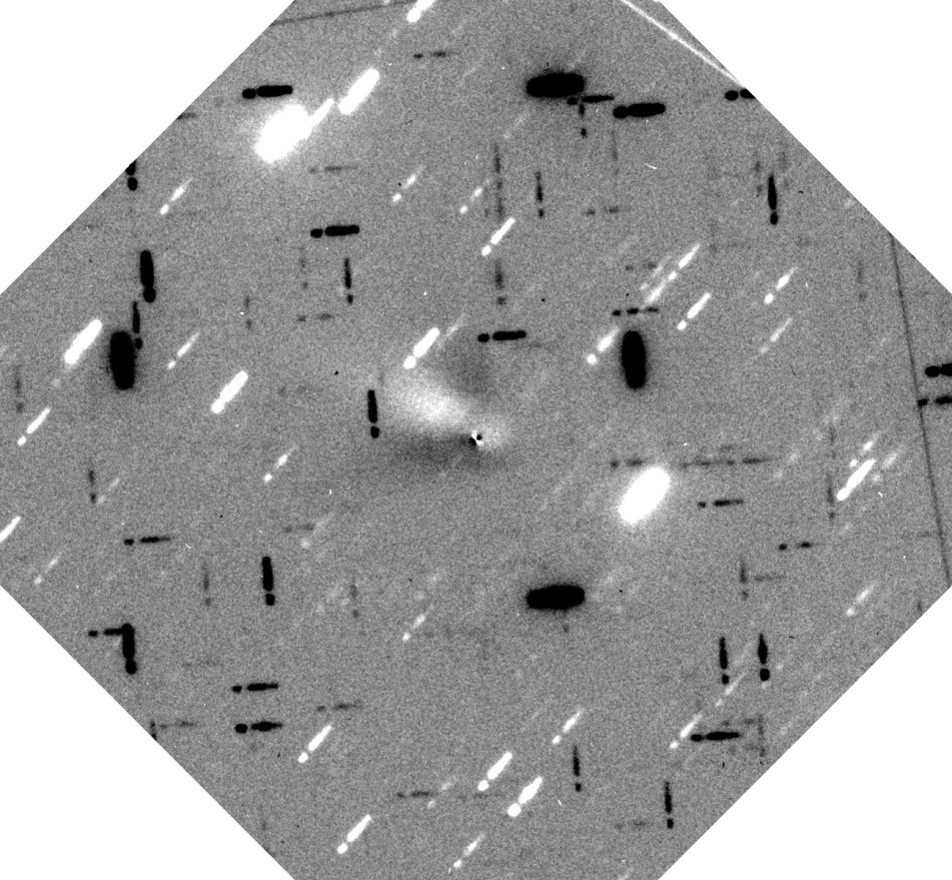}
    \caption{2022-03-09. The previous image is processed with a Larson-Sekanina filter with an angular shift ($\alpha=30${\textdegree}) in order to underline details of the inner coma. Two jets are clearly visible moving away from the nucleus. These features suggest the presence of discrete active areas on the nucleus surface.}

\end{SCfigure}

\newpage

\subsection{Spectra}

\begin{table}[h!]
\centering
\begin{tabular}{|c|c|c|c|c|c|c|c|c|c|c|c|}
\hline
\multicolumn{12}{|c|}{Observation details}                      \\ \hline 
\hline
$\#$  & date          & r     & $\Delta$ & RA     & DEC     & elong & phase & PLang & config  &  FlAng & N \\
      & (yyyy-mm-dd)  &  (AU) & (AU)     & (h)    & (°)     & (°)   & (°)   &  (°)   &       &  (°)  & \\ \hline 
1 & 2023-03-28 & 3.624 & 2.633 & 11.99 & $-$0.51 & 172.2 & 02.1 & $-$01.7 & A & $-$0 & 3 \\

\hline
\end{tabular}
\end{table}

\begin{figure}[h!]
    \centering
    \includegraphics[scale=0.368]{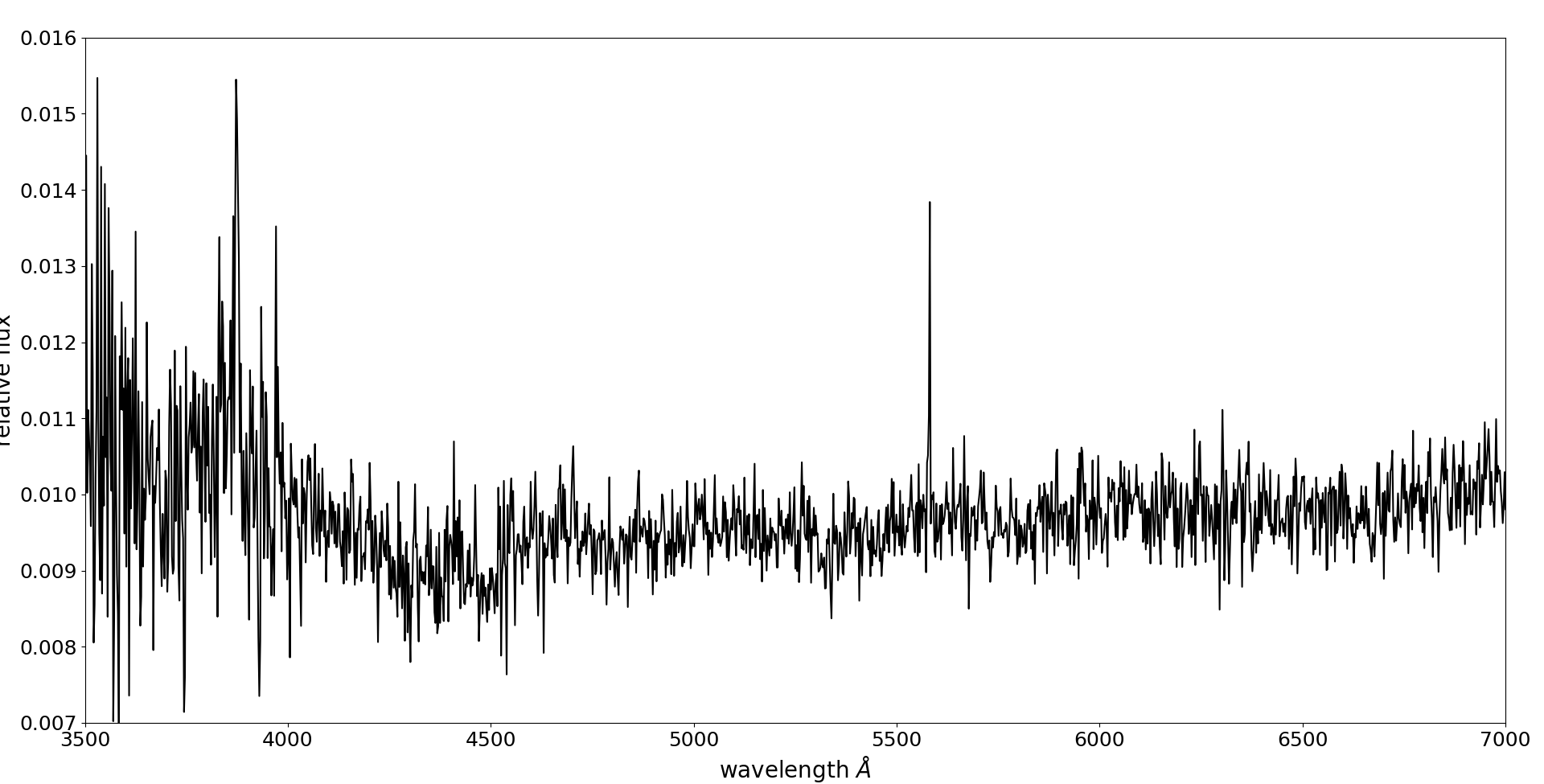}
    \caption{Spectrum of 2023-04-17; configuration A.}
\end{figure}

\begin{figure}[h!]
    \centering
    \includegraphics[scale=0.368]{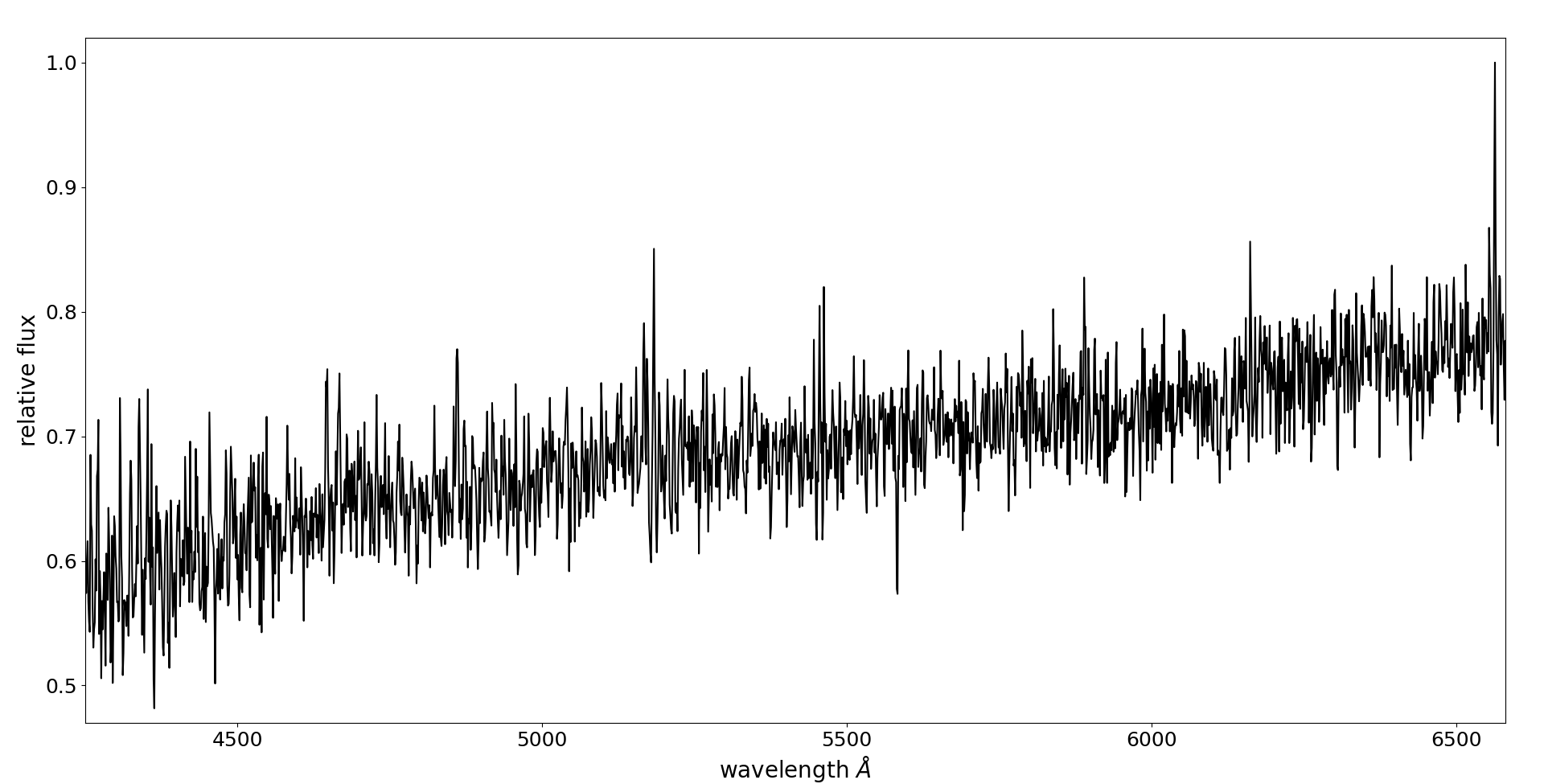}
    \caption{Spectrum of 2023-04-25; configuration B. }
\end{figure}

\newpage
\clearpage

\section{C/2019 Y4 (ATLAS)}
\label{cometa:C2019Y4}
\subsection{Description}

C/2019 Y4 (ATLAS) is a Long Period comet with a period of 5893 years and an absolute magnitude of 8.8$\pm$0.7.\footnote{\url{https://ssd.jpl.nasa.gov/tools/sbdb_lookup.html\#/?sstr=2019\%20Y4} visited on July 20, 2024}
It was first spotted by the 0.5m Asteroid Terrestrial-impact Last Alert System (ATLAS) on December 28, 2019.
Comet C/2019 Y4 received widespread media coverage due to its dramatic increase in brightness and an orbit similar to that of the Great Comet of 1844.
On March 22, 2020, the comet started disintegrating and continued to fade, without reaching the naked-eye visibility.

\noindent
We observed the comet at magnitude 8.\footnote{\url{https://cobs.si/comet/1861/ }, visited on July 20, 2024}
The Earth crossed the comet orbital plane on on February 16, 2019.

\begin{table}[h!]
\centering
\begin{tabular}{|c|c|c|}
\hline
\multicolumn{3}{|c|}{Orbital elements (epoch: February 24, 2020)}                      \\ \hline \hline
\textit{e} = 0.999225 & \textit{q} = 0.2529 & \textit{T} = 2459000.5166 \\ \hline
$\Omega$ = 120.5686 & $\omega$ = 177.4098  & \textit{i} = 45.3802 \\ \hline  
\end{tabular}
\end{table}

\begin{table}[h!]
\centering
\begin{tabular}{|c|c|c|c|c|c|c|c|c|}
\hline
\multicolumn{9}{|c|}{Comet ephemerides for key dates}                      \\ \hline 
\hline
& date         & r    & $\Delta$  & RA      & DEC      & elong  & phase  & PLang  \\
& (yyyy-mm-dd) & (AU) & (AU)      & (h)     & (°)      & (°)    & (°)    & (°) \\ \hline 

Perihelion       & \multicolumn{8}{|c|}{Not reached (calculated 2020-05-31)}   \\ 
Nearest approach & \multicolumn{8}{|c|}{Not reached}    \\ \hline
\end{tabular}

\end{table}

\vspace{0.5 cm}

\begin{figure}[h!]
    \centering
    \includegraphics[scale=0.38]{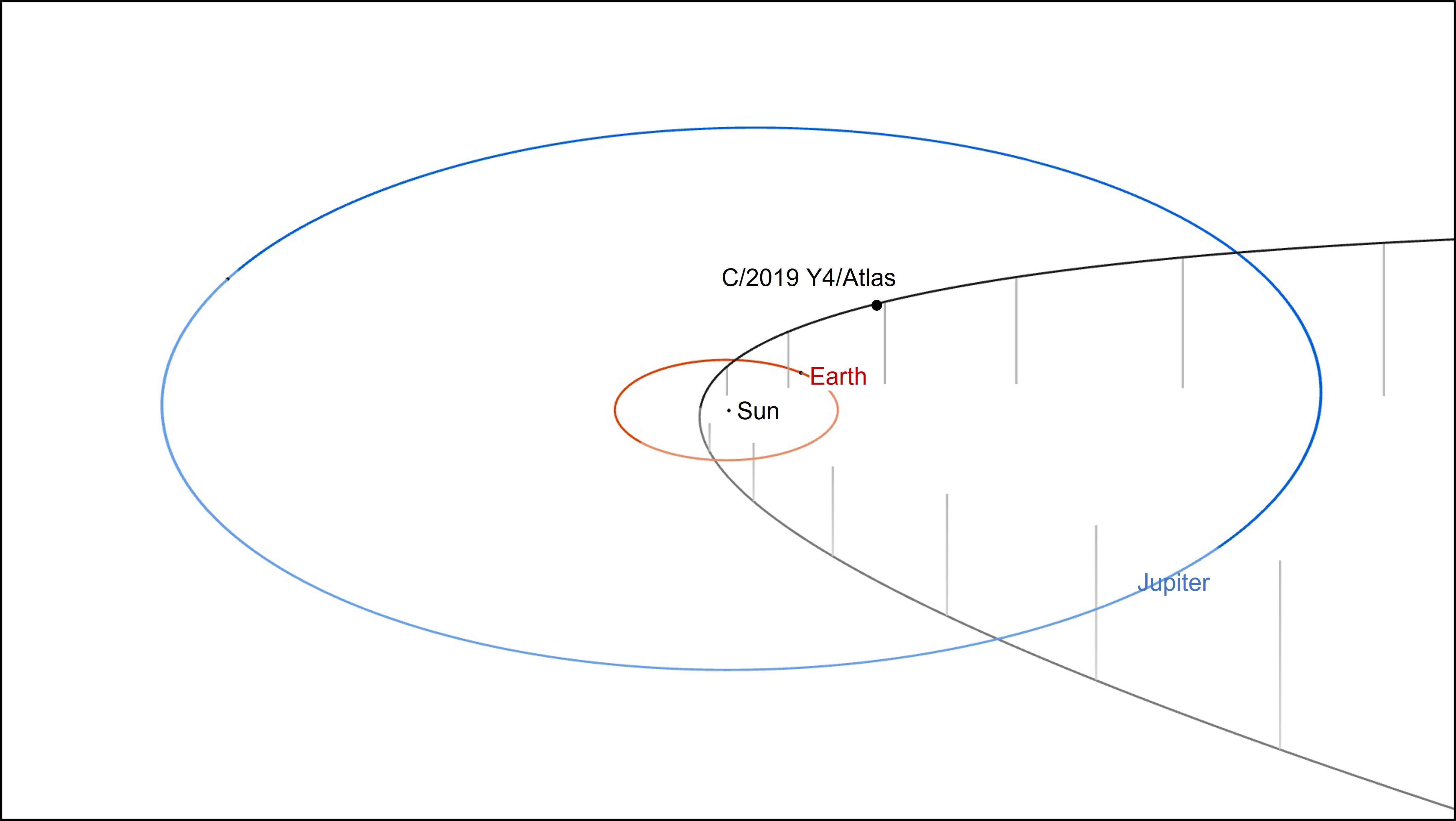}
    \caption{Orbit of comet C/2019 Y4 and position on March 22, 2020, when it started disintegrating, about two months before reaching perihelion. The field of view is set to the orbit of Jupiter for size comparison. Courtesy of NASA/JPL-Caltech.}
\end{figure} 


\newpage

\subsection{Images}

\begin{SCfigure}[0.8][h!]
    \centering
    \includegraphics[scale=0.4]{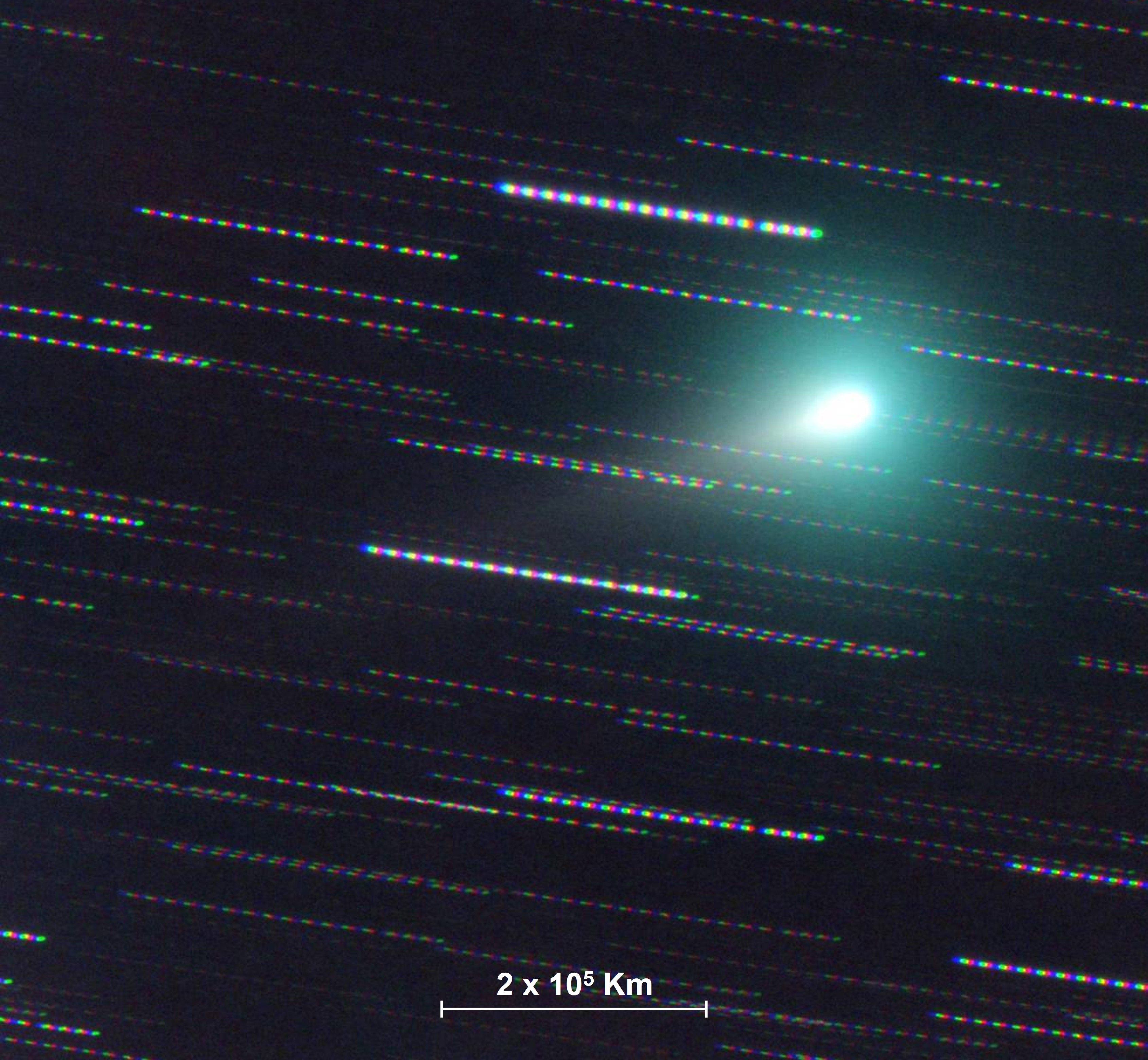}
     \caption{ 2020-04-02. Three-color Bgr composite from images taken with the Asiago Schmidt telescope.
     The comet was already in an advanced phase of disintegration and the coma appears very bright and green in color due to the fluorescence of atoms of diatomic carbon C$_2$.}
\end{SCfigure} 

\begin{SCfigure}[0.8][h!]
    \centering
    \includegraphics[scale=0.4]{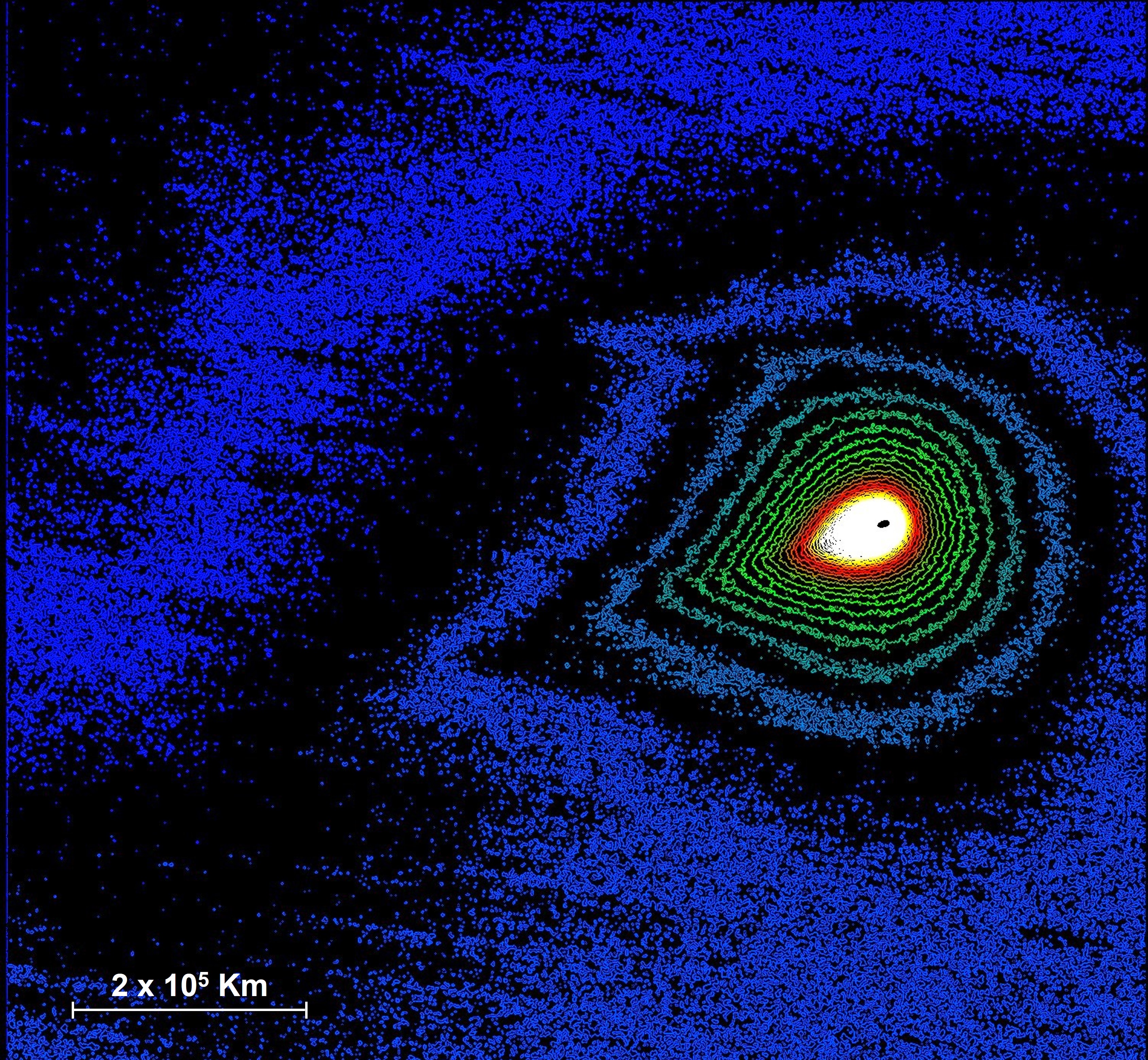}
    \caption{2020-04-02. A visualization in isophotes highlights the elongation of the coma and its merging into a weak tail. The nuclear region (in black) appears elongated due to the disintegration of the nucleus and the progressive distancing of the secondary bodies after the fragmentation.}
\end{SCfigure}

\newpage

\subsection{Spectra}

\begin{table}[h!]
\centering
\begin{tabular}{|c|c|c|c|c|c|c|c|c|c|c|c|}
\hline
\multicolumn{12}{|c|}{Observation details}                      \\ \hline 
\hline
$\#$  & date          & r     & $\Delta$ & RA     & DEC     & elong & phase & PLang& config  & FlAng & N \\
      & (yyyy-mm-dd)  &  (AU) & (AU)     & (h)    & (°)     & (°)   & (°)   &  (°)   &       &  (°)  & \\ \hline 

1 & 2020-03-11 & 1.824  & 1.137 & 10.33 & $+$63.78 & 117.6 & 28.9 & $-$28.9  & A & $+$90 & 3 \\
2 & 2020-03-19 & 1.689  & 1.093	& 09.52 & $+$66.87 & 107.8 & 34.2 & $-$33.7 & A & $+$0 & 3 \\
3 & 2020-03-28 & 1.532  & 1.056	& 08.42 & $+$68.43	& 96.4 & 40.4 & $-$38.5 & A & $+$50 & 5 \\
4 & 2020-04-01 & 1.458  & 1.042	& 07.93 & $+$68.47 & 91.1 & 43.3 & $-$40.4 & A & $+$90 & 3 \\
5 & 2020-04-06 & 1.365  & 1.026	& 07.35 & $+$67.97 & 84.7 & 46.9 & $-$42.5 & A & $+$25 & 3 \\
6 & 2020-04-07 & 1.348  & 1.023	& 07.25 & $+$67.82 & 83.5 & 47.6 & $-$42.9 & A & $+$5 & 1 \\
7 & 2020-04-10 & 1.204 & 1.013 & 06.93	& $+$67.25 & 79.7 & 49.8 & $-$44.0 & A & $-$0 & 5 \\
8 & 2020-04-17 & 1.154 & 0.987 & 06.30 & $+$65.42 & 70.8 & 55.3 & $-$46.3 & A & $-$80 & 1 \\

\hline
\end{tabular}
\end{table}

\begin{figure}[h!]

    \centering
    \includegraphics[scale=0.368]{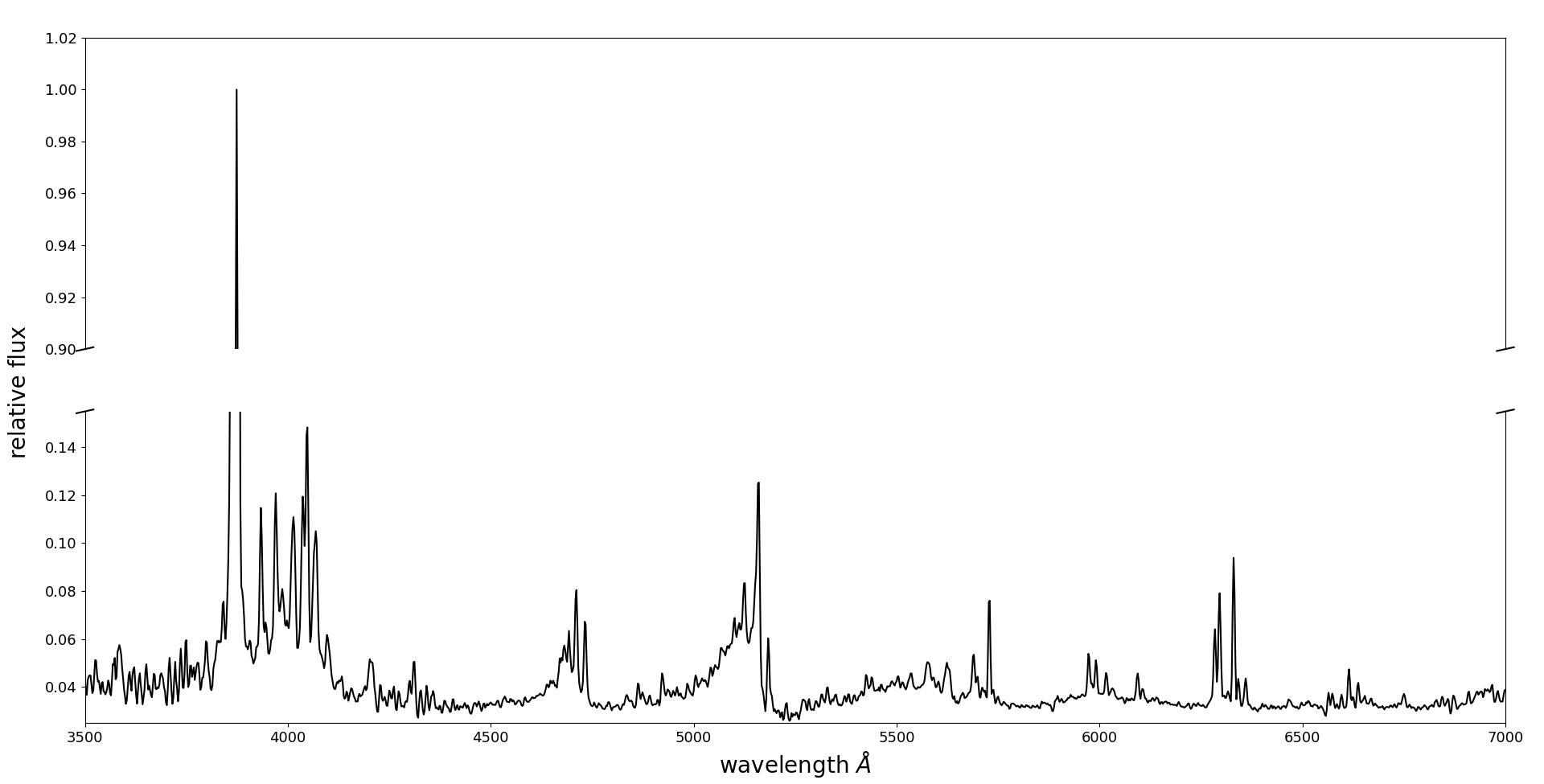}
    \caption{Spectrum of 2020-03-19; configuration A}

\end{figure}

\newpage
\clearpage

\section{C/2020 F3 (NEOWISE)}
\label{cometa:C2020F3}
\subsection{Description}

C/2020 F3 (NEOWISE) is a Long Period comet with a period of 6787 years and an absolute magnitude of 12.1$\pm$1.0.\footnote{\url{https://ssd.jpl.nasa.gov/tools/sbdb_lookup.html\#/?sstr=2020\%20F3} visited on July 20, 2024}
It was first spotted by the NEOWISE spacecraft on March 27, 2020.
The comet was an exceptional sight in the skies during the months of July and August 2020, with a long tail of up to twenty degrees.
Due to the proximity to the Sun at perihelion, it developed intense spectral lines from the sodium doublet.
Comet Neowise has an aphelic distance of more than 716 AU.

\noindent
We observed the comet at a visual magnitude of 6.\footnote{\url{https://cobs.si/comet/1875/ }, visited on July 20, 2024}
The Earth crossed the comet orbital plane on November 22, 2020.

\begin{table}[h!]
\centering
\begin{tabular}{|c|c|c|}
\hline
\multicolumn{3}{|c|}{Orbital elements (epoch: July 06, 2020)}                      \\ \hline \hline
\textit{e} = 0.999178 & \textit{q} = 0.2947 & \textit{T} = 2459034.1789 \\ \hline
$\Omega$ = 61.0104 & $\omega$ = 37.2787  & \textit{i} = 128.9375  \\ \hline  
\end{tabular}
\end{table}

\begin{table}[h!]
\centering
\begin{tabular}{|c|c|c|c|c|c|c|c|c|}
\hline
\multicolumn{9}{|c|}{Comet ephemerides for key dates}                      \\ \hline 
\hline
& date & r & $\Delta$ & RA & DEC & elong & phase & PLang \\
& (yyyy-mm-dd) & (AU) & (AU) & (h) & (°) & (°) & (°) & (°) \\ \hline 

Perihelion       & 2020-07-03 & 0.295  & 1.160 & 05.99  & $+$30.05  & 13.6 & 54.3  & $-$25.6  \\
Nearest approach & 2020-07-23 & 0.640  & 0.692 & 10.61  & $+$44.00  & 38.5 & 99.3  & $-$81.1 \\ \hline
\end{tabular}

\end{table}

\vspace{0.5 cm}

\begin{figure}[h!]
    \centering
    \includegraphics[scale=0.38]{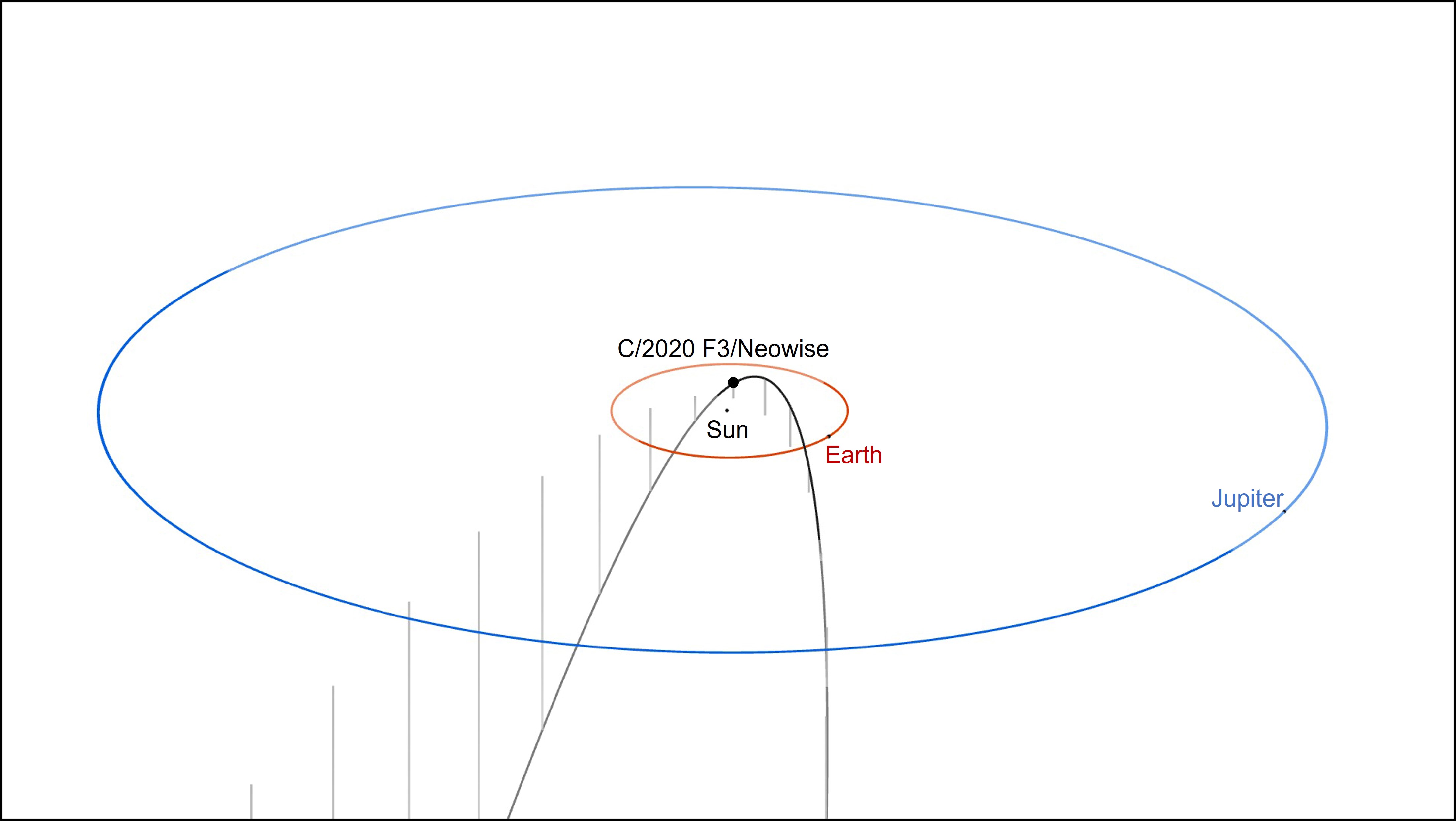}
    \caption{Orbit of comet C/2020 F3 and position on perihelion date. The field of view is set to the orbit of Jupiter for size comparison. Courtesy of NASA/JPL-Caltech.}
\end{figure} 


\newpage

\subsection{Images}

\begin{SCfigure}[0.8][h!]
 \centering
 \includegraphics[scale=0.4]{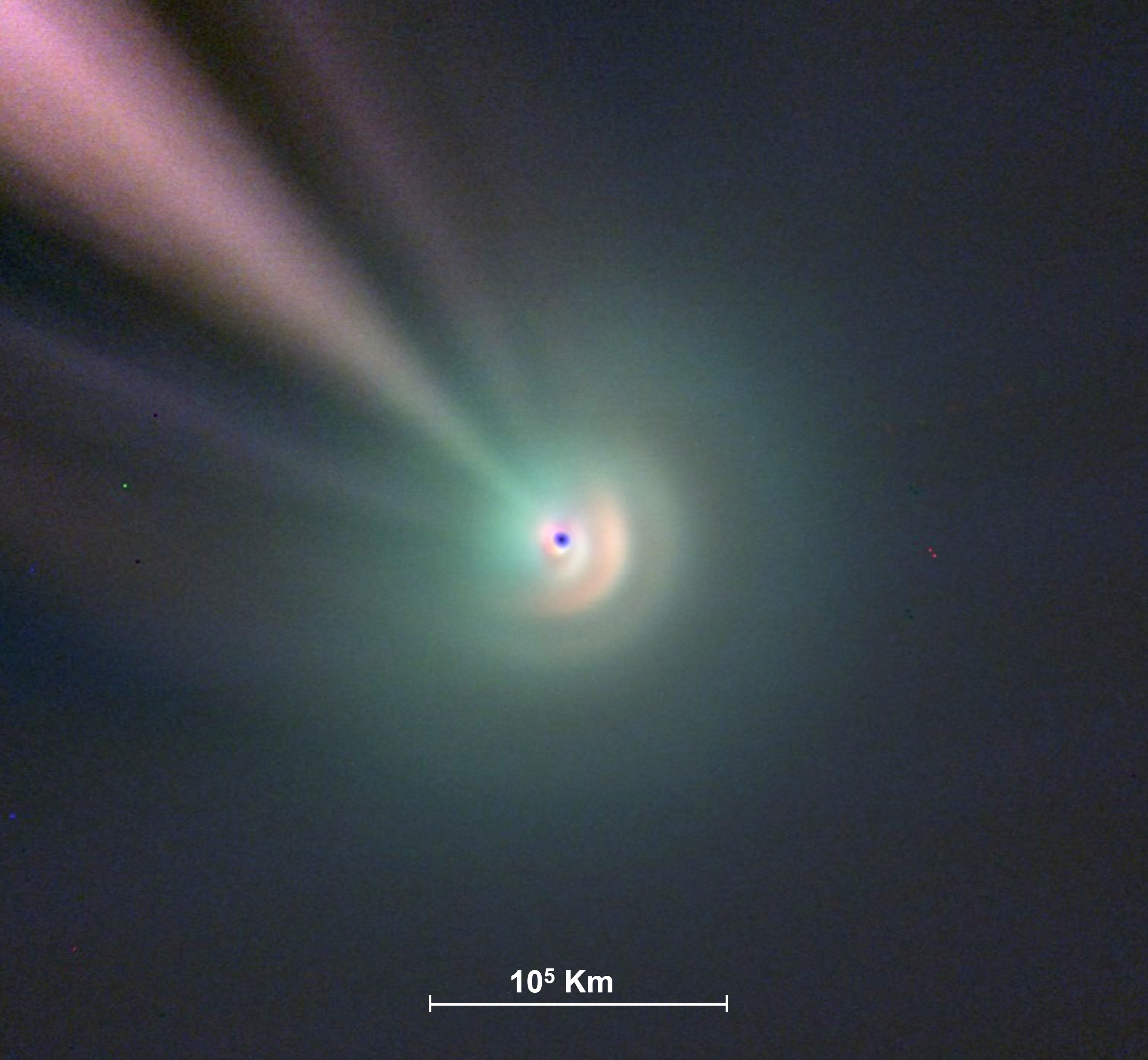}
 \caption{2020-07-21. Three-color BVR composite from images taken with the Asiago Copernico telescope. The comet was close to the perigee at the time of the imaging. The different colors allow to identify gases (green), dust (yellow-red) and ions (blue), which are the main components of coma and tails of a comet. The straight blue tail is oriented exactly in the anti-solar direction.}
\end{SCfigure} 

\begin{SCfigure}[0.8][h!]
 \centering
 \includegraphics[scale=0.4]{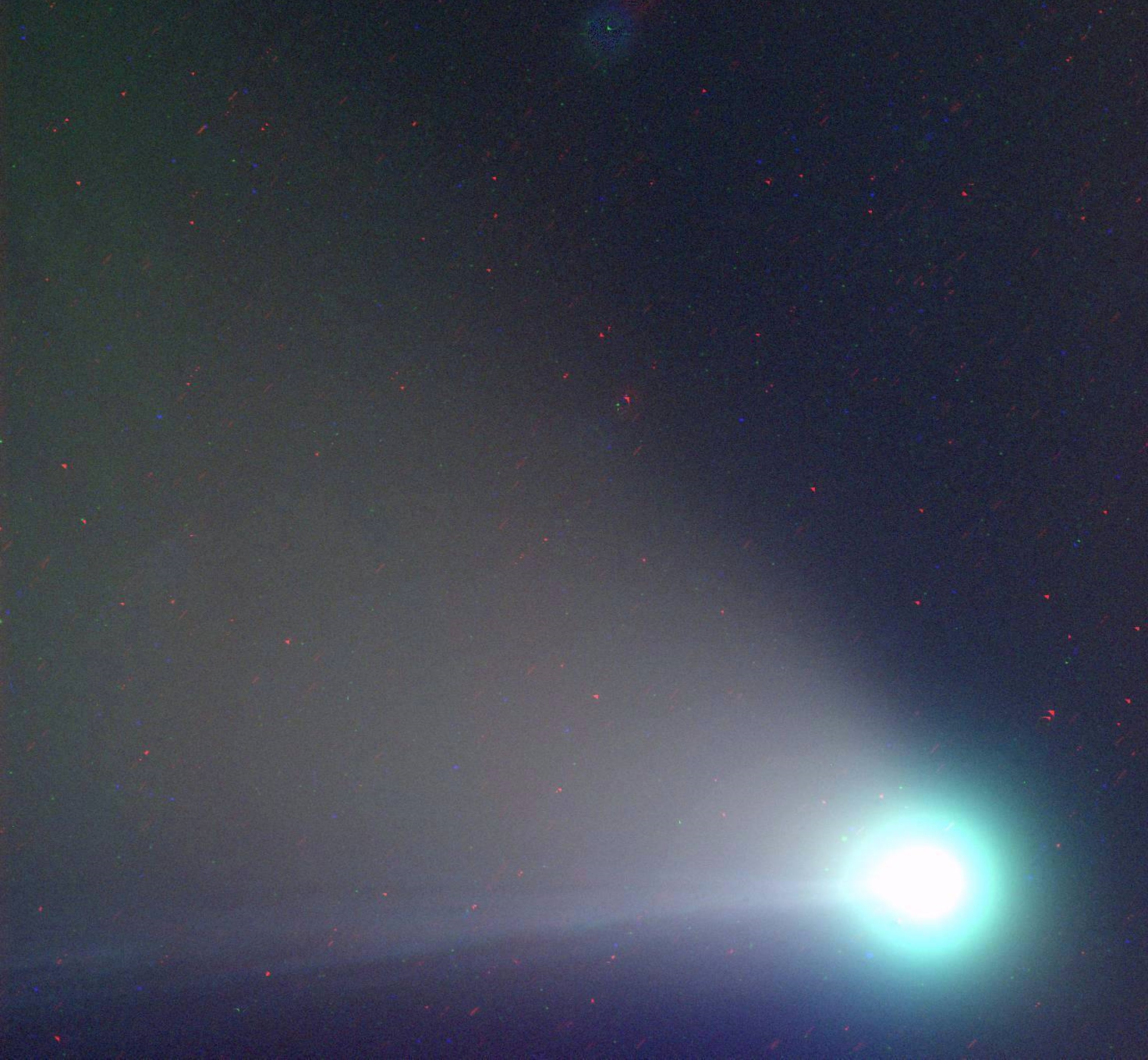}
 \caption{2020-08-05. Three-color BVR composite from images taken with the Asiago Schmidt telescope. Comet C/2020 F3 Neowise at its best, with the dust tail arranged in a fan and the thin ion tail (blue) that is projected straight towards East. The coma and the nucleus are even overexposed due to the high brightness of this comet. }
\end{SCfigure}

\newpage

\subsection{Spectra}

\begin{table}[h!]
\centering
\begin{tabular}{|c|c|c|c|c|c|c|c|c|c|c|c|}
\hline
\multicolumn{12}{|c|}{Observation details}                      \\ \hline 
\hline
$\#$ & date & r & $\Delta$ & RA & DEC & elong & phase & PLang& config & FlAng & N \\
 & (yyyy-mm-dd) & (AU) & (AU) & (h) & (°) & (°) & (°) & (°) & & (°) &  \\ \hline

1               & 2020-07-30 & 0.802  & 0.771 & 12.40  & $+$29.57 & 51.2 & 80.3 & $-$70.6 & A & $+$0 & 1 \\
2               & 2020-08-04 & 0.908  & 0.881 & 13.05  & $+$20.62  & 56.7 & 69.0 & $-$58.3 & A & $-$0 & 2 \\
 \hline
\end{tabular}
\end{table}

\begin{figure}[h!]
    \centering
    \includegraphics[scale=0.368]{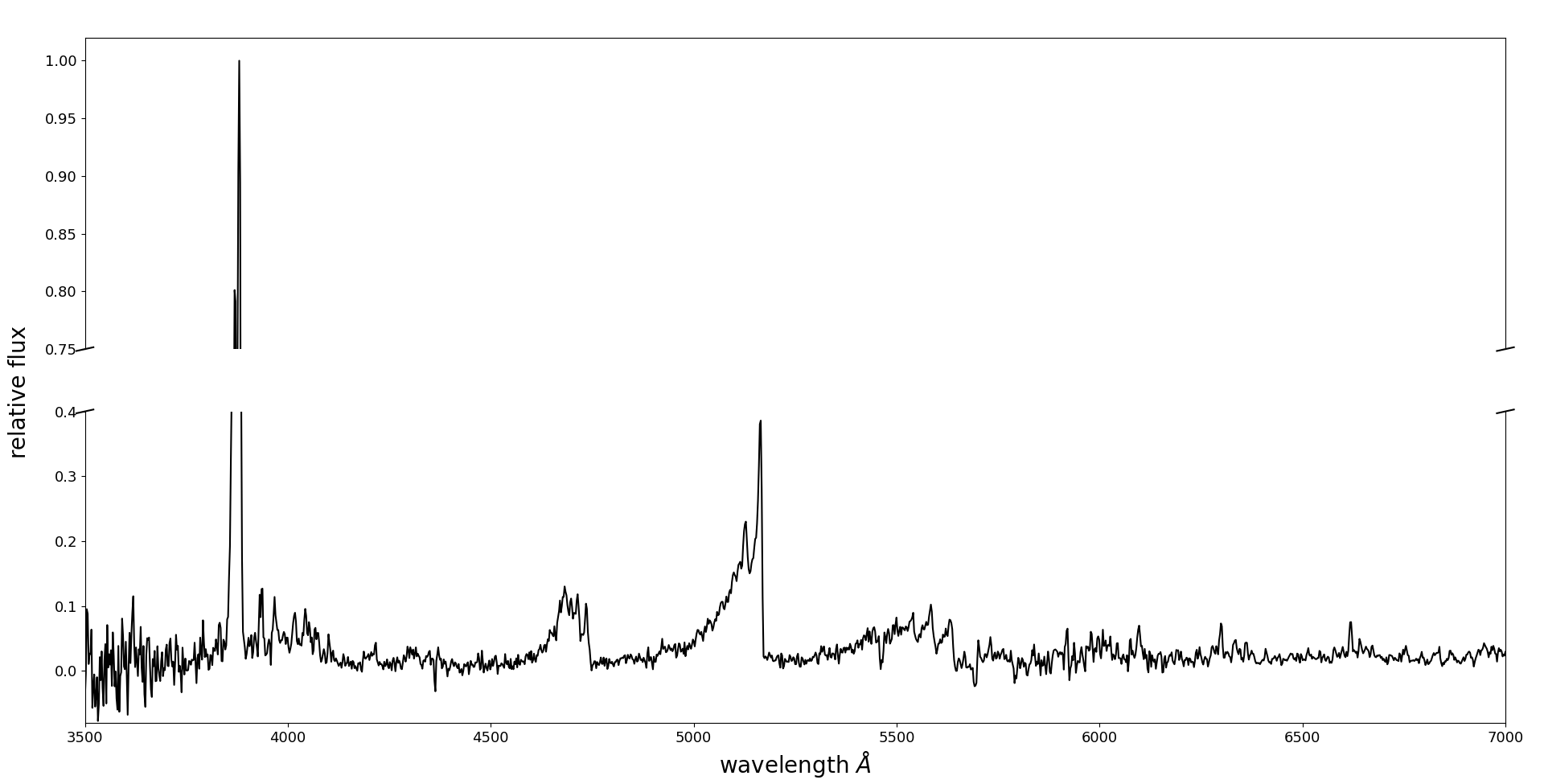}
    \caption{Spectrum of 2020-07-30; configuration A}

\end{figure}

\newpage
\clearpage

\section{C/2020 M3 (ATLAS)}
\label{cometa:C2020M3}
\subsection{Description}

C/2020 M3 (ATLAS) is a Halley-type comet with a period of 139 years and an absolute magnitude of 14.5$\pm$0.9.\footnote{\url{https://ssd.jpl.nasa.gov/tools/sbdb_lookup.html\#/?sstr=2020\%20M3} visited on July 20, 2024}
It was first spotted by the 0.5m Asteroid Terrestrial-impact Last Alert System (ATLAS) on June 27, 2020.
The absolute coma diameter increased from 70,000 km to 260,000 km between early September and mid-October 2020.
Comet C/2020 M3 has an aphelic distance of more than 52 AU. 

\noindent
We observed the comet at magnitude 7.\footnote{\url{https://cobs.si/comet/1932/ }, visited on July 20, 2024}
The Earth crossed the comet orbital plane on December 3, 2020.

\begin{table}[h!]
\centering
\begin{tabular}{|c|c|c|}
\hline
\multicolumn{3}{|c|}{Orbital elements (epoch: November 26, 2020)}                      \\ \hline \hline
\textit{e} = 0.9527 & \textit{q} = 1.2682 & \textit{T} = 2459148.1239 \\ \hline
$\Omega$ = 71.2502 & $\omega$ = 328.4463  & \textit{i} = 23.4735  \\ \hline  
\end{tabular}
\end{table}

\begin{table}[h!]
\centering
\begin{tabular}{|c|c|c|c|c|c|c|c|c|}
\hline
\multicolumn{9}{|c|}{Comet ephemerides for key dates}                      \\ \hline 
\hline
& date         & r    & $\Delta$  & RA      & DEC      & elong  & phase  & PLang  \\
& (yyyy-mm-dd) & (AU) & (AU)      & (h)     & (°)      & (°)    & (°)    & (°) \\ \hline 

Perihelion       & 2020-10-25 & 1.268 & 0.406 & 05.16 & $-$18.09 & 124.4 & 40.3 & $+$37.8 \\ 
Nearest approach & 2020-11-13 & 1.299 & 0.359 & 05.43 & $+$03.27 & 144.8 & 26.1 & $+$21.9 \\ \hline
\end{tabular}

\end{table}

\vspace{0.5 cm}

\begin{figure}[h!]
    \centering
    \includegraphics[scale=0.38]{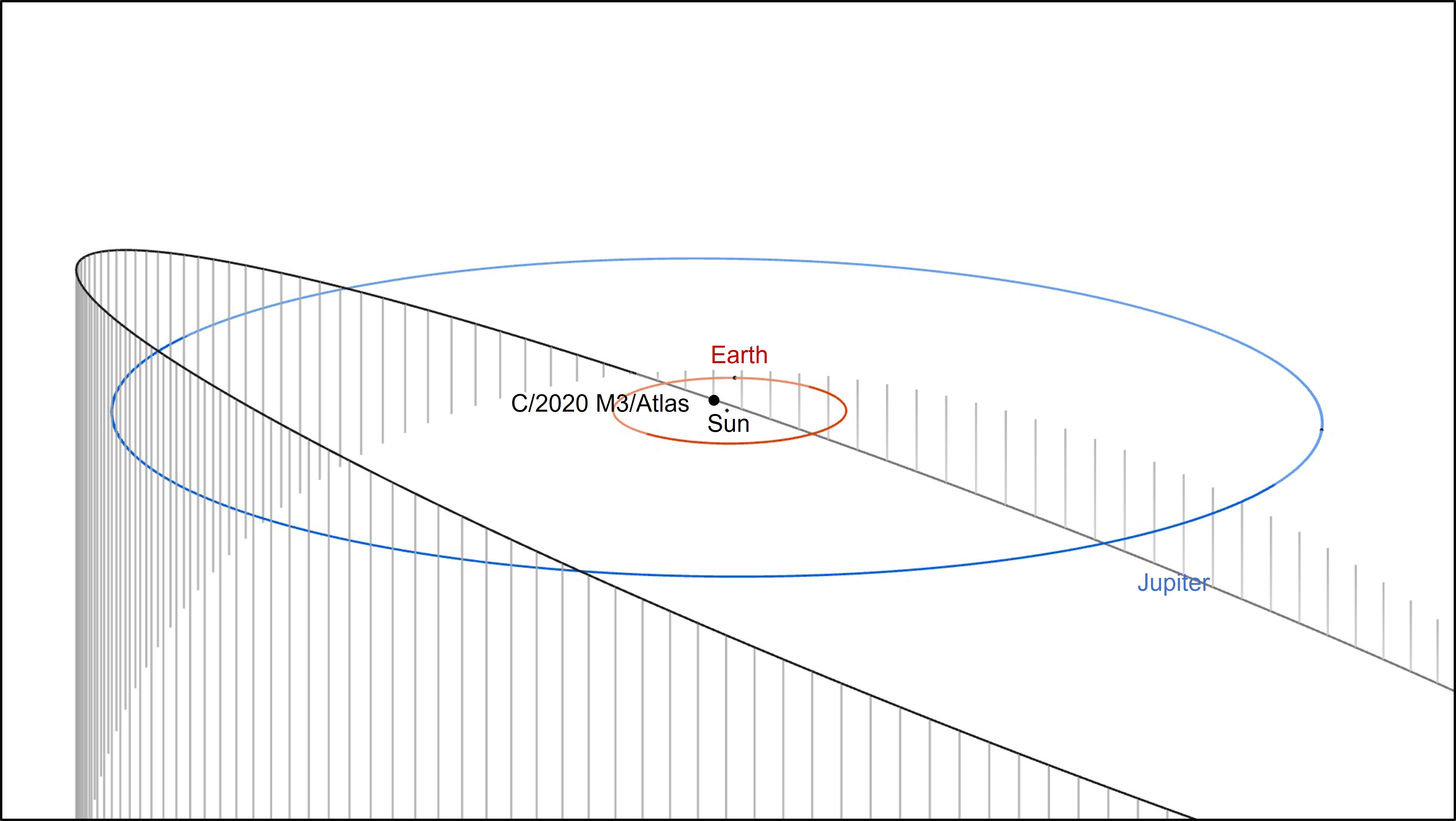}
    \caption{Orbit of comet C/2020 M3 and position on perihelion date. The field of view is set to the orbit of Jupiter for size comparison. Courtesy of NASA/JPL-Caltech.}
\end{figure} 


\newpage

\subsection{Images}

\begin{SCfigure}[0.8][h!]
    \centering
    \includegraphics[scale=0.4]{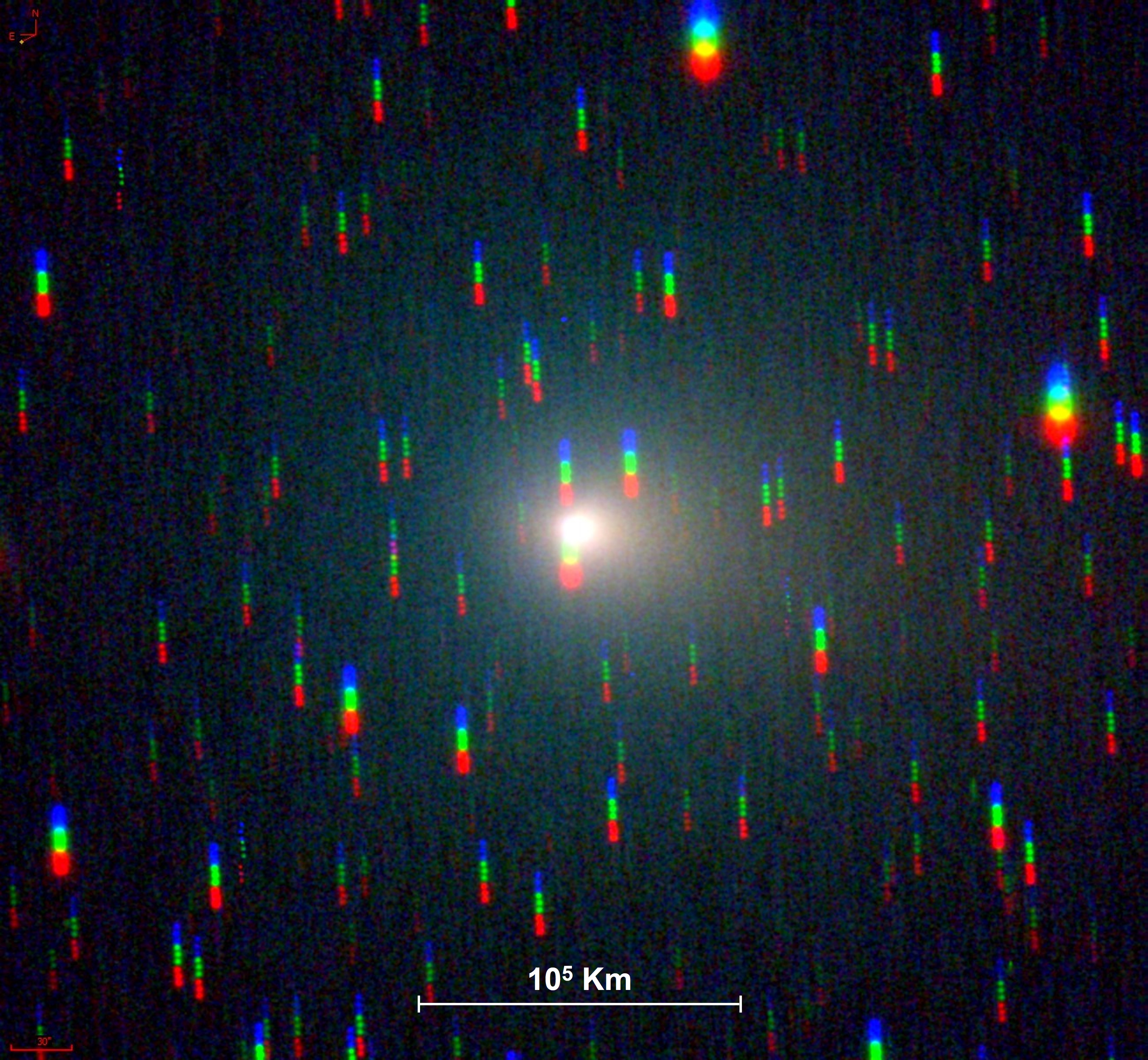}
     \caption{2021-01-15. Three-color BVR composite from images taken with the Schmidt telescope. The inner coma and the outer coma have different colors: the latter is composed of a higher concentration of gases emitting in the green region of the spectrum.}
\end{SCfigure} 

\begin{SCfigure}[0.8][h!]
    \centering

    \includegraphics[scale=0.4]{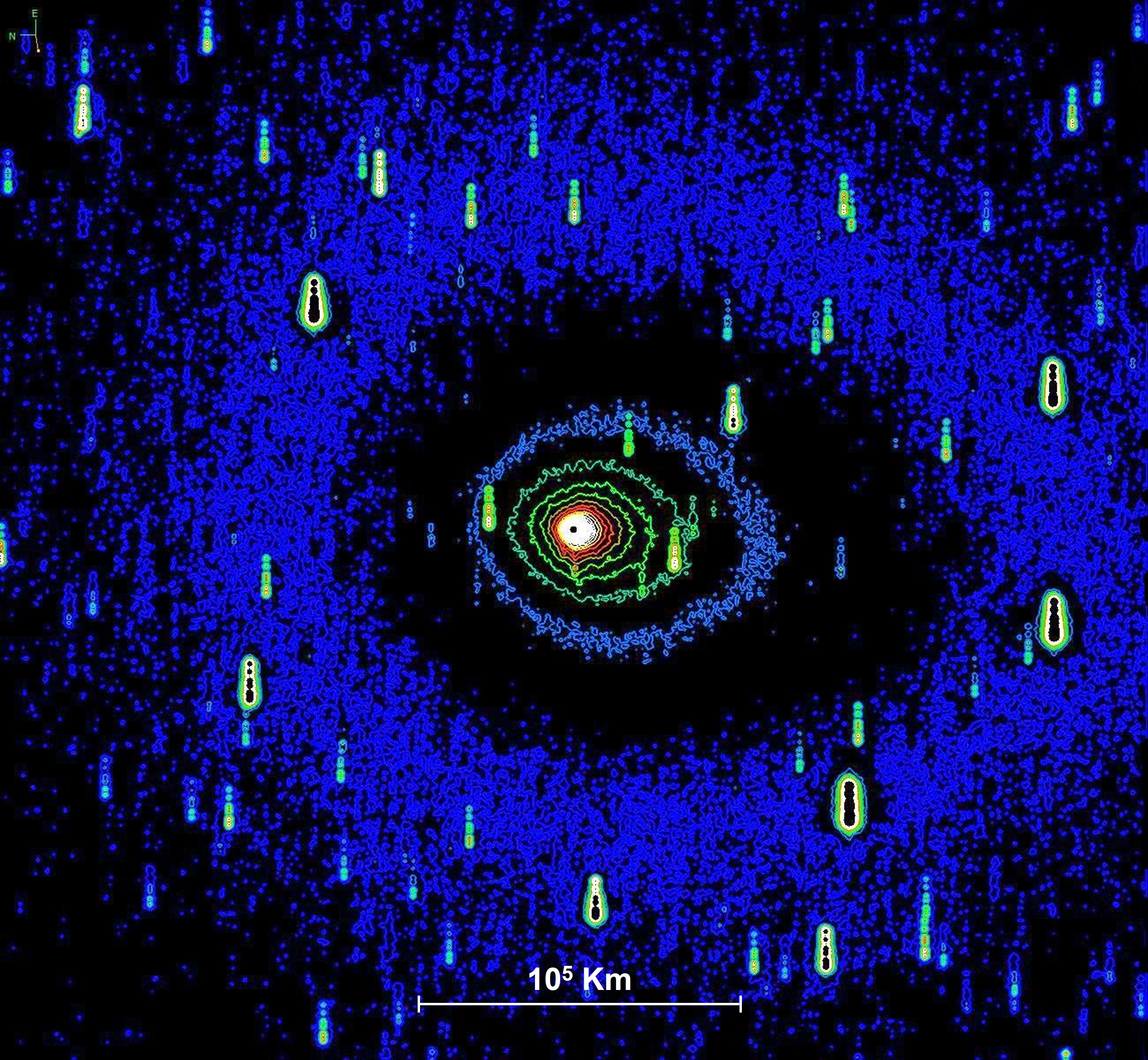}

    \caption{2021-01-18. Image taken three days later visualized as isophotes, showing the extension of the coma around the nucleus (identified with a black dot). The respective orbital positions of the comet and the Earth cause the tail to develop behind the coma and therefore to be almost completely hidden.}
\end{SCfigure}

\newpage

\subsection{Spectra}

\begin{table}[h!]
\centering
\begin{tabular}{|c|c|c|c|c|c|c|c|c|c|c|c|}
\hline
\multicolumn{12}{|c|}{Observation details}                      \\ \hline 
\hline
$\#$  & date          & r     & $\Delta$ & RA     & DEC     & elong & phase & PLang& config  & FlAng & N \\
      & (yyyy-mm-dd)  &  (AU) & (AU)     & (h)    & (°)     & (°)   & (°)   &  (°)   &       &  (°)  &  \\ \hline 
      
1 & 2020-11-12 & 1.298 & 0.359 & 05.43 & $+$01.33 & 144.2 & 26.5 & $+$22.4 & A & $+$90 & 1 \\
2 & 2020-11-14 & 1.301 & 0.358 & 05.45 & $+$03.95 & 145.4 & 25.6 & $+$21.4 & C & $+$90 & 2 \\
3 & 2020-11-16 & 1.304 & 0.358 & 05.45 & $+$06.58 & 146.9 & 24.4 & $+$20.1 & C & $+$90 & 2 \\
4 & 2020-11-17 & 1.315 & 0.360 & 05.45 & $+$07.92 & 150.9 & 21.4 & $+$16.8 & C & $-$0 & 2 \\
5 & 2020-11-21 & 1.332 & 0.367 & 05.47 & $+$13.20 & 156.5 & 17.2 & $+$11.9 & C & $+$50 & 1 \\

\hline
\end{tabular}
\end{table}

\begin{figure}[h!]

    \centering
    \includegraphics[scale=0.368]{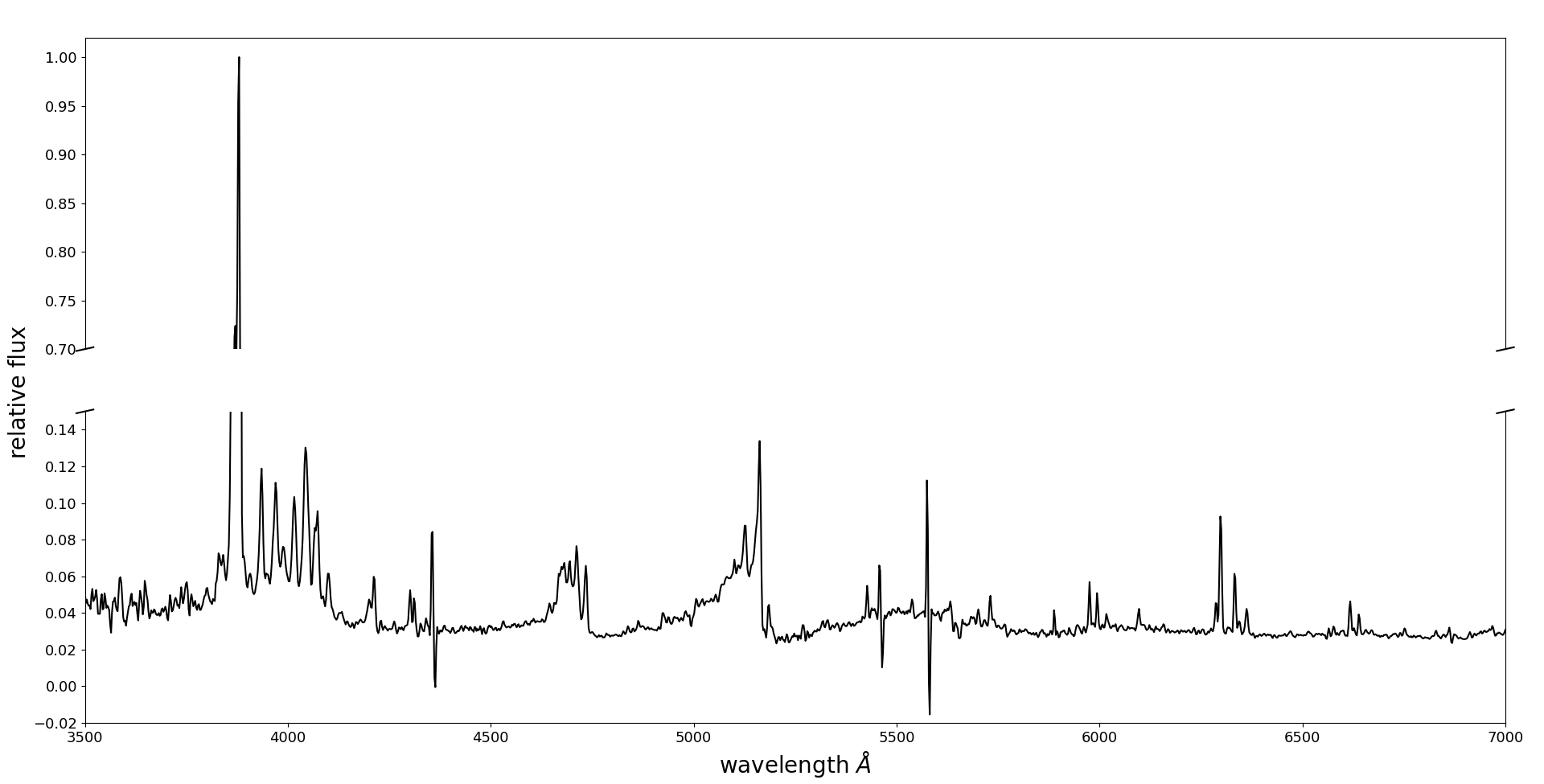}
    \caption{Spectrum of 2020-11-12; configuration A}

\end{figure}

\begin{figure}[h!]

    \centering
    \includegraphics[scale=0.368]{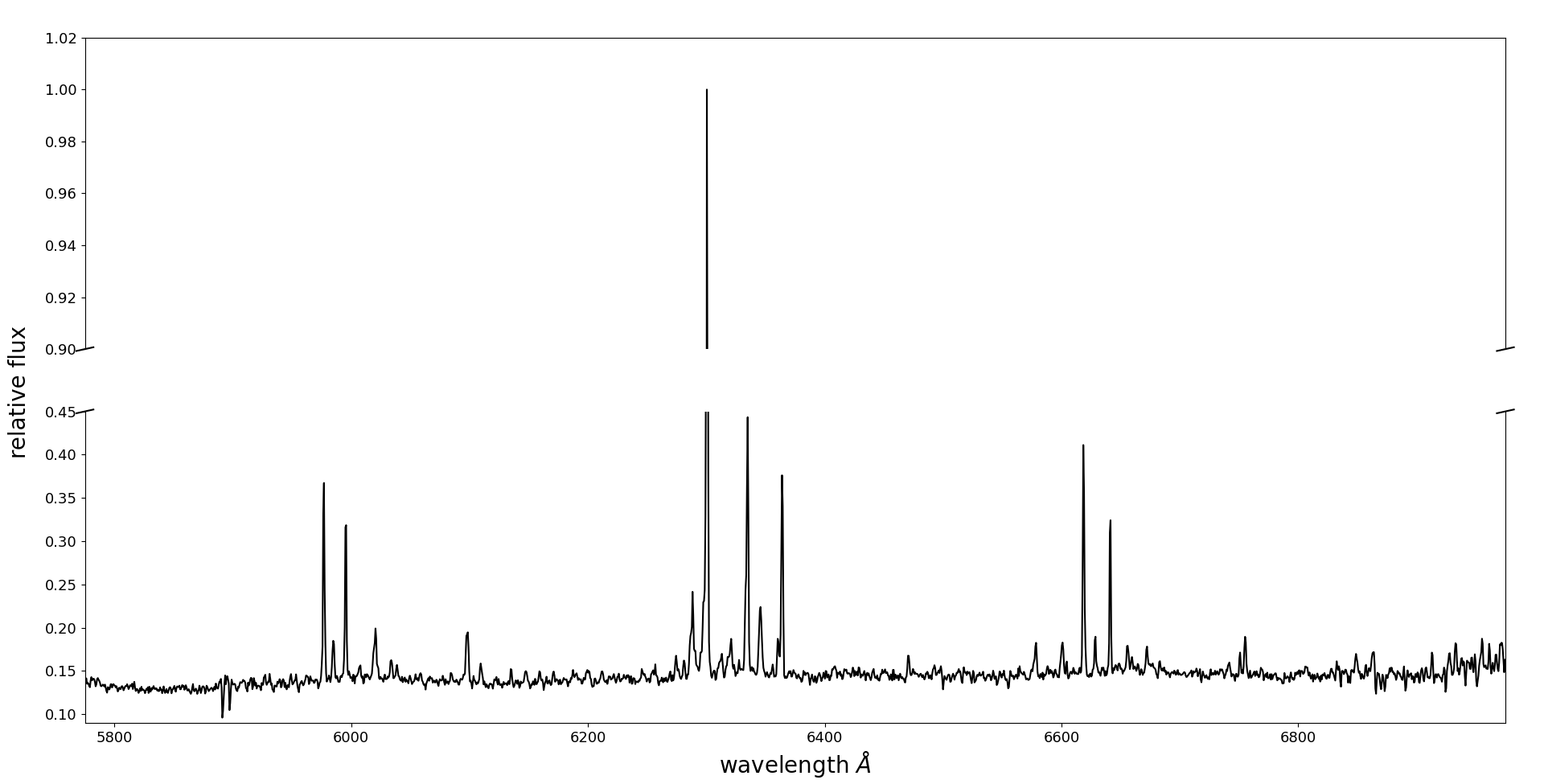}
    \caption{Spectrum of 2020-11-14; configuration C}

\end{figure}

\newpage
\clearpage

\section{C/2020 PV6 (PanSTARRS)}
\label{cometa:C2020PV6}
\subsection{Description}

C/2020 PV6 (PanSTARRS) is a comet with a period of 269 years and an absolute magnitude of 7.5$\pm$0.6.\footnote{\url{https://ssd.jpl.nasa.gov/tools/sbdb_lookup.html\#/?sstr=2020\%20PV6} visited on July 20, 2024}
It was discovered by the 1.8m Pan-STARRS 1 telescope located in Haleakala, Hawaii, on Aug 13, 2020.

\noindent
We observed the comet at magnitude 14.\footnote{\url{https://cobs.si/comet/3/  }, visited on July 20, 2024}
The Earth crossed the comet orbital plane on February 17 and August 23, 2021.

\begin{table}[h!]
\centering
\begin{tabular}{|c|c|c|}
\hline
\multicolumn{3}{|c|}{Orbital elements (epoch: September 29, 2021)}                     \\ \hline \hline
\textit{e} = 0.9450 & \textit{q} = 2.2956 & \textit{T} = 2459483.4993 \\ \hline
$\Omega$ = 329.1439 & $\omega$ = 71.3806  & \textit{i} = 128.2393  \\ \hline  
\end{tabular}
\end{table}

\begin{table}[h!]
\centering
\begin{tabular}{|c|c|c|c|c|c|c|c|c|}
\hline
\multicolumn{9}{|c|}{Comet ephemerides for key dates}                      \\ \hline 
\hline
& date         & r    & $\Delta$  & RA      & DEC      & elong  & phase  & PLang  \\
& (yyyy-mm-dd) & (AU) & (AU)      & (h)     & (°)      & (°)    & (°)    & (°) \\ \hline 

Perihelion       & 2021-09-26 & 2.296 & 2.567 & 16.23 & $+$21.43 & 63.1 & 22.9 & $+$09.9  \\ 
Nearest approach & 2021-07-19 & 2.416 & 1.671 & 19.26 & $+$32.06 & 126.3 & 19.8 & $-$14.8 \\ 
\hline
\end{tabular}

\end{table}

\vspace{0.5 cm}

\begin{figure}[h!]
    \centering
    \includegraphics[scale=0.38]{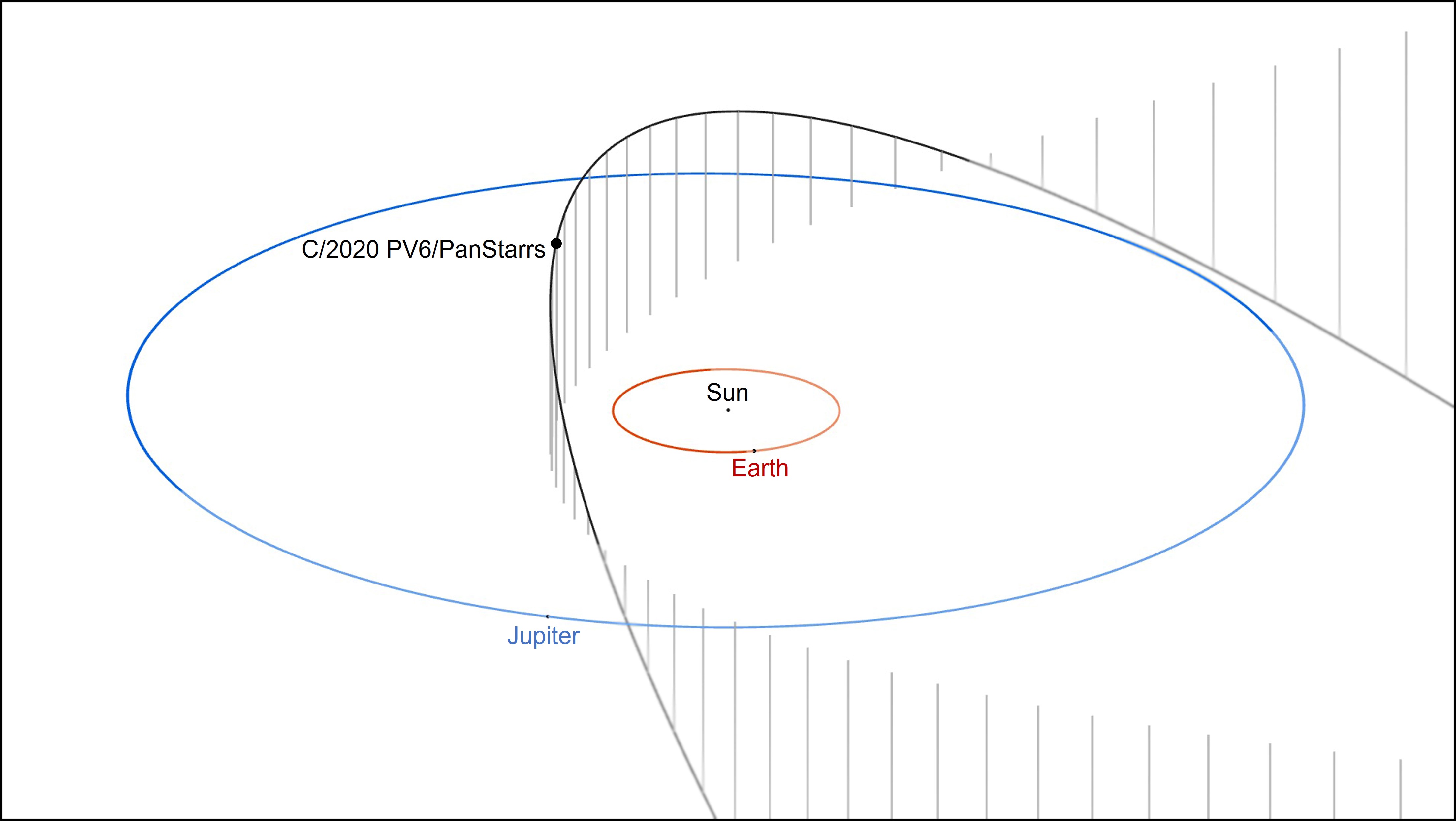}
    \caption{Orbit of comet C/2020 PV6 and position on perihelion date. The field of view is set to the orbit of Jupiter for size comparison. Courtesy of NASA/JPL-Caltech.}
\end{figure}

\newpage

\subsection{Images}

\begin{SCfigure}[0.8][h!]
    \centering
    \includegraphics[scale=0.4]{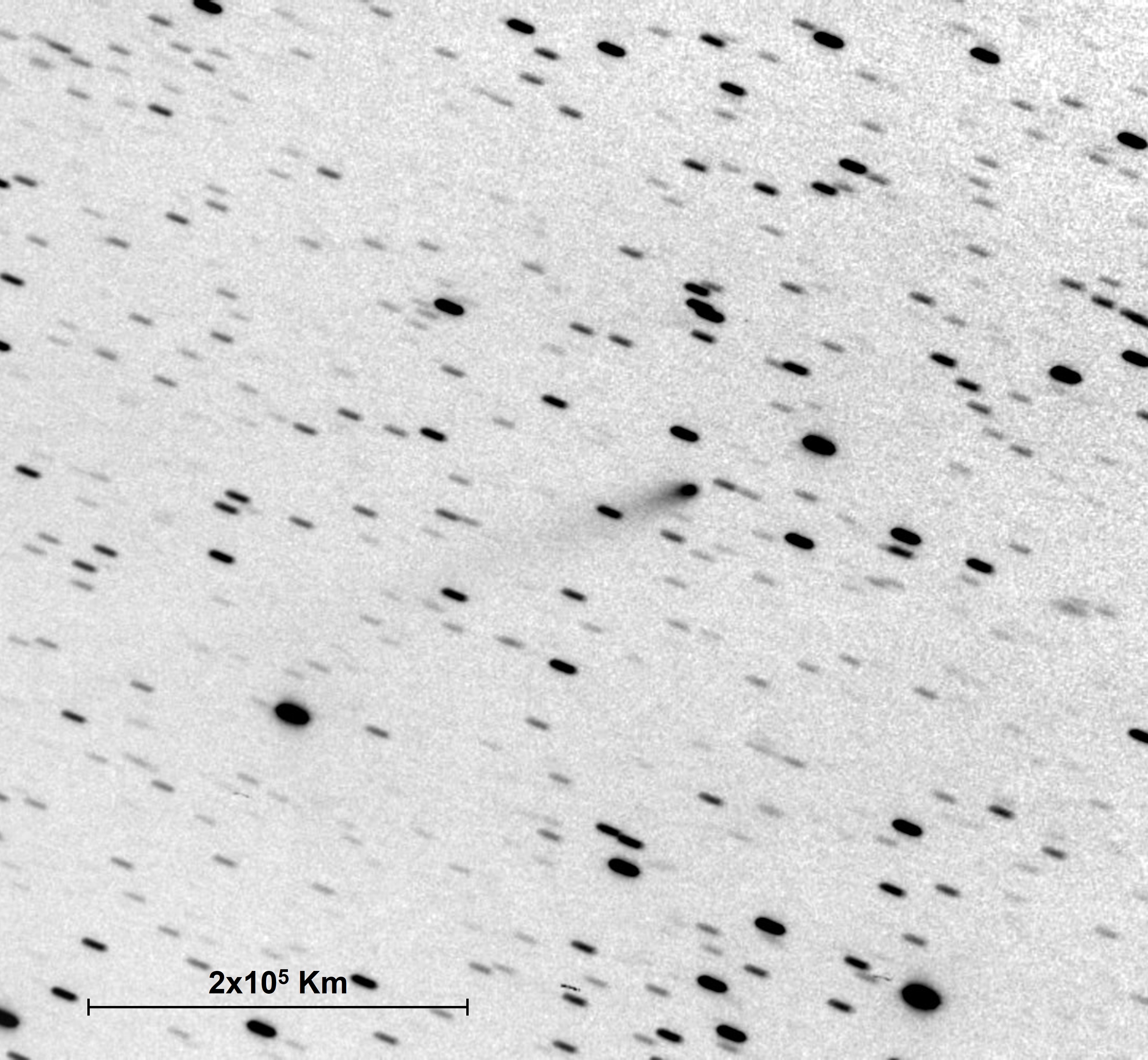}
     \caption{2021-08-20. Comet C/2020 PV6 exhibited low activity throughout its entire apparition. Image taken by J.G. Bosch with a 0.4m telescope as part of a collaboration within the "Asyago" initiative.
     The Earth was about to cross the orbital plane of the comet.}
\end{SCfigure} 

\begin{SCfigure}[0.8][h!]
    \centering
    \includegraphics[scale=0.4]{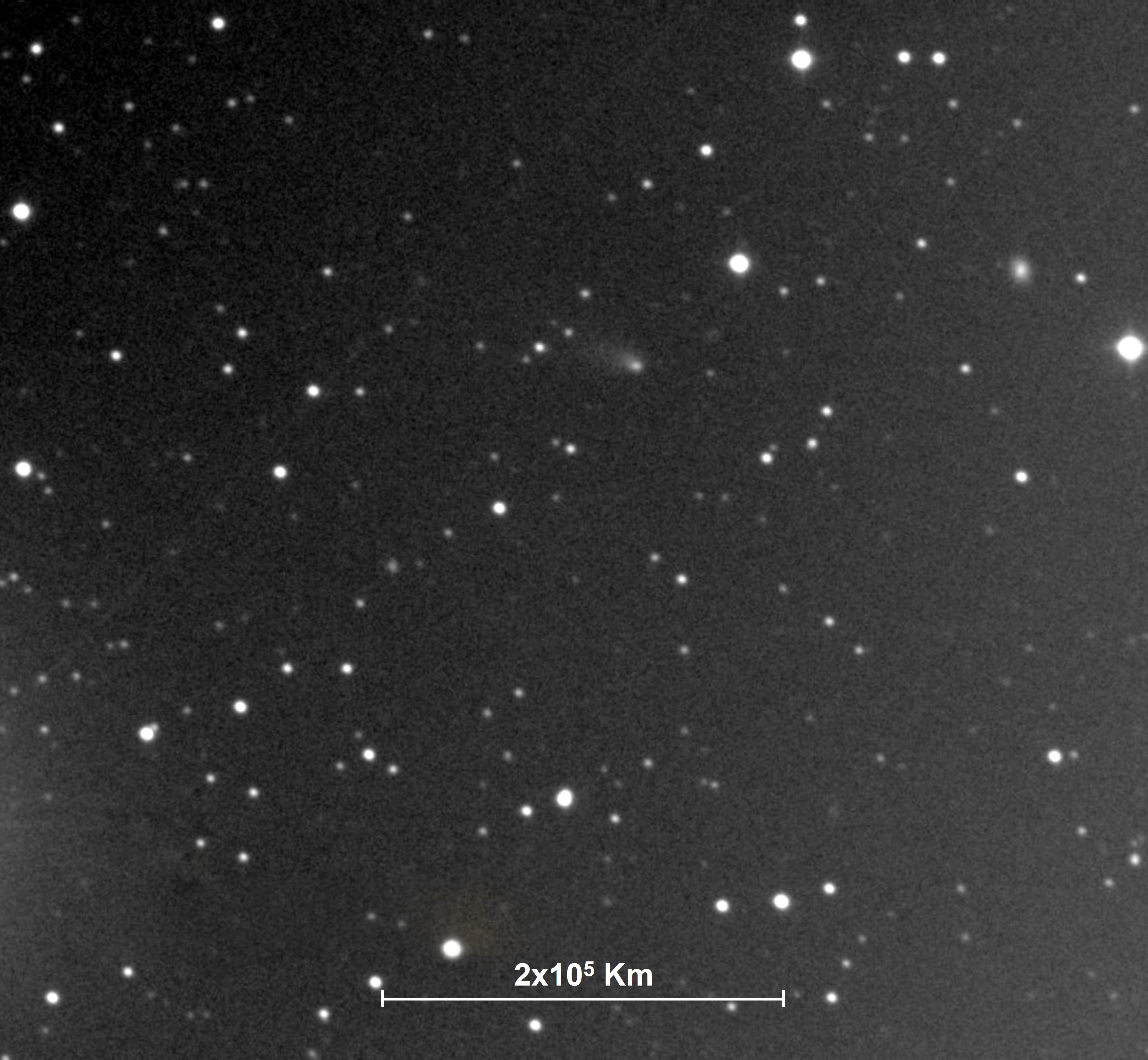}
    \caption{2021-11-07. C/2020 PV6 moved rapidly away from Earth approaching a heliacal conjunction that hid it from our view. Image taken by J.G. Bosch with a 0.4m telescope with the comet at an elongation of only 38° from the Sun.}
\end{SCfigure}

\newpage

\subsection{Spectra}

\begin{table}[h!]
\centering
\begin{tabular}{|c|c|c|c|c|c|c|c|c|c|c|c|}
\hline
\multicolumn{12}{|c|}{Observation details}                      \\ \hline 
\hline
$\#$  & date          & r     & $\Delta$ & RA     & DEC     & elong & phase & PLang& config  & FlAng & N \\
      & (yyyy-mm-dd)  &  (AU) & (AU)     & (h)    & (°)     & (°)   & (°)   &  (°)   &       &  (°)  &  \\ \hline 

1 & 2021-08-06 & 2.361 & 1.775 & 17.93 & $+$32.02 & 112.8 & 23.3 & $-$06.6 & A & $-$0 & 1 \\
2 & 2021-08-07 & 2.358 & 1.786 & 17.86 & $+$31.88 & 111.8 & 23.5 & $-$06.1 & A & $-$0 & 1 \\
3 & 2021-08-08 & 2.358 & 1.786 & 17.80 & $+$31.73 & 111.8 & 23.5 & $-$06.1 & A & $-$0 & 1 \\
\hline
\end{tabular}
\end{table}

\begin{figure}[h!]

    \centering
    \includegraphics[scale=0.368]{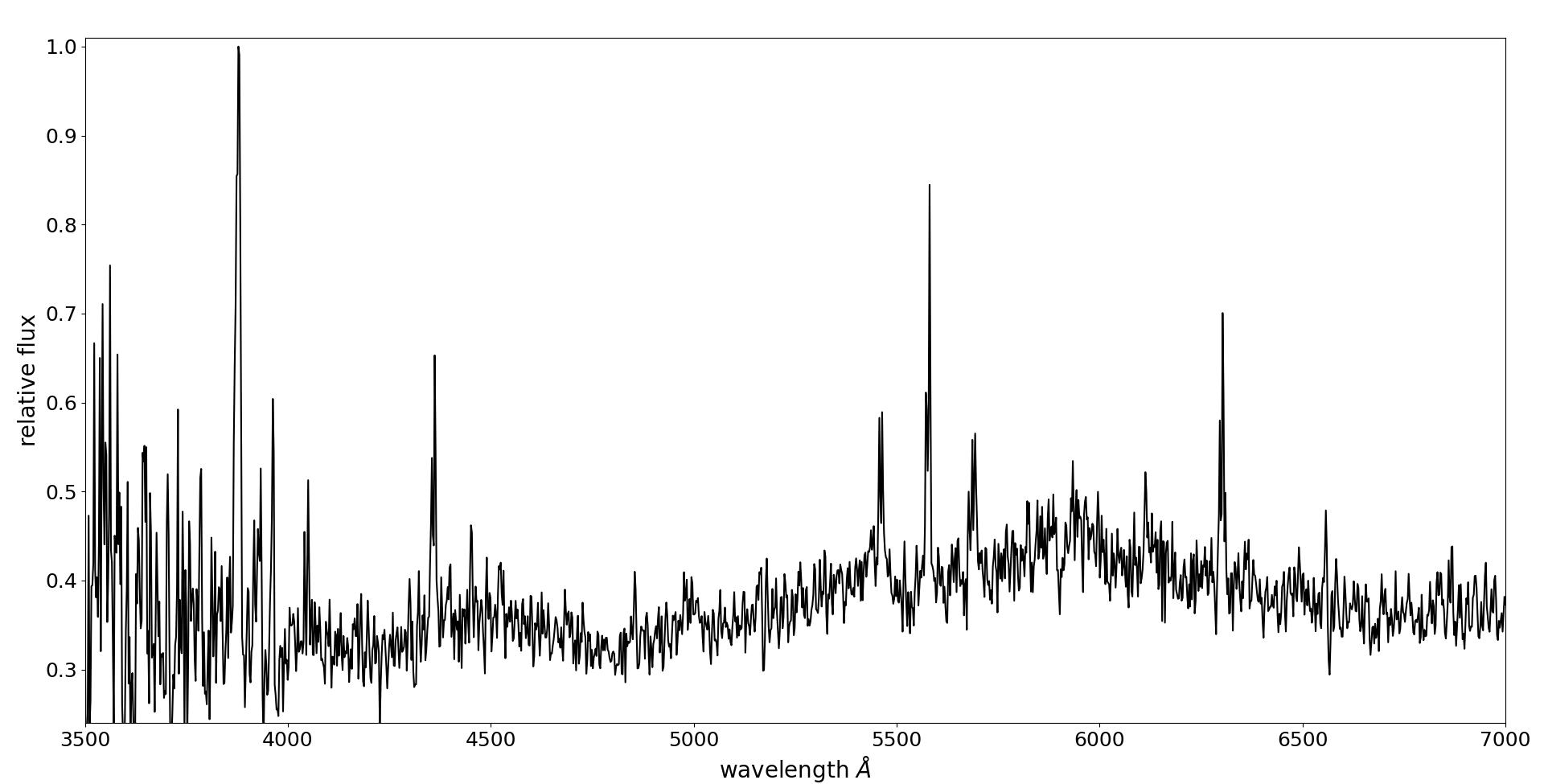}
    \caption{Spectrum of 2021-08-06; configuration A}

\end{figure}

\newpage
\clearpage

\section{C/2020 R4 (ATLAS)}
\label{cometa:C2020R4}
\subsection{Description}

C/2020 R4 (ATLAS) is a Long Period comet with a period of 951 years and an absolute magnitude of 12$\pm$1.0.\footnote{\url{https://ssd.jpl.nasa.gov/tools/sbdb_lookup.html\#/?sstr=2020\%20R4} visited on July 20, 2024}
It was first spotted by James Robinson using the 0.5m Asteroid Terrestrial-impact Last Alert System (ATLAS) on September 12, 2020.
Comet C/2020 R4 has an aphelic distance of more than 192 AU.
The Earth crossed the comet orbital plane on February 12 and August 16, 2021.

\noindent
We observed the comet at magnitude 10.\footnote{\url{https://cobs.si/comet/1954/ }, visited on July 20, 2024}

\begin{table}[h!]
\centering
\begin{tabular}{|c|c|c|}
\hline
\multicolumn{3}{|c|}{Orbital elements (epoch: March 13, 2021)}                      \\ \hline \hline
\textit{e} = 0.9894 & \textit{q} = 1.0286 & \textit{T} = 2459275.4355 \\ \hline
$\Omega$ = 323.2724 & $\omega$ =46.7079  & \textit{i} = 164.4633  \\ \hline  
\end{tabular}
\end{table}

\begin{table}[h!]
\centering
\begin{tabular}{|c|c|c|c|c|c|c|c|c|}
\hline
\multicolumn{9}{|c|}{Comet ephemerides for key dates}                      \\ \hline 
\hline
& date         & r    & $\Delta$  & RA      & DEC      & elong  & phase  & PLang  \\
& (yyyy-mm-dd) & (AU) & (AU)      & (h)     & (°)      & (°)    & (°)    & (°) \\ \hline 

Perihelion       & 2021-03-02 & 1.029 & 1.699 & 20.64 & $-$11.37 & 33.4 & 32.0 & $-$02.8  \\ 
Nearest approach & 2021-04-23 & 1.343 & 0.464 & 16.23 & $+$28.47 & 128.3 & 36.0 & $-$33.1 \\ 
\hline
\end{tabular}

\end{table}

\vspace{0.5 cm}

\begin{figure}[h!]
    \centering
    \includegraphics[scale=0.38]{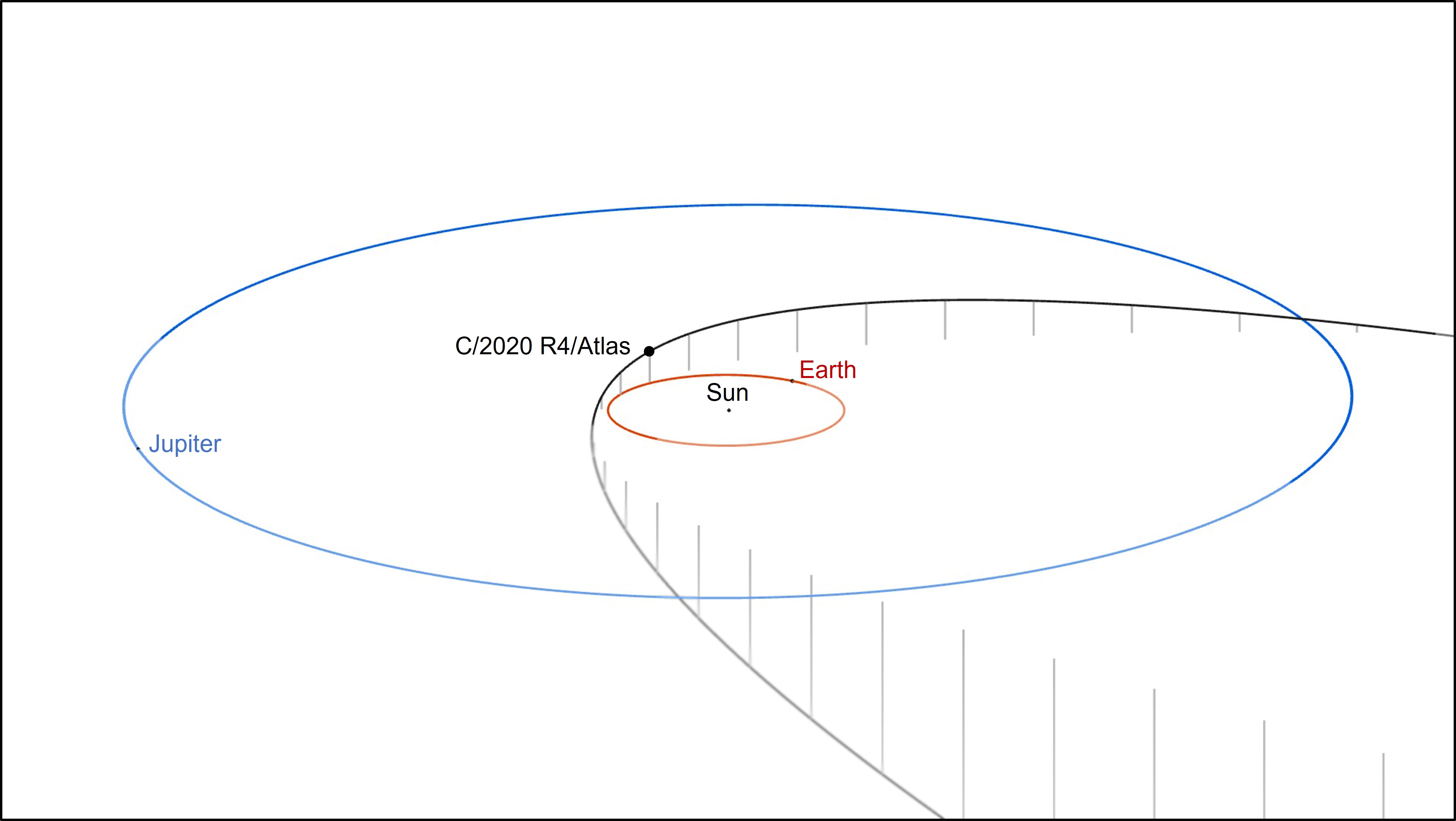}
    \caption{Orbit of comet C/2020 R4 and position on perihelion date. The field of view is set to the orbit of Jupiter for size comparison. Courtesy of NASA/JPL-Caltech.}
\end{figure}

\newpage

\subsection{Images}

\begin{SCfigure}[0.8][h!]
    \centering
    \includegraphics[scale=0.4]{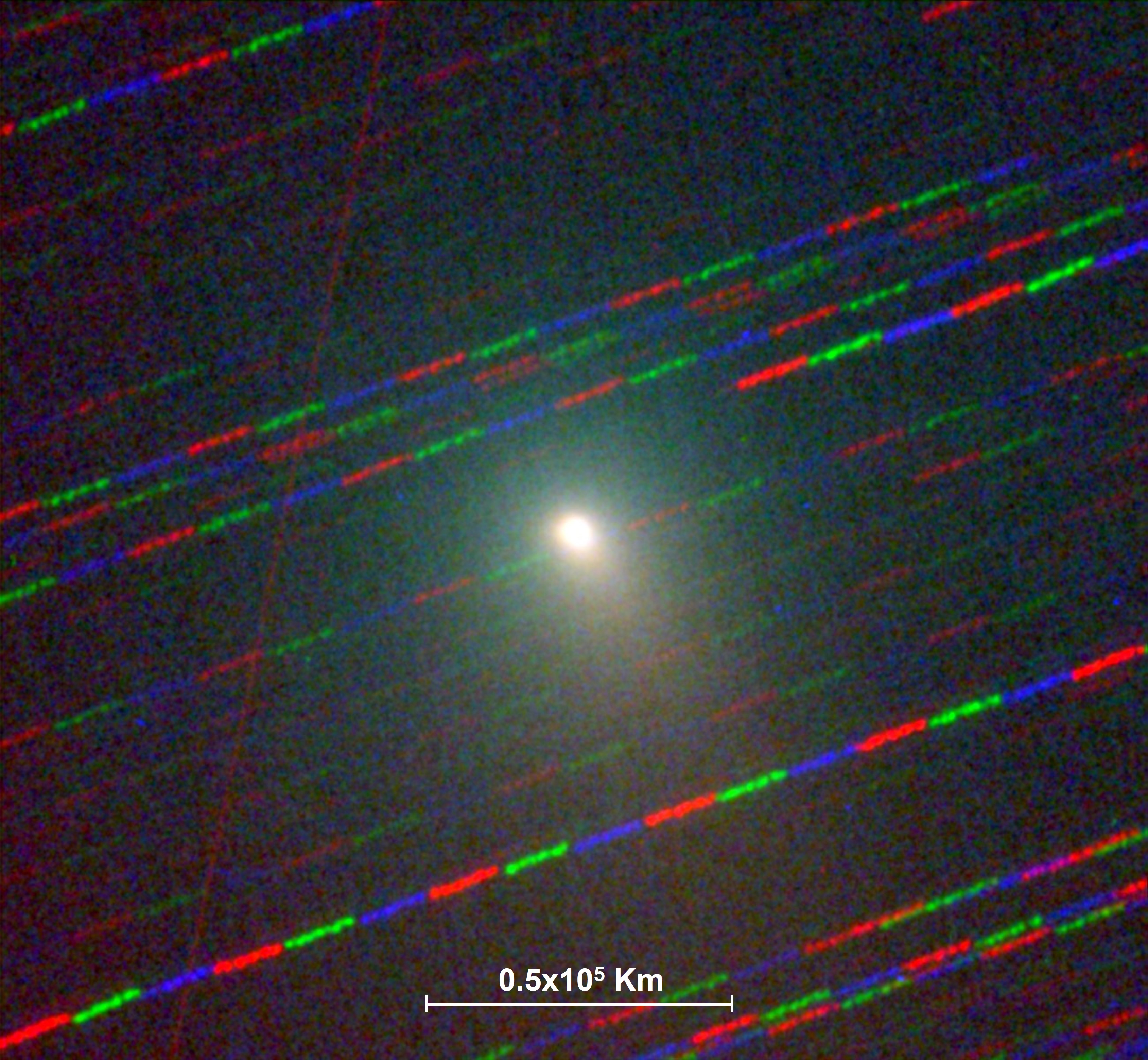}
     \caption{2021-04-22. Trichrome composite image taken with B, V, R filters at the Stazione Astronomica di Sozzago with the 0.4m Savonarola telescope.
The coma and tail have different colors: the first has a higher concentration of gases emitting in the green region of the spectrum, while the dust tail is reddened.}
\end{SCfigure} 

\begin{SCfigure}[0.8][h!]
    \centering
    \includegraphics[scale=0.4]{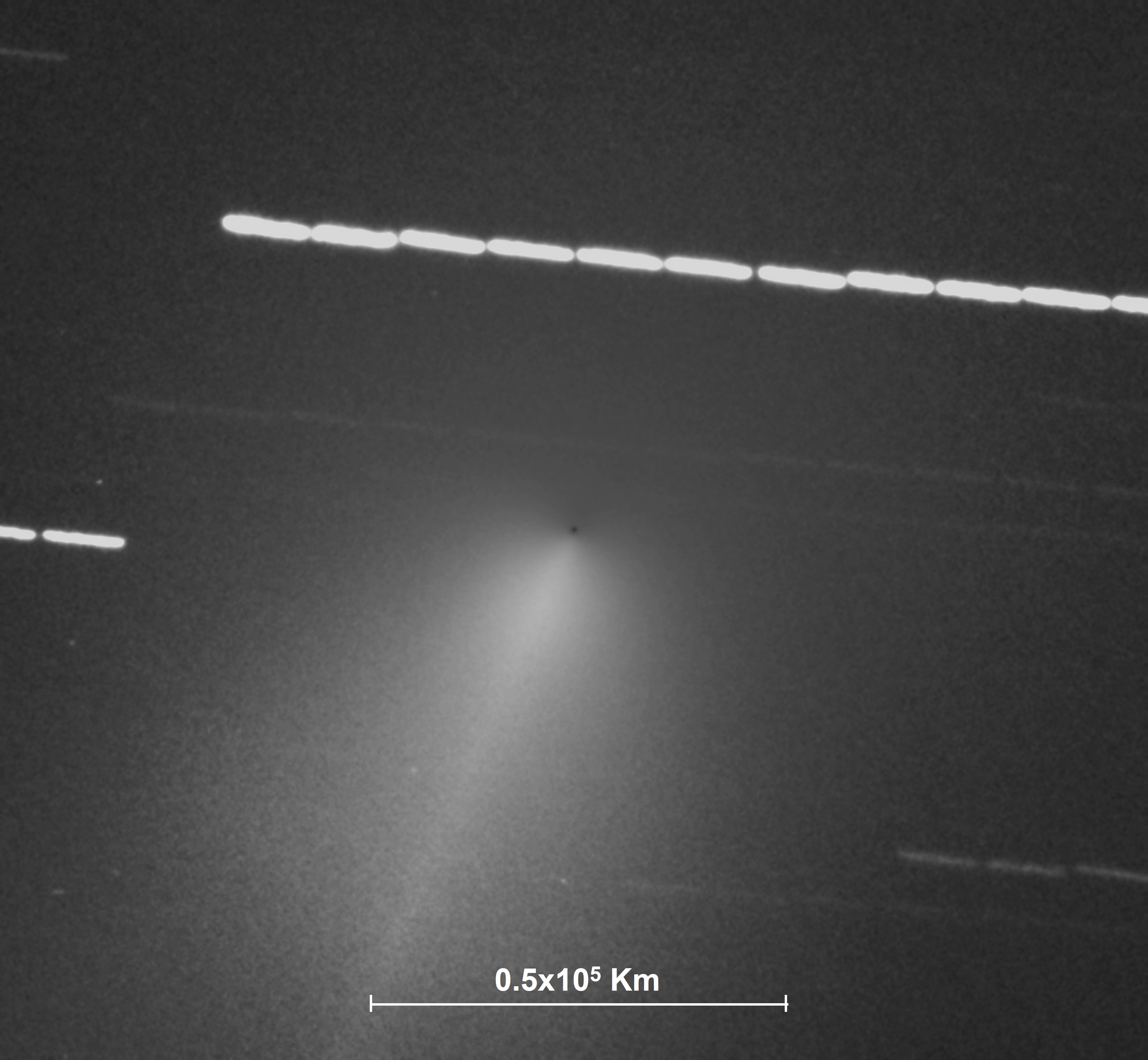}
    \caption{2021-05-01. When the comet got closer to Earth, the geometric conditions of observation changed rapidly. In just 10 days, the orientation of the tail changed by more than 40°.
    The image has been processed for a better visualization of the rectilinear ion tail, which is projected in the anti-solar direction. The cometary nucleus coincides with the black dot. }
\end{SCfigure}

\newpage

\subsection{Spectra}

\begin{table}[h!]
\centering
\begin{tabular}{|c|c|c|c|c|c|c|c|c|c|c|c|}
\hline
\multicolumn{12}{|c|}{Observation details}                      \\ \hline 
\hline
$\#$  & date          & r     & $\Delta$ & RA     & DEC     & elong & phase & PLang& config  & FlAng & N \\
      & (yyyy-mm-dd)  &  (AU) & (AU)     & (h)    & (°)     & (°)   & (°)   &  (°)   &       &  (°)  &  \\ \hline 

1 & 2021-04-23 & 1.338 & 0.464 & 16.40 & $+$27.68 & 127.2 & 36.7 & $-$32.9 & A & $+$0 & 2 \\

\hline
\end{tabular}
\end{table}

\begin{figure}[h!]

    \centering
    \includegraphics[scale=0.368]{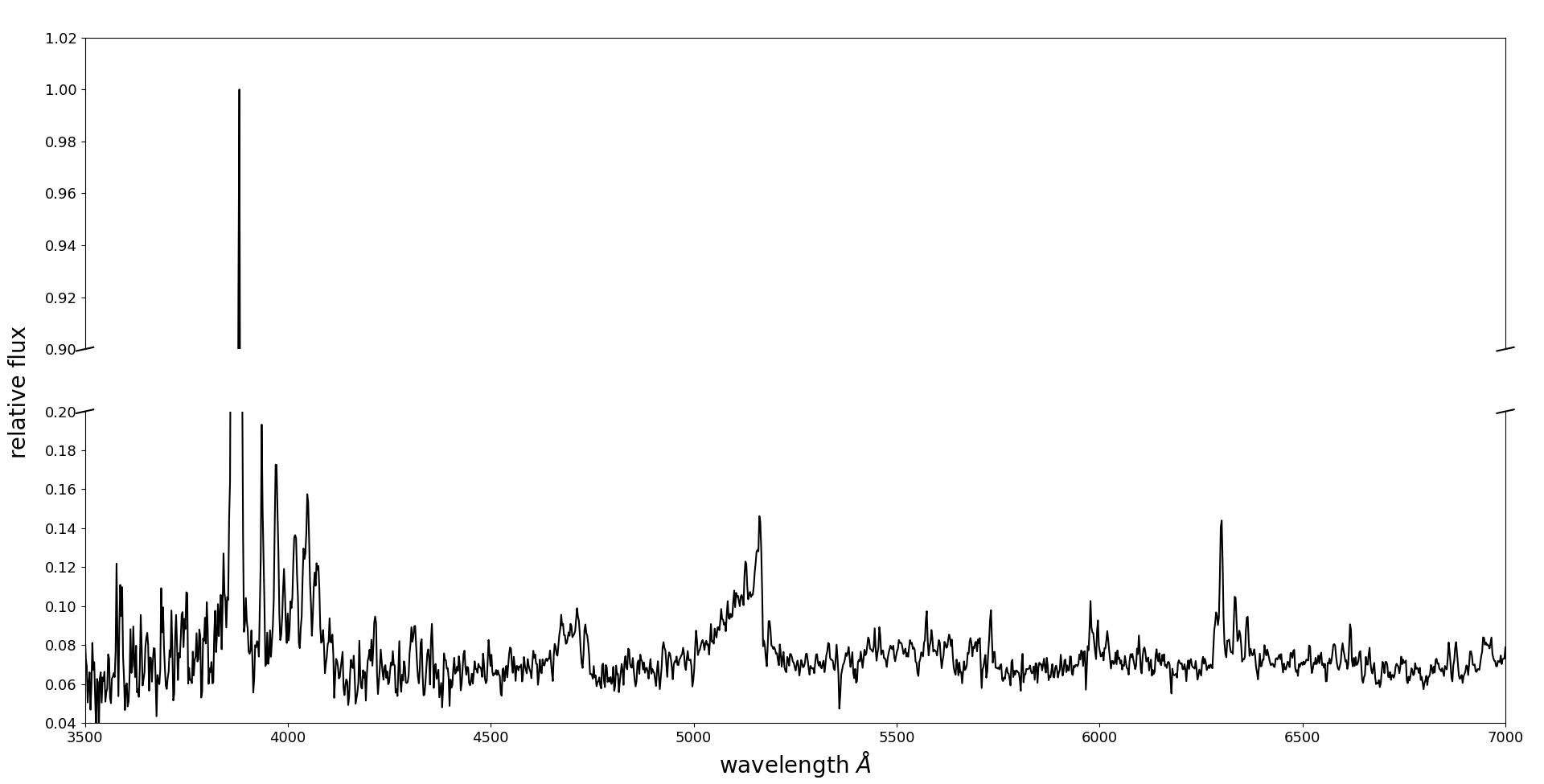}
    \caption{}
    
\end{figure}

\newpage
\clearpage

\section{C/2020 T2 (Palomar)}
\label{cometa:C2020T2}
\subsection{Description}

C/2020 T2 (Palomar) is a Long Period comet with a period of 5529 years and an absolute magnitude of 9.4$\pm$0.8.\footnote{\url{https://ssd.jpl.nasa.gov/tools/sbdb_lookup.html\#/?sstr=2020\%20T2} visited on July 20, 2024}
It was first spotted by Dimitry Duev using the 1.2m Mount Palomar Zwicky Transient Facility on October 7, 2020.
The Earth crossed the comet orbital plane on June 14, 2021.

\noindent
We observed the comet around visual magnitude 10.\footnote{\url{https://cobs.si/comet/2179/ }, visited on July 20, 2024}

\begin{table}[h!]
\centering
\begin{tabular}{|c|c|c|}
\hline
\multicolumn{3}{|c|}{Orbital elements (epoch: May 15, 2021)}                      \\ \hline \hline
\textit{e} = 0.9934 & \textit{q} = 2.0547 & \textit{T} = 2459406.6448 \\ \hline
$\Omega$ = 83.0484 & $\omega$ = 150.3815  & \textit{i} = 27.8729  \\ \hline  
\end{tabular}
\end{table}

\begin{table}[h!]
\centering
\begin{tabular}{|c|c|c|c|c|c|c|c|c|}
\hline
\multicolumn{9}{|c|}{Comet ephemerides for key dates}                      \\ \hline 
\hline
& date & r & $\Delta$ & RA & DEC & elong & phase & PLang \\
& (yyyy-mm-dd) & (AU) & (AU) & (h) & (°) & (°) & (°) & (°) \\ \hline 

Perihelion       & 2021-07-14 & 2.055  & 1.694 & 14.07 & $+$03.86  & 95.4  & 29.5  & $+$07.7  \\
Nearest approach & 2021-05-12 & 2.174  & 1.414 & 13.66 & $+$29.75  & 126.6 & 21.9  & $-$10.1 \\ \hline
\end{tabular}

\end{table}

\vspace{0.5 cm}

\begin{figure}[h!]
    \centering
    \includegraphics[scale=0.38]{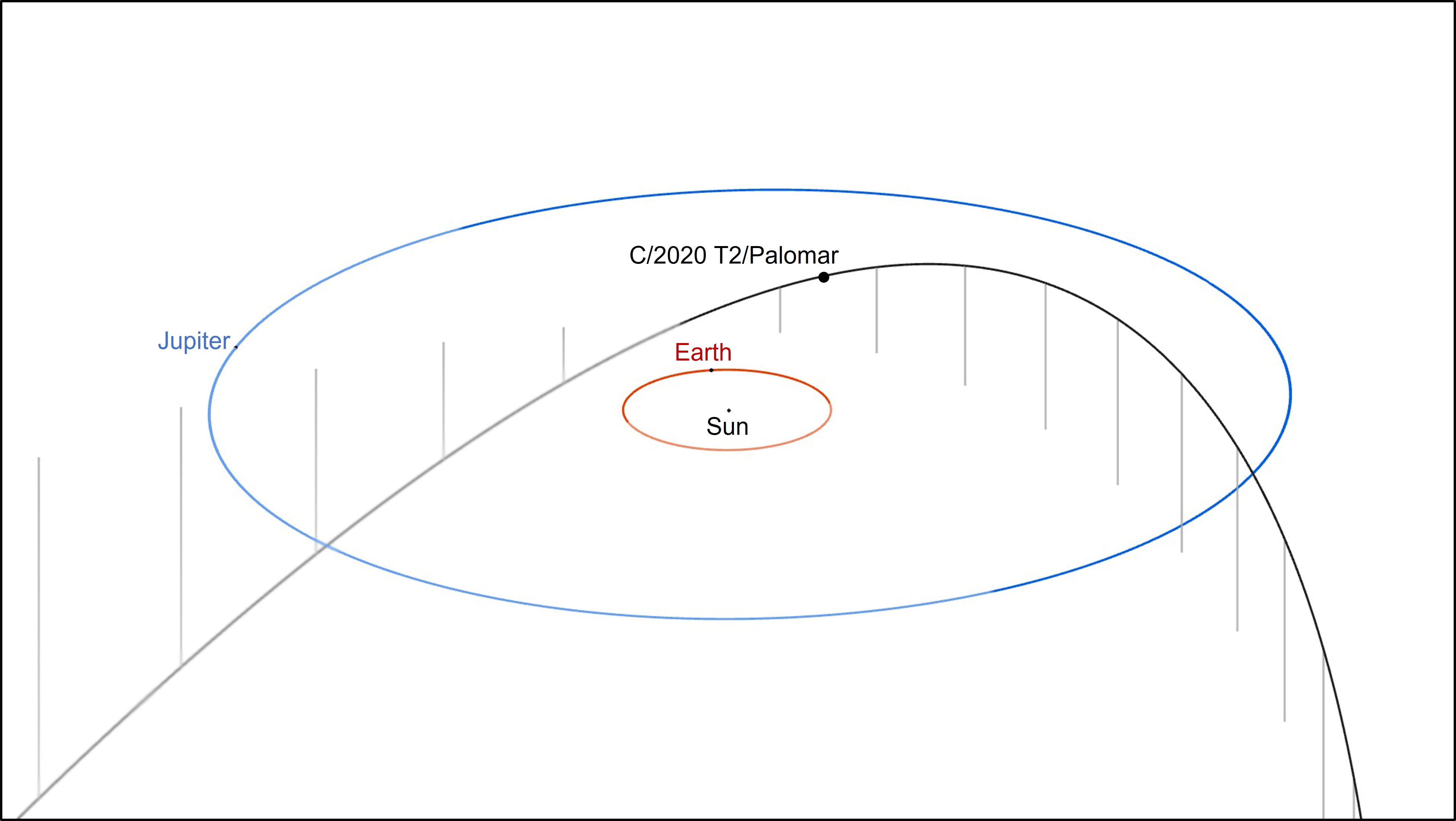}
    \caption{Orbit of comet C/2020 T2 and position on perihelion date. The field of view is set to the orbit of Jupiter for size comparison. Courtesy of NASA/JPL-Caltech.}
\end{figure}

\newpage

\subsection{Images}
\begin{SCfigure}[0.8][h!]
 \centering
 \includegraphics[scale=0.4]{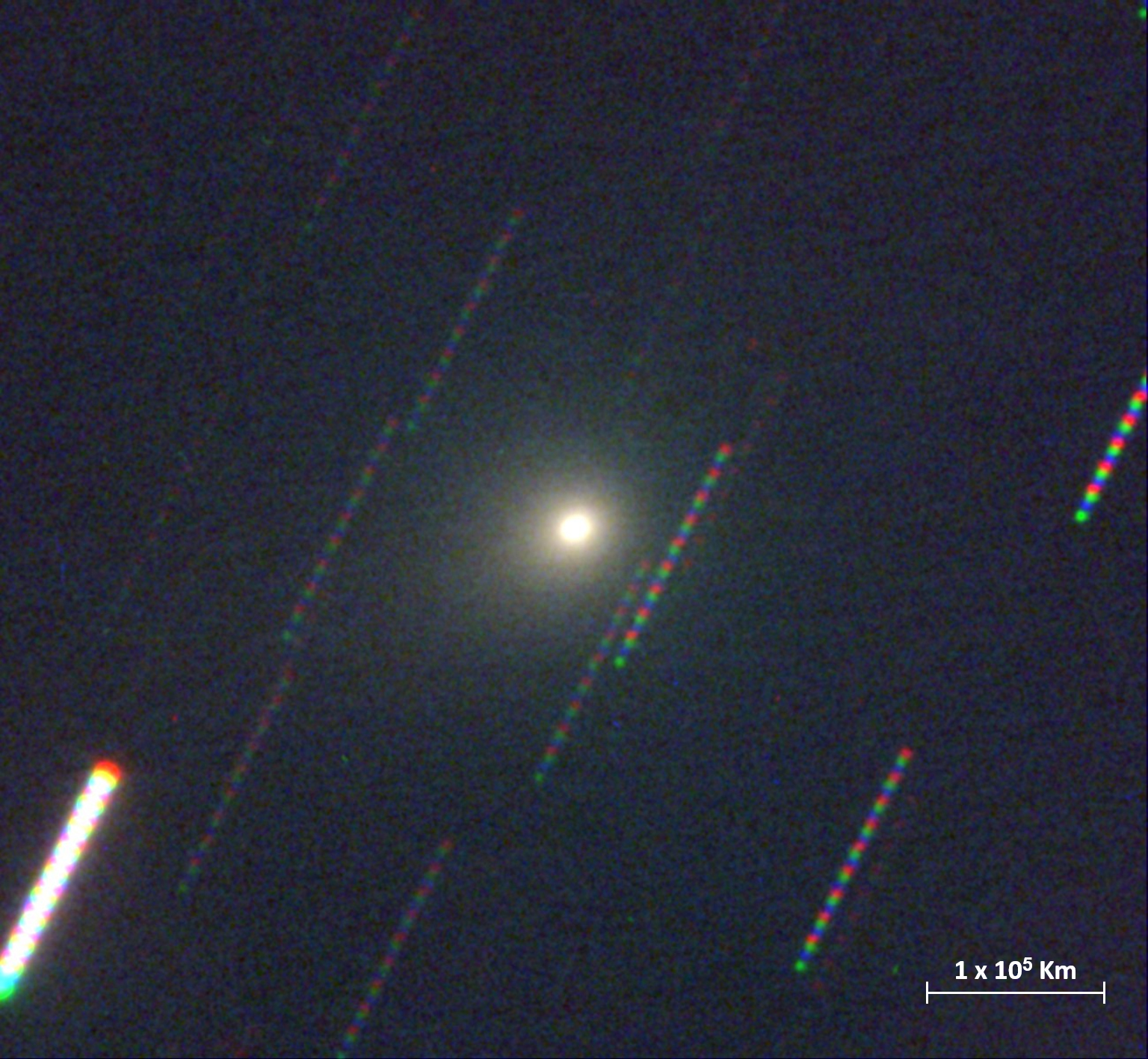}
 \caption{2021-07-05. Trichrome composite image taken with B, V, R filters at the Stazione Astronomica di Sozzago with the 0.4m Savonarola telescope.}
\end{SCfigure} 
\begin{SCfigure}[0.8][h!]
 \centering

 \includegraphics[scale=0.4]{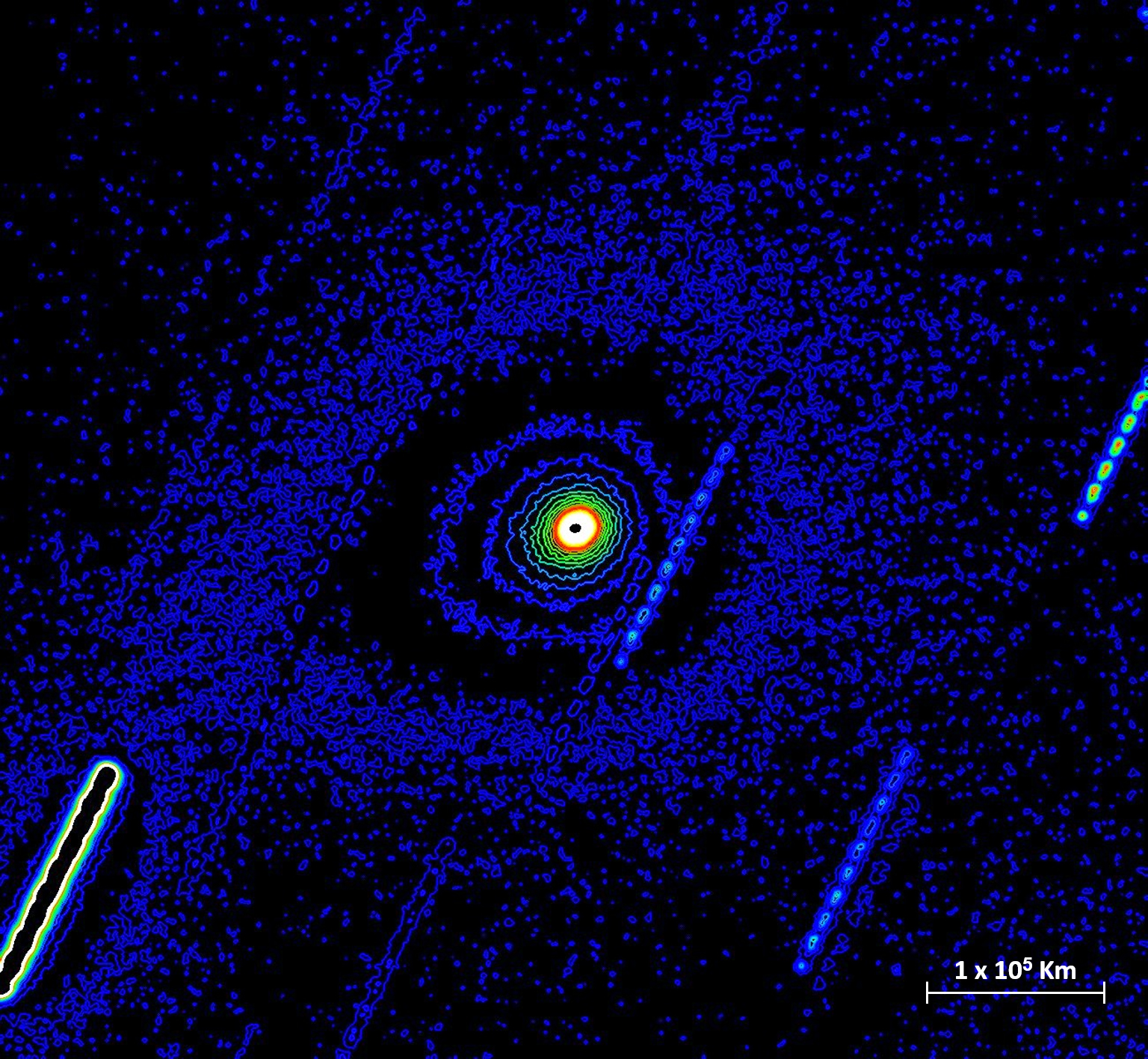}

 \caption{2021-07-05. Same image as above visualized in 500-ADU spaced isophotes.}
\end{SCfigure}
\newpage

\subsection{Spectra}

\begin{table}[h!]
\centering
\begin{tabular}{|c|c|c|c|c|c|c|c|c|c|c|c|}
\hline
\multicolumn{12}{|c|}{Observation details} \\ \hline 
\hline
$\#$ & date & r & $\Delta$ & RA & DEC & elong & phase & PLang& config & FlAng & N \\
 & (yyyy-mm-dd) & (AU) & (AU) & (h) & (°) & (°) & (°) & (°) & & (°) &  \\ \hline 

1   & 2021-03-31 & 2.376  & 1.548 & 14.05 & $+$34.25 & 136.7 & 16.8 & $-$16.7 & A & $+$0  & 2  \\
2   & 2021-06-03 & 2.102  & 1.453 & 13.60 & $+$22.00 & 116.3 & 25.7 & $-$03.3  & A & $+$0  & 3 \\
3   & 2021-06-10 & 2.086  & 1.478 & 13.63 & $+$19.08 & 112.2 & 26.8 & $-$01.1  & A & $+$0  & 1 \\
4   & 2021-07-30 & 2.058  & 1.588 & 13.83 & $+$10.08 & 102.1 & 28.8 &  $+$04.7  & A & $+$20 & 2 \\ \hline

\end{tabular}
\end{table}

\begin{figure}[h!]

    \centering
    \includegraphics[scale=0.368]{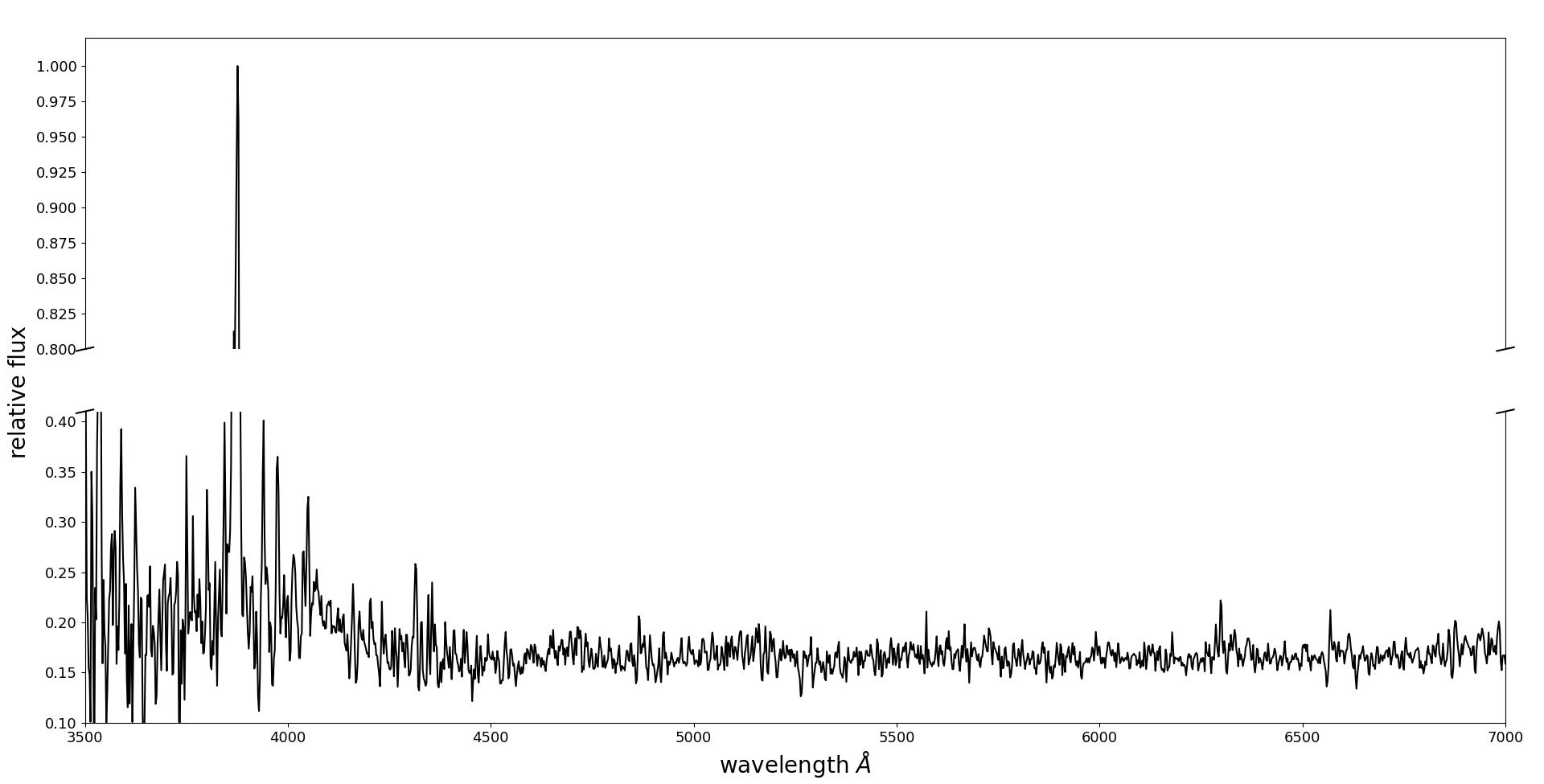}
    \caption{Spectrum of 2021-03-31; configuration A}

\end{figure}

\newpage
\clearpage

\section{C/2020 V2 (ZTF)}
\label{cometa:C2020V2}
\subsection{Description}

C/2020 V2 (ZTF) is a hyperbolic comet with an absolute magnitude of 8.7$\pm$0.8.\footnote{\url{https://ssd.jpl.nasa.gov/tools/sbdb_lookup.html\#/?sstr=2020\%20V2} visited on July 20, 2024}
It was first spotted by the 1.2m Mount Palomar Zwicky Transient Facility on November 2, 2020.

\noindent
We observed the comet at magnitude 10.\footnote{\url{https://cobs.si/comet/2194/} visited on July 21, 2024}
The Earth crossed the comet orbital plane on October 26, 2023.

\begin{table}[h!]
\centering
\begin{tabular}{|c|c|c|}
\hline
\multicolumn{3}{|c|}{Orbital elements (epoch: July 16, 2022)}                      \\ \hline \hline
\textit{e} = 1.0010 & \textit{q} = 2.2281 & \textit{T} = 2460073.0379 \\ \hline
$\Omega$ = 212.3694 & $\omega$ = 162.4155  & \textit{i} = 131.6109  \\ \hline  
\end{tabular}
\end{table}

\begin{table}[h!]
\centering
\begin{tabular}{|c|c|c|c|c|c|c|c|c|}
\hline
\multicolumn{9}{|c|}{Comet ephemerides for key dates}                      \\ \hline 
\hline
& date         & r    & $\Delta$  & RA      & DEC      & elong  & phase  & PLang  \\
& (yyyy-mm-dd) & (AU) & (AU)      & (h)     & (°)      & (°)    & (°)    & (°) \\ \hline 

Perihelion       & 2023-05-05 & 2.281 & 3.216 & 2.62 & $+$25.50 & 09.7 & 04.4 & $+$02.7  \\ 
Nearest approach & 2023-09-17 & 2.683 & 1.855 & 2.19 & $-$23.10 & 137.5 & 14.7 & $+$14.5  \\ 
\hline
\end{tabular}

\end{table}

\vspace{0.5 cm}

\begin{figure}[h!]
    \centering
    \includegraphics[scale=0.38]{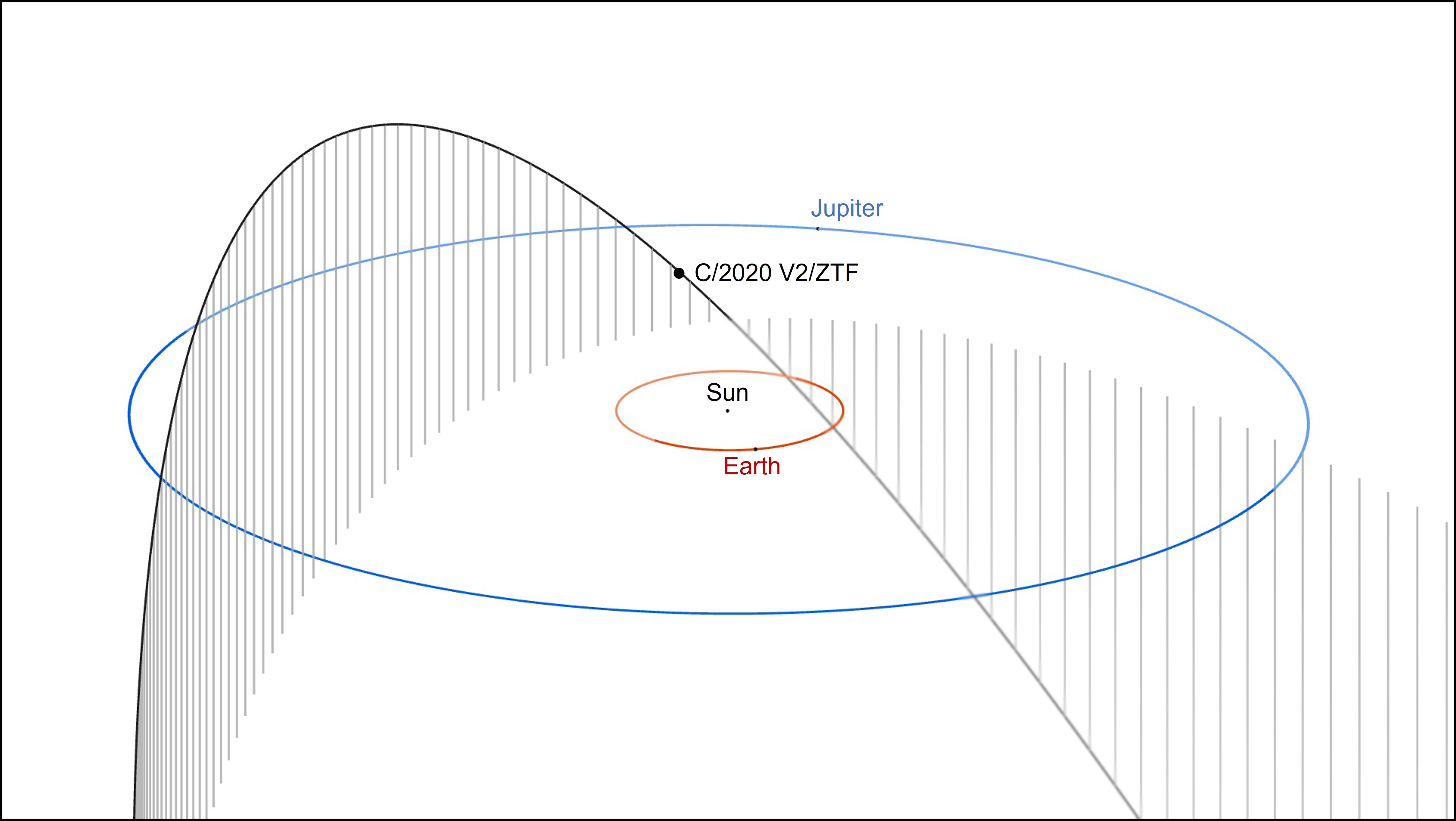}
    \caption{Orbit of comet C/2020 V2 and position on perihelion date. The field of view is set to the orbit of Jupiter for size comparison. Courtesy of NASA/JPL-Caltech.}
\end{figure}

\newpage

\subsection{Images}

\begin{SCfigure}[0.8][h!]
    \centering
    \includegraphics[scale=0.4]{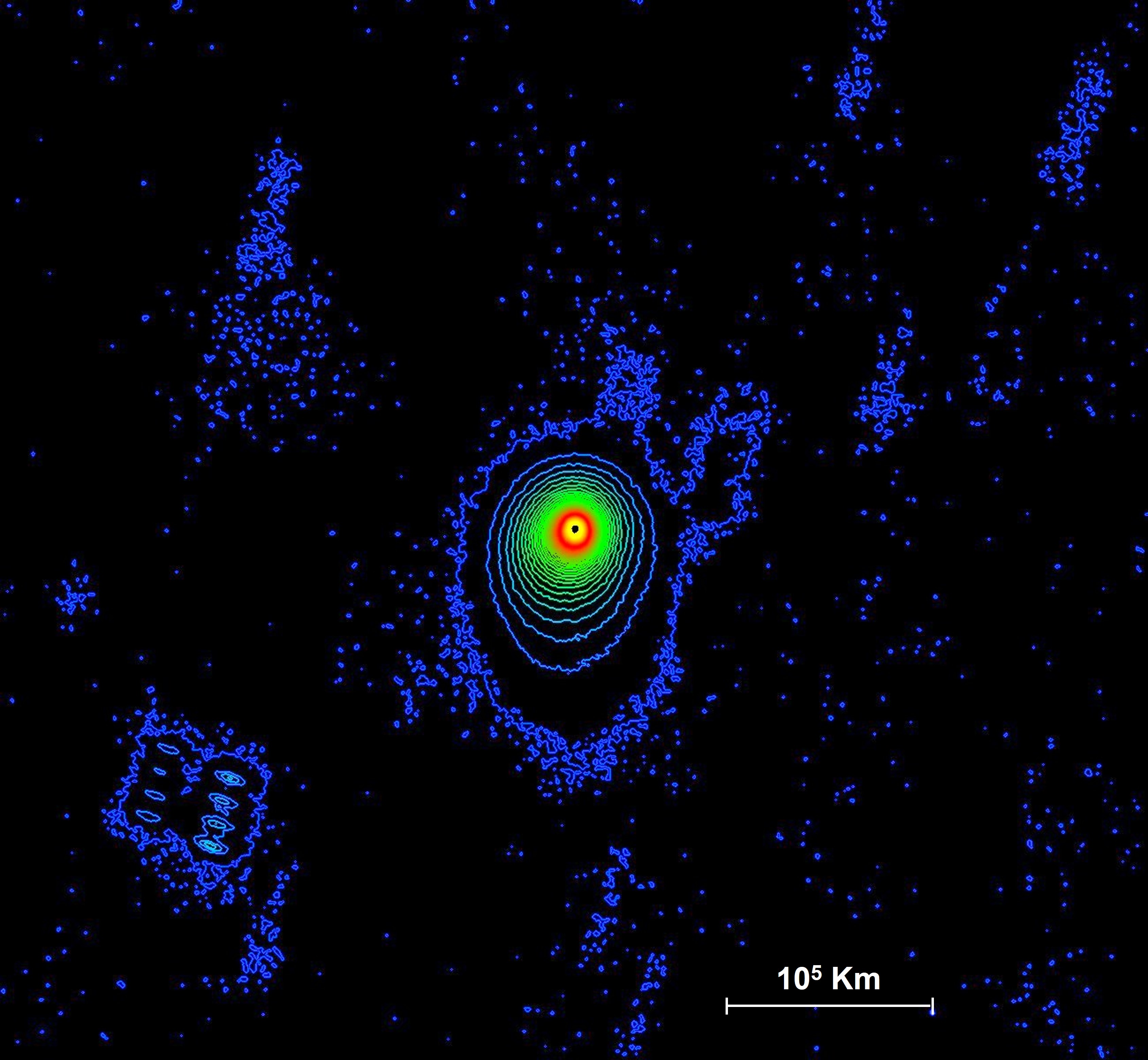}
     \caption{2022-01-25. Image taken with the Asiago Copernico telescope visualized with isophotes, which show a non-uniform coma. The position of the cometary nucleus coincides with the black dot The elongated trails are due to field stars that moved during the exposure.}
\end{SCfigure} 

\begin{SCfigure}[0.8][h!]
    \centering
    \includegraphics[scale=0.4]{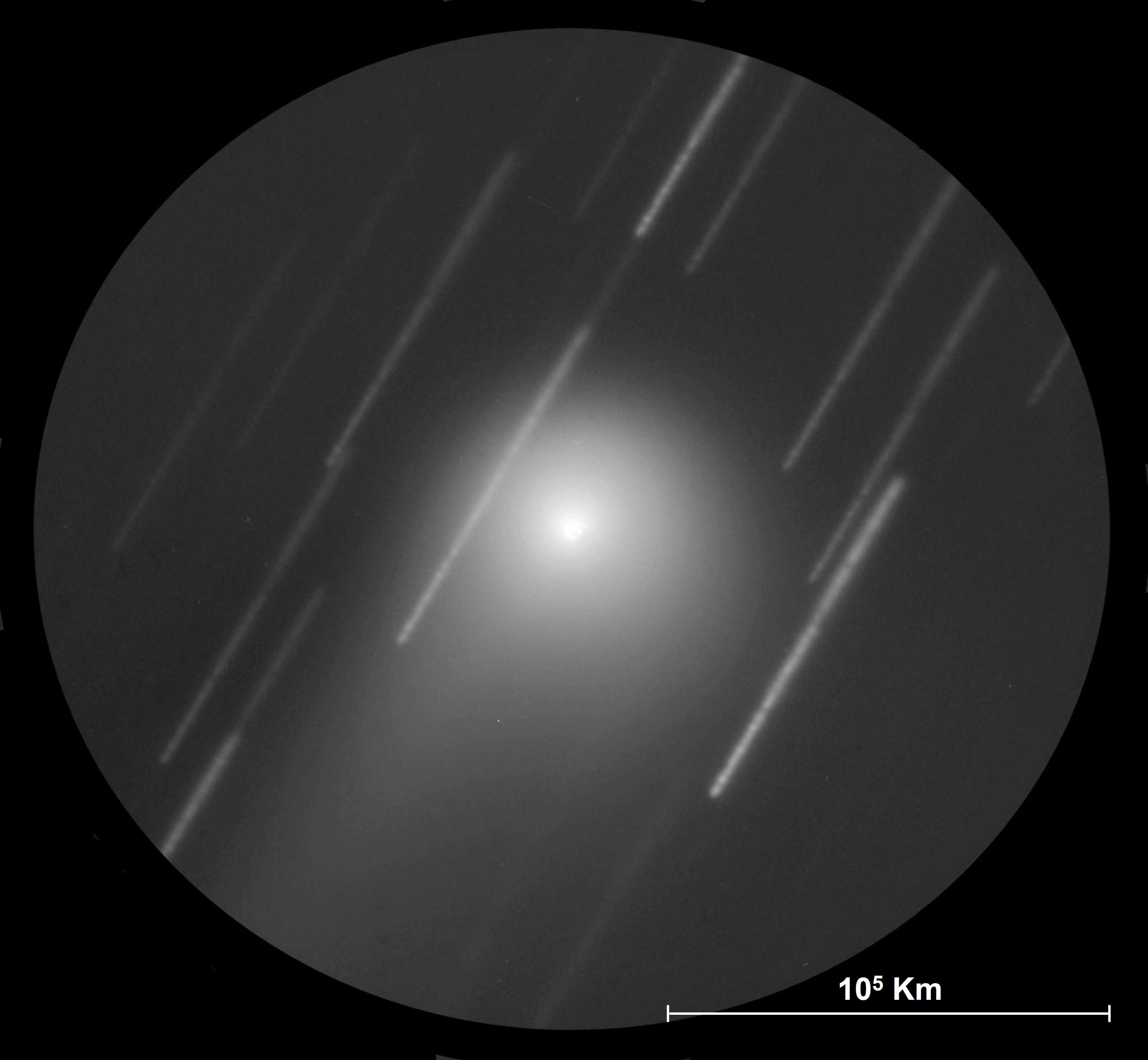}
    \caption{2022-12-14. Comet C/2020 V2 (ZTF) is approaching perihelion: continuous insulation causes it to develop a large coma and its tail to lengthen in an antisolar direction. Image taken with the Asiago Copernico telescope. The use of an r filter highlights the signal due to dust.}
\end{SCfigure}

\newpage

\subsection{Spectra}

\begin{table}[h!]
\centering
\begin{tabular}{|c|c|c|c|c|c|c|c|c|c|c|c|}
\hline
\multicolumn{12}{|c|}{Observation details}                      \\ \hline 
\hline
$\#$  & date          & r     & $\Delta$ & RA     & DEC     & elong & phase & PLang& config  & FlAng & N \\
      & (yyyy-mm-dd)  &  (AU) & (AU)     & (h)    & (°)     & (°)   & (°)   &  (°)   &       &  (°)  &  \\ \hline 

1 & 2022-10-20 & 3.145 & 3.236 & 11.02 & $+$57.62 & 75.9 & 17.9 & $+$01.1 & A & $-$0 & 4 \\
\hline
\end{tabular}
\end{table}

\begin{figure}[h!]

    \centering
    \includegraphics[scale=0.368]{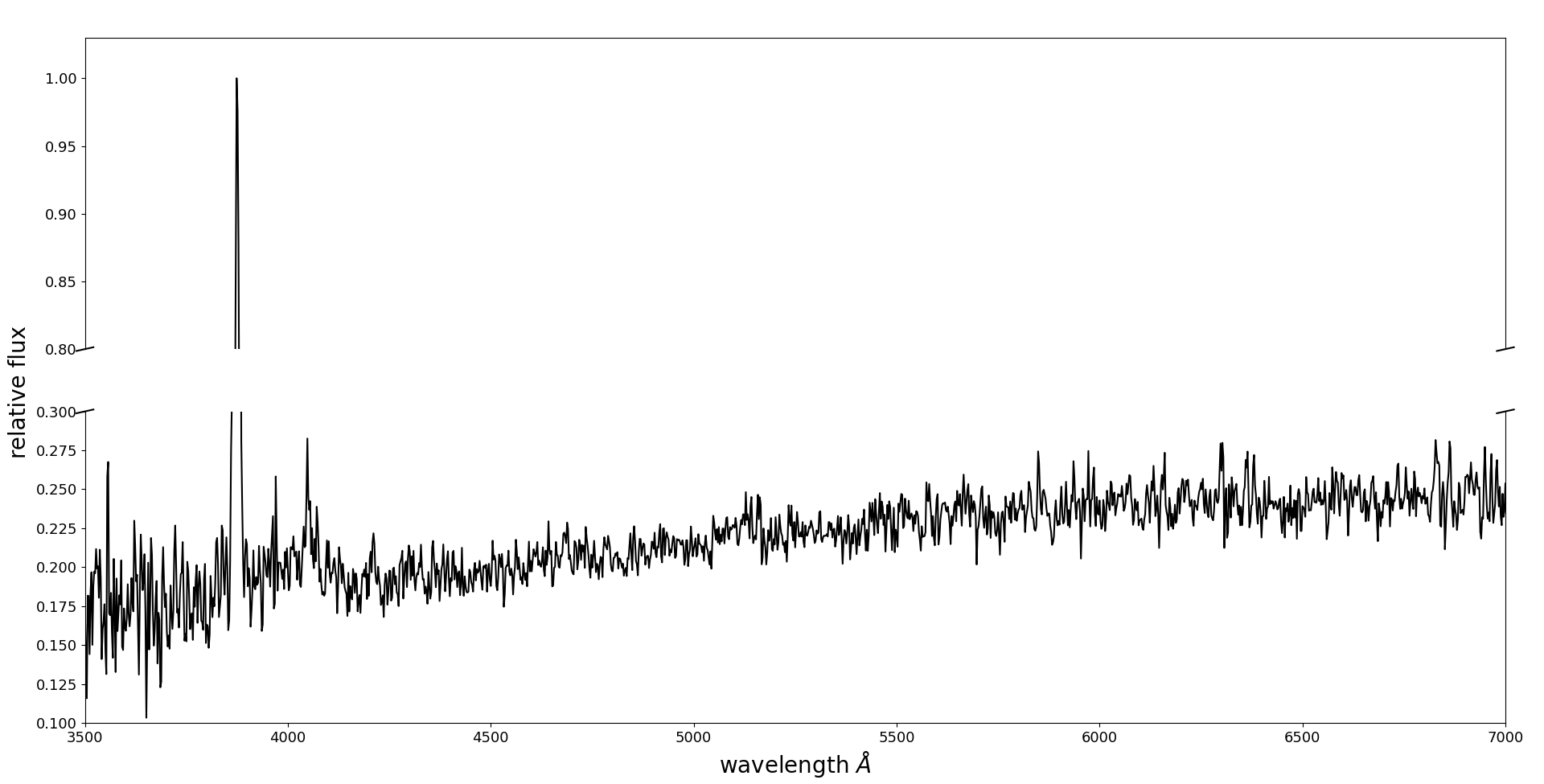}

\end{figure}

\newpage
\clearpage

\section{C/2021 X1 (Maury-Attard)}
\label{cometa:C2021X1}
\subsection{Description}

C/2021 X1 (Maury-Attard) is a hyperbolic comet with an absolute magnitude of 9.5$\pm$0.7.\footnote{\url{https://ssd.jpl.nasa.gov/tools/sbdb_lookup.html\#/?sstr=2021\%20X1} visited on July 21, 2024}
It was first spotted by Alain Maury with a 0.28m reflector at the San Pedro de Atacama Observatory, Chile, on December 2, 2021.
Initially, the provisional designation A/2021 X1 was assigned.

\noindent
We observed the comet around magnitude 14.\footnote{\url{https://cobs.si/comet/2328/ }, visited on July 21, 2024}
The Earth crossed the comet orbital plane on October 4, 2023.

\begin{table}[h!]
\centering
\begin{tabular}{|c|c|c|}
\hline
\multicolumn{3}{|c|}{Orbital elements (epoch: June 10, 2023)}                      \\ \hline \hline
\textit{e} = 1.0012 & \textit{q} = 3.2336 & \textit{T} = 2460091.8822 \\ \hline
$\Omega$ = 10.5886 & $\omega$ = 334.6106  & \textit{i} = 140.1171 \\ \hline  
\end{tabular}
\end{table}

\begin{table}[h!]
\centering
\begin{tabular}{|c|c|c|c|c|c|c|c|c|}
\hline
\multicolumn{9}{|c|}{Comet ephemerides for key dates}                      \\ \hline 
\hline
& date         & r    & $\Delta$  & RA      & DEC      & elong  & phase  & PLang  \\
& (yyyy-mm-dd) & (AU) & (AU)      & (h)     & (°)      & (°)    & (°)    & (°) \\ \hline 

Perihelion       & 2023-05-27 & 3.234 & 4.079 & 02.71 & $+$02.55 & 29.3 & 08.8 & $-$07.5  \\ 
Nearest approach & 2023-09-25 & 3.433 & 2.441 & 00.64 & $+$06.61 & 170.1 & 02.9 & $-$02.3\\ \hline
\end{tabular}

\end{table}

\vspace{0.5 cm}

\begin{figure}[h!]
    \centering
    {\includegraphics[scale=0.38]{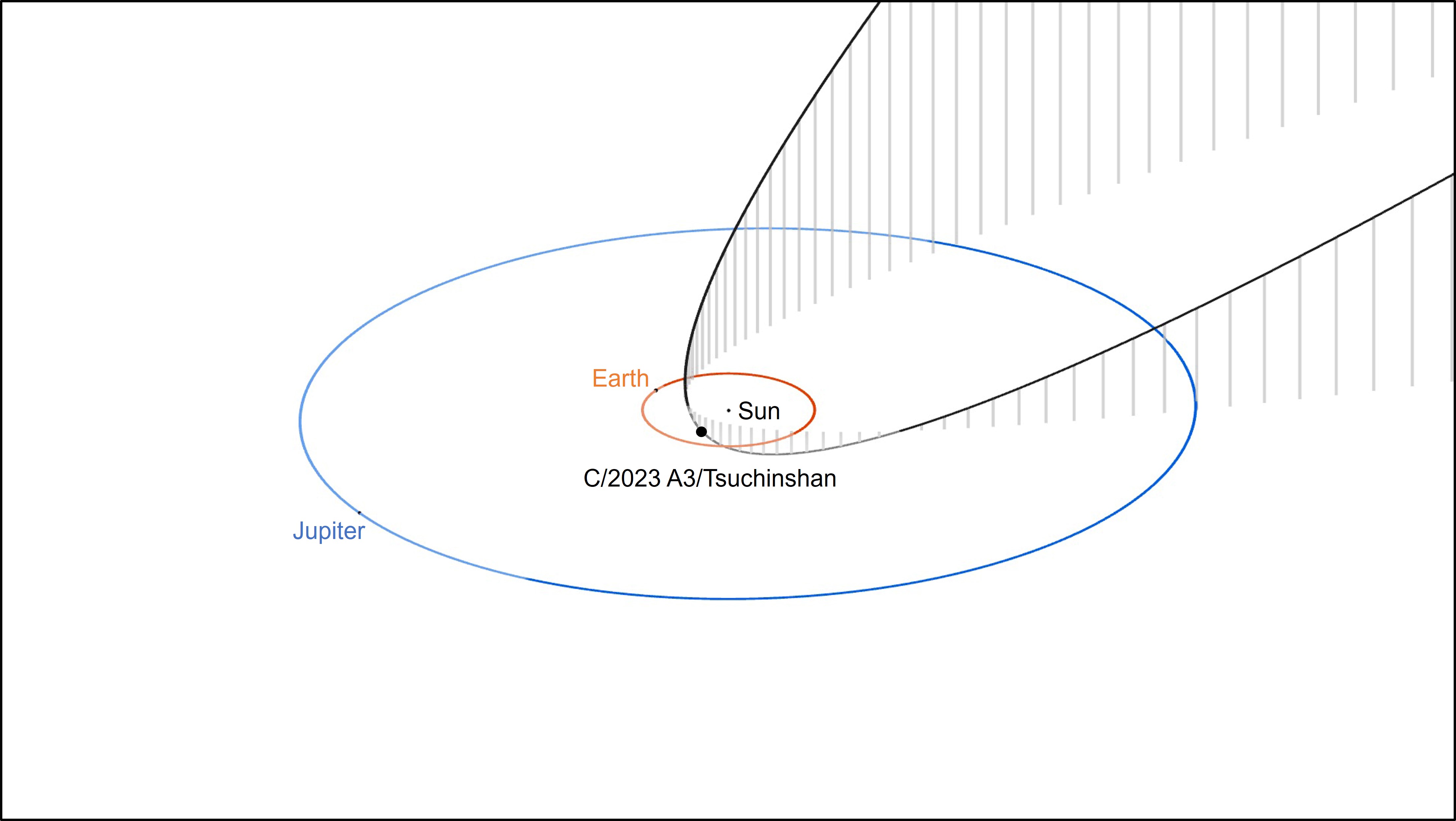}}
    \caption{Orbit of comet C/2021 X1 and position on perihelion date. The field of view is set to the orbits of Jupiter and Uranus for size comparison. Courtesy of NASA/JPL-Caltech.}
\end{figure}

\newpage

\subsection{Images}

\begin{SCfigure}[0.8][h!]
    \centering
    \includegraphics[scale=0.4]{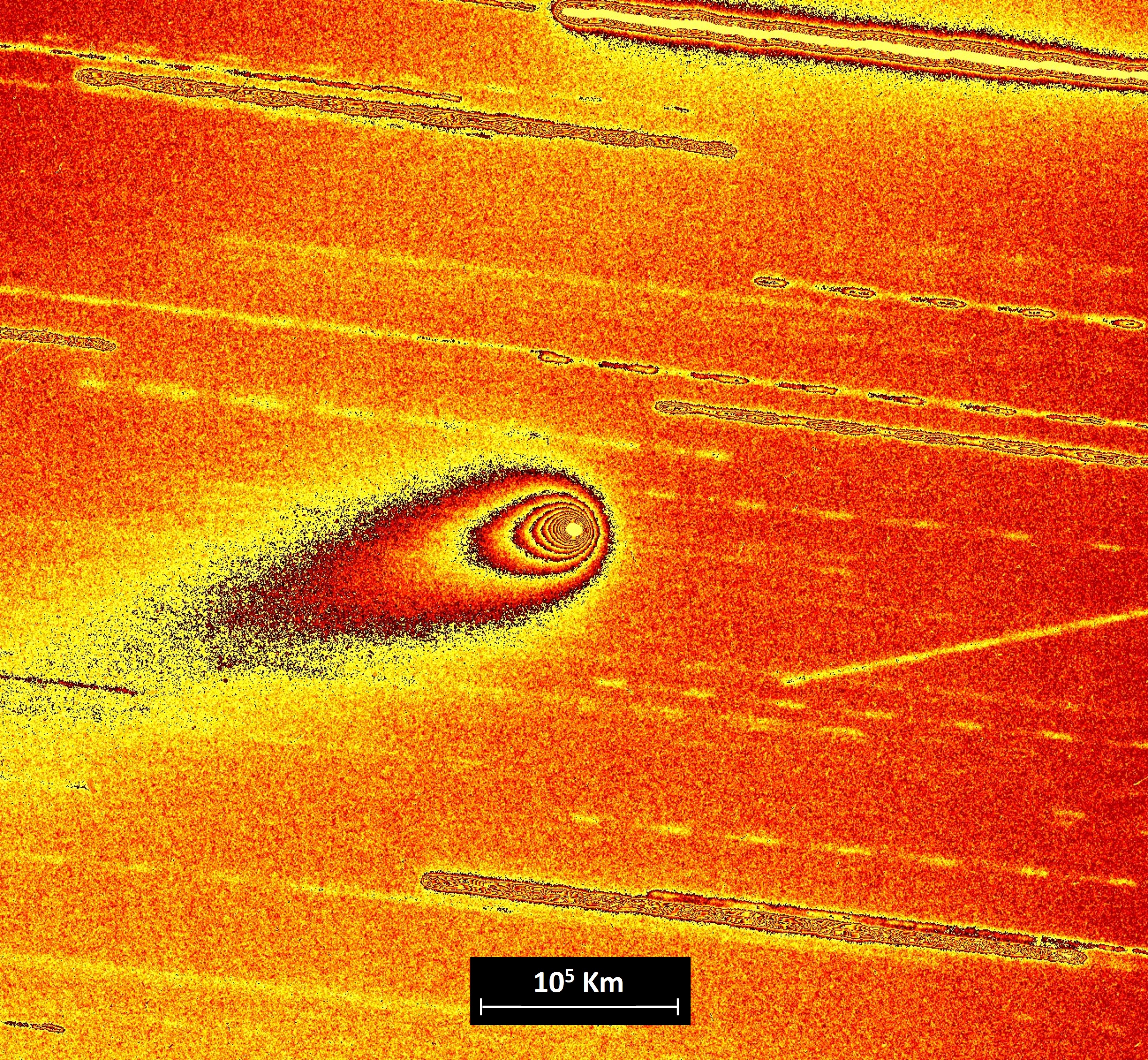}
     \caption{2023-10-06. Comet C/2021 X1 imaged at the Asiago Copernico telescope with r and i filters. At that epoch, comet Maury-Attard was 2.482 AU from Earth and 3.471 AU from the Sun. The color saw-tooth visualization highlights the typical characteristics of a comet: small nucleus inside the coma and tail that develops in an anti-solar direction.}

\end{SCfigure} 

\begin{SCfigure}[0.8][h!]
    \centering
    \includegraphics[scale=0.4]{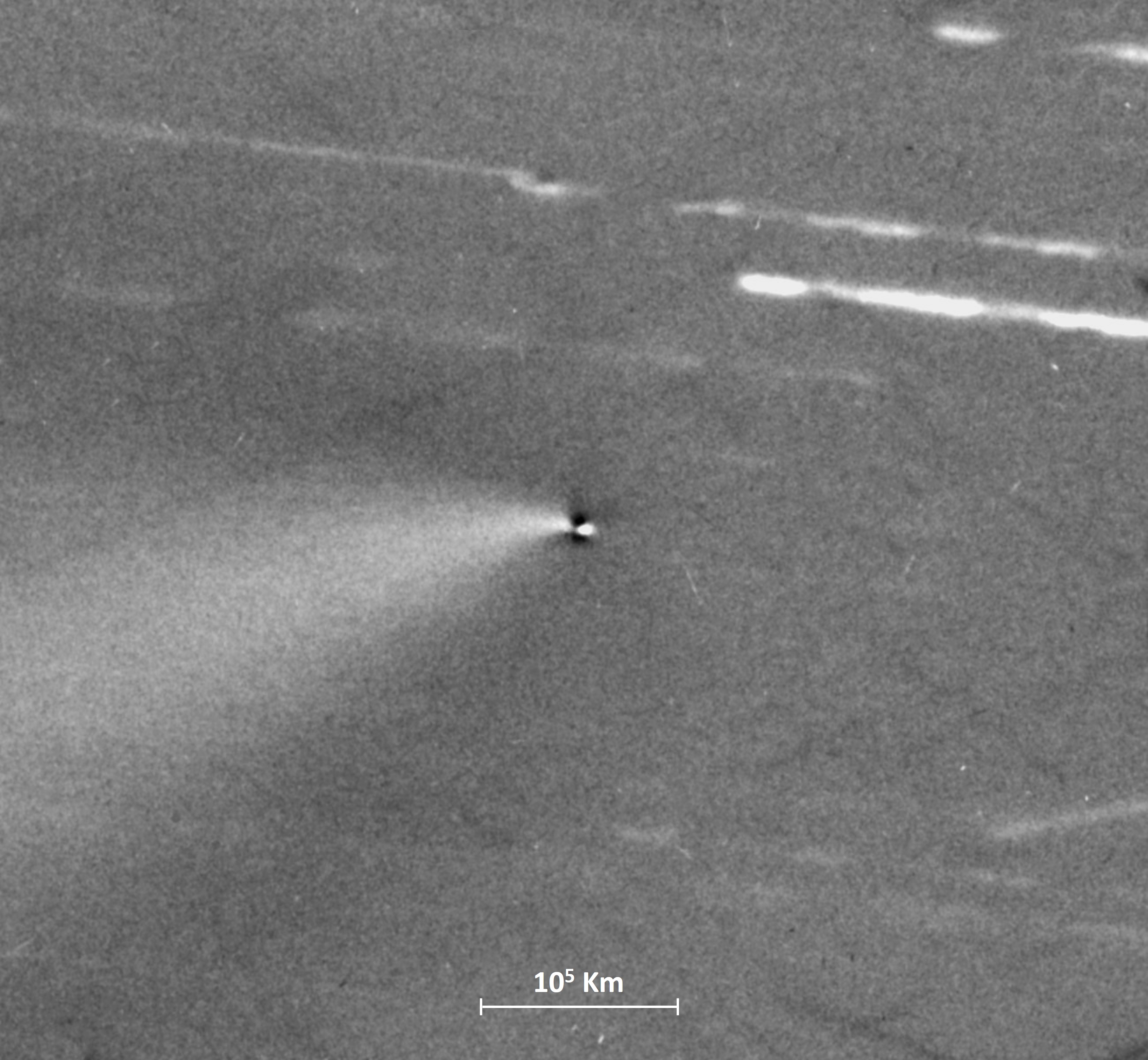}
    \caption{Processing the same image with the Larson-Sekanina filter allows to detect the presence of a prominent jet emitted from the nucleus towards PA 270°. The tail does not seem to "leave" the nucleus in an antisolar direction at a short distance from it.}

\end{SCfigure}

\newpage

\subsection{Spectra}

\begin{table}[h!]
\centering
\begin{tabular}{|c|c|c|c|c|c|c|c|c|c|c|c|}
\hline
\multicolumn{12}{|c|}{Observation details}                      \\ \hline 
\hline
$\#$  & date          & r     & $\Delta$ & RA     & DEC     & elong & phase & PLang & config  & FlAng & N \\
      & (yyyy-mm-dd)  &  (AU) & (AU)     & (h)    & (°)     & (°)   & (°)   &  (°)   &       &  (°)  &  \\ \hline 
1 & 2023-09-23 & 3.429 & 2.441 & 00.70 & $+$6.72 & 167.8 & 03.5 & $-$02.7 & A & $+$0 & 2 \\

\hline
\end{tabular}
\end{table}

\begin{figure}[h!]
    \centering
    \includegraphics[scale=0.58]{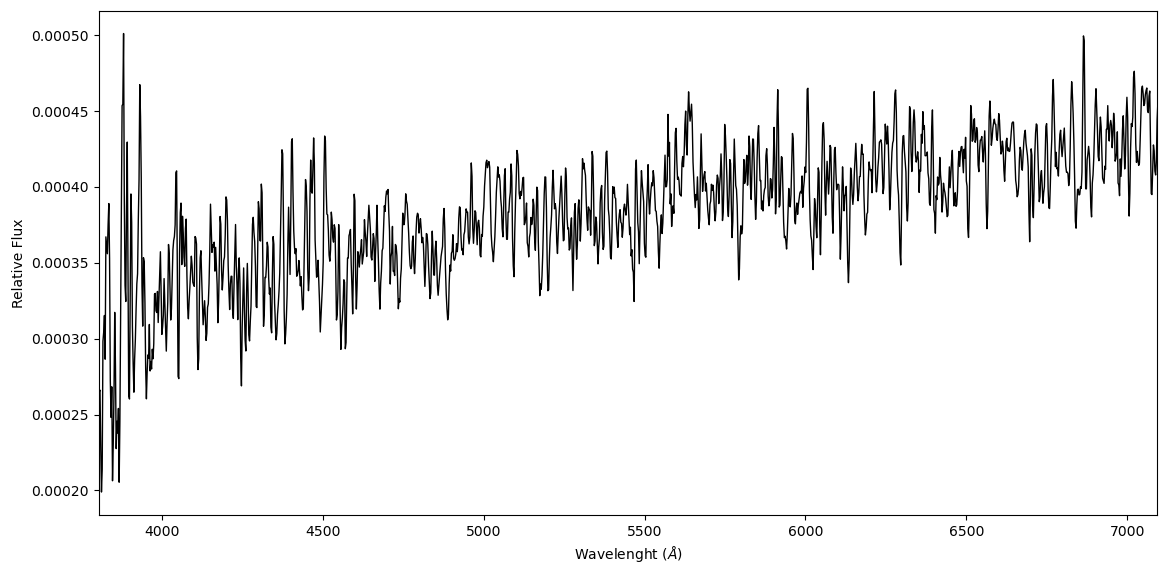}
    \caption{Spectrum of 2023-09-23; configuration A.}
\end{figure}

\newpage
\clearpage

\section{C/2022 E2 (ATLAS)}
\label{cometa:C2022E2}
\subsection{Description}

C/2022 E2 (ATLAS) is a hyperbolic comet with an absolute magnitude of 7.5$\pm$0.5.\footnote{\url{https://ssd.jpl.nasa.gov/tools/sbdb_lookup.html\#/?sstr=2022\%20E2} visited on July 21, 2024}
It was first spotted by the 0.5m Asteroid Terrestrial-impact Last Alert System (ATLAS) on March 7, 2022.

\noindent
We observed the comet around magnitude 13.\footnote{\url{https://cobs.si/comet/2322/ }, visited on July 21, 2024}
The Earth crossed the comet orbital plane on July 26, 2022 and on January 23, 2023.

\begin{table}[h!]
\centering
\begin{tabular}{|c|c|c|}
\hline
\multicolumn{3}{|c|}{Orbital elements (epoch: September 1, 2023)}                      \\ \hline \hline
\textit{e} = 1.0009 & \textit{q} = 3.6667 & \textit{T} = 2460567.5120 \\ \hline
$\Omega$ = 125.3688 & $\omega$ = 41.6940  & \textit{i} = 137.12349 \\ \hline  
\end{tabular}
\end{table}

\begin{table}[h!]
\centering
\begin{tabular}{|c|c|c|c|c|c|c|c|c|}
\hline
\multicolumn{9}{|c|}{Comet ephemerides for key dates}                      \\ \hline 
\hline
& date         & r    & $\Delta$  & RA      & DEC      & elong  & phase  & PLang  \\
& (yyyy-mm-dd) & (AU) & (AU)      & (h)     & (°)      & (°)    & (°)    & (°) \\ \hline 

Perihelion       & 2024-09-07 & 3.667 & 4.069 & 07.61 & $+$45.34 & 59.8 & 13.7 & $-$06.1  \\ 
Nearest approach & 2024-12-01 & 3.733 & 2.986 & 04.75 & $+$68.38 & 133.4 & 11.1 & $-$10.6\\ \hline
\end{tabular}

\end{table}

\vspace{0.5 cm}

\begin{figure}[h!]
    \centering
    \includegraphics[scale=0.38]{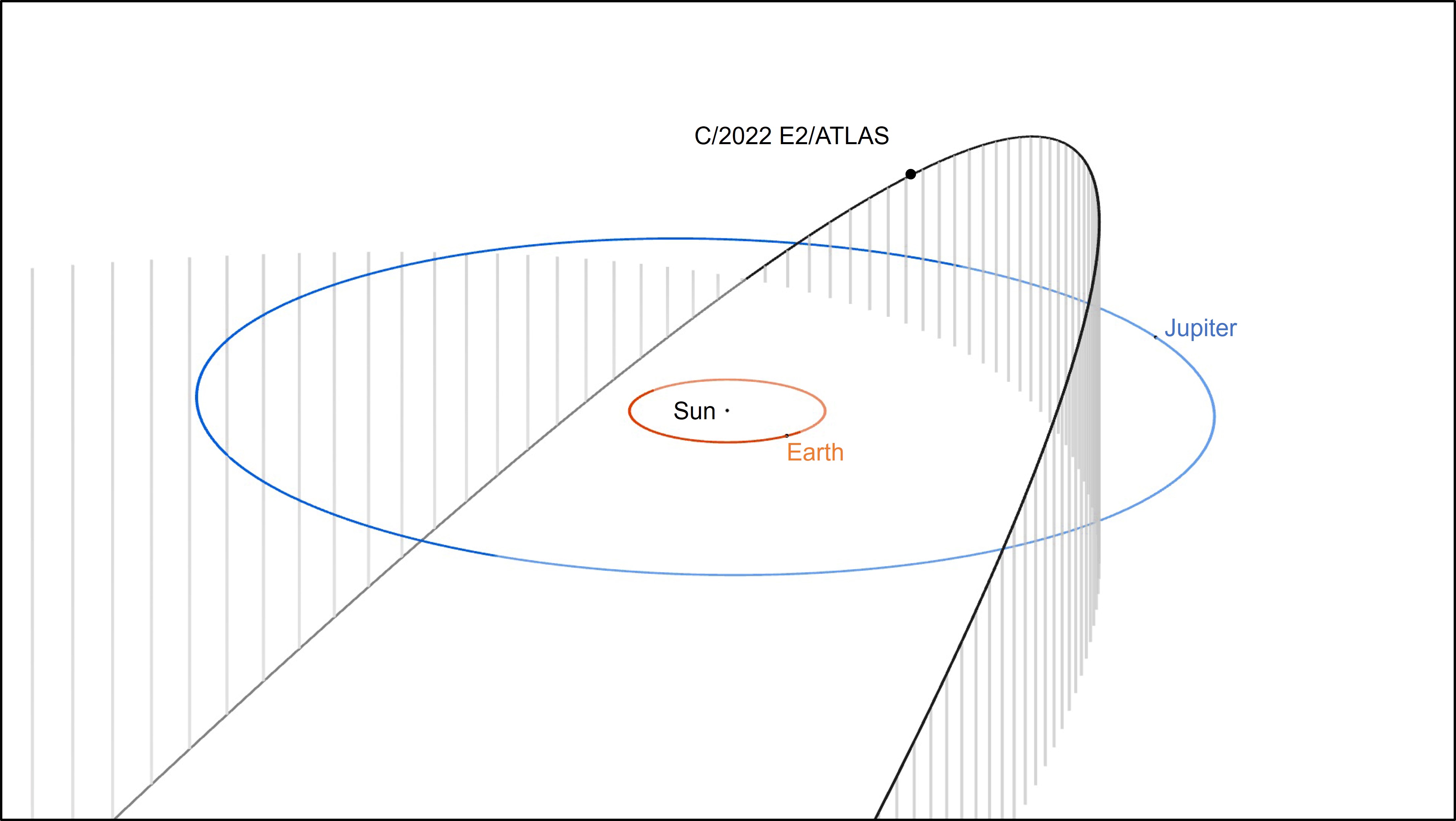}
    \caption{Orbit of comet C/2022 E2 and position on perihelion date. The field of view is set to the orbit of Jupiter for size comparison. Courtesy of NASA/JPL-Caltech.}
\end{figure}

\newpage

\subsection{Images}

\begin{SCfigure}[0.8][h!]
    \centering
    \includegraphics[scale=0.4]{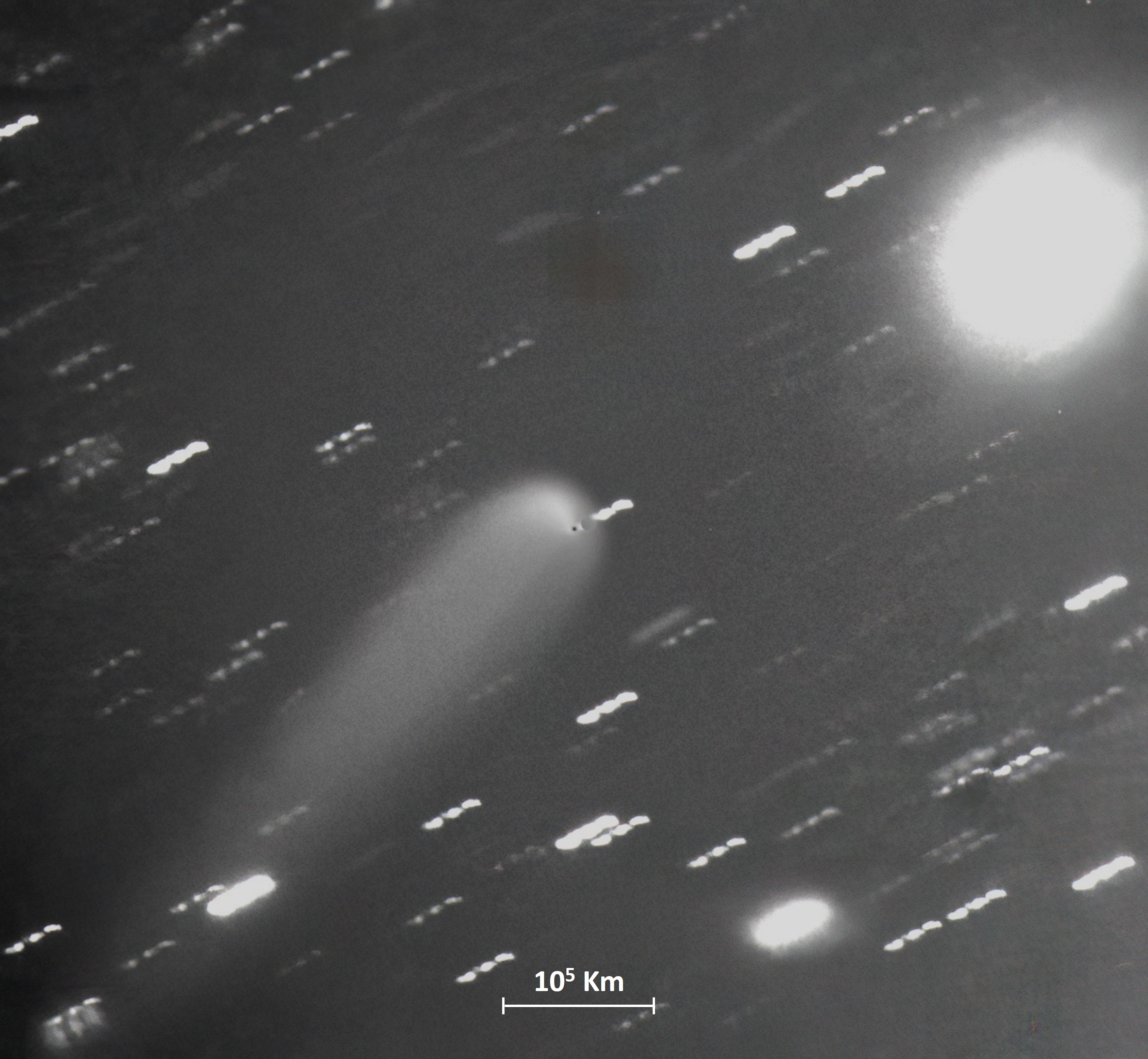}
     \caption{2024-02-14. Comet C/2022 E2 imaged at the Asiago Copernico telescope with r filter. At that epoch, comet ATLAS was 3.231 AU from Earth and 4.123 AU from the Sun. The tail develops in a South-East direction. Processing it with a 1/$\rho$ attenuating filter highlights the presence of a marked morphology in the inner coma.
}

\end{SCfigure} 

\begin{SCfigure}[0.8][h!]
    \centering
    \includegraphics[scale=0.4]{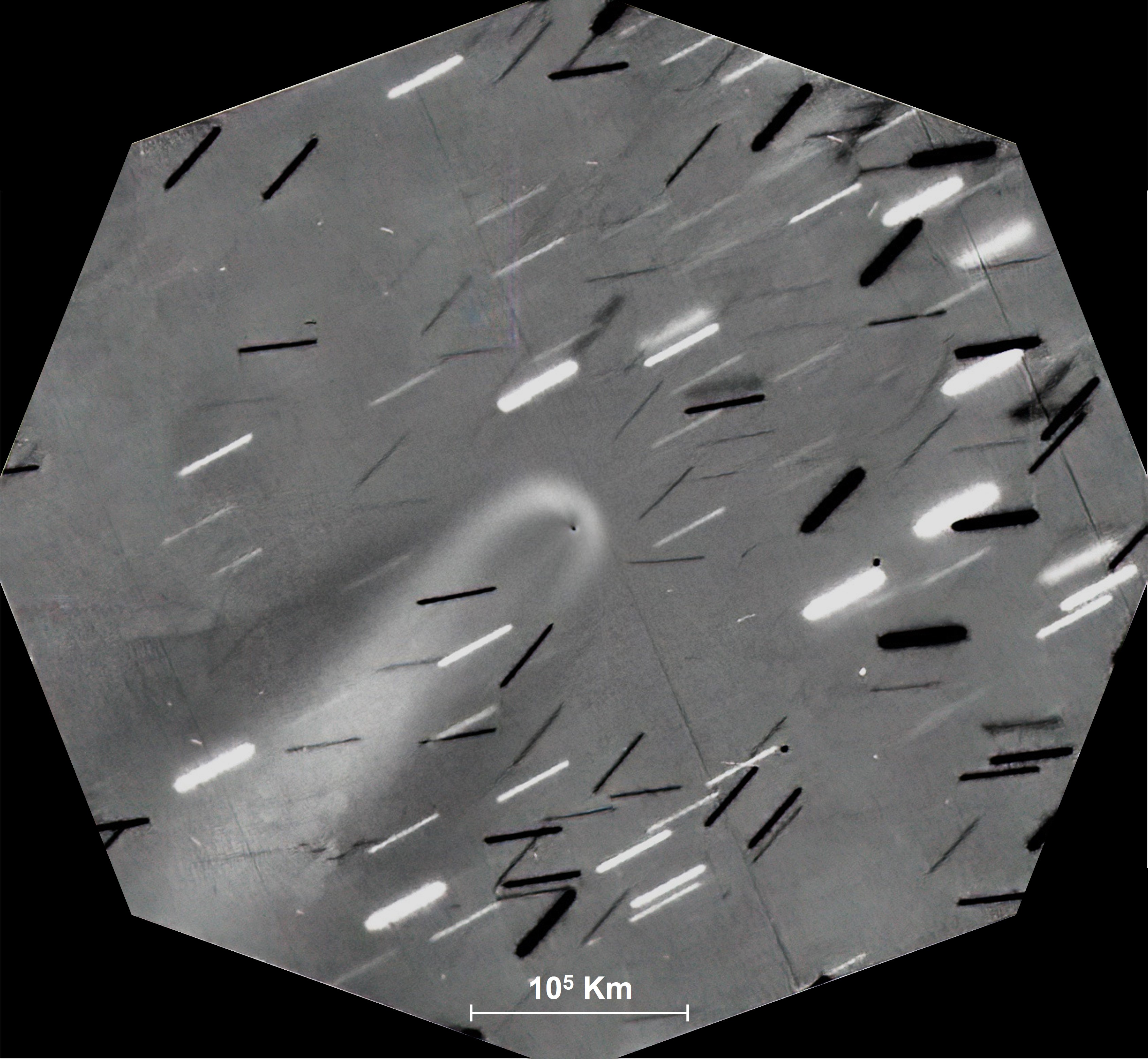}
    \caption{2024-02-15. Processing with the Larson-Sekanina filter of images of February 15, 2024, again taken with the Asiago Copernico telescope, highlights a long curvilinear structure (North of the nucleus), which then merges into the tail. The white streaks are due to the stars, the dark streaks are generated by the use of the L-S spatial filter.
}

\end{SCfigure}

\newpage

\subsection{Spectra}

\begin{table}[h!]
\centering
\begin{tabular}{|c|c|c|c|c|c|c|c|c|c|c|c|}
\hline
\multicolumn{12}{|c|}{Observation details}                      \\ \hline 
\hline
$\#$  & date          & r     & $\Delta$ & RA     & DEC     & elong & phase & PLang & config  & FlAng & N \\
      & (yyyy-mm-dd)  &  (AU) & (AU)     & (h)    & (°)     & (°)   & (°)   &  (°)   &       &  (°)  & \\ \hline 
1 & 2024-02-14 & 4.123 & 3.231 & 07.95 & $+$23.23 & 151.1 & 06.6 & $+$04.2 & A & $-$0 & 5 \\
2 & 2024-12-23 & 1.851 & 2.559 & 2.74 & $+$65.43 & 35.65 & 18.05 & 13.7 & A & $-$90 & 3 \\
3 & 2024-12-25 & 1.88 & 2.601 & 2.62 & $+$64.97 & 34.86 & 17.4 & 13.63 & A & $+$0 & 7 \\

\hline
\end{tabular}
\end{table}

\begin{figure}[h!]
    \centering
    \includegraphics[scale=0.58]{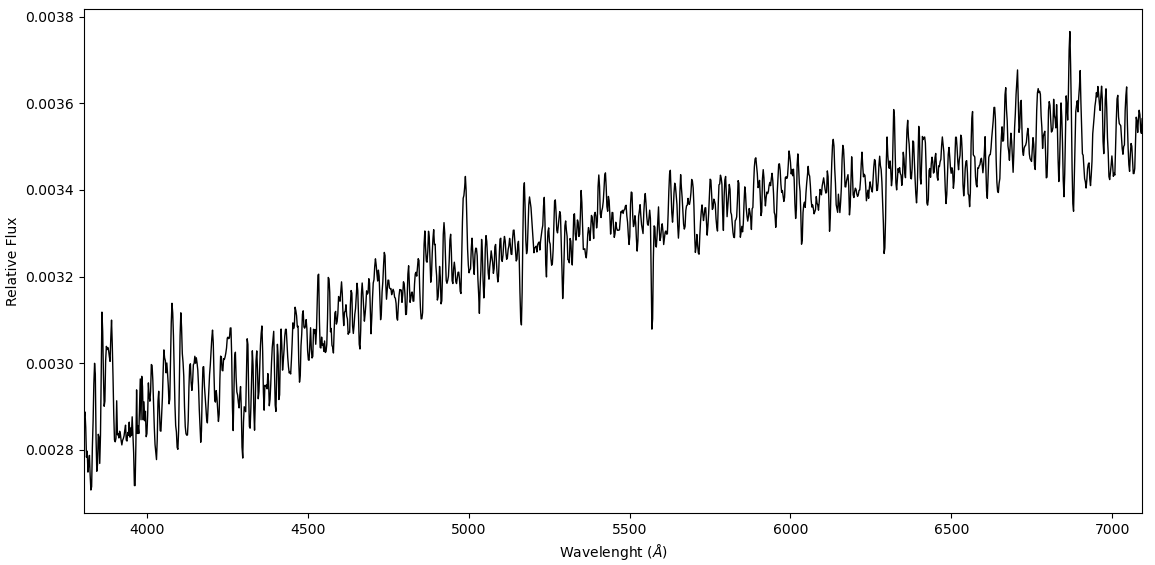}
    \caption{Spectrum of 2024-02-14; configuration A}
\end{figure}

\newpage
\clearpage

\section{C/2022 E3 (ZTF)}
\label{cometa:C2022E3}
\subsection{Description}

C/2022 E3 (ZTF) is a hyperbolic comet with an absolute magnitude of 10.8$\pm$0.9.\footnote{\url{https://ssd.jpl.nasa.gov/tools/sbdb_lookup.html\#/?sstr=2022\%20E3} visited on July 20, 2024}
It was first spotted by the 1.2m Mount Palomar Zwicky Transient Facility on February 2, 2022.
We observed the comet between magnitudes 13 and 10.\footnote{\url{https://cobs.si/comet/2323/} visited on July 21, 2024} However, this comet reached the visual magnitude 4 in January 2023.
The Earth crossed the comet orbital plane on July 26, 2022 and on January 23, 2023.

\begin{table}[h!]
\centering
\begin{tabular}{|c|c|c|}
\hline
\multicolumn{3}{|c|}{Orbital elements (epoch: October 21, 2022)}                      \\ \hline \hline
\textit{e} = 1.0003 & \textit{q} = 1.1122 & \textit{T} = 2459957.2852 \\ \hline
$\Omega$ = 302.5550 & $\omega$ = 145.8149  & \textit{i} = 109.1695  \\ \hline  
\end{tabular}
\end{table}

\begin{table}[h!]
\centering
\begin{tabular}{|c|c|c|c|c|c|c|c|c|}
\hline
\multicolumn{9}{|c|}{Comet ephemerides for key dates}                      \\ \hline 
\hline
& date         & r    & $\Delta$  & RA      & DEC      & elong  & phase  & PLang  \\
& (yyyy-mm-dd) & (AU) & (AU)      & (h)     & (°)      & (°)    & (°)    & (°) \\ \hline 

Perihelion       & 2023-01-12 & 1.112  & 0.715 & 15.80  & $+$38.57 &  80.1 & 60.6  & $+$14.0  \\ 
Nearest approach & 2023-02-01 & 1.158  & 0.284 & 06.64  & $+$72.70 & 121.0  & 46.8 & $-$33.0 \\ \hline
\end{tabular}

\end{table}

\vspace{0.5 cm}

\begin{figure}[h!]
    \centering
    \includegraphics[scale=0.38]{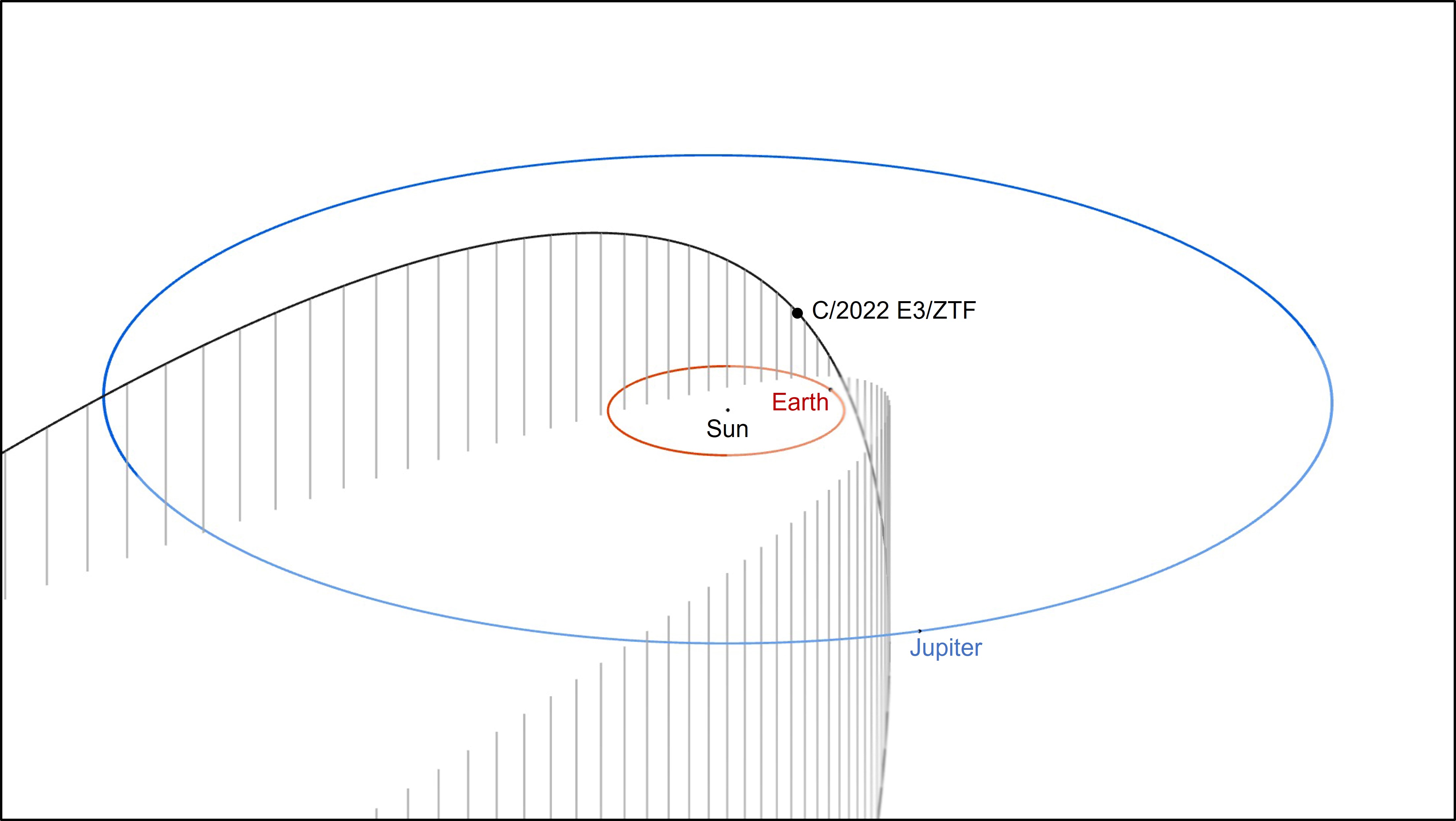}
    \caption{Orbit of comet C/2022 E3 and position on perihelion date. The field of view is set to the orbit of Jupiter for size comparison. Courtesy of NASA/JPL-Caltech.}
\end{figure}

\newpage

\subsection{Images}

\begin{SCfigure}[0.8][h!]
    \centering
    \includegraphics[scale=0.4]{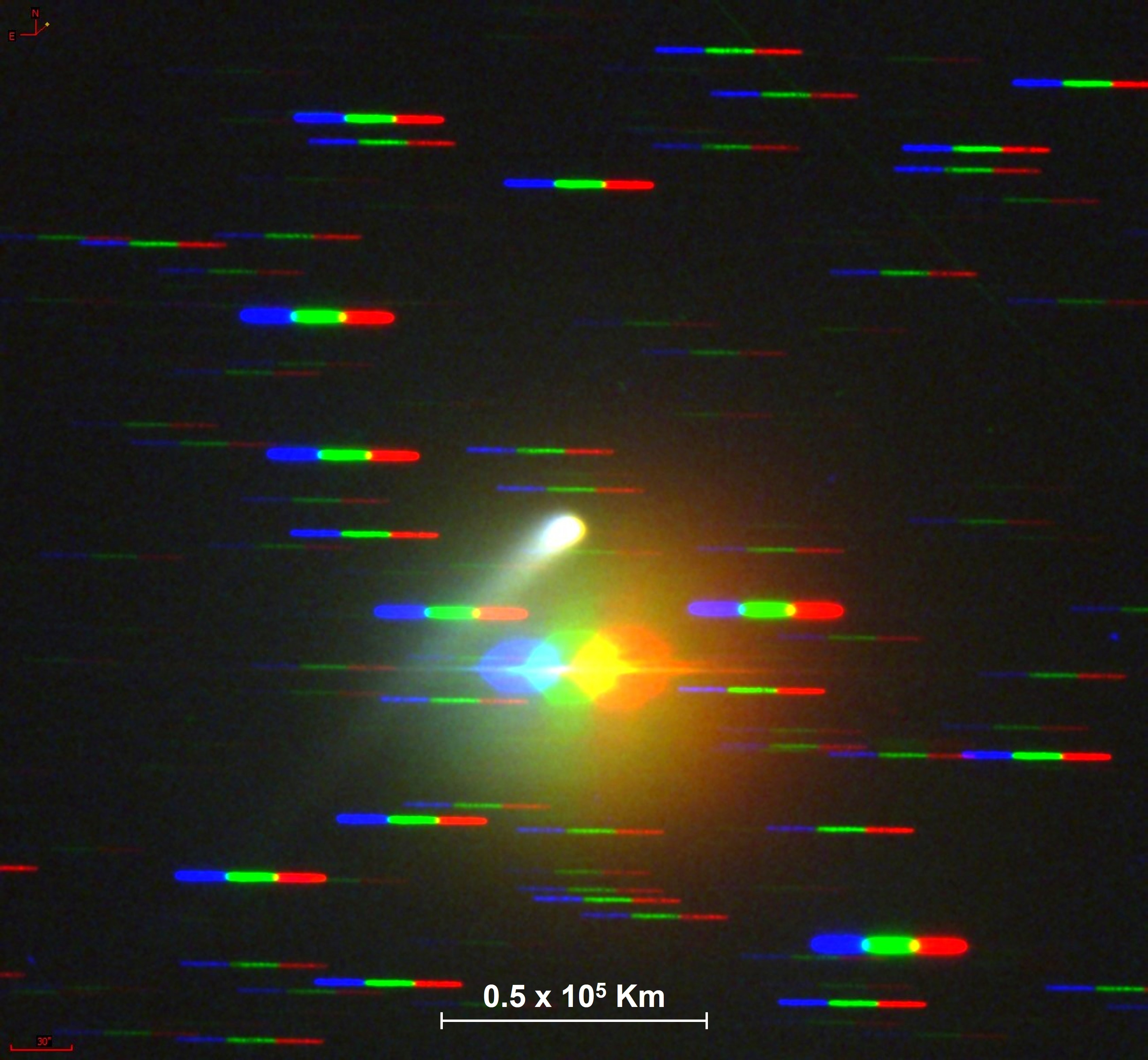}
     \caption{2022-07-31. Three-color BVR composite from images taken with the Schmidt telescope. Comet ZTF is hosted in a star-rich field in the constellation Hercules, close to the K0 star HD 159732 of magnitude 7. The tail develops in the south-west direction. }
\end{SCfigure} 

\begin{SCfigure}[0.8][h!]
    \centering

    \includegraphics[scale=0.4]{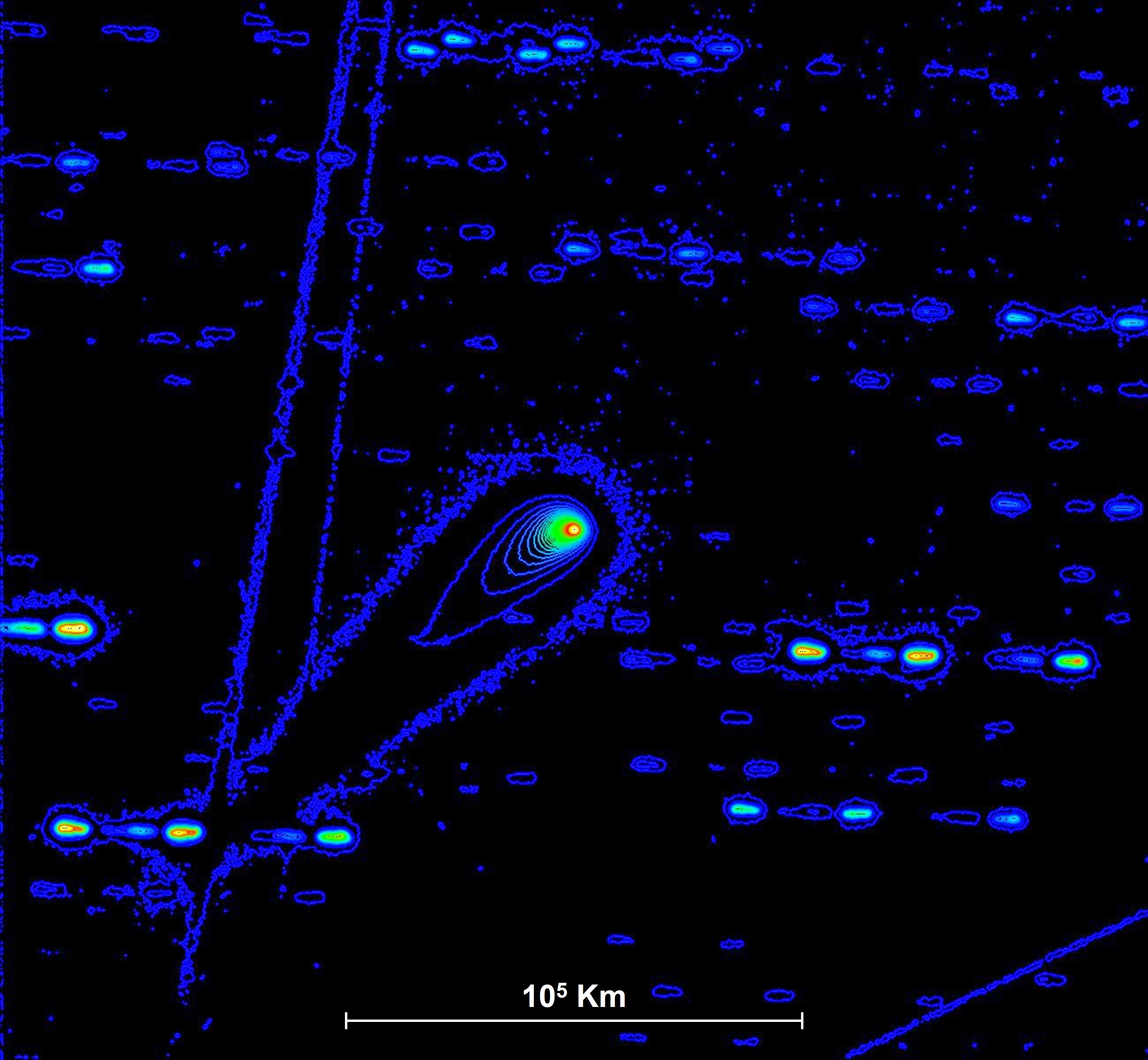}

    \caption{2022-08-02. The visualization with isophotes of an image taken with the Asiago Copernico telescope highlights the long tail, but also an asymmetry of the coma. }
\end{SCfigure}

\newpage

\subsection{Spectra}

\begin{table}[h!]
\centering
\begin{tabular}{|c|c|c|c|c|c|c|c|c|c|c|c|}
\hline
\multicolumn{12}{|c|}{Observation details}                      \\ \hline 
\hline
$\#$  & date          & r     & $\Delta$ & RA     & DEC     & elong & phase & PLang& config  & FlAng & N \\
      & (yyyy-mm-dd)  &  (AU) & (AU)     & (h)    & (°)     & (°)   & (°)   &  (°)   &       &  (°)  &  \\ \hline 

1 & 2022-07-18 & 2.789 & 2.150 & 18.20 & $+$35.22 & 119.1 & 18.6 & $-$03.1 & A & $-$0 & 3 \\
2 & 2022-07-22 & 2.744  & 2.134 & 18.07 & $+$35.52 & 116.6  & 19.3  & $-$01.5 & A & $+$0 & 5 \\
3 & 2022-08-16 & 2.459 & 2.136 & 16.909 & $+$34.85 & 96.1 & 24.2 & $+$9.1 & A & $-$0 & 6 \\
4 & 2022-08-19 & 2.425 & 2.144 & 16.83 & $+$34.55 & 93.4 & 24.6 & $+$10.2 & A  & $-$0 & 2 \\
5 & 2022-08-20 & 2.413 & 2.147 & 16.80 & $+$34.42 & 92.5 & 24.8 & $+$10.6 & A  & $-$0 & 1\\
6 & 2022-08-21 & 2.402 & 2.150 & 16.77 & $+$34.28 & 91.6 & 24.9 & $+$11.0 & A  & $-$0 & 2 \\
7 & 2022-08-22 & 2.390 & 2.153 & 16.73 & $+$34.15 & 90.7 & 25.0 & $+$11.3 & A  & $-$0 & 6 \\
8 & 2022-08-26 & 2.344 & 2.166 & 16.60 & $+$33.57 & 87.1 & 25.5 & $+$12.8 & B  & $+$0 & 3\\
9 & 2022-10-19 & 1.729 & 2.221 & 15.85 & $+$25.73 & 48.4 & 25.5 & $+$24.6 & A & $-$0 & 1 \\
10 & 2023-01-02 & 1.126 & 1.006 & 15.88 & $+$31.61 & 68.9 & 54.6 & $+$19.5 & C & $-$35 & 7 \\
11 & 2023-01-03 & 1.121 & 0.953 & 15.87 & $+$32.54 & 70.8 & 55.9 & $+$18.7 & C & $-$40 & 2 \\
12 & 2023-01-20 & 1.120 & 0.479 & 15.54 & $+$51.19 & 93.5 & 61.3 & $+$04.1 & D & $+$90 & 2 \\
13 & 2023-02-03 & 1.168	& 0.292 & 05.78 & $+$63.85 & 122.4 & 45.4 & $-$41.5 & D & $+$0 & 4 \\
14 & 2023-02-07 & 1.190	& 0.350 & 05.03 & $+$41.57 & 117.7 & 47.2 & $-$47.4 & D & $+$33 & 2\\
15 & 2023-02-10 & 1.207	& 0.412 & 04.90 & $+$33.02 & 112.6 & 49.0 & $-$47.2 & D & $-$0 & 2 \\
16 & 2023-02-15 & 1.241	& 0.549 & 04.70 & $+$16.45 & 104.1 & 50.5 & $-$44.1 & D & $+$50 & 2 \\ 
17 & 2023-02-28 & 1.344	& 0.932 & 04.63 & $+$01.98 & 88.6 &	47.5 & $-$37.4 & C & $+$0 & 8 \\

\hline
\end{tabular}
\end{table}

\begin{figure}[h!]

    \centering
    \includegraphics[scale=0.368]{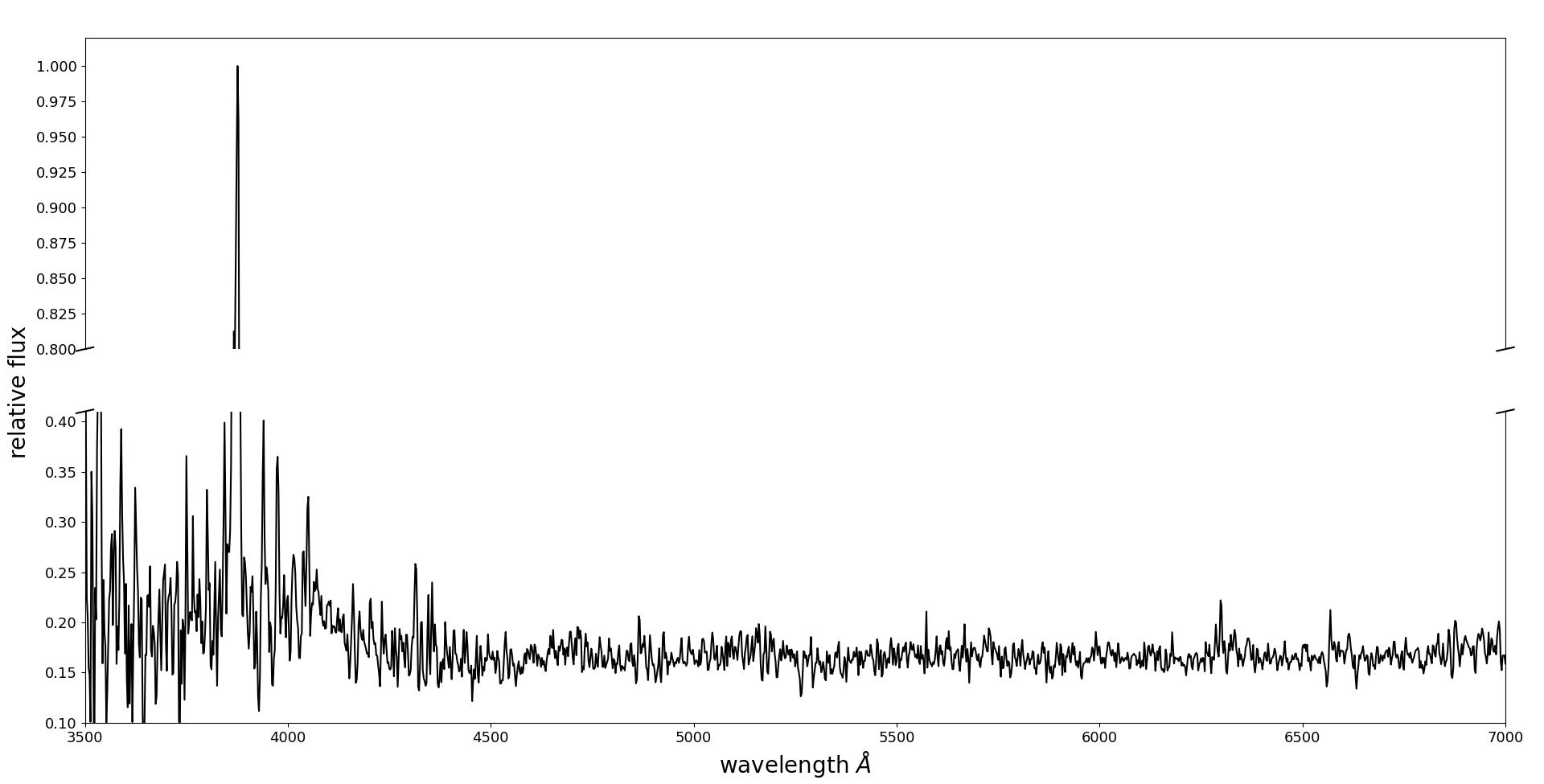}

\end{figure}

\begin{figure}[h!]
    \centering
    \includegraphics[scale=0.368]{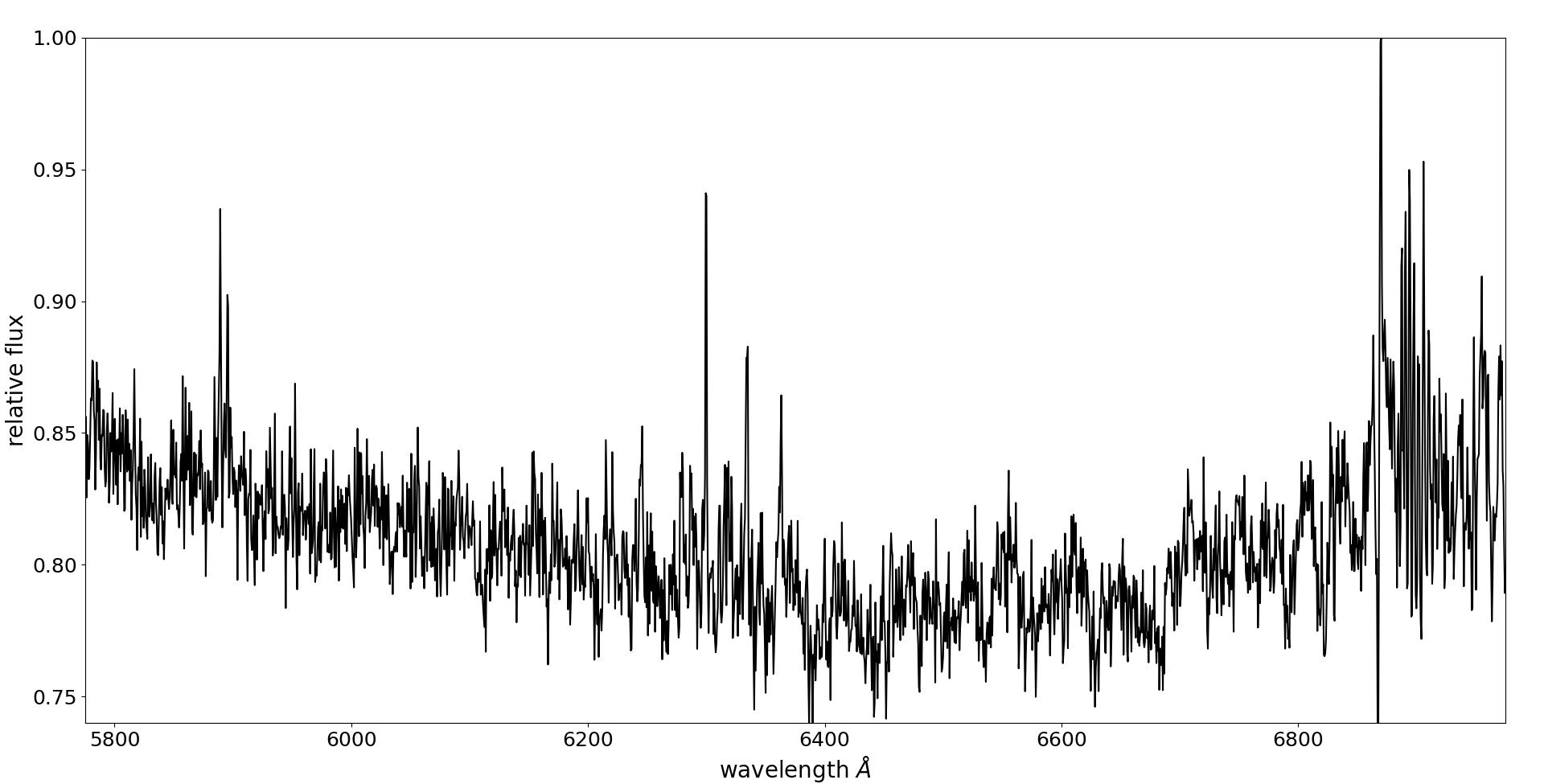}

\end{figure}

\newpage
\clearpage

\section{C/2023 E1 (ATLAS)}
\label{cometa:C2023E1}
\subsection{Description}

C/2023 E1 (ATLAS) is a Halley-type comet with a period of 85 years and an absolute magnitude of 16.2$\pm$0.8.\footnote{\url{https://ssd.jpl.nasa.gov/tools/sbdb_lookup.html\#/?sstr=2023\%20E1} visited on July 20, 2024}
It was first spotted by the 0.5m Asteroid Terrestrial-impact Last Alert System (ATLAS) on March 1, 2023.
Given its orbital period and inclination, this object is classified as a Halley-type comet captured by Neptune.
Observations show low content of dust and a coma rich of CN, \ch{C2}, \ch{C3} and [O I]. The spectra are characterized by strong emission lines of the molecules in the coma and an almost absent continuum. The coma exhibits blue/green color due to the fluorescence of \ch{C2} and \ch{C3} molecules.
We observed the comet between magnitude 9.5 and 8.\footnote{\url{https://cobs.si/comet/2411/ }, visited on July 20, 2024}

\begin{table}[h!]
\centering
\begin{tabular}{|c|c|c|}
\hline
\multicolumn{3}{|c|}{Orbital elements (epoch: July 2, 2023)}                      \\ \hline \hline
\textit{e} = 0.9469 & \textit{q} = 1.0266 & \textit{T} = 2460126.6075 \\ \hline
$\Omega$ = 164.5742 & $\omega$ = 105.8946  & \textit{i} = 38.3130 \\ \hline  
\end{tabular}
\end{table}

\begin{table}[h!]
\centering
\begin{tabular}{|c|c|c|c|c|c|c|c|c|}
\hline
\multicolumn{9}{|c|}{Comet ephemerides for key dates}                      \\ \hline 
\hline
& date         & r    & $\Delta$  & RA      & DEC      & elong  & phase  & PLang  \\
& (yyyy-mm-dd) & (AU) & (AU)      & (h)     & (°)      & (°)    & (°)    & (°) \\ \hline 

Perihelion       & 2023-07-01 & 1.027 & 0.643 & 14.75 & $+$79.66 & 72.5 & 70.8 & $-$63.2  \\ 
Nearest approach & 2023-08-18 & 1.290 & 0.375 & 21.74 & +35.62 & 131.2 & 36.2 & $-$34.3\\ \hline
\end{tabular}

\end{table}

\vspace{0.5 cm}

\begin{figure}[h!]
    \centering
    \includegraphics[scale=0.38]{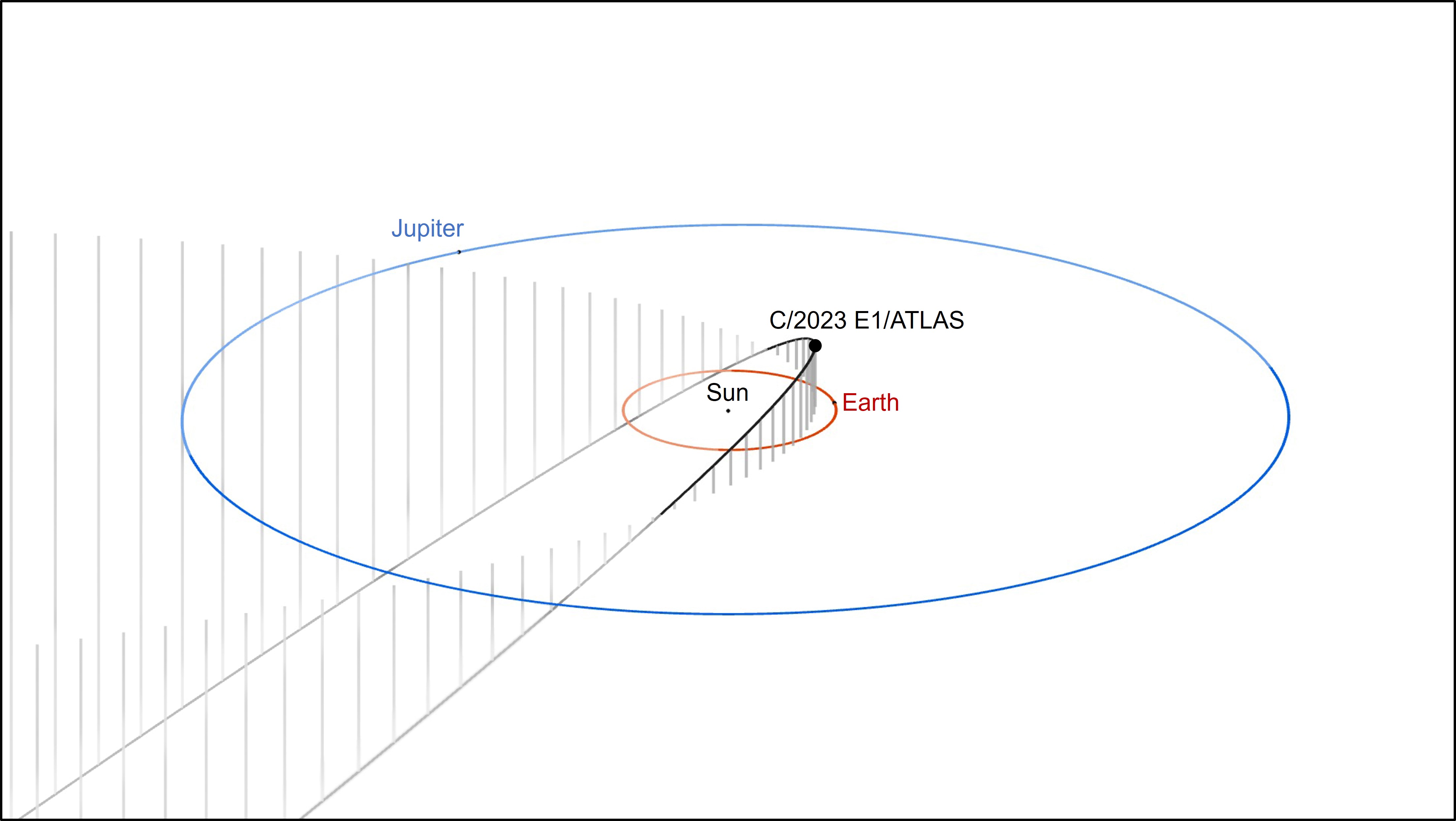}
    \caption{Orbit of comet C/2023 E1 and position on perihelion date. The field of view is set to the orbit of Jupiter for size comparison. Courtesy of NASA/JPL-Caltech.}
\end{figure}

\newpage

\subsection{Images}

\begin{SCfigure}[0.8][h!]
    \centering
    \includegraphics[scale=0.4]{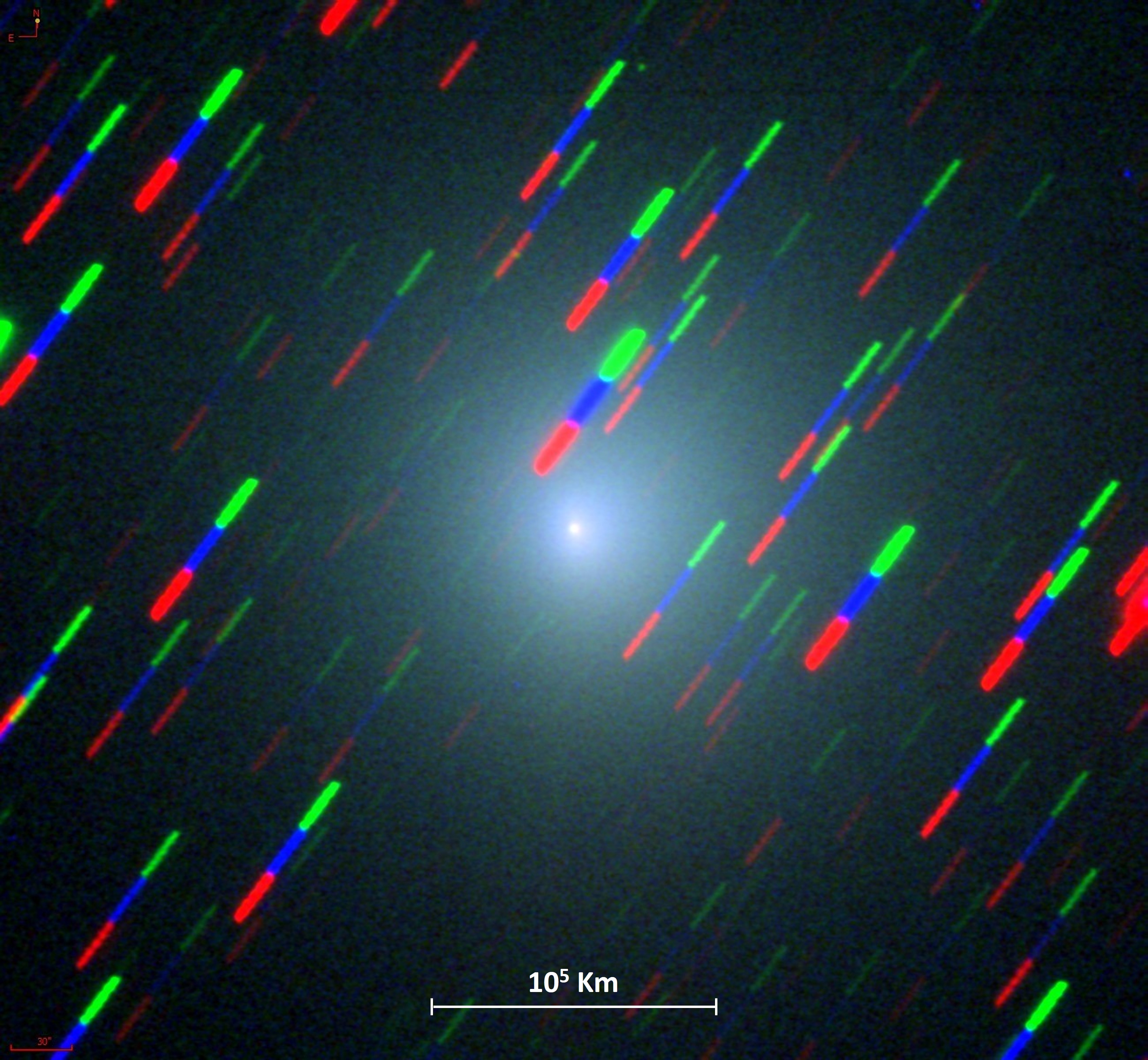}
     \caption{2023-07-23. Three-color BVR composite from images taken with the Schmidt telescope. The coma appears blue-green due to the fluorescence of CN and C$_2$ molecules.}

\end{SCfigure} 

\begin{SCfigure}[0.8][h!]
    \centering
    \includegraphics[scale=0.4]{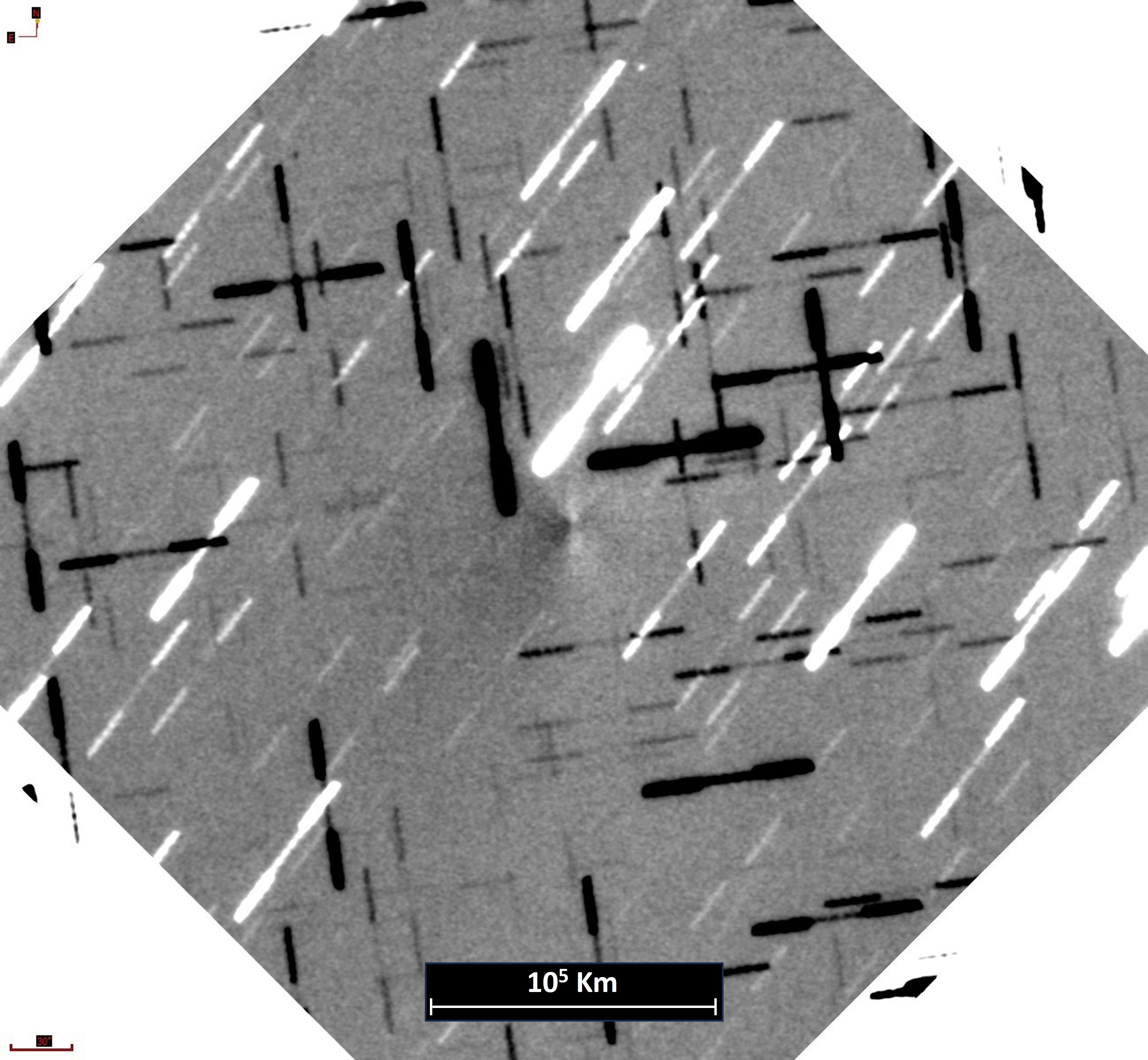}
    \caption{2023 07 23. Schmidt images processed with Larson-Sekanina filter ($\alpha=45$\textdegree). To show details of the inner coma. A structure with two jets is visible in a near-equatorial position, suggesting the existence of a single active area. Due to the rotation of the comet, this active area produces an emission cone. The two jets we see are actually the sides of the same emission cone.}

\end{SCfigure}

\newpage

\subsection{Spectra}

\begin{table}[h!]
\centering
\begin{tabular}{|c|c|c|c|c|c|c|c|c|c|c|c|}
\hline
\multicolumn{12}{|c|}{Observation details}                      \\ \hline 
\hline
$\#$  & date          & r     & $\Delta$ & RA     & DEC     & elong & phase & PLang & config  & FlAng & N \\
      & (yyyy-mm-dd)  &  (AU) & (AU)     & (h)    & (°)     & (°)   & (°)   &  (°)   &       &  (°)  & \\ \hline 

1 & 2023-06-25 & 1.030 & 0.673 & 13.88 & +77.97 & 71.90 & 69.70 & $-$62.16 & A & $-$77 & 4 \\
2 & 2023-06-29 & 1.027 & 0.652 & 14.47 & +79.20 & 72.25 & 70.55 & $-$62.93 & A & $+$90 & 1 \\
3* & 2023-07-01 & 1.027 & 0.640 & 14.83 & +79.70 & 72.59 & 70.88 & $-$63.26 & C & $+$0 & 1 \\
4 & 2023-07-23 & 1.093 & 0.492 & 19.99 & +74.45 & 85.43 & 67.90 & $-$62.99 & A & $+$90 & 2 \\
5 & 2023-08-05 & 1.180 & 0.410 & 21.23 & +59.25 & 103.76 & 56.54 & $-$53.84 & A & $+$0 & 1 \\
6 & 2023-08-06 & 1.188 & 0.405 & 21.28 & +57.77 & 105.50  & 55.34 & $-$52.74 & A & $+$30 & 1\\
7 & 2023-08-11 & 1.231 & 0.384 & 21.53 & +48.71 & 116.01 & 47.71 & $-$45.58 & A & $+$90 & 6 \\
8 & 2023-08-12 & 1.239 & 0.381 & 21.56 & +47.07 & 117.96 & 46.25 & $-$44.15 & A & $+$0 & 1 \\
9 & 2023-08-13 & 1.248 & 0.379 & 21.59 & +45.19 & 120.14 & 44.61 & $-$42.59 & A & $+$0 & 3 \\
10 & 2023-08-13 & 1.249 & 0.379 & 21.60 & +44.94 & 120.42 & 44.41 & $-$42.39 & C & $+$0 & 3 \\
11 & 2023-08-18 & 1.294 & 0.375 & 21.75 & +34.71 & 132.22 & 35.39 & $-$33.48 & A & $+$90 & 13 \\
12 & 2023-08-25 & 1.363 & 0.393 & 21.91 & +19.83 & 148.97 & 22.47 & $-$20.10 & A & $+$0 & 1 \\
13 & 2023-09-10 & 1.530 & 0.537 & 23.52 & +11.61 & 163.69 & 10.65 & $+$03.41 & C & $+$90 & 1 \\ \hline

\hline
\end{tabular}
\end{table}

\begin{figure}[h!]
    \centering
    \includegraphics[scale=0.58]{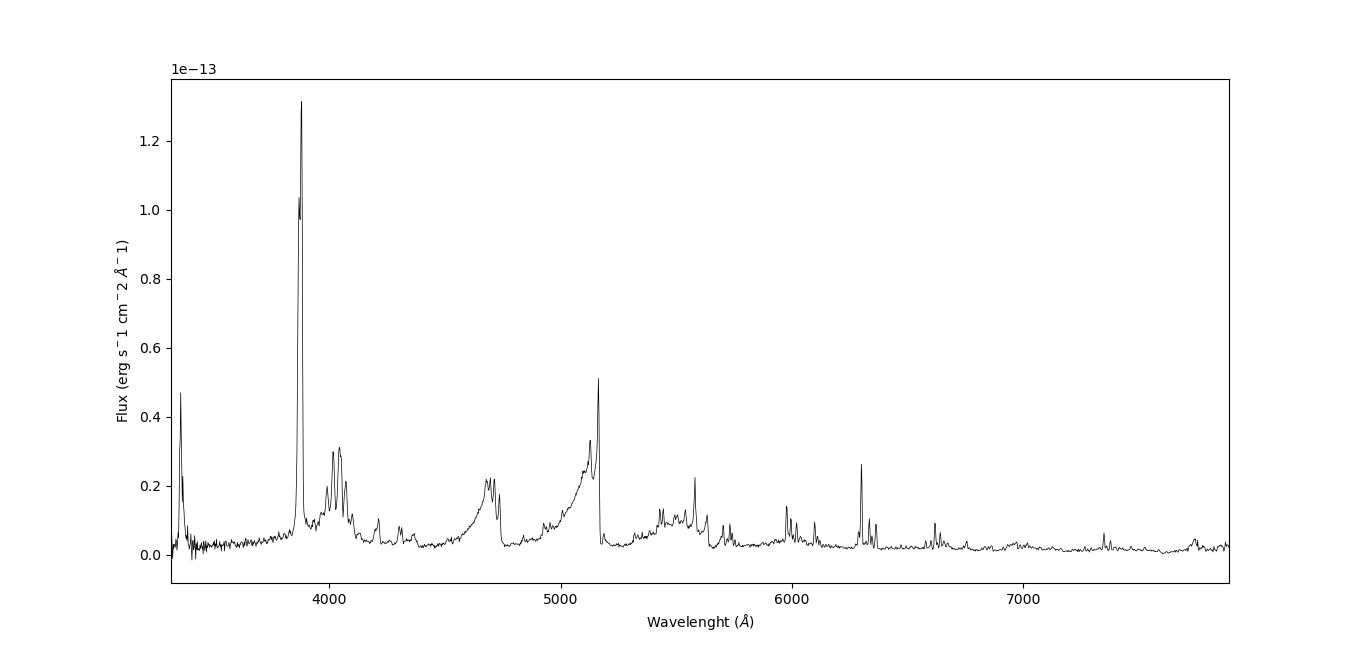}
    \caption{Spectrum of 2023-06-25; configuration A}
\end{figure}

\begin{figure}[h!]
    \centering
    \includegraphics[scale=0.58]{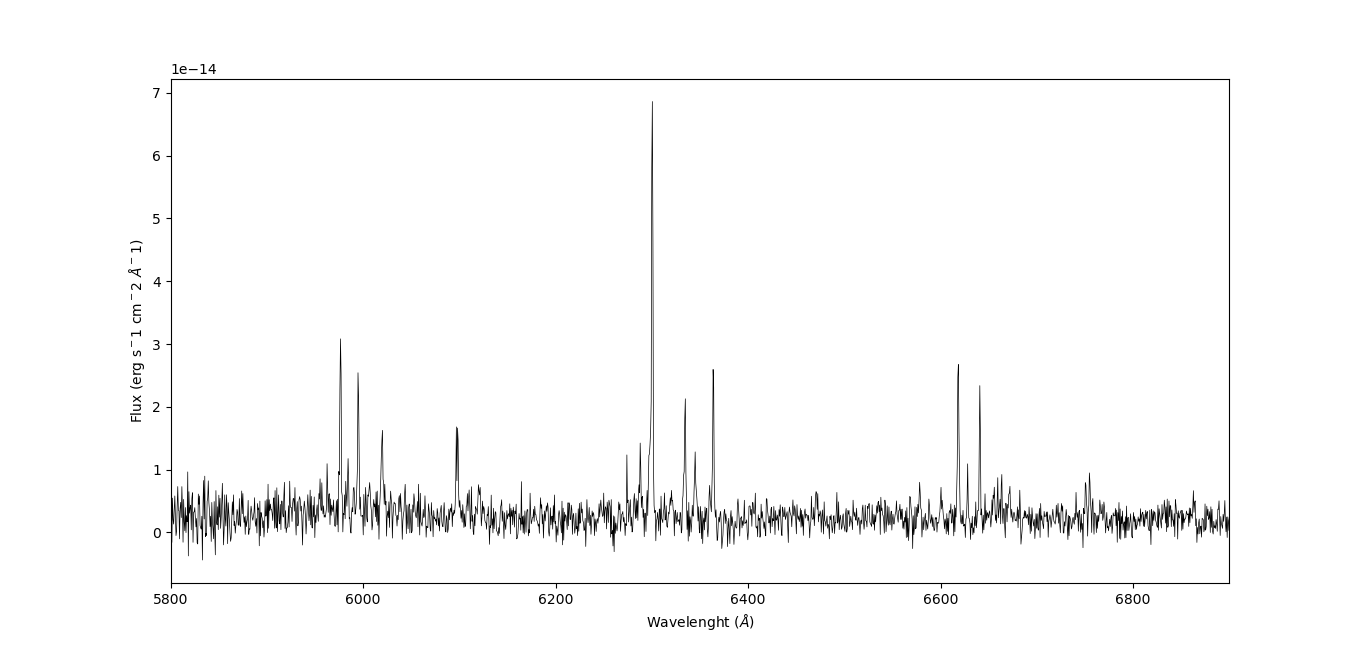}
    \caption{Spectrum of 2023-07-01; configuration C}
\end{figure}

\footnotetext{* No Solar Analog}

\newpage
\clearpage

\section{C/2023 H2 (Lemmon)}
\label{cometa:C2023H2}
\subsection{Description}

C/2023 H2 (Lemmon) is a Long Period comet with a period of 3731 years and an absolute magnitude of 16.1$\pm$1.0.\footnote{\url{https://ssd.jpl.nasa.gov/tools/sbdb_lookup.html\#/?sstr=2023\%20H2} visited on July 21, 2024}
It was first spotted by the Mt. Lemmon program on April 23, 2023.
The Earth crossed the comet orbital plane on April 29, 2023 and on October 31, 2023.

\noindent
We observed this comet between magnitude 11 and 7.\footnote{\url{https://cobs.si/comet/2431/}, visited on July 21, 2024}

\begin{table}[h!]
\centering
\begin{tabular}{|c|c|c|}
\hline
\multicolumn{3}{|c|}{Orbital elements (epoch: October 7, 2023)}                      \\ \hline \hline
\textit{e} = 0.9963 & \textit{q} = 0.8944 & \textit{T} = 2460246.6900 \\ \hline
$\Omega$ = 217.0448 & $\omega$ = 150.6504  & \textit{i} = 113.7535 \\ \hline  
\end{tabular}
\end{table}

\begin{table}[h!]
\centering
\begin{tabular}{|c|c|c|c|c|c|c|c|c|}
\hline
\multicolumn{9}{|c|}{Comet ephemerides for key dates}                      \\ \hline 
\hline
& date         & r    & $\Delta$  & RA      & DEC      & elong  & phase  & PLang  \\
& (yyyy-mm-dd) & (AU) & (AU)      & (h)     & (°)      & (°)    & (°)    & (°) \\ \hline 

Perihelion       & 2023-10-29 & 0.894 & 0.490 & 13.57 & $+$50.08 & 64.0 & 86.5 & $+$02.9  \\ 
Nearest approach & 2023-11-10 & 0.922 & 0.194 & 18.42 & $+$22.60 & 63.9 & 105.2 & $-$58.0\\ \hline
\end{tabular}

\end{table}

\vspace{0.5 cm}

\begin{figure}[h!]
    \centering
    \includegraphics[scale=0.38]{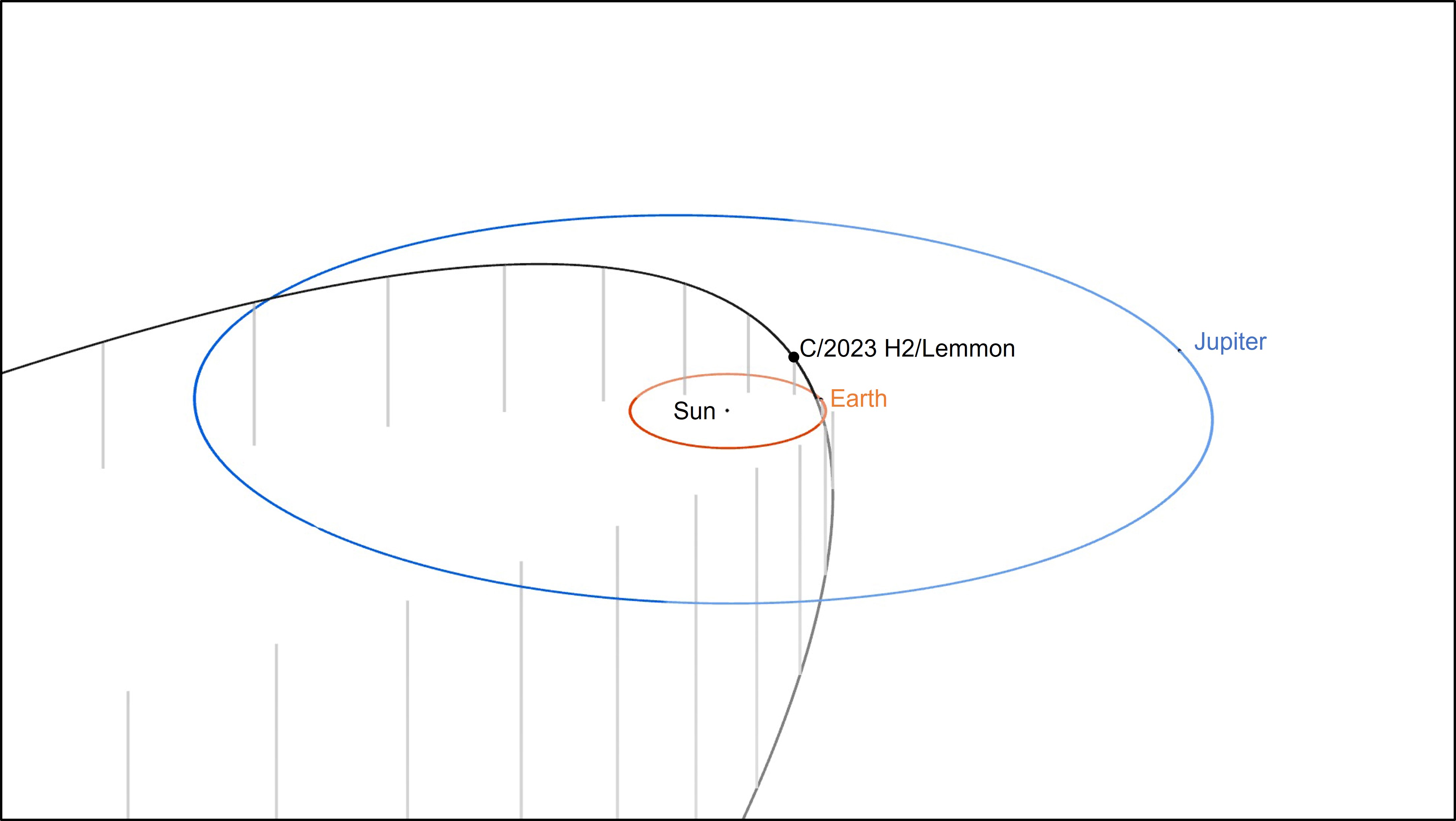}
    \caption{Orbit of comet C/2023 H2 and position on perihelion date. The field of view is set to the orbit of Jupiter for size comparison. Courtesy of NASA/JPL-Caltech.}
\end{figure}

\newpage

\subsection{Images}

\begin{SCfigure}[0.8][h!]
    \centering
    \includegraphics[scale=0.40]{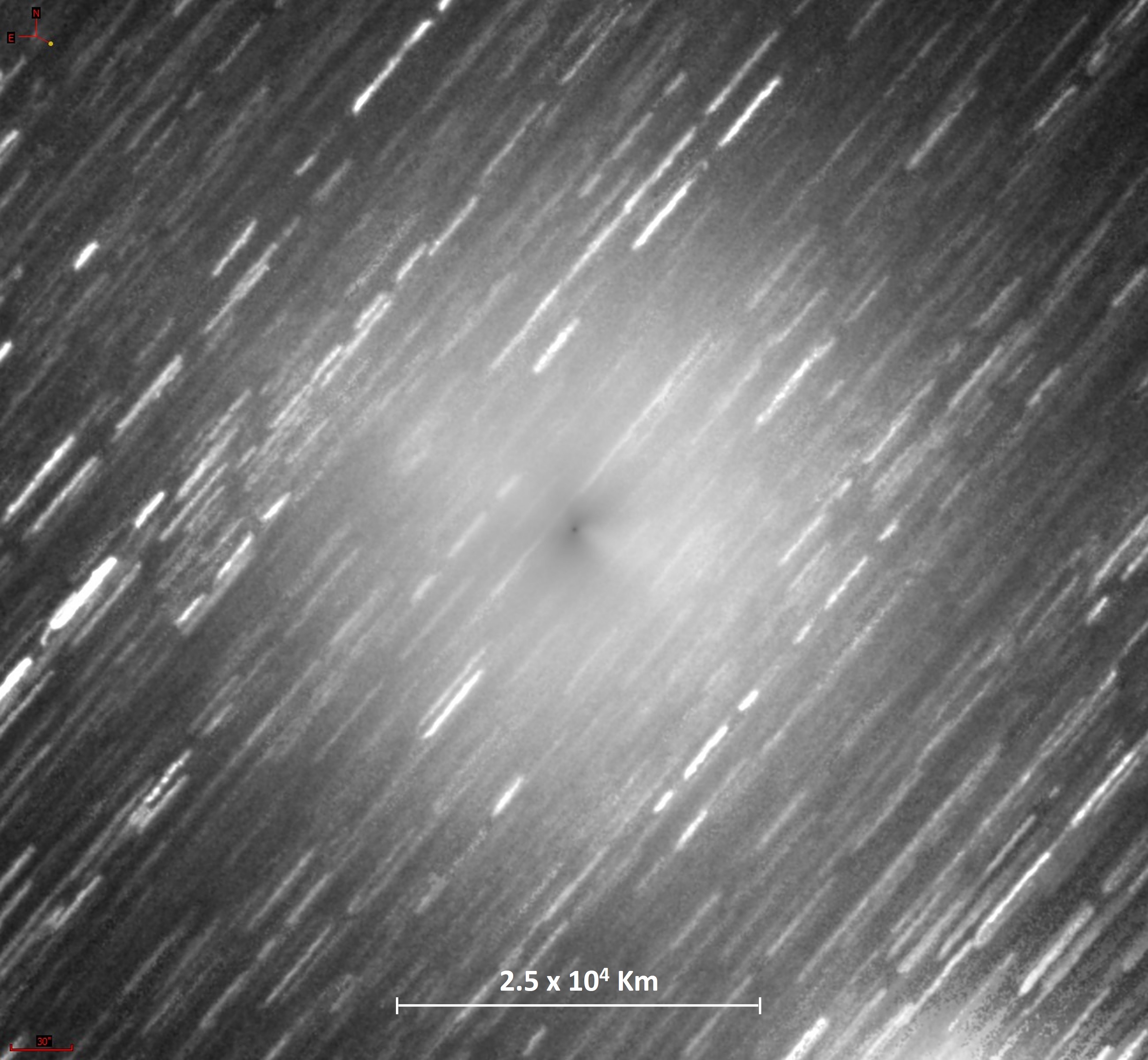}
     \caption{2023-11-11. Images taken at the Asiago Schmidt telescope. To observe the morphological details of the inner coma, the comet image was processed with an attenuator filter (1/R). C/2023 H2 came very close to Earth; this image was taken with the comet less than 30 million km away. }

\end{SCfigure} 

\begin{SCfigure}[0.8][h!]
    \centering
    \includegraphics[scale=0.40]{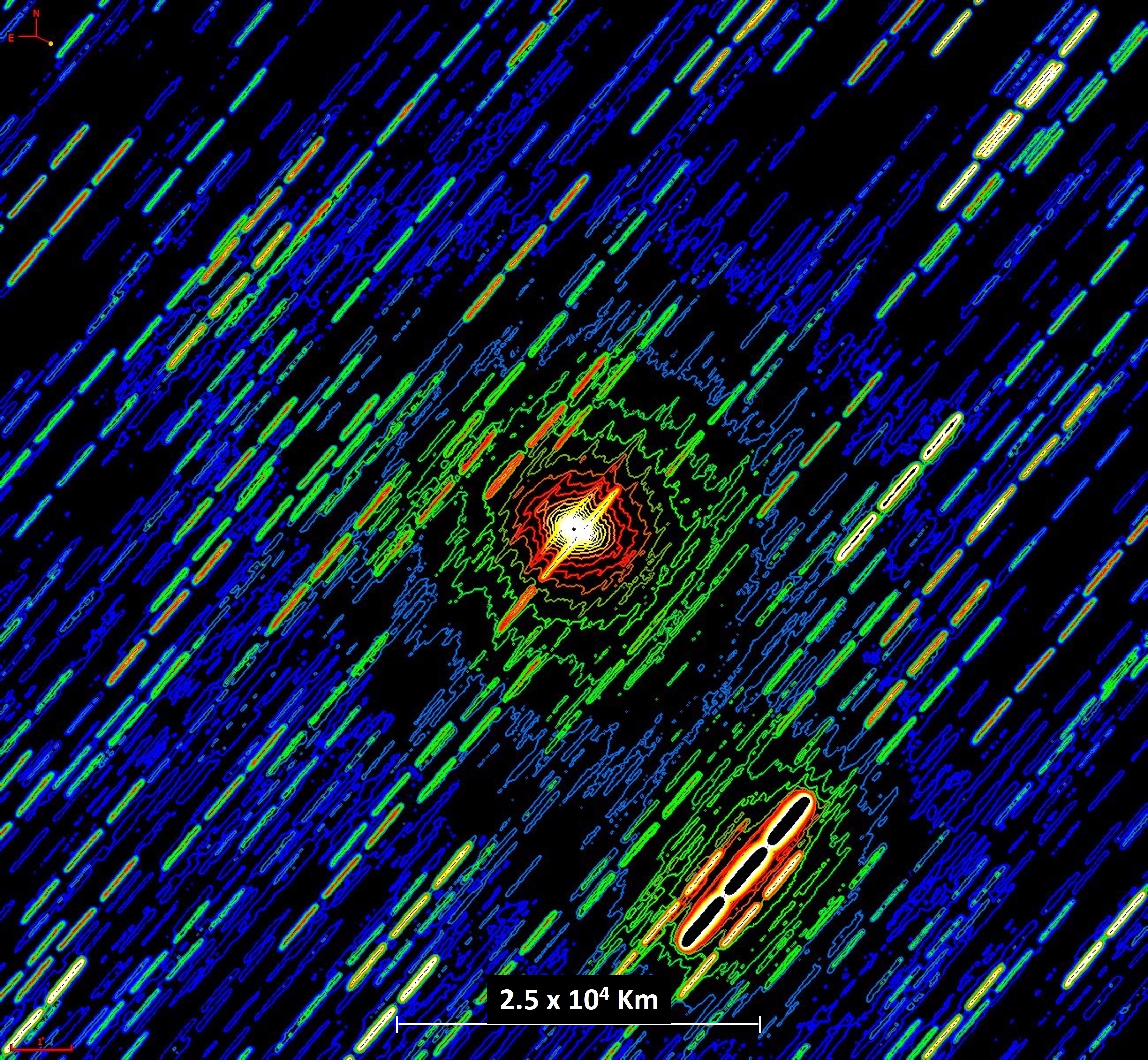}
    \caption{2023-11-11. The same image is visualized in isophotes in steps of 500 ADU. The nucleus of the comet is indicated by the black dot in the inner coma, which appears significantly elongated from NE to SW. The presence of many stars in the Milky Way makes 'reading' the image rather difficult. The outer coma occupies almost the entire field of view.}

\end{SCfigure}

\newpage

\subsection{Spectra}

\begin{table}[h!]
\centering
\begin{tabular}{|c|c|c|c|c|c|c|c|c|c|c|c|}
\hline
\multicolumn{12}{|c|}{Observation details}                      \\ \hline 
\hline
$\#$  & date          & r     & $\Delta$ & RA     & DEC     & elong & phase & PLang & config  & FlAng & N \\
      & (yyyy-mm-dd)  &  (AU) & (AU)     & (h)    & (°)     & (°)   & (°)   &  (°)   &       &  (°)  &  \\ \hline 

1 & 2023-10-12 & 0.947 & 1.052  & 12.27 & $+$46.53 & 54.9 & 59.7 & $+$16.2 & A & $-$64 & 5 \\ 
2 & 2023-10-25 & 0.897 & 0.633  & 13.00 & $+$49.22 & 62.3 & 78.9 & $+$08.5 & A & $-$60 & 8 \\ 
3 & 2023-11-12 & 0.932 & 0.203 & 19.48 & $+$04.87 & 67.8 & 100.5 & $-$78.7 & A & $+$90 & 13 \\
4 & 2023-11-17 & 0.961 & 0.311 & 20.941 & $-$21.04 & 75.9 & 85.8 & $-$62.3 & D & $+$90 & 7 \\

\hline
\end{tabular}
\end{table}

\begin{figure}[h!]
    \centering
    \includegraphics[scale=0.58]{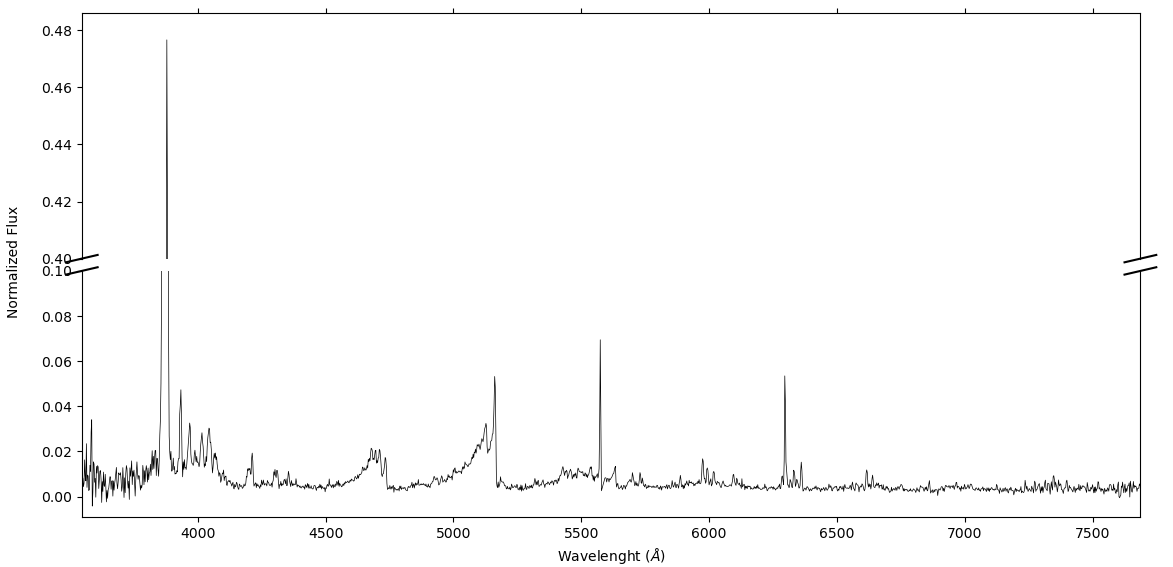}
    \caption{Spectrum of 2023-11-12; configuration A}
\end{figure}

\begin{figure}[h!]
    \centering
    \includegraphics[scale=0.58]{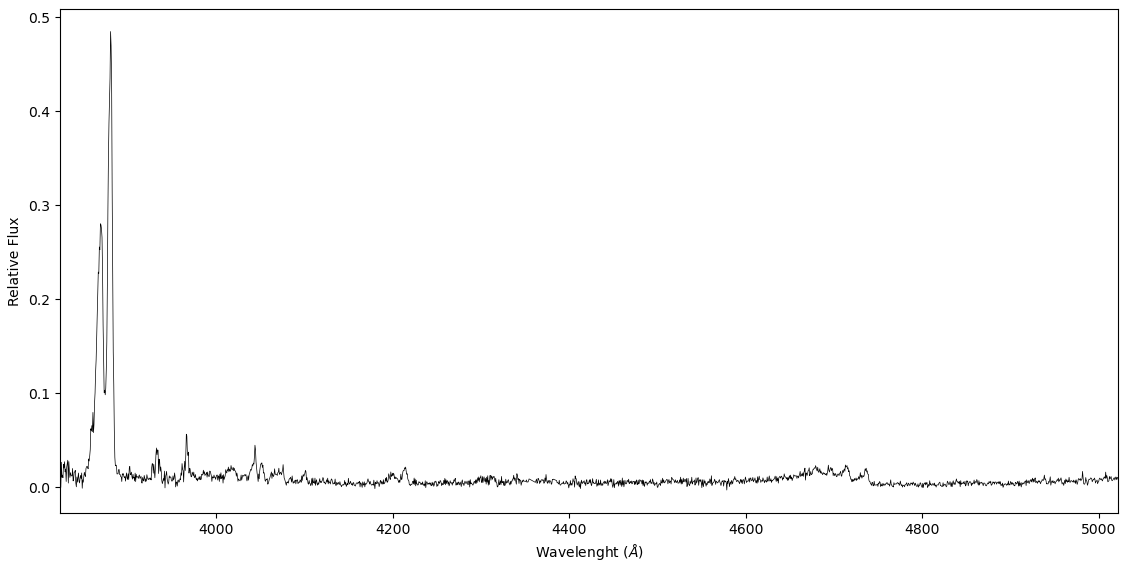}
    \caption{Spectrum of 2023-11-17; configuration D}
\end{figure}

\newpage
\clearpage

\section{C/2023 A3 (Tsuchinshan-ATLAS)}
\label{cometa:C2023A3}
\subsection{Description}

C/2023 A3 (Tsuchinshan-ATLAS) is a hyperbolic comet with an absolute magnitude of 8.6$\pm$0.7.\footnote{\url{https://ssd.jpl.nasa.gov/tools/sbdb_lookup.html\#/?sstr=2023\%20A3} visited on July 21, 2024}
It was first spotted by the 0.5m Asteroid Terrestrial-impact Last Alert System (ATLAS) in Sutherland (South Africa) on February 22, 2023.
The traditional name for PMO comets is Tsuchinshan.\footnote{\url{http://www.cbat.eps.harvard.edu/iau/cbet/005200/CBET005228.txt} visited on July 21, 2024}

\noindent
We observed the comet between magnitude 11 and 10.\footnote{\url{https://cobs.si/comet/2410/ }, visited on July 20, 2024}
The Earth crossed the comet orbital plane on October 15, 2024.

\begin{table}[h!]
\centering
\begin{tabular}{|c|c|c|}
\hline
\multicolumn{3}{|c|}{Orbital elements (epoch: November 24, 2023)}                      \\ \hline \hline
\textit{e} = 1.0001  & \textit{q} = 0.3915 & \textit{T} = 2460581.2288 \\ \hline
$\Omega$ = 21.5573 & $\omega$ = 308.4843  & \textit{i} = 139.1180 \\ \hline  
\end{tabular}
\end{table}

\begin{table}[h!]
\centering
\begin{tabular}{|c|c|c|c|c|c|c|c|c|}
\hline
\multicolumn{9}{|c|}{Comet ephemerides for key dates}                      \\ \hline 
\hline
& date         & r    & $\Delta$  & RA      & DEC      & elong  & phase  & PLang  \\
& (yyyy-mm-dd) & (AU) & (AU)      & (h)     & (°)      & (°)    & (°)    & (°) \\ \hline 

Perihelion       & 2024-09-27 & 0.391 & 0.919 & 10.77 & $-$6.07 & 23.0 & 90.6 & $-$12.1\\ 
Nearest approach & 2024-10-12 & 0.554 & 0.472 & 14.07 & $-$1.17 & 14.6 & 152.9 & $-$3.0\\ \hline
\end{tabular}

\end{table}

\vspace{0.5 cm}

\begin{figure}[h!]
    \centering
    \includegraphics[scale=0.38]{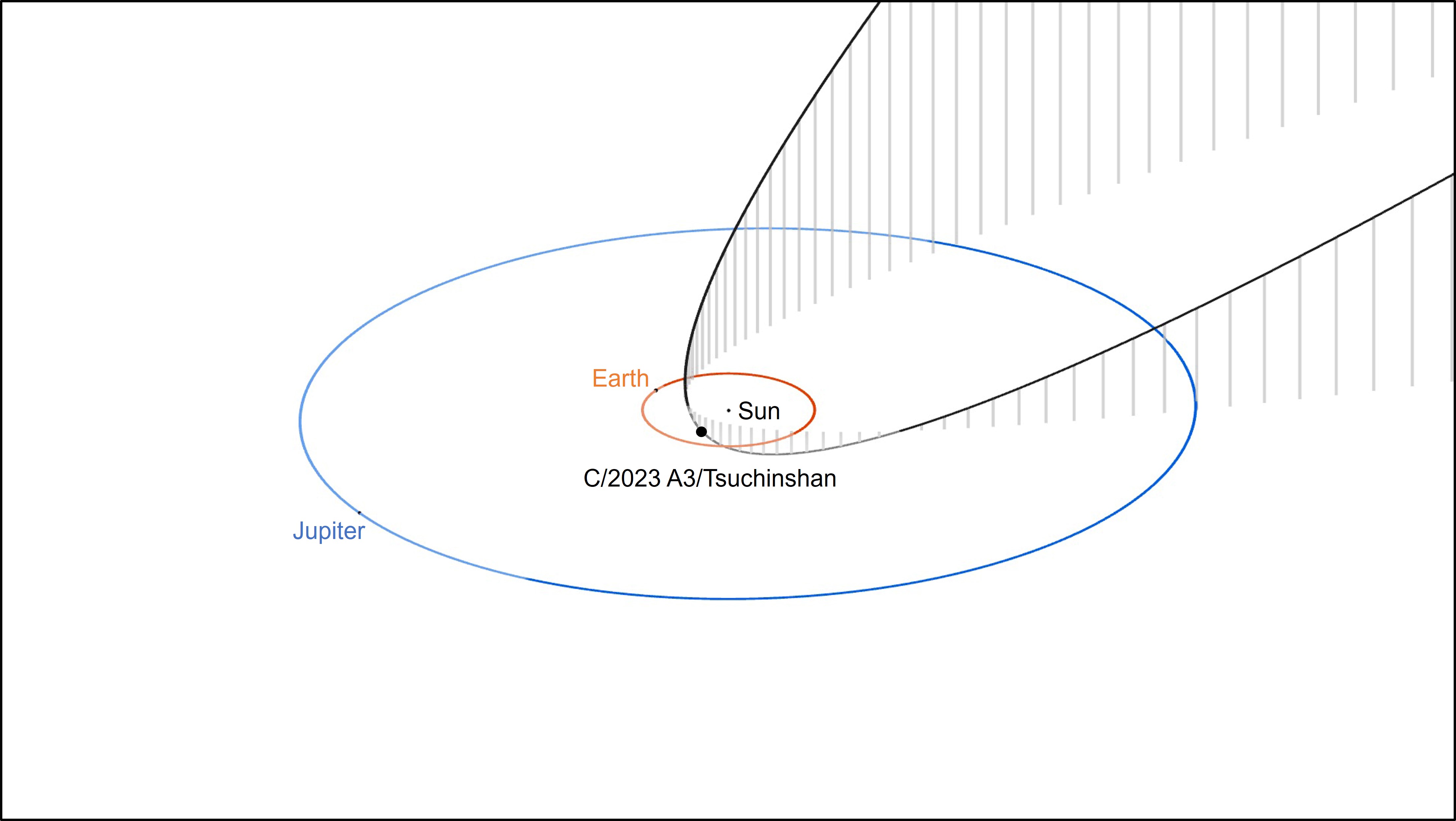}
    \caption{Orbit of comet C/2023 A3 and position on perihelion date. The field of view is set to the orbit of Jupiter for size comparison. Courtesy of NASA/JPL-Caltech.}
\end{figure}

\newpage

\subsection{Images}

\begin{SCfigure}[0.8][h!]
    \centering
    \includegraphics[scale=0.4]{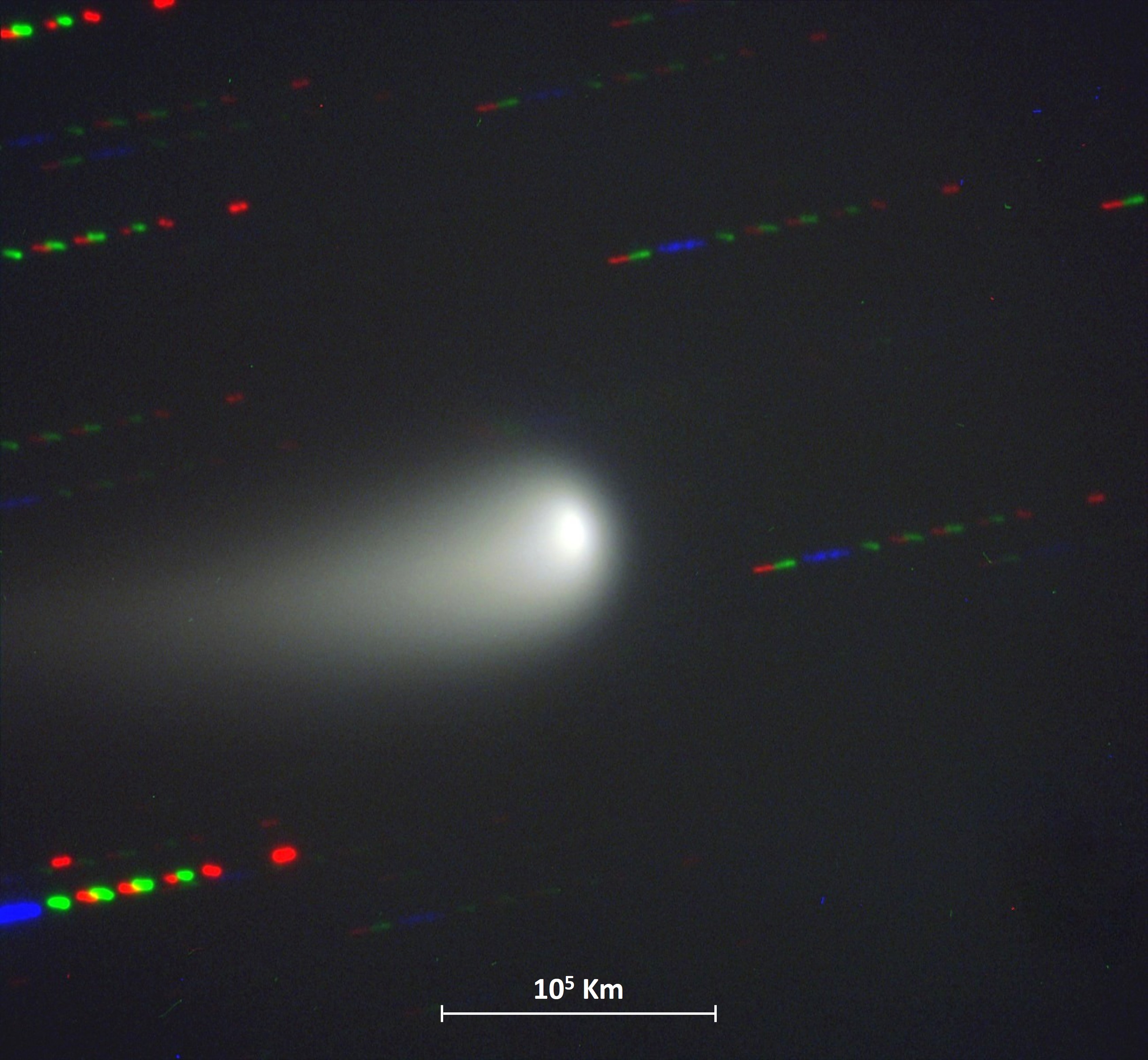}
     \caption{2024-05-05. Three-color BVR composite from images taken with the Asiago Copernico telescope.
     The inner coma of comet Tsuchinshan-ATLAS looks elongated from PA 30° to PA 210°. The tail extends in antisolar direction.}

\end{SCfigure} 

\begin{SCfigure}[0.8][h!]
    \centering
    \includegraphics[scale=0.4]{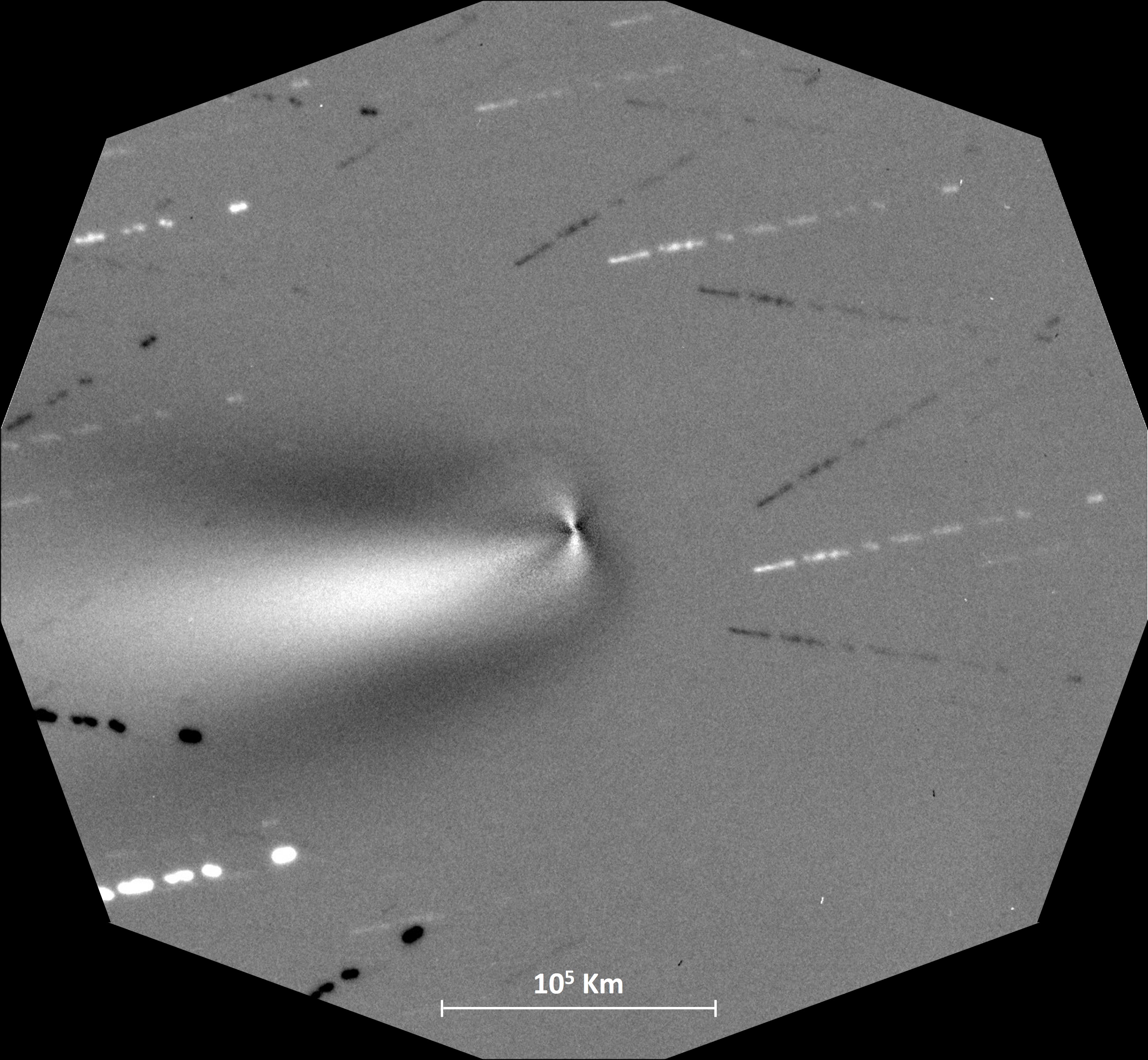}
    \caption{2023-05-05. Sum of the above BVR images shown in grayscale. Processing with the Larson-Sekanina spatial filter highlights the presence of two dust jets coming out of the nucleus in opposite directions. The force of solar radiation pressure, when it exceeds the ejection speed, pushes the dust backwards towards the tail. }

\end{SCfigure}

\newpage

\subsection{Spectra}

\begin{table}[h!]
\centering
\begin{tabular}{|c|c|c|c|c|c|c|c|c|c|c|c|}
\hline
\multicolumn{12}{|c|}{Observation details}                      \\ \hline 
\hline
$\#$  & date          & r     & $\Delta$ & RA     & DEC     & elong & phase & PLang & config  & FlAng & N \\
      & (yyyy-mm-dd)  &  (AU) & (AU)     & (h)    & (°)     & (°)   & (°)   &  (°)   &       &  (°)  &  \\ \hline 

1 & 2024-04-30 & 2.767 & 1.808 & 13.39 & $-$0.86 & 157.81 & 7.91 & $-$06.84 & A & $+$90 & 2 \\
2 & 2024-05-04 & 2.712 & 1.783 & 13.20 & $-$0.28 & 151.37 & 10.27 & $-$08.34 & A & $+$90 & 3 \\
3 & 2024-05-25 & 2.418 & 1.770 & 12.28 & $+$2.16 & 118.03 & 21.70 & $-$14.90 & A & $+$60 & 3 \\
4 & 2024-05-28 & 2.375 & 1.781 & 12.16 & $+$2.38 & 113.53 & 23.03 & $-$15.60 & A & $+$90 & 1 \\
5 & 2024-11-24 & 1.363 & 1.769 & 18.98 & $+$4.25 & 49.99 & 33.69 & 13.86 & A & $-$53 & 8 \\
6 & 2024-11-30 & 1.468 & 1.95 & 19.12 & $+$4.47 & 47.04 & 29.45 & 14.02 & D & $-$50 & 4 \\
7 & 2024-12-03 & 1.519 & 2.037 & 19.19 & $+$4.57 & 45.52 & 27.59 & 14.05 & D & $-$50 & 2 \\
8 & 2024-12-11 & 1.653 & 2.257 & 19.35 & $+$4.94 & 41.47 & 23.24 & 14.03 & C & $-$50 & 1 \\
9 & 2024-12-13 & 1.686 & 2.309 & 19.39 & $+$5.08 & 40.47 & 22.27 & 14.0 & A & $-$49 & 4 \\

\hline

\end{tabular}
\end{table}

\begin{figure}[h!]
    \centering
    \includegraphics[scale=0.58]{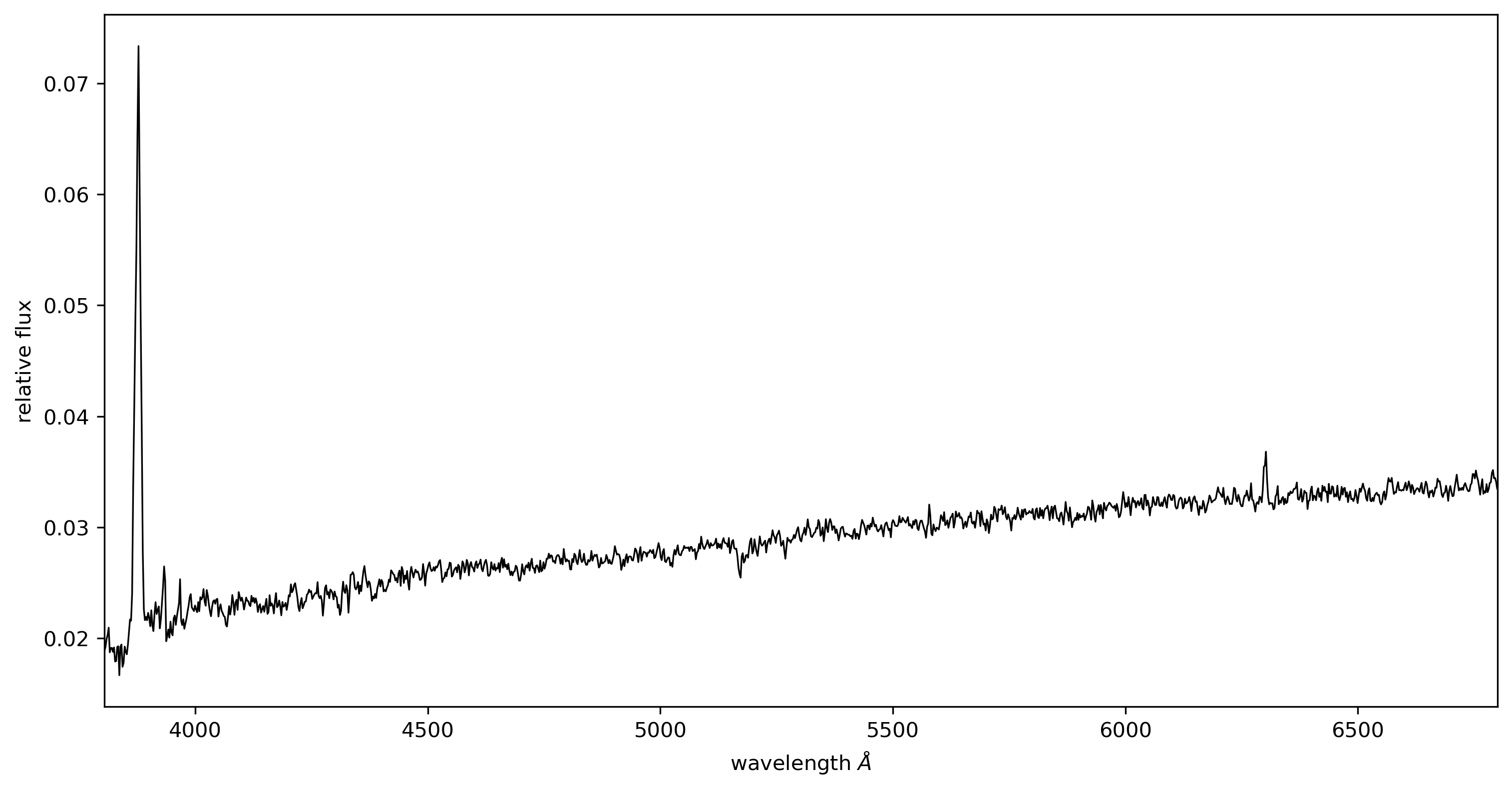}
    \caption{Spectrum of 2024-05-25; configuration A}
\end{figure}


\renewcommand{\thefootnote}{}

\footnotetext{* No Solar Analog}

\newpage
\clearpage

\section{Spectral line table}
Table of the most important emission lines observed in the Galileo's spectra 
(identification by Brown et al., AJ, 1996; Cremonese et al., A\&A, 2007; Cambianica et al., A\&A, 2021).

\begin{table}[h!]
\centering
\begin{tabular}{|c|c|c|c|c|c|c|c|}
\cline{1-2} \cline{4-5} \cline{7-8}
molecule & wavelength [\r{A}] & & molecule & wavelength [\r{A}] & &molecule & wavelength [\r{A}] \\
\cline{1-2} \cline{4-5} \cline{7-8}
\ch{NH}   & 3360 &  & \ch{NH2}  & 5687 &  & \ch{OI}   & 6300 \\ \cline{1-2} \cline{4-5} \cline{7-8} 
\ch{CO2+} & 3510 &  & \ch{NH2}  & 5700 &  & \ch{NH2}  & 6320 \\ \cline{1-2} \cline{4-5} \cline{7-8} 
\ch{OH+}  & 3590 &  & \ch{NH2}  & 5730 &  & \ch{NH2}  & 6335 \\ \cline{1-2} \cline{4-5} \cline{7-8} 
\ch{CN}   & 3880 &  & \ch{H2O+} & 5814 &  & \ch{OI}   & 6364 \\ \cline{1-2} \cline{4-5} \cline{7-8} 
\ch{C3}   & 3929 &  & \ch{Na}   & 5896 &  & \ch{NH2}  & 6434 \\ \cline{1-2} \cline{4-5} \cline{7-8} 
\ch{C3}   & 3936 &  & \ch{Na}   & 5900 &  & \ch{NH2}  & 6470 \\ \cline{1-2} \cline{4-5} \cline{7-8} 
\ch{C3}   & 3972 &  & \ch{C2}   & 5972 &  & \ch{NH2}  & 6498 \\ \cline{1-2} \cline{4-5} \cline{7-8} 
\ch{C3}   & 4012 &  & \ch{NH2}  & 5976 &  & \ch{C2}   & 6533 \\ \cline{1-2} \cline{4-5} \cline{7-8} 
\ch{C3}   & 4019 &  & \ch{NH2}  & 5984 &  & \ch{NH2}  & 6563 \\ \cline{1-2} \cline{4-5} \cline{7-8} 
\ch{C3}   & 4045 &  & \ch{NH2}  & 5994 &  & unid & 6577 \\ \cline{1-2} \cline{4-5} \cline{7-8} 
\ch{C3}   & 4067 &  & \ch{C2}   & 5988 &  & \ch{H2O+} & 6577 \\ \cline{1-2} \cline{4-5} \cline{7-8} 
\ch{CN}   & 4200 &  & \ch{NH2}  & 6006 &  & unid & 6588 \\ \cline{1-2} \cline{4-5} \cline{7-8} 
\ch{CN}   & 4213 &  & \ch{NH2}  & 6020 &  & unid & 6600 \\ \cline{1-2} \cline{4-5} \cline{7-8} 
\ch{CH}   & 4300 &  & \ch{C2}   & 6013 &  & \ch{NH2}  & 6619 \\ \cline{1-2} \cline{4-5} \cline{7-8} 
\ch{CH}   & 4312 &  & \ch{NH2}  & 6097 &  & \ch{NH2}  & 6618 \\ \cline{1-2} \cline{4-5} \cline{7-8} 
\ch{C2}   & 4365 &  & \ch{NH2}  & 6098 &  & \ch{NH2}  & 6627 \\ \cline{1-2} \cline{4-5} \cline{7-8} 
\ch{C2}   & 4515 &  & \ch{C2}   & 6100 &  & \ch{NH2}  & 6634 \\ \cline{1-2} \cline{4-5} \cline{7-8} 
\ch{C2}   & 4706 &  & \ch{NH2}  & 6108 &  & \ch{NH2}  & 6640 \\ \cline{1-2} \cline{4-5} \cline{7-8} 
\ch{C2}   & 4732 &  & \ch{H2O+} & 6135 &  & \ch{NH2}  & 6655 \\ \cline{1-2} \cline{4-5} \cline{7-8} 
\ch{C2}   & 4737 &  & \ch{H2O+} & 6147 &  & \ch{NH2}  & 6671 \\ \cline{1-2} \cline{4-5} \cline{7-8} 
\ch{C2}   & 4920 &  & \ch{H2O+} & 6158 &  & \ch{C2}   & 6740 \\ \cline{1-2} \cline{4-5} \cline{7-8} 
\ch{C2}   & 5128 &  & \ch{C2}   & 6161 &  & \ch{NH2}  & 6750 \\ \cline{1-2} \cline{4-5} \cline{7-8} 
\ch{C2}   & 5165 &  & \ch{C2}   & 6170 &  & \ch{NH2}  & 6754 \\ \cline{1-2} \cline{4-5} \cline{7-8} 
\ch{NH2}  & 5183 &  & \ch{H2O+} & 6198 &  & \ch{NH2}  & 6783 \\ \cline{1-2} \cline{4-5} \cline{7-8} 
\ch{C2}   & 5424 &  & \ch{H2O+} & 6210 &  & unid & 6829 \\ \cline{1-2} \cline{4-5} \cline{7-8} 
\ch{NH2}  & 5429 &  & \ch{NH2}  & 6273 &  & unid & 6834 \\ \cline{1-2} \cline{4-5} \cline{7-8} 
\ch{C2}   & 5436 &  & \ch{NH2}  & 6281 &  & unid & 6864 \\ \cline{1-2} \cline{4-5} \cline{7-8} 
unid & 5460 &  & \ch{NH2}  & 6288 &  & \ch{CN}   & 6949 \\ \cline{1-2} \cline{4-5} \cline{7-8} 
\ch{OI}   & 5577 &  & \ch{NH2}  & 6295 &  & \ch{NH2}  & 7012 \\ \cline{1-2} \cline{4-5} \cline{7-8} 
\ch{C2}   & 5620 &  & \ch{NH2}  & 6327 &  & \ch{NH2}  & 7348 \\ \cline{1-2} \cline{4-5} \cline{7-8} 
\end{tabular}
\end{table}

\newpage
\clearpage

\end{document}